\documentstyle[aps,prb,eqsecnum,psfig,floats]{revtex}

\tighten

\begin{document}

\draft

\title{Random Matrix Theories in Quantum Physics: Common Concepts}
\author{Thomas Guhr, Axel M\"uller--Groeling, and Hans A. Weidenm\"uller}
\address{Max--Planck--Institut f\"ur Kernphysik, Postfach 103980,
  69029 Heidelberg, Germany} 

\date{July 10, 1997}

\maketitle

\begin{abstract}
  We review the development of random--matrix theory (RMT) during the
  last decade. We emphasize both the theoretical aspects, and the
  application of the theory to a number of fields. These comprise
  chaotic and disordered systems, the localization problem, many--body
  quantum systems, the Calogero--Sutherland model, chiral symmetry
  breaking in QCD, and quantum gravity in two dimensions. The review
  is preceded by a brief historical survey of the developments of RMT
  and of localization theory since their inception. We emphasize the
  concepts common to the above--mentioned fields as well as the great
  diversity of RMT. In view of the universality of RMT, we suggest
  that the current development signals the emergence of a new 
  ``statistical mechanics'': Stochasticity and general symmetry
  requirements lead to universal laws not based on dynamical
  principles.  
\end{abstract}

\pacs{PACS numbers: 02.50.Ey, 05.45.+b, 21.10.-k, 24.60.Lz, 72.80.Ng\\
      Keywords: Random matrix theory, Chaos, Statistical many--body theory,
      Disordered solids}

\vskip 1cm
\centerline{{\it MPI preprint H V27 1997, submitted to Physics Reports}}
\vskip 3cm
\tableofcontents
\newpage

\setcounter{equation}{0}
\section{Introduction}
\label{intro}

During the last ten years, Random Matrix Theory (RMT) underwent an 
unexpected and rapid development: RMT has been successfully applied 
to an ever increasing variety of physical problems. 

\begin{figure}
\centerline{
\psfig{file=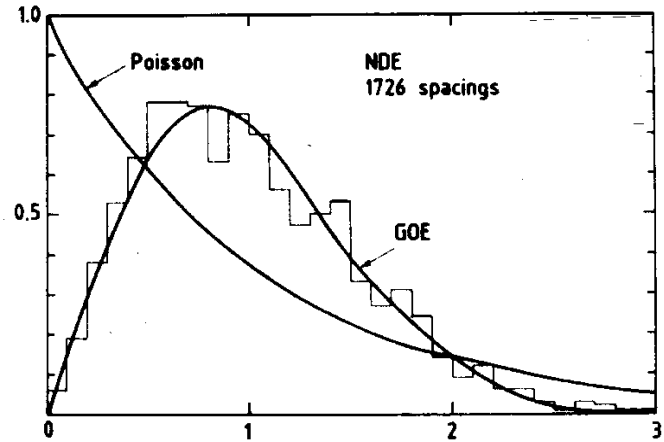,width=3.5in}
}
\caption{
Nearest neighbor spacing distribution for the ``Nuclear Data
Ensemble''  comprising 1726 spacings (histogram) versus $s=S/D$ with
$D$ the mean level spacing and $S$ the actual spacing. For comparison,
the RMT prediction labelled GOE and the result for a Poisson
distribution are also shown as solid lines. Taken from
Ref.~\protect\onlinecite{Boh83}.
} 
\label{fig1}
\end{figure}

\begin{figure}
\centerline{
\psfig{file=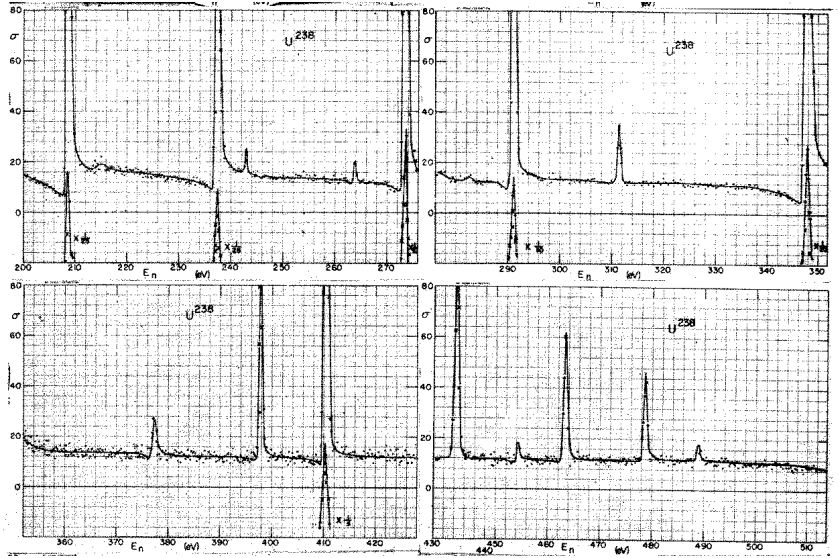,width=4.5in}
 }
\caption{
Total cross section versus neutron energy for scattering of neutrons on 
$^{238}$U. The resonances all have the same spin $1/2$ and positive parity.
Taken from Ref.~\protect\onlinecite{Garg64}.
} 
\label{fig2}
\end{figure}

\begin{figure}
\centerline{
\psfig{file=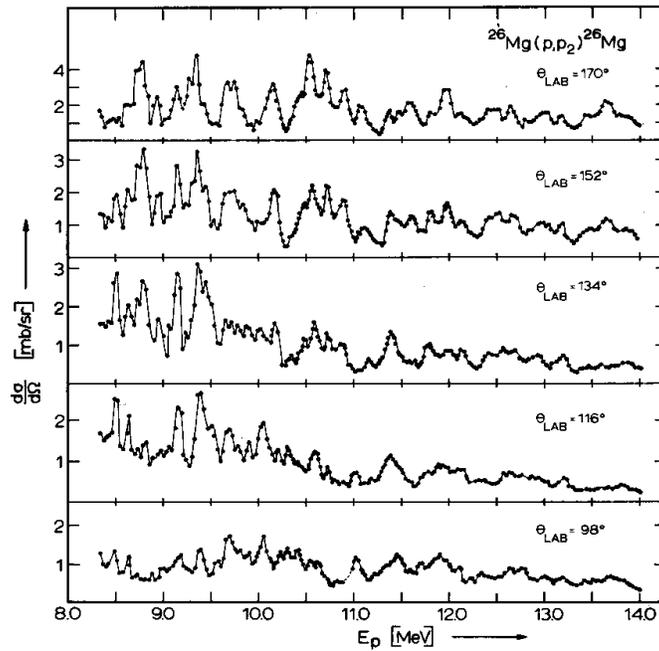,width=3.5in}
 }
\caption{
Differential cross section at several lab angles versus proton
c.m. energy (in MeV) for the reaction  
$^{26}$Mg$(p,p)^{26}$Mg leaving $^{26}$Mg in its second excited state.
Taken from Ref.~\protect\onlinecite{Hauss68}.
} 
\label{fig3}
\end{figure}

Originally, RMT was designed by Wigner to deal with the statistics 
of eigenvalues and eigenfunctions of complex many--body quantum 
systems. In this domain, RMT has been successfully applied to the 
description of spectral fluctuation properties of atomic nuclei, 
of complex atoms, and of complex molecules. The statistical 
fluctuations of scattering processes on such systems were also 
investigated. We demonstrate these statements in 
Figs.~\ref{fig1}, \ref{fig2} and \ref{fig3},
using examples taken from nuclear physics. The histogram in
Fig.~\ref{fig1} \cite{Boh83} shows the distribution of spacings of
nuclear levels versus the variable $s$, the actual spacing in units of
the mean level spacing $D$. The data set comprises 1726 spacings of
levels of the same spin and parity from a number of different nuclei. 
These data were obtained from neutron time--of--flight spectroscopy
and from high--resolution proton scattering. Thus, they refer to
spacings far from the ground--state region. The solid curve shows the 
random--matrix prediction for this ``nearest neighbor spacing (NNS) 
distribution''. This prediction is parameter--free and the agreement 
is, therefore, impressive. Typical data used in this analysis are
shown in Fig.~\ref{fig2} \cite{Garg64}. The data shown are only part
of the total data set measured for the target nucleus $^{238}$U. 
In the energy range between neutron threshold and about 2000 eV, the
total neutron scattering cross section on $^{238}$U displays a number
of well--separated (``isolated'') resonances. Each resonance is
interpreted as a quasibound state of the nucleus $^{239}$U.
The energies of these quasibound states provide the input for the
statistical analysis leading to Fig.~\ref{fig1}. We note the scale: At
neutron threshold, i.e. about 8 MeV above the ground state, the 
average spacing of the s--wave resonances shown in Fig.~\ref{fig2} is 
typically 10 eV! What happens as the energy $E$ increases? As is the
case for any many--body system, the average compound 
nuclear level spacing $D$ decreases nearly exponentially with 
energy. For the same reason, the number of states in the residual 
nuclei (which are available for decay of the compound nucleus)  
grows strongly with $E$. The net result is that the average width 
$\Gamma$ of the compound--nucleus resonances (which is very small 
compared to $D$ at neutron threshold) grows nearly exponentially with 
$E$. In heavy nuclei, $\Gamma \geq D$ already a few MeV above neutron
threshold, and the compound--nucleus resonances begin to overlap. A 
few MeV above this domain, we have $\Gamma \gg D$, and the
resonances overlap very strongly. At each bombarding energy, the
scattering amplitude is a linear superposition 
of contributions from many (roughly $\Gamma / D$) resonances. But the
low--energy scattering data show that these resonances behave
stochastically. This must also apply at higher energies. 
Figure~\ref{fig3} \cite{Hauss68} confirms this expectation. It shows
an example for the statistical fluctuations (``Ericson fluctuations''
\cite{Erics63}) seen in nuclear cross sections a few MeV above neutron
threshold. These fluctuations are stochastic but reproducible. The
width of the fluctuations grows with energy, since ever more decay
channels of the compound nucleus open up. Deriving the characteristic
features of these fluctuations as measured in terms of their variances
and correlation functions from RMT posed a challenge for the nuclear
physics community.       

\begin{figure}
\centerline{
\psfig{file=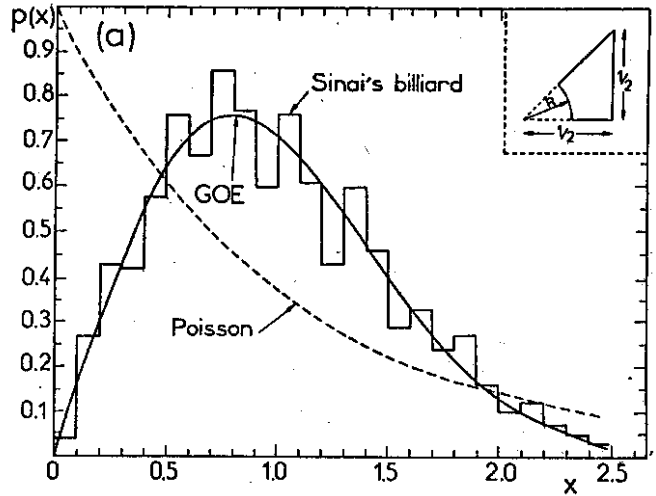,width=3.5in}
 }
\caption{
The nearest neighbor spacing distribution versus $s$ (defined as in
Fig.~\protect\ref{fig1}) for the Sinai billiard. The histogram
comprises about 1000 consecutive eigenvalues. Taken from
Ref.~\protect\onlinecite{Boh84b}.
} 
\label{fig4}
\end{figure}

\begin{figure}
\centerline{
\psfig{file=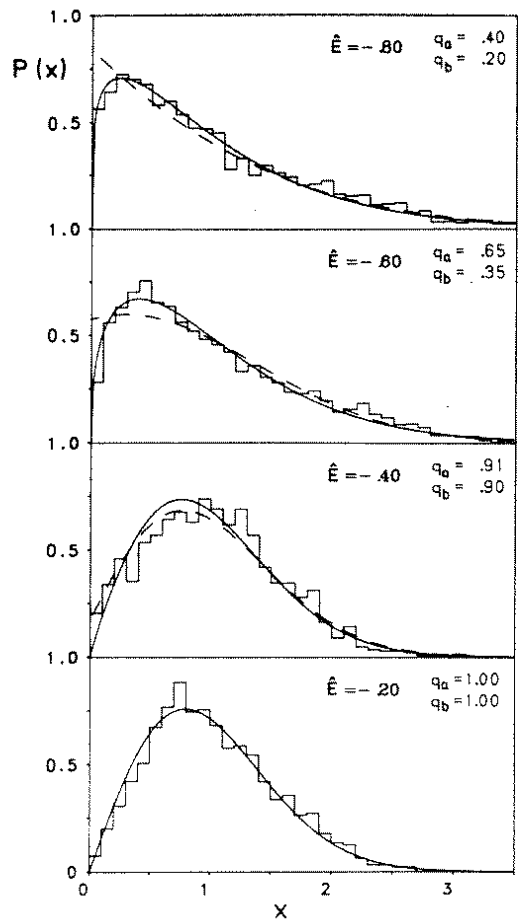,width=3.5in}
 }
\caption{
Nearest neighbor spacing distribution versus $s$ (as in
Fig.~\protect\ref{fig1}) for 
the hydrogen atom in a strong magnetic field. The levels are taken
from a vicinity of the scaled binding energy $\tilde{E}$. Solid and
dashed lines are fits, except for the bottom figure which represents
the GOE. The transition Poisson $\rightarrow$ GOE is clearly visible.
Taken from Ref.~\protect\onlinecite{Win87b}.
} 
\label{fig5}
\end{figure}

\begin{figure}
\centerline{
\psfig{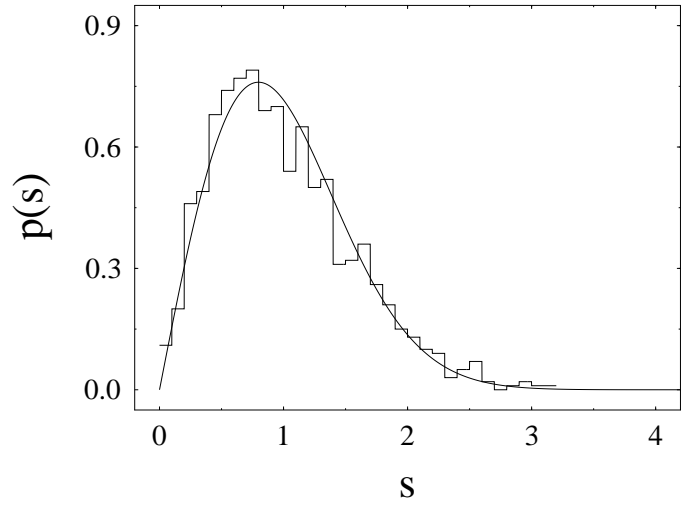}
 }
\vskip 5mm
\caption{
Nearest neighbour spacing distribution for elastomechanical modes in 
an irregularly shaped quartz crystal. Taken from
Ref.~\protect\onlinecite{Ell96}. 
} 
\label{fig6}
\end{figure}

\begin{figure}
\centerline{
\psfig{file=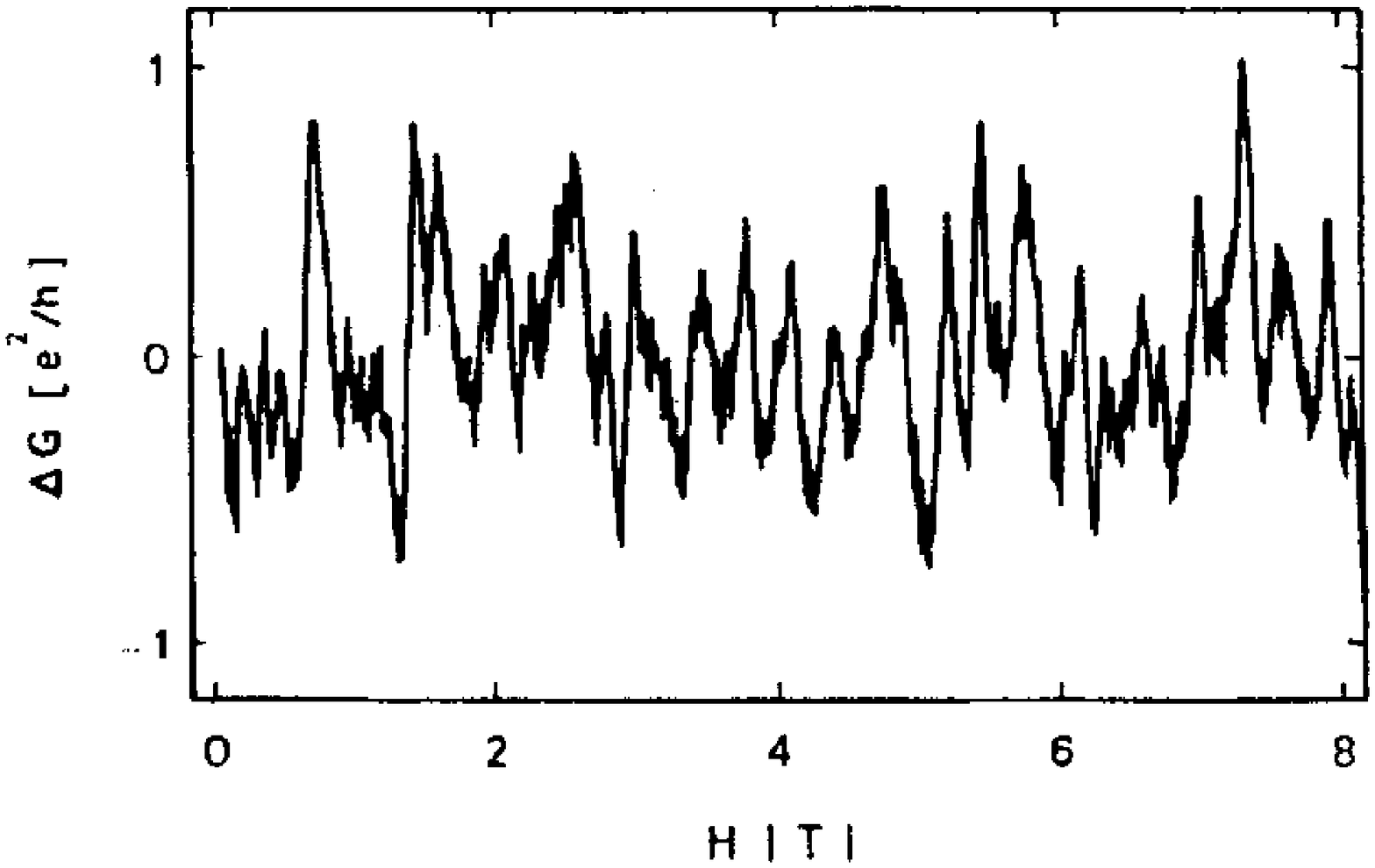,width=3.5in}
 }
\caption{
The difference $\Delta G$ of the conductance and its mean value, in
units of $e^2/h$, of a 310 nm wire at a temperature of 0.01 K versus
magnetic field strength $H$ in Tesla. 
Taken from Ref.~\protect\onlinecite{Web86}.
} 
\label{fig7}
\end{figure}

These applications of RMT were all in the spirit of Wigner's original
proposal. More recently, RMT has found a somewhat unexpected extension
of its domain of application. RMT has become an important tool in the
study of systems which are seemingly quite different from complex
many--body systems. Examples are: Equilibrium and transport properties
of disordered quantum systems and of classically chaotic quantum
systems with few degrees of freedom, two--dimensional gravity,
conformal field theory, and the chiral phase transition in quantum
chromodynamics. Figures~4 to 7 show several cases of interest. The
Sinai billiard is the prime example of a fully chaotic classical
system. Within statistics, the NNS distribution for the quantum
version shown \cite{Boh84b} as the histogram in Fig.~\ref{fig4} agrees
perfectly with the RMT prediction (solid line). The successful
application of RMT is not confined to toy models like the Sinai
billiard. Rydberg levels of the hydrogen atom in a strong magnetic
field have a spacing distribution \cite{Win87b} which once again
agrees with RMT (Fig.~\ref{fig5}). The same is true \cite{Ell96} of
the elastomechanical eigenfrequencies of irregularly shaped quartz
blocks (Fig.~\ref{fig6}). And the transmission of (classical or
quantum) waves through a disordered medium shows patterns similar to
Ericson fluctuations. Figure~\ref{fig7} \cite{Web86} shows the
fluctuations of the conductance of a wire of micrometer size caused by
applying an external magnetic field. Such observations and independent
theoretical progress paved the way for the rapid development of RMT.

Some of these developments were triggered or at least significantly
advanced by the introduction of a new tool into RMT: Efetov's
supersymmetry technique and the ensuing mapping of the random--matrix
problem onto a non--linear supersymmetric $\sigma$ model. This advance
has, in turn, spurred developments in mathematical physics relating,
among other items, to interacting Fermion systems in one dimension
(the Calogero--Sutherland model), to supersymmetric Lie algebras, to
Fourier transformation on graded manifolds, and to extensions of the
Itzykson--Zuber integral.

As the title suggests, we use the term ``Random Matrix Theory'' in its
broadest sense. It stands not only for the classical ensembles (the
three Gaussian ensembles introduced by Wigner and the three circular
ensembles introduced by Dyson) but for any stochastic modelling of a
Hamiltonian using a matrix representation. This comprises the embedded
ensembles of French {\it et al}. as well as the random band matrices
used for extended systems and several other cases. We attempt to
exhibit the common concepts underlying these Random Matrix Theories
and their application to physical problems.

To set the tone, and to introduce basic concepts, we begin with a
historical survey (Sec. \ref{hist}). Until the year 1983, RMT and
localization theory (i.e., the theory of disordered solids) developed
quite independently. For this reason, we devote separate subsections
to the two fields. In the years 1983 and 1984, two developments took
place which widened the scope of RMT enormously.  First, Efetov's
supersymmetric functional integrals, originally developed for
disordered solids, proved also applicable to and useful for problems
in RMT. This was a technical breakthrough. At the same time, it led to
a coalescence of RMT and of localization theory. Second, the ``Bohigas
conjecture'' established a generic link between RMT and the spectral
fluctuation properties of classically chaotic quantum systems with few
degrees of freedom. It was these two developments which in our view
largely triggered the explosive growth of RMT during the last decade.

Subsequent sections of the review are devoted to a description of this
growth.  In Sec.~\ref{thasp} we focus attention on formal developments
in RMT. We recall classic results, define essential spectral
observables, give a short introduction into scattering theory and
supersymmetry, and discuss wavefunction statistics, transitions and
parametric correlations. It is our aim to emphasize that RMT, in spite
of all its applications, has always been an independent field of
research in its own right. In Sec.~\ref{mbs} we summarize the role
played by RMT in describing statistical aspects of systems with many
degrees of freedom. Examples of such systems were mentioned above
(atoms, molecules, and nuclei) but also include interacting electrons
in ballistic mesoscopic devices.  Section~\ref{chaos} deals with the
application of RMT to ``quantum chaology'', i.e. to the quantum
manifestations of classical chaos. Here, we restrict ourselves to
systems with few degrees of freedom. Important examples are mesoscopic
systems wherein the electrons move independently and ballistically
(i.e., without impurity scattering).  Section \ref{disorder} deals
with the application of RMT to disordered systems, and to localization
theory. This application has had major repercussions in the theory of
mesoscopic systems with diffusive electron transport. Moreover, it has
focussed theoretical attention on novel issues like the spectral
fluctuation properties at the mobility edge. Treating the Coulomb
interaction between electrons in such systems forms a challenge which
has been addressed only recently. The need to handle the crossover
from ballistic to diffusive electron transport in mesoscopic systems
has led to a common view of both regimes, and to the discovery of new
and universal laws for parametric correlation functions in such
systems. However, we neither cover the integer nor the fractional
quantum Hall effect.  Section \ref{field} deals with the numerous
applications of RMT in one--dimensional systems of interacting
Fermions, in QCD, and in field theory and quantum gravity.
Section~\ref{univ} is devoted to a discussion of the universality of
RMT. This concerns the question whether or not certain statistical
properties are independent of the specified probability distribution
function in matrix space.

In our view, the enormous development of RMT during the last decade
signals the birth of a ``new kind of statistical mechanics'' (Dyson).
The present review can be seen as a survey of this emergent field. The
evidence is growing that not only disordered but also strongly
interacting quantum systems behave stochastically. The combination of
stochasticity and general symmetry principles leads to the emergence
of general laws.  Although not derived from first dynamical
principles, these laws lay claim to universal validity for (almost)
all quantum systems. RMT is the main tool to discover these universal
laws. In Sec. \ref{coco}, we end with general considerations and
speculations on the origins and possible implications of this ``new
statistical mechanics''.

The breadth of the field is such that a detailed account would be
completely beyond the scope of a review paper. As indicated in the
title, we focus attention on the concepts, both physical and
mathematical, which are either basic to the field, or are common to
many (if not all) of its branches.  While we cannot aim at
completeness, we attempt to provide for the last decade a bibliography
which comprises at least the most important contributions. For earlier
works, and for more comprehensive surveys of individual subfields, we
refer the reader to review articles or reprint collections at
appropriate places in the text.

We are painfully aware of the difficulty to give a balanced view of
the field. Although we tried hard, we probably could not avoid
misinterpretations, imbalances, and outright oversights and mistakes.
We do apologize to all those who feel that their work did not receive
enough attention, was misinterpreted, or was unjustifyably omitted
alltogether.

In each of the nine sections, we have adopted a notation which has
maximum overlap with the usage of the literature covered in that
section, often at the expense of consistency between different
sections. We felt that this approach would best serve our readers.
We could not always avoid using the same symbol for different
quantities. This is the case, for instance, for the symbol $P$. Most
often, $P$ denotes a probability density. Context and argument of the
symbol should identify the quantity unambiguously.

We could not have written this review without the continuous advice
and help of many people. Special thanks are due to David Campbell,
editor of {\it Physics Letters}, for having triggered the writing of
this paper. We are particularly grateful to J. Ambj\o rn, C.
Beenakker, L. Benet, R. Berkovits, G. Casati, Y. Gefen, R. Hofferbert,
H. K\"oppel, I. Lerner, K. Lindemann, E. Louis, C. Marcus, A. D.
Mirlin, R. Nazmitdinov, J. Nyg\aa rd, A. Richter, T. Seligman, M.
Simbel, H. J. St\"ockmann, G. Tanner, J.J.M. Verbaarschot, and T.
Wettig for reading parts or all of this review, and/or for many
helpful comments and suggestions. Part of this work was done while the
authors were visiting CIC, UNAM, Cuernavaca, Mexico.

\setcounter{equation}{0}
\section{Historical survey: the period between 1951 and 1983}
\label{hist}     

As mentioned above, in this period RMT and localization theory
developed virtually independently, and we therefore treat their
histories separately. 

After a period of rapid growth during the 1950's, RMT was almost
dormant until it virtually exploded about ten years ago. Our history
of RMT therefore consists of three parts. These parts describe (i) the
early period (1951 till 1963) in which the basic ideas and concepts
were formulated, and the classical results were obtained
(Sec.~\ref{early}); (ii) the period 1963 till 1983 in which the theory
was consolidated, relevant data were gathered, and some fundamental
open problems came to the surface (Sec.~\ref{cons}); (iii) the almost
simultaneous introduction of the supersymmetry method (Sec.~\ref{sup})
and of the Bohigas conjecture (Sec.~\ref{chao}) around 1983.

For a long time, applications of RMT have essentially been 
confined to nuclear physics. The reason is historical: This was the 
first area in physics where the available energy resolution was fine 
enough, i.e. of the order of $D$, and the data set large enough, to 
display spectral fluctuation properties relevant for tests of RMT. 
This is why --- aside from a description of the technical development
of RMT --- Secs.~\ref{early} and \ref{cons} deal almost entirely
with problems in nuclear theory. We will not mention again that here 
and in other systems, spectroscopic tests of the theory always involve
levels of the same spin and parity or, more generally, of the same
symmetry class.

After its inception by Anderson in 1958, localization theory received
a major boost by the work of Mott and, later, by the application of 
scaling concepts. This is the history described in 
Sec.~\ref{loc}. For pedagogical reasons, this section precedes the ones 
on supersymmetry and chaos.   

In this historical review, we refer to review papers wherever
possible. We do so at the expense of giving explicit references to
individual papers. 

\subsection{RMT: the early period}
\label{early}

Most references can be
found in Porter's book \cite{Por65} and are not given explicitly. 
Our account is not entirely historical: History serves as an
introduction to the relevant concepts.  

In 1951, Wigner proposed the use of RMT to describe certain
properties of excited states of atomic nuclei. This was the first time
RMT was used to model physical reality. To understand Wigner's 
motivation for taking such a daring step, it is well to recall the
conceptual development of nuclear theory preceding it.

\begin{figure}
\centerline{
\psfig{file=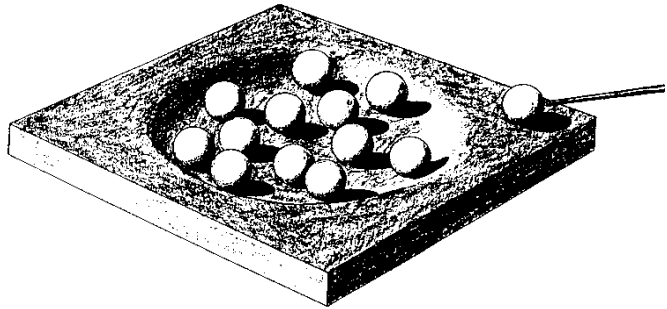,width=3.5in}
 }
\caption{
Photograph of Niels Bohr's wooden toy model for compound--nucleus
scattering. 
Taken from Ref.\protect\onlinecite{Bor36}.
} 
\label{fig8}
\end{figure}

In the scattering of slow neutrons by medium--weight and heavy nuclei,
narrow resonances had been observed. Each of these resonances 
corresponds to a long--lived ``compound state'' of the system formed by 
target nucleus and neutron. In his famous 1936 paper, N. Bohr
\cite{Bor36} had described 
the compound nucleus as a system of strongly interacting neutrons and 
protons. In a neutron--induced nuclear reaction, the strong
interaction was thought to lead to an almost equal sharing of the 
available energy between all constituents, i.e. to the attainment of
quasi--equilibrium. As a consequence of equilibration, formation
and decay of the compound nucleus should be almost independent 
processes. In his appeal to such statistical concepts, Bohr prepared
the ground for Wigner's work. In fact, RMT may be seen as a formal
implementation of Bohr's compound nucleus hypothesis. At the same
time, it is remarkable that the concepts and ideas formulated by Bohr 
have a strong kinship to ideas of classical chaotic motion which in
turn are now known to be strongly linked to RMT. This is most clearly 
seen in Fig.~\ref{fig8} \cite{Nat36}
which is a photograph of a wooden model used by Bohr
to illustrate his idea. The trough stands for the nuclear potential 
of the target nucleus. This potential binds the individual nucleons, 
the constituents of the target, represented as small spheres. An 
incoming nucleon with some kinetic energy (symbolized by the billard 
queue) hits the target. The collision is viewed as a sequence of 
nucleon--nucleon collisions which have nearly the character of 
hard--sphere scattering.

In the absence of a {\it dynamical} nuclear theory (the nuclear shell 
model had only just been discovered, and had not yet found universal
acceptance), Wigner focussed emphasis on the
{\em statistical} aspects of nuclear spectra as revealed in neutron
scattering data. At first sight, such a statistical approach to nuclear 
spectroscopy may seem bewildering. Indeed, the spectrum of any nucleus
(and, for that matter, of any conservative dynamical system) is 
determined unambiguously by the underlying Hamiltonian, leaving
seemingly no room for statistical concepts. Nonetheless, such concepts
may be a useful and perhaps even the only tool available to deal with 
spectral properties of systems for which the spectrum is sufficiently 
complex. An analogous situation occurs in number theory. The sequence 
of prime numbers is perfectly well defined in terms of a
deterministic set of rules. Nevertheless, the pattern of occurrence of
primes among the integers is so complex that statistical concepts
provide a very successful means of gaining information on the
distribution of primes. This applies, for instance, to the average 
density of primes (the average number of primes per unit interval), to
the root--mean--square deviation from this average, to the
distribution of spacings between consecutive primes, and to other 
relevant information which can be couched in statistical terms. 

The approach introduced by Wigner differs in a fundamental way from 
the standard application of statistical concepts in physics, and from
the example from number theory just described. In standard statistical
mechanics, one considers an ensemble of {\it identical} physical
systems, all governed by the {\it same} Hamiltonian but differing in
initial conditions, and calculates 
thermodynamic functions by averaging over this ensemble. In number
theory, one considers a single specimen --- the sequence of primes ---
and introduces statistical concepts by performing a running average
over this sequence. Wigner proceeded differently: He considered 
ensembles of dynamical systems governed by {\it different}
Hamiltonians with some common symmetry property. This novel
statistical approach focusses attention on the {\em generic} 
properties which are common to (almost) all members of the ensemble
and which are determined by the underlying fundamental symmetries. The 
application of the results obtained within this approach to individual
physical systems is justified provided there exists a suitable ergodic
theorem. We return to this point later.  

Actually, the approach taken by Wigner was not quite as general as
suggested in the previous paragraph. The ensembles of Hamiltonian 
matrices considered by Wigner are defined in terms of invariance 
requirements: With every Hamiltonian matrix belonging to the ensemble,
all matrices generated by suitable unitary transformations of Hilbert 
space are likewise members of the ensemble. This postulate guarantees
that there is no preferred basis in Hilbert space. Many recent
applications of RMT use extensions of Wigner's original approach and
violate this invariance principle. Such extensions will be discussed
later in this paper. 

It is always assumed in the sequel that all conserved quantum numbers
like spin or parity are utilized in such a way that the Hamiltonian 
matrix becomes block--diagonal, each block being characterized by a 
fixed set of such quantum numbers. We deal with only one such block 
in many cases. This block has dimension $N$. The basis states in
Hilbert space relating to this block are labelled by greek indices
like $\mu$ and $\nu$ which run from $1$ to $N$. Since Hilbert space 
is infinite--dimensional, the limit $N \rightarrow \infty$ is taken 
at some later stage. Taking this limit signals that we do not address 
quantum systems having a complete set of commuting observables. Taking
this limit also emphasises the generic aspects of the random--matrix 
approach. Inasmuch as RMT as a ``new kind of statistical mechanics''
bears some analogy to standard statistical mechanics, the limit $N
\rightarrow \infty$ is kin to the thermodynamic limit.  

Using early group--theoretical results by Wigner \cite{Wig59}, Dyson 
showed that in the framework of standard Schr\"odinger theory, there
are three generic ensembles of random matrices,  
defined in terms of the symmetry properties of the Hamiltonian.   

(i) Time--reversal invariant systems with rotational symmetry. For
such systems, the Hamiltonian matrix can be chosen real and symmetric,  
\begin{equation}
H_{mn}= H_{nm} = H^{*}_{mn}.
\label{H1} 
\end{equation}
Time--reversal invariant systems with integer spin and 
broken rotational symmetry also belong to this ensemble.

(ii) Systems in which time--reversal invariance is violated. This is
not an esoteric case but occurs frequently in applications. An example
is the Hamiltonian of an electron in a {\it fixed} external magnetic 
field. For such systems, the Hamiltonian matrices are Hermitean,   
\begin{equation}    
H_{mn} = [H^{\dagger}]_{nm}.
\label{H2} 
\end{equation}

(iii) Time--reversal invariant systems with half--integer spin and
broken rotational symmetry. The Hamiltonian matrix can be written in
terms of quaternions, or of the Pauli spin matrices $\sigma_{\gamma}$
with $\gamma = 1,2,3$. The Hamiltonian has the form
\begin{equation}
H^{(0)}_{nm} 1_2 - i \sum_{\gamma=1}^3 H^{(\gamma)}_{nm} \sigma_\gamma,
\label{H4}
\end{equation}
where all four matrices $H^{(\gamma)}$ with $\gamma=0,\ldots,3$ are
real and where $H^{(0)}$ is symmetric while $H^{(\gamma)}$ with
$\gamma=1,2,3$ are antisymmetric.

In all three cases the probability of finding a particular matrix 
is given by a weight function $P_{N\beta}(H)$ times
the product of the differentials of all independent matrix
elements. Both the symmetry properties (\ref{H1}), (\ref{H2}), and
(\ref{H4}), and the weight functions $P_{N\beta}(H)$ for $\beta =
1,2,4$ are invariant under orthogonal, unitary, and symplectic
transformations of the Hamiltonian, respectively. The index $\beta$
is often used to specify the ensemble
altogether. In a sense specified by Dyson, the three ensembles are
fundamentally irreducible and form the basis of all that follows. 
Novel ensembles not contained in this list may arise \cite{Gad93,And94}
when additional symmetries or constraints are imposed. Such ensembles 
have recently been discussed in the context of Andreev scattering 
\cite{Alt96}, and of chiral symmetry \cite{Shu93}.

The choice $P_{N\beta}(H)=1$ would be consistent with the symmetry
requirements but would lead to divergent integrals.
For the {\em Gaussian} ensembles considered by Wigner, the weight 
functions $P_{N\beta}$ are chosen to have Gaussian form,
\begin{equation}
P_{N\beta}(H) \propto \exp \left( -\frac{\beta N}{\lambda^2}\, 
                                {\rm tr} H^2 \right). 
\label{P3} 
\end{equation}
We have suppressed a normalization factor, see Sec.~\ref{qc1rcb}. 
The constant $\lambda$ 
is independent of $N$. The factor $N$ ascertains that the spectrum 
of the ensemble remains bounded in the limit $N \rightarrow \infty$. 
The independent elements of the Hamiltonian are independent random 
variables; the distributions factorize. For a more formal discussion
of the three Gaussion ensembles we refer to Sec.~\ref{qc1rc}.

The choice (\ref{P3}) defines the three canonical ensembles: The
Gaussian orthogonal ensemble (GOE) with $\beta = 1$. the Gaussian
unitary ensemble (GUE) with $\beta = 2$, and the Gaussian 
symplectic ensemble (GSE) with $\beta = 4$. These ensembles have 
similar weight functions $P_{N\beta}$ but different symmetries, and
consequently different volume elements in matrix space. As mentioned 
before, this review covers a variety of Random Matrix Theories. For
clarity, we refer to the ensembles introduced above jointly as to 
Gaussian Random Matrix Theory (GRMT). 

The introduction of RMT as a theory of physical systems poses two
questions. (a) How are predictions on observables obtained from RMT? 
(b) How are such predictions compared with physical reality? A third
question arises when we note that the requirements (i) to (iii) 
defining the three classical ensembles are based on general symmetry 
principles and their quantum--mechanical implementation, while the 
choice of the Gaussian functions (\ref{P3}) was dictated by
convenience. We must therefore ask: (c) Are the conclusions derived 
from the Gaussian ansatz generally valid, i.e. independent of the 
Gaussian form (which would then indeed serve only as a convenient 
vehicle)? 

Basic for most applications of GRMT is the distinction between 
{\it average} quantities and their {\it fluctuations}.
Because of the cutoff due to 
the Gaussian weight factors, all three Gaussian ensembles defined
above have a spectrum which in the limit $N \rightarrow \infty$ is 
bounded and has length $4 \lambda$. In the interval $-2 \lambda \leq E
\leq 2 \lambda$, the average level density has the shape of a 
semicircle (cf. Eq.~(\ref{R1}) below). For most physical systems, a 
bounded spectrum with semicircle shape is totally unrealistic. 
Therefore, GRMT is generically
useless for modelling {\it average} properties like the mean level
density. 
The situation is different for the statistical {\it fluctuations}
around mean values of observables. Since there are $N$ 
levels in the spectrum, $D \sim N^{-1}$ tends to zero as $N
\rightarrow \infty$. 
In this limit, fluctuation properties may become 
independent of the form of the global spectrum and of the choice of 
the Gaussian weight factors, and may attain universal validity. This 
is the expectation held upon employing GRMT, see in particular
Sec.~\ref{univ}. The wide range of successful 
applications of RMT in its Gaussian form has for a long time suggested
that this expectation is justified. Balian \cite{Bal68} derived the 
three Gaussian ensembles from a maximum entropy principle, postulating
the existence of a second moment of the Hamiltonian. His
derivation shows that in the absence of any further constraints, the
Gaussian weight factors are the must natural ones to use. This result 
and numerical studies gave further credence to the belief that
fluctuation properties should not depend on the Gaussian form of the
weight factors. There is now also direct evidence for this
assertion. It is presented in Sec.~\ref{univ}.

GRMT can thus be used to predict fluctuations of certain observables. The
general scheme is this: The average of the observable over the
ensemble serves as input, and the fluctuation properties are derived
from GRMT. Examples are local level correlation functions (with the
average level density as input to define the physical value of
$N / \lambda$), distribution laws and local correlation functions for
eigenvectors (which are usually measured by coupling to some external
field, the average of the square of the coupling matrix element
serving as input), etc. We will encounter numerous examples of this
type as we proceed. In some select cases, it is possible to work out 
the fluctuation properties of the observable completely, and to 
calculate the entire distribution function over the ensemble. In most
cases, however, this goal is too ambitious, and one has to settle for
the first few moments of the distribution. 

In comparing such predictions with experiment, we must relate the 
average over the (fictitious) ensemble given by GRMT with information 
on a given system. Here, an {\it ergodic hypothesis} is used. It says 
that the ensemble average is equal to the {\it running average}, taken
over a sufficiently large section of the spectrum, of almost any 
member of the ensemble. In specific cases, this ergodic hypothesis has
been proved \cite{Bro81}. 

It is clear that RMT cannot ever reproduce a given data set in its 
full detail. It can only yield the distribution function of, and the 
correlations between, the data points. An example is the nuclear 
level spectrum at neutron threshold, the first case ever studied in a 
statistically significant fashion. The actual sequence of levels 
observed experimentally remains inaccessible to RMT predictions, while
the distribution of spacings and the correlations between levels can 
be predicted. The same holds for the stochastic fluctuations of
nuclear cross sections, for universal conductance fluctuations of
mesoscopic probes, and for all other applications of RMT.

GRMT is essentially a parameter--free theory. Indeed, the only
parameter ($\lambda$) is fixed in terms of the local mean level
spacing of the system under study. The successful application of GRMT
to data shows that these data fluctuate stochastically in a manner
consistent with GRMT. Since GRMT can be derived from a maximum entropy
principle \cite{Bal68}, this statement is tantamount to saying that
the data under study carry no information content. It is an amazing
fact that out of this sheer stochasticity, which is only confined by
the general symmetry principles embodied in GRMT, the universal laws
alluded to above do emerge.

The distribution of the eigenvalues and eigenvectors is obviously a
central issue in applications of GRMT. To derive this distribution, it
is useful to introduce the eigenvalues $E_{\mu}$ and the eigenvectors
of the matrices as new independent variables. The group--theoretical
aspects of this procedure were probably first investigated by Hua
\cite{Hua63}, cf. also Ref.~\onlinecite{Cart35}. After this
transformation, the probability measures for the three Gaussian
ensembles factorize. One term in the product depends only on the
eigenvalues, the other, only on the eigenvectors (angles). Moreover,
because of the assumed invariance properties, the functions
$P_{N\beta}$ depend on the eigenvalues only. Therefore, eigenvectors
and eigenvalues are uncorrelated random variables. The form of the
invariant measure for the eigenvectors implies that in the limit $N
\rightarrow \infty$, these quantities are Gaussian-- distributed
random variables.

Following earlier work by Scott, Porter and Thomas suggested this
Gaussian distribution for the eigenvectors in 1956. Their paper became
important because it triggered many comparisons between GRMT
predictions and empirical data. In nuclear reaction theory, the
quantities which determine the strength of the coupling of a resonance
to a particular channel are the ``reduced partial width amplitudes'',
essentially given by the projection of the resonance eigenfunction
onto the channel surface. Identification of the resonance
eigenfunctions with the eigenfunctions of the GOE implies that the
reduced partial width amplitudes have a Gaussian probability
distribution centered at zero, and that the reduced partial widths
(the squares of the partial width amplitudes) have a $\chi^2$
distribution with one degree of freedom (the ``Porter--Thomas
distribution''). This distribution obtains directly from the volume
element in matrix space and holds irrespective of the choice of the
weight function $P_{N1}(H)$. It is given in terms of the average
partial width which serves as input parameter.

For the eigenvalues, the invariant measure takes the form $(\beta =
1,2,4)$ 
\begin{equation}
P_{N\beta}(E_1,\ldots,E_N) \prod_{m>n}|E_m-E_n|^{\beta}
\prod_{l=1}^N dE_l.
\label{E1} 
\end{equation} 
This expression displays the famous repulsion of eigenvalues which
becomes stronger with increasing $\beta$. (Eigenvalue repulsion as a
generic feature of quantum systems was first discussed by von Neumann
and Wigner \cite{Von29}. The arguments used in this paper explain the
${\beta}$--dependence of the form (\ref{E1})). We note that this level 
repulsion is due entirely to the volume element in matrix space, and
independent of the weight function: It influences local rather than 
global properties. It turned out to be very difficult to deduce 
information relevant for a comparison with experimental data from 
Eq.~(\ref{E1}). Such a comparison typically involves the correlation 
function between pairs of levels, or the distribution of spacings 
between nearest neighbors. The integrations over all but a few 
eigenvalues needed to derive such information were found to be very hard.  

Lacking exact results, and guided by the case $N = 2$, Wigner proposed 
in 1957 a form for the distribution $p(s)$ of spacings of
neighboring eigenvalues, i.e. for the nearest neighbor spacing
distribution mentioned in the Introduction. 
Here, $s$ is the level spacing in units of the local mean level
spacing $D$. This ``Wigner surmise'', originally stated for $\beta =
1$, has the form 
\begin{equation}
p_\beta(s) = a_{\beta} s^{\beta} \exp(-b_{\beta} s^2). 
\label{W1} 
\end{equation} 
The constants $a_{\beta}$ and $b_{\beta}$ are given in
Eq.~(\ref{1ba13}) of Sec.~\ref{qc1bab}. The Wigner surmise
shows a strong, $\beta$--dependent level repulsion at small
spacings (this is a reflection of the form (\ref{E1})), and a Gaussian
falloff at large spacings. This falloff has nothing to do with the 
assumed Gausian distribution of the ensembles and stems directly from 
the assumed form of the volume element in matrix space. The solid line
in Fig.~\ref{fig1} shows the $s$--dependence of $p(s)$ for the
orthogonal case. 

In the same paper, Wigner also derived the semicircle law mentioned
above. As a function of energy $E$, the average level density 
$\rho = D^{-1}$ is given by 
\begin{equation}
\rho(E) = \frac{N}{\pi \lambda} \sqrt{1 - (E/(2 \lambda))^2}.
\label{R1} 
\end{equation} 
Later, Pastur~\cite{Pas72} derived a quadratic equation for the
averaged retarded or advanced Green function. This equation coincides
in form with the scalar version of the saddle point 
equation~(\ref{super9}) of the non--linear $\sigma$--model and yields
for $\rho(E)$ the semicircle law~(\ref{R1}).

The difficulties encountered in calculating spectral properties of GRMT 
were overcome in 1960 when M. L. Mehta introduced the method of 
orthogonal polynomials, summarized in his book \cite{Meh91}. 
Mathematically, his methods are related to Selberg's integral. Mehta's
work provided the long--missing tool needed to calculate the spectral
fluctuation properties of the three canonical ensembles, and therefore
had an enormous impact on the field. Mehta could not only prove
earlier assertions but also obtained many new results. For instance, 
Gaudin and Mehta derived the exact form of the nearest neighbour 
spacing distribution. The Wigner surmise turns out to be an excellent 
approximation to this distribution. The method of orthogonal
polynomials has continued to find applications beyond the canonical
ensembles.   

GRMT is conceived as a generic theory and should apply beyond the
domain of nuclear physics. Early evidence in favor of this assertion,
and perhaps the earliest evidence ever in favor of the nearest neighbor
spacing distribution, was provided in 1960 by Porter and Rosenzweig. 
These authors analyzed spectra in complex atoms.

In a series of papers written between 1960 and 1962, Dyson
developed RMT much beyond the original ideas of Wigner. Aside from
showing that there are three fundamental ensembles (the unitary,
orthogonal and symplectic one), he contributed several novel ideas. 
As an alternative to the Gaussian ensembles defined above, he 
introduced the three ``circular ensembles'', the circular unitary
ensemble (CUE), the circular orthogonal ensemble (COE), and the
circular symplectic ensemble (CSE). In each of these ensembles, the
elements are unitary matrices of dimension $N$ rather than the 
Hamiltonian matrices used in Wigner's ensembles. Since the spectra 
of all three ensembles are automatically confined to a compact 
manifold (the eigenvalues exp$[i \theta_{\mu}]$ with $\mu = 1, 
\ldots,N$ of unitary matrices are located on the unit circle in the 
complex plane), there is no need to introduce the (arbitrary) Gaussian
weight factors of Eq.~(\ref{P3}). This removes an ambiguity of the 
Gaussian ensembles and simplifies the mathematical problems.
Unfortunately, these advantages are partly compensated by the fact
that the physical interpretation of the eigenvalues of the circular
ensembles is not obvious. 

We briefly give the mathematical definitions of the three
circular ensembles here because they will not be discussed in detail
in Sec.~\ref{thasp}. Each circular ensemble is defined by the 
same invariance postulate as the corresponding Gaussian ensemble. 

(i) The COE consists of symmetric unitary 
matrices $S$ of dimension $N$.  Each such matrix $S$ can be written as
$S = U^T U$ where $U$ is unitary and $T$ denotes the transpose. To
define the measure, we consider an infinitesimal neighborhood of $S$ 
given by $S + dS = U^T (1_N + i dM) U$ where $dM$ is infinitesimal,
real and symmetric. Let $d \mu_{ij}$ with $1 \leq i \leq j \leq N$ denote
the differentials of the elements of the matrix $M$. Then the measure
$\mu(dS)$ is defined by $\mu(dS) = \prod_{i \leq j} d \mu_{ij}$, and
the COE has the probability measure 
\begin{equation}
P(dS) = {\mu(dS) \over \int \mu(dS)}.
\label{Dysono}
\end{equation}  
The COE is invariant under every automorphism $S \rightarrow 
U'^{T}SU'$ where $U'$ is unitary.

(ii) The CUE consists of unitary matrices $S$ of dimension $N$. Each 
such matrix can be written as the product $S = UV$ of two unitary 
matrices $U$ and $V$ so that $S + dS = U(1_N + i dH)V$ where $dH$ is
infinitesimal Hermitean. The measure $\mu(dS)$ is defined by $\mu(dS) 
= \prod_{i \leq j} d {\rm Re} \mu_{ij} \prod_{i < j} d {\rm Im}
\mu_{ij}$ where  
$d \mu_{ij}$ denotes the differentials of the elements of the matrix
$H$. With this changed definition of $\mu(dS)$, the probability
measure again has the form of Eq.~(\ref{Dysono}). The CUE is invariant
under every automorphism $S \rightarrow U'SV'$ 
where $U',V'$ are unitary. 

(iii) The CSE consists of unitary matrices $S$ of dimension $N$ which
have the form $S = U^R U$ where $U$ is unitary, and where $R$ denotes
the dual, defined by the combined operation of time--reversal ($K$)
and transposition, $U^R = K U^T K^{-1}$. We have $S + dS = U^R(1_N + i
dM)U$ where $dM$ is infinitesimal, quaternion--real and self--dual. 
The measure $\mu(dS)$ is defined in terms of the differentials $d
\mu_{ij}^{\gamma}, \ \gamma = 0,\ldots,3$ of the four matrices
$M^{\gamma}$ of the quaternion decomposition of $M$ as $\mu(dS) = 
\prod_i d\mu_{ii}^0 \prod_{\gamma = 0}^3 \prod_{i \leq j} d
\mu_{ij}^{\gamma}$. With this changed definition of $\mu(dS)$, the
probability measure again has the form of Eq.~(\ref{Dysono}). The CSE
is invariant under every automorphism $S \rightarrow W^RSW$ 
where $W$ is unitary. 

The phase angles $\theta_{\mu}$ of Dyson's ensembles are expected to 
have the same local fluctuation properties as the eigenvalues 
$E_{\mu}$ of the corresponding Gaussian ensembles. This expectation 
is borne out by calculating the invariant measure. It carries the 
factor $\prod_{\mu > \nu} |\exp (i \theta_{\mu}) - \exp (i 
\theta_{\nu})|^{\beta}$, in full analogy to Eq.~(\ref{E1}). Using the 
Gaudin--Mehta method, Dyson was able to derive the $k$--level
correlation function (i.e. the probability density for $k$ phase
angles $\theta_{\mu}$ with $\mu = 1,\ldots,k \ll N$, irrespective of
the position of all others). These functions coincide
with the corresponding expressions for the Gaussian ensembles found by
Mehta, strengthening the belief that the Gaussian weight factor is 
immaterial for local fluctuation properties. 
The explicit formulas for the two--level correlation functions can be
found in Sec.~\ref{qc1rce}.

Aside from furnishing useful alternatives 
to the Gaussian ensembles, the circular ensembles have found direct 
application in stochastic scattering theory since the elements of 
the ensembles --- the unitary matrices $S$ --- can be viewed as 
scattering matrices.

Dyson also noticed an interesting connection between GRMT and a 
classical Coulomb gas. For the Gaussian ensembles, the factors 
appearing in Eq.~(\ref{E1}) can be written in the form 
\begin{equation}
  \exp \left(-\beta N \sum_{n=1}^N \left[\frac{E_n}{\lambda}\right]^2
  + \frac{\beta}{2} \sum_{m>n} \ln
  \left[\frac{E_m-E_n}{\lambda}\right]^2\right).
\label{E2} 
\end{equation} 
{}From a thermodynamic point of view, this expression is the free energy 
of a static Coulomb gas of $N$ particles in one dimension with
positions $E_1,\ldots,E_N$ and temperature $\beta^{-1}$. The gas is 
confined by a harmonic oscillator potential, which is absent for the
circular ensembles originally considered by Dyson.
The parameter $\beta$ plays the role of an
inverse temperature. This analogy helps to understand the fluctuation 
properties of the spectrum. Both the level repulsion at short distance
and the long--range stiffness of the spectrum (see Eq.~(\ref{D1}) and
the remarks following it) characteristic of GRMT become intuitively 
obvious. 

Another important idea in Dyson's work relates to cases of slightly 
broken symmetries or invariance properties. The ensembles considered 
so far all correspond to exactly obeyed symmetries and invariances. 
In many applications of RMT, it is necessary to consider cases of
slight symmetry breaking. This is true, for instance, for
isospin--mixing in nuclei, for parity violation in nuclei, or for
time--reversal invariance breaking in solids. These effects are caused
by the Coulomb interaction, the weak interaction, and an external
magnetic field acting on the electrons, respectively.
Such cases can be modeled by adding 
to the ensemble describing the conserved symmetry (or invariance)
another one, which violates the symmetry (or invariance) and has
relative strength $t$ with respect to the first. More precisely, 
the variances of the matrix elements in the second ensemble differ
by a factor $t^2$ from those in the first. By generalizing the
static Coulomb gas model of the last paragraph to a dynamical one,
where the $N$ particles, in addition to their mutual repulsion,
are also subject to dissipative forces, Dyson arrived at a Brownian 
motion model for the eigenvalues of a random matrix. Starting at time 
$t = 0$ with the case of pure symmetry, and letting the Brownian 
motion proceed, the eigenvalues move under the influence of the
symmetry--breaking ensemble. The Brownian--motion model has found
important applications in recent years. 

We refer to the Gaussian and the circular ensembles jointly as 
to classical Random Matrix Theory (cRMT).

In comparing cRMT results with experiment, it is useful to have
available statistical measures tailored to the fact that the data set
always comprises a finite sequence only. In 1963, Dyson and Mehta
introduced several statistical measures (the ``Dyson--Mehta
statistics'') to test a given sequence of levels for agreement with
cRMT. These measures have found wide application. One important
example is the $\Delta_3$ statistic, which will be discussed in detail
in Sec.~\ref{qc1bac}. Let $\eta(E)$ be the number of energy levels in
the interval $[-\infty,E]$, and let us assume that the spectrum has
been ``unfolded'' so that the local average level density is constant
(independent of $E$). On this unfolded, dimensionless energy scale
$\xi$, the number of levels in the interval $[-\infty,\xi]$ is 
given by $\widehat{\eta}(\xi)$. The $\Delta_3$ statistic
\begin{equation}
  \Delta_3(L) = \left\langle \min_{A,B} \frac{1}{L}
  \int_{\xi_s}^{\xi_s+L} \left(\widehat{\eta}(\xi) - A\xi - B\right)^2
                d\xi \right\rangle_{\xi_s}.
\label{D1} 
\end{equation}
measures how well the staircase function $\widehat{\eta}(\xi)$ can be
locally approximated by a straight line.  The angular brackets in
Eq.~(\ref{D1}) indicate an average over $\xi_s$.  If subsequent spacings
of nearest neighbors were uncorrelated, $\Delta_3(L)$ would be linear
in $L$, $\Delta_3(L) = L / 15$. An essential property of RMT is the
{\em logarithmic} dependence of $\Delta_3(L)$ on $L$. It is often
referred to as the `stiffness' of the spectrum (cf. the remarks below
Eq.~(\ref{E1})). Explicit expressions for $\Delta_3(L)$ in all three
symmetry classes are given in Sec.~\ref{qc1bac}.

As remarked above, the compound--nucleus resonances observed in slow
neutron scattering are prime candidates for applications of GRMT. By
implication, the neutron scattering cross section in this energy range
is expected to be a random process. In the Introduction, it was
pointed out that already a few MeV above neutron threshold, the
resonances overlap strongly. Nonetheless, the cross section should
still be stochastic. Using a simple statistical model for the nuclear
scattering matrix, Ericson proposed in 1963 that in this domain,
cross--section fluctuations with well--defined properties (``Ericson
fluctuations'') should exist. This proposal was made before the
necessary experimental resolution and the data shown in
Fig.~\ref{fig3} became available. The proposal gave a big boost to
nuclear reaction studies, see the review in Ref.~\onlinecite{Eri79}.
For quite some time, the connection between Ericson's model and cRMT
remained an open question, however.

\subsection{From 1963 to 1982: consolidation and application}
\label{cons}

In this period, nuclear reaction data became available which permitted
the first statistically significant test of GRMT predictions on level
fluctuations. Ericson's prediction of random cross--section
fluctuations was confirmed, and many reaction data in the domain of
strongly overlapping resonances were analyzed using his
model. Theorists tackled the problem of connecting Ericson
fluctuations with GRMT. Two--body random ensembles and embedded
ensembles were defined as a meaningful extension of GRMT. Symmetry
breaking became a theoretical issue in GRMT. The results were used to
establish upper bounds on the breaking of time--reversal symmetry in
nuclei.

Much of this work is reviewed in Refs.~\onlinecite{Bro81,Mah80}.  

A significant test of GRMT predictions requires a sufficiently large
data set. Collecting such a set for the sole purpose of testing
stochasticity may seem a thankless task. The opposite is the case. 
It is important to establish whether stochasticity as formulated in 
GRMT does apply in a given system within some domain of excitation 
energy, and to what extent this is the case. Moreover, once the 
applicability of GRMT is established, there is no room left for 
spectroscopic studies involving a level--by--level analysis. Indeed, 
Balian's derivation of GRMT from a maximum--entropy principle 
\cite{Bal68} shows that spectroscopic data taken in the domain of 
validity of GRMT carry no information content. 

In the 1970's, a determined experimental effort produced data on 
low--energy neutron scattering by a number of heavy nuclei, and on 
proton scattering near the Coulomb barrier by several nuclei with 
mass numbers around $50$. The number of levels with the same spin 
and parity seen for any one of these target nuclei ranged typically 
from $100$ to $200$. In 1982, the totality of these data was combined 
into the ``Nuclear Data Ensemble'' comprising $1726$ level
spacings. This ensemble was tested by Haq, Pandey and Bohigas 
\cite{Haq82,Boh83} for agreement with the GOE (the relevant ensemble for 
these cases). They used the nearest neighbor spacing distribution 
and the $\Delta_3$ statistic. This analysis produced the first 
statistically significant evidence for the agreement of GOE 
predictions and spectral properties of nuclei, see Fig.~\ref{fig1}.

The advent of electrostatic accelerators of sufficiently high energy
and of sufficiently good energy resolution (which must be better than
the average width $\Gamma$ of the compound--nucleus resonances) made 
it possible at that time to detect Ericson fluctuations \cite{Eri79} 
in nuclear cross sections, see Fig.~\ref{fig3}.
As a function of incident energy, the intensity of particles produced 
in a nuclear reaction at fixed scattering angle shows random
fluctuations. An evaluation of the intensity autocorrelation versus
incident energy yields the average lifetime $\hbar / \Gamma$ of the
compound nucleus, and the autocorrelation versus scattering angle
yields information on the angular momenta relevant for the reaction. 
>From the variance one finds the ratio of ``direct reactions'', i.e. of
particles emitted without delay, to the typically long--delayed
compound--nucleus reaction contribution. The ``elastic enhancement
factor'' favors elastic over inelastic compound--nucleus reactions
and is the forerunner of the ``weak localization correction'' in
mesoscopic physics. 
The authors of reference \cite{Eri79} were well aware of the fact that
the phenomena discovered in nuclear reactions are generic. And indeed,
most of the concepts developed within the theory of Ericson 
fluctuations have later resurfaced in different guise when the theory 
of wave propagation through disordered media was developed, and was
applied to chaotic and disordered mesoscopic conductors, and to light 
propagation through media with a randomly varying index of refraction.

At the same time, intense theoretical efforts were undertaken to 
give a solid foundation to Ericson's model, and to connect it with 
cRMT \cite{Mah80}. This was necessary in order to obtain a unified 
model for fluctuation properties of nuclear cross sections in the 
entire range of excitation energies extending from $\Gamma \ll D$ 
(i.e. neutron threshold where GRMT was known to work) till the 
Ericson domain $\Gamma \gg D$. The approaches started from two 
different hypotheses. 

(i) {\it The random Hamiltonian approach}.

Formal theories of nuclear resonance reactions express the
scattering matrix $S$ in terms of the nuclear Hamiltonian $H$, so that
$S = S(H)$. Since RMT is supposed to model the stochastic properties
of $H$, an ensemble of scattering matrices was obtained by replacing
the nulear Hamiltonian by the GOE, i.e. by writing $S = S$(GOE). For
example, the shell--model approach to nuclear reactions \cite{Mah69}
yields for the elements $S_{a b}(E)$ of the scattering matrix
\begin{equation}
S_{a b}(E) = \delta_{a b} - 2 i \pi \sum_{\mu \nu} W_{a \mu}
(D^{-1}(E))_{\mu \nu} W_{\nu b}.
\label{shell1}
\end{equation}
Here, $E$ is the energy, and $a,b$ refer to the physical channels. The
indices $\mu$ and $\nu$ of the inverse propagator $D_{\mu \nu}$ refer
to a complete set of orthonormal compound nucleus states, and $D$ 
(not to be confused with the mean level spacing) has the form
\begin{equation}
D_{\mu \nu}(E) = E \delta_{\mu \nu} - H_{\mu \nu} + i \pi \sum_c W_{\mu
  c} W_{c \nu}.
\label{shell2}
\end{equation}
We have omitted an irrelevant shift function. The symbol $H_{\mu \nu}$
stands for the projection of the nuclear Hamiltonian onto the set of
compound nucleus states, and the matrix elements $W_{\mu a}(E)$
describe the coupling between these states and the channels
$a$. Equations~(\ref{shell1},\ref{shell2}) are generically valid. By
assuming that $H$ is a member of a random--matrix ensemble, $S$
becomes an ensemble of scattering matrices, and the challenge consists
in calculating ensemble--averaged cross sections and correlation
functions. Work along these lines was carried out by Moldauer, by 
Agassi {\it et al}., by Feshbach {\it et al}., and by many others. {\em Inter 
alia}, this approach led to an asymptotic expansion kin to impurity 
perturbation theory in condensed matter physics. It is valid for 
$\Gamma \gg D$ and uses the assumption that the GRMT eigenvalue 
distribution can be replaced by a model with fixed nearest neighbor 
spacings (``picket fence model''). In this framework, Ericson's 
results could be derived. However, all attempts failed to obtain a 
unified theoretical GRMT treatment valid in the entire regime from 
$\Gamma \ll D$ till $\Gamma \gg D$. The methods developed by Mehta 
which had proven so successful for level correlations did not seem 
to work for the more complex problem of cross--section fluctuations. 

Much theoretical attention was also paid to reactions at energies 
beyond the Ericson regime where an extension of RMT is needed to 
describe physical reality. Indeed, the application of RMT to 
transport processes (like cross sections) implies some sort of 
equilibrium assumption: By virtue of the orthogonal invariance of 
the GOE, all states of the compound system are equally accessible 
to an incident particle, and a preferred basis in Hilbert space 
does not exist. This assumption is justified whenever the 
``internal equilibration time''
$\tau_{eq}$ needed to mix the states actually populated in the first 
encounter of projectile and target with the rest of the system is
small compared with the time $\hbar / \Gamma$ for particle emission.
But $\tau_{eq}$ changes little with energy while $\hbar / \Gamma$
decreases nearly exponentially over some energy interval. At energies
above the Ericson regime there 
exists a domain where decay happens while the system
equilibrates. This is the  domain of ``precompound'' or
``preequilibrium'' reactions. It requires
an extension of RMT using a preferred basis in Hilbert space,
reflecting the ever increasing complexity of nuclear configurations 
reached from the incident channel in a series of two--body
collisions. Formally, this extension of RMT is a forerunner of and
bears a close relationship to the modelling of quasi one--dimensional 
conductors in terms of random band matrices. In
both cases, we deal with non--ergodic systems where a novel energy (or
time) scale --- the equilibration time or, in disordered solids, 
the diffusion time --- comes into play.   

(ii) {\it The random $S$ matrix approach}. 

A maximum--entropy approach for the scattering matrix kin to
Balian's derivation of GRMT was developed by Mello, Seligman and
others. A closed expression for the probability distribution of the 
elements of the scattering matrix was derived for any set of input
parameters (the average elements of $S$). Unfortunately, the
result was too unwieldy to be evaluated except in the limiting cases
of few or very many open channels. 
Moreover, parametric correlation
functions, e.g. the correlation between two elements of the 
scattering matrix at different energies, seem not accessible to this 
approach. Under the name ``random transfer matrix approach'', a
similar approach has later found wide application in the theory 
of disordered quasi one--dimensional conductors, cf. Sec.~\ref{quasi1d_2}. 
Here again, the calculation of parametric correlation functions has 
been fraught with difficulties. 

The random $S$ matrix approach has the virtue of dealing directly
with the elements of the $S$ matrix as stochastic variables. Thereby
it avoids introducing a random Hamiltonian in the propagator as done
in Eqs.~(\ref{shell1},\ref{shell2}) and the ensuing difficulties in
calculating ensemble averages. This very appealing feature is partly
offset by the difficulties in calculating parametric correlations. 

A basic criticism leveled against GRMT relates to the fact
that in most physical systems, the fundamental interaction is a
two--body interaction. In a shell--model basis, this interaction
has vanishing matrix elements between all states differing in
the occupation numbers of more than two single--particle
states. Therefore, the interaction matrix is sparse. In an arbitrary 
(non--shell model) basis, this implies that the number of 
{\it independent} matrix elements is much smaller than in GRMT, where 
the coupling matrix elements of any pair of states are uncorrelated
random variables. This poses the question whether GRMT predictions 
apply to systems with two--body forces. It is also necessary to determine
whether the two--body matrix elements are actually random. Another
problem consists in understanding how a random two--body force acts in
a many--Fermion Hilbert space. This problem leads to the question of
how information propagates in spaces of increasing complexity. These 
problems were tackled mainly by French and collaborators
\cite{Bro81} and led to ``statistical nuclear spectroscopy'' as 
an approach to understand the workings of the two--body residual 
interaction of the nuclear shell model in large, but finite,
shell--model spaces. Specifically, it
was shown that in the shell model, the matrix elements of the 
two--body force are random and have a Gaussian distribution.
The mathematical problem of handling such random two--body forces
was formulated by introducing the ``embedded ensembles'': A random 
$m$--body interaction operates in a shell--model space with $K$ 
Fermions where $m < K$. The propagation of the two--body force in
complex spaces was investigated with the help of moments methods. It
was shown numerically that for $m = 2$, the spectral fluctuation
properties of such an ensemble with orthogonal symmetry are 
the same as for the GOE. The last result shows that GOE can
meaningfully be used in predicting spectral fluctation properties of
nuclei and other systems governed by two--body interactions (atoms and
molecules). Nonetheless, embedded ensembles rather than GRTM would
offer the proper way of formulating statistical nuclear
spectroscopy. Unfortunately, an analytical 
treatment of the embedded ensembles is still missing.

In the early 80's, Mehta and Pandey \cite{Pan83,Pan84} made a
significant advance in the theoretical treatment of GRMT. Their work
established for the first time the connection between GRMT, field
theory, and the Itzykson--Zuber integral. They
succeeded in extending the orthogonal polynomial method to the problem
of symmetry breaking. In this way, they showed that the nearest neighbor
spacing distribution (NNS) could be used as a test of time--reversal 
invariance in nuclei. The basic idea is that for small spacings, the
presence of an interaction breaking time--reversal symmetry would
change the linear slope of NNS characteristic of the GOE into a 
quadratic one typical for the GUE, cf. Eq.~(\ref{W1}). To work out this
idea, it is necessary to define a random ensemble which allows for the
GOE $\rightarrow$ GUE crossover transition. The ensemble defined by
Mehta and Pandey for this purpose has the form 
\begin{equation}
H_{nm} = H_{nm}^{\rm GOE} + i \sqrt{t/N} H_{nm}^{A}.
\label{Me}
\end{equation}   
Here, $H^{GOE}$ is the GOE, and $H^A$ is the ensemble of real
antisymmetric Gaussian distributed matrices having the same variance
as the GOE. For $t = 0$, the ensemble (\ref{Me}) coincides with the 
GOE, while for $t = N$, it coincides with the GUE. Writing the
strength parameter in the form $(t/N)^{1/2}$ is motivated by the following
consideration. 
The perturbation $(t/N)^{1/2} H^A$ is expected to influence the {\it 
local} fluctuation properties of the spectrum (those on the scale $D$)
for values of $t$ such that the mean--square matrix element of 
the perturbation is of order $D^2$. From Eq.~(\ref{R1}), this 
implies $(t/N) (\lambda^2/N) \sim D^2 \sim 
(\pi \lambda / N)^2$, or $t \sim 1$. This argument shows that the 
local fluctuation
properties are extremely sensitive to a symmetry--breaking perturbation. 
The results obtained by Mehta and Pandey were used by French {\it et al}. to
obtain an upper bound on time--reversal symmetry breaking in nuclei. 

A similar problem arises from the Coulomb interaction. This
interaction is weak compared to the nuclear force and leads to a small
breaking of the isospin quantum number. The random--matrix model
needed for this case is different from Eq.~(\ref{Me}),
however. Indeed, without Coulomb interaction the Hamiltonian is 
block diagonal with respect to the isospin quantum number. In the 
simplest but realistic case of the mixing of states with two different
isospins, it consists of two independent GOE blocks. The
isospin--breaking interaction couples the two blocks. The strength
parameter of this coupling again scales with $N^{-1/2}$. For spectral
fluctuations, this crossover transition problem has not yet been
solved analytically (except for the GUE case) whereas an analytical
treatment has been possible for nuclear reactions involving isospin
symmetry breaking. This case is, in fact, the simplest example for
the precompound reaction referred to above. The results provided an 
ideal tool for the study of symmetry breaking in statistical nuclear 
reactions. The data gave strong support to the underlying statistical 
model \cite{Har82}.  

We conclude this historical review with a synopsis. RMT may be viewed
as a new kind of statistical mechanics, and there are several formal 
aspects which support such a view. First, RMT is based on universal 
symmetry arguments. In this respect it differs from ordinary
statistical mechanics which is based on dynamical principles; this is
the fundamental novel feature of RMT. It is linked to the fact
that in RMT, the ensembles consist of physical systems with {\it
  different} Hamiltonians. Second, the Gaussian ensembles 
of RMT can be derived from a maximum entropy approach. The derivation
is very similar to the way in which the three standard ensembles 
(microcanonical, canonical and grand canonical ensemble) can be
obtained in statistical mechanics. Third, the limit $N \rightarrow 
\infty$ is kin to the thermodynamic limit. Fourth, in relating results
obtained in the framework of RMT to data, an ergodic theorem is
used. It says that the ensemble average of an observable is --- for
almost all members of the ensemble --- equal to the running average of
the same observable taken over the spectrum of a single member of the
ensemble. In statistical mechanics, the ergodic theorem states the
equality of the phase--space average of an observable and the average
of the same observable taken along a single trajectory over sufficiently 
long time. Finally, the elements of the GRMT matrices generically 
connect all states with each other. This is a consequence of the 
underlying (orthogonal, unitary or symplectic) symmetry of GRMT and is 
reminiscent of the equal {\it a priori} occupation probability of 
accessible states in statistical mechanics. The assumption is valid if
the time scale for internal equilibration is smaller than all other time
scales.  

In the period from 1963 till 1982, the evidence grew strongly that 
both in nuclear spectrocopy and in nuclear reaction theory, concepts 
related to RMT are very successful. Two elements were missing, 
however: (i) A compelling physical argument was lacking why GRMT was 
such a successful model. Of course, GRMT is a generic theory. But why
does it apply to nuclei? More precisely, and more generally: What are
the dynamical properties needed to ensure the applicability of GRMT to
a given physical system? (ii) Both in nuclear spectroscopy and in 
nuclear reaction theory, the mathematical tools available were 
insufficient to answer all the relevant questions analytically, and 
novel techniques were needed.  

\subsection{Localization theory}
\label{loc} 

It may be well to recall our motivation for including localization theory 
in this historical survey. Localization theory deals with the properties 
of electrons in disordered materials. Disorder is simulated by a random 
potential. The resulting random single--particle Hamiltonian shares many
features with an ensemble of random matrices having the same symmetry 
properties. In fact, after 1983 the two fields -- RMT and localization 
theory -- began to coalesce. A review of localization theory was given by 
Lee and Ramakrishnan \cite{Lee85} in 1985. Here, we focus on those elements 
of the development which are pertinent to our context, without any claim of
completeness.   

The traditional quantum--mechanical description of the resistance 
of an electron in a metal starts from Bloch waves, the eigenfunctions 
of a particle moving freely through an ideal crystal. The resistance 
is due to corrections to this idealized picture. One such correction 
is caused by impurity scattering. The actual distribution of impurity
scatterers in any given sample is never known. Theoretical models for
impurity scattering therefore use statistical concepts which are 
quite similar to the ones used in RMT. The actual impurity potential 
is replaced by an ensemble of impurity potentials with some assumed
probability distribution, and observables are calculated as averages 
over this ensemble. It is frequently assumed, for instance, that the 
impurity potential $V(\vec r)$ is a Gaussian random process with mean 
value zero and second moment 
\begin{equation}
\overline{V(\vec r_1) V(\vec r_2)} = \frac{1}{2 \pi \nu \tau} 
\delta(\vec r_1 - \vec r_2). 
\label{imp}
\end{equation}
Here, the ensemble average is indicated by a bar, $\nu =
(V\Delta)^{-1}$ is the density of single--particle states
$\Delta^{-1}$ per volume $V$, and $\tau$ is the 
elastic mean free time, i.e. $\tau^{-1}$ measures
the strength of the impurity potential. In 1958, Anderson \cite{And58}
realized that under certain conditions, the standard perturbative 
approach to impurity scattering fails. He introduced the concept of 
localization. Consider the eigenfunctions of the single--particle 
Hamiltonian $H = T + V$ with $T$ the kinetic energy operator and $V$ 
the impurity potential defined above. For sufficiently strong disorder, 
these eigenfunctions, although oscillatory, may be confined to a 
finite domain of space. More precisely, their envelope $\chi \sim 
\exp( - r/\xi)$ falls off exponentially at the border of the domain. 
The scale is given by the localization length $\xi$, and $r$ is the 
distance from the center of the domain. For probes larger than the
localization length, the contribution to the conductance from such 
localized eigenfunctions is exponentially small. This important 
discovery was followed by the proof \cite{Mot61} that in one dimension
{\it all} states are localized no matter how small the disorder. In 
higher dimensions, localization occurs preferentially in the tails of 
the bands where the electrons are bound in deep pockets of the 
impurity potential. If there are extended states in the band center,
they are energetically separated from the localized states in the
tails by the  mobility edges. These characteristic energies mark the
location of the metal--insulator transition.

In the 1970's, Thouless and collaborators applied scaling ideas to the
localization problem \cite{Tho74}. Information about localization
properties can be obtained by the following Gedankenexperiment.
Consider two cubic blocks of length $L$ in $d$ dimensions of a
conductor with impurity scattering. Connect the two blocks to each
other to form a single bigger conductor, and ask how the
eigenfunctions change. The answer will depend on the ratio $\xi/L$.
For $L \gg \xi$, the localized eigenfunctions in either of the smaller
blocks are affected very little, and the opposite is true for $L \ll
\xi$.  Thouless realized that there is a single parameter which
controls the behavior of the wave functions. It is given by the
dimensionless conductance $g$, defined in terms of the actual
conductance $G$ as
\begin{equation}
G = \frac{e^2}{h} \ g.
\label{condu}
\end{equation}
Thouless expressed $g$ in terms of the ratio of two energies, $g = g
(E_c/\Delta$). Here, $\Delta$ is the mean {\it single--particle} level
spacing in each of the smaller blocks, not to be confused with the
mean level spacing of the {\it many--body} problem used in previous
sections.  The Thouless energy $E_c$ is defined as the energy interval
covered by the single--particle energies when the boundary conditions
on opposite surfaces of the block are changed from periodic to
antiperiodic. The Thouless energy is given by
\begin{equation}
E_c =  \frac{\hbar {\cal D}}{L^2}.  
\label{thoul}
\end{equation} 
where ${\cal D}$ is the diffusion constant.
We note that $L^2/{\cal D}$ is the classical diffusion time through the
block. It is remarkable that the Thouless energy, defined in terms of 
a classical time scale, plays a central role in localization,
manifestly a pure quantum phenomenon. For length scales $L$ such that $E_c
\gg \Delta$, an electron is multiply scattered by the 
impurity potential and moves {\it diffusively} through the probe. 
(This is true provided $L$ is larger than the elastic mean free path 
$l$. Otherwise, the motion of the electron is ballistic). In this 
diffusive regime, we have 
\begin{equation}
g = E_C / \Delta = \xi / L.
\label{conductance}
\end{equation}
The last equality applies in quasi one--dimensional systems only. For
$E_c \sim \Delta$, $L$ is of the order of the localization length
$\xi$, and the conductance is of order unity. For even larger values
of $L$, $\xi \ll L$, the multiple scattering of the electron by the 
impurity potential leads to annihilation of the wave function by 
interference (localization), and the conductance falls off
exponentially with $L$. This picture was developed further by Wegner
who used the analogy to the scaling theory of critical phenomena.    

Quantitative scaling theory asks for the change of $g$ with the length
$L$ of the probe in $d$ dimensions and expresses the answer in terms
of the $\beta$ function $\beta(L) = d \ln g(L)/ d \ln(L)$. The calculation
of $\beta(L)$ makes use of diagrammatic perturbation theory, or of
field--theoretical methods (renormalization theory, loop expansion). 
The answer depends strongly on $d$. In one dimension, all states are 
localized, there is no diffusive regime, and $g$ falls off 
exponentially over all length scales $L \gg l$. Abrahams {\it et al}.
argued that $\beta(L)$ is a function of $g(L)$ only (``one--parameter 
scaling''). This implied that not only in one but also in two 
dimensions all states should be localized. The view of these authors 
was largely supported later by perturbative and renormalization 
calculations, cf., however, Sec.~\ref{akl}. For $d \geq
2$, the perturbative approach to localization made use of an expansion
of the observable in powers of the impurity potential. The resulting 
series is simplified with the help of the small parameter $(k_F
l)^{-1} \ll 1$ where $k_F$ is the Fermi wave length and $l$ is the 
elastic mean free path due to impurity scattering. This is the essence
of ``diagrammatic impurity perturbation theory''. In the presence of a
magnetic field, some diagrams are not affected (those containing
``diffusons'') while others (containing ``cooperons'') become
suppressed ever more strongly as the strength of the magnetic field
increases. This discussion carries over, of course, to the 
field--theoretical approach. We do not discuss here the results of 
these approaches.

Prior to the development of the supersymmetry method, the 
field--theoretical approach to the localization problem made use of
the replica trick, introduced by Edwards and Anderson in 1975 
\cite{Edw75}. The replica trick was originally introduced to
calculate the average of $\rm{ln} Z$ where $Z$ is the partition function 
of a disordered system, for instance, a spin glass. It was applied 
later also to the calculation of the ensemble average of a product 
of resolvents (or propagators). 

Observables like the density of states or the scattering matrix depend
on the impurity Hamiltonian $H$ typically through advanced or retarded
propagators $(E^{\pm} - H)^{-1}$ where the energy $E^{\pm}$ carries an
infinitesimalIy small positive or negative imaginary part. It is
notoriously difficult to calculate the ensemble average of a product
of such propagators directly. Instead, one writes the (trace or matrix
element of) the propagator as the logarithmic derivative of a suitable
generating function $F$. For instance, in case $H$ is a simple
Hermitean matrix of dimension $N$ rather than a space--dependent
operator,
\begin{equation}
{\rm tr}\frac{1}{E^{\pm} - H} = 
 \frac{\partial \ln F_C(J_C)}{\partial J_C}\Bigg|_{J_C=0}
\label{replica1}
\end{equation}
where the generating function $F_C$ has the form
\begin{equation} 
  F_C(J_C) = \int d[S] \exp \left(\pm i S^\dagger (E^{\pm}1_N - H +
                                J_C1_N)S\right).
\label{replica2}
\end{equation}
We have introduced an $N$ component vector $S$ with complex entries. The
integration is performed over real and imaginary part of each of the
$N$ complex variables $S_n \ n=1,\ldots,N$.  This procedure, analogous
to standard methods in statistical mechanics, has the advantage of
bringing $H$ into the exponent of $F_C$. Therefore, it is easy to
average $F_C$ if $H$ has a Gaussian distribution.  Unfortunately,
calculating the observable requires averaging the logarithm of $F_C$
(or of a product of such logarithms). This cannot be done without some
further trick. We write
\begin{equation}
\ln F_C = \lim_{n \rightarrow 0}\, \frac{1}{n} (F_C^n-1). 
\label{replica3}
\end{equation}
The replica trick consists in calculating the average of $F_C^n$ for
integer $n$ only and in using the result to perform the limit in
Eq.~(\ref{replica3}).  Many of the subsequent steps used in the
replica trick have later been applied analogously in the supersymmetry
method, see Sec.~\ref{sup}.  We refrain from giving further details
here. Suffice it to say that the trick of confining the variable $n$
in Eq.~(\ref{replica3}) to integer values often limits the scope of
the method: It yields asymptotic expansions rather than exact results.
A comparison between the replica trick and the supersymmetry method
\cite{Ver85a} has shown why this is so.

An important step in the development of the field--theoretical method
was the 
introduction of the $N$--orbital model by Wegner and its subsequent
analysis \cite{Weg79a,Opp79,Sca80}. In this discrete lattice model,
each site carries $N$ orbitals, and an asymptotic expansion in inverse
powers of $N$ is possible for $N \rightarrow \infty$. In the replica
formulation, this limit defines a saddle point of the functional
integral. It was found that for observables involving a product of
retarded and advanced propagators, the saddle point takes the form of
a saddle--point manifold with hyperbolic symmetry. This fact reflects
the occurrence of the mean level density which carries opposite signs
in the retarded and the advanced sector, respectively, and breaks the
symmetry between retarded and advanced fields. This implies the
existence of a massless or Goldstone mode which causes singularities
in the perturbation series. The resulting field theory is closely
related to a non--linear $\sigma$ model. All of these features
reappear in the supersymmetric formulation, except that the topology
of the saddle--point manifold is modified because of the simultaneous
presence of bosonic and fermionic degrees of freedom.

Prior to many of the developments described above, Gor'kov and
Eliashberg \cite{Gor65} realized that the Wigner--Dyson approach (RMT)
could be applied to the description of level spectra of small metallic
particles, and to the response of such particles 
to external electromagnetic fields. The success of this description
implied the question whether a derivation of RMT from first principles
was possible. This question played an important role in Efetov's work
on supersymmetry to which we turn next. 
 
\subsection{The supersymmetry method}
\label{sup}

At the end of Sec.~\ref{cons}, it was mentioned that in order to apply
RMT to a number of problems of practical interest, novel methods were
needed. What was the difficulty? An observable relating to transport
properties typically consists of the square of a quantum--mechanical
amplitude, or of a product of such squares. For example, the cross
section is a sum of squares of $S$ matrix elements, and the
correlation function of the cross section is a product of sums of
squares. Each quantum--mechanical amplitude occurring in a transport
observable contains a propagator, i.e. a factor $(E - H \pm
i\gamma)^{-1}$. Here $H$ stands for the random matrix, and $\gamma$
denotes a width matrix which describes the coupling to external
channels. An example is given by Eqs.~(\ref{shell1}, \ \ref{shell2}).
The occurrence of $\gamma$ is typical for transport problems. No such
matrix arises in spectral fluctuation problems (i.e., in the
calculation of the $n$--level correlation functions). While Mehta's
orthogonal polynomial method makes it possible to calculate the
latter, his method apparently fails for the former problem. (At least
we are not aware of a successful application of the method in the
former case). The reason probably is that the matrix $\gamma$
massively breaks the symmetry which would otherwise exist between
retarded and advanced propagators. Hence, a novel method is needed to
calculate ensemble averages over observables relating to transport.

The replica trick mentioned in Sec.~\ref{loc} seemed to offer such a
novel method. It had been applied to technically very similar problems
(scattering of electrons in a random impurity potential).  It had led
to an understanding of the singularities encountered in diagrammatic
perturbation theory, and to the hope of overcoming these problems in
terms of an exact integration over the Goldstone mode (the
saddle--point manifold). The first applications of this method to RMT
scattering problems were disappointing, however. A calculation of the
$S$--matrix autocorrelation function \cite{Wei84} gave results which
were in agreement with but did not go beyond those obtained from an
asymptotic expansion in powers of $D / \Gamma$ mentioned in
Sec.~\ref{cons} under (i). This was later understood as a generic
shortcoming of the replica trick \cite{Ver85a}. Remedy came from the
supersymmetry method.

The fundamental problem in using a generating functional of the form
of Eq.~(\ref{replica2}) or its analogue in extended solids is a
problem of normalization. Indeed, for $J_C = 0$, we have $F_C(0) =
\det[ \pm 2i \pi / (E^{\pm} - H)]$. The occurrence of the logarithm in
Eq.~(\ref{replica1}) is due to the need to remove this normalization
factor $F_C(0)$. The replica trick accomplishes the same aim by taking
the limit $n \rightarrow 0$ of $\det^n[ \pm 2i \pi / (E^{\pm} - H)]$.
Several authors \cite{Mac80,Par81} noted that this factor can also be
removed, and $F_C(0)$ thereby normalized to unity, by integrating in
Eq.~(\ref{replica1}) over both normal (commuting or bosonic) and
anticommuting (Grassmann or fermionic) variables. This approach was
developed in a series of papers by Efetov and coworkers and led to the
supersymmetry method, summarized in Refs.~\onlinecite{Efe83,Efe96}. A
short technical introduction to supersymmetry is given in
Sec.~\ref{susy}.

In developing the supersymmetry method, Efetov was mainly interested
in the theory of disordered metals. It was soon realized, however,
that the method is eminently suitable also for all scattering and
transport problems of RMT type \cite{Ver85}. In the supersymmetry
formulation, the width matrix $\gamma$ which generically appears in
these problems does not lead to difficulties. The supersymmetry
approach to scattering and transport problems applies likewise to
problems of disorder in $d > 1$ dimensions and has played an important
role in making the supersymmetry method such an ubiquitous tool for
stochastic quantum problems. With the help of this method it was
possible, for instance, to calculate average nuclear cross sections
and the $S$--matrix autocorrelation function versus energy in the
entire domain $\Gamma \ll D$ till $\Gamma \gg D$ \cite{Ver85}. This
provided the solution of a long--standing problem in statistical
nuclear reaction theory described in Sec.~\ref{cons} and, for $\Gamma
\gg D$, confirmed Bohr's picture of the compound nucleus.

It was mentioned above that the supersymmetry method is not applicable
to Hamiltonians containing explicitly interacting Bosons or Fermions
in the language of second quantization. This is because the symmetry
properties of the fermionic or bosonic creation and annihilation
operators clash with the introduction of supersymmetry. There is a
natural way to avoid this problem: Introduce a complete set of
(properly symmetrized) many--body states, and write the Hamiltonian as
a matrix in the associated Hilbert space. Nothing prevents us from
applying the supersymmetry method to the resulting matrix problem.
However, stochastic properties of one--body or two--body operators
lead to stochastic properties of the resulting Hamiltonian matrix
which are very difficult to handle. These difficulties are similar to
the ones encountered in the treatment of the embedded ensembles, see
Sec.~\ref{cons}. Thus it is not the supersymmetry method as such which
fails in these cases but rather the complexity of the statistical
properties of the Hamiltonian which renders the problem unmanageable.

Aside from this shortcoming, the supersymmetry method suffers from two
additional technical restrictions.  (i) The dimension of the
supermatrices in the non--linear $\sigma$ model grows linearly with
the number $k$ of propagators in the $k$--point function. This makes
the exact (rather than asymptotic) calculation for $k \geq 4$ so
cumbersome that until now very few calculations for $k = 4$ and none
for larger $k$, have been published. (ii) For a long time,
calculations for disordered solids based on the non--linear $\sigma$
model were restricted to quasi one--dimensional probes. Only recently
has it been possible to overcome this restriction.

\subsection{RMT and classical chaos}
\label{chao}

As mentioned several times in previous sections, RMT received a
significant boost by the discovery of its connection with classical
chaos. A review may be found in Ref.~\onlinecite{Boh84}.

In the late 1970's and early 1980's, several authors investigated the
quantum spectra of conservative systems which behave chaotically in
the classical limit (i.e., are $K$ systems in this limit). The
interest in this question arose naturally from the great attention
paid at that time to classical chaotic motion. Sequences of levels
pertaining to the same symmetry class were generated numerically for
the Bunimovich stadium by McDonald and Kaufman \cite{McD79} and by
Casati {\it et al}.  \cite{Cas80}, and for the Sinai billiard by Berry
\cite{Ber81}. These data suggested agreement of the NNS distribution
with the Wigner surmise. Interest in this question was further kindled
by a paper by Zaslavsky \cite{Zas81} which also contains references to
earlier work.  In 1984, Bohigas, Giannoni and Schmit \cite{Boh84b}
produced a statistically significant set of data, consisting of more
than 700 eigenenergies of the Sinai billiard, larger than what had
been available before. Perhaps more importantly, these authors
employed the methods of statistical analysis developed for nuclear and
atomic spectra to this data set. The good agreement between both the
NNS distribution and the $\Delta_3$ statistic calculated from the data
and the corresponding GOE results (see Fig.~\ref{fig8}) caused these
authors to formulate the following conjecture: ``{\it Spectra of
  time--reversal invariant systems whose classical analogues are $K$
  systems show the same fluctuation properties as predicted by the
  GOE.}'' The $K$ systems mentioned in the conjecture are the most
strongly mixing classical systems. The most unpredictable $K$ systems
form a sub--class, they are referred to as Bernouilli systems.  An
alternative, stronger version of the conjecture, also formulated in
Ref.~\onlinecite{Boh84b}, replaces $K$ systems by less chaotic systems
provided they are ergodic. In both its forms, this conjecture is
commonly referred to as the Bohigas conjecture. For systems without
time--reversal invariance, the GOE is replaced by the GUE.  replaced
by the GUE. Originally, Bohigas {\it et al.} formulated the conjecture
without referring to the semiclassical regime, i.e.~to the limit
$\hbar\to 0$. However, all attempts to proof the conjecture are based
on some kind of semiclassical approximation, see the discussion in
Sec.~\ref{chaos}.  Further evidence for this conjecture was soon
forthcoming from the numerical study of other chaotic two--degrees of
freedom systems, and from experimental and theoretical studies of the
hydrogen atom in a strong magnetic field.  The conjecture also
suggested a (perhaps naive) physical explanation why RMT works in
nuclei and other complex quantum systems. Such an explanation, which
refers to chaotic classical motion, goes beyond the mere statement of
generic properties which were used to introduce RMT by Wigner. In
fact, in the light of the Bohigas conjecture and later findings,
systems with classically mixed phase space, where layers of chaotic
and regular motion are intimately interwoven, would not be expected to
display RMT properties in a clean fashion.

There are two problems connected with the Bohigas conjecture. The
analytical understanding of the connection between GOE and fully
developed classical chaos is still incomplete, and the conjecture must
be complemented by a number of caveats. As for the first point: A
proof of the Bohigas conjecture is still missing.  Apart from very
recent attempts to use a $\sigma$--model approach \cite{Muz95a,And96}
or structural invariance \cite{Ley92,Ley96a} to substantiate this
conjecture (see Secs.~\ref{qc4ss} and \ref{qc4si}), the only extant
analytical link between classical chaos and RMT is based on the
semiclassical approximation. Using this approximation,
Berry~\cite{Ber85} showed that for classically chaotic systems, the
$\Delta_3$ statistic defined in Eq.~(\ref{D1}) above has the same
logarithmic dependence on the interval length $L$ as for RMT. His
argument also showed a limitation of RMT: The correspondence between
chaotic systems and RMT applies only up to a maximum $L$ value defined
by the shortest classical periodic orbit. Beyond this value,
$\Delta_3$ becomes constant for dynamical systems. Unfortunately,
Berry's arguments could so far not be extended to address other RMT
predictions such as the NNS distribution. This is because short energy
intervals relate to long periodic orbits. In chaotic systems these
orbits increase exponentially in density and thus are difficult to
handle. Progress on this problem has recently been made, however, see
Sec.~\ref{qc4po}.  The caveat that GOE properties apply only up to a
maximum $L$ value is practically relevant only for systems with few
degrees of freedom.  Indeed, the semiclassical approximation is of
order $\hbar$ while the mean level spacing is of order $\hbar^{d}$
with $d$ the number of degrees of freedom. For large values of $d$,
the length of the level sequence needed to observe deviations from GOE
is beyond experimental or numerical possibilities. Another caveat
concerns the structure of classical phase space. Fully developed chaos
implies that in the course of time, a single chaotic trajectory is
ergodic (i.e., approaches every point in phase space arbitrarily
closely). But this requirement is not sufficient to guarantee
agreement with GOE fluctuations. It is necessary that, in addition,
all parts of phase space are equally accessible. A counterexample is
provided by two chaotic billiards connected by a thin channel. Here,
the conjecture would apply only in the semiclassical limit $\hbar
\rightarrow 0$. More generally, when geometry and/or Cantori cause the
system to have more than one intrinsic time scale, the connection with
RMT becomes more complicated, see Sec.~\ref{qc4mp}. Some of these
issues are reviewed in Ref.~\onlinecite{Haa91}.

\setcounter{equation}{0}
\section{Theoretical aspects}
\label{thasp}

Although developed in the framework of Nuclear Physics and with the
aim of providing a statistical approach to the analysis and
description of spectra of complex many--body systems, RMT has become a
branch {\it sui generis} of Theoretical Physics, with its own concepts
and mathematical methods. RMT studies analytically the fluctuation
properties and correlation functions of levels and wave functions of
ensembles of random matrices, quite independently of possible
applications to physical systems. The present section is written with
this fact in mind. Without going into details of the derivations, we
give a self--contained presentation of the central concepts of RMT,
and we summarize the most important formal results. We restrict
ourselves to the Gaussian ensembles, because on the physically
relevant unfolded scale, the circular ensembles give identical
results. The enormous wealth of the published material forced us to
restrict ourselves to those topics which are of particular importance
either for the analysis of data or for the understanding of the 
relationship between RMT and fields such as chaos theory and condensed
matter physics. Thus, some important contributions of mainly
mathematical interest had to be left out. The most prominent example
is the Pechukas gas, whose field--theoretical aspects are only briefly
discussed in Sec.~\ref{field}.

The section is organized as follows. In Sec.~\ref{qc1rc}, we begin 
with the ``classical'' analytical study of level fluctuation
properties. In Sec.~\ref{qc1ba}, we discuss the analysis of data 
and the spectral observables. The supersymmetry method which is used
in numerous modern applications of RMT is introduced
in Sec.~\ref{susy}. Scattering systems and the statistics of wave 
functions and widths are discussed in Secs.~\ref{qc1cs} 
and~\ref{qc1ww}, respectively. The study of crossover transitions 
in spectral correlations and other observables is presented
in Secs.~\ref{qc1tc} and~\ref{qc1to}, respectively. 
Parametric level motion and the ensuing correlations are studied
in Sec.~\ref{qc1pc}.

\subsection{Results of classical Random Matrix Theory}
\label{qc1rc}
 
Some of the results collected below were mentioned already in the
Historical Survey. We do not reiterate the intuitive arguments given
there and focus attention on the formal aspects.  In our notation we
follow as far as possible Mehta's book on Random
Matrices~\cite{Meh91}.  In Sec.~\ref{qc1rca}, we give the formal
definition of correlation functions and discuss some general aspects.
The choices for the probability density distributions and the
statistical implications and consequences thereof are discussed in
Sec.~\ref{qc1rcb}. Analytical results for the rotation invariant
Gaussian ensembles and the rotation non--invariant ensembles are
presented in Secs.~\ref{qc1rcc} and~\ref{qc1rcd}. Results for the
particularly important two--level correlations are summarized in
Sec.~\ref{qc1rce}.

\subsubsection{Correlation functions}
\label{qc1rca}

We consider a spectrum of $N$ energy levels $x_n, \ n=1,\ldots,N$ with
values on the entire real axis. An ensemble of infinitely many such
spectra is defined in terms of a normalized probability density
$P_N^{(E)}(x_1,\ldots,x_N)$. For the volume element, we use the flat 
measure $\prod_{n=1}^N d x_n$. It is assumed that the levels are
equivalent. This implies that $P_N^{(E)}(x_1,\ldots,x_N)$ is invariant
under permutations of the arguments. The correlation functions
describe the statistical properties of this ensemble. According to
Dyson~\cite{Dys62a,Dys62b,Dys62c}, the $k$--point correlation function
is obtained by integrating the probability density over $N-k$ arguments,
\begin{equation}
R_k(x_1,\ldots,x_k) = \frac{N!}{(N-k)!} 
       \int_{-\infty}^{+\infty} dx_{k+1} \cdots
       \int_{-\infty}^{+\infty} dx_N \, P_N^{(E)}(x_1,\ldots,x_N) \ .
\label{1rc1}
\end{equation} 
The functions $R_k$ in Eq.~(\ref{1rc1}) measure the probability
density of finding a level around each of the positions
$x_1,\ldots,x_k$, the remaining levels not being observed. These
quantities are independent of the labeling of the levels. Since the
function $P_N^{(E)}(x_1,\ldots,x_N)$ is normalized to unity, the
$k$--level correlation functions $R_k$ are normalized to the combinatorial
factors $N!/(N-k)!$. 

Frequently, correlation functions are introduced in a different
way. Let $H$ denote a Hamiltonian defined in a Hilbert space with a
cutoff so that $H$ has finite dimension $N$ with $N \gg 1$. The
spectrum of $H$ contains $N$ levels and is given by the trace of the
imaginary part of the Green function, i.e.~of the matrix
$(x_p^--H)^{-1}$ where the energy $x_p^- = x_p - i \eta$ has a
small imaginary increment. We consider an infinite ensemble of such
Hamiltonians. The ensemble is defined by a probability density 
$P_N(H)$ and a flat measure in matrix space. The $k$--point correlation
functions can be defined as 
\begin{equation}
R_k(x_1,\ldots,x_k) = \frac{1}{\pi^k} \int P_N(H)
       \prod_{p=1}^k {\rm Im\,}{\rm tr\,}\frac{1}{x_p^--H} d[H]
\label{1rc2}
\end{equation} 
where the flat measure $d[H]$ is the Cartesian volume element of $H$,
i.e.~the product of the differentials of all independent elements of
$H$. With a proper identification of $P_N(H)$ and of $P_N^{(E)}$, the
two definitions~(\ref{1rc1}) and~(\ref{1rc2}) are basically
equivalent.  There is a minor difference, however. It is easily seen
that the definition~(\ref{1rc2}) yields terms containing factors
$\delta(x_p-x_q)$, independent of the form of the probability density.
This is due to identities of the type
\begin{eqnarray}
  & & \frac{1}{\pi^2}\, {\rm Im\,}{\rm tr\,}\frac{1}{x_p^--H} {\rm
    Im\,}{\rm tr\,}\frac{1}{x_q^--H} \nonumber\\ 
  & & \qquad =
    \sum_{n=1}^N \delta(x_p-\hat{x}_n)\sum_{m=1}^N \delta(x_q-\hat{x}_m)
    \nonumber\\ & & \qquad\qquad = \delta(x_p-x_q)\sum_{n=1}^N
    \delta(x_p-\hat{x}_n) + \sum_{n\neq m}
    \delta(x_p-\hat{x}_n)\delta(x_q-\hat{x}_m)
\label{1rc3}
\end{eqnarray} 
where the $\hat{x}_n$ are the eigenvalues of $H$. (We differ in this
one instance from Mehta's notation. Usually, the eigenvalues are also
denoted by $x_n$.) Inserting Eq.~(\ref{1rc3}) into the
definition~(\ref{1rc2}) for $k=2$, one obtains two terms. After
integration over the eigenvalues $\hat{x}_1$ and $\hat{x}_2$, the
second term has the structure of the definition~(\ref{1rc1}).  In the
general case for arbitrary $k$, identities of the form~(\ref{1rc3})
yield for $R_k$ a sum of terms. All terms but one contain at least one
$\delta$ function of the form $\delta(x_p-x_q)$. The argument of this
function does not depend on the eigenvalues $\hat{x}_n$.
Equations~(\ref{1rc2}) and (\ref{1rc1}) are fully equivalent only when
all these terms are omitted in Eq.~(\ref{1rc2}). Unfortunately, the
two definitions (\ref{1rc2}) and (\ref{1rc1}) are sometimes not
clearly distinguished in the literature.

Some general results apply without any specific assumptions on the
form of the probability density. For systems described by the
Schr\"odinger equation, it was shown by Wigner~\cite{Wig32} that there
are three symmetry classes. For systems which are invariant under time
reversal and under space rotation, the Hamiltonian matrix $H$ can be
chosen real symmetric. For systems with broken time--reversal
invariance, $H$ is complex Hermitean, irrespective of whether rotation
invariance holds or not. For time--reversal invariant systems with
broken rotation invariance and half--odd--integer spin, $H$ is
quaternion real, i.e.~self--dual Hermitean. Thus, $H$ can be viewed as
an $N \times N$ matrix with $2 \times 2$ entries or as an $2N \times
2N$ matrix. According to Kramers~\cite{Kra30}, all eigenvalues of this
last type of system are doubly degenerate. The label $\beta=1,2,4$
indicates the number of real parameters $H_{nm}^{(\gamma)}, \ 
\gamma=0,\ldots,\beta-1$ needed to specify a matrix element $H_{nm}$
in each of the three classes. For $\beta=1,2,4$, the volume element
has the form
\begin{equation}
d[H] = \prod_{n\ge m} dH_{nm}^{(0)} 
       \prod_{\gamma=1}^{\beta-1} \prod_{n>m} dH_{nm}^{(\gamma)} \ . 
\label{1rc4}
\end{equation} 

We emphasize two points. First, the limit $N\to\infty$ of infinite
dimension has to be taken. This compensates the effect of the
technically motivated cutoff in Hilbert space. Second, in order to 
eliminate the dependence on the mean level density $R_1(x_1)$, the
dimensionless scaled variables 
\begin{equation}
\xi_p = \xi_p(x_p) = 
\int_{-\infty}^{x_p} R_1(x_p^\prime) dx_p^\prime \ ,
                 \qquad p=1,\ldots,k 
\label{1rc5}
\end{equation} 
are introduced. The correlation functions $R_k$ with $k>1$ have to be
rescaled accordingly. In analytical calculations, this ``unfolding procedure''
is usually combined with the limit $N\to\infty$. The unfolded
correlation functions $X_k(\xi_1,\ldots,\xi_k)$ are then obtained by
equating the differential probabilities on both energy scales
\begin{equation}
X_k(\xi_1,\ldots,\xi_k) d\xi_1 \cdots d\xi_k 
     = R_k(x_1,\ldots,x_k) dx_1 \cdots dx_k \qquad (N\to\infty) .
\label{1rc6}
\end{equation} 
In particular we have $X_1(\xi_1)=1$ by construction. In a generic
situation, it is sufficient to perform the unfolding in a small region
of the spectrum, provided that in the limit $N\to\infty$ this region
contains infinitely many levels. Let the region be centered at zero.
We then put $\xi_p=x_p/D$. The mean level spacing $D$ is defined by
$D=1/R_1(0)$. We note that the mean level spacing $D$ depends on $N$.
The unfolded correlation functions can be written as
\begin{equation}
X_k(\xi_1,\ldots,\xi_k) 
     = \lim_{N\to\infty} D^k R_k(D\xi_1,\ldots,D\xi_k) \ . 
\label{1rc7}
\end{equation} 
The new energy variables $\xi_p$ are held fixed while the limit is taken.

\subsubsection{Form of the probability density}
\label{qc1rcb}

To specify the probability density, we confine ourselves to the
two ``classical'' cases which were of paramount importance for the
development of Random Matrix Theory. This will allow us to establish
the relation between the functions $P_N^{(E)}(x_1,\ldots,x_N)$ 
and $P_N(H)$. We use the definition~(\ref{1rc2}) which is based on the
Hamiltonian.  

(i) {\it Rotation invariant density}.
      
      The probability density $P_N(H)$ is taken to be invariant under
      arbitrary rotations of matrix space, $P_N(H) = P_N(U^{-1}HU)$
      where $U$ is in the global unitary group $U(N;\beta)$. 
      Physically, this means that all states in matrix space interact
      with each other and that no preferred direction exists. With 
      $H=U^{-1}XU$, we introduce the elements of the diagonal matrix
      $X={\rm diag}(x_1,\ldots,x_N)$ of the eigenvalues and the 
      independent elements of $U$ as new variables. Then the volume 
      element~(\ref{1rc4}) has the form~\cite{Meh91}
      \begin{eqnarray}
      d[H] &=& |\Delta_N(X)|^\beta d[X] d\mu(U) \nonumber\\
      \Delta_N(X) &=& \prod_{n>m} (x_n-x_m)
      \label{1rc8}
      \end{eqnarray} 
      where $\Delta_N(X)$ is the Vandermonde determinant. (To keep
      with the usual notation, the eigenvalues are denoted by $x_n$,
      in contrast to Eq.~(\ref{1rc3}).) The integration over the 
      group $U$ with the invariant Haar measure $d\mu(U)$ is trivial 
      and we can identify
      \begin{equation}
      P_N^{(E)}(x_1,\ldots,x_N) = 
                    \tilde{C}_{N\beta}P_N(X)|\Delta_N(X)|^\beta \ .
      \label{1rc9}
      \end{equation} 
      The constant $\tilde{C}_{N\beta}$ ensures the normalization.

(ii) {\it Rotation non--invariant density}.

      All states in matrix space are uncoupled, and the rotation
      invariance of the probability density is maximally broken. This
      situation is modeled by assigning the value zero to all
      off--diagonal matrix elements of $H$, and we can write
      \begin{equation}
      P_N(H) = \prod_{n=1}^N p^{(0)}(H_{nn}^{(0)}) 
             \prod_{\gamma=0}^{\beta-1} 
             \prod_{n>m} \delta(H_{nm}^{(\gamma)})  
      \label{1rc10}
      \end{equation} 
      with a normalized function $p^{(0)}(z)$.
      The matrices of this ensemble are diagonal by construction. The
      level positions $x_n$ in definition~(\ref{1rc1}) are given by
      the diagonal elements, $x_n=H_{nn}^{(0)}$. The probability
      density is 
      \begin{equation}
      P_N^{(E)}(x_1,\ldots,x_N) = \prod_{n=1}^N p^{(0)}(x_n) \ .
      \label{1rc11}
      \end{equation} 

We refer to the two cases (i,ii) as to the pure cases. Various
crossover transitions are discussed later. The rotation invariant case
always contains the Vandermonde determinant and thus the famous Wigner
repulsion $|x_n-x_m|^\beta$ between each pair of levels. The absence
of the level--repulsion factor is typical for the rotation
non--invariant case. The occurrence or absence of the Vandermonde
determinant is independent of the functional form of $P_N(X)$ or of
$p^{(0)}(z)$ but strongly influences the character of the correlation
functions.  They are Wigner--Dyson--like (Poisson--like) in the presence
(absence) of this determinant. This connects to the issue of
universality taken up in Sec.~\ref{univ}.

For the Gaussian ensembles, the functional form of the rotation
invariant density $P_N(H)=P_N(X)$ is determined by the following
condition. The matrix elements of $H$ not connected by symmetry
are assumed to be statistically independent. Together with rotation
invariance, this implies that $P_N(H)$ is a normalized
Gaussian~\cite{Meh91}, 
\begin{equation}
P_{N\beta}(H) = \sqrt{\frac{\beta}{2\pi}}^{N+N(N-1)\beta/2}
       \exp\left(-\frac{\beta}{2}{\rm tr\,}H^2\right) \ .
\label{1rc12}
\end{equation} 
This yields the Gaussian Orthogonal (GOE), Unitary (GUE) and
Symplectic (GSE) Ensembles. The joint probability density of the
eigenvalues has the form
\begin{equation}
P_{N\beta}^{(E)}(x_1,\ldots,x_N) = C_{N\beta} 
       \exp\left(-\frac{\beta}{2}\sum_{n=1}^Nx_n^2\right)
                |\Delta_N(X)|^\beta \ .
\label{1rc13}
\end{equation} 
The normalization constant is given by~\cite{Meh91}
\begin{equation}
C_{N\beta} = \frac{\beta^{N/2+\beta N(N-1)/4}}{(2\pi)^{N/2}}
             \frac{\Gamma^N(1+\beta/2)}
                  {\prod_{n=1}^N\Gamma(1+\beta n/2)} \ .
\label{1rc14}
\end{equation} 
The variances of the Gaussians in Eqs.~(\ref{1rc12}) and~(\ref{1rc13})
are simply given by $1/\beta$, and the energies $x_n$ are measured on
a dimensionless scale.  Sometimes, another scaling is employed in the
literature where energies $E_n$ and variances $2v^2/\beta$ are used.
Here $v$ has the dimension energy. Very often, the two cases can be
mapped onto each other with the simple substitution
$x_n=E_n/\sqrt{2v^2}$. In dealing with crossover transitions in
Sec.~\ref{qc1tc}, we use $2v^2/\beta$ as the variance of the Gaussian
distribution.  In Eq.~(\ref{P3}), we have used another form for
$P_{N\beta}(H)$.  This form implies the value $2\lambda^2/N\beta$ for
the variance of the Gaussian distributions. This $N$--dependent choice
is frequently made within the supersymmetry method to be discussed in
Sec.~\ref{susy}. It renders the radius of the Wigner semicircle (see
the Historical Survey) independent of $N$. This has some technical
advantages in the supersymmetry method if the limit of infinitely many
levels is taken through a saddle--point approximation, see
Sec.~\ref{susy}.  This choice yields for the mean level spacing at the
origin $D \propto 1/N$.  The $N$--independent choice of the variance
used throughout the present section makes the radius of the Wigner
semicircle grow like $\sqrt{N}$.  This implies for the mean level
spacing at the origin $D \propto 1/\sqrt{N}$.

\subsubsection{Analytical results for the Gaussian ensembles}
\label{qc1rcc}

The considerable mathematical difficulties involved in the complete
analytical calculation of the $k$--level correlation function for the
three Gaussian ensembles were overcome by Mehta, Gaudin and
Dyson~\cite{Meh91}.  The GUE is technically the simplest of the three
Gaussian ensembles.  The GOE and the GSE are much more difficult and,
somewhat surprisingly at first sight, closely related. We use the
definition~(\ref{1rc1}) of the $k$--level correlation functions. These
functions have a determinant structure. For all three ensembles, they
can be written as a quaternion determinant of $k \times k$ quaternion
matrices built on $2\times 2$ matrix entries
$\sigma_{N\beta}(x_p,x_q)$,
\begin{equation}
R_{\beta k}(x_1,\ldots,x_k) = 
        {\rm qdet\,}\left[\sigma_{N\beta}(x_p,x_q)
                  \right]_{p,q=1,\ldots,k} \ .
\label{1rc15}
\end{equation}
The quaternion determinant $\rm qdet$ of a $k\times k$ self--dual
quaternion matrix $Q$ is defined as
\begin{equation}
{\rm qdet\,}Q = \sqrt{{\rm det\,}C(Q)} \ .
\label{1rc16}
\end{equation}
Here, $C$ is the $2k\times 2k$ matrix constructed from $Q$ by using
the Pauli matrices as basis for the quaternion entries of $Q$. If $Q$
is self--dual, ${\rm det\,}C(Q)$ is a perfect square and the square
root can be taken. The determinant on the right hand side of
Eq.~(\ref{1rc15}) has precisely this property.  To further illustrate
the remarkably compact result~(\ref{1rc15}), we give explicit results
for the $2 \times 2$ matrices $\sigma_{N\beta}(x_p,x_q)$ which depend
on only two energies $x_p$ and $x_q$. All entries involve the
oscillator wave functions
\begin{equation}
\varphi_n(z) = \frac{1}{\sqrt{2^nn!\sqrt{\pi}}}
               \exp\left(-\frac{z^2}{2}\right)H_n(z)
\label{1rc17}
\end{equation}
where $H_n(z)$ are the standard Hermite polynomials. We define the
kernel 
\begin{equation}
K_N(x_p,x_q) = \sum_{n=0}^{N-1} \varphi_n(x_p) \varphi_n(x_q)
\label{1rc18}
\end{equation}
and the functions
\begin{eqnarray}
S_N(x_p,x_q) &=& K_N(x_p,x_q) + \frac{1}{2} \sqrt{\frac{N}{2}}
       \varphi_{N-1}(x_p) \int_{-\infty}^{+\infty} 
         {\rm sgn\,}(x_q-z)\varphi_N(z)dz
                           \nonumber\\
DS_N(x_p,x_q) &=& -\frac{\partial}{\partial x_q} S_N(x_p,x_q)
                           \nonumber\\
IS_N(x_p,x_q) &=&  \frac{1}{2} \int_{-\infty}^{+\infty} 
              {\rm sgn\,}(x_p-z)S_N(z,x_q)dz
                           \nonumber\\
JS_N(x_p,x_q) &=& IS_N(x_p,x_q) - \frac{1}{2}{\rm sgn\,}(x_p-x_q)
\label{1rc19}
\end{eqnarray}
where ${\rm sgn\,}(z)$ is the signum function. In the case of even 
$N$, the result for the GOE is
\begin{equation}
\sigma_{N1}(x_p,x_q) = \left[
         \begin{array}{cc}
                    S_N(x_p,x_q) & DS_N(x_p,x_q) \\
                    JS_N(x_p,x_q) & S_N(x_q,x_p)
         \end{array}
       \right] \ .
\label{1rc20}
\end{equation}
The (slight) modification for odd $N$ can be found in Mehta's
book~\cite{Meh91}. For the GUE one has for all $N$ 
\begin{equation}
\sigma_{N2}(x_p,x_q) = \left[
         \begin{array}{cc}
                    K_N(x_p,x_q) & 0 \\
                    0 & K_N(x_q,x_p)
         \end{array}   
       \right] \ .
\label{1rc21}
\end{equation}
The result for the GSE reads 
\begin{equation}
\sigma_{N4}(x_p,x_q) = \frac{1}{\sqrt{2}}\left[
         \begin{array}{cc}
             S_{2N+1}(\sqrt{2}x_p,\sqrt{2}x_q) & 
                     DS_{2N+1}(\sqrt{2}x_p,\sqrt{2}x_q) \\
             IS_{2N+1}(\sqrt{2}x_p,\sqrt{2}x_q) & 
                      S_{2N+1}(\sqrt{2}x_q,\sqrt{2}x_p)
         \end{array}
       \right] \ .
\label{1rc22}
\end{equation}
The similarity of the structures for the GOE and the GSE is apparent.
In the GUE case, the correlation functions can be written as
ordinary $k\times k$ determinants 
\begin{equation}
R_{2,k}(x_1,\ldots,x_k) = 
   {\rm det\,}\left[K_N(x_p,x_q)\right]_{p,q=1,\ldots,k} \ .
\label{1rc23}
\end{equation}
This is due to the simple structure of Eq.~(\ref{1rc21}) and implies
that the quaternion structure of Eq.~(\ref{1rc15}) is not essential
for $\beta=2$. 

For the unfolded correlation functions $X_k(\xi_1,\ldots,\xi_k)$
defined in Eqs.~(\ref{1rc6}) and~(\ref{1rc7}), the beautiful
determinant structure~(\ref{1rc15}) survives and the problem is
reduced to unfolding the $2 \times 2$ matrices
$\sigma_{N\beta}(x_p,x_q)$ and their entries~(\ref{1rc18})
and~(\ref{1rc19}). One finds on the unfolded scale in the limit of
infinitely many levels~\cite{Meh91} 
\begin{equation}
X_{\beta k}(\xi_1,\ldots,\xi_k) = 
        {\rm qdet\,}\left[\sigma_\beta(\xi_p-\xi_q)
                  \right]_{p,q=1,\ldots,k}  \ .
\label{1rc24}
\end{equation}
The unfolded correlation functions depend only on the differences
$\xi_p-\xi_q$ and are, thus, translation invariant. The entries of the
unfolded matrices $\sigma_\beta(r)$ are
\begin{eqnarray}
s(r) &=& \frac{\sin\pi r}{\pi r}
                           \nonumber\\
Ds(r) &=& \frac{d}{dr} s(r) \qquad {\rm and} \qquad
Is(r) = -\int_0^r s(r^\prime) dr^\prime \ .
\label{1rc25}
\end{eqnarray}
The function $s(\xi_p-\xi_q)=\lim_{N\to\infty}DK_N(x_p,x_q)$ is the 
unfolded kernel~(\ref{1rc18}). Explicitly, one has
\begin{eqnarray}
\sigma_1(r) &=& \left[
         \begin{array}{cc}
                    s(r) & Ds(r) \\
                    Is(r)-1/2 & s(r)
         \end{array}
                \right]
                            \nonumber\\ 
\sigma_2(r) &=& \left[
         \begin{array}{cc}
                    s(r) & 0 \\
                    0    & s(r)
         \end{array}
                \right]
                            \nonumber\\ 
\sigma_4(r) &=& \left[
         \begin{array}{cc}
                    s(2r) & Ds(2r) \\
                    Is(2r) & s(2r)
         \end{array}
                \right] \ .
\label{1rc26}
\end{eqnarray}
The result for the GUE can be written in the simple form
\begin{equation}
X_{2,k}(\xi_1,\ldots,\xi_k) = 
   {\rm det\,}\left[s(\xi_p-\xi_q)\right]_{p,q=1,\ldots,k} \ .
\label{1rc27}
\end{equation}
The determinant structure of Eqs.~(\ref{1rc15}) and~(\ref{1rc23})
and, therefore, also of Eqs.~(\ref{1rc24}) and~(\ref{1rc27}) is
entirely due to the Vandermonde determinant $|\Delta_N(X)|^\beta$ in 
Eq.~(\ref{1rc9}). Therefore, this structure prevails for all
probability densities $P_N(X)$ which factorize in the arguments $x_n, \
n=1,\ldots,N$. Moreover, in many cases, the unfolding will yield
exactly the same functional forms~(\ref{1rc25}). Actually, it is shown
in Sec.~\ref{univ} on universality that the results given above 
apply to a wide class of probability densities $P_N(X)$.

It is apparent from the determinant structure that the correlation
functions $R_{\beta k}$ with $k>1$ contain terms which reflect the
clustering of the $k$ levels into subgroups of $k^\prime=1,\ldots,k-1$
levels. In analogy to the Green functions or propagators in field
theory, it is possible to remove such terms by introducing a cumulant
or cluster expansion. The $k$--level cluster function $T_{\beta
k}(x_1,\ldots,x_k)$ is accordingly defined as the true or connected
part of $R_{\beta k}(x_1,\ldots,x_k)$. This means that it cannot be
written in terms of lower order correlation functions.
The completely disconnected part of $R_{\beta k}(x_1,\ldots,x_k)$ is
always the product of the $k$ mean level densities. For $k = 2$, we
have, in particular, 
\begin{equation}
R_{\beta 2}(x_1,x_2) = R_{\beta 1}(x_1)R_{\beta 1}(x_2) - 
                                  T_{\beta 2}(x_1,x_2) \ .
\label{1rc28}
\end{equation}
In general, the $k$--level cluster function can be written as
\begin{equation}
T_{\beta k}(x_1,\ldots,x_k) = \sum_\omega \frac{1}{2}{\rm tr\,}
     \prod_{p=1}^k \sigma_{N\beta}(x_{\omega(p)},x_{\omega(p+1)})
\label{1rc29}
\end{equation}
where the sum is over all $(k-1)!$ distinct cyclic permutations 
$\omega$ of the indices $1,2,\ldots,k$. Every cycle is closed by
the definition $\omega(k+1)=\omega(1)$. On the unfolded scale, the
$k$--level cluster functions are defined in
analogy to Eqs.~(\ref{1rc6}) and~(\ref{1rc7}), 
\begin{equation}
Y_{\beta k}(\xi_1,\ldots,\xi_k) 
     = \lim_{N\to\infty} D^k T_{\beta k}(D\xi_1,\ldots,D\xi_k) \ . 
\label{1rc30}
\end{equation} 
In the case of the Gaussian ensembles, these functions can be written
as 
\begin{equation}
Y_{\beta k}(\xi_1,\ldots,\xi_k) = \sum_\omega \frac{1}{2}{\rm tr\,}
     \prod_{p=1}^k \sigma_\beta(\xi_{\omega(p)}-\xi_{\omega(p+1)}) \ .
\label{1rc31}
\end{equation}
Again, there are simplifications for the GUE. The completely
disconnected part of $X_{\beta k}(\xi_1,\ldots,\xi_k)$ is simply
unity, i.e.~the product of all unfolded mean level densities.

\subsubsection{Analytical results for the 
                           rotation non--invariant case}
\label{qc1rcd}

In this case, the probability densities have the forms~(\ref{1rc10})
and~(\ref{1rc11}). The Vandermonde determinant is absent, and the
calculation of the correlation functions is almost trivial. As in the
case of the Gaussian ensembles, we use the definition~(\ref{1rc1}). 
For the level density one finds
\begin{equation}
R_1(x) = N  p^{(0)}(x) \ .
\label{1rc32}
\end{equation}
The $k$--level correlation functions are simply $k$--fold products of
mean level densities, 
\begin{eqnarray}
R_k(x_1,\ldots,x_k) &=& \frac{N!}{(N-k)!} \prod_{p=1}^k p^{(0)}(x_p) 
      = \frac{N!}{(N-k)!N^k} \prod_{p=1}^k R_1(x_p)
                                        \nonumber\\
     &=& \prod_{p=1}^k \left(1-\frac{p-1}{N}\right) R_1(x_p)
\label{1rc33}
\end{eqnarray}
The choice of the factor on the right hand side of Eq.~(\ref{1rc33})
guarantees the normalization of the $k$--level correlation function to
$N!/(N-k)!$.  By definition, this normalization applies to arbitrary
ensembles. For the Gaussian ensembles, the completely disconnected
part in the cumulant or cluster expansion of the $k$--level
correlation function is simply the product of all $k$ level densities.
The prefactor in front of this term is unity.  The functional form of
the expression~(\ref{1rc33}) coincides with this completely
disconnected part. However, to ensure the overall normalization, its
prefactor is different from unity. In the limit $N\to\infty$, all
factors $(1-(p-1)/N)$ in the last of Eqs.~(\ref{1rc33}) yield unity.
This leads to a normalization problem on the unfolded scale since the
limit $N\to\infty$ and the normalization do not commute.  We will
discuss this for the two--level cluster function in Sec.~\ref{qc1rce}.
Thus, it is not very convenient to define $k$--level cluster
functions.

If on the scale of the mean level spacing the function $p^{(0)}(z)$
has no structure, one has
\begin{equation}
X_k(\xi_1,\ldots,\xi_k) = 1 \qquad {\rm for \ all} \quad k
\label{1rc34}
\end{equation}
which coincides with the fully disconnected part in the case of
the Gaussian ensembles. The absence of all correlations in the 
spectrum is referred to as Poisson regularity and is formally
expressed in terms of the $k$--level cluster functions as
\begin{equation}
Y_k(\xi_1,\ldots,\xi_k) = 0 \qquad {\rm for} \quad k > 1  \ .
\label{1rc35}
\end{equation}
The opposite case of a regular but maximally correlated spectrum is
represented by the harmonic oscillator. There,
\begin{equation}
p^{(0)}(z) = \frac{1}{N} \sum_{n=-N/2}^{+N/2} \delta(z-nD) \ .
\label{1rc35a}
\end{equation}
The condition of smoothness of $p^{(0)}(z)$ is strongly violated. Hence 
Eqs.~(\ref{1rc34}) and~(\ref{1rc35}) do not apply. Instead one finds
\begin{equation}
X_1(\xi_1) = \sum_{n=-\infty}^{+\infty} \delta(\xi_1-n)
\label{1rc35b}
\end{equation}
for the level density of an infinite spectrum and
\begin{equation}
X_k(\xi_1,\ldots,\xi_k) = \prod_{p=1}^k X_1(\xi_p) 
\label{1rc35c}
\end{equation}
for the correlation functions. For $k>1$, these correlation
functions contain terms proportional to $\delta(\xi_p-\xi_q)$. This is
due to the form~(\ref{1rc35a}) of the density $p^{(0)}(z)$.

\subsubsection{Two--level correlation functions}
\label{qc1rce}

Two--level correlation functions are especially important for the
analysis of data. We summarize results for these functions on the
unfolded scale and for the pure ensembles. For translation invariant
spectra, these functions depend only on the energy difference
$r=\xi_1-\xi_2$ and one has 
\begin{equation}
X_2(r) = 1 - Y_2(r) \ .
\label{1rc36}
\end{equation}
In the Poisson case the connected part vanishes, $Y_2(r)=0$. For
the Gaussian ensembles, one has with $s(r)=\sin\pi r/\pi r$
\begin{eqnarray}
Y_{1,2}(r) &=& s^2(r) + 
       \frac{ds(r)}{dr}\int_r^\infty s(r^\prime)dr^\prime
                           \nonumber\\
Y_{2,2} (r) &=& s^2(r)
                           \nonumber\\
Y_{4,2}(r) &=& s^2(2r) - 
       \frac{ds(2r)}{dr}\int_0^r s(2r^\prime)dr^\prime
\label{1rc37}
\end{eqnarray}
where the first index labels the GOE, GUE and GSE, respectively.  In
all three cases, the normalization integral of $Y_{\beta 2}(r)$ over
the real axis yields unity. This ``sum rule'' reflects the
normalization mentioned in Sec.~\ref{qc1rcd}. In the Poisson case, on
the other hand, this sum rule yields zero since $Y_2(r)=0$.  This
``normalization problem'' arises because the normalization for {\it
  finite} $N$ and the limit $N\to\infty$ do not commute. In the
Poisson case, the sum rule is only satisfied for finite $N$. This
issue will be taken up in Sec.~\ref{spec_3}.
\begin{figure}
\centerline{
\psfig{file=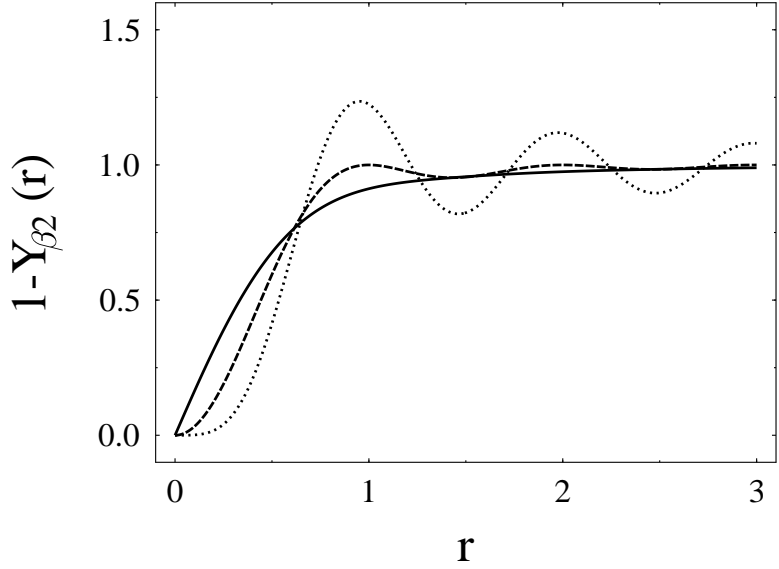,width=4.5in,angle=-90}
}
\caption{
  The two--level correlation functions $X_{\beta 2}(r)=1-Y_{\beta 2}(r)$ 
  on the unfolded energy scale. 
  The solid line is the GOE result ($\beta=1$), the
  dashed line is the GUE result ($\beta=2$) and the dotted line is the
  GSE result ($\beta=4$). We notice that $1-Y_{4,2}(r)$ overshoots 
  the value one due to strong oscillations.  
}
\label{figft1}
\end{figure}
In Fig.~\ref{figft1}, the two--level cluster function is shown for the
three Gaussian ensembles. In various applications, one needs the
Fourier transforms or ``two--level form factors''
\begin{equation}
b_2(t) = \int_{-\infty}^{+\infty} Y_2(r) \exp(i2\pi rt) dr \ .
\label{1rc38}
\end{equation}
In the Poisson case, one has $b_2(t)=0$. For the Gaussian ensembles,
one finds 
\begin{eqnarray}
b_{1,2}(t) &=& \left\{
         \begin{array}{cc}
         1-2|t|+|t|\ln(2|t|+1)          & 
                                \qquad {\rm if} \quad |t|\le 1 \\
         -1+|t|\ln\frac{2|t|+1}{2|t|-1} & 
                                \qquad {\rm if} \quad |t|> 1
         \end{array}
                \right.
                           \nonumber\\
b_{2,2}(t) &=& \left\{
         \begin{array}{cc}
         1-|t| & \qquad {\rm if} \quad |t|\le 1 \\
         0     & \qquad {\rm if} \quad |t|> 1
         \end{array}
                \right.
                           \nonumber\\
b_{4,2}(t) &=& \left\{
         \begin{array}{cc}
         1-\frac{1}{2}|t|+\frac{1}{4}|t|\ln\left||t|-1\right| 
                               & \qquad {\rm if} \quad |t|\le 2 \\
         0 & \qquad {\rm if} \quad |t|> 2
         \end{array}
                \right.
\label{1rc40}
\end{eqnarray}
The functions $1-b_{\beta 2}(t)$ are shown in Fig.~\ref{figft2}.  
\begin{figure}
\centerline{
\psfig{file=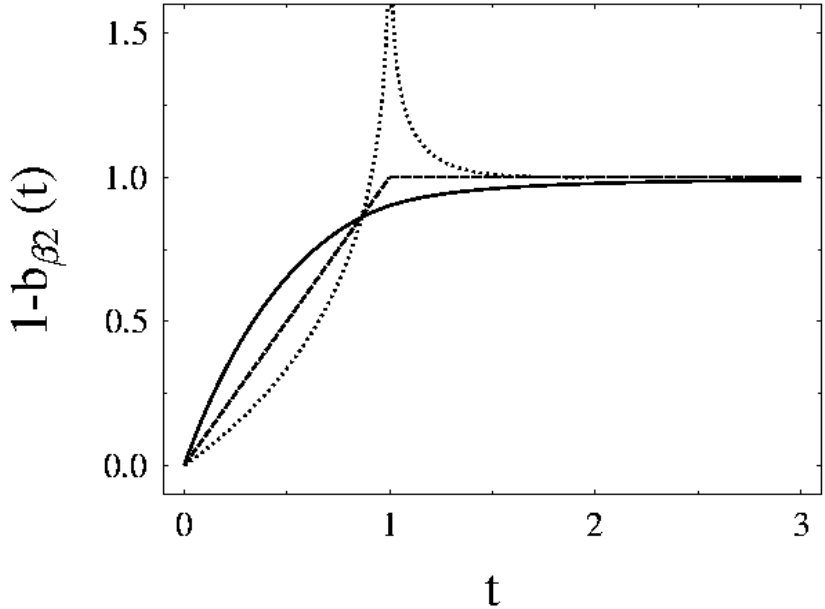,width=4.5in,angle=-90}
}
\caption{
  Relevant parts $1-b_{\beta 2}(t)$ of the Fourier transforms 
  of the unfolded two--level correlation functions 
  $X_{\beta 2}(r)=1-Y_{\beta 2}(r)$. The functions $b_{\beta 2}(t)$ 
  are referred to as the two--level form factors. The solid line is 
  the GOE result ($\beta=1$), the dashed line is the GUE result 
  ($\beta=2$) and the dotted line is the GSE result ($\beta=4$). 
  The different oscillatory structure of the two--level correlation 
  functions is most strikingly reflected by the behavior of the 
  Fourier transforms at $|t|=1$.
}
\label{figft2}
\end{figure}
The discontinuities of these functions or there derivatives reflect
the different oscillatory structures of the two--level correlation
functions.  We notice the singularity at $|t|=1$ in $b_{4,2}(t)$.

\subsection{Analysis of data and spectral observables}
\label{qc1ba}

In this section, we review some methods used in the statistical analysis
of experimental data, and we summarize the relevant predictions of Random
Matrix Theory. The unfolding procedure is described in Sec.~\ref{qc1baa}. 
Then, we discuss several statistical observables: The nearest neighbor
spacing distribution (Sec.~\ref{qc1bab}), long--range spectral
observables (Sec.~\ref{qc1bac}), and Fourier transform methods
(Sec.~\ref{qc1bad}). The superposition of several independent spectra
is treated in Sec.~\ref{qc1bae}. Specific properties of Random Matrix
Theory may yield additional tests of data. In Sec.~\ref{qc1baf}, we
present the GSE test of GOE data as an example. The predictions of Random
Matrix Theory are based on ensemble averages while experimental
results are obtained from a running average over the spectrum of a
single sample. The justification for this procedure is non--trivial
and raises the question of ergodicity. This question is briefly
discussed in Sec.~\ref{qc1bag}.

\subsubsection{Unfolding procedure}
\label{qc1baa}

A measurement or a calculation yields an ordered sequence of energies
$\{E_1,E_2,\ldots,E_N\}$ which form the stick spectrum or spectral
function, 
\begin{equation}
S(E) = \sum_{n=1}^N \delta(E-E_n) \ .
\label{1ba1}
\end{equation}
To analyze the fluctuation properties, this spectrum has to be
unfolded, i.e. the system--specific mean level density $R_1(E)$ must
be removed from the data. We define the cumulative spectral function
\begin{equation}
\eta(E) = \int_{-\infty}^E S(E^\prime) dE^\prime 
        = \sum_{n=1}^N \Theta(E-E_n) \ .
\label{1ba2}
\end{equation}
This function counts the number of levels with energy less than or
\begin{figure}
\centerline{
\psfig{file=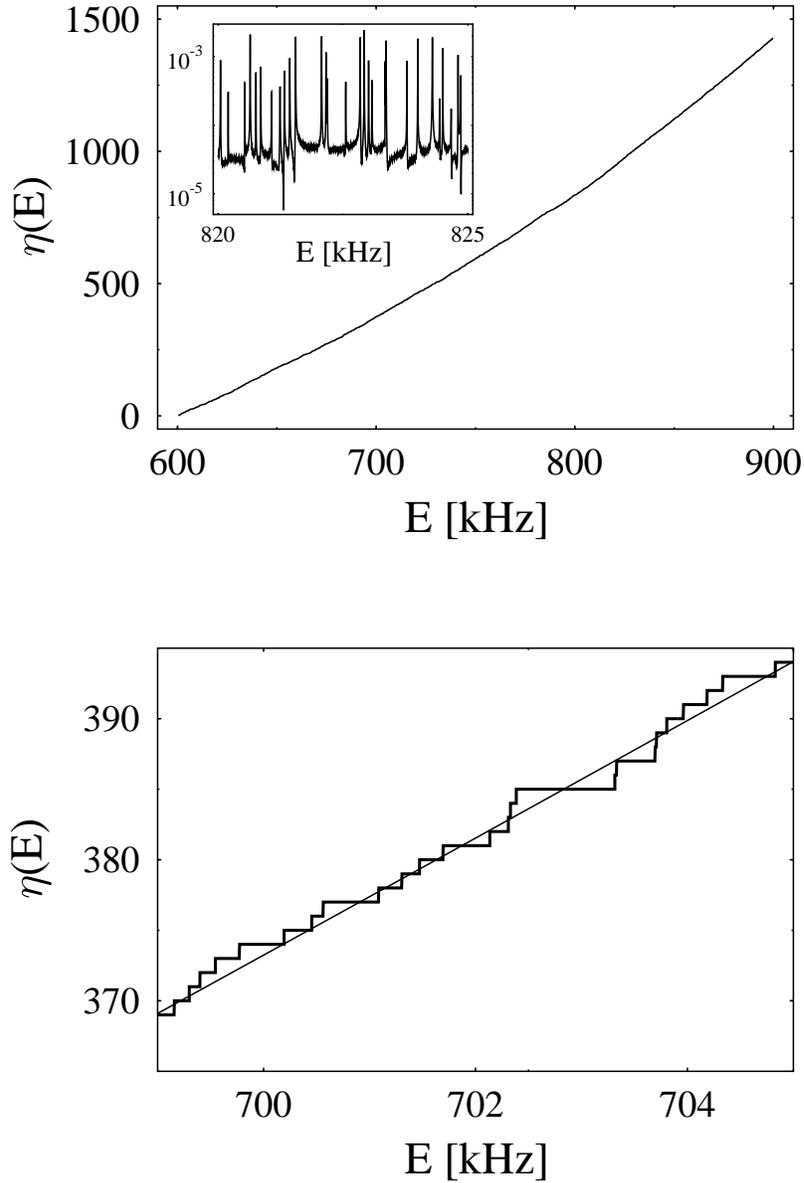,width=4.5in,angle=0}
}
\caption{
  Example of an experimentally obtained staircase function.
  The top figure shows the cumulative spectral function $\eta(E)$
  for a spectrum of 1428 elastomechanical eigenfrequencies of a
  resonating quartz block in the frequency range between 600~kHz
  and 900~kHz. To keep with the notation in the text, the frequency
  is denoted by $E$. The inset shows a small section of the measured
  spectrum between 820~kHz and 825~kHz. Due to the high number of levels,
  the staircase function appears as a smooth line. The smooth part
  $\xi(E)$ is a polynomial whose coefficients were found by a fit. 
  The bottom part shows a small section of the staircase 
  function between 699~kHz and 705~kHz, containing about 25 levels.
  The corresponding section of $\xi(E)$, i.e.~the polynomial fit, is drawn
  as a thin line. We notice that $\xi(E)$ is obtained by fitting to 
  the entire cumulative spectral function, not only to this section. 
  Adapted from Ref.~\protect\onlinecite{Ell96}.
}
\label{figft8}
\end{figure}
equal to $E$ and is also referred to as the staircase function. It is
decomposed into a smooth part $\xi(E)$ and a fluctuating part
$\eta_{\rm fl}(E)$,
\begin{equation}
\eta(E) = \xi(E) + \eta_{\rm fl}(E) \ .
\label{1ba3}
\end{equation}
The smooth part is given by the cumulative mean level density,
\begin{equation}
\xi(E) = \int_{-\infty}^E R_1(E^\prime) dE^\prime \ .
\label{1ba4}
\end{equation}
An example for an experimentally obtained staircase function and its
smooth part is shown in Fig.~\ref{figft8}.  To unfold the spectrum,
the sequence $\{E_1,E_2,\ldots,E_N\}$ is mapped onto the numbers
$\{\xi_1,\xi_2,\ldots,\xi_N\}$ with
\begin{equation}
\xi_n = \xi(E_n) \ , \qquad n=1,\ldots,N \ . 
\label{1ba5}
\end{equation}
In these new variables the cumulative spectral function simply reads
\begin{equation}
\widehat{\eta}(\xi) = \xi + \widehat{\eta}_{\rm fl}(\xi) \ .
\label{1ba6}
\end{equation}
The mean level density of the unfolded spectrum, i.e. the derivative
of the smooth part with respect to $\xi$, is unity, as required. 
Comparing Eqs.~(\ref{1rc5}) and~(\ref{1ba4}) we see that the present
unfolding procedure coincides with the analytical unfolding in Random
Matrix Theory. Therefore, measurement and theoretical prediction can
be directly compared with each other. In practice, the separation in
Eq.~(\ref{1ba3}) of a spectrum into a smooth and a fluctuating part can
be non--trivial task.  

\subsubsection{Nearest neighbor spacing distribution}
\label{qc1bab}

The nearest neighbor spacing distribution $p(s)$ is the observable
most commonly used to study the short--range fluctuations in the
spectrum. This function is equal to the probability density for two
{\it neighboring} levels $\xi_n$ and $\xi_{n+1}$ having the spacing
$s$. The function $p(s)$ involves all correlation functions $R_k$ with
$k\ge 2$ and differs from the two--level correlation function
$X_2(r)$.  The latter gives the probability of finding {\it any} two
levels at a distance $r$ from each other. Only for small arguments,
$p(s)$ is approximated by $X_2(s)$. The function $p(s)$ and its first
moment are normalized to unity,
\begin{equation}
\int_0^\infty p(s)ds = 1 \qquad {\rm and} \qquad
\int_0^\infty sp(s)ds = 1 \ .
\label{1ba7}
\end{equation}
For the uncorrelated or Poisson case, $p(s)=\exp(-s)$ while for the
harmonic oscillator, $p(s)=\delta(s-1)$. For the Gaussian ensembles,
the analytical calculation of $p(s)$ from the correlation functions
given in Sec.~\ref{qc1rc} is possible but highly
non--trivial~\cite{Meh91}. It leads to expressions involving infinite
\begin{figure}
\centerline{
\psfig{file=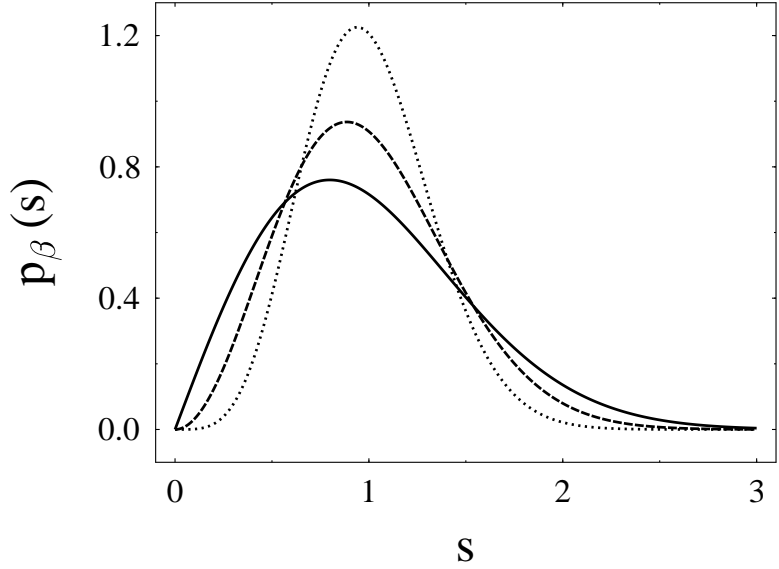,width=4.5in,angle=-90}
}
\caption{
  The Wigner surmises $p_\beta(s)$ for the nearest neighbor spacing 
  distribution. The solid line is the result for orthogonal symmetry
  ($\beta=1$), the dashed line is the result for unitary symmetry
  ($\beta=2$) and the dotted line is the result for symplectic symmetry
  ($\beta=4$). We notice the importance of the repulsion law $s^\beta$
  for small spacings. 
}
\label{figft3}
\end{figure}
products. An excellent approximation is given by the Wigner surmises
which are the exact ``spacing distribution'' for a $2\times 2$ matrix
model~\cite{Boh69} with Gaussian probability density function. In the
most general form the Wigner surmises reads 
\begin{equation}
p_\beta(s) = a_\beta s^\beta \exp\left(-b_\beta s^2\right)
\label{1ba12}
\end{equation}
for all three symmetry classes with $\beta=1,2,4$. The level repulsion
factor $s^{\beta}$ reflects the Vandermonde determinant in the
$N$--level probability density~(\ref{1rc13}). The constants
\begin{equation}
a_\beta = 2 \frac{\Gamma^{\beta+1}((\beta+2)/2)}
                 {\Gamma^{\beta+2}((\beta+1)/2)}
\qquad {\rm and} \qquad 
b_\beta = \frac{\Gamma^2((\beta+2)/2)}
               {\Gamma^2((\beta+1)/2)}
\label{1ba13}
\end{equation}
are explicitly given by $a_1=\pi/2$, $b_1=\pi/4$ (GOE),
$a_2=32/\pi^2$, $b_2=4/\pi$ (GUE), and $a_4=262144/729\pi^3$,
$b_4=64/9\pi$ (GSE), respectively. These functions $p_\beta(s)$
are shown in Fig.~\ref{figft3}. The Gaussian fall--off of $p(s)$
with large $s$ is unrelated to the assumed Gaussian probability
density of the three ensembles. Because of universality, the spacing
distribution~(\ref{1ba12}) is valid for a wide class of
probability densities, see Sec.~\ref{univ}.

A statistical argument due to Wigner~\cite{Wig56} leads to an
interesting heuristic formula for the spacing distribution,
\begin{equation}
p(s) = \mu(s) \exp\left(-\int_0^s\mu(s^\prime)ds^\prime\right) \ .
\label{1ba10}
\end{equation}
The repulsion function $\mu(s)$ models the presence or absence of the
Vandermonde determinant in the eigenvalue probability density. For a
linear repulsion $\mu(s)=\pi s/2$ one obtains the Wigner surmise, for a
constant value $\mu(s)=1$, the Poisson distribution. Choosing
$\mu(s)=c_qs^q$ with $0\le q \le 1$ leads to the Brody
distribution~\cite{Bro73},
\begin{equation}
p_q(s) = c_q s^q \exp\left(-\frac{c_q}{q+1}s^{q+1}\right)
\qquad {\rm with} \qquad c_q = \frac{\Gamma^{q+1}(1/(q+1))}{q+1}  \ .
\label{1ba11}
\end{equation}
In the case of orthogonal symmetry ($\beta=1$), this distribution
interpolates between the Poisson case and the Wigner surmise
(\ref{1ba12}).  The free parameter $q$ serves as a purely
phenomenological measure for the degree of mixing between Poisson and
GOE statistics.  Izrailev~\cite{Izra88,Izr90} proposed a different
phenomenological interpolation formula,
\begin{equation}
p_{\beta_{\rm eff}}(s) = A_{\beta_{\rm eff}} s^{\beta_{\rm eff}} 
         \exp\left(-\frac{\pi^2\beta_{\rm eff}}{16}s^2
         -\left(B_{\beta_{\rm eff}}-\frac{\pi\beta_{\rm eff}}{4}
         \right)s\right) \ .
\label{1ba11a}
\end{equation}
The parameter $\beta_{\rm eff}$ varies smoothly between zero, the
Poisson repulsion factor, and four, which applies to symplectic
symmetry. The normalization constants $A_{\beta_{\rm eff}}$ and
$B_{\beta_{\rm eff}}$ have to be determined from Eqs.~(\ref{1ba7}).
Caurier {\it et al.}~\cite{Cau90} and Lenz and Haake~\cite{Len91} 
calculated yet another interpolation formula for the spacing distribution 
starting from two--dimensional matrix models.  

Whenever the agreement of one of the distributions $p(s)$ with data or
numerical calculations is tested, it is helpful to use the integral
\begin{equation}
F(s) = \int_0^s p(s^\prime) ds^\prime 
\label{1ba14}
\end{equation}
which is referred to as the cumulative spacing distribution.

\subsubsection{Long--range spectral observables}
\label{qc1bac}

The nearest neighbor spacing distribution contains information about
the spectrum which involves short scales (a few mean level spacings). 
Long--range correlations are measured by quantities such as the level
number variance $\Sigma^2(L)$ and the spectral rigidity $\Delta_3(L)$. 
The level number variance is given by
\begin{equation}
\Sigma^2(L) = \langle \widehat{\eta}^2(L,\xi_s) \rangle
                 - \langle \widehat{\eta}(L,\xi_s) \rangle^2 \ .
\label{1ba15}
\end{equation}
Here, $\widehat{\eta}(L,\xi_s)$ counts the number of levels in the
interval [$\xi_s, \xi_s + L$] on the unfolded scale. The angular bracket
in Eq.~(\ref{1ba15}) denotes the average over starting points
$\xi_s$. By construction, i.e. unfolding, one has
$\langle\widehat{\eta}(L,\xi_s)\rangle=L$. Thus, in an interval of
length $L$ one expects on average $L\pm \sqrt{\Sigma^2(L)}$ levels. 
The spectral rigidity, introduced by Dyson and Mehta~\cite{Dys63}, is
closely related to $\Sigma^2(L)$. In an interval of length $L$. it is
defined as the least square deviation of the staircase function
$\widehat{\eta}(\xi)$ from the best fit to a straight line,
\begin{equation}
\Delta_3(L) = \frac{1}{L}\left\langle \min_{A,B} 
               \int_{\xi_s}^{\xi_s+L} 
                \left(\widehat{\eta}(\xi)-A\xi-B\right)^2 d\xi
                   \right\rangle \ .
\label{1ba16}
\end{equation}
The angular bracket is defined as in Eq.~(\ref{1ba15}). The
definition~(\ref{1ba16}) is very natural since by construction, the
smooth part of the staircase function $\widehat{\eta}(\xi)$ is $\xi$
itself.  The bottom part of Fig.~\ref{figft8} illustrates the meaning
of the spectral rigidity $\Delta_3(L)$.  Since the few stairs below
702~kHz come quite regularly, there is only a small root--mean--square
deviation from the thin line describing the smooth behavior.  This
implies a small contribution to $\Delta_3(L)$. Some of the levels
above 702~kHz are nearly degenerate which makes the succession of the
stairs rather irregular. Thus, the deviation from the smooth part is
larger, yielding a larger contribution to $\Delta_3(L)$. We notice
that the staircase function in Fig.~\ref{figft8} is shown on the
original scale prior to unfolding. However, since the section in the
bottom part is small, the smooth part is almost a straight line and
the statements just made apply qualitatively also to the unfolded
case. For fixed $\xi_s$, Bohigas and Giannoni~\cite{Boh75,Boh84} have
given a very useful procedure to obtain $\Delta_3(L)$ from a measured
or calculated spectrum.  A proper error analysis for $\Sigma^2(L)$ and
$\Delta_3(L)$ is non--trivial because the results for different values
of $L$ are strongly correlated. A consistent method of estimating
errors was given by Shriner and Mitchell~\cite{Shr92}.

In contrast to the spacing distribution, both the number variance and
the spectral rigidity probe only two--level correlations. Indeed,
$\Sigma^2(L)$ can be reduced to the form~\cite{Boh84}
\begin{equation}
\Sigma^2(L) = L - 2\int_0^L (L-r) Y_2(r) dr \ , 
\label{1ba18}
\end{equation}
provided the spectrum is translation invariant. As a consequence the
number variance is related to the two--level form factor $b_2(t)$
defined in Eq.~(\ref{1rc38}),
\begin{equation}
\lim_{L\to\infty}\frac{1}{L}\Sigma^2(L) = 1 - b_2(0) \ .
\label{1ba18a}
\end{equation}
For the Poisson spectrum without correlations and for the harmonic
oscillator one obtains $\Sigma^2(L) = L$ and $\Sigma^2(L) = 0$,
respectively. The results for the Gaussian ensembles lie between these
limiting cases. For large $L$ and $\beta=1,2,4$ one finds the following
approximations~\cite{Dys63,Meh91}, valid to order $1/L$,
\begin{eqnarray}
\Sigma^2_1(L) &=& \frac{2}{\pi^2}
       \left( \ln(2\pi L) + \gamma + 1 - \frac{\pi^2}{8} \right)
                           \nonumber\\
\Sigma^2_2(L) &=& \frac{1}{\pi^2}
       \left( \ln(2\pi L) + \gamma + 1 \right)
                           \nonumber\\
\Sigma^2_4(L) &=& \frac{1}{2\pi^2}
       \left( \ln(4\pi L) + \gamma + 1 + \frac{\pi^2}{8} \right) \ .
\label{1ba20}
\end{eqnarray}
Here $\gamma=0.5772\ldots$ is Euler's constant. These approximations
display the famous logarithmic behavior and the increasing stiffness
of the spectrum with increasing $\beta$. Exact expressions can be
found, for example in Sec.~V~C of Ref.~\onlinecite{Bro81}, they are shown in
\begin{figure}
\centerline{
\psfig{file=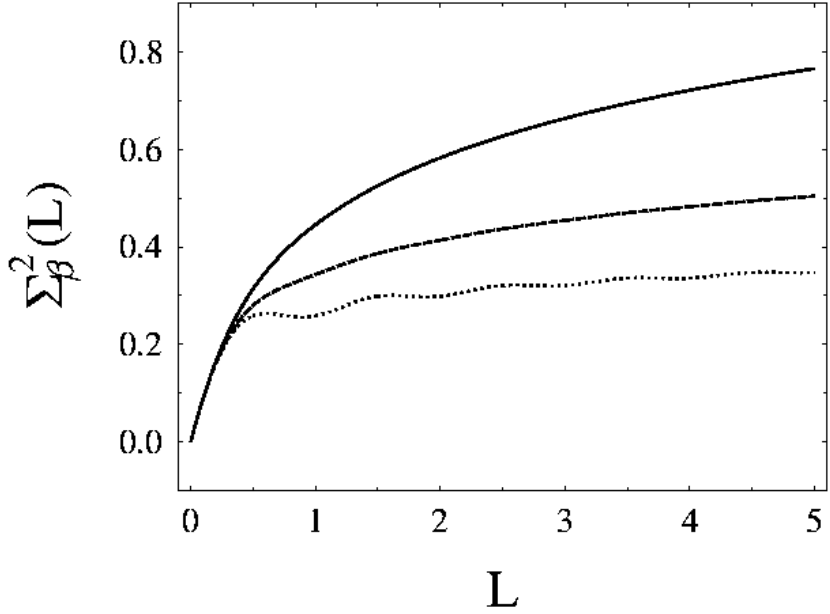,width=4.5in,angle=-90}
}
\caption{
  The level number variances $\Sigma^2_\beta(L)$ for the Gaussian
  ensembles. These are the exact results obtained from the 
  integral~(\protect\ref{1ba18}). The solid line is the GOE result 
  ($\beta=1$), the dashed line is the GUE result ($\beta=2$) and 
  the dotted line is the GSE result ($\beta=4$). We notice the 
  oscillations in $\Sigma^2_4(L)$ which are not contained in the 
  logarithmic approximation of Eq.~(\protect\ref{1ba20}).
}
\label{figft4}
\end{figure}
Fig~\ref{figft4}. There is a minor misprint in Eq.~(5.13) of
Ref.~\onlinecite{Bro81} for $\Sigma^2_4(L)$, the argument of
$\Sigma^2_2$ on the right hand side of this equation should read $2r$
instead of $r$.  Moreover, we notice that there are visible
oscillatory modulations in $\Sigma^2_4(L)$ which are not borne out in
the approximation~(\ref{1ba20}).

The spectral rigidity can similarly be expressed as an integral over
the two--point function, and the minimization in Eq.~(\ref{1ba16})
can be done analytically~\cite{Dys63,Pan79,Meh91}. For a spectrum
which is invariant under a continuous set of transformations, one
arrives at 
\begin{equation}
\Delta_3(L) = \frac{L}{15} - \frac{1}{15L^4} 
            \int_0^L(L-r)^3 (2L^2-9Lr-3r^2) Y_2(r) dr 
\label{1ba21}
\end{equation}
which shows a close formal similarity to Eq.~(\ref{1ba18}). 
Alternatively, one can write~\cite{Pan79}
\begin{equation}
\Delta_3(L) = \frac{2}{L^4} \int_0^L (L^3-2L^2r+r^3) \Sigma^2(r) dr \ .
\label{1ba22}
\end{equation}
This equation can be viewed as an integral transform of the level
number variance. Remarkably, the parabola is in the kernel of this
transform, i.e.~inserting $\Sigma^2(r)=r^2$ yields $\Delta_3(L)=0$.
For the limiting cases of the Poisson spectrum and the harmonic
oscillator, one has $\Delta_3(L) = L/15$ and $\Delta_3(L) = 1/12$,
respectively. The results for the Gaussian ensembles can be found by
numerical integration. For large $L$ and $\beta=1,2,4$, one has to
order $1/L$,
\begin{eqnarray}
\Delta_{3,1}(L) &=& \frac{1}{\pi^2}
       \left( \ln(2\pi L) + \gamma - \frac{5}{4} 
                                       - \frac{\pi^2}{8} \right)
                           \nonumber\\
\Delta_{3,2}(L) &=& \frac{1}{2\pi^2}
       \left( \ln(2\pi L) + \gamma - \frac{5}{4} \right)
                           \nonumber\\
\Delta_{3,4}(L) &=& \frac{1}{4\pi^2}
       \left( \ln(4\pi L) + \gamma - \frac{5}{4} 
                                       + \frac{\pi^2}{8} \right) \ .
\label{1ba24}
\end{eqnarray}
These approximations show once more the close relation between
$\Delta_{3,\beta}(L)$ and the level number variance 
$\Sigma^2_\beta(L)$. 

We do not address here observables which probe three--level,
four--level or even higher--order correlations. We refer the reader to
Refs.~\onlinecite{Dys63,Boh84}. In real systems, non--generic and
system--specific properties exist which are not described by Random
Matrix Theory. These show up on large scales and are discussed in
later sections.

\subsubsection{Fourier transforms}
\label{qc1bad}

Extracting the positions of the levels from a measured spectrum can be
difficult for at least two reasons. Either the resolution is so poor
that the extracted sequence is incomplete, or the data set is so large
that the task of finding all levels is simply too tedious. In these
cases it is desirable to introduce a statistical observable which is
directly related to the measured spectrum and in some way separates
the statistics of the level positions from the fluctuations due to the
line shapes of the levels. The correlation--hole method introduced by
Leviandier {\it et al.}~\cite{Lev86} is a first successful step in this
direction. 

We consider a measured spectrum $I(\xi)$ on the unfolded scale $\xi$. 
The observable of interest is the decay function, i.e.~the square
modulus of the Fourier transform of the spectrum
\begin{equation}
      |c(t)|^2 = \Bigg|\int\limits_{-\infty}^{+\infty} d\xi \,
                 I(\xi) \, \exp(2\pi i \xi t)\Bigg|^2 \ .
\label{1ba29}
\end{equation} 
The Fourier coordinate $t$ defines an unfolded time. 
Expression~(\ref{1ba29}) can be rewritten as the Fourier
transform 
\begin{equation}
|c(t)|^2 = \int\limits_{-\infty}^{+\infty} A(r) \exp(2\pi irt) dr
\label{1ba30}
\end{equation}
of the autocorrelation function
\begin{equation}
A(r) = \int\limits_{-\infty}^{+\infty} I(R-r/2) I(R+r/2) dR \ .
\label{1ba31}
\end{equation}
This autocorrelation function still contains non--generic information
about the system. To obtain the generic statistical properties, $A(r)$
or, equivalently, $|c(t)|^2$ have to be smoothed properly~\cite{Del91}. 
The resulting function $\langle A(r) \rangle$ is generically
translation invariant and can be related to the two--level cluster
function $Y_2(r)$ of Random Matrix Theory. The decay function
$|c(t)|^2$ is correspondingly related to the two--level form factor
$b_2(t)$ and at short times reflects long--range spectral
correlations, see Eq.~(\ref{1ba18a}). 

To be definite we discuss the model spectrum
\begin{equation}
I(\xi) = \sum_{n=1}^N y_n L(\xi-\xi_n) 
\label{1ba32}
\end{equation}
studied in Ref.~\onlinecite{Alt97} and investigated in a similar form
in Refs.~\onlinecite{Lev86,Guh90b,Lom91,Lom93}. Here, the levels
$\xi_n, \ n=1,\ldots,N$ on the unfolded scale have a common line shape
$L(\xi)$ and random intensities $y_n$, $N$ is assumed to be a very
large number. For non--negative times the decay function is given
by~\cite{Lev86,Guh90b,Lom91,Lom93,Alt97}
\begin{equation}
|c(t)|^2 = N\overline{y^2} \delta(t) 
  + N\overline{y^2} \left(1-\alpha b_2(t)\right) |\tilde{L}(t)|^2 \ ,
\label{1ba33}
\end{equation}
with $\alpha=\overline{y}^2/\overline{y^2}$ the ratio of the first two
moments of the intensity distribution. The function $\tilde{L}(t)$ is
the Fourier transform of the line shape $L(\xi)$. For a Lorentzian
line shape, $|\tilde{L}(t)|^2$ reflects the exponential decay of the
resonances. If $L(\xi)=\delta(\xi)$ and if all intensities $y_n$ are
unity, the decay function $|c(t)|^2$ equals $1-b_2(t)$ for non--zero
$t$. For an uncorrelated Poisson spectrum we have $b_2(t)=0$ and the
decay function is constant. In a correlated spectrum derived from one
of the Gaussian ensembles the decay function approaches zero with
vanishing $t$. This behavior is referred to as the correlation hole.
If the intensities are random numbers, the correlation hole is
described by the function $1-\alpha b_2(t)$ and does not approach
zero. In particular, if the intensity distribution is the
Porter--Thomas distribution (see Sec.~\ref{qc1wwa}), one has
$\alpha=1/3$. The $\delta$ function in Eq.~(\ref{1ba33}) occurs since
we assumed $N$ to be very large. For a finite number of levels, this
contribution will acquire a width, see
Refs.~\onlinecite{Guh90b,Lom93}.

The correlation--hole method separates the statistics of the level
positions from that of the intensities and line shapes. This is its
main advantage. For realistic spectra, i.e. with inclusion of the line
widths, the theory of the correlation hole was worked out in
Ref.~\onlinecite{Guh90b,Har91}. There, a scattering model was used.
The application of the Fourier transform to statistical spectra is
summarized and a qualitative discussion given in
Ref.~\onlinecite{Lom91}. A short version of the theory of the
correlation hole is presented in Appendix A of
Ref.~\onlinecite{Lom93}. In Ref.~\onlinecite{Alh93}, the correlation
hole is related to its classical analogue, the survival probability.

\subsubsection{Superposition of independent spectra}
\label{qc1bae}

We consider a system with a set of good quantum numbers such as spin
and parity. In such a system, the Hamiltonian $H$ assumes
block--diagonal form, $H={\rm diag}(H_1,H_2,\ldots,H_M)$. So far we
have always dealt with the spectrum generated by only one of the
sub--blocks $H_m$, $m=1,\ldots,M$. Now we consider the superposition
of $M$ such spectra.  The total level density $R_1(E)$ is simply the
sum of the $M$ sub--block densities, $R_1(E)=\sum_{m=1}^MR_{1m}(E)$.
We assume that these $M$ densities have roughly the same energy
dependence, and that the statistical measures for the $M$ individual
spectra are known. How can we calculate the corresponding statistical
measures for the total spectrum? For the spacing distribution, the
answer is already contained in the seminal article by Rosenzweig and
Porter~\cite{Rose60} of 1960, see also Mehta's book~\cite{Meh91}. The
unfolded observables are expressed in units of the total mean level
spacing $D$. Assuming that energy interval of interest is centered
around the origin $E=0$, we have $D=1/R_1(0)$. We also introduce the
fractional densities $g_m=R_{1m}(0)/R_1(0)$ with $\sum_{m=1}^Mg_m=1$.
Let $p_m(g_ms), \ m=1,\ldots,M$ be the nearest neighbor spacing
distribution of the $m$--th sub--spectrum. No assumption is made on the
form of the distributions $p_m(g_ms)$. The spacing distribution $p(s)$
of the superposition of the $M$ spectra is found to be
\begin{eqnarray}
p(s) &=& \frac{d^2}{ds^2} E(s)
   = E(s) \Bigg(\sum_{m=1}^M g_m^2\frac{p_m(g_ms)}{E_m(g_ms)} +
                                       \nonumber\\
 & & \qquad \left(\sum_{m=1}^M g_m\frac{1-F_m(g_ms)}{E_m(g_ms)}\right)^2
          - \sum_{m=1}^M\left( g_m\frac{1-F_m(g_ms)}{E_m(g_ms)}\right)^2
                                                          \Bigg) \ ,
\label{1ba25}
\end{eqnarray}
where $E(s) = \prod_{m=1}^M E_m(g_ms)$ and
\begin{equation}
F_m(g_ms) = \int_0^{g_ms} p_m(s^\prime) ds^\prime \ ,
\qquad 
         E_m(g_ms) = \int_{g_ms}^\infty (1-F_m(s^\prime)) ds^\prime \ .
\label{1ba26}
\end{equation}
Whenever the long--range spectral observables of interest are pure
two--point measures, it is easy to relate those of the full spectrum
and those of the $M$ individual spectra. Let $Y_{2m}(g_mr)$ be the
two--level cluster function, $\Sigma_m^2(g_mL)$ the level number
variance, and $\Delta_{3m}(g_mL)$ the spectral rigidity of the $m$-th
sub--spectrum, respectively. These observables are completely arbitrary.
Just like the $p_m(g_ms)$, they may coincide with but are not
restricted to any of the specific forms presented in previous
sections. {}From the definition of the two--point correlation
function, one finds immediately for the total two--level cluster
function
\begin{equation}
Y_2(r) = \sum_{m=1}^M g_m^2 Y_{2m}(g_mr) \ .
\label{1ba27}
\end{equation}
It then follows directly from Eqs.~(\ref{1ba18}),~(\ref{1ba21})
and~(\ref{1ba22}) that 
\begin{equation}
\Sigma^2(L) = \sum_{m=1}^M \Sigma_m^2(g_mL)
\qquad {\rm and} \qquad
\Delta_3(L) = \sum_{m=1}^M \Delta_{3m}(g_mL) \ .
\label{1ba28}
\end{equation}
The first property~(\ref{1ba28}) is not surprising because
$\Sigma^2(L)$ is a variance. All results given here apply only to $M$
strictly  {\it non--interacting} spectra. Crossover transitions due to
the breaking of a quantum number are discussed later.

Very important for the analysis of data is the superposition of a
large number $M$ of spectra. For all individual distributions
$p_m(g_ms)$ which play a role in realistic spectra the superposition
$p(s)$ tends rather quickly towards the Poisson limit.  The
superposition of only six or so spectra which individually follow the
Wigner surmises yields a spacing
\begin{figure}
\centerline{
\psfig{file=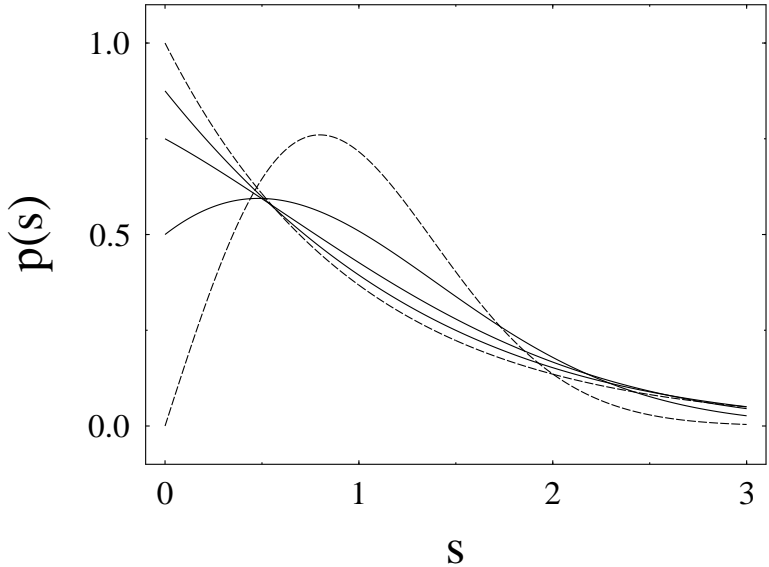,width=4.5in,angle=-90}
}
\caption{
  Nearest neighbor spacing distribution $p(s)$ as given in
  Eq.~(\protect\ref{1ba25}) for the
  superposition of $M$ spectra with equal fractional densities
  $g_m=1/M, \ m=1,\ldots,M$. Each individual spectrum follows the
  Wigner surmise. One has $p(0)=1-1/M$ at $s=0$. The cases $M=2,4,8$
  are shown as solid lines. The Wigner surmise and the Poisson 
  distribution are drawn as dashed lines.
}
\label{figft5}
\end{figure}
distribution which is experimentally hard to distinguish from a
Poisson distribution. This is illustrated in Fig.~\ref{figft5} for
spectra with equal fractional densities $g_m=1/M, \ m=1,\ldots,M$.  In
this case, formula~(\ref{1ba25}) implies $p(0)=1-1/M$, reflecting the
increasing number of degeneracies. For the interval lengths $L$ which
are usually available for the analysis of data, a similar statement
applies to $\Sigma^2(L)$ and $\Delta_3(L)$ as superpositions of the
relevant individual distributions $\Sigma_m^2(g_mL)$ and
$\Delta_{3m}(g_mL)$.  There are of course exceptions such as, for
example, the superposition of pure harmonic oscillator spectra.

The property~(\ref{1ba27}) of the two--level cluster function can be
used to estimate the number $M$ of superimposed independent spectra,
i.e.~of symmetries in a measured spectrum. This is important in
situations where an assignment of the set of quantum numbers to every
individual level in the spectrum is impossible or too tedious. In a
molecular physics context, Leviandier {\it et al.}~\cite{Lev86} argued that
the decay function $|c(t)|^2$ introduced in Sec.~\ref{qc1bad} is
rescaled by a factor $1/M$ so that the correlation hole becomes
narrower by a factor $M$. In a more general discussion \cite{Guh90b}, a
superposition of $M$ independent spectra was considered. Each spectrum
has a line shape $L_m(\xi)$ and a set of intensities with first and second
moments $\overline{y_m}$ and $\overline{y_m^2}$. One finds
\begin{equation}
|c(t)|^2 = f_0\delta(t) + f_1\sum_{m=1}^M g_m \overline{y_m^2} 
             \left(1-\alpha_m b_2(t/g_m)\right) |\tilde{L}_m(t)|^2 
\label{1ba28a}
\end{equation}
where $f_0$ and $f_1$ are system--specific constants, and where
$\alpha_m=\overline{y_m}^2/\overline{y_m^2}$. If all $M$ mean level
densities $R_{1 m}(E)$ are roughly equal, one has $g_m\simeq 1/M$, and
the correlation hole is indeed narrowed by a factor $M$.

Finally, we mention that Berry and Robnik proposed a spacing
distribution supposed to describe the transition from Poisson
behavior to the Wigner surmise.  This distribution is obtained from
formula~(\ref{1ba25}) by taking $M=2$ spectra, one satisfying the
Poisson spacing distribution, the other one, the Wigner surmise.
Strictly speaking this formula is only applicable to two
non--interacting spectra. It is nonetheless often used as a
phenomenological guess for mixed systems.

\subsubsection{GSE test of GOE data}
\label{qc1baf}

We consider a GOE spectrum $\{E_1,E_2,E_3,...\}$ with $E_n \leq
E_{n+1}$.  We divide the sequence $\{E_1,E_2,E_3,...\}$ into two
sequences with odd and even indices, i.e.~into $\{E_1,E_3,E_5,...\}$
and $\{E_2,E_4,E_6,...\}$ and consider each as a new spectrum. Then, a
theorem due to Mehta and Dyson~\cite{Meh63} states that after proper
unfolding, each of the two spectra obeys GSE statistics. This theorem
was recently used by Lombardi, Bohigas and Seligman~\cite{Lom94} for a
GSE test of GOE data. In this test, the two--level form factor
$b_2(t)$ and the correlation hole $|c(t)|^2$ are particularly useful
observables. In contrast to the GOE case, the GSE form factor has a
singularity at $t=1$ caused by the oscillatory structure of the
two--level cluster function $Y_2(r)$. Thus, one expects a
characteristic peak at $|t|=1$ which gives information on correlations
on the scale of about two mean level spacings. As pointed out in
Ref.~\onlinecite{Lom94}, the two--level observables calculated for the
two subsequences contain information on higher level correlations ($k
> 2$) of the original spectrum.

\subsubsection{Ergodicity}
\label{qc1bag}

In Random Matrix Theory, observables are calculated by averaging over
an ensemble of Hamiltonians. This average is indicated by a bar. 
Experimentally, the observables are calculated as running averages
over part of the spectrum of a given system, containing $N_I$ levels in
the interval $[E, I+E]$. This average is indicated by an angular
bracket. For any function $F(E)$ of energy $E$, it is given by  
\begin{equation}
\langle F(E) \rangle_I = \frac{1}{I} 
                         \int_E^{E+I} F(E^\prime) dE^\prime
                       = \frac{1}{N_I} 
                         \int_0^{N_I} F(E+Dr) dr
                       = \langle F \rangle_{N_I} 
\label{1ba34}
\end{equation}
where $D$ denotes the mean level spacing. A comparison of the results
of both procedures is meaningful only if they are equivalent, i.e.~if 
\begin{equation}
\overline{F(E)} = \langle F(E) \rangle
\label{1ba34a}
\end{equation} 
In this case, one speaks of ergodicity. Mathematical discussions of
ergodicity can be rather involved. Here, we only summarize a very
instructive discussion by French, Mello and Pandey~\cite{Fre78}. It is
assumed that the spectrum is generic so that after proper unfolding
averaging will give translation invariant results. As a test of
ergodicity, French {\it et al.} consider the variance
\begin{equation}
{\rm var}\langle F \rangle_{N_I} =
                 \overline{|\langle F -\overline{F}\rangle |^2} \ .
\label{1ba35}
\end{equation}
This quantity is the ensemble average of the squared difference
between running and ensemble average. It can be cast into a form
similar to Eq.~(\ref{1ba18}) for the level number variance,
\begin{equation}
{\rm var}\langle F \rangle_{N_I} =
   \frac{1}{I^2} \int_{-I}^{+I} (I-|\varepsilon|)
                          C_F(\varepsilon) d\varepsilon
   = \frac{1}{N_I^2} \int_{-N_I}^{+N_I} (N_I-|r|) C_F(Dr) dr \ .
\label{1ba36}
\end{equation}
The auto-covariance function 
$C_F(\varepsilon) =
\overline{(F^*(E)-\overline{F}^*) (F(E+\varepsilon)-\overline{F})}$
is, after proper unfolding, a translation invariant two--point
function independent of $E$. With the help of a result of the theory
of random functions~\cite{Yag62}, Eq.~(\ref{1ba36}) leads to
\begin{equation}
\lim_{r\to\infty} C_F(Dr) = 0
\qquad \Longrightarrow \qquad
\lim_{N_I\to\infty} {\rm var}\langle F \rangle_{N_I} = 0 \ .
\label{1ba38}
\end{equation}
or equivalently
\begin{equation}
\lim_{r\to\infty} C_F(Dr) = 0
\qquad \Longrightarrow \qquad
\lim_{N_I\to\infty} \langle F \rangle_{N_I} = \overline{F}  \ .
\label{1ba39}
\end{equation}
This is the desired ergodic behavior for almost all members of the
ensemble, except for a set of measure zero. We note that ergodicity is
attained only for a large, preferably infinite, number $N_I$ of
levels. The result~(\ref{1ba38}) can be generalized fairly easily: The
$k$--point function is ergodic if in the limit of infinite energy
differences. the ensemble--averaged $2k$--point function vanishes. We
have made no specific assumptions concerning $F(E)$. Thus, the
argument just presented also applies to correlations of scattering
matrix elements and all other level correlators of Random Matrix
Theory.

\subsection{Introduction to supersymmetry}
\label{susy}

The description of all technical aspects of supersymmetry would be 
far beyond the scope of this review. We attempt to give the interested 
reader an idea of the main elements of the method. For further reading, 
we mention Berezin's book \cite{Ber87} where mathematical aspects of  
superanalysis are described, see also Ref.~\onlinecite{Rot87}. Several
papers and review articles explain the use of the standard 
supersymmetry method in problems of chaos and disorder 
\cite{Efe83,Ver85,Plu95,Zuk96}, and Efetov \cite{Efe96} has 
recently devoted a book to the subject.  The term ``supersymmetry''
was first introduced by Wess and Zumino in relativistic field theory
\cite{Wes74}. In the present context, supersymmetry does involve a
symmetry between bosonic and fermionic integration variables but lacks
the invariance properties of a relativistic field theory. Perhaps more
importantly, the fermionic variables lack any physical meaning and are
introduced merely as bookkeeping devices. Wherever applicable, the
supersymmetry method has been found to be more powerful than the
replica trick: It yields exact results where the replica trick only
yields asymptotic expansions.  The price one has to pay consists in
using analysis on supermanifolds.  The replica trick, on the other
hand, finds a much wider range of applications. 

In Sec.~\ref{susya} we derive the supersymmetric representation 
of the generating function for level correlators. In Sec.~\ref{susyb}, 
the saddle--point approximation is used to derive the standard 
supersymmetric non--linear $\sigma$ model. An alternative technique 
which avoids the saddle--point approximation and is useful for purely 
spectral observables is discussed in Sec.~\ref{qc1tcb}.

\subsubsection{Supersymmetric representation of the
               generating function}
\label{susya}

We develop the method for an observable given by the product of the
retarded and advanced propagators tr$(E_1^{+} - H)^{-1}$ and
tr$(E_2^{-} - H)^{-1}$, with $H$ a member of the GUE, and $E_p^{\pm} =
E_p \pm i\eta$ with $\eta$ positive infinitesimal and $p = 1,2$.  It
is highly non--trivial to average a product of propagators over an
ensemble of matrices $H$ since $H$ appears in the denominators. In
Sec.~\ref{qc1rc}, we have shown that the Mehta--Dyson method
accomplishes this goal very elegantly in the case of spectral
correlations in the pure cases. If, however, an additional matrix
$\gamma$ appears in the propagator, rotation invariance in Hilbert
space is broken, and almost all of the convenient features of the
Mehta--Dyson method disappear. Scattering systems require such an
addition to the Hamiltonian to describe the coupling to the channels,
see Sec.~\ref{qc1cs}. Prior to the advent of supersymmetry, one was
restricted to perturbative or asymptotic expansions in these
situations, see Refs.~\onlinecite{Bro81,Mah80}.

To keep the presentation transparent, we do not discuss scattering
systems and restrict ourselves to the spectral two--point correlation
function, even though in this particular case, the supersymmetry
method only reproduces the results obtained earlier by Dyson and
Mehta.  We emphasize again that the supersymmetry method develops its
full power in those cases where the Mehta--Dyson method does not
apply. Later, we indicate the modifications for other cases (more than
two propagators, different symmetry of $H$, propagators in higher
dimension). Our choice of observable is dictated by the fact that the
occurrence of a pair of complex conjugate propagators is typical of
many problems. Our notation is that of Ref.~\onlinecite{Ver85}. We use
the form $\lambda^2/N$ for the variance of the Gaussian distribution
of $H$, see Sec.~\ref{qc1rcb}.

To perform the ensemble average exactly, i.e.~non--perturbatively,
one expresses the propagator as the derivative of a generating function
$F_p(J_p)$ with respect to a source variable $J_p$ such that
\begin{equation}
{\rm tr}\frac{1}{E_p^\pm - H} = 
             \frac{\partial}{\partial J_p}F_p(J_p)\Bigg|_{J_p=0} \ .
\label{susy1}
\end{equation}
It is easily seen that the choice 
\begin{equation}
F_p(J_p) = \frac{{\rm det}(E_p^\pm -H +J_p)}{{\rm det}(E_p^\pm -H -J_p)}
\label{susy2}
\end{equation}
represents such a generating function. We observe that the {\it
  logarithms} of the numerator and the denominator separately generate
the propagator, too. In contrast to the latter, however, the function
$F_p(J_p)$ has the very useful property of being normalized to unity,
i.e.~to a constant independent of $H$: We have $F_p(0)=1$. Still, an
exact average of $F_p(J_p)$ over the ensemble seems out of question.
This is the point where supersymmetry enters. The inverse of an
$N\times N$ determinant can be expressed as the Gaussian integral over
an $N$ component vector $S_p$ of complex commuting variables, while
the determinant itself is given as the Gaussian integral over an $N$
component vector $\chi_p$ of complex anticommuting variables. Thus we
have
\begin{eqnarray}
  F_p(J_p) &=& \int d[S_p] \exp\left( iS_p^\dagger(E_p^\pm -H
                    -J_p)S_p\right) \nonumber\\ 
           & & \qquad \int d[\chi_p] \exp\left(
  i\chi_p^\dagger(E_p^\pm -H +J_p)\chi_p\right) \ .
\label{susy3}
\end{eqnarray}
Calculating the average over the Gaussian ensemble is now straightforward
since $H$ appears in the exponent. 

With these tools, we can derive the generating function of the
two--point function. It is convenient to write $E=(E_1+E_2)/2$ and
$\varepsilon=E_1-E_2$. To be consistent with most of the literature,
we use the symbol $\varepsilon$ throughout this section and the symbol
$\omega$ in its stead in Sec.~\ref{disorder}. We already know that the
unfolded two--point function will only depend on $\varepsilon/D$ where
$D$ is the mean level spacing. We have to take into account that the
imaginary increment $i\eta$ lies on different sides of the real axis
for $p=1$ and $p=2$. To ensure convergence of the integrals one
proceeds as follows. We define a metric $L={\rm diag}(+1,+1,-1,-1)$,
reflecting the location of $i\eta$, and the direct product ${\bf
  L}=L\otimes 1_N$. Here, $1_N$ denotes the $N\times N$ unit matrix.
Moreover we set $J={\rm diag}(-J_1,+J_1,-J_2,+J_2)$ and introduce the
supervector $\Psi = (S_1,\chi_1,S_2,\chi_2)^T$. The proper generating
function is then given by
\begin{eqnarray}
F_1(J_1)F_2(J_2) &=& \int d[\Psi] \exp\biggl( i\Psi^\dagger 
                   {\bf L}^{1/2} \biggl(1_4\otimes(E-H)
                                         \nonumber\\ 
                 & & \qquad\qquad
       + \left(\frac{\varepsilon}{2} +i\eta\right)L\otimes 1_N
       + J\otimes 1_N\biggl) {\bf L}^{1/2}\Psi\biggl) \ .
\label{super3}
\end{eqnarray}
Some authors prefer a different form for expression~(\ref{super3}): 
The symmetric arrangement of the two factors $L^{1/2}$ is dropped in 
favor of a single factor $L$ appearing right behind $\Psi^{\dagger}$. 

For an observable containing $k > 2$ propagators (usually an equal
number of advanced and of retarded propagators), the {\it form} of
Eq.~(\ref{super3}) would be the same, but the dimension of the vectors
$\Psi$ and of the supermatrices would be $2kN$ rather than $4N$. The
dimensions of the supermatrices and supervectors and their internal
symmetries would change for real symmetric or quaternion (rather than
Hermitean) matrices $H$. The ``source term'' $J\otimes 1_N$ must be
altered, if matrix elements (rather than traces) of the propagators
are considered.  For Hamiltonians in $d > 0$ dimensions, the
four--component super vectors $\Psi_n^{\dagger}$ and $\Psi_n$ with
$n=1,\ldots,N$ are replaced by space--dependent vectors $\Psi(\vec
r)^{\dagger}$ and $\Psi(\vec r)$, respectively, and the bilinear forms
in the matrix space of $H$ are replaced by integrations over
$d$--dimensional space. The generating function becomes a generating
functional, and the theory turns into a field theory.

Only the $H$--dependent term in expression (\ref{super3}) is affected
by averaging over the GUE. It is easy to calculate
\begin{equation}
\int  d[H] P_{N2}(H) \exp\left( -i\Psi^\dagger (L\otimes H)\Psi \right)
     = \exp\left(\frac{1}{2N}{\rm trg} A^2 \right) 
\label{super4}
\end{equation}
where
\begin{equation}
A = i \lambda L^{1/2} \sum_{n=1}^N \Psi_n\Psi_n^{\dagger} L^{1/2}
\label{super7}
\end{equation}
is a four by four supermatrix. The
symbol ${\rm trg}$ denotes the supertrace. The term ${\rm trg} A^2$ in
the exponent is of fourth order in the integration variables contained
in $\Psi_n$, and a direct integration over these variables is,
therefore, impossible. This difficulty is overcome with the help of
the Hubbard--Stratonovich transformation. The fourth--order term in
the exponent is reduced to second order at the expense of introducing
a set of auxiliary integration variables,
\begin{equation}
\exp \left(\frac{1}{2N} {\rm trg} A^2 \right) = 
     \int d[\sigma] \exp \left(-\frac{N}{2}
        {\rm trg} (\sigma^2) - {\rm trg} (\sigma A) \right). 
\label{super5}
\end{equation}
The matrix $\sigma$ has essentially the same symmetries as the matrix
$iA$. The differential $d[\sigma]$ denotes the product of the
differentials of the independent matrix elements. The integration over
the $\Psi_n$ can now be done. The procedure sketched in 
Eqs.~(\ref{super4}) and~(\ref{super5}) can also be viewed as a double
Fourier transform, the first in ordinary, the second in 
superspace~\cite{Guh91}.

Collecting everything, we find for the ensemble average
$\overline{F_1(J_1)F_2(J_2)}$ of the generating function~(\ref{super3})
\begin{equation}
\overline{F_1(J_1)F_2(J_2)} = \int d[\sigma] \exp {\cal L}(\sigma)
\label{super6}
\end{equation}
where the ``Lagrangean'' ${\cal L}$ has the form
\begin{equation}
{\cal L}(\sigma) =  -\frac{N}{2} \ {\rm trg}
(\sigma^2) - N \ {\rm trg} \ln\left( E1_4 - \lambda  \sigma  + J +
                           \frac{\varepsilon}{2}L\right). 
\label{super8}
\end{equation}
In the case of a disordered system which might, for instance, be
described by a single--particle Hamiltonian $H = T + V$ where $T$ is
the kinetic energy operator, and $V$ is an impurity potential of the
form~(\ref{imp}), one obtains similarly a generating functional 
where $\sigma$ is a function of ${\vec r}$, and where the operator $T$
appears as argument under the logarithm. 

Equations~(\ref{super6},\ref{super8}) embody the first important
result of the supersymmetry technique: The ensemble average of
$F_1(J_1)F_2(J_2)$ has been calculated, and the resulting integral
expressed in terms of a small number of integration variables,
determined by the number of independent elements in the matrix
$\sigma$. The $4N$ original complex integration variables $\Psi$ have
disappeared.

The matrix $\sigma$ possesses the same dimension and the same
symmetries as the matrix $iA$. The matrix $iA$, in turn, comprises all 
the unitary invariants which can be constructed from the elements of 
each, equivalent, vector $\Psi_n$. Only such invariants survive the ensemble 
average. Hence, both $iA$ and $\sigma$ directly mirror the underlying 
unitary symmetry in superspace. Moreover, the dimension of ${\sigma}$ 
is as small as is consistent with both the form of the observable, and 
the symmetry of the problem. 

These statements apply equally to observables involving $k$--point
functions, see for example Ref.~\onlinecite{Guh91}, and to problems
with orthogonal~\cite{Ver85} or symplectic symmetry.  For Hamiltonians
in $d \geq 1$ dimensions, the matrix $\sigma$ attains a dependence on
a spatial variable $\vec r$. With proper modifications, the statements
just made carry over to this case, too, and the resulting expression
for the generating functional constitutes a genuine field
theory~\cite{Efe83}. In a conceptually important work,
Zirnbauer~\cite{Zirn96} recently extended the supersymmetry method to
to the circular ensembles, i.e.~to systems with unitary disorder.

\subsubsection{Saddle--point approximation and non--linear
  $\sigma$ model}
\label{susyb}

The second important step of the supersymmetry technique
consists in an approximate evaluation of the remaining integrations in
the ensemble--averaged supersymmetric generating function
in Eqs.~(\ref{super6},\ref{super8}), or at least in
the reduction of the problem to a non--linear $\sigma$ model. This is
accomplished by using the saddle--point approximation, suggested by
the occurrence of the factor $N$ multiplying all terms in the
exponent. We recall that in RMT, we always take the limit $N
\rightarrow \infty$. In some cases, it is advantageous to avoid
the saddle--point approximation, and to use an alternative 
technique, see Sec.~\ref{qc1tc}. 

The generating function~(\ref{super6},\ref{super8}) possesses an important
symmetry: Except for its dependence on $J$ and $\varepsilon$, it remains
invariant under the group $G$ of those global transformations $T$ with
$\sigma \rightarrow T^{-1} \sigma T$ which do not change the symmetry
of $\sigma$. The form of the supermatrices $T$ which fulfill this
requirement depends, of course, on the dimension of the matrix
$\sigma$ and on the symmetry (orthogonal, unitary or symplectic) of
the underlying Hamiltonian ensemble. 

The saddle point is found after neglecting in the expression
(\ref{super8}) for ${\cal L}$ both the term $\varepsilon$ and the
source term $J$. This is justified since $\varepsilon$ scales with
$1/N$. Indeed, the fluctuations depend on the unfolded energy
difference $\varepsilon/D$. Since $D$ vanishes like $1/N$, the same
must be true for $\varepsilon$. The source term $J$ may be viewed as
being infinitesimally small since the derivatives with respect to $J$
are taken at $J = 0$. The saddle--point equation is
\begin{equation}
\sigma = \frac{\lambda 1_4}{E - \lambda \sigma}.
\label{super9}
\end{equation}
This matrix equation is solved in two steps. First, a diagonal matrix
$Q_0$ is found which obeys Eq~(\ref{super9}). For $|E| \leq
2 \lambda$, i.e.~if $E$ lies within Wigner's semicircle, the
elements $Q^{\pm}$ of $Q_0$ have the values $Q^{\pm} =
E/(2 \lambda) \pm i \sqrt{1 - (E/(2 \lambda))^2}$. For 
a two--point function, i.e. a product of the trace of
an advanced and of a retarded propagator, it is necessary to
choose $Q_0 =$ diag$(Q^+,Q^+,Q^-,Q^-)$. 

This choice breaks the symmetry under the group $G$ defined above.
This is most easily seen by writing $Q_0$ in the form
\begin{equation}
Q_0 = \frac{E}{2 \lambda} 1_4 + \frac{\pi \lambda \rho(E)}{N}L
\label{super10}
\end{equation}
with $\rho$ defined in Eq.~(\ref{R1}). 
The matrix $L$ does not
commute with all elements $T$ of $G$: The symmetry is broken
by the mean level density $\rho$. We will see shortly that as a 
consequence of this broken symmetry, there occurs a Goldstone or 
zero mode. 

All matrices $Q = T^{-1} Q_0 T$ obtained from $Q_0$ by application of
a transformation $T \in G$ are also solutions of the saddle--point
equation, and the totality of all these matrices defines the
saddle--point manifold. Let $G_R$ with elements $R$ be the maximum
subgroup of $G$ which obeys $[R,L] = 0$ for all $R \in G_R$.  This
subgroup leaves $Q_0$ invariant. Let $G/G_R$ be the associated coset
space with elements $T_0$. Every $T \in G$ can be written in the form
$T=RT_0$ with $R \in G_R$ and $T_0 \in G/G_R$. The saddle--point
manifold is then given by $Q = T_0^{-1} Q_0 T_0$.  This manifold has
dimension $d > 0$ precisely because the symmetry under $G$ is broken
by $Q_0$, and it therefore constitutes the Goldstone mode caused by
the broken symmetry.

In a seminal paper, Sch\"afer and Wegner~\cite{Sca80} investigated the
structure of the saddle--point manifold for bosonic variables.  In
Refs.~\onlinecite{Efe83,Ver85}, their work was extended to the
supersymmetry formalism. It turns out that the saddle--point manifold
possesses mixed (hyperbolic and compact) symmetry. This property of
the saddle--point manifold is not reproduced in the replica trick, and
is the cause of its failure to produce exact results \cite{Ver85a}.
  
The integration over the massive modes (i.e. in directions orthogonal
to the saddle--point manifold) can be carried out. For Gaussian RMT,
these modes have a mass proportional to $N$, and yield corrections to
the saddle--point integral which vanish for $N \rightarrow \infty$.
The remaining integrations extend over the coset space $G/G_R$ and
yield a zero--dimensional non--linear $\sigma$ model (because the
saddle--point solutions $Q$ obey the non--linear Eq.~(\ref{super9})).
The generating function attains the form
\begin{equation}
  \overline{F_1(J_1)F_2(J_2)} = \int d \mu(t) (...)  \exp \left(
  i\frac{\pi\varepsilon}{4D} \ {\rm trg} (Q L) \right).
\label{super11}
\end{equation}
The dots indicate the $J$--dependent terms. Notice that the mean level
spacing is given by $D=\pi\lambda/N$ at $E=0$. The integration measure
$d \mu(t)$ is defined in terms of a suitable parameterization of the
elements $T_0$ of the coset space $G/G_R$.

{\it Mutatis mutandis}, these arguments carry over to the case of
disordered solids in $d \geq 1$ dimensions. For the potential model of
Eq.~(\ref{imp}), the large parameter which formally corresponds to $N$
is $k_Fl$ where $l = v_F \tau$ is the elastic mean free path,
$\tau$ is the elastic mean free time, and $k_F$ ($v_F$) is the Fermi
wave number (the Fermi velocity, respectively). The need to use this
parameter shows that the supersymmetry technique is confined to
problems with weak disorder.

Minimization of the Lagrangean in $d \geq 1$ leads to a diagonal
saddle--point solution $Q_0$ which has the same form as in RMT.  The
saddle--point manifold now comprises matrices $Q({\vec r}) =
T_0^{-1}({\vec r}) Q_0 T_0({\vec r})$ with a very slow ${\vec
  r}$--dependence (long wave--length limit). The resulting generating
functional $\overline{F(Q)}$ has the form
\begin{eqnarray}
  \overline{F(Q)} &=& \int d \mu(t({\vec r})) (...) \exp \biggl( 
  \frac{\pi \nu}{8} \int d^d r \biggl( \hbar \ {\cal D} \ {\rm trg}
  \left(\nabla Q({\vec r})\right)^2 \nonumber\\ & &
    \qquad\qquad\qquad + 2 i \varepsilon \ {\rm trg} (Q L) \biggl)
    \biggl).
\label{super12}
\end{eqnarray}
Here, ${\cal D} = v_F^2 \tau / d$ is the diffusion constant. Except
for the additional space dependence, all other symbols have the same
meaning as in Eq.~(\ref{super11}).

It is instructive to compare the terms in the exponents of
Eqs.~(\ref{super11},\ref{super12}). If we disregard the additional
space dependence in Eq.~(\ref{super12}), the ``symmetry--breaking
term'' proportional to $\varepsilon$ is identical in both cases.
Indeed, the space integration yields a factor $V$, so that $\nu V =
\Delta^{-1}$, and we surely must identify $\Delta$ with the Gaussian
RMT mean level spacing $D = \pi \lambda / N$.  The gradient term in
Eq.~(\ref{super12}) (obviously missing in Eq.~(\ref{super11})) has
roughly the magnitude $\hbar {\cal D} / L^2 = E_C$. For $E_C <
\varepsilon$, this term is more important than the symmetry--breaking
term, and conversely, see Sec.~\ref{disorder}.

We distinguish a diffusive regime, defined by the condition $g \gg 1$,
or $L \ll \xi$ with $\xi$ the localization length, and a localized
regime, where $L \gg \xi$. Since $g = E_C/ \Delta = \xi / L$ in the
diffusive regime, see Eq.~(\ref{conductance}), the conditions $L \ll
\xi$ and $E_C \gg \Delta$ are equivalent.  In the diffusive regime,
the symmetry--breaking term is the dominant zero mode. This implies
that {\it all correlation functions} of the disordered solid are
approximately of Gaussian RMT type in this regime! This statement
holds for energy differences $|\varepsilon| < E_C$ and establishes the
close link between Gaussian RMT and localization theory, see
Sec.~\ref{disorder}. In the localized regime, on the other hand, the
gradient term is the dominant zero mode. It may be surprising that in
this regime, Eq.~(\ref{super12}) retains its validity. The reason is
that the characteristic length scale $L_0$ over which $Q({\vec r})$
varies must obey the inequality $l \ll L_0 \ll \xi$. This is why a
non--linear $\sigma$ model does not apply for $d = 1$ where $l = \xi$.

In concluding this introduction to supersymmetry, we notice that the
supersymmetric representation of the generating function for the RMT
correlators provides an exact re--formulation of the RMT problem.
Therefore, all results derived from this function are fully equivalent
to the results of classical RMT. Starting from a white--noise
potential, on the other hand, one obtains a supersymmetric non--linear
$\sigma$ model. In the zero--dimensional limit, this model is fully
equivalent to th limit $N\to\infty$ of the supersymmetric
representation of RMT.  However, the supersymmetry method continues
yielding new results which, in many cases, could not be obtained with
other techniques.

\subsection{Scattering systems}
\label{qc1cs}

Scattering systems play a very important role in the application of
Random Matrix Theory, particularly in the context of transport
phenomena in mesoscopic systems. The theory of these systems requires
a generalization of the concepts introduced so far for bounded systems
and the associated spectral fluctuations. We keep our presentation
brief since various well written papers and reviews are available. We
mention the early reviews by Eckhardt~\cite{Eck88,Eck88b} and
Smilansky~\cite{Smi90,Smi91} which contain discussions of chaotic
scattering, especially from a semiclassical point of view. Likewise,
Doron, Smilansky and Frenkel~\cite{Dor91} mainly focus on
semiclassical aspects, while Lewenkopf and Weidenm\"uller~\cite{Lew91}
concentrate on aspects of Random Matrix Theory and on the comparison
with semiclassical results. Very recently, a review of chaotic
scattering was presented by Jung and Seligman~\cite{Jun97}, and one on
scattering and transport in mesoscopic and disordered systems by
Beenakker~\cite{Bee97}. In Sec.~\ref{qc1csa}, we summarize general
aspects of scattering theory, before we establish the link to Random
Matrix Theories in Sec.~\ref{qc1csb}.

\subsubsection{General aspects}
\label{qc1csa}

In the context of RMT, scattering processes are important in both,
cavities and/or mesoscopic wires connected to the outside world by
leads or antennae, and non--relativistic many--body systems as
encountered in nuclear, atomic and molecular physics. In all these
cases, the participants in the scattering process move freely at
asymptotically large distances but encounter strong interactions in
the interaction region. We idealize a scattering system as consisting
of a {\it compact} interaction region, i.e.~a region of finite volume,
and of channels through which this region is accessible and in which
the propagation is free. A general scattering theory based on this
idealization was constructed by Wigner and Eisenbud~\cite{Wig47}. 
These authors made the physically well justified further assumption
that only two--body fragmentation is allowed at low energies. They
also postulated short--range interactions. However, the formalism can
be extended to the Coulomb interaction, too. 

To illuminate the physical content of this approach, we first consider
a bound--state problem defined by restricting the motion of all
particles to the compact interaction region. Then, the spectrum is
discrete. As we allow for the coupling to the channels, many of the
bound eigenstates turn into resonances. These resonances may dominate
the scattering process. Following Ref.~\onlinecite{Mah69}, we present
here a variant of the general theory which focuses attention from the
outset onto such resonances. We consider $N$ orthonormal functions
$\varphi_\mu, \ \mu=1,\ldots,N$ localized on the interaction region,
where eventually the limit $N\to\infty$ has to be taken. These
functions represent a basis for the quasibound eigenstates. The
$\Lambda$ channels are represented by the channel wave functions
$\chi_c(E), c=1,\ldots,\Lambda$ where $E$ is the total energy. These
states obey the orthonormality condition
$\langle\chi_a(E)|\chi_b(E^\prime)\rangle=\delta_{ab}\delta(E-E^\prime)$.
They do not vanish in the interaction region but are orthogonal to the
bound--state wave functions, $\langle\chi_a(E)|\varphi_{\mu}\rangle =
0$ for all $E, a, \mu$. In this basis, the total Hamiltonian has the
form
\begin{eqnarray}
{\cal H} &=& \sum_{\mu,\nu=1}^N
     |\varphi_\mu\rangle H_{\mu\nu}\langle\varphi_\nu|
                                           \nonumber\\
   & & \qquad
   + \sum_{\mu=0}^N\sum_{c=1}^\Lambda\int_{\varepsilon_c}^\infty
      dE \left(|\chi_c(E)\rangle W_{c \mu} \langle\varphi_\mu|
        + |\varphi_\mu\rangle W_{\mu c}\langle\chi_c(E)|\right)
                                           \nonumber\\
   & & \qquad + \sum_{c=1}^\Lambda\int_{\varepsilon_c}^\infty
                         dE|\chi_c(E)\rangle E\langle\chi_c(E)| \ .
\label{1cs1}
\end{eqnarray}
Here $\varepsilon_c$ is the threshold energy in channel $c$ (for $E <
\varepsilon_c$ only evanescent waves exist, and for $E >
\varepsilon_c$ the channel is said to be open). The first term in
Eq.~(\ref{1cs1}) involves the $N\times N$ matrix $H$ which describes
the mutual coupling of the bound states. The second term contains the
$N \times \Lambda$ coupling matrix $W$ between channels and bound
states. This term causes the bound states to acquire widths, and to
become resonances. The third term is assumed to be diagonal in the
channels. This assumption can be lifted but provides a useful
simplification and is realistic in later applications. In the
framework of Eq.~(\ref{1cs1}), transitions between different channels
are possible only via intermediate population of the bound states. We
assume that elastic background phase shifts in the channels can be
neglected so that the functions $\chi_a(E)$ are essentially plane
waves.

The physical assumptions made in this model become even more
transparent as one works out~\cite{Mah69} the scattering matrix
$S_{ab}(E)$, 
\begin{equation}
S_{ab}(E) = \delta_{ab} - i2\pi W_a^\dagger D^{-1}(E) W_b
\label{1cs2}
\end{equation}
where the $N$ component vector $W_a$ is the $a$--th column of the 
matrix $W$. The propagator $D^{-1}(E)$ is the inverse of
\begin{equation}
D(E) = E 1_N - H +i\pi \sum_{c \ {\rm open}} W_cW_c^\dagger \ .
\label{1cs3}
\end{equation}
This notation should not be confused with that for the mean level
spacing $D$ between the eigenvalues of $H$. All threshold effects,
including the dependence of $W$ on $E$ have been neglected. The
scattering matrix element $S_{ab}(E)$ has $N$ poles corresponding to
$N$ resonances. For $N=1$, the scattering matrix has Breit--Wigner
form, with $W_{c 1}$ denoting the partial width amplitude. For $N>1$,
Eqs.~(\ref{1cs2}) and~(\ref{1cs3}) constitute $N$--level unitary
generalizations of the Breit--Wigner formula. In the limit
$\sum_{c=1}^\Lambda |W_{\mu c}|^2 \ll D$ for all $\mu$, the model
describes $N$ isolated resonances $\mu=1,\ldots,N$, with widths very
small compared to the spacing between resonances. Here, $D$ denotes
the mean level spacing. As the ratios $\sum_{c=1}^\Lambda |W_{\mu
  c}|^2/D$ increase, the resonances begin to overlap. Eventually, one
arrives at the Ericson regime of strongly overlapping resonances where
the fluctuations of the cross section are no longer related to
individual resonances. Increasing the ratios $\sum_{c=1}^\Lambda
|W_{\mu c}|^2/D$ even further, one returns to the regime of isolated
resonances.

\subsubsection{Scattering and Random Matrix Theory}
\label{qc1csb}

Apart from the general constraints mentioned above, no specific
assumptions on the bound--state Hamiltonian $H$ have been made in the
derivation of the formal result~(\ref{1cs2}). We now connect this
model to RMT by assuming that the Hamiltonian $H$ in the interaction
region can be represented by a random matrix drawn from one of the
Gaussian ensembles. In particular, for a time--reversal invariant
system, $H$ is a member of the GOE. With this assumption, the
fluctuation properties of $S$ can be predicted, given the ensemble
average of $S$ which serves as an input parameter. 

Anticipating the discussion in later sections, we argue that this
choice for $H$ is justified in three typical physical situations:
Scattering at complex many--body systems, chaotic scattering in
few--degrees--of--freedom systems, and scattering by a disordered
quantum dot. Indeed, the spectral fluctuation properties of a wide
class of complex many--body systems such as nuclei, atoms and
molecules are known to be well described by the spectral observables
discussed in Sec.~\ref{qc1ba}. Therefore, it is very natural to choose
$H$ from the GOE. For chaotic systems with few degrees of freedom,
such as billiards, the justification is based on the Bohigas
conjecture. For a disordered quantum dot, the random Hamiltonian
directly simulates the effect of the random potential.
 
The autocorrelation function of the scattering matrix $S$ is defined
as an ensemble average,
\begin{equation}
C_{abcd}(\varepsilon) = \overline{S_{ab}(E) 
                                S_{cd}^*(E+\varepsilon)}
                          - \overline{S_{ab}(E)}\,
                            \overline{S_{cd}^*(E)} \ .
\label{1cs4}
\end{equation}
In a generic situation, this correlator is, after proper unfolding,
independent of the energy $E$. The mathematical techniques developed
by Mehta and Dyson cannot be used to work out the correlator~(\ref{1cs4}).
Extending the supersymmetric method which Efetov~\cite{Efe83} had
developed in the framework of disordered systems, Verbaarschot,
Weidenm\"uller and Zirnbauer~\cite{Ver85} derived the exact result
\begin{eqnarray}
C_{abcd}(\varepsilon) &=& \frac{1}{8} \int_0^\infty d\lambda_1
          \int_0^\infty d\lambda_2\int_0^1 d\lambda
\frac{(1-\lambda)\lambda|\lambda_1-\lambda_2|}
     {\sqrt{(1+\lambda_1)\lambda_1(1+\lambda_2)\lambda_2}
            (\lambda+\lambda_1)^2(\lambda+\lambda_2)^2}
                                           \nonumber\\
   & & \qquad \exp\left(-i\pi\frac{\varepsilon}{D}
              (\lambda_1+\lambda_2+2\lambda)\right)
       \prod_{e=1}^\Lambda \frac{(1-T_e\lambda)}
                 {\sqrt{(1+T_e\lambda_1)(1+T_e\lambda_2)}}
                                           \nonumber\\
   & & \qquad \Biggl( \delta_{ab}\delta_{cd}
              \overline{S_{aa}}\overline{S_{cc}^*}T_aT_c
        \left(\frac{\lambda_1}{1+T_a\lambda_1}+
              \frac{\lambda_2}{1+T_a\lambda_2}+
                    \frac{2\lambda}{1-T_a\lambda}\right)
                                           \nonumber\\
   & & \qquad\qquad\qquad\qquad
        \left(\frac{\lambda_1}{1+T_c\lambda_1}+
              \frac{\lambda_2}{1+T_c\lambda_2}+
                    \frac{2\lambda}{1-T_c\lambda}\right)
                                           \nonumber\\
   & & \qquad + (\delta_{ac}\delta_{bd}+\delta_{ad}\delta_{bc})T_aT_b
        \Biggl(\frac{\lambda_1(1+\lambda_1)}
                   {(1+T_a\lambda_1)(1+T_b\lambda_1)}
                                           \nonumber\\
   & & \qquad\qquad + \frac{\lambda_2(1+\lambda_2)}
                   {(1+T_a\lambda_2)(1+T_b\lambda_2)}+
                   \frac{\lambda(1-\lambda)}
                   {(1+T_a\lambda)(1+T_b\lambda)} \Biggl) \Biggl) \ .
\label{1cs5}
\end{eqnarray}
The coefficients $T_c=1-|\overline{S_{cc}(E)}|^2$ are called
transmission coefficients and define the strength of the coupling
between resonances and channels. Each $T_c$ depends non--linearly on
the quantity $\sum_{\mu} W_{\mu c}^2$. If all coefficients $T_c$ are
unity, the fluctuations attain maximal size. An expansion of
$C_{abcd}(\varepsilon)$ to first order in the inverse of
$\sum_{c=1}^\Lambda T_c$ yields the Hauser--Feshbach
formula~\cite{Hau52}, see Eq.~(\ref{3nu1}) of Sec.~\ref{qc3nua}.

In the stochastic approach to scattering just described, the
bound--state Hamiltonian $H$ occurring in expressions~(\ref{1cs2})
and~(\ref{1cs3}) is replaced by an ensemble of random matrices.
Alternatively, it is possible to consider the $\Lambda \times \Lambda$
scattering matrix $S$ itself as a stochastic object, without the
intermediate step of choosing a random Hamiltonian. In this case, one
directly constructs an ensemble of scattering matrices $S$. This is
the approach by Dyson~\cite{Dys62b} who defined the three circular
ensembles: the Circular Orthogonal Ensemble (COE), the Circular
Unitary Ensemble (CUE), and the Circular Symplectic Ensemble (CSE),
see Sec.~\ref{early}.  These three ensembles correspond,
respectively, to the GOE, the GUE, and the GSE. The invariance
properties used by Dyson imply that the ensemble averages of the $S$
matrices vanish for all three ensembles, $\overline{S}=0$. This is
equivalent to saying that the transmission coefficients are unity, or
that the coupling to the channels is maximal. In that sense, the three
ensembles of Dyson are subsets of the more general ensembles
introduced above via the random Hamiltonian approach. We do not go
into further details here of the random scattering matrix approach. We
return to this point in Sec.~\ref{disorder}.

Using the random Hamiltonian approach and the supersymmetry method,
Fyodorov and Sommers~\cite{Fyod96a,Fyod96b,Fyod97} recently worked out
several other statistical measures. They studied the case of broken
time--reversal invariance (GUE Hamiltonian). In particular, they
established a link to the random scattering matrix approach by showing
that in both approaches the pair--correlation functions of phase
shifts at fixed energy coincide. Moreover, they also calculated
analytically the distribution of the poles of the scattering matrix.
Because of the last term in Eq.~(\ref{1cs3}) which represents the
coupling to the channels, these poles occur in the lower half of the
energy plane.  Their results apply to any breaking of Hermitecity
while previous studies are only valid the case of strongly broken
Hermitecity. We mention in passing that there is a relation to
Ginibre's~\cite{Gini65} ensemble of matrices with arbitrary complex
eigenvalues.  A discussion can be found in Haake's book~\cite{Haa91}.

\subsection{Wave functions and widths}
\label{qc1ww}

For the understanding of the connection between classical and quantum
chaos, the study of the stochastic properties of wave functions has
become very important. For an investigation of wave packet dynamics,
of scars and of the relationship with periodic orbits, systems with
few degrees of freedom are particularly well suited. A review can be
found in chapter~15 of Gutzwiller's book~\cite{Gut90}. Here, we focus
on the statistical properties of wave functions in chaotic systems and
on the way these properties are modeled in Random Matrix Theory
(Sec.~\ref{qc1wwa}), and on recent developments in mesoscopic physics,
see Sec.~\ref{qc1wwb}. The limits of the RMT description have been
explored by various authors. As one example we mention a conceptually
important contribution due to Tomsovic~\cite{Toms96}. He showed the
existence of parametric correlations (see Sec.~\ref{qc1pc}) between
states and level velocities in a particular system. Random Matrix
Theory cannot describe these correlations.

\subsubsection{Results derived in the framework of classical RMT}
\label{qc1wwa}

To characterize the distribution of the eigenfunctions of an ensemble
of random matrices, one considers a fixed basis and introduces the
probability density $P_N(a)$ of finding for the components a value
between $a$ and $a + da$, with $N$ the dimension of the matrices.
Rotation invariance of the ensemble implies that $P_N(a)$ is
determined by the Haar measure $d\mu(U)$ of the group of diagonalizing
matrices, $H=U^{-1}XU$. For the orthogonal case, a quick and
elementary derivation~\cite{Por65} of $P_N(a)$ proceeds as follows. We
consider one component of one of the $N$ eigenvectors of $H$. In polar
coordinates, it can be represented by $\cos\vartheta$. The associated
part of the measure is $\sin^{N-2}\vartheta$. For $-1 < a < +1$, we have
\begin{equation}
P_N(a) = C_N \int_0^\pi \delta(a-\cos\vartheta) 
             \sin^{N-2}\vartheta d\vartheta \ .
\label{1ww1}
\end{equation} 
The constant $C_N$ is determined by the normalization of $P_N(a)$ to
unity. A straightforward calculation yields
\begin{equation}
P_N(a) = \frac{\Gamma(N/2)}{\sqrt{\pi}\Gamma((N-1)/2)}
                   \left(1-a^2\right)^{(N-3)/2}
\label{1ww2}
\end{equation} 
which is symmetric in $a$. This expression is exact for every integer
$N$. For the second moment of this distribution, one finds
$\overline{a^2}=1/N$. In the limit of large $N$, the distribution
approaches the Gaussian
\begin{equation}
P_N(a) = \sqrt{\frac{N}{2\pi}}\exp\left(-\frac{N}{2}a^2\right)
\label{1ww3}
\end{equation} 
with variance $1/N$. 

In many cases, the distribution of the wave function components is not
accessible to a direct measurement. However, for isolated resonances
the reduced partial width amplitudes $\gamma_{nc}$ for level $n$ and
scattering channel $c$ can be determined from a scattering experiment.
Let $J_{mc}$ be the overlap integral of the channel wave function in
channel $c$ and the $m$--th wave function of the basis set. Then,
\begin{equation}
\gamma_{nc} = \sum_{m=1}^N U_{nm} J_{mc}
\label{1ww4}
\end{equation} 
with $U$ as defined above. Consider $L$ levels $n=1,\ldots,L$ where
$L<N$. As observed by Ullah~\cite{Ull67}, the joint probability
distribution $P(\gamma_{1c},\ldots,\gamma_{Lc})$ for the $L$ partial
width amplitudes can be written as
\begin{equation}
P(\gamma_{1c},\ldots,\gamma_{Lc}) = 
          \int \prod_{n=1}^L \delta\left(\gamma_{nc}-
              \sum_{m=1}^N U_{nm} J_{mc}\right) d\mu(U) \ .
\label{1ww5}
\end{equation} 
With the help of a proper orthogonal transformation~\cite{Ull67} of 
the eigenvectors, this expression can be reduced to a form which
generalizes Eq.~(\ref{1ww1}). One finds
\begin{eqnarray}
& & P(\gamma_{1c},\ldots,\gamma_{Lc}) = \nonumber\\
& & \qquad \qquad   \frac{\Gamma(N/2)}
         {\left(\pi N \overline{\gamma_c^2}\right)^{L/2}\Gamma((N-L)/2)}
             \left(1-\frac{1}{N\overline{\gamma_c^2}}
                   \sum_{n=1}^L \gamma_{nc}^2\right)^{(N-L-2)/2}
\label{1ww6}
\end{eqnarray} 
where $N\overline{\gamma_c^2}=N\overline{\gamma_{nc}^2} = \sum_{n=1}^N
J_{nc}^2$. Again, this is exact for any $N$. The functional forms of
Eqs.~(\ref{1ww2}) and~(\ref{1ww6}) agree for $L=1$ as expected. In the
sequel, we consider only the case $L=1$ and drop the level index. 

Typically, the quantity measured in an experiment is the partial
width $\Gamma_{c}$ for the decay of a level into a channel $c$, rather
than the partial width amplitude $\gamma_c$. Apart from
proportionality constants such as penetration factors etc., we have
$\Gamma_{c}= \gamma_{c}^2$. For large $N$, Eq.~(\ref{1ww6}) then
yields the famous Porter--Thomas law  
\begin{equation}
P(\Gamma_c) d\Gamma_c = \frac{1}{\sqrt{\pi}} 
                    \frac{\exp(-\Gamma_c/2\overline{\Gamma}_c)}
                         {\sqrt{\Gamma_c/2\overline{\Gamma}_c}}
                    \frac{d\Gamma_c}{2\overline{\Gamma}_c} 
\label{1ww7}
\end{equation} 
where $\overline{\Gamma}_c=\overline{\gamma_c^2}$. These
considerations can easily be generalized to unitary and symplectic
symmetry. The value of the symmetry parameter $\beta$ with
$\beta=1,2,4$ is equal to the dimension of the number space on which
the random matrices are constructed. The reduced width amplitudes
$\gamma_c$ accordingly possess $\beta$ components $\gamma_{cj}, \ 
j=1,\ldots,\beta$.  If all amplitudes $\gamma_{cj}$ are distributed
with the same Gaussian probability distribution, the absolute square
$\Gamma_c=\sum_{j=1}^\beta \gamma_{cj}^2$ of $\gamma_c$ is distributed
according to
\begin{equation}
P_\beta(\Gamma_c) d\Gamma_c = \frac{1}{\Gamma(\beta/2)} 
     \left(\frac{\beta\Gamma_c}{2\overline{\Gamma}_c}\right)^{\beta/2-1}
     \exp\left(-\frac{\beta\Gamma_c}{2\overline{\Gamma}_c}\right)
     \frac{\beta d\Gamma_c}{2\overline{\Gamma}_c} \ .
\label{1ww8}
\end{equation} 
This implies that for the three symmetry classes, the distributions
\begin{figure}
\centerline{
\psfig{file=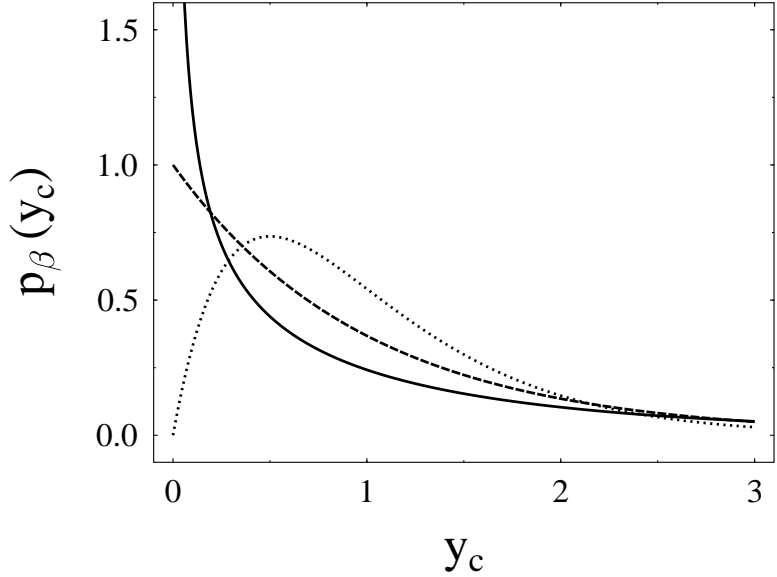,width=4.5in,angle=-90}
}
\caption{
  The distributions $p_\beta(y_c)$ of the normalized partial
  widths $y_c=\Gamma_c/\overline{\Gamma}_c$ for the Gaussian
  ensembles. The solid line is the GOE result ($\beta=1$), the
  dashed line is the GUE result ($\beta=2$) and the dotted line is the
  GSE result ($\beta=4$). A singularity at $y_c=0$ arises in the 
  orthogonal case ($\beta=1$) only.
}
\label{figft6}
\end{figure}
$P_{\beta}(\Gamma_c)$ are special cases of a $\chi_\nu^2$ distribution for
$\nu=\beta$ degrees of freedom. We note that only the orthogonal case has a
singularity for vanishing width $\Gamma_c$. One often introduces a
normalized partial width $y_c=\Gamma_c/\overline{\Gamma}_c$ and its
distribution $p_\beta(y_c)$ defined by 
$p_\beta(y_c)dy_c = P_\beta(\Gamma_c)d\Gamma_c$. This function is
shown in Fig.~\ref{figft6} for the three Gaussian ensembles. 

In a system with $\Lambda$ open channels, there may exist correlations
between the partial width amplitudes $\gamma_c, \ c=1,\ldots,\Lambda$. 
Typically, such correlations have a dynamical origin. An example is
provided by direct reactions in nuclear physics. The correlation
matrix of the partial width amplitudes is defined as 
\begin{equation}
M_{cc^\prime} = \overline{\gamma_c^*\gamma_{c^\prime}} \ .
\label{1ww8a}
\end{equation} 
In the limit of large level number $N$ and under the condition that
$\Lambda \ll N$, the distribution of the partial width amplitudes 
was found to be~\cite{Kri63,Ull63}
\begin{equation}
P(\vec{\gamma}) = \frac{1}{{\rm det}^{\beta/2}M}
       \exp\left(-\frac{\beta}{2}
          \vec{\gamma}^\dagger M^{-1}\vec{\gamma}\right) \ .
\label{1ww8b}
\end{equation} 
Here, $\vec{\gamma}=(\gamma_1,\ldots,\gamma_\Lambda)$ is a $\Lambda$
component vector containing the partial width amplitudes.  In
Ref.~\onlinecite{Ull67} it is also shown how to derive multi--level
and multi--channel distributions for finite level number $N$.

Having investigated the distribution of the partial widths $\Gamma_c$,
we now turn to the distribution function $P(\Gamma)$ of the {\it total}
width $\Gamma$. For isolated resonances, we have $\Gamma = \sum_{c =
1}^{\Lambda} \Gamma_c$ where $\Lambda$ is the number of open channels. 
Using and extending the methods described in Porter's
review~\cite{Por65} and Ullah's book~\cite{Ulla92}, Alhassid and
Lewenkopf~\cite{Alh95a} derived in a direct calculation a very general
result for the width distribution. Let $M$ be the correlation
matrix~(\ref{1ww8a}) of the partial width amplitudes. For
$\beta=1,2,4$, the distribution of the total width $\Gamma$ is given
by
\begin{equation}
P(\Gamma) = \frac{1}{2\pi} \int_{-\infty}^{+\infty}
                   \frac{\exp(-i\Gamma t)}
         {\det^{\beta/2}(1_\Lambda-i2tM/\beta)} dt
\label{1ww21}
\end{equation} 
where $1_\Lambda$ is the $\Lambda \times \Lambda$ unit matrix. We note
that $P(\Gamma)$ depends only on the eigenvalues $w_c^2$ of $M$. Since
$M$ is Hermitean and positive definite, the $w_c^2$'s are all positive. 
In the unitary case ($\beta = 2$), the integral~(\ref{1ww21}) can be
done and yields
\begin{equation}
P(\Gamma) =  \frac{1}{\prod_{c=1}^\Lambda w_c} 
             \sum_{c=1}^\Lambda \frac{\exp(-\Gamma/w_c^2)}
    {\prod_{c\neq c^\prime} \left(w_c^{-2}-w_{c^\prime}^{-2}\right)} \ .
\label{1ww22}
\end{equation} 
For two equivalent channels, i.e.~$\Lambda=2$ and $M_{11} = M_{22} =
\overline{\Gamma}/2$ one finds
\begin{equation}
P(\Gamma) = \frac{2}{|f|} \exp\left(-\frac{2\Gamma}
                          {(1-|f|^2)\overline{\Gamma}}\right)
            \sinh\left(-\frac{2|f|\Gamma}
                          {(1-|f|^2)\overline{\Gamma}}\right) \ .
\label{1ww23}
\end{equation} 
Here $f=M_{12}/\sqrt{M_{11}M_{22}}$ measures the correlation between
the two channels. This is in agreement with Ref.~\onlinecite{Muc95a}
where this result was derived using the supersymmetry method, see
Sec.~\ref{qc1wwb}.  For $\beta = 1$ one finds~\cite{Alh95a}
\begin{equation}
P(\Gamma) = \frac{1}{\sqrt{1-f^2}} \exp\left(-\frac{\Gamma}
                          {(1-f^2)\overline{\Gamma}}\right)
            I_0\left(-\frac{f\Gamma}
                          {(1-f^2)\overline{\Gamma}}\right) \ .
\label{1ww24}
\end{equation} 
where $I_0$ is the modified Bessel function of zeroth order.  The
methods described in Refs.~\onlinecite{Por65,Ull67} and used above
allow for a very efficient computation of the distribution of wave
functions and widths.

\subsubsection{Results derived in the framework of 
               the supersymmetry method}
\label{qc1wwb}

The results summarized in the previous section were mainly prompted by
research in scattering theory, and were obtained in the framework of
random matrix theory. Recently, condensed matter physicists became
also interested in distributions of wave functions and widths,
particularly in the context of transport properties of quantum
dots. Prigodin et al~\cite{Pri93} studied the statistics of 
conductance fluctuations in a quantum dot. {\it Inter alias}, this led
to a re--derivation of the Porter--Thomas distribution within the
supersymmetry formalism, cf. also the discussion by Efetov and
Prigodin~\cite{Efe93}. 

In many situations there is interest in probability measures which go
beyond the distribution of wave--function components. This applies, in
particular, to the spatial correlations of wave functions for {\it
chaotic} systems~\cite{Ber77,Ber90,Pri95a}. Berry~\cite{Ber77,Gut90}
assumed that the eigenfunction $\psi_E(\vec{r})$ at energy $E$ in a
time--reversal invariant, ergodic system is an
infinite superposition of random plane waves. For the space--averaged
correlation function of this eigenfunction in a $d$--dimensional
system, taken at two different points in space, he found
\begin{equation}
\frac{\langle \psi_E(\vec{r}_1)\psi_E^*(\vec{r}_2) 
                            \rangle_{\rm space}}
{\langle |\psi_E((\vec{r}_1+\vec{r}_2)/2)|^2\rangle_{\rm space}}
       = 2^{d/2-1} \Gamma \left(\frac{d}{2} \right) \,
                   \frac{J_{d/2-1}(kr)}{(kr)^{d/2-1}} \ .
\label{1ww9}
\end{equation} 
Here $\vec{k}$ is the wave vector, and $k$ the inverse de Broglie 
wave length, $r = |\vec{r}_1 - \vec{r}_2|$ is the distance between 
the two points. 

Closely related questions arise in {\it disordered} mesoscopic systems. 
For a single--particle model with random impurity potential,
cf. Sec.~\ref{disorder}, 
\begin{equation}
H = \frac{\vec{p}^2}{2m} + V(\vec{r}),
\label{gwn}
\end{equation}
Prigodin {\it et al.}~\cite{Pri94a} and Prigodin~\cite{Pri95a} 
recently calculated wave function
correlation functions with the help of the supersymmetry method. In a
system of volume $V$, they obtained 
\begin{equation}
\overline{|\psi_E(\vec{r}_1)\psi_E(\vec{r}_2)|^2} V^2
                = 1 + c_\beta |f(r)|^2 \ ,
\label{1ww12}
\end{equation} 
with $\beta = 1,2,4$ and $c_1=c_4=1$, and $c_2=2$. Under the
assumption of ergodicity, the function $f(r)$ is closely related to
the averaged spatial correlator~(\ref{1ww9}),
\begin{equation}
f(r) = \overline{\psi_E(\vec{r}_1)\psi_E^*(\vec{r}_2)} V \ .
\label{1ww13}
\end{equation} 
The existence of spatial correlations of chaotic wave functions
demonstrated in this work is essentially caused by the finite
(non--zero) wave length. 

Equations~(\ref{1ww12}) and~(\ref{1ww13}) connect the fourth moment of
the wave function with the square of the second moment, suggesting a
Gaussian--like probability density. To work out this probability
density in its most general form, let us consider $M$ local densities
of a given eigenfunction, 
$a_j=V|\psi_E(\vec{r_j})|^2, j = 1, \ldots,M$, 
and the spatial correlator~\cite{Pri95a},
\begin{equation}
P(a_1,\ldots,a_M,\vec{r}_1,\ldots,\vec{r}_M) = 
    \overline{\prod_{j=1}^M
      \delta\left(a_j-V|\psi_E(\vec{r}_j)|^2\right)} \ .
\label{1ww14}
\end{equation} 
The partial widths discussed in the previous section can be viewed as
special cases of the quantities $a_j$. Therefore, the correlation
function~(\ref{1ww14}) has to reproduce the results of that section
for $M = 1$. 

The distribution function~(\ref{1ww14}) is constructed from its
moments. These moments can be expressed in terms of retarded (and
advanced) Green functions~\cite{Weg80,Alt89} defined as
\begin{equation}
G_R(E,\vec{r}_1,\vec{r}_2) = \sum_{n=1}^N 
        \frac{\psi_{E_n}(\vec{r}_1)\psi_{E_n}^*(\vec{r}_2)}
             {E-E_n+i\eta}
\label{1ww15}
\end{equation}
where $\eta$ is infinitesimal. We consider the case $M = 2$. Let
$R_1(E)$ denote the mean level density. The moments can be written as 
\begin{eqnarray}
& &\overline{|\psi_E(\vec{r}_1)|^{2n}|\psi_E(\vec{r}_2)|^{2m}}
 = \frac{i^{n-m}(n-1)!(m-1)!}{2R_1(E)\pi^{n+m} (n+m-2)!}
                                        \nonumber\\
& & \qquad       \lim_{\eta\to 0} \eta^{n+m-1}
   \overline{\left(G_R(E+i\eta/2,\vec{r}_1,\vec{r}_1)
         \right)^n\left(G_A(E+i\eta/2,\vec{r}_2,\vec{r}_2)\right)^m} \ .
\label{1ww16}
\end{eqnarray} 
For vanishing $\eta$, the only contributions come from terms in which
the energy $E$ coincides with one of the eigenenergies $E_n$. 

In the unitary case $\beta=2$ one obtains~\cite{Pri95a} the
distribution function
\begin{equation}
P(a_1,a_2,r) = \frac{1}{1-f^2(r)} 
               \exp\left(-\frac{a_1+a_2}{1-f^2(r)}\right)
               I_0\left(\frac{2f(r)\sqrt{a_1a_2}}{1-f^2(r)}\right)
\label{1ww17}
\end{equation}
which contains the averaged spatial correlator $f(r)$, corresponding
to the quantity $f$ in Eq.~(\ref{1ww23}). The estimate~(\ref{1ww9})
shows that $f(r)$ takes values between $0$ and $1$. The function
$f(r)$ can be calculated in special situations. In a quantum
dot~\cite{Pri95a}, it is given by the Friedel function. In any
realistic physical system, there will be a critical distance $r_c$
beyond which the spatial correlator vanishes, $f(r)=0, \ r>r_c$. Then,
the two--point function becomes a product of two one--point functions,
$P(a_1,a_2,r)\simeq P(a_1)P(a_2)$, see also Ref.~\onlinecite{Pri93}.
For very short distances $r\to 0$, on the other hand, we have $f(r)\to
1$ so that $P(a_1,a_2,r)\simeq P(a_1) \delta(a_1-a_2)$.

The mathematically much harder but physically very important case of
orthogonal symmetry ($\beta = 1$) was also worked out by Prigodin {\it
  et al.}~\cite{Pri95b}. Srednicki~\cite{Sred96} showed that these
results can also be obtained in a direct calculation without
supersymmetry. Using Berry's conjecture~\cite{Ber77} he found a
distribution for the {\it wave functions} which formally coincides
with the distribution~(\ref{1ww8b}) for the partial width amplitudes.
{}From that, he immediatley obtains Eq.~(\ref{1ww17}) for $\beta=2$
and, in the notation of Ref.~\onlinecite{Pri95b},
\begin{equation}
P(a_1,a_2,r) = \frac{1}{2\pi\sqrt{1-f^2(r)}\sqrt{a_1a_2}} 
               \exp\left(-\frac{a_1+a_2}{2(1-f^2(r))}\right)
               \cosh\left(\frac{f(r)\sqrt{a_1a_2}}{1-f^2(r)}\right)
\label{1ww18}
\end{equation}
for $\beta=1$.  In the case of $\Lambda=2$ equivalent channels with
$\overline{\Gamma}_1=\overline{\Gamma}_2$, the distribution of the
total width is given by the integral~\cite{Muc95a}
\begin{equation}
P(\Gamma) = \int_0^\infty da_1 \int_0^\infty da_2 P(a_1,a_2,r) 
       \delta\left(\Gamma-\frac{a_1+a_2}{2}\overline{\Gamma}\right)
\label{1ww19}
\end{equation} 
where $\overline{\Gamma}$ is the average value. Inserting
Eqs.~(\ref{1ww17}) and~(\ref{1ww18}), one can obtain the
results~(\ref{1ww23}) and~(\ref{1ww24}).

This does not imply, however, that the supersymmetry method has become
obsolete in this context. It is not clear how to use the methods of
Refs.~\onlinecite{Por65,Ull67} in a crossover region, such as, for
example, the gradual breaking of time--reversal invariance. In the
supersymmetry method, such calculations are possible, see
Sec.~\ref{qc1to}.

\subsection{Crossover transitions in spectral correlations}
\label{qc1tc}

The concept of fully developed chaos and, likewise, the strict
applicability of the orthogonal, unitary or symplectic symmetry class
are often an idealization. Frequently, we deal with systems which
undergo a crossover transition between different statistics, or
different universality classes. The physically most interesting
situations are
(i) transitions from regular to chaotic fluctuation properties,
(ii) breaking of time--reversal invariance, and
(iii) breaking of symmetries.

We stress the difference between the breaking of time--reversal
invariance and symmetry breaking. In quantum mechanics, the
time--reversal operator is antiunitary while an operator corresponding
to a symmetry of the Hamiltonian is unitary. If a symmetry such as
angular momentum, parity etc.~holds, the Hamiltonian can be reduced
and can be written in block--diagonal form. Each block belongs to a
given quantum number, i.e.~a representation of the symmetry such as
positive or negative parity. On the other hand, the time--reversal
operator cannot be used to reduce the Hamiltonian.
 
In this section we study crossover properties of energy--correlation
functions. These functions depend on one or several external
parameters which represent the physical situation. The functions must
be distinguished from the ``parametric correlations'' which involve
two or more values of the same external parameter. The latter are
discussed in Sec.~\ref{qc1pc}.

In Sec.~\ref{qc1tca} we discuss transitions in terms of a diffusion
process for the joint probability density in matrix space. Thereby, we
summarize some of the results of Dyson's Brownian motion model. In
Sec.~\ref{qc1tcb} we show that a similar diffusion equation exists for
the generating functions of the correlators in the much smaller space
of supermatrices. In Sec.~\ref{qc1tcc} we summarize the numerical and
analytical results. 

\subsubsection{Diffusion in the space of ordinary matrices}
\label{qc1tca}

We recall the definition~(\ref{1rc12}) of the probability distribution
for the Gaussian ensembles. The volume element~(\ref{1rc8}) factorizes
in an eigenvalue part and an angular part. Because of rotation
invariance, the same statement holds for the joint probability density
of the eigenvalues which is given by Eq.~(\ref{1rc9}) and the
eigenvectors which is the Haar measure of the diagonalizing group.
The local fluctuation properties are governed by the Vandermonde
determinant in Eqs.~(\ref{1rc8}) and~(\ref{1rc9}). This quantity turns
out to be also crucial for the investigation of crossover transitions.
{}From a mathematical viewpoint, this is not surprising: The Jacobian
of a transformation reflects the topology of the spaces in question.
However, the various breakings of rotation invariance employed in
studying crossover transitions destroy the convenient factorization in
an eigenvalue part and an angular part.

Dyson~\cite{Dys62d,Dys72} was the first to realize that crossover
transitions can be formulated as Brownian motion of the eigenvalues.
He constructed the Smoluchowski equation for the joint probability
density. Recent reviews can be found in the books by
Haake~\cite{Haa91} and Mehta~\cite{Meh91}, see also the discussion by
Lenz and Haake~\cite{Len90}. The connection to the Pechukas gas and
the Calogero--Sutherland--Moser model~\cite{Haa91} will be discussed
in Sec.~\ref{field}. Here, we want to take a slightly different route
and use explicitly the universality of the unfolded correlations, see
Sec.~\ref{univ}. It allows us to restrict ourselves to the Gaussian
probability density~(\ref{1rc12}).

To formulate the problem, we consider a random Hamiltonian
\begin{equation}
H = H^{(0)} + \alpha H^{(1)}
\label{1tc4}
\end{equation}
as a function of the transition parameter $\alpha$. Each of the three
matrices belongs to the same symmetry class. The elements of $H^{(1)}$
are Gaussian distributed according to Eq.~(\ref{1rc12}). The
distribution $P_N^{(0)}(H^{(0)})$ of the elements of $H^{(0)}$,
however, is completely arbitrary. In particular, $P_N^{(0)}(H^{(0)})$
can restrict the symmetry of $H^{(0)}$ to a sub--symmetry of
$H^{(1)}$. In the case of a Hermitean $H^{(1)}$, for example,
$P_N^{(0)}(H^{(0)})$ can put all imaginary parts of the elements of
$H^{(0)}$ to zero, thereby effectively restricting $H^{(0)}$ to a real
symmetric matrix.  The general problem is: How does the distribution
$P_N^{(0)}$ develop into the full Gaussian~(\ref{1rc12}) as $\alpha$
increases?

We absorb the parameter $\alpha$ into the probability density by writing 
$\beta/4\alpha^2$ instead of $\beta/2$ in the exponent of
Eq.~(\ref{1rc12}). The resulting function is denoted by
$P_{N\beta}^{(1)}$. The squared transition parameter may be viewed as a
fictitious time $t = \alpha^2 / 2$ which governs the evolution from
the arbitrary into the Gaussian form. Introducing the matrix gradient
$\partial/\partial H$ and the Laplacean
\begin{equation}
\Delta = {\rm tr}\frac{\partial^2}{\partial H^2}
\label{1tc5}
\end{equation}
in the Cartesian spaces of the three symmetry classes, we can identify
the Gaussian $P_{N\beta}^{(1)}(H,t)$ as a diffusion kernel defined through
\begin{equation}
\frac{\beta}{4}\Delta P_{N\beta}^{(1)}(H,t) = 
         \frac{\partial}{\partial t} P_{N\beta}^{(1)}(H,t) 
\qquad {\rm and} \qquad \lim_{t\to 0} P_{N\beta}^{(1)}(H,t) = \delta(H) \ .
\label{1tc6}
\end{equation}
The evolution of the total probability density is given
as the solution of the diffusion equation
\begin{equation}
\frac{\beta}{4}\Delta P_{N\beta}(H,t) = 
         \frac{\partial}{\partial t} P_{N\beta}(H,t) 
\qquad {\rm with} \qquad 
         \lim_{t\to 0} P_{N\beta}(H,t) = P_N^{(0)}(H) \ . 
\label{1tc7}
\end{equation}
According to Eq.~(\ref{1tc6}), the solution can be written as the
convolution 
\begin{equation}
P_{N\beta}(H,t) = \int P_{N\beta}^{(1)}(H-H^{(0)},t) 
                           P_N^{(0)}(H^{(0)}) d[H^{(0)}] \ .
\label{1tc8}
\end{equation}
For initial conditions which depend only on the eigenvalues
$X^{(0)}$ of $H^{(0)}$, the diffusion process is restricted to the
curved space of the eigenvalues. For the joint probability density, one
obtains 
\begin{equation}
\frac{\beta}{4}\Delta_X P_{N\beta}(X,t) = 
         \frac{\partial}{\partial t} P_{N\beta}(X,t) 
\qquad {\rm and} \qquad 
         \lim_{t\to 0} P_{N\beta}(X,t) = P_N^{(0)}(X) \ .
\label{1tc9}
\end{equation}
The ``radial part'' of the Laplacean is given by
\begin{equation}
\Delta_X = \frac{1}{|\Delta_N(X)|^\beta} \sum_{n=1}^N 
           \frac{\partial}{\partial x_n} |\Delta_N(X)|^\beta
           \frac{\partial}{\partial x_n} \ .
\label{1tc10}
\end{equation}
The notation $\Delta_N(X)$ for the Vandermonde determinant should not
be confused with the symbol $\Delta_X$ standing for the radial part of
the Laplacian.  This diffusion equation~(\ref{1tc9}) is solved by the
convolution
\begin{equation}
P_{N\beta}(X,t) = \int \Gamma_{N\beta}(X,X^{(0)},t) 
           P_N^{(0)}(X^{(0)}) |\Delta_N(X^{(0)})|^\beta
                                  d[X^{(0)}] 
\label{1tc11}
\end{equation}
and the kernel 
$\Gamma_{N\beta}(X,X^{(0)},t)$ is given by the group integral
\begin{equation}
\Gamma_{N\beta}(X,X^{(0)},t) = 
              \int P_{N\beta}^{(1)}(U^{-1}XU-X^{(0)},t) d\mu(U) \ .
\label{1tc13}
\end{equation}
In the unitary case $\beta=2$, this is the famous 
Harish-Chandra-Itzykson-Zuber integral~\cite{Har58,Itz80}. For the
joint probability density of the eigenvalues, it yields
\begin{eqnarray}
P_{N2}^{(E)}(X,t) &=& P_{N2}(X,t) \Delta_N^2(X)
            = \frac{1}{\sqrt{2\pi t}^N}
                \int d[X^{(0)}] P_N^{(0)}(X^{(0)}) 
                                     \nonumber\\
            & & \qquad \qquad    
                \exp\left(-\frac{1}{2t}
                   {\rm tr} (X-X^{(0)})^2\right)
                \frac{\Delta_N(X)}{\Delta_N(X^{(0)})} \ .
\label{1tc14}
\end{eqnarray}
Since Eq.~(\ref{1rc1}) applies also in the case of transitions, we can
use Eq.~(\ref{1tc14}) to compute the $k$--level correlation functions
in the unitary case. An alternative procedure for the Circular Unitary
Ensemble was recently given by Pandey~\cite{Pan95}. He starts from an
exact formal solution of the diffusion equation in terms of
representation functions of the unitary group $U(N)$. His method is
equivalent to calculating the Harish-Chandra-Itzykson-Zuber integral.
As discussed in Sec.~\ref{qc1rca}, the correlation functions must
be properly unfolded by introducing~\cite{Pan81} an unfolded
transition parameter $\lambda=\alpha/D$ and an unfolded fictitious
time $\tau=\lambda^2/2$.

An important property of the correlation functions was obtained by
French {\it et al.}~\cite{Fre88a}. Starting from the Fokker--Planck or
Smoluchowski equation closely related to the diffusion equation
discussed above, these authors derived hierarchic relations among the
correlation functions.  The unfolded correlation functions are defined
by extending Eqs.~(\ref{1rc6}) and~(\ref{1rc7}). On this unfolded
scale, the hierarchic equations involve the fictitious time $\tau$.
They acquire the form
\begin{eqnarray}
\frac{\partial}{\partial \tau} X_{\beta k} (\xi_1,\ldots,\xi_k,\tau) 
      &=& \sum_{p=1}^k \frac{1}{|\Delta_k(\xi)|^\beta}
                     \frac{\partial}{\partial\xi_p}|\Delta_k(\xi)|^\beta
                     \frac{\partial}{\partial\xi_p}
                               X_{\beta k}(\xi_1,\ldots,\xi_k,\tau)
                                 \nonumber\\
      & & - \beta \sum_{p=1}^k \frac{\partial}{\partial\xi_p}
          \int_{-\infty}^{+\infty}
         \frac{X_{\beta(k+1)}(\xi_1,\ldots,\xi_k,\xi_{k+1},\tau)}
                    {\xi_p-\xi_{k+1}} d\xi_{k+1}
\label{1tc18}
\end{eqnarray}
where $\Delta_k(\xi)=\prod_{p<q}(\xi_p-\xi_q)$. For $k=1$ one
obtains a closed equation which coincides with the famous Pastur 
equation~\cite{Pas72}.

Als already mentioned, Brownian motion and diffusion also exist in the
circular ensembles~\cite{Dys62d,Pan95}. We mention this here only
briefly since on the unfolded scale, the fluctuation properties agree
with those of the Gaussian ensembles. Pandey and Shukla~\cite{Pan91}
have shown that the ensuing hierarchic relations coincide with
Eq.~(\ref{1tc18}).

\subsubsection{Diffusion in the space of supermatrices}
\label{qc1tcb}

Remarkably, the diffusion equation (\ref{1tc9}) for the joint
probability density in ordinary space has an analogue in
superspace~\cite{Guh96a}. We consider an ensemble of random
Hamiltonians of the form~(\ref{1tc4}). We are interested in the
correlation functions $\widehat{R}_{\beta k}(x_1,\ldots,x_k,t)$. In
contrast to Eq.~(\ref{1rc2}), these functions are defined in terms of
the full Green functions and do contain the principal value parts.
Following Refs.~\onlinecite{Efe83,Ver85}, we express these correlation
functions as derivatives
\begin{equation}
\widehat{R}_{\beta k}(x_1,\ldots,x_k,t) \ = \ \frac{1}{(2\pi)^k} \,  
      \frac{\partial^k}{\prod_{p=1}^k \partial J_p} \,
      Z_{\beta k}(x+J,t) \Bigg|_{J=0}
\label{1tc19}
\end{equation}
of a normalized generating function $Z_{\beta k}(x+J,t)$.  Here, $x$
and $J$ contain the $k$ energies and the $k$ source variables, $x =
{\rm diag}(x_1,x_1,x_1,x_1,\ldots,x_k,x_k,x_k,x_k)$, $J = {\rm
  diag}(-J_1,-J_1,+J_1,+J_1,\ldots,-J_k,-J_k,+J_k,+J_k)$ for $\beta=1$
and $\beta=4$, and $x = {\rm diag}(x_1,x_1,\ldots,x_k,x_k)$, $J = {\rm
  diag}(-J_1,+J_1,\ldots,-J_k,+J_k)$ for $\beta=2$. The physically
interesting functions $R_{\beta k}(x_1,\ldots,x_k,t)$ of
Eq.~(\ref{1rc2}) are constructed by the operation $\Im Z_{\beta
  k}(x+J,t)$ which produces the proper linear
combination~\cite{Guh96a,Guh91} without the principal value parts.
With the help of standard techniques of the supersymmetry
method~\cite{Efe83,Ver85}, the average over the Gaussian distributed
random part $H^{(1)}$ can be performed directly and the generating
function acquires the form
\begin{eqnarray}
Z_{\beta k}(x+J,\alpha) &=& \int d[H^{(0)}] P_N^{(0)}(H^{(0)})
                      \int d[\sigma] Q_{\beta k}(\sigma,t)  
                     \nonumber\\
& & \qquad \qquad {\rm detg}^{-1}\left((x^\pm+J-\sigma)\otimes 1_N - 
                            1_{\zeta_\beta}\otimes H^{(0)}\right) \ .
\label{1tc20}
\end{eqnarray}
Here, $1_N$ and $1_{\zeta_\beta k}$ are unit matrices of dimension $N$
and $\zeta_\beta k$, respectively, and $\zeta_\beta=4$ for the GOE and
the GSE while $\zeta_\beta=2$ for the GUE. The supermatrix $\sigma$
reflects the symmetries of $H^{(1)}$ and has dimension $\zeta_\beta
k$. With $c_{\beta k}$ a normalization constant, the graded
probability density is given by
\begin{equation}
Q_{\beta k}(\sigma,t) = c_{\beta k}
         \exp\left(-\frac{\beta}{4t}{\rm trg}\sigma^2\right)
\label{1tc21}
\end{equation}
In the limit $t\to 0$ it reduces to the superspace $\delta$ function
$\delta(\sigma)$. For vanishing transition parameter $t$, $Z_{\beta
  k}$ therefore becomes the generating function
\begin{eqnarray}
Z_k^{(0)}(x+J) &=& \int d[H^{(0)}] P_N^{(0)}(H^{(0)}) 
                                 \nonumber\\
& & \qquad \qquad {\rm detg}^{-1}\left((x^\pm+J)\otimes 1_N - 
                            1_{\zeta_\beta k}\otimes H^{(0)}\right) 
\label{1tc22}
\end{eqnarray}
for the correlations of the ensemble of matrices $H^{(0)}$.

Because of the Gaussian form of the graded probability
density~(\ref{1tc21}) we can formulate a diffusion process in
superspace. We shift $\sigma \to \sigma -x -J$ and diagonalize the
supermatrix $\sigma=u^{-1}su$ where $s$ is the diagonal matrix of the
eigenvalues of $\sigma$. There are $2k$ eigenvalues for the GUE and
$3k$ eigenvalues for the GOE and the GSE. We now replace the matrix
$x+J$ by an diagonal matrix $r$ having the same form as $s$. These
steps yield
\begin{equation}
Z_{\beta k}(r,t) \ = \ \int Q_{\beta k}(\sigma-r,t) \, 
                          Z_k^{(0)}(s) \, d[\sigma] \ .
\label{1tc25}
\end{equation}
There are only $2k$ energies and source variables in
$x+J$, whereas the matrix $r$ contains $3k$ variables for $\beta=1$
and $\beta=4$. The convolution integral~(\ref{1tc25}) is related to
a diffusion equation. Since $Z_k^{(0)}(s)$ depends only on $s$, the
diffusion takes place in the curved space of the eigenvalues of the
supermatrix. In analogy to the previous section, the diffusion
equation has the form
\begin{equation}
\frac{\beta}{4} \Delta_r \, Z_{\beta k}(r,t) \ = \ 
        \frac{\partial}{\partial t} \, Z_{\beta k}(r,t) \qquad
      {\rm with} \qquad \lim_{t\to 0} Z_{\beta k}(r,t) \ = \ 
                           Z_k^{(0)}(r) 
\label{1tc30}
\end{equation}
and holds for all three symmetry classes and for arbitrary initial
generating functions $Z_k^{(0)}(r)$. The Laplacean $\Delta_r$ is the
``radial part'' of the full Laplacean $\Delta = {\rm trg}\partial^2 /
\partial \sigma^2$, defined in terms of the matrix gradient $\partial
/\partial\sigma$. As usual, the radial part involves the Jacobians or
Berezinians $B_{\beta k}(r)$ of the transformation to radial
coordinates. Explicit expressions for these quantities are not given
here and may be found in Ref.~\onlinecite{Guh96a}. We notice that
Eq.~(\ref{1tc9}) describes a diffusion of the probability density of
the eigenvalues in the space of ordinary matrices whereas
Eq.~(\ref{1tc30}) describes a diffusion of the generating function for
the correlations in the space of supermatrices.

The corresponding diffusion kernel is given by the angular average
\begin{equation}
\gamma_{\beta k}(s,r,t) \ = \ 
               \int Q_{\beta k}(u^{-1}su-r,t) \, d\mu(u)
\label{1tc31}
\end{equation}
where $d \mu(u)$ is the Haar measure. It also satisfies the diffusion
Eq.~(\ref{1tc30}), and an initial condition at $t = 0$ not given here. 
This implies that we can write the solution of Eq.~(\ref{1tc30}) in
the form 
\begin{equation}
Z_{\beta k}(r,t) = \int \gamma_{\beta k}(s,r,t) 
                   Z_k^{(0)}(s)  B_{\beta k}(s) d[s] \ .
\label{1tc33}
\end{equation}
The diffusion process cannot be formulated for the $k$ variables $x$
but only for the $2k$ or $3k$ variables $r$, respectively. Therefore,
the role of the joint probability density in ordinary space is here
played by the generating function rather than the correlation
functions.

In the unitary case $\beta=2$, the kernel $\gamma_{\beta k}$ is
explicitly known: It is the supersymmetric Harish-Chandra-Itzykson-Zuber
integral~\cite{Guh91}. The solution of the diffusion equation reads 
\begin{equation}
Z_{2,k}(r,t) = \frac{1}{B_{2,k}(r)}
               \int G_k(r-s,t) Z_k^{(0)}(s) B_{2,k}(r) d[s] 
\label{1tc34}
\end{equation}
where the relevant part of the kernel is the Gaussian
\begin{equation}
G_k(s-r,t) = \frac{1}{\sqrt{2\pi t}^{2k}} 
                  \exp\left(-\frac{1}{2t}{\rm trg}(s-r)^2\right) \ .
\label{1tc35}
\end{equation}
The correlation functions can now be calculated
straightforwardly~\cite{Guh96a,Guh96b}. They are given by the 
$2k$--dimensional integral representation
\begin{equation}
R_{2,k}(x_1,\ldots,x_k,t) = 
        \frac{(-1)^k}{\pi^k}  \int G_k(s-x,t) 
                  \Im Z_k^{(0)}(s) B_{2,k}(r)d[s]
\label{1tc36}
\end{equation}
which is exact for all initial conditions and also exact for any
finite level number $N$.

It can easily be seen \cite{Guh96a} that the transition on the
unfolded scale is a diffusive process as well. With $\rho=r/D$,
$z_{\beta k}^{(0)}(\rho) = \lim_{N\to\infty} Z_{\beta k}^{(0)}(r)$,
and $z_{\beta k}(\rho,\tau) = \lim_{N\to\infty} Z_{\beta k}(r,t)$ one
finds
\begin{equation}
\frac{\beta}{4} \Delta_\rho z_{\beta k}(\rho,\tau) =
     \frac{\partial}{\partial\tau}  z_{\beta k}(\rho,\tau) \qquad
      {\rm with} \qquad \lim_{\tau\to 0} z_{\beta k}(\rho,\tau) \ = \ 
                           z_k^{(0)}(\rho) \ .
\label{1tc38}
\end{equation}
The diffusion processes on the original and the unfolded scale are the
same, only the initial conditions are different, see
Ref.~\onlinecite{Guh96a}. Therefore, the integral
representation~(\ref{1tc33}) must hold on the unfolded scale as well,
\begin{equation}
z_{\beta k}(\rho,\tau) \ = \ \int \gamma_{\beta k}(s,\rho,\tau) \,
        z_k^{(0)}(s) \, B_{\beta k}(s) d[s] \ .
\label{1tc39}
\end{equation}
In particular, in the unitary case $\beta=2$ we find
\begin{equation}
z_{2,k}(\rho,\tau) = \frac{1}{B_{2,k}(\rho)}
               \int G_k(\rho-s,t) z_k^{(0)}(s) B_{2,k}(r) d[s] 
\label{1tc40}
\end{equation}
for the generating function. {}From that, the $2k$--dimensional
integral representation
\begin{equation}
X_{2,k}(\xi_1,\ldots,\xi_k,\tau) = 
        \frac{(-1)^k}{\pi^k} \int G_k(s-\xi,\tau)
                  \Im z_k^{(0)}(s) B_{2,k}(r)d[s]
\label{1tc41}
\end{equation}
for the correlation functions on the unfolded scale is obtained. In
the case $k = 2$, a further simplification arises because the
two--level correlation function depends only on $r=\xi_1-\xi_2$. The
remaining four integrals can be trivially reduced to two integrals and
one has~\cite{Guh96a,Guh96b}
\begin{eqnarray}
X_{2,2}(r,\tau) &=& \frac{4}{\pi^3 \tau} 
       \int\limits_{-\infty}^{+\infty}
       \int\limits_{-\infty}^{+\infty}
       \exp\left(-\frac{1}{4\tau}(t_1^2+t_2^2)\right) 
                                              \nonumber\\
& & \qquad  \qquad 
           \frac{t_1t_2}{(t_1^2+t_2^2)^2} 
           \sinh\frac{rt_1}{2\tau} 
           \sin\frac{rt_2}{2\tau} 
           \Im z_2^{(0)}(t_1,t_2) dt_1 dt_2 \ .
\label{1tc42}
\end{eqnarray}
The trigonometric and the hyperbolic functions can be directly 
traced back to the eigenvalues in the Boson--Boson and the
Fermion--Fermion block.

We compare the hierarchic equations~(\ref{1tc18}) and the diffusion
equation~(\ref{1tc38}). The former couple the correlation functions
with index $k$ to all those with index up to $k+1$ and thus cannot be
viewed as describing a diffusion process. The diffusion
equation~(\ref{1tc38}) in superspace, on the other hand, does not
couple different values of $k$. In this sense, the diffusion
equation~(\ref{1tc38}) diagonalizes the hierarchic
equations~(\ref{1tc18}).

Starting from Eqs.~(\ref{1tc30}) and~(\ref{1tc38}), one can also
derive stationary equations for the pure ensembles on the original and
the unfolded scale~\cite{Guh96a}.

In the case of symmetry breaking, the standard model is that of a
block--diagonal Hamiltonian $H^{(0)}$ to which the perturbation $\alpha
H^{(1)}$ is added, 
\begin{equation}
H^{(0)} \, = \, \left[ \begin{array}{cc}
                        H^{(0,1)} & 0 \\
                             0    & H^{(0,2)}  
                       \end{array} \right]
\qquad {\rm and} \qquad
H^{(1)} \, = \, \left[ \begin{array}{cc}
                        0 & H^{(1,P)} \\
                        H^{(1,P)\dagger} & 0  
                       \end{array} \right] \ .
\label{1tc42z}
\end{equation}
The matrices $H^{(0,1)}$ and $H^{(0,2)}$ represent the system for two
fixed values $1$ and $2$ of some quantum number such as parity or
isospin. The perturbation $H^{(1)}$ breaks the block--diagonal
structure.  Although symmetry breaking is not fully compatible with
diffusion in superspace, formula~(\ref{1tc42}) was found to apply to
this case~\cite{Guh90}, too. Thus, it is valid beyond the diffusion
model.

\subsubsection{Numerical and analytical results}
\label{qc1tcc}

We follow the scheme mentioned in the introduction to 
Sec.~\ref{qc1tc} and begin with the transition from regularity to
chaos. Classical integrability may be destroyed when some system
parameter such as shape, external field, etc., varies. Then, the
structure of classical phase space changes and chaotic regions
appear. Such change affects the fluctuation properties of the
analogous quantum system. Modeling this process is not trivial,
however, since fluctuation properties of regular systems are not
generic. Therefore, the crossover transition may not have generic
properties either. Nonetheless, much work has been devoted to this
problem. Poisson regularity and harmonic oscillator regularity have
served as main examples of classically integrable systems, see
Secs.~\ref{qc1bab} and~\ref{qc1bac}. 

Several scenarios have been used to study the transition from Poisson
statistics to Wigner--Dyson (WD) statistics: (i) An ensemble of
block--diagonal Hamiltonian matrices consisting of two blocks, one
with Poisson statistics and the other taken from one of the three
classical ensembles, will show a transition from Poisson to WD
statistics as the ratio of the dimensions of the regular and the
random matrix block changes from infinity to zero. Following work by
Mehta~\cite{Meh91}, Berry and Robnik~\cite{Ber84} calculated a spacing
distribution which describes this transition. (ii) An ensemble of
Hamiltonians of the form $H^{(0)}+\alpha H^{(1)}$ discussed in the
previous section where $H^{(0)}$ is diagonal and the diagonal elements
obey Poisson statistics, while $H^{(1)}$ is a member of an ensemble of
random matrices of proper symmetry. The transition parameter $\alpha$
defines the relative strength of $H^{(1)}$. On the unfolded scale, the
parameter $\lambda=\alpha/D$ where $D$ is the mean level spacing has
to be used. The phenomenological Brody formula~\cite{Bro73}
interpolates between the two limiting cases $\lambda=0$ and
$\lambda\to\infty$ without being explicitly derived from the model,
see Sec.~\ref{qc1bab}. (iii) Random band matrices where all matrix
elements located beyond a certain distance from the main diagonal,
vanish. For fixed matrix dimension and decreasing (increasing) band
width, the fluctuation properties approach Poisson (WD)
statistics~\cite{Izr90,Fyo91}, respectively. These, the related chains
of Gaussian ensembles~\cite{Iid90,Zir92} and other models involving
crossover transitions are studied in Secs.~\ref{rbm}, \ref{quasi1d}
and~\ref{spec}, respectively. (iv) Moshe {\it et al.}~\cite{Mos94}
studied a rotation invariant ensemble which interpolates between
Poisson and WD statistics.

A large body of work exists on models of type (ii). Caurier {\it et
  al.}~\cite{Cau90} and Lenz and Haake~\cite{Len91} calculated the
spacing distribution of two--dimensional matrix models.  Numerical
studies for time--reversal symmetric systems were presented in
Refs.~\onlinecite{Guh89,Per95,Miz95,Wei96}. For a small chaotic
admixture, closed formulae were obtained analytically in different,
but equivalent, perturbation schemes by French {\it et
  al.}~\cite{Fre88a}, by Leyvraz and Seligman~\cite{Ley90}, and by
Leyvraz~\cite{Ley93} for the transition from Poisson regularity to all
three Gaussian ensembles. The exact computation of the two--point
function for the entire transition has up to now been possible only in
the unitary case. Unlike the GOE and the GSE, the GUE can be
understood as an interaction--free statistical model,
cf.~Eq.~(\ref{eq_2:24}) of Sec.~\ref{quasi1d_2}. This is reflected in
the fact that the diffusion kernel~(\ref{1tc13}) is known explicitly
as the Harish--Chandra--Itzykson--Zuber integral.  Lenz~\cite{Len92}
calculated the correlation functions with the Mehta--Dyson
method~\cite{Meh91}.  He used Eq.~(\ref{1tc14}) in Eq.~(\ref{1rc1})
for $k=2$ and found a four--dimensional integral representation. His
result contains a ratio of determinants which depend on the level
number. Unfortunately, this representation does not seem to be
amenable to further analytical treatment.  Pandey~\cite{Pan95} worked
out an exact expression for the two--point function on the unfolded
scale in terms of a double integral.  Avoiding formula~(\ref{1tc14})
he used an exact formal solution of the Fokker-Planck equation in
terms of representation functions of the unitary group $U(N)$. Another
exact derivation of the two--point function on both scales, the
original and the unfolded one, was given in
Refs.~\onlinecite{Guh96a,Guh96b} with the help of the supersymmetry
method. Equation (\ref{1tc42}) gives the two--point function as a
double integral for arbitrary initial conditions, and one simply needs
to compute the proper initial condition of the Poisson ensemble. This
can be done straightforwardly.  Rather compact exact expressions for
the $k$--point functions on the original and the unfolded scale were
given in terms of a $2k$--dimensional integral in
Refs.~\onlinecite{Guh96a,Guh96b}.

Most of the concepts just discussed can be adopted to the transition
from harmonic oscillator regularity to chaos. In the spirit of the
Berry--Robnik procedure, one can write down a formula for the spacing
distribution in a block--diagonal model. The transition in a model of
the type $H^{(0)}+\alpha H^{(1)}$ was simulated in
Ref.~\onlinecite{Guh89}. As in the case of the transition from Poisson
regularity, analytical results for the two--point function for the
entire transition are only available in the unitary
case~\cite{Pan95,Guh96c}.

For the breaking of time--reversal invariance, the random matrix
ensemble has to interpolate between the GOE and the GUE limits, see
Eq.~(\ref{Me}). This particular choice for the total Hamiltonian $H$
is often referred to as the Mehta--Pandey Hamiltonian.  In a
perturbative scheme, analytical results for the two--point
correlations and the level number variance were given by French {\it
  et al.}~\cite{Fre88a,Fre88b}. Using the
Harish-Chandra-Itzykson-Zuber integral, Mehta and
Pandey~\cite{Pan83,Pan84} derived exact analytical results on the
original and on the unfolded scale in a pioneering study as early as
1983. They employed Eq.~(\ref{1tc14}) in Eq.~(\ref{1rc1}) and found
that the $k$--level correlation function can be written as a $k \times
k$ quaternion determinant. This expression reproduces the quaternion
determinant in the GOE limit and reduces to the ordinary determinant
in the GUE limit. The calculation benefits from this determinant
structure which is absent in the Poisson to GUE transition. In
Ref.~\onlinecite{All93a} the result for the two--point function on the
unfolded scale was reproduced by using the supersymmetry method and
the saddle--point approximation which leads to Efetov's non--linear
$\sigma$ model. Pandey and Shukla~\cite{Pan91} calculated all
correlation functions of the corresponding transition in the circular
ensembles.

Symmetry breaking is stuidied in terms of the model defined in
Eq.~(\ref{1tc42z}).  Both numerical simulations and analytical results
were reported in Ref.~\onlinecite{Guh90}. The latter were derived for
the unitary case and for equal dimensions of the symmetry--conserving
blocks. The supersymmetry technique was used which led to an integral
representation of the type~(\ref{1tc42}).  A generalization to the
case where the blocks have different dimensions is due to
Pandey~\cite{Pan95}. A detailed discussion in the framework of
perturbation theory was given by Leitner~\cite{Lei93}.  His work
includes a surprisingly efficient formula for the spacing
distribution.

A remarkable speed with which the Wigner--Dyson fluctuations are
reached on short scales is common to all transitions discussed here.
More precisely, as the transition parameter $\lambda=\alpha/D$
increases, observables such as the spacing distribution $p(s)$ become in
almost all cases indistinguishable from the Wigner--Dyson expectation
when the transition parameter is in the regime $0.7 \le \lambda \le 1$
or so. In the spectral long range observables, such as the level
number variance $\Sigma^2(L)$ and the spectral rigidity $\Delta_3(L)$,
one can identify, as a function of $\lambda$, the maximal interval
length $L_{\rm max}$ within which the fluctuations are of
Wigner--Dyson type. This issue is very important in many--body systems
and in disordered systems, it relates to the spreading width and to
the Thouless energy. This discussion will be taken up in Secs.~\ref{mbs}
and~\ref{disorder}.

Finally, we mention a model which employs the concept of transitions
in random matrix theory for a deeper understanding of classical and
semiclassical systems.  Aiming at an understanding of how the the
classical phase space in a mixed system manifests itself in quantum
mechanics, Berry and Robnik~\cite{Ber84} and later, in much more
detail, Bohigas {\it et al.}~\cite{Boh93a} constructed random matrix
models whose block structure can be viewed as generalizations of
certain transition models. This is discussed in Sec.~\ref{qc4mp}.
Formally, these models are in the spirit of the Rosenzweig--Porter
model~\cite{Rose60} which was introduced as early as 1960.

\subsection{Crossover transitions in other observables}
\label{qc1to}

Crossover transitions also affect the distributions of wave functions
and widths. Most authors considered the transition from orthogonal to
unitary symmetry, i.e.~the breaking of time--reversal invariance. To
obtain a family of distributions which interpolates between the
$\chi_\beta^2$ distributions of the widths in Eq.~(\ref{1ww8}) for the
two limiting cases $\beta=1$ and $\beta=2$, one could simply view
$\beta$ as a continuous variable. This is mathematically possible. It
is not clear, however, how to relate the continuous variable $\beta$ to
the strength of an interaction which breaks time--reversal invariance. 
Another interpolating formula due to Zyczkowski and Lenz~\cite{Zyc91} 
has similar shortcomings. 

A meaningful interpolating formula has to be derived from a
Hamiltonian of the form $H=H^{(0)}+\alpha H^{(1)}$ where $\alpha$ must
be a physically well--defined parameter. Such a derivation is
non--trivial since the joint probability densities of eigenvalues and
eigenvectors of the random matrix $H$ factorize only in the two
limiting cases $\alpha = 0$ and $\alpha = \infty$. Sommers and
Iida~\cite{Som94} and Fal'ko and Efetov~\cite{Fal94} gave exact
integral representations for the distributions of wave functions and
widths which are valid for the entire transition regime. The
supersymmetry method is used in both works. Although equivalent, the
starting points are slightly different. In Ref.~\onlinecite{Som94},
the full distribution of eigenvalues and eigenvectors in the
transition regime was used, whereas in Ref.~\onlinecite{Fal94}, the moments
of the eigenvector distribution were calculated.

Fal'ko and Efetov~\cite{Fal96} calculated correlations of wave
functions of the type~(\ref{1ww14}) in the crossover transition from
orthogonal to unitary symmetry and found a surprising result. In the
two limiting cases, the correlations at distant points in space vanish
as expected. For $\beta=1$ and $\beta=2$, the correlation functions
become products of two $\chi_\beta^2$ distributions. However, the
correlations do not vanish in the transition regime. Recently, van
Langen {et al.}~\cite{vanL97} studied these unexpected long--range
correlations by relating them to phase--rigidity fluctuations of
chaotic wave functions.

\subsection{Parametric level motion and parametric level correlations}
\label{qc1pc}

Do chaotic systems possess observables other than energy correlators
or wave function and width distributions which also show a high degree
of universality? This question, answered in the affirmative in recent
years, is treated in the present section.  In Sec.~\ref{qc1pca}, we
summarize results on the motion of energy levels when a parameter of
the system varies. The correlations associated with this motion are
presented in Sec.~\ref{qc1pcb}. In Sec.~\ref{qc1pcc}, we discuss a
generalization of the Gaussian ensembles, the Gaussian processes,
which models these parametric correlations.

\subsubsection{Level motion and curvature}
\label{qc1pca}

Wilkinson~\cite{Wil88,Wil90} studied a classically chaotic quantum
system with Hamiltonian $H(X)$ which is in the same symmetry class for
all values of $X$. On the scale of the mean level spacing, the
spectral fluctuations of $H(X)$ have the same statistics for all $X$
and are given by one of the three Gaussian ensembles. However, both
eigenvalues $E_n=E_n(X)$ and eigenfunctions depend on $X$. For a
chaotic system, there is always level repulsion, in contrast to the
regular case. Hence, a plot of the spectrum of $H$ versus $X$ shows
\begin{figure}
\centerline{
\psfig{file=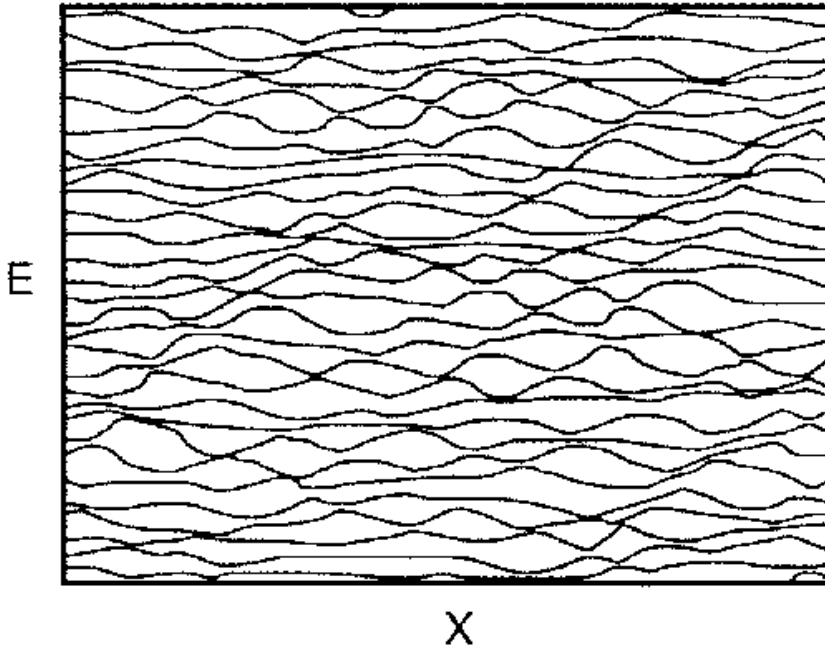,width=4.5in}
}
\caption{
  The motion of eigenenergies as a function of an external
  parameter $X$. Many avoided crossings due to level
  repulsion can be seen, indicating that the level statistics
  at any fixed value of $X$ is predominantly of Wigner--Dyson type.
  This picture illustrates results of a calculation for the
  hydrogen atom in a strong magnetic field by Goldberg 
  {\it et al.}~\protect\cite{Gol91}.
  Taken from Ref.~\protect\onlinecite{Been94}.
}
\label{figft9}
\end{figure}
many avoided crossings, see Fig.~\ref{figft9}.  We consider the
parameter $X$ as a function of a fictitious time $t$, $X=X(t)$.
Suppose that at time $t = 0$, the system is prepared in a high--lying
eigenstate of $H(X(0))$. Wilkinson showed that with increasing $t$,
the occupation probability of eigenstates of $H(X(t))$ spreads
diffusively from the initially prepared state. The parameter $\kappa
\propto dX/dt$ measures the non--adiabaticity of the change of $H$.
For small $\kappa$, the diffusion process is governed by Landau--Zener
transitions at the avoided crossings. The diffusion constant is
proportional to $\kappa^{(\beta+2)/2}$ where $\beta$ labels the three
symmetry classes. This reflects the power law $s^\beta$ of the nearest
neighbor spacing distributions for small spacings. It is argued that,
for large $\kappa$, the diffusion is governed by Ohmic dissipation and
the diffusion constant is proportional to $\kappa^2$, independent of
the symmetry class. The statistics of the avoided crossings was worked
out in Ref.~\onlinecite{Wil89}. In recent, related work, Wilkinson and
Austin~\cite{Wilk95} found a diffusion constant proportional to
$\kappa^2$ for $\beta=1$. This is in agreement with the Kubo formula
(see Sec.~\ref{quasi1d_1}). However, there is a limiting value of
$\kappa^2$ beyond which the rate is lower. To test these results,
these authors also performed numerical studies involving random band
matrices.  This study suggests that the transition mechanism is
different from the above mentioned discussion.  Simultaneously,
Bulgac, Do Dang and Kusnezov~\cite{Bulg97} performed closely related
studies.

A useful local measure of the parametric dependence of eigenvalues is
the cur\-va\-ture
\begin{equation}
K_n = \frac{d^2E_n(X)}{dX^2}\Bigg|_{X=0}
\label{1pc1}
\end{equation}
of the level motion. Indeed, $|K_n|$ takes large values at the avoided
crossings. Several authors~\cite{Pom74,Noi80,Bri85} used $K_n$ or
related quantities to investigate the transition from regular to
chaotic motion. Another measure is the distribution $P(K)$ of
curvatures. We drop the index $n$ because $P(K)$ must be the same for
all eigenvalues. Gaspard {\it et al.}~\cite{Gas89,Gas90a} investigated
$P(K)$ in the fully chaotic regime. These authors used arguments
involving the Pechukas gas~\cite{Pec83} and the generalized
Calogero--Moser system~\cite{Gib83,Nak86} to derive the asymptotic
behavior of $P(K)$ for large $K$. They found that this form is
universal and given by $1/|K|^{\beta+2}$.  As in the case of the
diffusion of occupation probabilities discussed above, this form is
due to level repulsion since roughly $|K| \propto 1/s$. The universal
character of this result was numerically confirmed in
Refs.~\onlinecite{Sah91,Tak92,Zak93,Bra94} for several different
systems. In studying the kicked top, Zakrzewski and
Delande~\cite{Zak93} conjectured that the full distribution is given
by
\begin{equation}
P(k) = \frac{c_\beta}{(1+k^2)^{(\beta+2)/2}}
\label{1pc2}
\end{equation}
where the normalization constants read $c_1=1/2$, $c_2=1/2\pi$ and
$c_4=8/3\pi$. The dimensionless curvature $k =
DK/\pi\beta\langle(dE_n/dX)^2\rangle$ is given in terms of the mean
level spacing $D$ and the mean square level velocity
$\langle(dE_n/dX)^2\rangle$. This quantity is independent of the index
$n$. Using the supersymmetry method, von Oppen~\cite{Opp94,Opp95a}
could prove that the conjecture~(\ref{1pc2}) is indeed true. He used a
random matrix model defined through the Hamiltonian
\begin{equation}
H(X) = H_1 \cos X + H_2 \sin X
\label{1pc4}
\end{equation}
in which both $H_1$ and $H_2$ belong to the same symmetry class. 
Several authors have stressed that at this point a caveat is in
order: Second order perturbation theory yields an exact
expression~\cite{Opp94} for the curvature~(\ref{1pc1}) in the
model~(\ref{1pc4}). Consequently, formula~(\ref{1pc2}) is universal
only for those models which yield an equivalent expression for the 
curvature. The class of, in this sense, equivalent models, however,
is obviously very large. Furthermore, the tail of the distribution
is expected to be rather insensitive to the particular model because
it is directly related to the spacing distribution. In numerical
simulations of billiard systems,
Li and Robnik~\cite{LiR96} found deviations from the formula
(\ref{1pc2}). Recently, Yurkevich and Kravtsov~\cite{Yurk97} 
calculated analytically corrections to this formula.

\subsubsection{Parametric correlation functions}
\label{qc1pcb}

In contrast to the previous section we consider now two different
points $X$ and $X^\prime$ in the space of the parameter which governs
the level motion. We study the correlation of $E_n(X)$ and
$E_m(X^\prime)$ as function of the separation in energy and in
parameter value. For $X=X'$ we have to recover the universal
spectral correlations of RMT. New universal features can only be
expected on scales on which the gross features of the system play no
role, i.e. on the scale of the mean level spacing $D$. We therefore
define the unfolded energies $\varepsilon_n(X)=E_n(X)/D$. We must also
unfold the level motion. The appropriate scale is given by the velocity
$d\varepsilon_n(X)/dX$. We define the new parameter
\begin{equation}
x = \sqrt{\left\langle\left(\frac{d\varepsilon_n(X)}{dX}\right)^2
                                 \right\rangle} \, X \ . 
\label{1pc6}
\end{equation}
The average is performed either over the spectrum or, in an
analytical calculation, over the ensemble. We recall that the same
average is used for the normalization of the curvature parameter $k$
below Eq.~(\ref{1pc2}).

Goldberg {\it et al.}~\cite{Gol91} introduced the parametric
number variance as a probe of correlations between different points in
parameter space. The cumulative spectral function or staircase function,
\begin{equation}
\widehat{\eta}(\varepsilon,x) = 
     \sum_{n=1}^N \Theta(\varepsilon-\varepsilon_n(x))
\label{1pc7}
\end{equation}
serves to define the parametric level number variance
\begin{equation}
v(x) = \left\langle\Big(\widehat{\eta}(\varepsilon,\overline{x}-x/2)-
    \widehat{\eta}(\varepsilon,\overline{x}+x/2)\Big)^2
             \right\rangle
\label{1pc8}
\end{equation}
which is obtained by averaging over both the spectrum (or the
ensemble) and over the position of the midpoint $\overline{x}$ between
two points in parameter space, keeping the distance $x$ fixed. Note
that $v(x)$ is not directly related to the spectral level number
variance $\Sigma^2(L)$: At $x=0$, the function $v(x)$ vanishes by
definition. With increasing $x$, the two staircase functions in
Eq.~(\ref{1pc8}) get more and more ``out of phase''. Hence, we expect
that $v(x)$ is a monotonically increasing function of $x$.

Another natural measure of parametric level correlations is the
correlation function of the velocity of a fixed eigenvalue
$\varepsilon_n$ taken at two different points $\overline{x}-x/2$
$\overline{x}+x/2$
\begin{equation}
c(x) = \left\langle\frac{\partial\varepsilon_n(\overline{x}-x/2)}
                                   {\partial\overline{x}}
         \frac{\partial\varepsilon_n(\overline{x}+x/2)}
                    {\partial\overline{x}}\right\rangle \ .
\label{1pc9}
\end{equation}
This function was introduced by Szafer and Altshuler~\cite{Sza93} who
studied a disordered electronic system in the form of a ring threaded
by a magnetic flux. Using diagrammatic perturbation theory, these
authors~\cite{Alt86} found that beyond a critical regime the
correlator~(\ref{1pc9}) becomes independent of any system properties
and inversely proportional to the squared flux ($x^2$ in the present
notation). Together with the investigations by Goldberg {\it et al.},
this gave the first clear indication that parametric correlations of
level motion possess universal features.

Beenakker~\cite{Bee93b} argued that it ought to be possible to extract
such universal features from the standard concepts of Random Matrix
Theory. Indeed, Dyson's Brownian motion~\cite{Dys62d,Dys72} and the
Pechukas gas~\cite{Pec83} both model the motion of levels as function
of a fictitious time $t$. Beenakker studied Dyson's Brownian motion
under the crucial assumption that $t$ can be identified with the
square of the parameter which governs the level motion. He found that
the smooth behavior in $x$ of the correlator~(\ref{1pc9}) is
universally given by 
\begin{equation}
c_\beta(x) \simeq -\frac{2}{\pi^2\beta x^2}  
                  \qquad {\rm for} \qquad x \gg 1 \ .
\label{1pc10}
\end{equation}
This confirmed the result of Szafer and Altshuler in the unitary case
$\beta=2$. The restriction $x \gg 1$ is not imposed by Dyson's
Brownian motion model itself, but is only due to the particular method
chosen in Ref.~\onlinecite{Bee93b} to derive Eq.~(\ref{1pc10}).
Beenakker pointed out that the universality of his result confirms the
fact that parametric correlations are dominated by level repulsion and
thus uniquely determined by the symmetry class. The specific physical
realization of the parameter $X$ is immaterial. In the unitary case,
the result~(\ref{1pc10}) was also obtained by Brezin and
Zee~\cite{Bre94} who used a diagrammatic expansion.

Within the supersymmetric $\sigma$ model, Simons and
Altshuler~\cite{Sim93a,Sim93b} simultaneously derived exact
expressions for the parametric density--density correlator
at two different energies $\overline{\varepsilon}-\varepsilon/2$ and
$\overline{\varepsilon}+\varepsilon/2$,
\begin{eqnarray}
k_\beta(\varepsilon,x) &=& \sum_{n,m=1}^N \Big\langle 
      \delta(\overline{\varepsilon}-\varepsilon/2-
             \varepsilon_n(\overline{x}-x/2))
                      \nonumber\\
& & \qquad \qquad \qquad \delta(\overline{\varepsilon}+\varepsilon/2-
               \varepsilon_m(\overline{x}+x/2)) \Big\rangle - 1
\label{1pc11}
\end{eqnarray}
and the parametric current--current correlator at fixed energy 
separation
\begin{eqnarray}
&&\widetilde{c}_\beta(\varepsilon,x) = \frac{1}{
      \sum_{n,m=1}^N \left\langle 
      \delta(\varepsilon_n(\overline{x}-x/2)
             -\varepsilon_m(\overline{x}+x/2)) - \varepsilon)
                                 \right\rangle} 
                                 \nonumber\\
 & &  \quad \sum_{n,m=1}^N \bigg\langle 
      \delta(\varepsilon_n(\overline{x}-x/2)
             -\varepsilon_m(\overline{x}+x/2)) - \varepsilon)
         \frac{\partial\varepsilon_n(\overline{x}-x/2)}
           {\partial\overline{x}}
      \frac{\partial\varepsilon_m(\overline{x}+x/2)}
           {\partial\overline{x}} \bigg\rangle \ . \nonumber\\
\label{1pc12}
\end{eqnarray}
The two correlators~(\ref{1pc11}) and~(\ref{1pc12}) are connected
through a Ward identity. Hence, we restrict ourselves to the
density--density correlator~(\ref{1pc11}). Simons and Altshuler
studied a disordered system. They used Efetov's supersymmetric
non--linear $\sigma$ model~\cite{Efe83}. They point out that the same
results can be derived starting from a random matrix model in the
spirit of Ref.~\onlinecite{Ver85}. The integral representations in the
three universality classes read
\begin{eqnarray}
k_1(\varepsilon,x) &=& {\rm Re}\int_{-1}^{+1} d\lambda
                             \int_1^\infty d\lambda_1
                             \int_1^\infty d\lambda_2
          \frac{(1-\lambda^2)(\lambda-\lambda_1\lambda_2)^2}
               {(2\lambda\lambda_1\lambda_2-\lambda^2-
                     \lambda_1^2-\lambda_2^2+1)^2}
                                   \nonumber\\
& & \quad 
    \exp\left(-\frac{\pi^2x^2}{4}(2\lambda_1^2\lambda_2^2-\lambda^2-
                     \lambda_1^2-\lambda_2^2+1) 
        -i\pi\varepsilon^+(\lambda-\lambda_1\lambda_2)\right)
                                   \nonumber\\
k_2(\varepsilon,x) &=& \frac{1}{2}{\rm Re}\int_{-1}^{+1} d\lambda
                             \int_1^\infty d\lambda_1
\exp\left(-\frac{\pi^2x^2}{2}(\lambda_1^2-\lambda^2)
                           -i\pi\varepsilon^+(\lambda-\lambda_1)\right)
                                   \nonumber\\
k_4(\varepsilon,x) &=& \frac{1}{2}{\rm Re}\int_1^\infty d\lambda
                             \int_{-1}^{+1} d\lambda_1
                             \int_{-1}^{+1} d\lambda_2
          \frac{(\lambda^2-1)(\lambda-\lambda_1\lambda_2)^2}
               {(2\lambda\lambda_1\lambda_2-\lambda^2-
                     \lambda_1^2-\lambda_2^2+1)^2}
                               \nonumber      \\
& & \quad
    \exp\left(-2\pi^2x^2(\lambda_1^2+\lambda_2^2+\lambda^2-
                2\lambda_1^2\lambda_2^2-1) 
        +i2\pi\varepsilon^+(\lambda-\lambda_1\lambda_2)\right)
\label{1pc15}
\end{eqnarray}
where the energy difference $\varepsilon$ is given an imaginary
increment. The symplectic case was worked out in
Ref.~\onlinecite{Sim93c}.  For $x=0$, Efetov's integral
representations of the spectral two--point functions~\cite{Efe83} are
recovered. For $\beta=2$, Beenakker and Rejai~\cite{Been94} performed
a calculation in the framework of Dyson's Brownian motion model which
yields full agreement with the results of Simons and Altshuler. {}From
the definition~(\ref{1pc12}) it is obvious that for large $x$, the
current--current correlator $\widetilde{c}(0,x)$ at zero energy
separation has to approximate the velocity correlator $c(x)$ defined
in Eq.~(\ref{1pc9}). Using the Ward identity mentioned above, the
result~(\ref{1pc10}) can indeed be rederived. Remarkably, no integral
representation of the form (\ref{1pc15}) valid for arbitrary $x$ could
\begin{figure}
\centerline{
\psfig{file=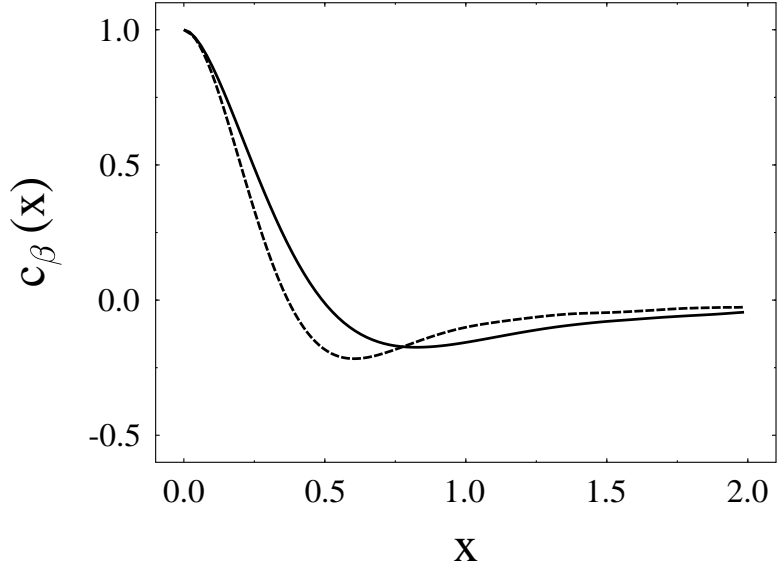,width=4.5in,angle=-90}
}
\caption{
  The parametric velocity correlators $c_\beta(x)$, obtained from
  numerical simulations. The solid line is the GOE result ($\beta=1$),
  the dashed line is the GUE result ($\beta=2$). The curves have a
  very small numerical error. 
  Taken from Ref.~\protect\onlinecite{Mucc96}.
}
\label{figft7}
\end{figure}
be derived for $c(x)$ yet. Mucciolo~\cite{Mucc96,Bruu96} obtained
$c_{\beta}(x)$ by extensive numerical simulations for orthogonal
($\beta=1$) and unitary ($\beta=2$) symmetry. He used
random matrices of the form~(\ref{1pc4}). The results are shown in 
Fig.~\ref{figft7}.

Just as its spectral analogue, the parametric level number variance
can be expressed in terms of the two--point function,
\begin{equation}
v_\beta(x) = 2 \lim_{x^\prime\to 0}
      \int_{-\infty}^0 d\varepsilon
      \int_{-\infty}^{\varepsilon} d\varepsilon^\prime
          \left(k_\beta(\varepsilon^\prime,x^\prime)-
                    k_\beta(\varepsilon^\prime,x)\right) \ .
\label{1pc16}
\end{equation}
For small values of $x$, these integrals can be worked out. For the
unitary and the orthogonal case, one finds
\begin{equation}
v_\beta(x) \simeq \sqrt{\frac{2}{\pi}} |x| 
            \qquad {\rm for} \qquad x \to 0 \ .
\label{1pc17}
\end{equation}
For larger values of $x$, the parametric level number variance can be
obtained by numerical integration of Eq.~(\ref{1pc16}). It turns out
that $v_1(x)$ and $v_2(x)$ begin to differ.  A discrepancy with the
semiclassical formula in Ref.~\onlinecite{Gol91} was attributed by
Simons and Altshuler to the failure of the semiclassical method for
small $x$.

The results derived within the supersymmetric non--linear $\sigma$
model give the most general proof of the universality of parametric
correlations. Although experiments and numerical simulations have
verified this universality, see Ref.~\onlinecite{Sim93d}, deviations
have also been found.

Naturally, the question arises whether observables involving the wave
functions also possess universal features. Using the Gaussian
processes discussed in the following section, Attias and
Alhassid~\cite{Att95} studied the normalized energy--dependent
parametric overlap of wave functions $\psi_n(\vec{r},x)$ at different
values of the dimensionless parameter $x$,
\begin{eqnarray}
\widetilde{o}_\beta(\varepsilon,x) &=&
      \frac{1}{\sum_{n,m=1}^N \left\langle 
      \delta(\varepsilon_n(\overline{x}-x/2)
             -\varepsilon_m(\overline{x}+x/2)) - \varepsilon)
                                 \right\rangle} 
                                      \nonumber\\
& & \qquad \sum_{n,m=1}^N \Big\langle 
  |\langle\psi_n(\vec{r},\overline{x}-x/2)
              |\psi_m(\vec{r},\overline{x}+x/2)\rangle|^2
                                      \nonumber\\       
& & \qquad \qquad \qquad \qquad \delta(\varepsilon_n(\overline{x}-x/2)
    -\varepsilon_m(\overline{x}+x/2)) - \varepsilon)\Big\rangle \ .
\label{1pc17a}
\end{eqnarray}
With the supersymmetry method, exact expressions were given for
$\beta=1$ and $\beta=2$ and compared to numerical simulations of the
Anderson model. In the orthogonal case and for zero energy separation,
the overlap is Lorentzian. As $\varepsilon$ increases, the shape
changes, and the peak moves from $x=0$ to larger values of $x$.

Mucciolo {\it et al.}~\cite{Muc95b} also studied the dependence of
wave functions in the chaotic regime on a parameter, using the full
joint distribution function
\begin{eqnarray}
w_\beta(\overline{a},a,X) &=& \Big\langle\delta(\overline{a}-a/2-
                         V|\psi_n(\vec{r},\overline{x}-x/2)|^2)
                                   \nonumber\\
& & \qquad \qquad \delta(\overline{a}+a/2-
             V|\psi_n(\vec{r},\overline{x}+x/2)|^2)\Big\rangle \ .
\label{1pc17b}
\end{eqnarray}
They showed that in the limit of small $x$, this distribution has
asymptotically the universal form 
\begin{equation}
w_\beta(\overline{a},a,X) \simeq P_\beta(\overline{a})
               \frac{1}{2\pi x\sqrt{\overline{a}}}
  B_\beta\left(\frac{a}{2\pi x\sqrt{\overline{a}}}\right)
            \qquad {\rm for} \qquad x \to 0
\label{1pc17c}
\end{equation}
where $P_\beta$ is the $\chi^2$ distribution defined in Eq.~(\ref{1ww8}).
The function $B_\beta$ is also universal and depends only on the
symmetries of the system. In the unitary case, the function
$B_2$ was given explicitly.

\subsubsection{Gaussian random processes}
\label{qc1pcc}

The universality of the parametric correlation functions suggests
the existence of a proper generalization of the Gaussian ensembles. 
This generalization should lead naturally to the parametric
correlations. In the context of parametric level motion, and for
numerical simulations of the diffusion constants discussed in
Sec.~\ref{qc1pca}, Wilkinson~\cite{Wil90} introduced a parameter
dependence in the GOE. Let the Gaussian distributed white--noise
matrices $W_{nm}(X)$ have zero mean values and a second moment given
by 
\begin{equation}
\overline{W_{nm}(X)W_{n^\prime m^\prime}(X^\prime)}
              = (\delta_{nn^\prime}\delta_{mm^\prime}
                    +\delta_{mn^\prime}\delta_{nm^\prime})
                 \delta(X-X^\prime) \ .
\label{1pc18}
\end{equation}
A parametric dependence of the random Hamiltonian $H$ can then be
defined by convoluting these white noise functions with a smoothly
varying function $h(X)$,
\begin{equation}
H_{nm}(X) = \int_{-\infty}^{+\infty} W_{nm}(X^\prime)
                             h(X-X^\prime) dX^\prime \ .
\label{1pc19}
\end{equation}
For the second moments of the matrix elements this yields
\begin{eqnarray}
\overline{H_{nm}(X)H_{n^\prime m^\prime}(X^\prime)}
              &=& (\delta_{nn^\prime}\delta_{mm^\prime}
                    +\delta_{mn^\prime}\delta_{nm^\prime})
                   \widehat{f}(X-X^\prime)
                         \nonumber\\
\widehat{f}(X) &=& \int_{-\infty}^{+\infty} h(X-X^\prime) h(-X^\prime) 
                                 dX^\prime \ .
\label{1pc20}
\end{eqnarray}
The function $\widehat{f}(X)$ is translation invariant by construction
but not necessarily symmetric in $X$. In studying related questions in
a different context, Ko {\it et al.}~\cite{KoPi76} and Brink {\it et
  al.}~\cite{Brin79} had earlier introduced distributions defined by
an equation similar to Eq.~(\ref{1pc20}). In these cases, however, the
function $f(X)$ had an additional dependence on the matrix indices,
leading to a band structure of the matrices. More recently, related
studies have been performed by Bulgac {\it et al.}~\cite{Bulg97}.
{}From the point of view of statistical mechanics, the introduction of
the dependence on such a parameter $X$ in Eqs.~(\ref{1pc18})
and~(\ref{1pc20}) is tantamount to the introduction of another
dimension: The zero--dimensional Random Matrix Theory becomes
one--dimensional. In Sec.~\ref{cs_1}, we discuss a related important
process, the free propagation of a time--dependent random matrix.

A suitable unifying generalization of Wilkinson's ideas was introduced
by Alhassid and Attias~\cite{Att95,Alh95b} in the form of the Gaussian
random process. Let $H(X)$ denote matrices of dimension $N$, with
volume element $d[H(X)]$, which depend on the parameter $X$. For all
$X$, the matrices belong to the same symmetry class labeled by $\beta$. 
The Gaussian Orthogonal, Unitary and Symplectic Processes, GOP, GUP
and GSP, respectively, are defined through the probability density
\begin{equation}
P(H(X)) \propto \exp\left(-\frac{\beta}{2v^2} 
   \int dX \int dX^\prime {\rm tr}H(X)K(X,X^\prime)H(X^\prime)\right)
\label{1pc24}
\end{equation}
where $K(X,X^\prime)$ is a scalar function. Since the distribution
is Gaussian, it is completely determined by the first two moments
\begin{eqnarray}
\overline{H_{nm}(X)} &=& 0
                                            \nonumber\\
\overline{H_{nm}(X)H_{n^\prime m^\prime}(X^\prime)}
              &=& \frac{v^2}{2\beta} f(X,X^\prime) 
                          g_{\beta,nmn^\prime m^\prime}
\label{1pc25}
\end{eqnarray}
where $g_{1,nmn^\prime m^\prime}=(\delta_{nn^\prime}\delta_{mm^\prime}
+\delta_{mn^\prime}\delta_{nm^\prime})$ in the orthogonal and
$g_{2,nmn^\prime m^\prime}=2\delta_{nm^\prime}\delta_{n^\prime m}$ in
the unitary case. Equation~(\ref{1pc25}) bears a close formal relationship
to Eq.~(\ref{1pc20}). It is assumed, however, that $f$ be both
translation invariant and symmetric, $f(X,X^\prime)=f(|X-X^\prime|)$,
and that $f(0)=1$. This implies that $K(X,X^\prime)$ also obeys
$K(X,X^\prime) = K(|X-X^\prime|)$. The kernel $K$ and the function $f$
are related via their Fourier transforms $\widetilde{K}(k)$ and
$\widetilde{f}(k)$ by  
\begin{equation}
\widetilde{K}(k) = 1 / \widetilde{f}(k) \ . 
\label{1pc26}
\end{equation}
All Gaussian processes can be constructed from the elementary Gaussian
process defined by the white noise weight functions
\begin{equation}
f(X-X^\prime) = \delta(X-X^\prime) \qquad {\rm and} \qquad 
K(X-X^\prime) = \delta(X-X^\prime) \ .
\label{1pc27a}
\end{equation}
The elementary Gaussian process formally includes also the case of a
Gaussian white noise potential. To see this, we interpret the
parameter $X$ as the $d$--dimensional position vector and insert
the Fourier transform of Eq.~(\ref{1pc26}) into Eq.~(\ref{1pc24}) 
for matrix dimension $N=1$.

As an illustration, we present one of the examples given in
Refs.~\onlinecite{Alh95b,Att95}. We consider the Ornstein--Uhlenbeck
Gaussian process~\cite{Uhl30}
\begin{eqnarray}
f(X) &=& \exp(-\gamma|X|) \qquad {\rm and} 
                    \nonumber\\
K(X-X^\prime) &=& \frac{1}{2\gamma}\left(\gamma^2 - 
        \frac{\partial^2}{\partial X^2}\right) \delta(X-X^\prime)
\label{1pc28}
\end{eqnarray}
where $\gamma$ is a positive constant. If the parameter $X$ is
identified with time $t$, the weight functions~(\ref{1pc28}) lead to
the action of a freely propagating matrix with quadratic potential,
see Eq.~(\ref{eq_cs_6}).  Naturally, both the physical and
mathematical significance of $X$ can be very different for different
choices of weight functions.  In Refs.~\onlinecite{Att95,Alh95b}, the
relationship between the Ornstein--Uhlenbeck process and Dyson's
Brownian motion model is also discussed from the viewpoint of Gaussian
processes.

\setcounter{equation}{0}
\section{Many--body systems}
\label{mbs}

We deal with self--bound systems consisting of several or 
many particles which interact through two--body forces.  In the
case of atomic nuclei, the particles are nucleons and in the 
case of atoms and molecules, they are ions and electrons. We can cover
only a small fraction of the enormous wealth of published
material. Atomic nuclei display considerable complexity and have
been studied extensively. This is why we put emphasis on this
field which is discussed in Sec.~\ref{qc3nu}, while atoms and
molecules are the subject of Sec.~\ref{qc3am}. Some general
questions arising in the context of many--body systems are
addressed in Sec.~\ref{qc3gd}.

\subsection{Atomic nuclei}
\label{qc3nu}

Medium--weight or heavy nuclei are complex many--body systems
{\it par excellence}~\cite{Boh69,Bohr75}. A nucleus with mass number
$A=N+Z$ contains $N$ neutrons and $Z$ protons. The size of a
nucleon and the average distance between the nucleons are both
about 1~fm. The range of the strong interaction is also of the order
of 1~fm. The energy scale is MeV (Mega electron volts). 

As outlined in the Historical Survey, the high complexity of nuclei
prompted the introduction of statistical concepts already in the early
days of nuclear physics. This eventually led to Random Matrix Theory.
This development has had enormous impact on statistical many--body
physics at large. To prepare the ground for the introduction of
statistical concepts, we present in Sec.~\ref{qc3nua} a very brief
introduction into the most essential aspects of nuclear physics. In
Secs.~\ref{qc3nub} and~\ref{qc3nuc}, we collect, respectively, the
evidence for random matrix properties of spectral fluctuations, and
data on the distribution of scattering parameters and widths. All
theoretical models for the nucleus are effective, i.e., do not start
from the basic nucleon--nucleon interaction. Thus, it is important to
see whether the fluctuation properties displayed by the data are also
borne out in model calculations. This is discussed in
Sec.~\ref{qc3nud}. In Sec.~\ref{qc3nue} we summarize some of the work
which employs Random Matrix Theory as a tool to obtain upper bounds or
values for the strengths of invariance-- and/or symmetry--breaking
forces.

The are numerous reviews on Random Matrix Theory in nuclear 
physics. We mention
Refs.~\onlinecite{Por65,Garg72,Mah80,Weid80,Bro81,Fren82,Seli86,Bohi88,Zele96a}.

\subsubsection {Basic features of nuclei}
\label{qc3nua}

The cross section for the elastic scattering of slow neutrons on
heavy nuclei displays sharp resonances with widths ranging
from about 3~meV to 1~eV, and with spacings of about 10~eV,
see Fig.~\ref{fig2}.  These resonances cannot be due to a single
interaction between the incident neutron and a target nucleon.
Indeed, a straightforward estimate shows that the inverse
reaction time for such a two--body collision yields a value for the
resonance width which is in the MeV range. The small observed
widths correspond to much longer interaction times. This requires
that many nucleons in the target participate in the reaction, and
that each resonance has a complex wave function. While it may
thus be difficult to account for the properties of individual
resonances, the average of the cross section over many
resonances can be obtained by the following statistical argument. The
available energy is assumed to be spread uniformly over all
participating nucleons in such a way that the target nucleus and the
incident neutron together form a new system, the compound nucleus. 
This system equilibrates and loses memory of its mode of formation
before it decays. As a consequence, formation and decay of the
compound nucleus are independent processes. This is the essence
of Bohr's picture~\cite{Bor36} of the compound nucleus. 

The picture yields a formula for the average cross section $\langle
\sigma_{ab} \rangle$. The incident channel (in our example, the slow
neutron) is denoted by $a$. The average probability for
formation of the compound nucleus from the incident channel is given
by the transmission coefficient $T_a$.  The decay of the compound
nucleus into other channels $c$ is likewise governed by associated
transmission coefficients $T_c$.  Depending on the available energy,
$c$ may stand for an inelastically scattered neutron, a proton, an
$\alpha$ particle, etc.  In Bohr's statistical picture, $\langle
\sigma_{ab} \rangle$ is proportional to $T_a G_b$ where $G_b$ is the
probability for the decay into channel $b$. Since the nucleus has to
decay into one of the $\Lambda$ open channels, we have
$\sum_{b=1}^\Lambda G_b=1$. On the other hand, $G_b$ must be
proportional to $T_b$.  Thus, $G_b=T_b/\sum_{c=1}^\Lambda T_c$. This
yields the Hauser--Feshbach formula~\cite{Hau52},
\begin{equation}
\langle\sigma_{ab}\rangle = C_{\rm kin}
                          \frac{T_aT_b}{\sum_{c=1}^\Lambda T_c} \ .
\label{3nu1}
\end{equation}
Kinematic factors have been absorbed in the constant 
$C_{\rm kin}$. 

These stochastic concepts received strong support when in 1966
Ericson~\cite{Eri79} showed that the nuclear cross section can be
viewed as a random variable. His ideas apply in the regime where
the average resonance width is much larger than the mean level
spacing, i.e. in the regime of strongly overlapping resonances. At
any one energy, the cross section is due to the superposition of
many resonances. If each resonance contributes a random
amplitude, the cross section fluctuates randomly versus energy.
Minima and maxima are not related to individual resonances.
Nevertheless, the cross section is fully reproducible. Therefore,
the fluctuations do not represent white noise, but internal
stochastic behavior of the compound system.

Prior to Ericson's observation, the discovery of shell structure in
nuclear binding energies had led to quite different a picture of the
nucleus. At certain values of the neutron and proton numbers $N$ and 
$Z$, the nuclear binding energy displays maxima. As in atomic
physics, these maxima can be explained in the framework of a
shell model~\cite{Goep55}. In contrast to the atomic case, a strong
spin--orbit force has to be invoked to describe the data. In the
nuclear shell model, each nucleon moves independently in a mean
field $U$ generated by all other nucleons. The Hamiltonian takes
the form
\begin{eqnarray}
H &=& \sum_{k=1}^A \left(T(k)+U(k)\right) + H_{\rm res}
                                                 \nonumber\\
H_{\rm res} &=& \sum_{k<l}W(k,l) - \sum_{k=1}^A U(k) \ .
\label{3nu1a}
\end{eqnarray}
Here, $T(k)$ is the kinetic energy and $W(k,l)$ is the two--body
interaction between nucleons $k$ and $l$. The difference 
$H_{\rm res}$ between the true two--body interaction and the
mean field is referred to as the residual interaction. (This picture
is oversimplified and does not account for the strong repulsion
between pairs of nucleons.) The model is very successful in
accounting for the low--energy spectra of many nuclei,
particularly when the residual interaction is chosen
phenomenologically. In the space of Slater determinants of
shell--model states, the Hamilton matrix acquires a highly complex
structure, making it a well--suited object for a statistical
approach. In this basis, the matrix elements of $H_{\rm res}$ have a
nearly Gaussian distribution. This fact links the shell model to the
random matrix approach to nuclear reactions.

Nuclei are capable of yet another type of motion. This was discovered
shortly after the emergence of the shell model. A nucleus can be
viewed as a piece of dense but elastic matter with spherical or
ellipsoidal shape. This object can vibrate and, in the case of
deformation, rotate. Collective modes, the quantal manifestations of
such motion, occur in many nuclei. The minimum rotational energy
$E(I)$ a nucleus must have for a given total spin $I$ defines a lower
bound for its total energy $E$. In the $(E,I)$ plane, the curve of
$E(I)$ versus $I$ has parabolic shape. It is called the yrast line.
This curve defines rotation without inner excitation (``cold
rotation''). Excitations heat up the system. On the yrast line, the
collective motion decays through the emission of electric quadrupole
($E2$) radiation. A state of spin $I$ on the yrast line decays into a
state of spin $I-2$, and so on.  States connected by a sequence of
such strong $E2$ transitions form a rotational band. Near the yrast
line, the members of a rotational band are characterized by the $K$
quantum number. This is the projection of the intrinsic angular
momentum (caused by single--particle excitations) onto the symmetry
axis of the deformed nucleus. Collective excitations often follow an
easily predictable pattern, leaving little room for the kind of
stochastic behavior modeled by random matrices.

Collective excitations at low energy and low angular momentum values
are often modeled in terms of the Interacting Boson Model
(IBM)~\cite{Iach87}. Due to the strong pairing force in the nuclear
Hamiltonian, pairs of protons or of neutrons couple to bosonic pairs
which form the basic ingredients of the model. In its basic version,
the model contains Bosons with spin zero ($s$ Bosons) and Bosons with
spin two ($d$ Bosons). For nuclei with even $A$, the IBM Hamiltonian
is solely expressed in terms of these Bosons. Nucleons do not appear.
The Hamiltonian has the form
\begin{equation}
H = c_0 n_d + c_2 Q^\chi\cdot Q^\chi + c_1 \vec{L}^2 \ .
\label{3ibm1}
\end{equation}
Here $\vec{L}$ is the total angular momentum operator and $c_0$, $c_1$
and $c_2$ are parameters. The operator $n_d$ counts the number of $d$
Bosons. The quadrupole operator $Q^\chi$ models the deformation of the
nucleus. It depends on a parameter $\chi$.  The
Hamiltonian~(\ref{3ibm1}) possesses a U(6) and an O(3) symmetry,
corresponding to the conservation of Boson number and angular
momentum. Three different group chains connect U(6) with O(3). Each
chain represents a dynamical symmetry which is realized in the
Hamiltonian~(\ref{3ibm1}) for a certain choice of parameters.
Physically, these dynamical symmetries correspond to vibrational,
rotational and deformation--instable nuclei. Every IBM Hamiltonian
describing a real nucleus corresponds to a point in the space of
parameters of the IBM.  The IBM can be viewed as a mapping of the
space of shell--model Slater determinants onto a --- drastically
smaller --- sub--space spanned by bosonic wave functions.

In summary, independent--particle motion with coupling through
the residual interaction, collective modes due to deformation, and
a stochastic approach form the main ingredients of nuclear theory.
These elements have different relative weights at different
excitation energies and mass numbers $A$. 

For the statistical analysis, the concept of the spreading width 
has proven very useful. We consider a nucleus modeled by a 
Hamiltonian $H=H_0+V$. The first part, $H_0$, describes the sub--system
which is pure with respect to symmetries, invariances or other
properties. The second part, $V$, is a perturbation which destroys
this property. For example, $H_0$ conserves time--reversal invariance
and $V$ breaks it, or, $H_0$ models purely collective motion while
$V$ contains single--particle excitations. The influence of the perturbation
can most conveniently be measured if the eigenbasis of $H_0$ is taken
as the ``reference frame''. As $V$ increases, an eigenstate of $H$ 
spreads out, starting from an initial eigenstate of $H_0$. The averaged
energy scale of this spreading in the spectrum can be shown to be
given by the spreading width
\begin{equation}
\Gamma = 2\pi \frac{\langle V^2 \rangle}{D_0}
\label{3nua2}
\end{equation}
where $D_0$ is the mean level spacing due to $H_0$ and
$\langle V^2 \rangle$ is the mean square perturbation matrix element.
The spreading width $\Gamma$ should not be confused with the
decay width of an open quantum system. It is conceptually
closely related to the Thouless energy which is used in condensed
matter physics, see Sec.~\ref{disorder}. The spreading width is
frequently used to characterize the strength of a perturbation $V$.
This is done because $\Gamma$, in contrast to both the matrix elements
of $V$ and $H_0$, depends only very little on excitation energy or
mass number. Indeed, $D_0$ decreases essentially exponentially with
excitation energy $E$, and the same is true for $\langle V^2 \rangle$
because the complexity of eigenstates of $H_0$ grows strongly with
$E$. It can be shown~\cite{Bro81} that these exponential dependences
cancel in the ratio $\Gamma$ of Eq.~(\ref{3nua2}).

\subsubsection{Spectral fluctuations}
\label{qc3nub}

Data of sufficiently high quality to test spectral fluctuations in
nuclei are available only in restricted energy windows. It is crucial
that sequences of levels of the same spin and parity be complete (no
missing levels) and pure (no levels with wrong quantum numbers).
Spectra which contain levels of all spins and parities yield a Poisson
distribution~\cite{Kats66,Huiz67}, see Sec.~\ref{qc1bae}. One such
window is defined by the scattering of slow
neutrons~\cite{Garg64,Liou72,Liou75,Jain75}, see Fig.~\ref{fig2}, and
of protons at the Coulomb barrier~\cite{Wats81}. Earlier than in other
areas of physics, the energy resolution and the size of the data set
attained here were sufficient for a statistical analysis. In 1982,
Haq, Pandey and Bohigas~\cite{Haq82} analyzed the set of 1726 nuclear
spacings then available from such data. This ``nuclear data ensemble''
is a combination of data on neutron~\cite{Hack78,Chri80} and
proton~\cite{Bilp76} resonances. For neutrons, about 150 to 170
$s$ wave resonances per nucleus located typically about 9~MeV above
the ground state, contribute to the ensemble. For protons, the
sequences consist of 60 to 80 levels per nucleus. After separate
unfolding, the sequences were lumped together and analyzed. Remarkable
agreement of both spacing distribution (see Fig.~\ref{fig1}) and
spectral rigidity with GOE predictions was found. Three-- and
four--level correlations also agree well~\cite{Bohi85}. Recently,
Lombardi, Bohigas and Seligman~\cite{Lom94} analyzed the spectral
two--level form factor of the nuclear data ensemble and applied the
GSE test discussed in Sec.~\ref{qc1baf}. Again, consistence with
random matrix predictions was found.

\begin{figure}
\centerline{
\psfig{file=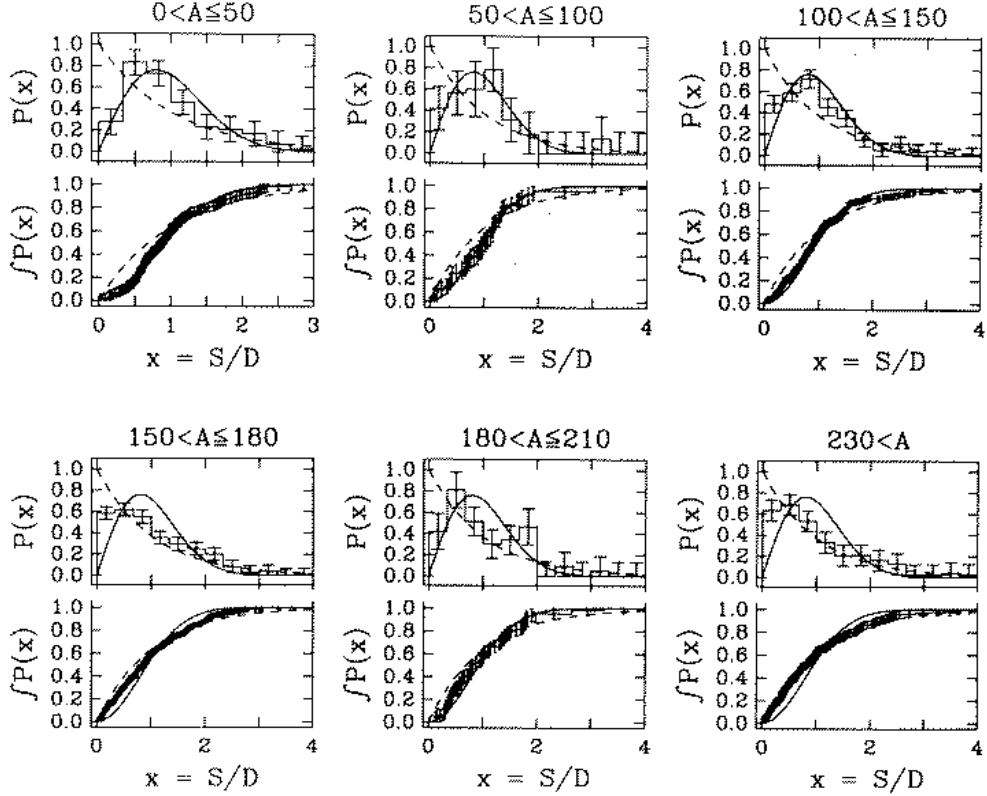,width=5.5in}
}
\caption{
  Nearest neighbor spacing distributions and cumulative spacing
  distributions for all levels in six different regions of the nuclear
  mass parameter $A$. The dashed lines show the behavior for the
  Poisson distribution, and the smooth solid lines show it for the
  Wigner surmise.
  Taken from Ref.~\protect\onlinecite{Shri91}.
}
\label{figmb5}
\end{figure}

Another window for a test of spectral fluctuations exists in the
ground--state domain. This region was first investigated by Brody {\it
  et al.}~\cite{Brod76} who found some evidence for GOE--like
behavior. A much more detailed study is due to von Egidy, Behkami and
Schmidt~\cite{vEgi85,vEgi87}. The data set of this analysis consists
of 1761 levels from 75 nuclides between $^{20}$F and $^{250}$Cf. The
set is obtained from the analysis of $(n,\gamma)$ reactions. The data
are separated into pure sequences. The combined nearest neighbor
spacing distribution lies between a Poisson distribution and the
Wigner surmise. Abul--Magd and Weidenm\"uller~\cite{Abul85} divided
the data set of Ref.~\onlinecite{vEgi85} into a number of groups and
analyzed the spacing distribution and the spectral rigidity within
each group.  It was found that the collective states with spin $J$ and
parity $\pi$ values of $J^\pi=2^+$ and $4^+$ in rotational nuclei show
Poisson characteristics whereas the fluctuations of the remaining
levels are much closer to GOE predictions. A more complete study along
similar lines is due to Shriner, Mitchell and von Egidy~\cite{Shri91}.
These authors analyzed a data set of 988 spacings from 60 nuclides for
which the level schemes are believed to be complete in an energy
interval starting with the ground state, and over some range of
angular momentum values.  The results show a strong mass dependence.
The spacing distributions of the lighter nuclides are close to GOE
while those of the heavier ones are closer to a Poisson distribution,
see Fig.~\ref{figmb5}.
This suggests that the coupling of independent particles by the
residual interaction generically leads to GOE behavior, and that
collectivity is the origin of Poisson statistics. Further support for
this hypothesis comes from the investigation of spin and shape effects
in Ref.~\onlinecite{Shri91}, although the number of states available
was rather small. The spacing distributions of the $2^+$ and $4^+$
states in heavier and deformed nuclides (where collectivity is
strongest) are close to Poisson while those of the $0^+$ and $3^+$
states in the same nuclides and the $2^+$ and $4^+$ states in
spherical nuclides are close to GOE. A Poisson distribution would be
expected for a superposition of states with different quantum
numbers~\cite{Meh91}, see Sec.~\ref{qc1bae}.  Candidates for such
unidentified conserved quantum numbers are isospin $T$ and the $K$
quantum number introduced in the previous section. However, even a
weak breaking of the symmetry associated with $T$ or $K$ quickly leads
to GOE statistics, see Sec.~\ref{qc1tcc}. In the data of Shriner {\it
  et al.}~there is a hint towards an interuption of the gradual
transition from Wigner--Dyson to Poisson statistics in the mass region
$180 < A \le 210$.  Abul--Magd and Simbel~\cite{Abul96} argued that
this could be related to the occurence of oblate deformation in this
mass region.

Another window exists in deformed rare--earth nuclei with mass numbers
$A$ between 155 and 185. In contrast to the cases discussed above, the
data are taken near the yrast line and contain much higher spin
values. Garrett and co--workers~\cite{Garr94a,Garr94b} analyzed 2522
level spacings. Very recently, the analysis was extended~\cite{Garr97}
and includes now 3130 spacings.  The average spacing was calculated
for the total ensemble, before dividing it into pure sequences of
fixed spin and parity, because the properties of the deformed nuclei
in this mass region are known to be similar. The authors state that,
for this and other reasons, the usual method to determine the average
spacing for individual pure sequences can be replaced by this
procedure of combining the sequences in the yrast region.  The
resulting spacing distribution is the closest to a Poisson
distribution yet obtained except for a statistically significant
suppression of very small spacings. It is argued that this suppression
indicates that $K$ is (nearly) a good quantum number for the
low--lying states in deformed rare--earth nuclei.

Little is known about the spectral fluctuations of individual nuclei.
The most complete data sets available are probably the ones for
$^{26}$Al, analyzed by Mitchell {\it et al.}~\cite{Mit88,Shri90}, and
for $^{116}$Sn, analyzed by Raman {\it et al.}~\cite{Rama91}. The
results for $^{26}$Al are important for the study of symmetry breaking
and are discussed in Sec.~\ref{qc3nue}. The authors of
Ref.~\onlinecite{Rama91} present an almost complete level scheme of
$^{116}$Sn from the ground state to a maximum excitation energy of
$4.3~{\rm MeV}$. The resulting spacing distribution lies between a
Poisson and a Wigner distribution.  Bybee {\it et al.}~\cite{Bybe96}
applied the correlation--hole method (see Sec.~\ref{qc1bad}) to
spectra of $^{57}$Co and $^{167}$Er.  However, these authors conclude,
that the limited sample size of their data yields ambiguous results
whereas the spacing distribution and the spectral rigidity are
statistically more significant.

\begin{figure}
\centerline{
\psfig{file=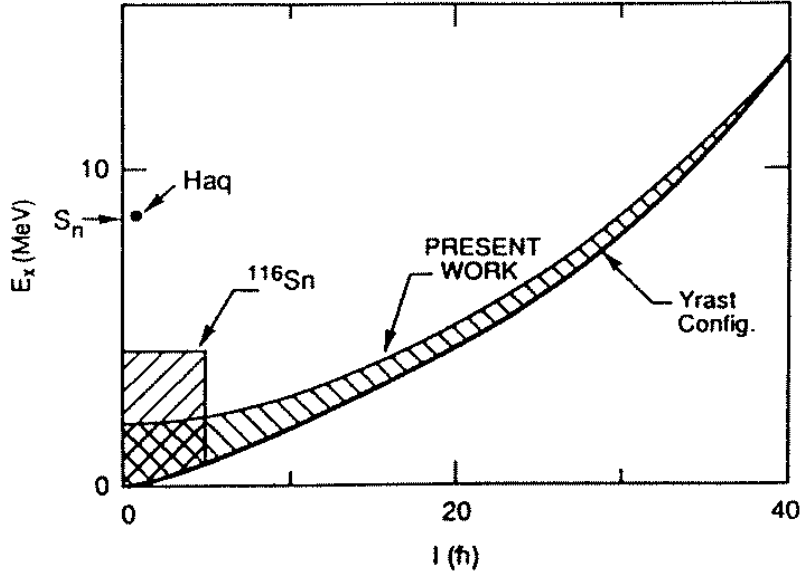,width=4.5in,angle=180}
}
\caption{
  The range of excitation energies $E_x$ and angular momenta $I$
  associated with the three data sets mentioned in the text.  The
  point in this diagram from which the levels for the nuclear data
  ensemble are taken is labeled with ``Haq''. The shaded area labeled
  ``$^{116}$Sn'' indicates the area in the plane from which the data
  for the analysis of Raman {\it et al.}~\protect\cite{Rama91} were
  gathered. The shaded area designated ``present work'' refers to the
  analysis of Garrett {\it et al.}~\protect\cite{Garr94a,Garr94b,Garr97}.  
  The yrast line is shown as a solid line.  
  Taken from Ref.~\protect\onlinecite{Garr97}.  
}
\label{figmb7}
\end{figure}

In Refs.~\onlinecite{Garr94b,Garr97} it is attempted to draw a
coherent picture of the results summarized above. In the work of
Refs.~\onlinecite{Garr94a,Garr94b}, the value of the mean level
spacing, $D=297~{\rm keV}$, considerably exceeds the typical size of
the interaction matrix elements, $V_{\rm int} < 35~{\rm keV}$, between
the states. Thus, the fluctuations are expected to be close to
Poisson.  For small spacings $s < 70~{\rm keV}/D$, a deviation from
the Poisson distribution is seen because here $s$ is comparable with
$V_{\rm int}/D$. The other extreme is provided by the nuclear data
ensemble analyzed in Refs.~\onlinecite{Haq82,Bohi85,Lom94}.  Here, the
mean level spacing $D$ is much smaller than a typical interaction
matrix element, $D\ll V_{\rm int}$. The states are strongly mixed and
the fluctuations are GOE. In the case of $^{116}$Sn, the mean level
spacing $D=110~{\rm keV}$ is comparable with the interaction matrix
element $V_{\rm int}$, causing the spacing distribution to lie between
the Poisson and the Wigner distribution. Figure~\ref{figmb7} shows the
three windows in which the aforementioned data were taken.  This
interpretation suggests a statistical model of the type
$H^{(0)}+\alpha H^{(1)}$ where $H^{(0)}$ and $H^{(1)}$ model Poisson
and GOE fluctuations, respectively, see Sec.~\ref{qc1tc}. The
parameter $\lambda=\alpha/D$ is of the order of $V_{\rm int}/D$.

\subsubsection{Resonances and scattering}
\label{qc3nuc}

Aside from the resonance positions, data for resonance reactions also
yield partial and total widths. RMT predicts the distribution of
partial widths to be of Porter--Thomas form, see Sec.~\ref{qc1wwa}.
Prior to a comparison of data with this prediction, it is necessary to
remove the penetration factors due to the Coulomb and/or angular
momentum barriers from the measured partial widths. Early experimental
evidence for the validity of the Porter--Thomas distribution for the
resulting ``reduced'' widths is summarized in Refs.~\onlinecite{Por65}
and~\onlinecite{Garg72}. Later, the distributions of reduced widths for
neutron~\cite{Chri80} and proton~\cite{Mitc85} resonance data were
shown to be consistent with the Porter--Thomas law. However,
Harney~\cite{Harn84} pointed out that a statistically significant test
requires many more data than were available at the time. Therefore,
Shriner, Mitchell and Bilpuch~\cite{Shri87,Shri89} used another
statistical observable. The Porter--Thomas distribution for the
reduced widths results from the Gaussian distribution for the reduced
width amplitudes, see Sec.~\ref{qc1wwa}, and that latter distribution
is tested. With $n$ the running index of a set of resonances, the
reduced width amplitudes $\gamma_{na}$ and $\gamma_{nb}$ for two
different channels $a$ and $b$ are taken as two data sets. If two sets
$\{x_1,\ldots,x_N\}$ and $\{y_1,\ldots,y_N\}$ have a Gaussian
distribution, the linear correlation coefficient
\begin{equation}
\rho(x,y) = \frac{\sum_{n=1}^N\left(x_n-\langle x \rangle\right)
                              \left(y_n-\langle y \rangle\right)}
           {\sqrt{\sum_{n=1}^N\left(x_n-\langle x \rangle\right)^2
                  \sum_{n=1}^N\left(y_n-\langle y \rangle\right)^2}}
\label{3nu2}
\end{equation}
has the property $\rho(x^2,y^2)=\rho^2(x,y)$. This test is far more
sensitive than a comparison with the Porter--Thomas law because it
involves the joint moments of the partial width amplitudes
$\gamma_{na}$ and $\gamma_{nb}$ and, thus, the relative signs of the
amplitudes.  This information was provided by the analysis of
interference phenomena in the exit channels in proton--induced
reactions. The analysis yields a statistically significant
confirmation of GOE predictions.

The successful analysis of the distribution of the positions and
widths of well--resolved isolated resonances in terms of the GOE
implies that compound nucleus scattering itself is a stochastic
process and can be modeled by the stochastic scattering problem
introduced in Sec.~\ref{qc1cs}. It is assumed that this modeling
applies not only in the domain of isolated resonances but also at
higher energies where resonances overlap. Apart from kinematic
factors, the compound nucleus cross section is then given by the
square modulus of the scattering matrix in Eq.~(\ref{1cs2}) of
Sec.~\ref{qc1cs}, and the fluctuation properties of compound nucleus
cross sections can be predicted. The history of this problem is
mirrored in the reviews in Refs.~\onlinecite{Garg72,Mah80,Bro81}. For
a long time, only perturbative solutions were available, valid either
for small or for large values of the ratio $\Gamma / D$. Here,
$\Gamma$ is the average total width of and $D$ the mean spacing
between resonances. Use of the supersymmetry method~\cite{Ver85} made
it possible to work out the fluctuations of the scattering matrix in
all regimes, from isolated resonances to the Ericson regime of
strongly overlapping resonances, see Sec.~\ref{qc1csb}.  The result is
also of practical use. An example is the application to reactor
physics at low energies, see Ref.~\onlinecite{Qaim92}. The leading
term of an expansion of the result in inverse powers of
$\sum_{c=1}^\Lambda T_c$ yields the Hauser--Feshbach
formula~(\ref{3nu1}). Nishioka and Weidenm\"uller~\cite{Nish85x}
showed that direct reactions can easily be included in the formalism.

The use of the GOE in this model for compound nucleus scattering
implies that the internal equilibration time of the nucleus is small
in comparison with the decay time $\hbar / \Gamma$. This is true at
low incident energies. However, in reactions induced by light ions,
$\hbar / \Gamma$ decreases strongly with energy and becomes comparable
to the equilibration time for incident energies higher than 10~MeV or
so. Then, the compound nucleus can decay before it is completely
equilibrated, and the model of Sec.~\ref{qc1csb} must be extended to
account for this fact. This is done by using the nuclear shell model
and dividing the compound system into classes of states. Each class
comprises states with fixed particle--hole number and is represented
by a random matrix. Owing to the two--body character of the residual
interaction, only neighboring classes are coupled. This coupling is
also modeled by random matrices. The total Hamiltonian is then a band
matrix whose entries are random matrices. Following earlier work by
Agassi, Weidenm\"uller and Mantzouranis~\cite{Agas75} and Feshbach,
Kerman and Koonin~\cite{Fesh80}, Nishioka {\it et al.}~\cite{Nish86}
succeeded in calculating the average cross section for this model.
They used the supersymmetry method and a loop expansion~\cite{Weid84}.
The strong breaking of rotation invariance in Hilbert space of this
random matrix model amounts to the introduction of a dimension. This
is why this model later became important for mesoscopic systems, see
Sec.~\ref{quasi1d_3}. The ideas used in the model have been tested
perhaps most stringently in nuclear reactions with isospin mixing, see
Sec.~\ref{qc3nue}.

Another approach to compound nucleus scattering considers the 
scattering matrix itself as the basic random entity and uses 
the maximum entropy approach to determine the fluctuation
properties, see Sec.~\ref{qc1csb}. The random transfer matrix
technique used in mesoscopic physics (see Sec.~\ref{quasi1d_2}) 
can be viewed as an extension of this approach.

\subsubsection{Model systems}
\label{qc3nud}

As mentioned above, it is important to test whether spectral
fluctuation properties predicted by various nuclear models correctly
reproduce the properties of nuclei reviewed in the last two sections.
This is true, in particular, because all models of the nuclear
many--body problem are effective models based on mean--field
approaches.

There are numerous studies of spectral fluctuation properties within
the nuclear shell model. Early work is reviewed in
Ref.~\onlinecite{Bro81}.  We only mention the studies of Soyeur and
Zuker~\cite{Soye72} and of Wong~\cite{Wong70} which date back to the
early seventies. The former authors found good agreement of the spin
$J=2$ states in $^{24}$Mg with the Wigner surmise. Wong calculated
spectra including several $J$ values whose spacing distributions
follow the Poisson law. We can also mention only a few of the recent
results. Ormand and Broglia~\cite{Orma92} performed realistic
shell--model calculations for light nuclei in the mass region
$20<A<35$ and for $^{46}$V. Very good agreement of the spacing
distribution and the spectral rigidity with the GOE prediction was
found in the region near the ground state.  However, in these
calculations isospin conservation was assumed, see Sec.~\ref{qc3nue}.

The influence of the mean field was investigated by Dro\.zd\.z {\it et
al.}~\cite{Droz94} in the spectrum of two--particle two--hole
states. As expected on general grounds, there is no level
repulsion within a pure mean--field approach. GOE type
fluctuations occur when the residual interaction is switched
on. Level repulsion arises from particle--hole re--scattering effects.
Zelevinsky {\it et al.}~\cite{Zele96a,Zele96b} calculated spectra of light
nuclei as a function of the strength of the residual interaction. A
relatively weak interaction suffices to produce a Wigner type
spacing distribution. This is agreement with the RMT expectations,
see Sec.~\ref{qc1tc}.

The distribution of the amplitudes of shell--model wave functions in a
fixed basis has also been compared to the GOE prediction~\cite{Bro81}.
In the ground--state domain, Whitehead {\it et al.}~\cite{Whit78} and
Verbaarschot and Brussaard~\cite{Verb79} found deviations from the
Gaussian behavior predicted by the GOE. The single--particle states
within a major shell are not completely degenerate. The splitting of
the single--particle energies counteracts the mixing of states by the
residual interaction. Interestingly, Whitehead {\it et
  al.}~\cite{Whit78} set up a random matrix with a block structure in
the spirit of the Rosenzweig--Porter model to be discussed in
Sec.~\ref{qc3ama}.  Brown and Bertsch~\cite{Brow84} showed that the
distribution of the amplitudes becomes Gaussian at higher excitations
where the level density is larger. It was also argued by Zelevinsky
{\it et al.}~\cite{Zele95} that calculations with degenerate
single--particle energies yield GOE features at lower energies than
models which are believed to describe the nucleus in a more realistic
fashion.  A discussion can be found in Zelevinsky's
review~\cite{Zele96a}. The decay in time of the compound nucleus has
also been the object of several studies. For many (few) open channels,
RMT predicts exponential (algebraic) decay~\cite{Lew91,Ditt92,Har92},
respectively.  This was compared with model calculations in various
regimes~\cite{Droz96,Pers96}.

The interplay between collective motion and single--particle effects
can ideally be studied in hot rotating nuclei, cf.  Sec.~\ref{qc3nua}.
The results reviewed in Sec.~\ref{qc3nub} suggest that collective
motion is likely to yield Poisson or harmonic oscillator statistics
whereas single--particle effects should lead to GOE behavior. Reviews
were given by \AA berg~\cite{Aber92} and D\o ssing {\it et
  al.}~\cite{Doss96}.  Rotational bands exist not only near the yrast
line (see Sec.~\ref{qc3nua}), but also at excitation energies of a few
MeV above it. At these energies, the rotational bands are mixed by the
residual interaction. Thus, a state with spin $I$ does not decay into
a single state of spin $I - 2$ within the same band, but into several
or even many states with spin $I-2$. This phenomenon is referred to as
rotational damping and is characterized by a spreading width, see
Sec.~\ref{qc3nua}. Experimental and theoretical evidence for
rotational damping was given by Leander~\cite{Lean82} as early as
1982. Numerous investigations have since been added. We only mention
the work of Lauritzen, D\o ssing and Broglia~\cite{Laur86}. The
picture was confirmed experimentally~\cite{Hers92a,Hers92b}. Guhr and
Weidenm\"uller~\cite{Guh89} interpreted rotational damping in terms of
the random matrix model $H^{(0)}+\alpha H^{(1)}$ discussed in
Sec.~\ref{qc1tc}. The collective part, modeled by $H^{(0)}$, is chosen
to have a Poisson or harmonic oscillator spectrum. The mixing is
modeled by $H^{(1)}$ which is drawn from the GOE. The influence of the
perturbation rises with excitation energy. \AA berg~\cite{Aber90}
addressed the same issue in a still schematic but more realistic model
and obtained similar results. Person and \AA berg~\cite{Per95} and
Mizutori and \AA berg~\cite{Miz95} studied a random matrix model
based on \AA berg's approach. They identified a critical scale $L_{\rm
  max}$ in the energy spectrum. Below this scale, the long--range
correlations of levels are of GOE type and above it, they become
Poisson--like. This length scale is related to the spreading width or
``energy localization length'' of the wave functions. The problem is
similar to problems in disordered systems, see Sec.~\ref{shp}. In the
random matrix model $H^{(0)} + \alpha H^{(1)}$ with $H^{(1)}$ drawn
from the GUE, Guhr and M\"uller--Groeling~\cite{Guh96} studied such a
critical length analytically. D\o ssing {\it et
  al.}~\cite{Doss96,Doss97} discussed microscopic models of rotational
damping and compared with GOE predictions.
 
Various studies address single--particle motion in deformed
potentials. This motion may itself be chaotic. Of the many works on
this subject, we mention, {\it inter alia}, the early papers by Arvieu
{\it et al.}~\cite{Arvi87,Arvi91} which focus on classical and
semiclassical aspects.  Milek, N\"orenberg and Rozmej~\cite{Mile89}
studied a two--center shell model. Such a model is used in the theory
of nuclear fission. It describes the separation of a spherical nucleus
via deformation and formation of a neck into two separated spherical
nuclei. Poisson--like spacing distributions were found in the two
limiting cases. When a neck is formed and rotational symmetry is
broken, the spacing distribution agrees with the Wigner surmise.
Heiss, Nazmitdinov and Radu~\cite{Heis94a,Heis95} investigated an
axially symmetric potential of quadrupole shape. Such a potential is
used to model deformed nuclei. They found that the addition of an
octupole deformation can render the motion chaotic. In the case of a
very strong deformation, the spacing distribution is quite close to
the Wigner surmise. In similar work, Heiss and
Nazmitdinov~\cite{Heis94b} studied other deformed potentials which are
often used to describe deformed nuclei and metallic clusters. Such and
other studies of {\it single--particle} motion cannot answer the
question whether the nuclear {\it many--body} problem is chaotic. In
fact, the influence of the single--particle motion on chaotic features
of nuclei is not yet understood, particularly since the residual
interaction is known to mix the single--particle states already at
rather low excitation energies.  We must also remember that shell
structure, one of the central elements of theoretical nuclear physics,
is due to regular motion.  We mention in passing that the study of
shell effects~\cite{Stru67,Brac72} led Strutinsky and
Magner~\cite{Stru76} as early as 1976 to an interpretation of shell
structure in terms of periodic orbits, a fact little known outside the
nuclear physics community.

Certain features of the Interacting Boson Model (IBM), see
Sec.~\ref{qc3nua}, which can be related to chaotic motion, have been
studied intensely by Alhassid, Novoselsky and Whelan~\cite{Alha90} and
Alhassid and Whelan~\cite{Alha91a,Alha91b}. A semiclassical
approximation to the Hamiltonian $H$ defined in Eq.~(\ref{3ibm1}) is
used to define the classical limit, and to check whether the classical
dynamics is chaotic. The inverse number of Bosons plays the role of
$\hbar$. The construction uses coherent states. We recall that $H$
depends on a number of parameters. For those values where one of the
three group chains applies, the classical motion is regular, and the
spectral fluctuations are close to Poisson. For values of the
parameters where the dynamical symmetries are broken, there is a
tendency toward chaotic classical motion and GOE statistics.
Remarkably, a regular strip was found within this domain which
probably corresponds to a previously unknown approximate symmetry.
However, the coordinates and momenta of the semiclassical
approximation to $H$ do not represent in any direct way the spatial
coordinates and momenta of the shell--model Hamiltonian which the IBM
Hamiltonian is supposed to approximate. Thus, no conclusion can be
drawn as to whether the nuclear many--body problem is regular or
chaotic.  Lopac, Brant and Paar~\cite{Lopa90} investigated numerically
the Interacting Boson Fermion Model. Here, the odd nucleon is coupled
to the Bosons of the IBM. Analyzing the spectral rigidity, they found
fluctuation properties between Poisson and GOE.

\subsubsection{Invariances and symmetries}
\label{qc3nue}

In previous sections, we summarized some of the evidence showing that
in certain domains of excitation energy and mass number, spectral
fluctuation properties of nuclei agree with RMT.  We also showed that
these random matrix features are consistent with certain nuclear
models. We now use these results as tools to address the breaking of
symmetry and/or invariance. We focus on a domain of excitation
energies and mass number where RMT concepts do apply, and we address
the breaking of time--reversal invariance, of the isospin quantum
number, and of parity. The important facts are: (i) In this domain,
there is a strong enhancement of symmetry breaking. Indeed, a
noticeable effect appears whenever the relevant matrix element is of
the order of the mean level spacing $D$, in agreement with the general
consideration of Sec.~\ref{qc1tcc}. We recall that at neutron
threshold, $D \simeq 10~{\rm eV}$ in heavy nuclei, whereas $D \simeq
100~{\rm keV}$ near the ground state. (ii) In this domain, the nuclear
wave functions are stochastic. Meaningful information on the strength
of the symmetry--breaking interaction can only be gained by
determining the root--mean--square (rms) matrix element from a large
number of measurements. This is done under the assumption that the
fluctuations are described by RMT. There is a bonus to this approach:
By definition, the rms matrix element does not depend on details of
the wave functions of the individual states. In the analysis,
analytical and numerical results on crossover transitions mentioned in
Sec.~\ref{qc1tc} are used.

The breaking of time--reversal invariance is modeled as a crossover
transition from GOE to GUE. Two tests have been used to obtain an
upper bound on the violation of this invariance. (i) In the nearest
neighbor spacing distribution, the slope for small $s$ changes from
linear to quadratic, and the level number variance changes
accordingly, as the GOE is replaced by the GUE. Absence of this effect
provides a test.  French {\it et al.}~\cite{Fren85,Fre88a} employed a
perturbative calculation of the level number variance $\Sigma^2(L)$
valid for small admixtures of the GUE. This is physically justified
because the breaking of time--reversal invariance is expected to be
weak.  They used the nuclear data ensemble discussed in
Sec.~\ref{qc3nua}, see Ref.~\onlinecite{Haq82}. The value of the level
number variance at $L=1$ is plotted as a function of the parameter
measuring the invariance--breaking GUE admixture and compared with the
data, and an upper bound for time--reversal invariance breaking is
found. (ii) The principle of detailed balance is tested in the regime
of Ericson fluctuations. The cross section for a nuclear reaction and
its inverse are compared. The average cross section is time--reversal
symmetric, and the test applies to the fluctuations of the cross
section. Blanke {\it et al.}~\cite{Blan83} measured the cross section
of the reaction $^{24}{\rm Mg} + \alpha \longrightarrow\, ^{27}{\rm
  Al} + {\rm p}$ and its inverse. Boos\'e, Harney and
Weidenm\"uller~\cite{Boos86a,Boos86b} used the random matrix model for
the GOE to GUE transition within the random $S$--matrix approach
described in Sec.~\ref{quasi1d_3}. The observable is the correlation
function $\langle\sigma_{a\to b}\sigma_{b\to a}\rangle/
\langle\sigma_{a\to b}\rangle\langle\sigma_{b\to a}\rangle$ relating
the cross sections of the two reactions. The analysis yielded an upper
bound for the spreading width of $10^{-2}~{\rm eV}$. This is
consistent with the value found in the analysis of French {\it et
  al.}~\cite{Fren85,Fre88a}. {}From this value an upper bound for the
relative strength of the time--reversal invariance breaking
interaction in the shell--model Hamiltonian of $10^{-3}$ can be
derived.
 
\begin{figure}
\centerline{
\psfig{file=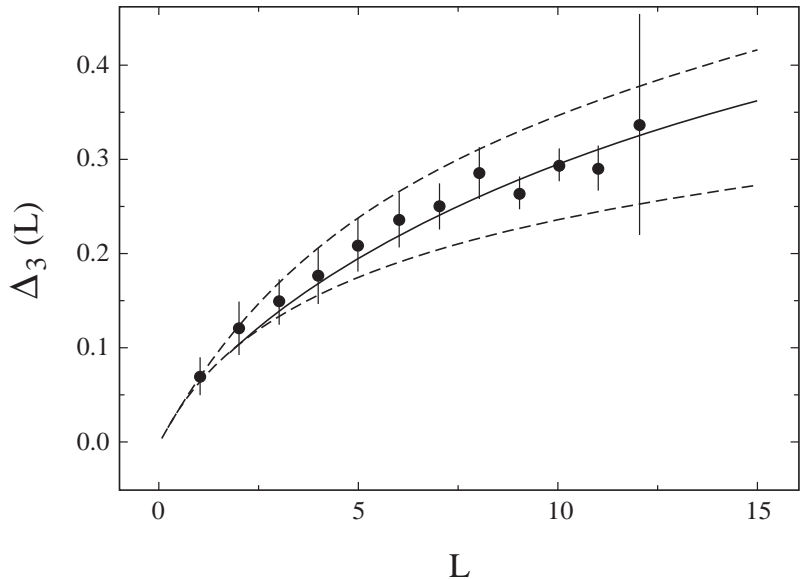,width=4.5in}
}
\caption{
  Experimental spectral rigidity $\Delta_3(L)$ in $^{26}$Al.  Solid
  dots represent the experimental results of the 100 levels between
  $0$ and $8$~MeV excitation energy. The solid line is the result of
  a numerical simulation of the random matrix model incorporating
  a symmetry breaking. This value of the symmetry breaking gave the
  best fit to the experimental data. For comparison the theoretical
  results for no symmetry and one fully conserved symmetry are
  shown as lower and upper dashed lines, respectively.
  Adapted from Ref.~\protect\onlinecite{Guh90}.
}
\label{figmb9}
\end{figure}

As shown in Sec.~\ref{qc1tc}, the breaking of symmetries (which are
always associated with quantum numbers) involves a structure of the
Hamiltonian matrices which differs from the one for time--reversal
violation. Nevertheless, certain features, particularly the remarkable
sensitivity of the fluctuations to the crossover transition, are the
same. In nuclei, isospin is only an approximate symmetry because the
Coulomb interaction, although much weaker than the nuclear force, is
not completely negligible.  Harney, Richter and
Weidenm\"uller~\cite{Har82} analyzed isospin breaking in compound
nucleus reactions by comparing the data with analytical calculations
based on a random matrix model mentioned in Sec.~\ref{qc1tc}. The
strength of symmetry breaking is measured in terms of the spreading
width introduced in Sec.~\ref{qc3nua}.  Over the entire mass range $25
< A < 240$ and for all energies investigated, the spreading widths for
isospin mixing resulting from this statistical analysis lie in a
narrow band between 1~keV and 100~keV. Mitchell {\it et
  al.}~\cite{Mit88,Shri90} tested isospin violation in the spectrum of
a single nucleus, $^{26}$Al. In this nucleus a nearly complete scheme
of about 100 levels is available ranging from the ground state up to
8~MeV. The levels are classified by spin, parity, and
isospin~\cite{Endt86,Endt88a,Endt88b}. Of these 100 levels, 75 have
isospin $T=0$ and 25 have isospin $T=1$. Mitchell {\it et
  al.}~obtained a spacing distribution and a spectral rigidity which
lie between the two limiting RMT cases of perfectly conserved and
fully broken isospin.  Guhr and Weidenm\"uller~\cite{Guh90} compared
the spectral rigidity found experimentally in Ref.~\onlinecite{Mit88}
with a numerical simulation of a full--fledged random matrix model,
see Fig.~\ref{figmb9}. Within
relatively large errors, the resulting rms Coulomb matrix element of
about 20~keV agrees with the data mentioned above.  In
Ref.~\onlinecite{Shri90} a slightly different experimental analysis
was presented. An interpretation in the framework of the RMT model for
symmetry breaking is still possible. Due to the relatively large
errors, the resulting rms Coulomb matrix did not change significantly.
However, improved experimental data are highly desirable.  Extensive
shell--model calculations by Endt {\it et al.}~\cite{Endt88b} and
Ormand and Broglia~\cite{Orma92} which included the Coulomb
interaction, yielded good agreement with both experiment and the
statistical analysis.  Very recently, Adams {\it et al.}~\cite{Adam97}
calculated the distribution of reduced transition probabilities in
$^{22}$Na using the shell model.  Magnetic dipole ($M1$) and electric
quadrupole ($E2$) transitions where studied among the first 25 states
with positive parity, spin values ranging from $J=0$ to $J=8$, and
isospin values $T=0$ and $T=1$. The results are consistent with a
Porter--Thomas distribution. Remarkably, no dependence on spin,
isospin, electromagnetic character or excitation energy was found.

The violation of parity conservation is modeled in the same way as
isospin symmetry breaking. Naively, one might expect a very weak
effect: The weak interaction is about $10^{-7}$ times weaker than the
strong interaction. Whereas the spreading width of nuclear states due
to the strong interaction is of the order of MeV, the spreading width
for the weak interaction is, therefore, expected to be of the order of
$10^{-7}$ eV. There are, however, two mechanisms which strongly
enhance parity violation in the regime of isolated compound nucleus
resonances~\cite{Sush80,Buna83}. First, in this regime resonances of
opposite parity have typical spacings $D$ of a few eV. A spreading
width of $10^{-7}$ eV thus implies a rms weak interaction matrix
element $V_{\rm weak}$ of about $10^{-4}$ eV and a ratio $V_{\rm
  weak}/D$ of about $10^{-4}$.  This value is much larger than the
naive estimate $10^{-7}$ and yields an enhancement factor of about
$10^3$. Second, if an incident neutron populates a $p$ wave resonance
which by parity violation is mixed with an $s$ wave resonance, decay
of the latter by $s$ wave neutron emission is not hindered by the
angular momentum barrier. Comparing this decay with the decay of the
$p$ wave resonance back into the entrance channel, one obtains a
second enhancement factor which is also about $10^3$. It is therefore
not surprising that signals for parity violation amounting to several
per cent were subsequently observed~\cite{Alfi83} in the
helicity--dependence of neutron transmission through several $p$ wave
resonances. A statistical analysis requires many such data, however.
The TRIPLE collaboration has furnished and analyzed such a data base,
see Refs.~\onlinecite{Bowm90,Fran91}.  The spreading width deduced
from the data is of the order of $10^{-7}~{\rm eV}$, in agreement with
theoretical expectations. We refer to a recent theoretical review on
parity and time--reversal violation by Flambaum and
Gribakin~\cite{Flam95}.

\subsection{Atoms and molecules}
\label{qc3am}

Complex atoms and polyatomic molecules are governed by the
Coulomb interaction. The relevant energy scale is very
different from the nuclear case: The ionization energies for
atoms and the dissociation energies for molecules are of the order
of several electron volts. In Secs.~\ref{qc3ama} and~\ref{qc3amb},
we discuss atoms and molecules, respectively.

\subsubsection{Atoms}
\label{qc3ama}

It is not clear a priori whether atomic spectra are expected to
show random matrix fluctuations. Complex atoms differ from nuclei 
in various ways. The nuclear charge $Ze$ is concentrated
in the center. The mean--field approximation, the basis of the
atomic shell model, is better justified and a better approximation 
than in the nuclear case. Nevertheless, the screened interaction
between the electrons is not negligible. But is it strong enough to
induce RMT behavior? Rosenzweig and Porter~\cite{Rose60} seemingly
were the first to investigate the fluctuation properties of spectra of
complex atoms. This seminal work was done as early as 1960, only
a few years after the first evidence for random matrix fluctuations
in nuclear spectra had been found, see Porter's review~\cite{Por65}. 

\begin{figure}
\centerline{
\psfig{file=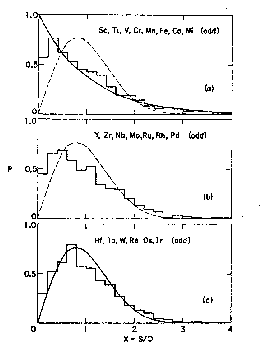,width=3.5in}
}
\caption{
  Nearest neighbor spacing distribution for the odd parity levels in
  the first, second and third chain (histograms (a), (b) and (c),
  respectively). Separate distributions were constructed for the $J$
  sequences of each element and then the results were combined.  For
  comparison, the Poisson distribution and the Wigner surmise are also
  drawn.
  Taken from Ref.~\protect\onlinecite{Rose60}.
}
\label{figmb3}
\end{figure}

Rosenzweig and Porter analyzed spectra of neutral atoms, members of
three chains. The first chain consists of the atoms Sc, Ti, V, Cr, Mn,
Fe, Co, Ni with $Z=21,22,\ldots,28$, the second one of Y, Zr, Nb, Mo,
Tc, Ru, Rh, Pd with $Z=39,40,\ldots,46$ and the third one of Lu, Hf,
Ta, W, Re, Os, Ir, Pt with $Z=71,72,\ldots,78$. The levels were
grouped according to parity $\pi$ and total angular momentum $J$. The
angular momentum values ranged from $J=0$ to $J=17/2$. No data were
available for Tc, Lu and Pt. In the three chains, there was a total of
between 1000 and 2000 spacings for each parity. For each pair of
$(J,\pi)$ values, the spacing distributions were analyzed in each of
the three chains. For the odd parity levels, the first chain gave a
spacing distribution close to Poisson, the third one a spacing
distribution close to the Wigner surmise, and the second one an
intermediate result, see Fig.~\ref{figmb3}. To explain their results,
Rosenzweig
and Porter used arguments related to the superposition of spectra, see
Sec.~\ref{qc1bae}. In the three chains, the sub--shells $3d$, $4d$ and
$5d$, respectively, are filled. The chains differ in the strength of
the spin--orbit force. In the first long chain, this force is very
weak, and the Russell--Saunders or $LS$ coupling scheme applies. For
larger $Z$ values, the spin--orbit force becomes stronger and the
coupling scheme changes. In the second long chain, $LS$ coupling gives
a relatively poor but still acceptable description. In the third
chain, the coupling scheme is very different. In the first chain, $L$
and $S$ can be viewed as almost good quantum numbers. The spacing
distribution for a given $J$ is a superposition of as many nearly
independent spectra as there are $(L,S)$ pairs for that $J$, and the
spacing distribution is very close to Poisson. This interpretation is
strongly supported by an analysis which separates levels also by $L$
and $S$ quantum numbers.  The ensuing spacing distribution is very
close to the Wigner surmise.  In the case of the third chain, there
are no nearly conserved symmetries, and the spacing distribution is
Wigner. In the intermediate case of the second chain, the symmetries
are not completely destroyed yet. The picture differs for levels with
even parity.  For various reasons, the even parity spectra are simpler
and closer to the ground state.  To support their argument, Rosenzweig
and Porter set up a random matrix model. For fixed $J$, they used a
block--diagonal Hamiltonian. Each block represents a possible $(L,S)$
pair and is modeled by a GOE. The remaining matrix elements are
independent Gaussian distributed random variables with rms value
$\mu$. The spacing distribution was calculated numerically.  As $\mu$
increases, the spacing distribution changes from Poisson to Wigner.
The relevant parameter is $N\mu^2$, where $N$ is the matrix dimension.
It was interpreted as the rms symmetry--breaking strength normalized
to the mean level spacing, see Sec.~\ref{qc1tc}.

We have discussed this work in some detail because in many
ways it was a truly pioneering contribution. Aside from its
technical aspects which later found application in many other
papers, this work --- and a preliminary study by the same
authors --- constitutes the first application of RMT outside the 
field of nuclear physics. It provided the first strong evidence for
the universal applicability of RMT, independently of the specific
interaction governing the physical system. In 1961, only a year later, 
Trees~\cite{Tree61} extended this work by analyzing spacing
distributions of calculated spectra.

In 1983, Camarda and Georgopulos~\cite{Cama83} analyzed new level 
schemes of atoms in the beginning of the third chain, from
La with $Z=57$ to Lu with $Z=71$, using the spectral rigidity and
the covariance of adjacent spacings as observables. (For each pair
of consecutive spacings, the latter is given by the
expression~(\ref{3nu2})). This test of long--range correlations is
also consistent with GOE predictions. 

Of the more recent work, we only mention the extensive theoretical
study of the rare--earth atom Ce by Flambaum 
{\it et al.}~\cite{Flam94}. A realistic model is used to calculate the
structure of complex atomic states. Above 1~eV excitation energy,
these states acquire a structure similar to that of compound states in
heavy nuclei (a linear superposition of very many basis states).
Various spectral observables were worked out in this model and
compared with GOE predictions.

\subsubsection{Molecules}
\label{qc3amb}

In a polyatomic molecule, there are three types of excitations:
Electronic excitations, vibrations, and rotations. The energy scale
of the electronic excitations is a few electron volt. Vibrational
states built on top of every electronic state have an energy scale
of 0.1~eV.  On top of the vibrational states, there are rotational
states with an energy scale of 0.1~meV. There is no spherical
symmetry. The spectra are characterized by quantum numbers
which differ from the ones used in atomic physics. Modern laser
spectroscopy is capable of resolving the enormously rich and
complex spectra of such molecules even at very high level density.
Traditional theoretical techniques often fail in this regime, a
description of individual levels is out of the question, and an
analysis of the data with the aid of statistical concepts is called for.

The molecule NO$_2$ provides a particularly interesting example
and has played a prominent role in the application of RMT
concepts. Between the ground state and the dissociation threshold
at about 3~eV, there are four electronic states. The 
Born--Oppenheimer approximation (based on an adiabatic
separation of nuclear and electronic motion) breaks down in
certain parts of the spectrum. It is this feature which makes
NO$_2$ such an interesting object of experimental and theoretical
studies of spectral fluctuations, although the molecule consists of
only three atoms. 

In 1983, Haller, K\"oppel and Cederbaum~\cite{Hall83} studied
experimental and theoretical vibronic spectra of the molecule NO$_2$
and of the acetylene kation C$_2$H$_4^+$ in a regime where the
Born--Oppenheimer approximation fails. Here, the term vibronic means
that the excitation mechanism is only vibrational and electronic.
Rotations are absent. Good evidence for GOE fluctuations was found in
both the short-- and long--range spectral observables. Zimmermann {\it
  et al.}~\cite{Zimm88} extended the analysis of NO$_2$ data,
including a comparison of the measured intensity distribution with the
Porter--Thomas law.  Theoretical studies were presented by Zimmermann,
K\"oppel and Cederbaum~\cite{Zimm89}. Leitner, K\"oppel and
Cederbaum~\cite{Leit96} compared the spectral statistics and the
classical dynamics of both NO$_2$ and C$_2$H$_4^+$. The dynamics of
those systems with coupled potential surfaces can only approximately
be described in classical terms since the electronic degrees of
freedoms remain quantized.  However, it turns out that the phase space
is dominated by one of the potential surfaces. This allowed the
authors of Ref.~\onlinecite{Leit96} to empirically establish a
correspondence between this classical dynamics and the spectral
fluctuation properties. In 1985, Mukamel, Sue and Pandey~\cite{Muka84}
investigated the spectral fluctuations and the distribution of
resonance intensities in data on vibrationally highly excited
acetylene. Deviations from the GOE predictions are attributed to
missing levels. Sundberg {\it et al.}~\cite{Sund85} extended this
analysis of acetylene spectra.

It is apparent from this discussion that NO$_2$ has played a very
important role in the investigation of spectral fluctuations in
molecular physics. This was possible because the Grenoble group
succeeded in improving the quality of experimental data substantially. 
Georges, Delon and Jost~\cite{Geor95} have given a
detailed account of the measurements on NO$_2$. They summarize
the evidence for GOE fluctuations obtained from short-- and
long--range spectral observables as follows. The vibronic states
built on the first and the second electronic state show GOE
fluctuations. This can be attributed to the mixture of these 
states due the breakdown of the Born--Oppenheimer approximation.
Higher up in the spectrum, GOE fluctuations occur for another 
reason in rovibronic states, i.e.~in states where rotations on top of
vibronic states become important. The mixing of these states is due to 
a symmetry breaking which has specific molecular physics
causes. Very recently, parametric correlations as discussed in
Sec.~\ref{qc1pc} were addressed experimentally and theoretically
in NO$_2$ by Nyg\aa rd, Delon and Jost~\cite{Nyga96}.

\begin{figure}
\centerline{
\psfig{file=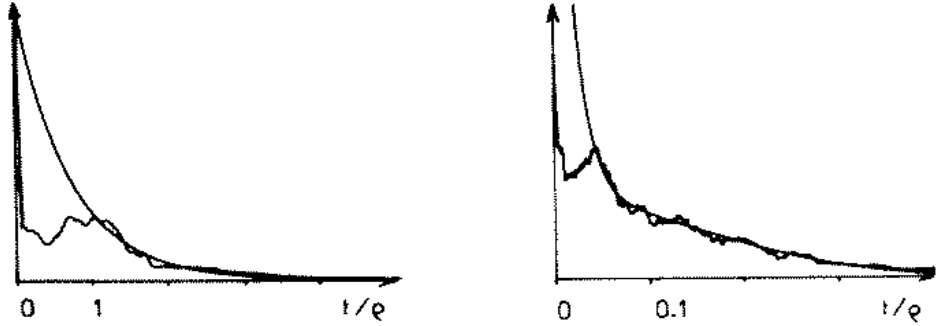,width=5.5in}
}
\caption{
  Correlation--hole analysis for a stimulated emission pumping
  spectrum of acetylene (left) and for an anticrossing spectrum of
  methylglyoxal (right). The properly smoothed decay function defined in 
  Sec.~\protect\ref{qc1bad} is shown. The variable $t/\rho$ should be 
  identified with the variable $t$ of Sec.~\protect\ref{qc1bad}.  
  The hole is due
  to the correlation of the level positions while the exponential
  decay is due to the resonance shapes. Notice the drastically
  different widths of the two holes, indicating a very different
  number of symmetries.  Taken from Ref.~\protect\onlinecite{Lev86}.  
}
\label{figmb4}
\end{figure}

The correlation--hole method (Sec.~\ref{qc1bad}) introduced by
Leviandier {\it et al.}~\cite{Lev86} brought a conceptually new aspect
into the analysis of spectral fluctuations, see also the
review~\cite{Jost86}. This technique is designed for the analysis of
raw spectra. It is less sensitive to the effect of missing levels than
other observables. It does not require a line--by--line identification
of the resonances. Hence, it can deal with the enormous amount of data
supplied by modern laser spectroscopy. The correlation--hole method
effectively separates the statistical fluctuations of the level
positions and of the resonance shapes, see Fig.~\ref{figmb4}.
Leviandier {\it et al.}~applied this technique to anticrossing spectra
of methylglyoxal.  Consider a singlet state $|s\rangle$ populated from
the singlet ground state $|s_0\rangle$.  A triplet state $|t\rangle$
couples to $|s\rangle$ by some interaction $V_{st}$. If the energy
difference between the two states obeys $|E_t-E_s|\gg V_{st}$, the
interaction can be neglected and the resonance fluorescence light back
into the ground state $|s_0\rangle$ attains a maximum.  If, however,
$|E_t-E_s|$ and $V_{st}$ are comparable, the admixture of the triplet
state leads to a suppression of the fluorescence yield, since
transitions from $|t\rangle$ to $|s_0\rangle$ are forbidden. The
energy of the triplet states is shifted by a magnetic field $B$, and
the resonance fluorescence yield $I(B)$ is measured as a function of
$B$. For each isolated triplet state, $I(B)$ has Lorentzian shape. In
methylglyoxal, there are many such triplet states with overlapping
Lorentzians, and the measured resonance fluorescence intensity $I(B)$
has a complicated dependence on $B$.  Leviandier {\it et al.}~found a
deep correlation hole in the decay function constructed from $I(B)$.
The fluctuations of the anticrossing spectrum must therefore at least
to some extent reflect GOE behavior. The exponential fall--off of the
decay function is due to the Lorentzian shapes of the individual
resonances. The arguments presented in Sec.~\ref{qc1bad} were used to
conclude that the measured $I(B)$ must be a superposition of several
non--interacting spectra.  Further experimental work is due to Delon,
Jost and Lombardi~\cite{Del91}. Pique {\it et al.}~\cite{Piqu87}
applied the same type of analysis to spectra on acetylene. In the
framework of scattering theory, a full--fledged random matrix model
simulating these experiments was studied analytically~\cite{Guh90b}
and numerically~\cite{Har91}. Detailed accounts of the
correlation--hole method which also include other aspects of the
statistical analysis were given by Lombardi {\it et al.}~\cite{Lom91}
and by Lombardi and Seligman~\cite{Lom93}.

The distribution of decay rates and related questions were
investigated theoretically by Zimmermann, K\"oppel and
Cederbaum~\cite{Zimm89}, by Miller {\it et al.}~\cite{Mill90} and by
Peskin, Reisler and Miller~\cite{Pesk94}, cf. also the work by
Persson, Gorin and Rotter~\cite{Pers96}, see Sec.~\ref{qc3nud}.

Davis~\cite{Davi93} introduced the interesting concept of
hierarchical analysis: The structure of peaks in a spectrum is
studied as a function of the resolution. In this way, hierarchical
trees are generated which correspond to states of ever higher 
complexity. This analysis is intimately related to the constraints
imposed by the experimental set--up and can yield significant
statistical information. It seems worthwhile to see whether links
can be established to RMT.

\subsection{General aspects}
\label{qc3gd}

This short presentation of the evidence for random matrix
fluctuations in many--body systems would be incomplete without
a brief discussion of two questions posed by the results: (i) To
what extent do the Gaussian ensembles properly model the
properties of many--body systems? This question, asked 
in Sec.~\ref{qc3gda}, addresses the very foundations of the
application of Random Matrix Theory to real systems. (ii) Are
many--body systems chaotic? This question connects to the
discussion of chaos in Sec.~\ref{chaos} and is treated in
Sec.~\ref{qc3gdb}.

\subsubsection{To what extent do the Gaussian ensembles model 
               many--body systems properly?}
\label{qc3gda}

We recall that the distribution $P(H)$ for the Gaussian ensembles is
uniquely derived from two assumptions: (i) $P(H)$ is invariant under
orthogonal, unitary or symplectic transformations, respectively, and
(ii) the independent matrix elements are uncorrelated random
variables. There is evidence that both in atoms and in nuclei, the
distribution of matrix elements differs from what is assumed in the
Gaussian ensembles. The key argument is based on the observation that
in both systems, the residual interaction is of two--body type. Thus,
in a shell--model basis, the Hamiltonian matrix is sparse. Put
differently, in an arbitrary basis, the elements of the Hamiltonian
matrix are strongly correlated. In contradistinction, the Gaussian
ensembles may be said to contain matrix elements of $k$--body type,
with $k$ ranging all the way up to $N$. This is because in these
ensembles, {\it all} matrix elements not connected by symmetry are
uncorrelated.  To what extend does this difference affect spectral
fluctuations?  The question can be formulated more precisely as
follows. We consider a system of Fermions with a
two--body~\cite{Fren70,Bohi71} and, more generally, a $k$--body random
interaction. The matrix elements of the $k$--body operator between
$k$--particle states are assumed to be independent Gaussian
distributed random variables. There is evidence from nuclear
physics~\cite{Bro81} that for $k = 2$, this is a realistic assumption.
There are $m$ Fermions which occupy $N$ degenerate single--particle
levels.  Here, $k \leq m \ll N$. This defines the embedded Gaussian
$k$--body ensembles~\cite{Mon75}. Which are the spectral fluctuation
properties of these ensembles? Very little is known analytically. This
is why the embedded ensembles receive scarce attention in the present
review.  Numerical simulations~\cite{Bro81} have shown, however, that
the local fluctuation properties of the embedded two--body random
ensembles do agree with RMT predictions.

The Gaussian ensembles do not only fail to take account of the
two--body character of the residual interaction. They also
misrepresent the actual distribution of the matrix elements in a
non--shell--model basis. This is shown by work of Flambaum {\it et
al.}~\cite{Flam94} who investigated the distribution of the 
non--diagonal matrix elements of the Hamiltonian for the $J=4$
states of odd and even parity in Ce. The basis is given by all possible
single--determinant states which are first constructed for a
given value of the total azimuthal quantum number $M$ of the atom
and then rotated such that the Hamiltonian is block diagonal
in the total angular momentum $J$. These states are labeled
by an index $n$. For $m \neq n$, Flambaum {\it et al}.~found
the empirical law
\begin{equation}
P(H_{nm}) = C \frac{\exp\left(-|H_{nm}|/V\right)}{\sqrt{|H_{nm}|}}
\label{3gd1}
\end{equation}
where $C$ and $V$ are constants. For $m \simeq n$, a peak occurs
on top of this distribution. Nonetheless, the spectral
fluctuations calculated/measured in this atom agree with the GOE.

In spite of these shortcomings, the Gaussian ensembles generically
model the fluctuation properties correctly. The reason is that
fluctuations are studied on the {\it unfolded}, i.e. {\it local},
scale defined by the mean level spacing. This is in contrast to
average or global properties which cannot be modeled by these
ensembles. The local fluctuation properties are very insensitive to
details of the distribution $P(H)$. For a class of ensembles, this is
shown in Sec.~\ref{univ}. More can be found in the review by Brody
{\it et al.}~\cite{Bro81}. This issues have received renewed interest
in recent years, cf. the review by Flambaum and Gribakin~\cite{Flam95}
and the work of M\"uller and Harney~\cite{Mull87,Mull88} and Granzow,
Harney and Kalka~\cite{Gran95}.

\subsubsection{Are many--body systems chaotic?}
\label{qc3gdb}

We have presented evidence to the effect that the complex structure of
the many--body system often yields fluctuation properties which are
consistent with RMT. There is also evidence for Poisson--type
fluctuations, especially in systems capable of collective motion. In
the description of these phenomena, we have tried to avoid the terms
``chaotic'' and ``regular'', respectively, for these two scenarios. It
is shown in Sec.~\ref{chaos} that for systems with few degrees of
freedom, there is strong and growing evidence for the correctness of
the Bohigas conjecture. As stated in the Introduction, this conjecture
links classical chaos to the spectral fluctuations described by the
Gaussian ensembles. Does this conjecture also apply to many--body
systems? We are not aware of any arguments which would establish a
link between many--body systems and classical chaos. Various studies
have shown chaotic features of single--particle motion in realistic
potentials, see Sec.~\ref{qc3nud}. Therefore, it is intriguing and may
even be justified to think of complex many--body systems such as
nuclei, atoms or molecules as chaotic. Such thinking must come to
terms, however, with the existence of regular collective and
single--particle properties in the same systems which may show up as
gross structures at the very energies where the underlying
fluctuations are of GOE type. Thus, a theoretical understanding of why
RMT is so successful in many--body systems has yet to be found. The
evidence we have at hand is either experimental or phenomenological.
RMT was developed and used in the context of many--body physics long
before the connection to classical chaos in few--degrees--of--freedom
systems had been suggested. Thus, in retrospective, it is no surprise
that phrases such as ``unreasonable effectiveness'' were used to
characterize the predictive power of Random Matrix Theory in
many--body physics.

\setcounter{equation}{0}
\section{Quantum chaos}
\label{chaos}

RMT originated in nuclear physics and was conceived as a statistical
approach to systems with many degrees of freedom. However, RMT applies
also to few--degrees--of--freedom systems with chaotic classical
dynamics. This observation is of recent origin, and it forms the
content of the present section.  In 1984, Bohigas, Giannoni and
Schmit~\cite{Boh84b} stated the famous conjecture: ``{\it Spectra of
  time--reversal invariant systems whose classical analogues are $K$
  systems show the same fluctuation properties as predicted by the
  GOE.}'' We recall that $K$ systems are the most strongly mixing
classical systems. The most unpredictable $K$ systems are referred to
as Bernouilli systems.  An alternative, stronger version of the
conjecture, also formulated in Ref.~\onlinecite{Boh84b}, replaces $K$
systems by less chaotic systems provided they are ergodic. In both its
forms, this conjecture is commonly referred to as the Bohigas
conjecture. For systems without time--reversal invariance, the GOE is
replaced by the GUE. In its original version, the Bohigas conjecture
was stated without a reference to the semiclassical regime, i.e.~to
the limit $\hbar\to 0$. However, all attempts to proof the conjecture
are based on some kind of semiclassical approximation, see
Sec.~\ref{qc4psi}.

We begin with a brief review of the historical development prior and
up to the formulation of the Bohigas conjecture (Sec.~\ref{qc4ea}).
In Sec.~\ref{qc2dl}, we briefly outline the main concepts of dynamical
localization.  We then turn to systems which were crucial to the
development (billiards, the hydrogen atom in a strong magnetic field,
and others). These systems form the first major part of the present
section (Secs.~\ref{qc2bi} to \ref{qc2ms}). Although representatives
of mesoscopic systems, quantum dots are included here
(Sec.~\ref{qc2qd}) because they share many properties with billiards.
The examples given and many others strongly support the Bohigas
conjecture. This accumulation of convincing evidence has led to
several attempts at establishing a theoretical link between classical
chaos and RMT. These are reviewed in the second part of this section,
i.e.~in Secs.~\ref{qc4mp} to~\ref{qc4psi}. Lack of space forces us to
present only the key elements of this discussion. In Sec.~\ref{qc4mp},
we outline a hypothesis which has particular bearing on RMT. It
concerns the way in which the structure of classical phase space
influences quantum properties. Three approaches towards a formal proof
of the Bohigas conjecture are presented in Sec.~\ref{qc4psi}: Periodic
orbit theory, a field--theoretic approach using the supersymmetry
method, and a probabilistic argument based on structural invariance.
In Sec.~\ref{qc4su}, a brief summary of the present section is given.
There are many excellent reviews on classical and quantum chaos.  With
respect to classical and semiclassical aspects, we mention only, in
the order of appearance, the books by Moser~\cite{Mos73}, Lichtenberg
and Liebermann~\cite{Lic83}, Schuster~\cite{Sch84} and
Gutzwiller~\cite{Gut90}.

\subsection{Historical development}
\label{qc4ea}

In its beginnings, the development of quantum chaos was strongly
influenced by chemists who tried to understand simple molecules. The
phenomenon of avoided level crossings was of particular interest. We
follow partly the survey given by Robnik~\cite{Rob84}.

In 1929, von Neumann and Wigner~\cite{Von29} had shown that level
crossings cannot occur in generic systems and that avoided crossings
appear instead. Teller~\cite{Tel37} stated the same theorem again in
1937. Percival~\cite{Per73} argued in 1973 that, in the semiclassical
limit, the energy spectrum ought to consist of a regular and an
irregular part reflecting the regular and irregular parts of classical
phase space. In the early eighties Ramaswamy and Marcus~\cite{Ram81},
Berry~\cite{Ber83}. Zaslavsky~\cite{Zas81} and
Marcus~\cite{Mar80a,Mar80b} showed that level repulsion and avoided
crossings are typical for non--integrable systems. Already in the late
seventies and early eighties, Zaslavsky~\cite{Zas77,Zas79,Zas81},
Berry~\cite{Ber77a,Ber77b,Ber81,Ber83} and Berry and
Tabor~\cite{Ber77c} expected a drastic change in the spacing
distribution to accompany the transition from integrability to
non--integrability. In 1979, McDonald and Kaufman~\cite{McD79}
published their pioneering numerical study of the quantum analogues of
the classically integrable circular billiard and of the classically
chaotic Bunimovich stadium. This work was closely followed by
numerical work of Casati {\it et al.}~\cite{Cas80} on the Bunimovich
stadium. In these papers, the above--mentioned expectations came true.
Further and improved evidence, also in billiard systems, was presented
by Berry~\cite{Ber81} and by Bohigas {\it et
  al.}~\cite{Boh84,Boh84b,Boh84c}. All this work is reviewed in more
detail in Sec.~\ref{qc2bi}. The behavior of eigenfunctions near
avoided crossings was simultaneously studied by Noid {\it et
al.}~\cite{Noi79,Noi80}, Marcus~\cite{Mar80a,Mar80b} and Ramaswamy and
Marcus~\cite{Ram81} who showed that the eigenfunctions are strongly
mixed. According to McDonald and Kaufman~\cite{McD79} and
Noid~\cite{Noi79} this causes a random pattern of the wave functions
whose nodal lines do not cross. In these papers, evidence was found
that the low--lying eigenvalues and the corresponding eigenfunctions
do not follow this pattern and have mainly regular features.

For small spacings, level repulsion was expected to lead to a spacing
distribution proportional to $s^{\nu}$. Using geometrical arguments,
Berry~\cite{Ber81} suggested the value unity for the ``critical
exponent'' $\nu$, in keeping with the Wigner surmise for the spacing
distribution in time--reversal invariant many--body systems.
Zaslavsky~\cite{Zas81} related $\nu$ to the Kolmogorov entropy for the
classical system. His work was a pioneering attempt to find a
quantitative relationship between classical chaos and spectral
fluctuations, although no convincing evidence for his assertion was
found.

More formal arguments were used in the framework of statistical
mechanics. Pechukas~\cite{Pec83}, Yukawa~\cite{Yuk85} and Nakamura and
Lakshmanan~\cite{Nak86} showed that the probability distribution of
the energy eigenvalues of the ``Pechukas gas'' coincides with the GOE
probability density. We do not follow this important line here because
there is no easy link to chaos in few--degrees--of--freedom systems.

\subsection{Dynamical localization}
\label{qc2dl}

Dynamical localization is a completely unexpected and novel feature of
quantum chaos. In a classical system, driven by some external
time--dependent force, the occupation probability in phase space
spreads forever diffusively. Not so in the quantum analogue: After
some characteristic time, the wave packet stops spreading in {\it
  momentum} space. This phenomenon is referred to as dynamical
localization. It is ``dual'' to the localization in {\it
  configuration} space typical of disordered systems, see
Sec.~\ref{disorder}. The close relationship between both types of
localization phenomena was discovered by Fishman {\it et
  al.}~\cite{Fis82,Grem82,Grem84}. In both cases, there is quantum
suppression of classical diffusion. The wave packet relaxes into a
non--ergodic localized state. For the case of the kicked rotor, a step
towards formally establishing the correspondence between both types of
localization was recently made by using the supersymmetric non--linear
$\sigma$ model~\cite{All96}, see Sec.~\ref{rbm}.  A detailed account
of dynamical localization was given by Casati and
Chirikov~\cite{Casa95a,Casa95b}, see also Ref.~\onlinecite{Heis92}. We
partly follow Ref.~\onlinecite{Casa95a}.

{}From a conceptual point of view, the study of wave functions is
quite different from that of level statistics.  Levels are eigenvalues
of the Hamiltonian and can be studied experimentally.  The spectral
fluctuation properties of the quantum analogues of classically chaotic
systems have accordingly received much interest, and they form the
content of the bulk of the present section. The wave function $\psi$,
on the other hand, is not an observable.  Investigating the evolution
of $\psi(t)$ in time $t$, one disregards all aspects of quantum
mechanics related to the measurement process, and the collapse of the
wave function. However, the wave function $\psi(t)$ obeys the
Schr\"odinger equation, a deterministic equation.  Thus, direct
comparison of the time evolution of $\psi(t)$ with the dynamics of the
corresponding classical system is useful. Some authors prefer to
reserve the term quantum chaos for investigations conducted in this
spirit. 

It should be emphasized that quantum dynamical localization in
classically conservative systems is due to quantum suppression of
classically chaotic diffusion. Many semiclassical studies address
questions such as scars around periodic orbits or similar phenomena
which are often also referred to as localization. These two effects,
however, should clearly be distinguished.

Two time scales govern dynamical localization.  The relaxation time
scale $t_R$ is proportional to the density $\rho_0$ of those
eigenstates (quasi energy levels) which are admixed to the original
wave packet at $t=0$. This is the time scale on which periodic orbit
theory can be applied. The time scale $t_e$ is characteristic of the
{\it classical} relaxation to the ergodic steady state. The quantity
\begin{equation}
\lambda = \sqrt{\frac{t_R}{t_e}}
\label{2dl1}
\end{equation}
is a measure for the ergodicity of the state. Shnirelman~\cite{Shni74}
argued that for $\lambda \gg 1$ the final steady state and all
eigenfunctions are ergodic, and the Wigner functions are close to the
classical microcanonical distribution in phase space.  If, however,
$\lambda \ll 1$, all states are non--ergodic, and their structure is
governed by quantum mechanics.

On a very short time scale (which, according to the uncertainty
principle, involves very high energies) the wave packet spreads quite
like a classical cloud of particles in phase space would.  Then quantum
effects set in, the wave packet spreads in a manner not allowed by the
constraints of Liouville's theorem in classical mechanics, and
eventually may stop spreading and remain in a non--ergodic localized
state. To measure the size of localized states, Izrailev~\cite{Izr90}
introduced the entropy localization length
\begin{equation}
l_h = \exp(h) 
\label{2dl2}
\end{equation}
where the entropy is given by
\begin{equation}
h = - \sum_n |\varphi(n)|^2 \ln |\varphi(n)|^2 \ .
\label{2dl3}
\end{equation}
Here, $\varphi(n)$ are the wave functions in momentum space. In
various cases, the wave functions are exponentially localized,
i.e.~one has $\varphi(n) \rightarrow \exp(-|n|/l), \ |n| \rightarrow
\infty$.  In those cases, $l_h$ turns out to be close to $l$.  The
spacing distribution of localized states obeys, in the extreme case,
the Poisson law whereas the Wigner surmise describes that of fully
ergodic systems. Thus, the spacing distribution can give information
about the degree of ergodicity and localization.  Remarkably, it has
been found in numerical studies that the repulsion parameter
$\beta_{\rm eff}$ in Izrailev's phenomenological crossover 
formula~(\ref{1ba11a}) can be approximated by
\begin{equation}
\beta_{\rm eff} = \beta \frac{l_{\langle h \rangle}}{l_e}
\label{2dl5}
\end{equation}
where $l_{\langle h \rangle}$ is the average of $l_h$ over all states
and $l_e \ge l_{\langle h \rangle}$ is the maximal entropy localization
length corresponding to ergodic states. As usual, $\beta$ labels the
three symmetry classes. Another measure is the inverse participation ratio
\begin{equation}
\xi_m^{-1} = \sum_n |\varphi_m(n)|^4
\label{2dl5a}
\end{equation}
which is related but not identical to $l_h$.

One of the first examples to show dynamical localization was the
kicked rotor. Another case, which is experimentally accessible, is
that of the hydrogen atom in a microwave field.  Experimental results
by Bayfield and Koch~\cite{Bayf74} led to a series of activities.
Casati {\it et al.}~\cite{Casat84} predicted localization effects in
this system which were experimentally confirmed by Bayfield {\it et
  al.}~\cite{Bayf89}.  We refer the reader to the review in
Ref.~\onlinecite{Heis92}. Recently, these studies have been extended
to the hydrogen atom in both magnetic and microwave fields by
Benvenuto {\it et al.}~\cite{Benv97}.

\subsection{Billiards}
\label{qc2bi}

A classical billiard consists of a point particle which moves freely
in a compact domain of $d$--dimensional space and which is reflected
elastically by the boundary of this domain. The energy and the modulus
of the momentum of the particle are constants of the motion which,
therefore, is determined by geometrical optics. Hence, the dynamics
(regular, mixed, or chaotic) is the same for all energies. We restrict
ourselves to $d = 2$. This case was and is conceptually of great
importance. Examples of integrable (regular) billiards are furnished
by the circle and the ellipse. Among the many examples of chaotic
billiards, two have received particular attention. The Sinai billiard
consists of a square and an inserted concentric circle. The particle
moves in the domain between square and circle. The Bunimovich stadium
\begin{figure}
\centerline{
\psfig{file=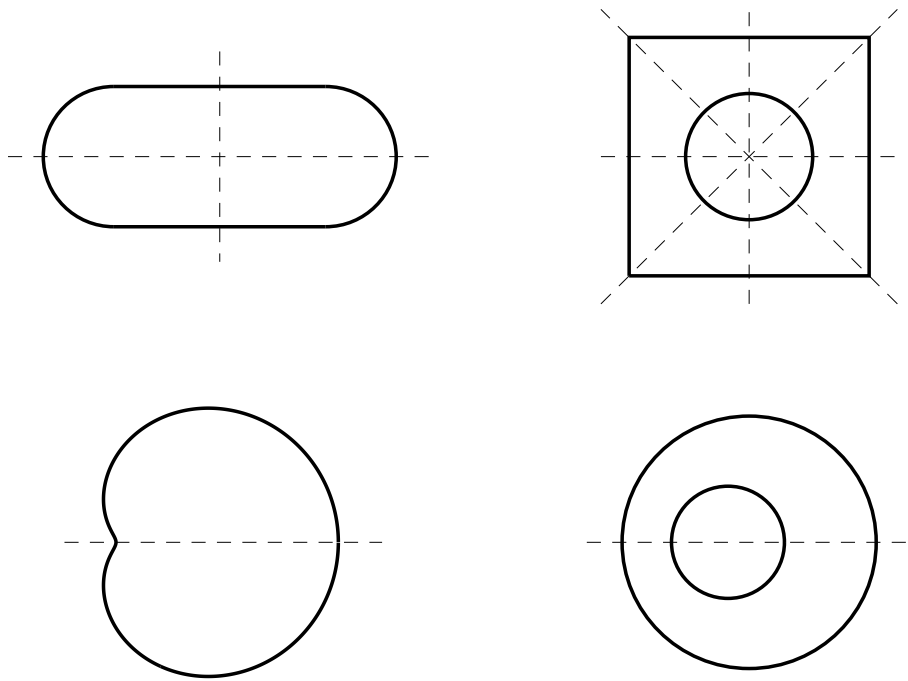,width=5.5in,angle=-90}
}
\caption{
  Four important billiards: the Bunimovich stadium
  (top left), the Sinai billiard (top right), the Pascalian
  snail (bottom left) and the annular billiard (bottom right).
  To reduce the symmetries, these billiard have to be cut
  along the symmetry axes which are drawn as dashed lines.
}
\label{figch11}
\end{figure}
is of oval shape and consists of two equally long sections of parallel
straight lines which at both ends are joined by two semicircles. In
the Sinai billiard, chaos is due to the defocusing effect of
reflection at the inner circle. In the Bunimovich stadium, chaos
arises because the straight lines destroy the rotation invariance of
the two semicircles. Alternatively, chaotic dynamics can be attributed
to an ``over--focussing'' effect in this geometry.  Four important
billiards are shown in Fig.~\ref{figch11}.  Sinai and
Bunimovich~\cite{Sin68,Bun74,Bun80} have shown that both billiards are
fully chaotic and belong to the Bernoulli class, i.e.~to the strongest
form of chaotic dynamics.  Mathematical aspects were discussed by
Katok and Strelcyn~\cite{Kat86}. Physics--oriented reviews on
classical chaos in billiards can be found in the article by Bohigas
{\it et al.}~\cite{Boh84} and in Gutzwiller's book~\cite{Gut90}.

The quantum analogue of a classical billiard is defined by the
stationary Schr\"odinger equation with Dirichlet boundary conditions
(the wave function vanishes at the boundary). Apart from factors, the
Hamilton operator for a free particle is simply the Laplacean.
Therefore, the problem is mathematically equivalent to determining the
vibrations of a membrane. Independently of the connection to quantum
mechanics, such and more general systems are of considerable interest
and have been studied by mathematicians for a long time. For example,
the Dirichlet condition may be replaced by the Neumann boundary
condition (the normal derivative of the wave function vanishes at the
boundary).  In these systems, the smoothed part of the level density
and the smoothed part of the cumulative density $\xi(k)$, see
Sec.~\ref{qc1ba}, show some very general features. As a function of
the wavenumber $k$, $\xi(k)$ is given by
\begin{equation}
\xi(k) = \frac{A}{4\pi} k^2 \mp \frac{L}{4\pi} k + C \ .
\label{2bi1}
\end{equation}
Here $A$ is the area and $L$ the length of the perimeter of the
billiard. The constant $C$ describes curvature corrections, cusps and
other topological properties. The minus and the plus sign in front of
the perimeter term apply in the case of Dirichlet and Neumann boundary
conditions, respectively. In the case of a particle of mass $m$ in a
quantum billiard, the energy is $E=(\hbar k)^2/2m$.
Formula~(\ref{2bi1}) is valid for arbitrary geometries. The area term
was found by Weyl~\cite{Wey11} in 1911 and 1912, the perimeter term by
Kac~\cite{Kac66} in 1966. More general results for arbitrary linear
boundary conditions were derived by Balian and Bloch~\cite{Bal70}, cf.
also Ref.~\onlinecite{Bal75}. A review of the interesting history of
these results can be found in Refs.~\onlinecite{Boh84,Gut90,Boh91a}.

We focus attention on the spectral fluctuation properties of quantum
billiards. In sharp contrast to the level density, these properties
depend sensitively on the shape of the boundary. The spectrum of a
billiard can be unfolded very efficiently with the help of
formula~(\ref{2bi1}). To analyze the spectral fluctuations, it is
important to use only levels which pertain to the same set of quantum
numbers. This is done by removing all symmetries from the problem. The
Sinai billiard, for instance, has four symmetry axes. One considers
instead a billiard comprising 1/8th of the area. In the case of the
Bunimovich billiard, a billiard of 1/4th of the area is considered,
see Fig.~\ref{figch11}.

\begin{figure}
\centerline{
\psfig{file=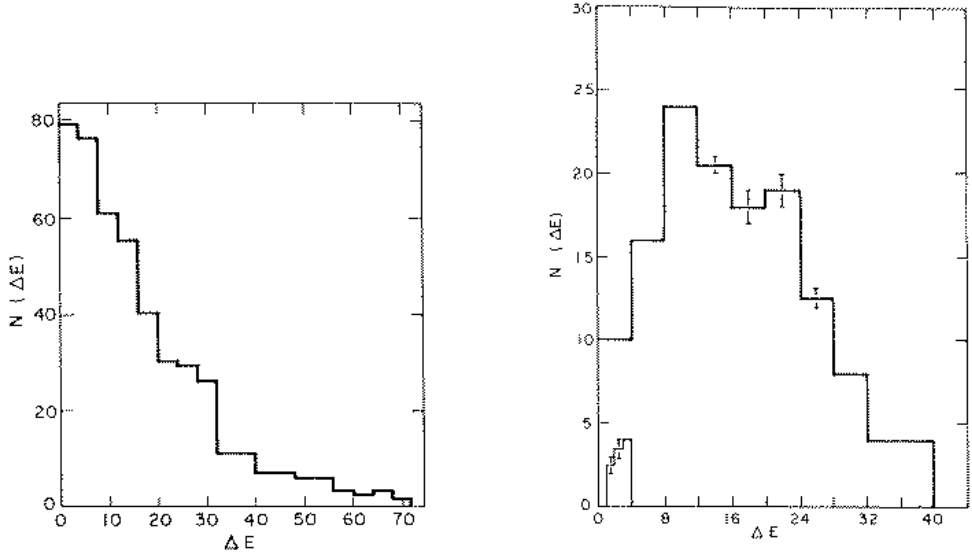,width=5.5in}
}
\caption{
  Nearest neighbor spacing distributions of odd parity levels in the
  circle (left) and the Bunimovich stadium (right). We notice that the
  spacing $\Delta E$ is not normalized to the mean level spacing and
  that the distributions $N(\Delta E)$ are normalized to the total
  number of spacings. The energy ranges are $2500<E<10000$ for the
  circle and $2500<E<4900$ for the stadium where the units are fixed
  by the choice $\hbar^2/2m=1$ in the Schr\"odinger equation. 
  Taken from Ref.~\protect\onlinecite{McD79}. 
}
\label{figch6}
\end{figure}

McDonald and Kaufman~\cite{McD79} investigated numerically the
spectrum of the circle, a classically integrable system. The results
for the spacing distribution are shown in Fig.~\ref{figch6}. Level
clustering and degeneracies due to the absence of level repulsion
could clearly be seen. The rectangle, also regular, was studied with
comparably high statistics (more than 10 000 eigenvalues) by Casati
{\it et al.}~\cite{Cas85}. To avoid degeneracies, the squared ratio of
the side lengths was chosen as an irrational number. Simultaneously,
Feingold~\cite{Fei85} performed similar numerical work. The spacing
distribution probes correlations on the scale of up to three or four
mean level spacings and agrees very well with the Poisson
distribution.  The spectral rigidity $\Delta_3(L)$ for the system
considered agrees with the Poisson behavior $L/15$ only up to a
critical value of $L \simeq 100$ and becomes constant for larger
interval lengths $L$. The spectrum of the rectangle is fully random
only on scales below this critical value. On larger scales
non--trivial correlations appear which cannot be seen in the spacing
distribution. Casati {\it et al.}~\cite{Cas84} argue that such
correlations occur generically in integrable systems.  The very large
number of levels needed to see this saturation of the spectral
rigidity is often not accessible in numerical simulations, let alone
experiments.

Chaotic billiards behave very differently. In their pioneering work of
1979, McDonald and Kaufman~\cite{McD79} calculated eigenenergies and
eigenfunctions of the Bunimovich billiard numerically. They showed
that the spacing distribution is of Wigner type and stressed the
qualitative difference to the spacing distribution of the circle
calculated in the same paper, see Fig.~\ref{figch6}. McDonald and
Kaufman related this difference to the classical integrability or
non--integrability of the two cases. They also observed that the nodal
lines of eigenfunctions in the Bunimovich stadium do not cross. The
stadium was also investigated by Casati {\it et al.}~\cite{Cas80}.
Berry~\cite{Ber81} studied the Sinai billiard in 1981, calculated
about 500 eigenvalues and found linear level repulsion for small
spacings. He gave a simple analytical argument for this behavior and
mentioned the connection to RMT.  Berry's work demonstrates that
accurate numerical methods for the calculation of a large number of
levels are not only technically important, they often prompt new
conceptual developments. In 1984, Bohigas {\it et
  al.}~\cite{Boh84b,Boh84} computed more than 700 eigenvalues of the
Sinai billiard and found very good agreement of the spacing
distribution with the Wigner surmise, see Fig.~\ref{fig4}. This very
accurate calculation provided the first statistically significant
evidence of the validity of the Wigner surmise not only for small, but
for all spacings. Bohigas {\it et al.} also showed that the spectral
rigidity $\Delta_3(L)$ agrees very well with the GOE prediction up to
$L \simeq 15$. Bohigas {\it et al.}~\cite{Boh84c} also calculated
about 800 eigenvalues of the Bunimovich stadium and found the same
excellent agreement of the data with GOE predictions.  Mainly on the
basis of these results, they formulated the famous conjecture quoted
in the introduction to the present section.  Beginning at that time,
many more examples were studied numerically, comprising billiards and
other systems with few degrees of freedom.  Some of these cases are
treated in other parts of this review. Most examples essentially
support the conjecture, some mark its limitations.

The first study of the transition from regular to chaotic spectral
fluctuations in a billiard is probably due to Robnik~\cite{Rob84} in
1984. His billiard is defined by the conformal quadratic map
$z^\prime=az+bz^2$ of the unit circle, described by the complex
coordinate $z$.  This is also referred to as the Pascalian snail, see
Fig.~\ref{figch11}. The parameter of interest is $\lambda=b/a$ with $0
\leq \lambda \leq 1/2$.  For $\lambda=0$ the billiard is integrable.
For $\lambda=1/2$, a cusp occurs and the map is no longer conformal.
For $1/4 \leq \lambda \leq 1/2$, the classical billiard shows strong
chaos~\cite{Rob83}. The quantum analogue displays a
transition~\cite{Rob84} from Poisson to Wigner--Dyson statistics
as $\lambda$ increases from $0$ to $1/2$.  Robnik also introduced
another billiard defined by the cubic map
\begin{equation}
z^\prime = \frac{z+bz^2+cz^3\exp(i\delta)}{\sqrt{1+2b^2+3c^2}} \ .
\label{2bi1a}
\end{equation}
It is called the Africa billiard since for some values of the
parameters $b$, $c$ and $\delta$, the shape resembles that of the
continent Africa. Ishikawa and Yukawa~\cite{Ish85} studied the
transition from Poisson to Wigner--Dyson statistics in yet another
billiard.

Of the numerous studies of billiards since 1984, we can mention but a
few. A strong line of research uses semiclassical quantization and the
theory of periodic orbits to investigate quantum properties of
non--integrable billiards. Reviews are found in
Refs.~\onlinecite{Eck88,Gut90,Boh91a}. Sieber {\it et
  al.}~\cite{Sie93} investigated the effect of the bouncing ball
orbits on the spectral fluctuations of the Bunimovich stadium. The
family of orbits which are perpendicular to the parallel straight
sections has a strong effect, such that the long--range spectral
fluctuations appear more regular.  Sieber {\it et al.} constructed a
semiclassical method to remove the contributions of these orbits from
the spectral fluctuations.  The general aspects of these effects had
earlier been studied by Keating and Berry~\cite{Keat87}.  Bohigas {\it
  et al.}~\cite{Boh93b} introduced the annular billiard. It consists
of the area between two circles with different radii, see
Fig.~\ref{figch11}. The billiard is integrable only when the circles
are concentric. The properties of the system depend on two parameters,
the ratio of the two radii and the distance between the two centers.
The phase space is mixed. In the quantum system, tunneling from one
regular region in phase space to another may be enhanced by the
intermediate chaotic domain (``chaos--assisted tunneling''). This
prediction of the Bohigas--Tomsovic--Ullmo model for mixed
systems~\cite{Boh93a} has been confirmed analytically by Doron and
Frischat~\cite{Dor95}.  Dittrich and Smilansky~\cite{Ditxx,Dityy} and
Dittrich {\it et al.}~\cite{Ditzz} studied a chain of pairwise coupled
billiards, with the aim to study localization, see also the review in
Ref.~\onlinecite{Dit96}. The universal form of the curvature
distribution (Sec.~\ref{qc1pca}) was numerically confirmed for the
Bunimovich stadium by Takami and Hasegawa~\cite{Tak92}, and for
Robnik's billiard defined above by Li and Robnik~\cite{LiR96}. In the
latter case, deviations for small curvatures from the analytical
form~(\ref{1pc2}) were attributed to peculiarities of the billiard.
Tomsovic~\cite{Toms96} studied parametric correlations in the
Bunimovich stadium to investigate phase space localization properties
of chaotic eigenstates. He introduced a correlator between state
overlap intensities and level velocities.  Tomsovic found sizable
correlations due to scarring of specific states Those correlations are
not described by RMT.  Sieber {\it et al.}~\cite{Siexx} discussed
parametric statistics in billiards with mixed boundary conditions.
Schanz and Smilansky~\cite{Schxx} studied the Sinai billiard from a
scattering point of view. They found good agreement of the spacing
distribution of the eigenphases of the scattering matrix with Random
Matrix predictions, see also the review in Ref.~\onlinecite{Smila95}.
Primack and Smilansky~\cite{Prim95} performed numerical simulations
for the three--dimensional Sinai billiard. They found good agreement
of the level statistics with RMT.

Pseudo--integrable billiards form a separate class, with specific
fluctuation properties.  A pseudo--integrable system is defined as a
dynamical system whose phase--space trajectories live on surfaces with
genus higher than one. The classical dynamics is not
chaotic~\cite{Zeml75}.  Billiards in the form of certain polygons fall
into this class.  Detailed studies of the spectral fluctuations in
pseudo--integrable models can be found, for example, in the works of
Richens and Berry~\cite{Rich81}, Cheon and Cohen~\cite{Che89a} and
Shudo {\it et al.}~\cite{Shud94}.  The spectral fluctuation properties
of those systems have confusing features: on shorter scales,
fluctuations properties such as the spacing distribution are often
indistinguishable from the Wigner surmise. Only on longer scales, in
the spectral rigidity for example, sizeable deviations from the
properties of a chaotic system become visible.  Balazs {\it et
  al.}~\cite{NBa95,Chat96} introduced another class of billiards:
These authors demonstrated that the free motion of a two--dimensional
rigid body bouncing between two walls is equivalent to a billiard
system. Thus, coin tossing can be viewed as a billiard problem.
Billiards involving a non--Euclidean metric are of great interest for
the problem of quantization. Schmit~\cite{CSc91} has given a review on
billiards on the hyperbolic plane. In chapter 19 of his
book~\cite{Gut90}, Gutzwiller discusses the motion on a surface of
constant negative curvature. The structure of wave functions can give
rich information on the relationship between properties of the
classical and the quantum billiard, see Heller's review~\cite{Hel91}
and Gutzwiller's book~\cite{Gut90}. In some billiard systems,
geometrical optics does not suffice to understand the motion. Examples
are ray splitting effects~\cite{Pra96}, see Sec.~\ref{qc2deb}, and
Andreev billiards~\cite{All96,Fra96z,Mel96,All97} which prompted the
introduction of new random matrix ensembles, see Sec.~\ref{spec_2}.
Recently, Borgonovi {\it et al.}~\cite{Borg96} and Frahm and
Shepelyansky~\cite{Frah97a,Frah97b} studied localization effects of
wave functions in billiards which partly destroy level repulsion. In
such regimes, the fluctuations lie between Poisson and GOE. These
authors studied diffusive effects in the Bunimovich stadium and rough
billiards, respectively. Again, in the semiclassical limit $\hbar\to
0$, pure GOE statistics is found. This relates to the general
discussion of dynamical localization in Sec.~\ref{qc2dl}, i.e.~to the
quantum suppression of classically chaotic diffussion.

\subsection{Demonstrative experiments}
\label{qc2de}

The experiments discussed below show that billiards are not purely
theoretical constructs but can be realized experimentally. For
instance, quantum billiards can be simulated by microwave cavities.
Such simulations are valuable because parameters such as the size and
the shape of the cavity, the range of frequencies etc. can be altered,
and they have been used to study specific aspects of quantum chaos.
Simulations are not as fundamental, of course, as experiments on
systems such as atoms, molecules, or nuclei where entirely new and
unexpected phenomena may occur. Still, in some cases the simulation
experiments exceed their original purpose and turn into interesting
research objects of their own right. Moreover, some experiments (such
as the ones on elastomechanical resonances in three--dimensional
solids) open new frontiers: The wave equation and the boundary
conditions in these bodies differs qualitatively from the
Schr\"odinger equation.

The field was pioneered by Schr\"oder~\cite{MSc54} who, about 40 years
ago and long before the connection to quantum chaos was established,
realized the connection between acoustic waves and microwaves. He
experimented with microwaves in order to study the acoustic properties
of rooms. He used rectangular cavities and, interestingly, analyzed
the nearest neighbor spacing distribution. For spacings larger than a
mean level spacing or so, he found agreement with the exponential law
now referred to as the Poisson distribution. He also discussed
connections of his experiments with number theory. This and much more
can be found in a fascinating book~\cite{MSc86} he wrote on number
theory in science and communication.

Demonstrative experiments with microwaves and elastomechanical waves
are discussed in Secs.~\ref{qc2dea} and~\ref{qc2deb}, respectively.

\subsubsection{Microwave cavities}
\label{qc2dea}

The stationary Helmholtz wave equations for the electric and the
magnetic field $\vec{E}(\vec{r})$ and $\vec{B}(\vec{r})$,
respectively, in a three--dimensional metal cavity read
\begin{equation}
(\Delta + \vec{k}^2)\vec{E}(\vec{r}) = 0 
\qquad {\rm and} \qquad
(\Delta + \vec{k}^2)\vec{B}(\vec{r}) = 0 
\label{2de1}
\end{equation}
where $\vec{k}$ is the wave vector. If the boundary of the cavity is
perfectly conducting, the tangential component of the electric and the
normal component of the magnetic field vector vanish at the boundary.
We consider a flat cavity formed by two plane, congruent and parallel
metal plates. The cavity is uniform in the direction perpendicular to
the planes, the $z$ axis. With $h$ the distance between the plates,
modes with wave number $k < \pi/h$ cannot have a node in the $z$
direction and are therefore effectively two--dimensional. These modes
are transverse magnetic, i.e.~the electric field vector is parallel to
the $z$ direction. For such modes, the relevant part of the wave
equation reduces to a two--dimensional equation for the $z$ component
of the electric field,
\begin{equation}
\left(\frac{\partial^2}{\partial x^2}+
      \frac{\partial^2}{\partial y^2}+k_x^2+k_y^2\right) 
                              E_z(x,y) = 0 \ .
\label{2de2}
\end{equation}
The tangential component of $\vec{B}$ can be calculated from $E_z$. If
we identify $E_z(x,y)$ with the wave function $\psi(x,y)$ and $k^2$ with
$2mE/\hbar^2$, Eq.~(\ref{2de2}) becomes identical to the Schr\"odinger
equation for a two--dimensional billiard. In the electromagnetic case,
the ray limit is attained if the wave length is small compared to the
dimension of the cavity. This corresponds to the classical limit in
quantum mechanics.

Almost simultaneously but independently, St\"ockmann and
Stein~\cite{Sto90} and Doron {\it et al.}~\cite{Dor90a} in 1990
published the first experimental studies of microwaves in such flat
metal cavities and interpreted their results from the viewpoint of
classical chaos. St\"ockmann and Stein measured about 1000 eigenmodes
of a Bunimovich stadium and of a Sinai billiard. Both cavities were
about half a meter in size and 8~mm thick. The experimental spacing
distribution agrees very well with the Wigner surmise. In a rectangle
of similar size, St\"ockmann and Stein obtained a Poisson
distribution.  Doron {\it et al.}~investigated scattering in an
``elbow'' shaped cavity and compared the autocorrelation function of
the scattering matrix versus frequency to a semiclassical
formula~\cite{Blu88} which predicts a Lorentzian shape. This shape
applies when many channels are open. The autocorrelation function is
non--Lorentzian in the case of few open channels~\cite{Lew91,Lew92}.
Lewenkopf {\it et al.}~\cite{Lew92} improved the analysis of the data
in the light of this fact. Doron {\it et al.}~also discussed the
enhancement of the Wigner time delay due to time--reversal symmetry.
Another scattering experiment in microwave cavities was performed by
Schultz {\it et al.}~\cite{SSc91}.

Sridhar~\cite{Sri91}, Sridhar and Heller~\cite{Sri92a} and Sridhar
{\it et al.}~\cite{Sri92b} used a cavity to study wave functions of
the Sinai billiard. To measure the distribution of the electrical
field strength within the cavity, Sridhar {\it et al.}~\cite{Sri92b}
used a perturbative technique. These authors demonstrated the
connection between the wave functions and classical periodic orbits
and identified scarred states predicted by Heller~\cite{Hel91}. They
also showed that a symmetry breaking in the Sinai billiard leads to
localized eigenfunctions. Related experiments were simultaneously done
by Stein and St\"ockmann~\cite{Ste92} in the Bunimovich stadium. The
measured microwave power is proportional to the imaginary part of the
Green function. The latter can be expressed semiclassically in terms
of periodic orbits by Gutzwiller's trace formula~\cite{Gut90}. Thus,
via Fourier transform, the measured power gives information about the
classical orbits.  Stein {\it et al.}~\cite{Stei95} extended these
investigations and thereby experimentally verified an insight due to
Tomsovic and Heller~\cite{Toms91,Toms93}. These authors had found that
the semiclassical dynamics has a strong impact on the quantum
mechanical behavior in billiards over very long time scales.

The statistics of wave functions was measured directly by Kudrolli
{\it et al.}~\cite{Kud95}. For the Bunimovich stadium, slight
deviations from the Porter--Thomas law are caused by the bouncing ball
orbits. The correlator of wave functions at different space points
(see Sec.~\ref{qc1wwb}) agrees well with theoretical predictions for
chaotic systems. Tiles in the cavity which act as hard scatterers
change the cavity into a disordered system. Deviations from the
predictions for chaotic systems due to localization effects are found
both for wave function statistics and correlators. The results agree
well with a formula by Fyodorov and Mirlin~\cite{Mir93} which accounts
for localization effects. Prigodin {\it et al.}~\cite{Pri95b} also
compared experimental and theoretical results for wave functions, see
Sec.~\ref{qc1wwb}.

All these experiments suffer from a broadening of the resonances due
to absorption by the walls of the cavity. This makes it impossible to
separate close--lying resonances. Missing levels pose a serious
problem for the analysis of spectral fluctuations. To cure this
problem, Gr\"af {\it et al.}~\cite{Gra92} used superconducting niobium
cavities at a temperature of 2~K. This results in a dramatic
improvement of the resolution. About 1000 well resolved resonances
were measured in the Bunimovich stadium. The effect of the
bouncing--ball orbits (Sec.~\ref{qc2bi}) is clearly visible in the
spectral rigidity, and can be removed with the help of the
semiclassical analysis of Sieber {\it et al.}~\cite{Sie93}. With the
same superconducting technique, Alt {\it et al.}~\cite{Alt94} measured
eigenmodes in the hyperbola billiard and compared the results with
various transition formulae for the spectral fluctuations. Because of
the high resolution, Alt {\it et al.}~\cite{Alt93} could compare the shape
of individual resonances with the Breit--Wigner formula and found
excellent agreement over the full dynamical range. Likewise, Alt et.
al~\cite{Alt95} found good agreement of the distribution of partial
widths with the Porter--Thomas law, see Sec.~\ref{qc1wwa}. These
authors also confirmed the above--mentioned theoretical predictions of
Lewenkopf and Weidenm\"uller~\cite{Lew91} and Harney {\it et
al.}~\cite{Har92} for the autocorrelation function of the scattering
matrix in the limit of isolated resonances. The non--Lorentzian shape
of this function implies that the Fourier transform decays
algebraically and not exponentially. A detailed investigation of the
decay properties of the Bunimovich stadium billiard was performed by
Alt {\it et al.}~\cite{Alt96y}.

Measurements of spectral observables at a crossover transition are
particularly interesting. For the case of gradual breaking of
time--reversal invariance, such experiments were conducted by So {\it
  et al.}~\cite{SoA95} and Stoffregen {\it et al.}~\cite{USt95}.  So
{\it et al.}~used a cavity with a thin ferrite strip adjacent to one
of the walls.  Time--reversal invariance is broken when a magnetic
field is applied to the ferrite. In the $\Delta_3$ statistic, the GOE
$\to$ GUE transition was observed with surprisingly good resolution.
Stoffregen {\it et al.}~placed a wave guide inside a microwave cavity.
The wave guide contained a ``microwave isolator'' controlled by a
magnetic field.  One side of the cavity could be moved. This made it
possible to study the motion of levels versus the length of the
cavity. Stoffregen {\it et al.}~determined the nearest neighbor
spacing distribution, the tail of the curvature distribution of the
eigenvalues, and the distance of closest approach at avoided crossings
and found agreement of all three observables with GUE predictions. In
an earlier study, St\"ockmann {\it et al.}~\cite{Sto94a} had already
shown experimentally that in the case of time--reversal invariance,
the tail of the curvature distribution agrees well with theoretical
predictions, see Sec.~\ref{qc1pca}. A detailed theoretical study
of microwave billiards with broken time--reversal invariance was
given by Haake {\it al.}~\cite{Haak96}. Recently, Kollmann {\it et
  al.}~\cite{Kollm94} and St\"ockmann {\it et al.}~\cite{Stock96}
compared experimental results on the level motion with the predictions
of the Pechukas gas~\cite{Pec83}.  They also study velocity and
curvature distributions, see Sec.~\ref{qc1pc}.  Moreover, a link to
periodic orbit theory is discussed.

Richter~\cite{Rich97} discusses numerous further projects with
microwaves: The Pascalian snail for the study of the regularity
chaos transition, the annular billiard for the investigation of
chaos assisted tunneling and experiments addressing localization.

In addition to the spectral rigidity, there exists another statistical
observable, the correlation hole, to measure long--range correlations,
see Sec.~\ref{qc1bad}. The Fourier transform of the spectral
two--point correlation function maps long--range properties onto
short--range ones in Fourier space. For RMT correlations, this
function has a hole at small values of the variable. Kudrolli {\it et
  al.}~\cite{Kud94} found good agreement with the GOE prediction. Alt
{\it et al.}~\cite{Alt97} investigated the correlation hole in more
detail, including an analysis of scattering data.

As early as 1991, Haake {\it et al.}~\cite{Haak91} used microwave
experiments to study a certain pseudo--integrable billiard, see
Sec.~\ref{qc2bi}.  Shudo {\it et al.}~\cite{Shud94} investigated
pseudo--integrable billiards in the shape of polygons.

Cavities have been used also for another purpose. Cavities with simple
boundaries show good agreement with the Weyl formula for the mean
density of eigenvalues. This changes when tiles are placed within the
cavity to roughen it. Deviations from the Weyl formula can be viewed
as indicating a change of dimensionality~\cite{Srix1}. In another
study, Sridhar and Kudrolli~\cite{Srix2} confirm experimentally the
theorem of isospectrality. It states that pairs of billiards with
different, but related, shapes exist which have identical spectra.

RMT was designed to describe fluctuation properties of quantum
systems.  Does it also work for the spectral fluctuation properties of
other wave equations? Deus {\it et al.}~\cite{Deu95}, Alt {\it et
  al.}~\cite{Alt96x,Alt97x} and D\"orr {\it et al.}~\cite{Doer97}
studied the statistics of eigenmodes in three--dimensional microwave
cavities. In a truly three--dimensional system, the vector Helmholtz
equations~(\ref{2de1}) are structurally different from the scalar
Schr\"odinger equation. Deus et al. found very good agreement of the
spacing distribution with the Wigner surmise but deviations of the
long--range correlations from the GOE prediction. They attributed the
difference to a remnant of regularity in the irregularly shaped
cavity. Alt et al.~\cite{Alt96x} used a more regular cavity and found
statistics intermediate between Poisson and GOE even for the spacing
distribution.  Alt {\it et al.}~\cite{Alt97x} study a
three--dimensional Sinai billiard. In both investigations, periodic
orbit theory is applied and bouncing ball modes are extracted. The
same geometry was investigated by D\"orr {\it et al.}~\cite{Doer97}
who studied scarring and various statistical observables of the
electromagnetic field distributions.

\subsubsection{Acoustics and elastomechanics}
\label{qc2deb}

Weaver~\cite{Wea89} was the first author to study the question raised
in the previous paragraph: Does RMT apply to the spectral fluctuations
of wave equations different from the Schr\"odinger equation? In 1989,
a year before the first microwave experiments were reported, he
studied the elastomechanical eigenmodes of aluminum blocks. The motion
of the displacement vector $\vec{u}(\vec{r},t)$ of a mass element at
position $\vec{r}$ is governed by Navier's equation~\cite{Lan59}. For
an isotropic material, the stationary wave equation~\cite{Lan59} for
the displacement vector $\vec{u}(\vec{r})$ can be decomposed into two
Helmholtz equations for the longitudinal ($L$) and transverse ($T$)
parts. These equations read
\begin{equation}
\left(\Delta + \frac{\omega^2}{c_X^2}\right)\vec{u}_X(\vec{r}) = 0
                \qquad {\rm for} \quad X=L,T \ .
\label{2de5}
\end{equation}
Here $c_L$ and $c_T$ are the longitudinal and the transverse
velocities, respectively, and $\omega$ is the frequency.  Free
boundary conditions (vanishing stress at the surface) apply. The
longitudinal mode represents a pressure wave, the two transverse
modes, shear waves. These modes are coupled through reflection at the
boundary. Since always $c_L > c_T$, Snell's reflection law leads to
ray splitting: The reflection of an incoming longitudinal or
transverse wave at the boundary yields both a longitudinal and a
transverse wave traveling in different directions. This phenomenon is
also referred to as mode conversion. Thus, elastomechanical wave
equations are structurally very different from the Schr\"odinger
equation and even more complicated than the wave equation for the
electromagnetic field in three dimensions.

Weaver used rectangular aluminum blocks a few centimeters in size with
angled slits. The slits break the symmetry of the blocks. Dropping
steel balls onto the blocks and measuring the response, he found about
150 eigenmodes. The spacing distribution agrees well with the Wigner
surmise, but the long--range correlations deviate from GOE
statistics. Why are the results for these nearly regular blocks so
close to GOE? Bohigas {\it et al.}~\cite{Boh91b} attributed this fact to the
angled slits which act as defocusing structures. Delande {\it et
al.}~\cite{Del94} used periodic orbit theory to analyze the problem and
reached a similar conclusion. Weaver and Sornette~\cite{Wea95} showed
that these systems can be viewed as being pseudo--integrable. On short
frequency scales, the correlations are indistinguishable from GOE
predictions. These authors supported their conclusion by numerical
simulations of a rectangular membrane with a point scatterer. Cheon
and Cohen~\cite{Che89a} reached a similar conclusion for polygons and
pseudo--integrable billiards.

Using Weaver's technique on brick--sized aluminum blocks, Ellegaard
{\it et al.}~\cite{Ell95} measured about 400 eigenmodes in the
interval from 0 to 100 kHz. Rectangular blocks display Poisson
statistics. The spectrum is a superposition of eight spectra
pertaining to different symmetries. In such a case, Poisson statistics
is known~\cite{Meh91} to apply almost irrespective of the statistics
of the individual spectra, see Sec.~\ref{qc1bae}.  Using blocks in the
shape of a three--dimensional Sinai billiard, Ellegaard {\it et
  al.}~observed the Poisson $\to$ GOE transition for the first time in
a three--dimensional system.

\begin{figure}
\centerline{
\psfig{file=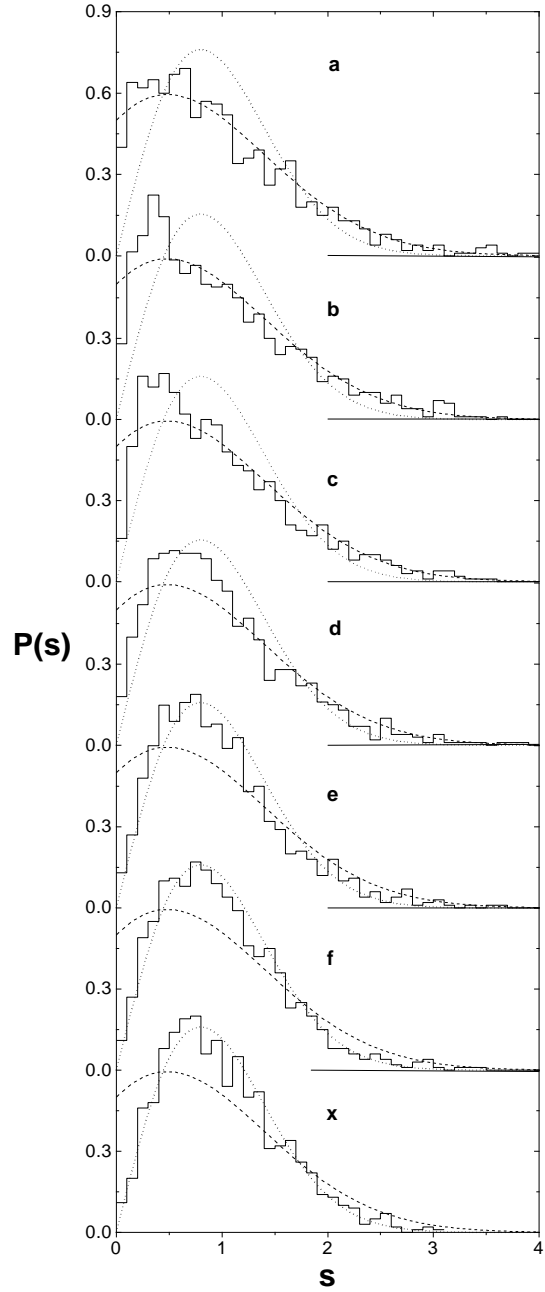,width=3.5in}
}
\caption{
  Nearest neighbor spacing distribution $P(s)$ for the different
  radii $r$ of the removed octant: (a) $r=0$~mm, (b) $r=0.5$~mm, (c)
  $r=0.8$~mm, (d) $r=1.1$~mm, (e) $r=1.4$~mm, (f) $r=1.7$~mm,
  (x) a huge radius of $r\simeq 10$~mm. The dotted and the dashed
  curves are the theoretical predictions for a chaotic system
  containing no or one symmetry, respectively. 
  Taken from Ref.~\protect\onlinecite{Ell96}. 
} 
\label{figch7}
\end{figure}

Monocrystalline quartz blocks a few centimeters in size are even
better suited objects because about 1500 eigenvalues can be
measured~\cite{Ell96} in the range from 600 to 900 kHz.  Unlike
aluminum, quartz is an anisotropic substance. Both the longitudinal
and transverse velocities differ in the direction of different axes.
Thus, the wave equation is even more complicated than
Eqs.~(\ref{2de5}) which apply to isotropic materials.  Quartz belongs
to the discrete crystal group D$_3$. The rectangular blocks used by
Ellegaard {\it et al.}~\cite{Ell96} were cut in such a way that all
symmetries are fully broken except a ``flip'' symmetry which has two
representations, just like parity. The spectra were measured in
transmission. The resolution was close to that of superconducting
microwave cavities. The existence of the flip symmetry makes it
possible to measure a gradual symmetry breaking which is statistically
fully equivalent to the breaking of a quantum number such as parity or
isospin~\cite{Mit88,Guh90}, see Secs.~\ref{qc1tc} and~\ref{qc3nue}.
The symmetry violation is achieved by removing successively bigger
octants of a sphere from one corner, thereby again creating a
three--dimensional Sinai billiard. The results for the spacing
distribution are shown in Fig.~\ref{figch7}. They are in agreement
with the random matrix simulations of Ref.~\onlinecite{Guh90}. The tiny
symmetry violation induced by small octants immediately lifts all
degeneracies. The spectral rigidity deviates from the GOE prediction.
This is probably because the blocks used have features similar to
those of a scalar pseudo--integrable system, see Sec.~\ref{qc2bi}.  On
short scales, however, the spectral fluctuations are indistinguishable
from RMT.  

The experiments show that RMT applies far beyond quantum mechanics to
classical wave equations which are very different from the
Schr\"odinger equation. This is true even for anisotropic materials
and for non--trivial effects such as crossover transitions. As Weaver
points out in his original paper~\cite{Wea89}, statistical analysis
has had a long history in acoustics, especially room acoustics and
instrument calibration, see Ref.~\onlinecite{Ebe84}. Vibrations of
complex structures have likewise been analyzed for a long time with
statistical methods~\cite{Lyo75}. We are under the impression,
however, that Weaver was the first author to recognize the importance
of RMT for the understanding of fluctuation properties in acoustics
and elastomechanics. (Schr\"oder's work~\cite{MSc54} addressed a
regular system).

We turn to two--dimensional systems. Berry~\cite{Ber81} cites
unpublished experimental work by Ede on the vibrations of metal plates
in the shape of a Sinai billiard. Sapoval {\it et al.}~\cite{Sap91}
investigated numerically and experimentally vibrations of membranes
and found that fractal boundaries lead to localization effects. This
is related to the work of Sridhar {\it et al.}~\cite{Srix1} mentioned
above.  In 1992, Legrande {\it et al.}~\cite{Leg92} studied numerically
the flexural vibrations (bending modes) of thin metal plates in the
shape of a Bunimovich stadium and extended in the $(x,y)$ plane. The
stationary wave equation for the displacement $u_z(x,y)$ perpendicular
to the plate is of fourth order and can be written as
\begin{equation}
(\Delta + k^2) (\Delta - k^2) u_z(x,y) = 0
\label{2de6}
\end{equation}
where $\Delta$ is the Laplacean in two dimensions. The wave number is
given by $k^4=3\omega^2\rho(1-\nu^2)/Eh^2$ where $\omega$ is the
frequency, $\rho$ the density, $h$ the thickness, $\nu$ the Poisson
number and $E$ the elasticity module. Longitudinal modes obeying the
second order Helmholtz equation exist as well in elastic plates, but
are not studied in Ref.~\onlinecite{Leg92}. Legrande {\it et al.}~used
the boundary condition of a clamped plate. Because of the
form~(\ref{2de6}) the waves are a superposition of Helmholtz waves,
and of exponentially decreasing and increasing solutions due to the
operator $\Delta-k^2$. Nevertheless, the spectral fluctuations agree
well with GOE predictions.  Recently, Bogomolny and
Hugues~\cite{Bogo96} calculated the smooth part of the cumulative
level density, i.e.~the Weyl formula, for flexural vibrations in
elastic plates. They used a ``semiclassical'' approximation.
Moreover, they worked out trace formulae for the integrable and the
chaotic case.

Ray splitting occurs also in two--dimensional systems and causes
problems similar to those of elastomechanics. Prange {\it et
al.}~\cite{Pra96} calculated the cumulative eigenfrequency density in a
two--dimensional system with ray splitting. In three dimensions, the
technical difficulties are such that the cumulative eigenvalue density
of a solid with free boundaries has not yet been calculated. Dupois
{\it et al.}~\cite{Dup60} solved a simplified problem by introducing
periodic boundary conditions.

\subsection{Quantum dots}
\label{qc2qd}

During the last decade, it has become possible to manufacture
microstructures of $\mu$m size. In the case of semiconductors, the
electrons form a two--dimensional gas located between the two layers
of a heterostructure. The potential confining this two--dimensional
gas can be controlled extremely well, and it is possible to fabricate
devices with widely varying shapes. At temperatures below 1~K or so,
the phase coherence length is larger than the system size, and quantum
coherence plays an important role. The elastic mean free path is
typically $10~\mu$m or larger. The electrons move ``ballistically'',
i.e.~they scatter only at the boundaries defined by the confining
potential. The device can be viewed as a billiard. When coupled to
external leads, it is called a quantum dot. The number of electrons on
the dot varies typically between a few and several hundred. The
coupling between dot and leads may be controlled by gates which create
a potential barrier. A review of electron transport in quantum dots
was recently given by Kouwenhoven {\it et al.}~\cite{Kouw97}.  Here,
we are concerned with level statistics and wave function correlations
in ballistic quantum dots. In our discussion, we disregard the Coulomb
interaction between electrons, save for the charging energy which is a
mean field effect. Interaction effects in quantum dots have drawn
interest only recently and are not discussed here, a detailed
discussion was presented by Marcus {\it et al.}~\cite{Marc97}.

It is very useful to discuss quantum dots in the framework of
scattering theory.  Let $G$ denote the conductance and $\Lambda$ the
total number of channels in each lead, defined by the number of
transverse modes below the Fermi energy. The dimensionless conductance
$g = (h/e^2)G$ is the central object of theoretical interest. 
The Landauer formula 
\begin{equation}
g = \sum_{a=1}^\Lambda \sum_{b=1}^\Lambda
         \left(|S_{ab}^{LR}|^2 + |S_{ba}^{RL}|^2\right) 
\label{2qd1}
\end{equation}
expresses $g$ as the total transmission. The latter can be written in
terms of the scattering matrix $S$ describing the electron transport
through the device. Here, $S^{LR}$ and $S^{RL}$ are those blocks of
the scattering matrix $S$ which describe scattering from the left to
the right lead and from the right to the left lead, respectively. A
more detailed discussion of this formula is given in
Sec.~\ref{quasi1d_1}.  The dimensionless conductance $g$ is bounded by
twice the number of channels in one lead, $g \leq 2\Lambda$. The
properties of the dot depend on the strength of its coupling to the
leads. Without barriers, we have $g \geq 2$. The electrons move freely
between dot and leads, the quasibound states in the dot become
strongly overlapping resonances, and $g$ shows conductance
fluctuations. This regime is addressed in Sec.~\ref{qc2qda}. For $g <
1$ or so, the electrons must tunnel through the barriers between dot
and leads, and $g$ displays isolated resonances.  This ``Coulomb
blockade regime'' forms the topic of Sec.~\ref{qc2qdb}.

\subsubsection{Magnetoconductance and level correlations}
\label{qc2qda}

In 1990, Jalabert {\it et al.}~\cite{Jal90} suggested that
fluctuations of observables such as the conductance might contain
information about the shape of quantum dots. The spectral fluctuations
of a regular system such as the circle differ qualitatively from those
of a chaotic system such as the Bunimovich stadium. This difference
might also affect the conductance fluctuations of quantum dots with
either shape. It was known that magnetotransport properties in small
two--dimensional conductors do depend on the
geometry~\cite{Ree89,Bar89a,Bee89}, and Doron {\it et
  al.}~\cite{Dor90a} investigated almost simultaneously the open
``elbow'' billiard with a similar aim, cf. Sec.~\ref{qc2dea}.  But
Jalabert {\it et al.}~established the connection between classical
chaos and ballistic quantum dots. These authors investigated both
semiclassically and numerically the fluctuations of $g$ versus an
external magnetic field (``magnetoconductance fluctuations'').
Jalabert {\it et al.}~used the Landauer formula~(\ref{2qd1}) and
calculated semiclassically the autocorrelation function of $g$ versus
magnetic field strength $B$,
\begin{equation}
c(B) = \langle \delta g(\overline{B}-B/2)
                     \delta g(\overline{B}+B/2) \rangle
\label{2qd2}
\end{equation}
where $\delta g = g - \langle g\rangle$. The average is taken over an
appropriate interval of values of $\overline{B}$. For a chaotic dot,
they found
\begin{equation}
c(B) = \frac{c(0)}{\left(1+(eB/hc\alpha_{\rm cl})^2\right)^2}
\label{2qd3}
\end{equation}
where the constant $1/\alpha_{\rm cl}$ denotes the root--mean--square
area enclosed by trajectories traversing the dot. The power spectrum
(Fourier transform) of $c(B)$ decays exponentially,
\begin{equation}
s(\omega) = s(0) (1+hc\alpha_{\rm cl}\omega) 
                    \exp(-hc\alpha_{\rm cl}\omega) \ .
\label{2qd4}
\end{equation}
As mentioned in Sec.~\ref{qc2dea}, Bl\"umel and Smilansky~\cite{Blu88}
had computed a related quantity, the energy correlator of the
scattering matrix, also semiclassically. The result~(\ref{2qd4}) holds
only for weak fields where the cyclotron radius is larger than the
size of the dot. Moreover, the semiclassical approximation used in the
derivation is justified only for large channel number, $\Lambda \gg
1$. Jalabert {\it et al.}~compare formula~(\ref{2qd3}) with numerical
simulations in two classically chaotic scattering systems: a
Bunimovich stadium with two external leads and a four disk junction.
The latter consists of four identical disks whose centers form the
corners of a square. The distance between these disks is a quarter of
the disk radius. Thus the four disks form an open scattering
system~\cite{Eckh90,Hans95}, extending the famous three disk
scattering system~\cite{Cvi89}. Good agreement is found.  Deviations
in the tail of the Lorentzian are ascribed to non--universal behavior
due to short trajectories.  Moreover, to check the uniqueness of their
findings for chaotic systems, Jalabert {\it et al.}~perform numerical
simulations on a rectangle which is integrable. There are
fluctuations, but the Fourier transform indicates that they are
qualitatively different and have non--universal shape. Related
theoretical work was presented by Doron {\it et al.}~\cite{Dor91},
Jensen~\cite{Jen91} and Oakeshott and MacKinnon~\cite{Oak92}.

In 1992, Marcus {\it et al.}~\cite{Mar92} performed an experimental
\begin{figure}
\centerline{
\psfig{file=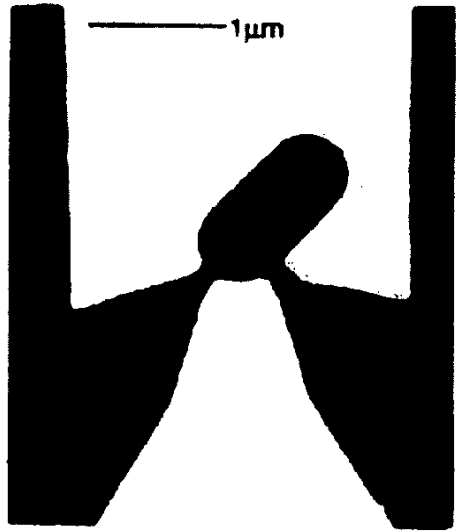,width=2.5in}
}
\caption{
  Electron micrograph of a quantum dot in the shape of a Bunimovich
  stadium, with $1~\mu$m bar for scale. This device was used in the 
  experiments of Ref.~\protect\onlinecite{Mar92}. The electrons can move 
  in the black area. Two leads are coupled to the stadium.
  Adapted from Ref.~\protect\onlinecite{Mar92}.  
}
\label{figch10}
\end{figure}
test of the difference between regular and chaotic quantum dots.
These authors fabricated two quantum dots in the form of a circle of
radius $0.44~\mu$m, and two quantum dots in the shape of a Bunimovich
stadium of length $1.2~\mu$m and width $0.6~\mu$m, see
Fig.~\ref{figch10}. All dots had the same area of $0.41~\mu{\rm m}^2$.
Both the electron density of $3.8\cdot 10^{11}~{\rm cm}^{-2}$ and the
mobility were measured and yielded an elastic mean free path of
$2.6~\mu$m, several times the sizes of the dots, so that the electrons
move ballistically.  The power spectra show a striking difference for
the two geometries.  The semiclassical prediction~(\ref{2qd4}) for
chaotic motion agrees with the data for the Bunimovich stadium and not
with those for the circle.  The results are shown in
Fig.~\ref{figch2}.
\begin{figure}
\centerline{
\psfig{file=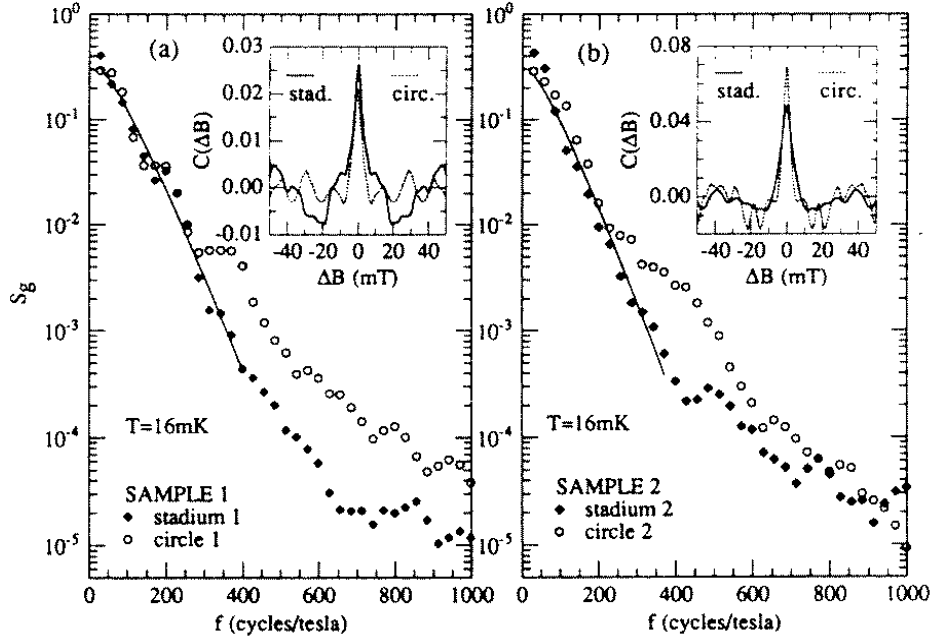,width=5.5in}
}
\caption{
  Averaged power spectra (here denoted by $S_g(f)$) of conductance
  fluctuations $\delta g(B)$ for stadium (solid diamonds) and circle
  (open circles) with approximately three transverse modes. The left
  and the right figures display the results for two different samples.
  Solid curves are fits of the semiclassical theory.  The
  autocorrelation functions (here denoted by $C(\Delta B)$) of stadium
  (solid) and circle (dashed) are shown as insets.  
  Taken from Ref.~\protect\onlinecite{Mar92}.  
}
\label{figch2}
\end{figure}
The number of channels in this experiment was rather small, $\Lambda
\simeq 3$. Hence the semiclassical analysis is not very well
justified.  However, Marcus {\it et al.}~provided convincing
experimental evidence that, as Jalabert {\it et al.}~had predicted,
the difference between regular and chaotic motion reveals itself in
the conductance fluctuations of quantum dots.

At zero magnetic field, the resistance displays sizeable peaks. The
widths differ for the two geometries. At 20~mK, Marcus {\it et
  al.}~found a few milli Tesla for the width of the peak of the
Stadium and a much smaller value for that of the circle. A connection
between these peaks and the weak localization effect~\cite{Lee85} in
disordered systems was suggested. Baranger {\it et al.}~\cite{Bar93}
investigated this phenomenon semiclassically. At zero magnetic field,
the system is time--reversal invariant. The amplitudes which
contribute to the reflection coefficient come in pairs.
Semiclassically, the members of a pair correspond to time--reversed
classical trajectories and interfere constructively. As the magnetic
field is turned on and time--reversal symmetry is broken, the two
amplitudes get out of phase. Hence at zero field, the reflection
coefficient has a maximum and the transmission coefficient has a
minimum. This leads to a maximum of the resistance.  Baranger {\it et
  al.}~showed that this weak localization effect depends on whether
the dynamics in the dot is regular or chaotic. The calculation
accounts for the existence of the peaks and for the difference in
widths. The authors pointed out that the ``diagonal
approximation''~\cite{Gut90} typically used in the semiclassical
approximation (Sec.~\ref{qc4po}) does not yield all relevant
contributions, in contrast to a full--fledged random matrix model such
as that of Iida {\it et al.}~\cite{Iid90}.

In further experimental work, the weak localization dip of the
conductance in the Bunimovich stadium was measured~\cite{Ber94} for
temperatures between 1.5~K and 4.25~K. In keeping with semiclassical
arguments, the dip becomes less pronounced as the temperature
increases, see also Ref.~\onlinecite{Kel94}.

The experiments on chaotic quantum dots had shown that the shape of
the weak localization peak is Lorentzian. Pluhar {\it et
  al.}~\cite{Plu94,Plu95} reproduced this behavior in a random matrix
model, using the random Hamiltonian approach to the scattering matrix
as discussed in Sec.~\ref{qc1cs}. The dependence on magnetic field was
modeled as a GOE $\to$ GUE crossover transition. The fictitious time
$t$ in Eq.~(\ref{Me}) of Sec.~\ref{cons} is proportional to the square
of the magnetic flux through the dot. In the notation of
Sec.~\ref{qc1tc} one has $t \propto \lambda^2$. Assuming ideal
coupling to the leads, these authors obtain an exact expression in
terms of a three--dimensional integral for $\langle g(t,\Lambda)
\rangle$.  Numerical integration shows that the expression
\begin{equation}
\langle g(t,\Lambda) \rangle \simeq \Lambda \, - \,
   \frac{\Lambda}{2\Lambda+1} \, \frac{1}{1+2(2\Lambda-1)t/\Lambda^2}
\label{2qd8}
\end{equation}
is an excellent approximation to $\langle g(t,\Lambda) \rangle$ for
all $\Lambda \geq 2$. This expression has Lorentzian shape. Gerland
and Weidenm\"uller~\cite{Ger96} extended this calculation to the case
of non--ideal coupling to the leads (barriers). They show that the
peak shape is always Lorentzian but that the parameters depend
sensitively on the transmission coefficients. Since electron--electron
interaction is not included in this model, these results are not valid
in the Coulomb blockade regime.

Using a Brownian motion model in the space of scattering matrices,
Frahm and Pichard~\cite{FrP95a} and Rau~\cite{Rau95} investigated the
same problem, now formulated as a crossover transition from COE to CUE
in the random $S$--matrix approach (cf. Sec.~\ref{quasi1d}). Frahm and
Pichard calculated all correlators of of $g$ as functions of the
eigenvalues of the transmission matrix and of a fictitious time $t$.
Unfortunately, it has not been possible to relate $t$ with the flux
through the dot. Moreover, the analytical dependence of $\langle
g(t,\Lambda) \rangle$ on $t$ differs from Eq.~(\ref{2qd8}). This
suggests that the Brownian motion model in the space of $S$ matrices
may not be a suitable means to model the crossover transition. This is
in line with a difficulty often encountered in attempts to model
parametric correlations in the random $S$ matrix (or transfer matrix)
approach, cf. Sec.~\ref{quasi1d}.

The calculation of the correlator $c(B)$ of the magnetoconductance
defined in Eq.~(\ref{2qd2}) within a random matrix approach poses a
much harder problem. Frahm and Pichard~\cite{FrP95a} point out that it
is highly doubtful whether $c(B)$ can be worked out at all in the
random $S$ matrix approach. Within an appropriate extension of the
model of Pluhar {\it et al.}, Frahm~\cite{Fra95e} calculated $c(B)$
using an asymptotic expansion in the inverse channel number $\Lambda$
and obtained the result~(\ref{2qd2}). Recently, Gossiaux {\it et
al.}~\cite{Gos96} tackled the full problem for arbitrary channel
number.

Prigodin {\it et al.}~\cite{Pri93} investigated the conductance
distribution $P(g)$ for a chaotic quantum dot weakly coupled to leads.
The Hamiltonian has the form
\begin{equation}
H = H_c + \frac{i\alpha}{2\pi}\Delta + 
    \frac{iV\Delta}{2\pi}\left(\alpha_1\delta(\vec{r}-\vec{r}_1)
                    +\alpha_2\delta(\vec{r}-\vec{r}_2)\right) \ .
\label{2qd5}
\end{equation}
Here, $H_c$ is the disorder Hamiltonian of the chaotic dot. The
second, non--Hermitean term on the right hand side of Eq.~(\ref{2qd5})
accounts for incoherent processes such as electron--phonon scattering
which produce a width $\overline{\Gamma}=\alpha\Delta / 2 \pi$
measured in units of the mean single--particle level spacing $\Delta$.
Prigodin {\it et al.}~focus on the regime $\Delta \ll
\overline{\Gamma} \ll E_C$ where $E_C$ is the Thouless energy, see
Sec.~\ref{disorder}. In this regime, an interaction--free theory is
appropriate.  The last term on the right hand side of Eq.~(\ref{2qd5})
describes~\cite{Zir92} the coupling to the leads via point contacts at
$\vec{r}_1$, $\vec{r}_2$ with $V$ the volume and $\alpha_i, \ i=1,2$
dimensionless parameters.  Prigodin {\it et al.}~study this model with
Efetov's supersymmetry formalism. In the zero mode approximation and
in the weak tunneling limit $\alpha \gg \alpha_i, \ i=1,2$, they
compute the moments
\begin{equation}
\langle g^n \rangle = \int P(g) g^n dg \ , \qquad n=0,1,2,\ldots \ .
\label{2qd6}
\end{equation}
of the conductance. If the magnetic field is sufficiently large, weak
localization effects can be neglected and the unitary symmetry class
is appropriate. The distribution $P(g)$ decreases monotonically with
$g$ and $g=0$ is the most probable value. This is in contrast to the
case of systems strongly connected with leads. Prigodin {\it et
  al.}~also re--derive the Porter--Thomas distribution for the unitary
case, see Sec.~\ref{qc1wwb}. Marcus {\it et al.}~\cite{Mar94}
measured the magnetoconductance in the tunneling regime where the dot
is nearly isolated. The power spectrum is enhanced at high magnetic
frequencies.  This is consistent with the results of Prigodin {\it et
  al.}~\cite{Pri93}.  Moreover, the magnetoconductance displays a
crossover from aperiodic fluctuations at lower fields to periodic,
Aharonov--Bohm--like fluctuations at higher fields. At the crossover
point, the cyclotron radius of the electron motion has roughly the
size of the Bunimovich stadium. For fields larger than this critical
value, the electron bounces along the boundary of the stadium, and the
motion becomes regular.
 
Baranger and Mello~\cite{Bar94} and Jalabert {\it et al.}~\cite{Jal94}
used the random $S$ matrix approach to calculate statistical
observables, particularly the distribution of the dimensionless
conductance $g$. They discuss their results as functions of the
channel number $\Lambda$ and the universality class $\beta$. In the
case of only one channel in each of the two leads, they find
\begin{equation}
P(g) =  \frac{\beta}{2} g^{\beta/2-1}
\label{2qd7}
\end{equation}
which shows a qualitative difference for the three cases
$\beta=1,2,4$.  With increasing channel number, $P(g)$ quickly
approaches a Gaussian with non--zero mean. These results were obtained
for ideal leads and for maximal coupling $\Gamma$ between dot and
leads, and do not contradict those of Prigodin {\it et al}. Brouwer
and Beenakker~\cite{Bro94} bridged the gap between these two
approaches by calculating, in the random $S$ matrix model, the
distribution of $g$ for arbitrary $\Gamma$ and for all $\beta$. For
$\beta = 2$ and small $\Gamma$, the result of Prigodin {\it et al.}~is
reproduced.

\subsubsection{Coulomb blockade regime and wave function statistics}
\label{qc2qdb}

In this regime, the barriers between dot and leads are so high that
the intrinsic mean width $\overline{\Gamma}$ of the quasibound states
in the dot is much smaller than the resonance spacing. The conductance
is measured as function of the gate voltage applied to the dot. At low
temperature, a series of well--defined isolated resonances is seen.
Such measurements were made by Meirav {\it et al.}~\cite{Mei90},
Kouwenhoven {\it et al.}~\cite{Kou91}, McEuen {\it et
  al.}~\cite{McE91}, Weis {\it et al.}~\cite{Wei92} and Foxman {\it et
  al.}~\cite{Fox93}, see also the review by Kastner~\cite{Kas92}. At
each resonance, another quasibound level of the quantum dot passes
through the Fermi surface from above as the gate voltage is increased,
and the number of electrons on the dot increases by one.
Beenakker~\cite{Bee91} has shown that the resonance spacing
$E_n-E_{n-1}+e^2/C$ is the sum of the spacing $E_n-E_{n-1}$ between
adjacent quasibound single--particle levels in the dot, and the
charging energy $e^2/C$. This charging energy suppresses tunneling
processes between resonance peaks and causes the Coulomb blockade. For
sufficiently small values of the capacity $C$, the charging energy
dominates the expression $E_n-E_{n-1}+e^2/C$. In a plot of conductance
versus gate voltage, equally spaced peaks appear. Several theoretical
investigations~\cite{McE91,Bee91,Mei91} address the Coulomb blockade
regime. Here we focus on a feature which is of particular interest for
RMT. In many of the experiments~\cite{Mei90,McE91}, the peak widths
are dominated by temperature broadening, so that $\overline{\Gamma}
\ll k_BT \ll e^2/C$, and all peaks have the same width. Nevertheless,
the peak heights fluctuate by orders of magnitude. This behavior can
be modeled within a random matrix approach. Several aspects of this
problem are of general interest and are presented in Sec.~\ref{qc1ww}.
Here we discuss the applications specific to quantum dots.
   
We are interested in the regime $\overline{\Gamma} \ll k_BT <
(E_n-E_{n-1})$. The height of a resonance peak is given by
\begin{equation}
  g_{\rm max} = \frac{\overline{\Gamma}}{4\pi k_BT} \alpha \qquad {\rm
    with} \qquad \alpha = \frac{1}{\overline{\Gamma}}
  \frac{\Gamma^{(L)}\Gamma^{(R)}} {\Gamma^{(L)}+\Gamma^{(R)}}
\label{2qd9}
\end{equation}
where $\Gamma^{(L)}$ and $\Gamma^{(R)}$ are the partial decay widths
of the resonance into the channels of the left and the right lead,
respectively. RMT had been shown to give the correct description for
the spectral fluctuations in irregularly shaped quantum dots. In 1992,
Jalabert {\it et al.}~\cite{Jal92} used RMT to determine the statistical
properties of the quantity $\alpha$ in Eq.~(\ref{2qd9}). For an
asymmetric dot with a single decay channel per lead,
the distribution function $P_{\beta}(\alpha)$ is obtained as
\begin{eqnarray}
P_{1}(\alpha) &=& \sqrt{\frac{2}{\pi\alpha}} \exp(-2\alpha)
                                               \nonumber\\
P_{2}(\alpha) &=& 4\alpha\Big(K_0(2\alpha)+K_1(2\alpha)\Big)
                                             \exp(-2\alpha)
\label{2qd10}
\end{eqnarray}
for GOE and GUE, respectively. Here, asymmetric means that possible
reflection symmetries have been removed by proper deformations of the
system geometries. If such symmetries are present, one has
$\Gamma^{(L)}=\Gamma^{(R)}$ and $\alpha = \Gamma/2\overline{\Gamma}$.
Then the distribution function coincides with the Porter--Thomas law.
In Eq.~(\ref{2qd10}), $K_n$ is the $n$--th modified Bessel function of
the second kind. Stone and Bruus~\cite{Sto94,Bru94} performed
extensive numerical tests of these results. They used the Africa
billiard~(\ref{2bi1a}) as a model for the quantum dot and found good
agreement with both of the predictions~(\ref{2qd10}).  Experimental
verifications were presented by Chang {\it et al.}~\cite{Chan96} and
Folk {\it et al.}~\cite{Folk96}. Stone and Bruus~\cite{Sto94,Bru94}
also computed the parametric velocity correlator~(\ref{1pc9}) and
showed its universality. This work was extended both analytically and
numerically by several other authors~\cite{Pri93,Muc95a,Alh95a}.
Prigodin {\it et al.}~\cite{Pri93} discuss to which extend their
results can be applied to the Coulomb blockade regime. Bruus {\it et
  al.}~\cite{Bruu96} and Alhassid and Attias~\cite{Alha96} studied
parametric correlations (see Sec.~\ref{qc1pcb}) of the conductance
peaks as a function of the magnetic field separation. They predict a
squared Lorentzian in the case of fully broken time--reversal
invariance. This was also experimentally verified in
Ref.~\onlinecite{Folk96}.

\subsection{Hydrogen atom in a strong magnetic field}
\label{qc2hh}

The hydrogen atom in a strong magnetic field has played a central role
in the conceptual development of quantum chaos. This is mainly due to
three properties of this system. (i) It is a ``real'' system for which
experimental data were available when theorists became interested in
quantum chaos. (ii) Effectively, it has two degrees of freedom and is,
therefore, calculable. (iii) It has an extremely useful scaling
property which simplified the calculations and helped in the
understanding of experimental results. Our brief discussion centers on
the random matrix aspects of this system. Detailed presentations,
particularly on the periodic orbit aspects, can be found in the
reviews by Friedrich and Wintgen~\cite{Fri89}, Hasegawa {\it et
  al.}~\cite{Has89}, and in Gutzwiller's book~\cite{Gut90}. A detailed
discussion of the present status of experimental work on the hydrogen
atom in magnetic and electric fields was recently given by
Neumann~\cite{Neu96}.

\begin{figure}
\centerline{
\psfig{file=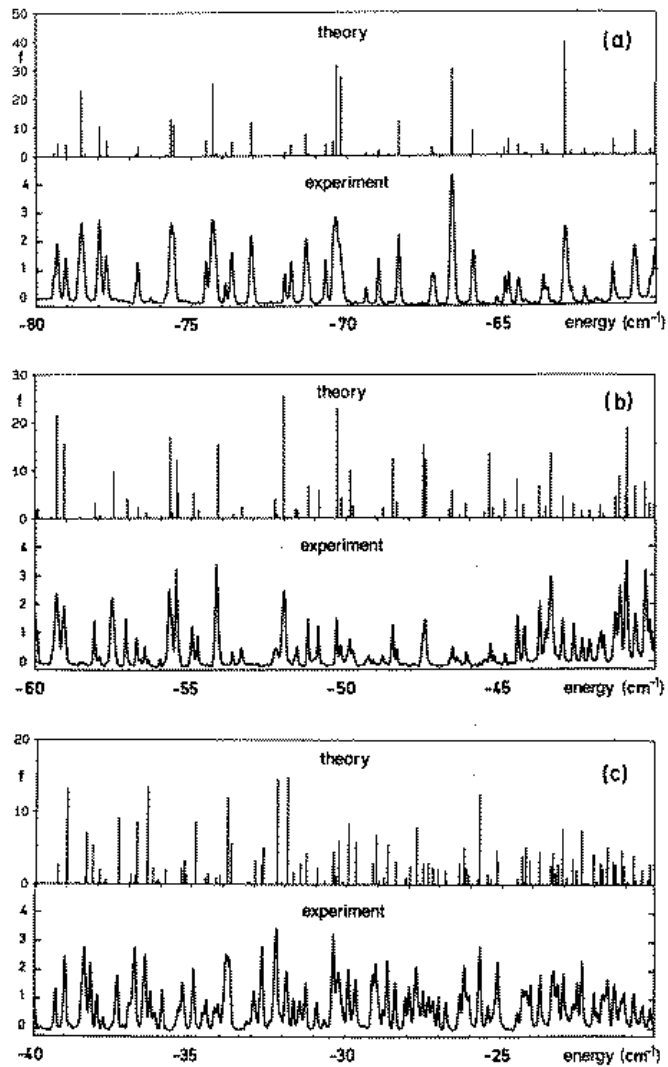,width=3.5in}
}
\caption{
  Comparison of theoretical and experimental results for the 
  deuterium Rydberg atom. The magnetic field strength is 5.96~T.
  In spectroscopic units, the energies of the Rydberg states 
  are between -80~cm$^{-1}$ and -20~cm$^{-1}$, where 1~cm$^{-1}$ 
  corresponds to 0.12~meV. At the end of the energy range, 
  i.e.~above -25~cm$^{-1}$, the corresponding classical system
  becomes completely chaotic. 
  Taken from Ref.~\protect\onlinecite{Holl87}. 
}
\label{figch9}
\end{figure}

The Hamiltonian of the hydrogen atom in a homogeneous constant
magnetic field $B$ in $z$ direction reads
\begin{equation}
H = \frac{\vec{p}^2}{2m} - \frac{e^2}{r} - \omega_L L_z
                   + \frac{1}{2}\omega_L^2(x^2+y^2)
\label{2hh1}
\end{equation}
with $r=\sqrt{x^2+y^2+z^2}$ and where $\omega_L=eB/2mc$ is the Larmor
frequency. The first two terms on the right hand side correspond to
the Hamiltonian of the hydrogen atom without magnetic field. The third
term represents the paramagnetic Zeeman contribution, the fourth one
is the diamagnetic term. The last two terms break rotation invariance
and destroy full integrability. With increasing magnetic field
strength, ever larger parts of classical phase space become chaotic.
We express the magnetic field in natural units defined by dividing the
Rydberg energy $me^4/2\hbar^2$ by the Larmor energy $\hbar\omega_L$.
Then, $B=\gamma\cdot 2.35\cdot 10^5{\rm T}$ where the magnetic field
strength parameter $\gamma$ is dimensionless. We use atomic units.
Because of axial symmetry with respect to the $z$ axis, the azimuthal
quantum number $M$ is a good quantum number, and so is parity $\pi$.
It is convenient to rotate the frame with the Larmor frequency
$\omega_L$. This removes the paramagnetic term. In cylindrical
coordinates, the momenta in $\rho=\sqrt{x^2+y^2}$ and $z$ direction
are denoted by $p_\rho$ and $p_z$, respectively. For fixed $M^\pi$
values, the Hamiltonian reads
\begin{equation}
H = \frac{1}{2}\left(p_\rho^2+p_z^2\right) + 
            \frac{M^2}{2\rho^2} - \frac{1}{\sqrt{\rho^2+z^2}} 
                                  +\frac{1}{8}\gamma^2\rho^2 \ .
\label{2hh2}
\end{equation}
Precise calculations which yield an excellent description of the
experimental data were performed by various groups. As an example, we
show results obtained by Holle {\it et al.}~\cite{Holl87} in
Fig.~\ref{figch9}.
The Hamiltonian~(\ref{2hh2}) has the remarkable scaling
property~\cite{Rob82}
\begin{equation}
H(\vec{p},\vec{r},\gamma) = \gamma^{2/3} 
         H(\gamma^{-1/3}\vec{p},\gamma^{2/3}\vec{r},1) \ .
\label{2hh3}
\end{equation}
Wintgen~\cite{Win87a} and Wintgen and Friedrich~\cite{Win87b,Win87c}
realized that the scaling property~(\ref{2hh3}) can be used for an
efficient computation of the eigenvalues. This step was crucial. It
was used later also by Holle {\it et al.}~\cite{Hol88} for a much
improved understanding of experimental data, see the
review~\cite{Fri89}. Wintgen and Friedrich~\cite{Win87c} calculated
the spectrum some meV below the ionization threshold at magnetic field
strengths of a few Tesla. All properties depend only on the scaled
energy $\varepsilon = \gamma^{-2/3}E$. This variable increases with
both excitation energy and magnetic field strength.  One has
$\varepsilon < 0$ for bound states and $\varepsilon > 0$ for
resonances.  Between $\varepsilon=-0.8$ and $\varepsilon=-0.2$, the
spacing distribution for bound states changes from Poisson to Wigner,
see Fig.~\ref{fig5}. In the same range, the fraction of classical
phase space filled by regular trajectories decreases drastically. The
coincidence of these facts strongly promoted the understanding of
quantum chaos. As a function of $\varepsilon$ and for $L$ values below
a critical value $L_{\rm max}$, the spectral rigidity follows the same
pattern but saturates for $L > L_{\rm max}$. This is in agreement with
Berry's~\cite{Ber85} general argument based on periodic orbit theory,
see Sec.~\ref{qc4po}. The hydrogen atom in a strong field has been of
great importance for the development of periodic orbit theory.
Although the relevance of classical periodic orbits for the quantum
spectrum was already known~\cite{Mai87}, Wintgen~\cite{Win88a} was the
first author to realize the full applicability of Gutzwiller's theory.
For reviews see Refs.~\onlinecite{Fri89,Has89,Gut90}.

Although a ``real'' system, the hydrogen atom in a strong magnetic
field is intimately connected to one of the toy models studied
theoretically. Upon the introduction of parabolic coordinates
$\mu_{\pm}$ with $\mu_\pm^2=r\pm z$, the Hamiltonian~(\ref{2hh1}) can
be written as the sum of three terms.  Each of the first two terms
represents a harmonic oscillator with angular momentum barrier, while
the third term provides a sixth order coupling in $\mu_\pm$ between
the two. This is actually the form used by Delande and
Gay~\cite{Dela81} and Wintgen and Friedrich~\cite{Fri89} to calculate
eigenstates.

Simons {\it et al.}~\cite{Sim93g} verified the existence of universal
parametric correlations, see Sec.~\ref{qc1pcb}, in the spectrum of the
hydrogen atom versus magnetic field strength, see also
Ref.~\onlinecite{Gol91}. These authors demonstrated the universality
of the parametric level number variance and of the velocity
correlator.

Zakrzewski {\it et al.}~\cite{Zak95} made the following interesting
observation. Very close to ionization threshold, the nearest neighbor
spacing distribution deviates from the Wigner surmise. This is caused
by an almost complete absence of spacings larger than about 1.5 mean
level spacings. The classical motion is chaotic. However, the
diamagnetic term in Eq.~(\ref{2hh2}) confines the motion only in the
$(x,y)$ plane. In the $z$ direction, the electron can move very far
away from the proton. At such large distances, the Hamiltonian is the
sum of two integrable parts. Zakrzewski {\it et al.}~showed that this
effect can be described by a regular Hamiltonian coupled to a chaotic
one modeled by a random matrix.

\subsection{Model systems}
\label{qc2ms}

Among a large variety of theoretical models, we subjectively select a
few which, in our judgement, had impact on RMT. Regarding other
systems, we refer the reader to the reviews by
Bohigas~\cite{Boh84,Boh91a}, Eckhardt~\cite{Eck88}, 
Gutzwiller~\cite{Gut90}, and Haake~\cite{Haa91}. Molecular chemists
introduced many such models systems which later became
interesting for quantum chaos. In Secs.~\ref{qc2msa} and~\ref{qc2msb},
we discuss coupled oscillators and the anisotropic Kepler problem,
respectively.

\subsubsection{Coupled oscillators}
\label{qc2msa}

In 1973, Percival~\cite{Per73} predicted that, in the semiclassical
limit, the energy spectrum of coupled oscillators consists of a
regular and an irregular part which reflect the classically regular
and irregular motion, respectively, see Sec.~\ref{qc4ea}. To test this
prediction, Pullen and Edmonds~\cite{Pul81} in 1981 investigated
numerically the classical and the quantum properties of two
one--dimensional harmonic oscillators coupled by a fourth--order
interaction,
\begin{eqnarray}
H &=& H_1 + H_2 + 4kx_1^2x_2^2   \nonumber\\
H_i &=& \frac{1}{2}(p_i^2+x_i^2), \qquad i=1,2
\label{2ms1}
\end{eqnarray}
where $k$ is the coupling parameter. This system has several
advantages over other model systems. For example, it is bound at all
energies. Pullen and Edmonds studied the levels $E_n(k)$ as functions
of the transition parameter $k$, and worked out the second
differences
\begin{equation}
\Delta_n^2 = |(E_n(k+\Delta k)-E_n(k))-(E_n(k)-E_n(k-\Delta k))| \ .
\label{2ms2}
\end{equation}
Large values of the $\Delta_n^2$'s are due to avoided crossings and
occur in the regime where the classical dynamics is chaotic. Today we
would say that $\Delta_n^2$ measures the curvature of the 
parametric level motion and
indicates chaotic dynamics, see Sec.~\ref{qc1pca}. Aside from their
interest for chaotic systems, pairs of coupled harmonic oscillators
became interesting also as models for a new class of integrable
systems with a second hidden integral of motion~\cite{Dor83,Gra83}.

A year later, Haller {\it et al.}~\cite{Hal84} studied the
system~(\ref{2ms1}) numerically. This system possesses a C$_{4v}$
point--group symmetry. It also has a scaling property, and the levels
can be computed efficiently. Haller {\it et al.}~worked out the
spacing distribution for a set of $k$ values and found a transition
from Poisson to Wigner behavior which was fitted with the
Berry--Robnik formula. The classical analogue of this system was known
to undergo a transition from predominantly regular to chaotic motion.
The work of Haller {\it et al.}~was one of the first studies to show
the close relationship between the transitions from regular to chaotic
motion in classical mechanics, and from Poisson to Wigner--Dyson
statistics in the quantum case. Meyer {\it et al.}~\cite{Mey84} added
further evidence in the frame of classical mechanics.

In a conceptually important contribution, Seligman {\it et
  al.}~\cite{Sel84} investigated a related system which has a more
\begin{figure}
\centerline{
\psfig{file=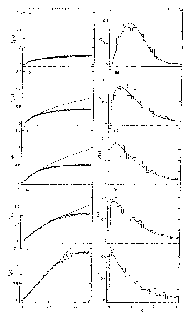,width=3.5in}
}
\caption{
  Numerically calculated spectral rigidities and nearest neighbor spacing 
  distributions for the two coupled oscillators defined in 
  Eq.~(\protect\ref{2ms3}). The parameters are $\alpha_1=0$, 
  $\beta_1=122.1$, $\gamma_1=0$, $\alpha_2=0$, $\beta_2=24.1$, 
  $\gamma_2=0$, $\alpha_{12}=0$, $\beta_{12}=-50\lambda$ and 
  $\gamma_{12}=0$. The transition is studied as a function of 
  the interaction strength $\lambda$. 
  Figures (a) to (e) correspond to interaction strengths
  $\lambda=0.10,0.04,0.02,0.01$ and $0$. We notice the saturation
  of the spectral rigidities. The lines were obtained from 
  a model of random band matrices. 
  Taken from Ref.~\protect\onlinecite{Sel84}.
}
\label{figch5}
\end{figure}
complex structure and shows richer features. They chose two
one--dimensional oscillators with fourth--order potentials coupled by
an interaction of the same form,
\begin{eqnarray}
H &=& \frac{1}{2}(p_1^2+p_2^2) + V_1(x_1) + V_2(x_2)
                               + V_{12}(|x_1-x_2|) \nonumber\\
V_i(x) &=& \alpha_i x^2 + \beta_i x^4 + \gamma_i x^6, \qquad i=1,2,12 \ .
\label{2ms3}
\end{eqnarray}
The extensive numerical simulations show that, controlled by the
parameters $\alpha_i$, $\beta_i$ and $\gamma_i$, the spacing
distribution exhibits a transition from Poisson to GOE statistics when
the classical system makes the related crossover from predominantly
regular to chaotic motion. Seligman {\it et al.}~also worked out the
spectral rigidity and found a saturation beyond some interval length
$L_{\rm max}$.  Their results are shown in Fig.~\ref{figch5} for 
a certain choice of parameters.
They also
investigated the behavior of the corresponding classical system.
Shortly thereafter, Berry~\cite{Ber85} expressed the spectral rigidity
in terms of periodic orbits and explained the saturation, see
Sec.~\ref{qc4po}. To explain the transition, Seligman {\it et
  al.}~also constructed a model of random band matrices. Very similar
models are nowadays used in localization theory, see Sec.~\ref{rbm}.

Zimmermann {\it et al.}~\cite{Zim86} calculated the spectral rigidity
for the system~(\ref{2ms1}) and compared the saturation effect
quantitatively to Berry's prediction. Berry argues that the maximal
interval length $L_{\rm max}$ for universal GOE behavior is given by
$L_{\rm max}=2\pi\hbar /DT_{\rm min}$ where $D$ is the mean level
spacing and $T_{\rm min}$ is the period of the shortest periodic
orbit, see Sec.~\ref{qc4po}. Zimmermann {\it et al.}~found that this
estimate is consistent with the tendency of the numerically obtained
$L_{\rm max}$ as function of the coupling parameter.

Bohigas {\it et al.}~\cite{Boh90a,Boh90b,Boh93a} used two coupled
quartic oscillators,
\begin{equation}
H = \frac{1}{2}(p_1^2+p_2^2) 
 +a(\lambda)\left(\frac{1}{b}x_1^4 +bx_2^4 +2\lambda x_1^2x_2^2\right)
\label{2ms5}
\end{equation}
with parameters $\lambda$, $b$ and $a(\lambda)$ as a model system to
study the effect of classical transport on quantum properties. In a
parameter range where the system is mixed, there exists a classical
flux in phase space through imperfect barriers which are due to
Cantori. Cantori are remnants of tori in the classical phase space of
a mixed system.  A semiclassical theory for the level number variance
was derived and applied to this system by Smilansky {\it et
  al.}~\cite{Smi92a}.  Because of its conceptual importance, we
discuss this system and the concepts emerging from this study in
Sec.~\ref{qc4mp}.

\subsubsection{Anisotropic Kepler problem}
\label{qc2msb}

Another model system introduced by condensed--matter physicists is the
anisotropic Kepler problem. The Hamiltonian is
\begin{equation}
H = \frac{p_\rho^2}{2m_\rho}+\frac{p_z^2}{2m_z} - 
             \frac{e^2}{\sqrt{\rho^2+z^2}} 
\label{2ms6}
\end{equation}
with $\rho=\sqrt{x^2+y^2}$.  The anisotropy is caused by the
difference between the masses $m_\rho$ and $m_z$. This model describes
donor--impurity levels in semiconductors.  In 1971,
Gutzwiller~\cite{Gut71,Gut73} had shown that the classical problem
exhibits hard chaos. Later this problem became very important in
periodic orbit theory. For most classical systems, the
Kolmogorov--Arnold--Moser theorem prevents the abrupt transition from
integrable to ergodic behavior: The structure of most invariant
surfaces in phase space changes smoothly under small perturbations.
However, this theorem does not apply to the pure Coulomb system.
Therefore, it was not clear how the system~(\ref{2ms6}) behaves if the
mass anisotropy $1-m_\rho/m_z$ differs slightly from zero.
Gutzwiller~\cite{Gut80} had found strong evidence for an abrupt
transition from regular to ergodic motion. His results and the
question whether the spectral fluctuations change abruptly, too, led
Wintgen and Marxer~\cite{Wint88} to a detailed numerical study of the
classical anisotropic Kepler problem and its quantum analogue. At a
mass anisotropy of $0.2$, classical phase space is densely filled with
remnants of tori, i.e.~with Cantori. Thus, the system is only weakly
chaotic.  At sufficiently high excitation energies, the nearest
neighbor spacing distribution agrees perfectly with the Wigner
surmise. But the spectral rigidity follows the GOE prediction only up
to an interval length of $L\simeq 7$ and then grows linearly with $L$
in a Poisson--like fashion. The speed of convergence to the GOE
prediction is different in different regions of the spectrum.  It is
conjectured that the intermediate scale deviations and the slow
convergence to the GOE predictions are connected to the pronounced
Cantori structure of classical phase space.  This example shows that,
in its gross features, even a somewhat special system such as the
anisotropic Kepler problem is consistent with the overall picture
developed throughout this review: RMT type correlations become visible
very quickly on short scales in the spectrum.  On larger scales,
however, there is stronger resistance to RMT correlations.

\subsection{Classical phase space and quantum mechanics}
\label{qc4mp}

We now leave the discussion of special systems and return to the
general aspects: The connection between quantum chaos and RMT.  How
does the structure of classical phase space manifest itself in quantum
mechanics? About ten years ago, Berry and Robnik~\cite{Ber84}
conjectured that the spectrum of a mixed system should be a
superposition of independent spectra, each associated with a chaotic
(or regular) region in phase space. This is related to
Percival's~\cite{Per73} picture, see Sec.~\ref{qc4ea}.  Under the
crucial assumption of independence of all these spectra, the nearest
neighbor spacing distribution and other fluctuation measures can be
predicted. In the simplest case, one obtains the Berry--Robnik formula
for the spacing distribution (Sec.~\ref{qc1ba}). In a more realistic
picture, boundaries partitioning the chaotic part of classical phase
space must be taken into account. Such boundaries may, for instance,
be due to Cantori. Classical trajectories do pass such boundaries.
Nevertheless, the existence of the boundaries defines a new time scale
for the system.  Bohigas, Tomsovic and Ullmo~\cite{Boh93a} extended
the Berry--Robnik surmise by accounting for this time scale in terms
of the classical flux through the boundary. The ensuing random matrix
model acquires block structure. Each region of phase space is
represented by a Gaussian or Poissonian ensemble, as the case may be.
The corresponding matrix blocks are located on the diagonal of the
total Hamiltonian matrix. The diagonal blocks are coupled through
blocks of random matrices. The strength (second moment) of the latter
is determined by the classical flux between the regions. The random
matrix model for symmetry breaking (Sec.~\ref{qc1tc}) is a special
case of this model. The mathematical concept is very similar to early
work by Rosenzweig and Porter on the spacing distribution in complex
atoms, see Sec.~\ref{qc3ama}. In the model of Bohigas, Tomsovic and
Ullmo, there exists, in addition to the coupling between different
regions of the chaotic part of phase space due to classical flux, also
a coupling between the regular and the chaotic blocks and among the
latter which is due to quantum--mechanical tunneling processes. 
This relates to the idea of ``chaos--assisted tunneling''
(see Sec.~\ref{qc2bi}) where separated regular regions in phase space
are coupled via a chaotic region.

One can view the Bohigas--Tomsovic--Ullmo model as an extension of the
original Bohigas conjecture: Since Cantori lead to more than one
intrinsic time scale the ensuing random matrix model acquires a more
complicated structure.

\subsection{Towards a proof of the Bohigas conjecture}
\label{qc4psi}

The most remarkable feature of RMT is the prediction of universal
statistical behavior of quantum systems. After the formulation of the
Bohigas conjecture, it became a challenge to find analytical arguments
linking classical chaos with RMT. Three approaches towards a formal
proof of the Bohigas conjecture are presented here: Periodic orbit
theory, Berry's argument, and subsequent theoretical work
(Sec.~\ref{qc4po}), a field--theoretic approach using the
supersymmetry method (Sec.~\ref{qc4ss}), and a group--theoretical and
probabilistic argument based on structural invariance
(Sec.~\ref{qc4si}). All three approaches involve, in one way or the
other, a semiclassical approximation to quantum mechanics which links
quantum and classical behavior, i.e.~formally, the limit $\hbar\to 0$
is taken.

\subsubsection{Periodic orbit theory}
\label{qc4po}

A natural framework for limking RMT and classical chaos is provided by
periodic orbit theory.  Developed by
Gutzwiller~\cite{Gut67,Gut69,Gut70,Gut71,Gut78} and by Balian and
Bloch~\cite{Bal72,Bal74}, periodic orbit theory is based on the
saddle--point approximation to Feynman's path integral for $\hbar \to
0$ and is one of the means to implement the semiclassical
approximation. In an important paper, Berry~\cite{Ber85} used periodic
orbit theory to investigate universal aspects of the $\Delta_3$
statistics (Sec.~\ref{qc1bac}). We recall that $\Delta_3(L)$ is
defined in terms of the staircase function $\widehat{\eta}(\xi)$. This
function is the unfolded integral over the level density $\rho(E)$,
and $\Delta_3(L)$ is obtained by minimizing the mean square deviation
of $\widehat{\eta}(\xi)$ from a straight line over an energy interval
of length $L$, see Eq.~(\ref{1ba16}). Berry writes $\rho(E)$ as the
sum of a smooth and a fluctuating part,
\begin{equation}
\rho(E) = \langle\rho(E)\rangle + \rho_{\rm fl}(E) \ .
\label{4po1}
\end{equation}
The smooth part $\langle\rho(E)\rangle$ is identical to the mean level
density $R_1(E)$, see Sec.~\ref{qc1baa}. Periodic orbit theory can be
used to write the fluctuating part as the sum over classical
periodic orbits (``trace formula''),
\begin{equation}
\rho_{\rm fl}(E) = \frac{1}{\hbar^{\mu+1}}\sum_{j}A_j(E)
                        \exp\left(\frac{i}{\hbar}S_j(E)\right)
\label{4po2}
\end{equation}
where $j$ labels all distinct orbits, including multiple traversals.
The phase factor in Eq.~(\ref{4po2}) is determined by the classical
action $S_j(E)$ of the orbits. The amplitude $A_j(E)$ is related to
the monodromy matrix of the orbit. In an integrable system, periodic
orbits form $d-1$ parameter families filling $d$--dimensional
phase--space tori. A chaotic system is defined as an ergodic system in
which all periodic orbits are isolated and unstable. It can be shown
that in Eq.~(\ref{4po2}), this implies $\mu=(d-1)/2$ for integrable
and $\mu=0$ for chaotic systems. The length of the averaging interval
needed for a separation of $\langle\rho(E)\rangle$ and of $\rho_{\rm
  fl}(E)$ must obey two conditions. (i) It must be small on the
classical scale, i.e.~small compared to the energy $E$ of the system.
(ii) It must be large compared to two intrinsic energy scales of the
system, the mean level spacing $D=1/R_1(E)$ and the energy $\hbar /
T_{\rm min}$ given by the period $T_{\rm min}$ of the shortest
periodic orbit. The associated energy $DL_{\rm max}=h/T_{\rm min}$
sets an important scale.  Since $D$ is of order $\hbar^d$ and $T_{\rm
  min}$ is of order $\hbar$, we have $D \ll \hbar / T_{\rm min}$ in
the semiclassical regime. The largest fluctuations in Eq.~(\ref{4po2})
stem from $\hbar / T_{\rm min}$. The energy range $DL$ is classically
small, and we can write $S_j(E + \varepsilon)\simeq S_j(E)+\varepsilon
T_j(E)$ where $T_j(E) = dS_j(E)/dE$ is the period of the orbit labeled
$j$. Ignoring the $\varepsilon$ dependence of the amplitudes $A_j$ and
of the mean level spacing $D$, the energy average in the
definition~(\ref{1ba16}) of $\Delta_3(L)$ can be performed trivially,
resulting in a double sum over periodic orbits. Inspection of the
energy scales allowed Berry to simplify the result further and to
write it as an integral over all periods,
\begin{equation}
\Delta_3(L) = \frac{2}{\hbar^{2\mu}} \int_0^\infty \frac{dT}{T^2}
                      \varphi(T) G\left(\frac{DT}{2\hbar}L\right)
\label{4po3}
\end{equation}
where the function 
\begin{equation}
\varphi(T) = \left\langle \sum_i\sum_{j,+} A_iA_j
                 \cos\left(\frac{S_i-S_j}{\hbar}\right)
                 \delta\left(T-\frac{1}{2}(T_i+T_j)\right)
                                                \right\rangle
\label{4po4}
\end{equation}
is the averaged double sum over the periodic orbit contributions. The
plus sign in the sum over $j$ denotes the restriction to positive
traversals, $T_j>0$. The function
\begin{equation}
G(y) = 1 - \left(\frac{\sin y}{y}\right)^2
         - 3\left(\frac{d}{dy}\frac{\sin y}{y}\right)^2
\label{4po5}
\end{equation}
does not depend on individual orbits $j$. The shape of $G(y)$ is very
similar to (but $G(y)$ is different from) the two--level correlation
function $X_2(r)$ of the GUE. Because of its shape, $G(y)$ selects
from $\varphi(T)$ only those pairs of orbits whose average period
exceeds $2\hbar/DL$. According to Berry, this feature reflects the
fact that $\Delta_3(L)$ measures the deviations from the linear
behavior of the staircase. Closer inspection of the steps leading to
Eq.~(\ref{4po4}) shows that the linear behavior is determined by
orbits with $T_j<\hbar/DL$ while the deviations are due to orbits with
$T_j>\hbar/DL$. Berry then argues that the two--point function
$\varphi(T)$ can be written in the form
\begin{equation}
\varphi(T) = \frac{\hbar^{2\mu+1}}{2\pi D} 
             \left(1-b_2^{\rm sc}\left(\frac{DT}{h}\right)\right) \ .
\label{4po6}
\end{equation}
Here $b_2^{\rm sc}(t)$ is the semiclassical approximation to the
two--level form factor defined in Eq.~(\ref{1rc38}) of
Sec.~\ref{qc1rce}, i.e.~the Fourier transform of the unfolded
two--level cluster function $Y_2(r)$. Collecting everything and
introducing the dimensionless time $t=DT/h$ as new integration
variable, one arrives at the semiclassical result
\begin{equation}
\Delta_3(L) = \frac{1}{2\pi^2} \int_0^\infty \frac{dt}{t^2}
                      (1-b_2^{\rm sc}(t)) G(\pi Lt) \ .
\label{4po7}
\end{equation}
According to Eq.~(\ref{4po7}), the spectral rigidity can be obtained
from a semiclassical approximation to the two--level form factor 
$b_2^{\rm sc}(t)$.

It remains to show that the semiclassical approximation to $b_2^{\rm
  sc}(t)$ leads to an expression for $\Delta_3(L)$ which is consistent
with RMT.  To this end, Berry uses sum rules. The first sum rule for
the diagonal part of the sum in Eq.~(\ref{4po4}),
\begin{equation}
\varphi_{\rm dia}(T) = \left\langle\sum_{j,+} A_j^2 
                            \delta(T-T_j)\right\rangle
\label{4po8}
\end{equation}
is due to Hannay and Ozorio de Almeida~\cite{Han84}. The sum rule
applies for large values of $T$. Periodic orbit theory shows that, in
this limit, the density of periodic orbits increases, while the
intensities $A_j^2$ decrease algebraically for integrable and
exponentially for chaotic systems. Hence the limiting relations
\begin{equation}
\varphi_{\rm dia}(T) \longrightarrow \frac{1}{2\pi D} 
                \left\{
         \begin{array}{cc}
         \hbar^d          &  \qquad {\rm integrable} \\
         \hbar t          &  \qquad {\rm chaotic}
         \end{array}
                \right.
\label{4po9}
\end{equation}
follow for large $T$ where $t=DT/h$ is the dimensionless time of 
the periods. The second sum rule due to Berry~\cite{Ber85} says
\begin{equation}
\varphi(T) \longrightarrow \frac{\hbar^{2\mu+1}}{2\pi D}
\label{4po10}
\end{equation}
in the regime $DT\gg\hbar$ or $t\gg 1$. This sum rule guarantees that
the amplitudes and phases of very long orbits in the trace formula
correctly generate the mean level density. Comparison with
Eq.~(\ref{4po6}) shows that the second sum rule implies $b_2^{\rm
  sc}(t)\to 0$ for $t\gg 1$. This is equivalent to the vanishing of
the two--level correlations on scales much larger than the mean level
spacing.

One further ingredient from periodic orbit theory is still needed. It
can be shown that for $T\gg T_{\rm min}$ and because of destructive
interference, the diagonal sum~(\ref{4po8}) is a good approximation to
the averaged double sum (\ref{4po4}). We recall that $T_{\rm min}$ is
the period of the shortest classical periodic orbit. In the regime
$t_{\rm min} \ll t \ll 1$ where $t_{\rm min}=DT_{\rm min}/h$, the
limit relations~(\ref{4po9}) of the diagonal sum can be used.  In the
case of integrable systems, the limit relation~(\ref{4po9}) coincides
with the sum rule~(\ref{4po10}) implying that the diagonal
approximation is also valid in the regime $t\gg 1$. Thus, $b_2^{\rm
  sc}(t)=0$ is a good approximation for the two--level form factor
down to values of $t_{\rm min}$ and Eq.~(\ref{4po7}) yields
\begin{equation}
\Delta_3(L) = \frac{1}{2\pi^2} \int_0^\infty \frac{dt}{t^2}
                       G(\pi Lt)
            = \frac{L}{15} \ ,
\label{4po11}
\end{equation}
the random matrix result for Poisson regularity. For fully chaotic
systems, the diagonal approximation must break down in the regime
$t\gg 1$, because the sum rule~(\ref{4po10}) is in conflict with the
limit relation~(\ref{4po9}). The off-diagonal terms enforce the
physically required cutoff of the limit relation~(\ref{4po9}),
expressed by $b_2^{\rm sc}(t)=0$ for $t \gg 1$. In order to work out
the integral~(\ref{4po7}), Berry chooses a linear interpolation
$1-b_2^{\rm sc}(t) \propto t$ for $t_{\rm min} < t < 1$ as suggested
by the limit relation~(\ref{4po9}), implying
\begin{equation}
\Delta_3(L) = \frac{1}{\beta\pi^2}\ln L + c_\beta
\label{4po12}
\end{equation}
in the regime $1\ll L \ll L_{\rm max}=1/t_{\rm min}$. As usual we have
$\beta=1$ and $\beta=2$ for time--reversal invariant and
non--invariant systems, respectively. This logarithmic behavior is in
agreement with the random matrix prediction~(\ref{1ba24}) of
Sec.~\ref{qc1bac}. Berry also calculates the numerical values of the
constants $c_\beta$. For $\beta=2$, he finds exact agreement with the
GUE prediction. This is so because the linear interpolation coincides
with the GUE result~(\ref{1rc40}) for the two--level form factor. For
$\beta=1$, there is no numerical agreement with the GOE prediction.
This is because the GOE result~(\ref{1rc40}) differs form the linear
interpolation. Periodic orbit theory predicts universal behavior of
the spectral rigidity only in the regime $L\ll L_{\rm max}$. Beyond
this scale, the spectral rigidity saturates at a value
$\Delta_3(\infty)$. Berry~\cite{Ber84} works out $\Delta_3(\infty)$
and discusses examples.

It is very difficult to go beyond Berry's semiclassical approximation
to the two--level form factor, i.e.~beyond the diagonal approximation
to the periodic orbit double sum $\varphi(T)$. In some special,
non--generic cases~\cite{Kea91,Luo94,Bog96a}, and for the Riemann
$\zeta$ function~\cite{Kea95,Bog95}, such a step has been possible.
Recently, progress has been made by Bogomolny and
Keating~\cite{Bog96b} in the general case. These authors use a
relationship discovered by Andreev and Altshuler~\cite{And95} to
connect the non--diagonal contribution in the double sum~(\ref{4po4})
to the diagonal one. The key assumption is that in generic systems and
modulo exact degeneracies, the orbits up to a cut--off period of the
order of $h / D$ can be treated as statistically independent. Bogomolny
and Keating point out that their results are closely related to the
ones found by using the supersymmetric non--linear $\sigma$ model
treated in the following section.

\subsubsection{Supersymmetric field theory}
\label{qc4ss}

We discuss two approaches \cite{Muz95a,And96} based on semiclassical
field theory. The central aim in both approaches is the same: To
derive, within a semiclassical framework, a supersymmetric generating
functional for conservative systems. For systems with fully chaotic
classical phase space and in the long wave--length limit, this
functional should yield RMT statistics for the eigenvalues. In
addition, the derivation should display the limits of validity of RMT
level statistics. Although based on a semiclassical approximation,
both approaches avoid the use of periodic orbits or similar entities.
The two approaches are technically demanding. Moreover, they touch
upon difficult issues in ergodic theory and are still under
discussion. Therefore, we restrict ourselves here to a discussion of
the central ideas.

The approach by Muzykantskii and Khmelnitzkii \cite{Muz95a}, related
to work by the same authors \cite{Muz95} discussed in
Sec.~\ref{akl_3}, starts from a disordered system and considers the
limit of vanishing disorder.  While it retains the usual semiclassical
condition $k_F l \gg 1$ of Efetov's supersymmetry approach (where
$k_F$ is the Fermi wave number and $l$ the elastic mean free path), it
does not use the long wave--length limit in its usual form, $q l \gg
1$. The wave number $q$ is typically given by the inverse of the
linear dimension $L_s$ of the sample, and the paper focuses on the
limit $l \gg L_s$. (In Sec.~\ref{disorder}, the symbol $L$ is used for
the sample size. Here we use $L_s$ to distinguish it from the symbol
$L$ for the metric used in the supersymmetry method.  Apart from this,
we use the same notation as in Sec.~\ref{disorder}.)

The authors consider a supersymmetric generating functional (averaged
over disorder) of the form of Eq.~(\ref{super8}). For a disordered
system, the Lagrangean takes the form
\begin{equation}
  {\cal L}(\sigma) = \frac{\pi \nu}{8 \tau} \int d^d r \ {\rm trg}
  (\sigma^2) - \frac{1}{2} \int d^d r \ {\rm trg} \ln(-i K)
\label{field1}
\end{equation}
with $K = E - T - \lambda \sigma + \varepsilon L/2$, where $T$ is the
operator of the kinetic energy and $L$ is the metric defined in
Sec.~\ref{susya}. The source term is omitted. The authors circumvent
the usual saddle--point approximation. Instead, they construct the
Green function $G(\vec{r}, \vec{r}' | \sigma)$ of the operator $K$, a
matrix--valued function in superspace. Let ${\tilde
  G}(\vec{r},\vec{p})$ denote the Wigner transform of $G$. Integrating
${\tilde G}(\vec{r},\vec{p})$ over the modulus $|\vec{p}|$ of the
momentum $\vec{p} = \vec{n}|\vec{p}|$ defines a function
$g_{\vec{n}}(\vec{r})$ for which a dynamical equation is derived. With
$v_F$ the Fermi velocity, this equation reads
\begin{equation}
2 v_F \vec{n} \frac{\partial g_{\vec{n}}(\vec{r})}{\partial\vec{r}}
= i \left[\frac{\varepsilon}{2}L  - \lambda \sigma, g_{\vec{n}} \right].
\label{field2}
\end{equation}
A functional $\Phi$ of $g_{\vec{n}}(\vec{r})$ is introduced with the
property that the stationary points of $\Phi$ (with respect to a
variation of $g_{\vec{n}}(\vec{r})$) are the solutions of
Eq.~(\ref{field2}). It is shown that in the limit $k_F l \gg 1$, the
action of Eq.~(\ref{field1}) can be rewritten in terms of the
functional $\Phi$. This technically somewhat complex procedure allows
taking the limit $k_F l \gg 1$ without implying $L_s \gg l$. Further
confidence in the derivation stems from the fact that in the long
wave--length limit, the resulting action ${\cal L}(\Phi)$ reduces to
Efetov's form, Eq.~(\ref{super12}). Moreover, it is found that ${\cal
  L}(\Phi)$ remains well--defined in the limit $l \rightarrow \infty$.
This suggests that in this limit, ${\cal L}(\Phi)$ yields the correct
level statistics for systems without disorder. It is shown that for
fully chaotic systems, limitations of the range of validity of
Wigner--Dyson statistics are determined by properties of the classical
Liouville operator.

The approach by Andreev {\it et al.}~\cite{And96,Simo97} constructs a
field theory where the effective action is associated with flow in
classical phase space.  Starting point is a supersymmetric generating
functional containing the Hamiltonian operator $H$ for a system
without disorder. A Gaussian average over energy centered at $E$ for
the single system described by $H$ replaces the ensemble average
typical for disordered systems. It leads naturally to the
Hubbard--Stratonovich transformation, and to the introduction of the
$\sigma$ matrix, see Sec.~\ref{susy}. The resulting 
Lagrangean is given by (we again omit the source terms)
\begin{equation}
  {\cal L}(\sigma) = \frac{1}{2} {\rm trg}_{\vec{q}} \sigma^2 + {\rm
    trg}_{\vec{q}} \ln K
\label{field3}
\end{equation}
where $K = E - H -N \sigma  + \varepsilon L/2$. Here, $\sigma$
depends on two sets of coordinates, ${\rm trg}_{\vec{q}}$ indicates a
graded trace as well as an integration over both sets, while $N$
denotes the number of levels (eigenvalues of $H$) in the averaging
interval. We note the formal similarity of Eq.~(\ref{field1}) and of
Eq.~(\ref{field3}). A saddle--point approximation based on $N \gg 1$
leads to a saddle--point manifold and the occurrence of Goldstone
modes. The resulting non--linear $\sigma$ model contains the classical
Liouville operator. It is equivalent \cite{And96} to the model obtained
in the zero disorder limit by Muzykantskii and Khmelnitzkii, although
the two theories differ in form. 

If $H$ is the Hamiltonian of a classically chaotic system, the
non--linear $\sigma$ model requires regularization. This is achieved
by adding a small noise to the classical Liouville operator. This
procedure renders the classical motion irreversible. In the limit of
vanishing strength of the noise, the spectrum of the resulting
time--evolution operator (known as the Perron--Frobenius operator)
reflects intrinsic irreversiblity properties of the underlying
classical dynamics. The spectral properties of this operator determine
the statistics of the eigenvalues of $H$. If the Perron--Frobenius
operator describes an exponential relaxation towards equilibrium, then
its spectrum has a gap between the lowest and the next eigenvalue, and
the spectrum of $H$ displays Wigner--Dyson statistics in an energy
interval determined by the size of the gap. Corrections to RMT can be
obtained using the spectral properties of the Perron--Frobenius
operator.

\subsubsection{Structural invariance}
\label{qc4si}

In order to furnish the Bohigas conjecture with a formal
justification, Leyvraz and Seligman~\cite{Ley92,Ley96a} take an
alternative approach reminiscent of the use of probability theory in
statistical mechanics. The three key steps of the argument are: (i) A
given object is identified as a typical representative of some
ensemble. (ii) This ensemble is known to have a particular property
$p$ with probability one. (iii) As a member of the ensemble, the given
object possesses property $p$ with probability one. Leyvraz and
Seligman use this general probabilistic reasoning to establish a
formal link between chaos and RMT. In this context, the three steps
read: (i) A given classical system is identified as a typical member
of a set $\Sigma$ of classical systems which can be embedded in an
ensemble $E$. (ii) Upon quantization, the members of $E$ are known to
possess random matrix fluctuations with probability one. (iii)
Therefore, the given system also possesses random matrix fluctuations
with probability one. We note that the term ensemble here has a
meaning which differs from its use in RMT.

The set $\Sigma$ of classical systems is defined in terms of a number
of characteristic properties common to all members of $\Sigma$. Such
properties may be dynamical symmetries, behavior under time reversal,
etc. The key concept in the construction of the ensemble $E$ is
structural invariance. It is based on a group--theoretical notion. Let
$G$ be the group of transformations which map a system in $\Sigma$
onto another such system. If $G$ possesses an invariant measure, then
$G$ induces an invariant measure on $\Sigma$, and this defines the
ensemble $E$. The properties defining $\Sigma$ must be ergodic,
i.e.~hold with probability one for any element of $E$.

Examples show that, in general, $G$ has infinite dimension and,
therefore, does not possess an invariant measure. To overcome this
difficulty, Leyvraz and Seligman consider systems with compact phase
space $\Gamma$ and with phase--space volume $|\Gamma|$. The associated
Hilbert space has finite dimension $N = |\Gamma| / h^d$ where $d$ is
the number of degrees of freedom of the system. For systems without
any discrete symmetries, invariant tori etc., i.e.~for the fully
chaotic systems, the elements of $G$ are given by the set $\Xi$ of
all canonical maps of $\Gamma$ onto itself. According to an
observation by Dirac~\cite{Dir47}, a unitary transformation can be
assigned to every element in $\Xi$. Thus, $\Xi$ is mapped onto the
unitary group $U(N)$ in $N$ dimensions. This group $U(N)$ possesses an
invariant measure $d \mu (U(N))$ which is precisely the probability
density of the Circular Unitary Ensemble (CUE). The third step is
trivial and amounts to saying that quantization of any map $C$ in
$\Xi$ yields the random matrix fluctuations of the CUE for the
eigenphases of the quantum system. Leyvraz and Seligman also show that
the inclusion of time--reversal invariance as a relevant property
leads to the Circular Orthogonal Ensemble (COE). Moreover, they apply
the same line of reasoning to time--independent Hamiltonians and
obtain random matrix fluctuations for the eigenvalues of the quantum
system. A critical discussion of the difficulties related to the
construction of the ensemble $E$, the quantization procedure, and the
necessity to consider a finite dimensional Hilbert space, is given in
Ref.~\onlinecite{Ley96a}.

By constructing a surprising example, Leyvraz, Schmit and
Seligman~\cite{Ley96b} show that in this approach, symmetries are
easily dealt with. They consider a billiard in the shape of an
equilateral triangle. It has three--fold rotation symmetry and three
mirror symmetries with respect to the three symmetry axes. The mirror
symmetries can be removed by a proper non--symmetric rounding of the
edges. The resulting billiard still possesses the three--fold rotation
symmetry. The system is time--reversal invariant. This suggests that
the spectral fluctuations are described by the GOE. However,
surprisingly, the GUE applies. Leyvraz, Schmit and
Seligman~\cite{Ley96b} show this with the help of structural
invariance.  A semiclassical explanation was given by Keating and
Robbins~\cite{Keat97}: The characters of the corresponding symmetry
group are complex which leads to phase contributions in the traces of
the Green function in the irreducible representation. Formally similar
to a breaking of time--reversal invariance due to an Aharanov--Bohm
flux, this yields GUE statistics.

\subsection{Summary: quantum chaos}
\label{qc4su}

There exists overwhelming evidence that for fully chaotic systems
governed by a single time scale, the Bohigas conjecture applies. A
body of analytical arguments lends theoretical plausibility to this
statement. However, a fully satisfactory proof of the conjecture is
still lacking. For strongly chaotic systems with more than one
intrinsic time scale (example: a chain of pairwise weakly coupled
billiards, each fully chaotic), RMT in its pure form cannot apply.
Such systems are kin to quasi one--dimensional disordered mesoscopic
systems and possess the same statistical properties. A detailed
discussion of such systems and their spectral statistics is given in
Sec.~\ref{disorder}. The situation is more difficult and less clear
for the generic classical systems with mixed phase space. Here, the
hypothesis of Sec.~\ref{qc4mp} is likely to apply. However, a
derivation of this hypothesis is lacking, and so is a
non--phenomenological determination of the coupling between blocks.

\setcounter{equation}{0}
\section{Disordered mesoscopic systems}
\label{disorder}

\begin{figure}
\centerline{
\psfig{file=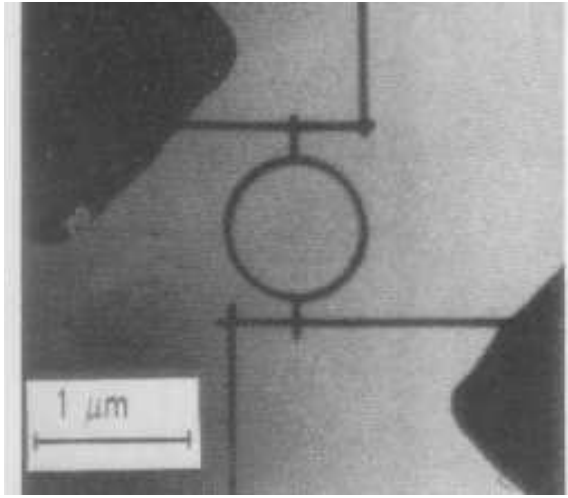,width=3.0in}
 }
\vskip 0.5cm
\caption{Picture of a mesoscopic gold loop taken with a scanning
  transmission electron microscope (STEM).
  Taken from Ref.~\protect\onlinecite{Wash91}.
} 
\label{afig1}
\end{figure}

The investigation of disordered systems in one, two or three dimensions 
is almost exclusively a domain of condensed matter physics. The motion of 
electrons in a crystal with random impurities provides a key example. 
Particularly important for our purposes is the {\it mesoscopic regime} to 
which we will restrict ourselves in the following. In this regime, the 
phase coherence length $L_\phi$ is the largest length scale of the problem. 
Phase coherence is lost by inelastic scattering (in the case of electrons, 
through scattering by phonons or by other electrons), while elastic 
scattering does {\it not} destroy phase coherence. Thus, $L_\phi$ is 
essentially given by the {\it inelastic} mean free path. For electrons, 
$L_\phi$ grows strongly with decreasing temperature. For typical sample
sizes $L$ in the $\mu$m regime (see Fig.~\ref{afig1}), 
the condition $L_{\Phi} > L$ can be met at 
temperatures below $100$ mK or so. Here, quantum coherence is a dominant 
feature of experiment and theory. Closely related coherence phenomena
occur when classical waves (light or sound) pass through a disordered medium. 
The theoretical treatments of phenomena in these different fields of physics 
bear a close analogy. Unless otherwise stated, we confine ourselves to the 
case of electrons.

Stochastic features enter into mesoscopic physics in two ways. The
first source of randomness is provided by real spatial disorder.
Schr\"odinger waves propagate through disordered media diffusively
and/or show localization.  Disorder is theoretically described in
terms of a random {\it disorder potential} (see Eq.~(\ref{imp})) with
a suitably chosen correlation length and probability distribution.
Such stochastic modeling leads directly to a theoretical description
in terms of an ensemble of Hamiltonians and constitutes a natural
extension of classical RMT. The essential difference to the latter is
due to the central roles played by dimension $d$, and by the spatial
extension of the sample. The second, more implicit source of
randomness arises from elastic wave reflection at suitably shaped
boundaries which in the classical limit would lead to chaotic motion.
For electrons, this situation is realized in ``ballistic systems''
where the elastic mean free path $\ell$ is larger than the sample size
$L$. This source of randomness is the same as described in the
preceding section. Again, dimension $d$ and spatial extension of the
sample play a key role.

In Sec.~\ref{qc1bag} we have stressed the distinction between two
averaging symbols,  the angular brackets $\langle\ldots\rangle$ for
spectral averages and the bar $\overline{(\ldots)}$ for ensemble
averages. The equality of these two averages can be proven in certain
cases, see Sec.~\ref{qc1bag}. In many other cases, however, the
equality of an experimental average (obtained by averaging over the
spectrum, an external magnetic field, or the Fermi energy)
and a theoretical ensemble average remains a hypothesis,
cf. Sec.~\ref{quasi1d_1}. In this section on 
disordered systems we mainly use the symbol $\langle\ldots\rangle$ for
average quantities, in spite of the fact that these quantities have
typically been calculated by averaging over the random potential.
In the few cases where we deal with a purely theoretical quantity
which cannot be observed in an experiment (like, e.g., an average
generating functional) we employ the bar.

For a systematic approach to the properties of spatially extended
mesoscopic systems we have to consider various relevant length and
energy (or time) scales. We assume that the Fermi energy $E_F$ is
always the largest energy scale. The length scales $\ell$ and $L$ were
defined above. A third important length scale is provided by the
localization length $\xi$. Except for strictly one--dimensional
samples, we always have $\ell < \xi$, and we assume in the following
that $\ell \ll \xi$.  Related to the two length scales $\ell$ and $L$
are two time (and corresponding energy) scales, the elastic scattering
time $\tau = \ell / v_F$ and the time--of--flight $t_f = L / v_F$
through the sample, with $v_F$ the Fermi velocity. Further relevant
energy (and related time) scales are the single--particle level
spacing $\Delta$ at the Fermi surface and the Thouless energy $E_C =
\hbar {\cal D}/L^2 = \hbar t_d^{-1}$, where ${\cal D}$ is the
diffusion constant and $t_d$ is the classical diffusion time through
the sample. It was already remarked in Sec.~\ref{loc} that $E_C$ plays
a central role in mesoscopic physics.

Four different regimes are distinguished. 

(i) The {\it localized} regime with $L \gg \xi$. The system size
exceeds the localization length, and the amplitude for propagation of
an electron through the probe is exponentially small.

(ii) The {\it diffusive} regime with $\ell\ll L\ll \xi$. The system size 
is intermediate between the elastic mean free path and the localization 
length. Electrons are multiply scattered by random impurities and propagate 
diffusively. (Almost) all states of the system are extended and Ohm's
law applies.  
In this regime, we have a further inequality. The dimensionless conductance 
$g$ is given by $g = E_C / \Delta \gg 1$, see
Eq.~(\ref{conductance}). Hence, we  
also have $\Delta \ll E_C$ and, moreover, $E_C \ll \hbar \tau^{-1}$ (this 
defines the diffusive regime). Thus, the diffusion time through the sample 
is too short to resolve individual levels but (trivially) long enough for 
the electron to be multiply scattered. The particular significance of the 
inequality $\Delta \ll E_C$ is discussed further below.

(iii) The condition $L \ll \ell$ or, equivalently $\tau^{-1} \ll t_f^{-1}$ 
characterizes both the {\it ballistic} and the {\it nearly clean} regimes. 
The system size is smaller than the elastic mean free path. The Thouless 
energy is replaced by the inverse time of flight $t_f^{-1}$ through the 
sample. The two regimes are distinguished by the degree of disorder.

(iii a) The {\it ballistic regime} where $\Delta \ll \hbar \tau^{-1} \ll 
\hbar t_f^{-1}$. The disorder, characterized by $\tau^{-1}$, is strong enough 
to thoroughly mix many energy levels of the system. 

(iii b) The {\it nearly clean} regime where $\hbar \tau^{-1} \ll \Delta, \hbar 
t_f^{-1}$. The disorder is so weak that it can be taken into account by 
low--order perturbation theory. 

All subsequent discussions of mesoscopic systems make use of this 
classification scheme. It should be evident already that the existence 
of four mesoscopic regimes vastly extends the range of problems dealt with 
in classical RMT. For instance, spectral fluctuation properties may 
differ in the four regimes. And the existence in $d > 2$ of a mobility 
edge separating localized and extended states requires special attention.

In the following we discuss equilibrium (Secs.~\ref{spec} and \ref{pc})
and transport (Secs.~\ref{quasi1d},~\ref{rbm},~\ref{akl}, and~\ref{shp})
properties of
mesoscopic systems, respectively. Sec.~\ref{spec} deals 
with the statistics of energy eigenvalues. Sec.~\ref{pc} is
devoted to persistent currents. Sec.~\ref{quasi1d} discusses the important
case of quasi one--dimensional wires. This leads naturally to the
investigation of random band matrices in Sec.~\ref{rbm}. In Sec.~\ref{akl}
we turn to systems in two and higher dimensions with special emphasis
on the distribution function of conductance fluctuations in mesoscopic
systems. Sec.~\ref{shp} deals with the rather recent topic of two
interacting electrons in a random impurity potential. Some concluding
remarks are contained in Sec.~\ref{sum}.
We emphasize those facts, phenomena, concepts, and 
theoretical developments which are closely related to Random Matrix Theory 
in its widest sense. We pay special attention to the role of dimension, 
and of the spatial extension of the systems under study. We do so at the 
expense of a systematic introduction into and account of mesoscopic physics.

\subsection{Energy eigenvalue statistics}
\label{spec}

This and the following
section are devoted to the {\it spectral} properties of isolated
mesoscopic systems. Why has this topic found such wide interest? After all,
{\it transport} properties have so far been the main source of experimental 
information on mesoscopic systems. And the study of the electrical
polarizability of an ensemble of small metallic particles by Gorkov and 
Eliashberg \cite{Gor65} in terms of spectral correlations was a rather 
isolated occurrence.
 
The very strong interest in spectral fluctuation properties results from 
the intimate connection between transport properties and spectral properties
of mesoscopic  systems. In a seminal paper, this connection was pointed out 
already in 1977 by Thouless \cite{Tho77}, see Sec.~\ref{loc}. His
insight lies at  
the heart of the modern scaling theory of localization and helped to understand
and interpret numerous mesoscopic fluctuation phenomena. Altshuler and 
Shklovskii \cite{Alt86} used the Thouless energy and a postulated connection 
to RMT to obtain a semiquantitative understanding of universal 
conductance fluctuations, one of the main manifestations of quantum coherence 
in mesoscopic systems. Finally, persistent currents in mesoscopic
rings, which are the topic of Sec.~\ref{pc},
measure the sensitivity of the spectrum to an external magnetic flux and 
provide a direct link between experiment and spectral fluctuation properties.

In Ref.~\onlinecite{Gor65}, it was suggested that GRMT and
Wigner--Dyson (WD) statistics in the form described in
Sec.~\ref{qc1rc} might be used to also address the spectral statistics
in small disordered metallic particles at low temperatures. This
suggestion could, however, not have been correct in general since WD
statistics contains no parameter related to the dimensionality (or
even the spatial extension) of the system. Research on spectral
fluctuations in disordered metals has therefore to a large extent been
driven by the following questions: Under which circumstances does WD
statistics apply, where should one expect deviations, and how can
these be calculated?

We approach the subject as follows. First, we recall 
in Sec.~\ref{spec_1} how the validity of 
WD statistics was established in the diffusive regime in some limiting case, 
and how corrections to this limit were calculated. Second,
in Sec.~\ref{spec_2}, we summarize 
attempts to go beyond diffusive systems, in the direction of both the 
ballistic and the localized regime. Third, we deal with spectral statistics 
in the proximity of the metal--insulator transition in Sec.~\ref{spec_3}.

A recent review by Dittrich~\cite{Dit96} nicely complements our
discussion in the present section in that it focuses on the
semiclassical approach to spectral statistics in (quasi)
one--dimensional disordered systems.

\subsubsection{Diffusive regime and Wigner--Dyson statistics}
\label{spec_1}

The above--mentioned conjecture by Gorkov and Eliashberg \cite{Gor65} 
was proved by Efetov \cite{Efe83}. Efetov showed that many statistical 
fluctuation measures of small disordered metallic particles can be calculated 
in terms of a functional integral over a field of supermatrices $Q(\vec{r})$. 
This ``non--linear $\sigma$ model'' was described in Sec.~\ref{susy}. For 
instance, the two--point correlation function $R(s)$ with $s = \omega / 
\Delta$ is -- apart from an additive constant -- given by
\begin{equation}
R(s) \propto {\rm Re} \int d[Q] (\ldots) \exp\left( -S[Q] \right)
\label{eq_4:01}
\end{equation}
where (with $\hbar=1$)
\begin{equation}
S[Q] = - \frac{\pi\nu}{8} \int {\rm trg} \left[ {\cal D}(\nabla Q)^2 +
2i\omega L Q\right] d^dr \ .
\label{eq_4:02}
\end{equation}
Here, $\nu = 1/(V\Delta)$ is the density of states (with $V$ the
volume and $\Delta$ the mean single--particle level spacing of the
system), ${\cal D}$ the diffusion constant, $ L$ a diagonal
supermatrix, and $Q$ a supermatrix whose detailed structure depends on
the symmetry class. The precise definitions can be found in
Sec.~\ref{susy}, cf. also Ref.~\onlinecite{Efe83}. The quantity
$\omega$ is the difference between two eigenvalues. The
pre--exponential terms in Eq.~(\ref{eq_4:01}) are not explicitly
specified. They typically involve certain components of $Q$. In the
context of Random Matrix Theory the correlation function $R(s)$ is
typically denoted by $X_2(r\equiv s)$, see Eq.~(\ref{1rc36}).

The ``effective Lagrangean'' $S[Q]$ comprises two terms, one with and one 
without spatial derivative. The gradient term owes its existence to spatial
fluctuations in the $Q$ field and would disappear if the system under
consideration were point--like. The second or ``symmetry--breaking'' term 
$\propto {\rm trg}[ L Q]$ survives even in the zero--dimensional limit 
$Q(\vec{r}) =  Q_0 = {\rm const}$. 

Efetov showed that in the zero--dimensional limit, his non--linear 
$\sigma$ model yields WD statistics for all three symmetry classes. He did
so by reproducing the two--level correlation functions $R(s)$ of GRMT
from the symmetry--breaking term of the non--linear $\sigma$ model. 
This result established the equivalence of the zero--dimensional 
$\sigma$ model and RMT (here in the narrow sense of the three Gaussian 
ensembles). Deviations from this limit are due to the spatial extension of 
the system. 

To determine the range of validity of GRMT, one has to compare the energy 
scales associated with the two terms in $S[Q]$. For the gradient term this 
scale is given by the Thouless energy $E_C = {\cal D} / L^2$, while the 
symmetry--breaking term is proportional to the difference $\omega$ of two 
eigenvalues. For $\omega \ll E_C$ the symmetry--breaking term gives the 
dominant zero mode, and one expects WD statistics, while for $\omega \geq 
E_C$ corrections to WD statistics should arise.  

First corrections of this type were calculated by Altshuler and Shklovskii
\cite{Alt86} within the perturbative approach of the impurity diagram
technique. (We briefly turn to this method in our discussion of transport in 
quasi one--dimensional systems, see Sec.~\ref{quasi1d}). Their study
was partly  
motivated by the phenomenon of universal conductance fluctuations, i.e. by 
the fact that the variance $\langle \delta g^2\rangle$ of the dimensionless 
conductance $g$ is of order unity in mesoscopic systems, see
Sec.~\ref{quasi1d_1}. 
The argument by Thouless \cite{Tho77} (to which we have already alluded in 
the beginning of this section) linked spectral statistics and the statistics 
of the conductance through the relation $g = E_C / \Delta$. This relation is 
rewritten in the form
\begin{equation}
g = \langle N(E_C) \rangle \ ,
\label{eq_4:03}
\end{equation}
where $\langle N(\omega)\rangle$ is the mean number of levels within
an energy  
interval of size $\omega$. In the diffusive regime, where $\Delta \ll
E_C$, we 
have $g\gg 1$ (good conductor). Through Eq.~(\ref{eq_4:03}), the variance 
$\langle \delta g^2\rangle$ became linked with the number variance
$\Sigma^2$ (see Sec.~\ref{qc1bac}). Following common usage in much of the
condensed matter literature we denote the number variance by
$\langle \delta N^2 \rangle$ in this section. 
Since Eq.~(\ref{eq_4:03}) limits the spectral 
range of interest to the Thouless energy $E_C$, WD statistics seemed 
appropriate. However, WD statistics implies $\langle \delta N^2 \rangle 
\propto \ln \langle N\rangle$, and the finite width of the energy levels 
due to the coupling to the leads had to be invoked \cite{Alt86} to reduce 
this value to $\langle \delta N^2\rangle \approx 1$ as required. For 
$\omega \gg E_C$ the result of the perturbative analysis in
Ref.~\onlinecite{Alt86} was 
\begin{equation}
\langle \delta N(\omega)^2 \rangle = \frac{c_d k \tilde{s}^2}{\beta}
\left(\frac{\omega}{E_C}\right)^{d/2} \ ,
\label{eq_4:04}
\end{equation}
where $c_d$ is a constant which depends on dimension $d$, where $\beta = 1,2$, 
and $4$ for orthogonal, unitary and symplectic symmetry, respectively, where 
$k$ is the number of independent spectra considered, and where $\tilde{s}$ 
characterizes the degeneracy of the levels. This result shows that with 
increasing energy $\omega$, the stiffness in the spectra of disordered
metals is  
reduced and the fluctuations in the number of levels increases. The physical 
picture behind these results is the following. For energies smaller than
$E_C$, i.e. for time scales larger than the diffusion time through the sample,
the wave packet of a diffusing electron has reached the boundary of
the sample and is, in a sense, equilibrated. The spatial extension of
the sample is then irrelevant and, therefore, WD statistics applies. This is 
precisely the same situation as in compound nucleus scattering, see 
Sec.~\ref{qc3nu}. In the opposite limit of energies larger than $E_C$
we deal with  
a wave packet that has not yet reached the boundaries of the sample. The 
corresponding level statistics has therefore not yet attained the universal 
limit and depends on the dimensionality of the system.

\begin{figure}
\centerline{
\psfig{file=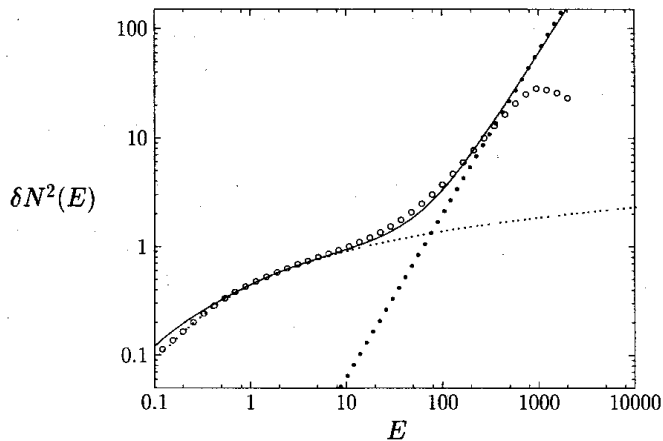,width=3.5in}
}
\caption{The number variance
  $\delta N^2$ for a system with $20^3$ sites versus the energy
  $\omega\equiv E$ in units of the mean level spacing (open circles).
  The Thouless energy is approximately $E_C\approx 2.5$.
  For comparison, the RMT behavior
  (dotted line), the asymptotic $E^{3/2}$ behavior of
  Eq.~(\protect\ref{eq_4:04}) (full circles), and the full
  perturbative result derived in Ref.~\protect\onlinecite{Alt86} are also
  shown. Taken from Ref.~\protect\onlinecite{Bra95}.
}
\label{fig12}
\end{figure}

Numerically, the crossover to the regime described by
Eq.~(\ref{eq_4:04}) was observed by Braun and Montambaux~\cite{Bra95},
see Fig.~\ref{fig12}. 

The corrections found by Altshuler and Shklovskii \cite{Alt86} in the 
diffusive regime could be refined in later investigations, in particular by
avoiding their perturbative analysis. Their calculation had focussed on the 
two--level correlation function $R(s)$ as a necessary prerequisite to 
determine the number variance. 
In the regime $1\ll s\ll g$, the perturbative approach \cite{Alt86} was 
only capable of yielding the smooth, non--oscillating behavior of $R(s)$.
In two papers \cite{Krv94a,And95}, corrections to $R(s)$ were 
calculated in the non--perturbative framework of the non--linear $\sigma$ model.
In the method used by Kravtsov and Mirlin, the parametrization
$Q=T_0^{-1}LT_0$ for a constant $Q$ matrix was modified by the
substitution $L\rightarrow \tilde{Q}(\vec{r})$ with an ensuing
non--trivial change in the integration measure of the supersymmetric
functional \cite{Mir95a}. Fluctuations of
$\tilde{Q}(\vec{r})$ around the origin $L$ of the coset space (leading to
contributions from the gradient term in Eq.~(\ref{eq_4:02})) were
taken into account perturbatively and were integrated out in the spirit of a
renormalization group treatment, see Sec.~\ref{akl_2}. For $g\gg s$ this
led to the result
\begin{equation}
R(s) = 1 - \frac{\sin^2(\pi s)}{(\pi s)^2} + 
       \frac{a_d}{\pi^2 g^2} \sin^2(\pi s) 
\label{eq_4:05}
\end{equation}
with $a_d$ a constant depending on dimension. Equation~(\ref{eq_4:05})
holds for the unitary case. Expressions for orthogonal and symplectic
symmetry are also given in Ref.~\onlinecite{Krv94a}. The first two
terms in Eq.~(\ref{eq_4:05}) represent the result for the GUE, i.e.
the universal limit. Since $g \approx E_C / \Delta$, see
Eq.~(\ref{eq_4:03}), the condition $g \gg s$ means $E_C \gg \omega$.
Hence, the third term in Eq.~(\ref{eq_4:05}) is a very small
correction to the zero--dimensional limit of Eq.~(\ref{eq_4:02})
originating from the comparatively massive modes of the gradient term.
Therefore, corrections to the result of cRMT are quite small in the
entire diffusive regime. The smooth part of Eq.~(\ref{eq_4:05}),
\begin{equation}
R(s) = 1- \frac{1}{2\pi^2 s^2} + \frac{a_d}{2\pi^2 g^2} \ ,
\label{eq_4:06}
\end{equation}
had already been found in the perturbative calculation \cite{Alt86}. 

Andreev and Altshuler \cite{And95} rederived Eq.~(\ref{eq_4:05}) for $1\ll s 
\ll g$. They also showed that for $s\gg g$ the oscillations in $R(s)$
are exponentially damped and only the perturbative corrections of
Ref.~\onlinecite{Alt86} survive.
Their calculation was based on a new and interesting technical point. 
Following Ref.~\onlinecite{And95} we consider the generalization of 
Eq.~(\ref{eq_4:01}) and Eq.~(\ref{eq_4:02}) to parametric perturbations of 
the impurity potential characterized by the strength parameter $x$,
\begin{eqnarray}
R(\omega, x) &\propto& \frac{\partial^2}{\partial J^2}
\int d[Q] \exp(-S_J[Q]) \Bigg\vert_{J=0} \ ,   \nonumber\\
S_J[Q] &=& - \frac{\pi\nu}{8} \int {\rm trg}
\left[ D(\nabla Q)^2 + 2i\omega L Q + iJ L k Q
  -x^2\Delta( L Q)^2/2 \right] \ .
\label{eq_4:07}
\end{eqnarray}
The derivative with respect to $J$ produces the pre--exponential terms
omitted in Eq.~(\ref{eq_4:01}), and $k$ is a diagonal matrix with elements
equal to $1$ and $-1$ in the Boson--Boson block and Fermion--Fermion
block, respectively. 

Let us focus attention on the case of unitary symmetry.
The conventional treatment takes into account the usual saddle point
$\propto L$ of
the effective Lagrangean. There exists, however, one additional saddle
point \cite{And95}. It has the form
$-k L$ and is well defined for any finite non--zero $x$. This saddle
point breaks  
supersymmetry and had previously been overlooked. Taking into account the 
contributions of both saddle points leads to Eq.~(\ref{eq_4:05}). The 
orthogonal and symplectic cases could be treated similarly \cite{And95}.

An interesting observation concerning the corrections to
WD statistics for $s\gg g$ was made by Kravtsov and Lerner
\cite{Krv95a}. We recall that $s\gg g$ corresponds to $\omega\gg E_C$,
i.e. to the limit in which Eq.~(\ref{eq_4:04}) is valid.
In this regime, the general (smooth) result for the
two--point correlation function derived in Ref.~\onlinecite{Alt86} reads
\begin{equation}
R(s) = \frac{b_d}{\beta g^{d/2} |s|^{2-d/2}} \ .
\label{eq_4:08}
\end{equation}
But in two dimensions we have $b_d = 0$, and the next order in $g^{-1}$
has to be invoked (weak localization corrections)! As a result, $R(s)$
for $d = 2$ is given by
\begin{equation}
R(s) = \frac{\eta(g,\beta)}{\beta} \frac{1}{g^2} \frac{1}{|s|} \ ,
\label{eq_4:09}
\end{equation}
where $\eta(g,\beta) = (\beta-2)^{-1}$ for $\beta = 1,4$ and
$\eta(g,\beta) = -1/2g$ for $\beta=2$. Hence, in contrast to the number
variance (\ref{eq_4:04}), the level correlation function was found 
\cite{Krv95a} to be totally governed by weak localization corrections in 
this regime. 

Motivated by the validity of WD statistics in the metallic regime, 
Akkermans and Montambaux \cite{Akk92} re--examined Thouless' arguments 
\cite{Edw72,Tho77} connecting dissipation with level statistics, and 
the Thouless relation (\ref{eq_4:03}). (In the original work, the
influence of WD statistics had not been taken into account, and this had 
meanwhile been found to be manifestly incorrect). Let the closed
system with  
eigenvalues $E_n$ be subject to an external perturbation $\varphi$, and 
define the dimensionless quantity $g_C$ by
\begin{equation}
g_C = \frac{1}{\Delta} \left\langle \left( \frac{\partial^2
  E_n}{\partial \varphi^2} \right)^2_{\varphi = 0} \right\rangle
  ^{1/2}. 
\label{eq_4:013}
\end{equation}
Obviously, $g_C$ measures the sensitivity of the levels $E_n$ to an 
external perturbation. (The perturbation can, for example, be realized by
a magnetic flux $\phi$ with $\varphi = 2\pi\phi/\phi_0$). Therefore, we may
view Eq.~(\ref{eq_4:013}) as a definition of the Thouless energy $E_C$ such 
that $ g_C = E_C / \Delta$. The statement equivalent to Eq.~(\ref{eq_4:03}) 
is then simply $g = g_C$ \cite{Edw72} where $g$ is the dimensionless 
conductance as given by the Kubo formula. Akkermans and Montambaux 
succeeded in deriving this ``Thouless formula''. Using scattering theory, 
the Friedel sum rule, and very general assumptions concerning the number 
correlator $\langle \delta N(E_F,\varphi) \delta N(E_F,\varphi') \rangle$, 
they could write $g$ in the form
\begin{equation}
g = -\frac{1}{4} \frac{\partial^2}{\partial\varphi^2} \langle \delta
N^2(E_F,\varphi) \rangle \Bigg\vert_{\varphi=0} \ .
\label{eq_4:014}
\end{equation}
Assuming WD statistics for the energy levels gave the central result
\begin{equation}
g \propto g_C  \ .
\label{eq_4:015}
\end{equation}
Thus, the Thouless formula was derived in a much more rigorous fashion. In
addition, it was found that \cite{Akk92}
\begin{equation}
\overline{
\left\langle
\left(
\frac{\partial E}{\partial\varphi}
\right)^2
\right\rangle
}^\varphi
\propto \Delta \left\langle \left(
\frac{\partial^2 E}{\partial\varphi^2} \right)^2_{\varphi=0}
\right\rangle^{1/2}
\label{eq_4:016}
\end{equation}
(with $E$ a typical energy level). The bar on the l.h.s.  of
Eq.~(\ref{eq_4:016})  
denotes an average over all flux values. We note that Eq.~(\ref{eq_4:016}) 
connects a flux average with a local quantity at $\varphi = 0$.

As the last item, we address the effect of time--reversal symmetry
breaking in the diffusive regime. This problem arises in isolated
mesoscopic rings threaded by a magnetic flux $\phi$. In
Refs.~\onlinecite{Dup91} and \onlinecite{All93a}, this problem was
investigated and compared to the GOE $\rightarrow$ GUE crossover
transition first considered by Pandey and Mehta \cite{Pan83}, 
who analytically calculated all spectral correlation functions for a
certain random matrix model, see
Sec.~\ref{qc1tcc}. Dupuis and Montambaux \cite{Dup91} established
numerically that the crossover transition is governed by the parameter
$(E_C / \Delta) (\phi/\phi_0)^2$. In the ergodic regime $\omega <
E_C$, the transition is well described by the Pandey--Mehta random
matrix model \cite{Pan83}, see Eq.~(\ref{Me}), provided the parameter
$t$ used there is identified with the parameter
$4\pi(E_C/\Delta)(\phi/\phi_0)^2$. Altland, Iida and Efetov
\cite{All93a} calculated the two--point correlation function $R(s)$ in
the crossover regime analytically. The calculation used the
supersymmetry method and a novel parametrization of the $Q$ matrix, $Q
= T^{-1} L T$, tailored to the breaking of time--reversal symmetry.
The matrices $T$ were decomposed into two (Cooperon and Diffuson)
parts,
\begin{equation}
T = T_CT_D  \ ,
\label{eq_4:023}
\end{equation}
with
\begin{equation}
[T_D,\tau_3] = 0.
\label{eq_4:024}
\end{equation}
where $\tau_3={\rm diag}(1,-1)$ is the matrix which describes time--reversal 
symmetry breaking. With this construction, the symmetry--breaking term in 
the effective Lagrangean depends only on $T_C$. This fact simplifies the 
calculation considerably. As a result, the effective Lagrangean in the 
supersymmetric functional (\ref{eq_4:01}) for the correlation function 
$R(s)$ has the form
\begin{equation}
S[Q] = -\frac{\pi\nu A}{8} \int {\rm trg} 
\left[ {\cal D}(\nabla_x-i(2\pi/L)(\phi/\phi_0)[\tau_3,Q])^2 +
2i\omega L Q\right] dx \ .
\label{eq_4:022}
\end{equation}
Here, $x$ is the longitudinal coordinate in the ring and $A$ the cross
section of the ring. In the regime $\omega \ll E_C$, i.e in zero--mode
approximation, an explicit result for $R(s)$ was derived, which
coincides with the findings in Ref.~\onlinecite{Pan83}.

This concludes our discussion of the diffusive regime. We have seen that in 
an energy range defined by the Thouless energy $E_C$, the spectral fluctuation 
properties (including the crossover induced by the breaking of time--reversal 
symmetry) are well described by the Gaussian ensembles. Deviations from this 
universal limit can be systematically calculated and are found to be 
generically small.

\subsubsection{Ballistic and localized regimes}
\label{spec_2}

We summarize some important work on the crossover transition from the 
diffusive to either the ballistic or the localized regime, cf. the 
classification given at the beginning of this section. This crossover can 
be realized by changing either the system size or the strength of disorder. 
The extreme examples of a practically clean and a strongly localized system 
show that the regime chosen has a drastic influence on the level statistics. 
In the former case the levels are determined by the quantization conditions 
dictated by the boundary of the system while in the latter case the system 
is composed of essentially independent localization volumes, and level 
repulsion is suppressed.

Altland and Gefen \cite{All93} addressed the crossover from the diffusive 
to the ballistic regime. They studied non--interacting electrons in a 
two--dimensional square geometry with a random white noise potential.
Within a suitable perturbative framework an equation for the two--point level
correlation function $R(s;B_1,B_2)$ depending on two magnetic fields $B_1$ and 
$B_2$ was derived, 
\begin{eqnarray}
R(s;B_1,B_2) &=& -\frac{\Delta^2}{2\pi^2} \partial^2_\omega
\Bigg(
{\rm tr} \Bigg\{ \hat{\ln}[1-\zeta^{(D)}(\omega,B_-)] \nonumber\\
 & & \qquad\qquad\quad +  \hat{\ln}[1-\zeta^{(C)}(\omega,B_+)] - S_1(\omega) 
\Bigg\}\Bigg) \ .
\label{eq_4:017}
\end{eqnarray}
Here, $B_\pm = B_1\pm B_2$, 
$\hat{\ln}(1-x)$ is defined by the power series of $\ln(1-x)+x$,
$S_1(\omega)$ is an essentially unimportant contribution and
the coordinate representation of the (Cooperon and Diffuson)
operators $\zeta^{(C,D)}$ reads
\begin{eqnarray}
\zeta^{(D)}(\omega,B_-;r,r') &=& \frac{\Delta}{2\pi\tau}
G^{+}(E_1,B_1;r,r')G^{-}(E_2,B_2;r',r) \ , \nonumber\\
\zeta^{(C)}(\omega,B_-;r,r') &=& \frac{\Delta}{2\pi\tau}
G^{+}(E_1,B_1;r,r')G^{-}(E_2,B_2;r,r') \ ,
\label{eq_4:018}
\end{eqnarray}
where $G^{\pm}$ are averaged single--particle Green functions. The
spectral properties are entirely governed by the eigenvalues of the
operators $\zeta^{(C,D)}$. Both the diffusive and the ballistic regime
can now be investigated. In the diffusive regime, three energy ranges
were considered, D1 ($0\le\omega<E_C$), D2 ($E_C < \omega<1/\tau$),
and D3 ($1/\tau<\omega$).  It turns out that the range D1 coincides
with the domain of WD statistics established in
Ref.~\onlinecite{Efe83}. In the range D2, the results of Altshuler and
Shklovskii \cite{Alt86} apply. These two cases have been treated in
detail in the preceding section. In the range D3, we probe the
dynamics of the system for times smaller than the elastic scattering
time $\tau$. In this range the direction of the particle momentum is
not yet completely randomized and the microscopic features of the
random potential matter. For a standard Gaussian white noise potential
the number variance was found to increase logarithmically,
\begin{equation}
\langle \delta N^2 \rangle = \frac{1}{\pi}
\left(\frac{L}{2\pi\ell}\right)^2 \ln(E\tau) \ .
\label{eq_4:019}
\end{equation}
This result is non--universal since it depends on the details of the random 
potential. In the ballistic regime, the Thouless energy $E_C$ loses its 
physical significance because the electron typically traverses the sample 
non--diffusively. The extension of the sample enters the description now 
via the time of flight $t_f$ through the system. Again, three energy ranges 
were considered: B1 ($0\le\omega<1/\tau$), B2 ($1/\tau<\omega<1/t_f$), and 
B3 ($1/t_f<\omega$). The range B1 corresponds to very large times. The
electron traverses the system many times and is multiply scattered in the 
process. This range is therefore similar to the diffusive range D1. The range 
B3, on the other hand, corresponds to the case where the electron does not 
have the time to randomize its momentum or even to reach the boundary of the 
system. Therefore, B3 is analogous to D3. Finally, the range B2 is typical 
for the ballistic regime and has no analogue in the diffusive regime. Here, 
the electron has sufficient time to explore the whole system {\it without} 
undergoing complete momentum relaxation via scattering by the random 
potential. Therefore, the boundary of the system plays an important role. 
In the case of a two--dimensional square (or rectangle) \cite{All93} the 
number variance for the range B2 is given by
\begin{equation}
\langle\delta N^2\rangle = -\frac{2}{\pi^2} \ln (\gamma\tau) -
\frac{2}{\pi^2} \frac{(E\tau)^2}{1+(E\tau)^2} \ ,
\label{eq_4:020}
\end{equation}
where $\gamma$ is a phenomenological level broadening of the order of
$\Delta$. Interestingly, in this regime the fluctuations in the level
number decrease upon increasing the energy window $E$. 
A discussion of both, the difference between energy-- and
disorder--averaging in ballistic systems, and the effect of finite
magnetic fields is also contained in Ref.~\onlinecite{All93}. 

Another interesting point was made in Ref.~\onlinecite{All94}.  There,
it was shown that several observables are still characterized by the
Thouless energy $E_C$ in the ballistic regime, although this energy
scale can no longer be interpreted as the inverse transport time
through the sample.

Building on ideas by Oppermann~\cite{Opp90}, Altland and Zirnbauer
showed~\cite{All96,All97} that new universality classes arise in
mesoscopic structures with normally conducting--superconducting (NS)
interfaces. They found that the ergodic limit of a quantum dot in
contact with superconducting regions (cf. regimes D1 and B1 above) is
not characterized by the three Gaussian ensembles GOE, GUE, and GSE,
which define WD spectral statistics. Instead, the NS system is
described by four new ensembles dubbed C, D, CI, and DIII after the
associated four symmetric spaces in Cartan's classification scheme,
see Ref.~\onlinecite{Zir96}. These four ensembles correspond to the
four possibilities of combining good or broken time reversal
invariance with good or broken spin rotation invariance.  Shortly
after Ref.~\onlinecite{All96} became available a microscopic model for
class C was derived by Frahm {\it et al.}~\cite{Fra96z}.  To justify
their claims the authors of Refs.~\onlinecite{All96,All97} started
from the Bogoliubov--de Gennes Hamiltonian
\begin{equation}
{\cal H} = \left[ \begin{array}{cc}
                  h      &   \Delta   \\
               -\Delta^* &  -h^T   
                 \end{array}  \right]
\label{eq_4:020a}
\end{equation}
with the diagonal block $h_{\alpha\beta} = h^*_{\beta\alpha}$ and the
pairing field $\Delta_{\alpha\beta} = -\Delta_{\beta\alpha}$.
Depending on the symmetry class considered, further constraints on $h$
and $\Delta$ are necessary. With the two additional assumptions that
the phase shift for Andreev scattering vanishes on average over the NS
interface, and that the classical dynamics of the N region is chaotic,
the Hamiltonian (\ref{eq_4:020a}) could be replaced by a random matrix
of appropriate symmetry, whose independent entries are Gaussian
distributed. Several properties of the ensuing four random matrix
ensembles were derived in Ref.~\onlinecite{All97}, in particular the
spectral $n$--point correlation functions for classes C and D, and ---
for an open NS dot --- the weak localization corrections to the
conductance which had earlier been considered by Brouwer and
Beenakker~\cite{Bro95}. Surprisingly, some of these corrections
persist in an external magnetic field~\cite{Bro95,All96}.  Whether or
not universal conductance fluctuations in the presence of Andreev
scattering depend on time--reversal symmetry breaking seems to depend
on the precise physical situation~\cite{Bro95a,All97}.

The authors of Ref.~\onlinecite{All97} argue that the four new
ensembles together with the three Gaussian and the three chiral
ensembles (see Sec.~\ref{conn}) exhaust Cartan's classification scheme
for symmetric spaces. Therefore they do not expect that additional
universality classes will be found.

The four ensembles just discussed apply to the case where the
proximity effect, i.e. an excitation gap in the spectrum of the normal
metal induced by the superconductor, is suppressed. A RMT treatment of
the proximity effect has been given by Melsen {\it et al.}~\cite{Mel96}.

\begin{figure}
\centerline{
\psfig{file=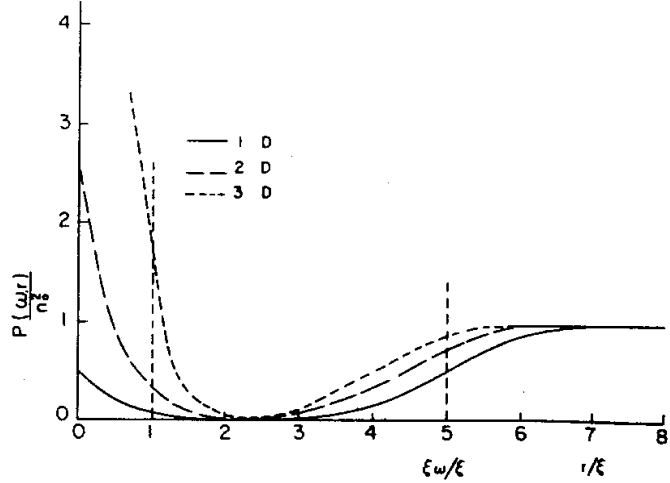,width=3.5in}
 }
\caption{
  Schematic drawing of the local density of
  states correlation function (normalized by the product $n_0^2$ of
  the density of states) as a function of $r=|\vec{r}|$ (in units of
  $\xi$) for $\omega\ll\Delta$. 
  Taken from Ref.~\protect\onlinecite{Siv87}.
} 
\label{afig2}
\end{figure}

Having explored the properties of various ballistic or nearly 
ballistic systems, we now turn our attention to the transition
to localization.
In the localized regime, wave functions of nearly degenerate states may 
have an exponentially small overlap. Therefore, level repulsion is 
suppressed and in the strongly localized regime, one expects Poisson 
statistics for the eigenvalues. The study of the crossover to the 
localized regime requires non--perturbative methods. In a numerical study, 
Sivan and Imry \cite{Siv87} employed a tight--binding model with diagonal 
disorder,
\begin{equation}
H = \sum_i h_i |i\rangle\langle i| + \sum_{ij} V_{ij} |i\rangle\langle
j| \ ,
\label{eq_4:011}
\end{equation}
where $h_i$ is Gaussian distributed and $V_{ij}$ connects only nearest
neighbors. According to the analytical results of
Ref.~\onlinecite{Efe83}, it is expected that in the diffusive regime,
the nearest neighbor spacing distribution $P(s)$ should agree (in a
certain energy range) with WD statistics. This was explicitly
confirmed in one and three dimensions.  In the localized regime, the
authors investigated the local density of states correlation function
\begin{equation}
P_E(\omega, \vec{r}) = \left\langle \sum_{ij}
\delta(E-E_i)\delta(E+\omega-E_j) \int_V d^3x \, \psi_i^2(\vec{x})
\psi_j^2(\vec{x}+\vec{r}) \right\rangle,
\label{eq_4:012}
\end{equation}
which had earlier been calculated by Gorkov, Dorokhov, and Prigara
\cite{Gor83} for one--dimensional chains.  This function contains
information about the correlations of electronic states which are
energetically separated by $\omega$ and spatially separated by
$r=|\vec{r}|$. The general behavior of $P_E(\omega, \vec{r})$ for
$d=1,2,3$ can be characterized as follows \cite{Gor83,Siv87}, see
Fig.~\ref{afig2}. For $r\to 0$ $P_E$ assumes a finite,
$\omega$--independent value, from which it decays to exponentially
small values for $\ell < r < \xi_\omega$ (level repulsion).  At the
scale $\xi_\omega$ the function crosses over to its disconnected part
(no correlations). It was pointed out in Ref.~\onlinecite{Siv87} that
this behavior follows from earlier ideas by Mott \cite{Mot70}, who had
already introduced the frequency--dependent length scale $\xi_\omega =
\xi \ln(c\Delta/\omega)$ as the typical tunneling range between almost
degenerate states. Here, $\xi$ is the localization length and $c$ a
constant.

Analytical progress in understanding level statistics in the localized
regime is possible with the help of suitable model systems. Here, we
consider the quasi one--dimensional mesoscopic wire with unitary
symmetry for which transfer matrix techniques should be useful. (This
system is further discussed in the context of transport properties,
cf. Sec.~\ref{quasi1d}). Altland and Fuchs \cite{All95} used the
results of Ref.~\onlinecite{Efe83} and, in the spirit of the transfer
matrix approach, reduced the calculation of $R(\omega,L)$ (see
Eq.~(\ref{eq_4:01})) to the solution of the differential equations
\begin{eqnarray}
\left[ -\partial_t + {\cal O} \right] Y_0(\lambda,t) &=& 0,
\qquad Y_0(\lambda,0) = 1 \ ,  \nonumber\\
\left[ -\partial_t + {\cal O} \right] Y_-(\lambda,t) &=&
  Y_0(\lambda,t), \qquad Y_-(\lambda,0) = 0 \ .
\label{eq_4:021}
\end{eqnarray}
Here, $\lambda = (\lambda_1,\lambda_2)$ are the radial coordinates for the 
$Q$ matrix in the unitary case, $t=r/\xi$ with $r$ the coordinate along the 
wire, and ${\cal O} = \Delta_r/16 + V(\lambda)$ with $\Delta_r$ the radial
part of the Laplacian on the manifold of $Q$ matrices and $V(\lambda)
= -i(\omega/\pi\Delta_\xi)(\lambda_1-\lambda_2)$ ($\Delta_\xi$ is the
level spacing in one localization volume). The ratio $\omega/\Delta_\xi$ 
governs the relative importance of the two terms in ${\cal O}$. The 
correlation function $R(\omega,L)$ can be expressed in terms of the 
eigenvalues and eigenfunctions of ${\cal O}$ \cite{All95}. In general, 
Eqs.~(\ref{eq_4:021}) cannot be solved analytically, and one has to calculate 
these eigenvalues and eigenfunctions on a computer. The main results were
as follows. For $t<1$ the WD and the Altshuler--Shklovskii regimes
were recovered, albeit with superimposed oscillations on the scale
$\Delta$ in the latter. This non--perturbative effect had not been
calculated earlier. For $t>1$ a gradual crossover to the Poisson limit
($R(\omega,L)=0$ as $L\to \infty$) was observed, obeying the scaling law
$R(\omega,L) \approx (\xi/L)^d f(\omega/\Delta_\xi)$. The function $f(x)$ 
is proportional to $\ln x$ for very small frequencies ($x\ll 1$) and 
proportional to $x^{-3/2}$ for larger frequencies ($x>1$). The latter 
behavior is reminiscent of the Altshuler--Shklovskii regime, see 
Eq.~(\ref{eq_4:08}).

We have seen that in the various regimes relevant for disordered mesoscopic 
systems, a fairly thorough understanding of spectral fluctuations has been 
attained. The use of WD statistics frequently served as a reference standard 
which helped to identify and interpret relevant length and energy scales.
The greatest challenge is probably still posed by the phenomenon of 
localization, in particular by the existence of a metal insulator transition 
to which we turn next.

\subsubsection{Critical distribution at the metal insulator
  transition} 
\label{spec_3}

For $d > 2$ (or, in the symplectic symmetry class, for  $d\ge 2$) there
occurs generically a metal insulator transition (MIT). It is characterized 
by the existence of a {\it mobility edge} $E_M$ which separates extended and 
localized states. With the help of
the Anderson model \cite{And58} the mobility
edge can be investigated as follows. For disorder values $W$ exceeding
a certain critical disorder $W_M$ all states in the spectrum are
localized. For $W<W_M$, however, extended states appear in the center
of the band. This means that states in the band center are critical for
$W=W_M$. We address the question: What are the 
properties of the energy eigenvalue statistics in the vicinity of or, in the 
thermodynamic limit, at the MIT? We use the index $M$ to label quantities 
which refer to the mobility edge.

An early argument due to Altshuler {\it et al.} \cite{Alt88}
suggesting that the MIT might significantly influence the energy
eigenvalue statistics used the following interpretation of the result
(\ref{eq_4:04}) obtained in Ref.~\onlinecite{Alt86} for $\omega >
E_C$. In the time $\hbar / \omega$ an electron propagates over a
distance $L_\omega=({\cal D} / \omega)^{1/2}$. This distance is
smaller than the system size $L=( {\cal D} / E_C)^{1/2}$. Therefore,
cubes of size $L_\omega$ possess independently fluctuating spectra,
and $\langle \delta N^2 \rangle$ is proportional to the number of such
cubes. For $d=3$, this yields
\begin{equation}
\langle \delta N^2 \rangle \propto \left( \frac{L}{L_\omega} \right)^3
= \left( \frac{\omega}{E_C} \right)^{3/2} \ ,
\label{eq_4:1}
\end{equation}
in keeping with Eq.~(\ref{eq_4:04}). At the MIT, two important effects 
have to be taken into account. First, we have $E_C \approx \Delta$. 
Therefore, the range of validity of WD statistics is confined to a few 
levels at most, and the argument leading to Eq.~(\ref{eq_4:1}) applies
practically  
to all energies. Second, at the MIT the dimensionless conductance $g$ is 
expected to approach a critical value $g_M$ wich is {\it independent of the 
system size} $L$. Imposing this condition, one finds that the diffusion 
constant ${\cal D}$ becomes scale--dependent, ${\cal D} = {\cal D}(L)$ with 
${\cal D}(L) \propto 1 / L$ for $d = 3$. Taking these points into account, 
Altshuler {\it et al.} \cite{Alt88} could estimate the behavior of $\langle 
\delta N^2 \rangle$,
\begin{equation}
\frac{\langle \delta N^2 \rangle}{\langle N(\omega) \rangle} = \chi
\approx 0.25 \ .
\label{eq_4:2}
\end{equation}
This result was closer to the Poisson distribution (which gives $\langle\delta 
N^2\rangle = \langle N\rangle$) than to Random Matrix Theory (which gives 
$\langle\delta N^2\rangle\approx \ln\langle N\rangle$). The corresponding 
estimate for the nearest neighbor spacing distribution $P(s)$ with 
$s=\omega/\Delta$ gave
\begin{equation}
P(s)\propto \exp(-s/\chi) \ ,
\label{eq_4:3}
\end{equation}
which again is reminiscent of the Poisson distribution. It turned out, 
however, that these arguments were oversimplified and that the estimates 
(\ref{eq_4:2}) and (\ref{eq_4:3}) had to be revised. 

Shklovskii {\it et al.} \cite{Shk93} were the first to formulate the
idea that the nearest neighbor spacing distribution $P_M(s)$ at the
MIT might be universal, thus representing a third possibility besides
the Wigner--Dyson statistics $P_{WD}(s)$ and the Poisson law $P_P(s)$.
In the band center of a three--dimensional Anderson model, and in the
thermodynamic limit $L\to\infty$, one expects that $P(s)=P_{WD}(s)$
below the critical disorder ($W < W_M$) and $P(s)=P_P(s)$ above it ($W
> W_M$). For $W > W_M$, this is because $P(s)$ is the superposition of
the contributions from infinitely many and statistically independent
localization volumes. For $W < W_M$, on the other hand, the range of
validity of Wigner--Dyson statistics is extended to infinity since
$E_C/\Delta$ diverges in the thermodynamic limit. Precisely at the MIT
these arguments do not apply since the localization length $\xi(W)$
diverges and $E_C \approx \Delta$. This fact led to the postulate
\cite{Shk93} of a third ``critical'' distribution, which was believed
to be a hybrid of $P_{WD}(s)$ and $P_P(s)$. Linear short--range level
repulsion was expected to be followed by Poisson--like behavior, in
agreement with Ref.~\onlinecite{Alt88}.

\begin{figure}
\centerline{
\psfig{file=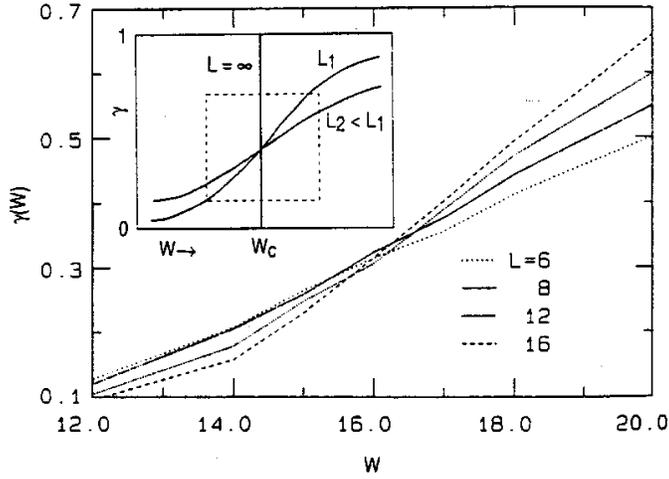,width=3.5in}
 }
\vskip 0.3cm
\caption{
The quantity $\gamma(W,L)$ as a function of $W$ for different system
sizes $L$ from numerical diagonalization. Inset: schematic figure
illustrating the smooth crossover behavior for finite system
sizes. Taken from Ref.~\protect\onlinecite{Shk93}.
} 
\label{afig3}
\end{figure}

To check the hypothesis of a third universal distribution, an
interesting test was developed and performed in
Ref.~\onlinecite{Shk93}. We consider the integral $A=\int_2^\infty
P(s)ds$ over the tail of $P(s)$. We recall that $s$ is the energy
differenc $\omega$ in units of the level spacing $\Delta$. Moreover,
$P_{WD}(s)$ and $P_P(s)$ coincide near $s = 2$. The quantity
\begin{equation}
\gamma(W,L) = \frac{A-A_{WD}}{A_P-A_{WD}}
\label{eq_4:4}
\end{equation}
vanishes for WD statistics, and equals unity for the Poisson
distribution. In the thermodynamic limit, 
$\gamma(W,L)$ exhibits a sharp crossover from $\gamma=0$ to $\gamma=1$ as $W$ 
crosses the critical value $W_M$. (For finite $L$ the crossover is
smooth). If a  
critical, size--independent distribution at the MIT were to exist, the 
curves of $\gamma(W,L)$ as functions of $W$ and for fixed but different 
$L$ should all intersect at one critical point $W_M$ (in 
Fig.~\ref{afig3} this point is denoted by $W_c$). Furthermore these  
curves should obey the scaling law
\begin{equation}
\gamma(W,L) = f(L/\xi(W)) 
\label{eq_4:5}
\end{equation}
with $\xi(W)$ the correlation length of the metal insulator
transition.  These expectations were numerically verified in
Ref.~\onlinecite{Shk93}, and estimates for both $W_M$ and the critical
exponent $\nu$ of the localization length were given.

Further numerical confirmation for the scenario developed in
Ref.~\onlinecite{Shk93} was provided by Hofstetter and Schreiber
\cite{Hof93}. In particular it was verified that while for $W < W_M$
and $W > W_M$ the function $P(s)$ tends to the Wigner--Dyson and the
Poisson distribution, respectively, as the system size increases, a
size--independent third distribution emerges for $W = W_M$. This fact
and the critical behavior of the quantity $\gamma$ defined in
Eq.~(\ref{eq_4:4}) were exploited in Refs.~\onlinecite{Hof94,Zha95} to
determine, as done in Ref.~\onlinecite{Shk93}, $W_M$ and the critical
exponent $\nu$ from the spectral statistics alone. The results ($W_M =
16.5$, $\nu = 1.34 \pm .10$ \cite{Hof94} and $W_M = 16.35$, $\nu =
1.45 \pm .08$ \cite{Zha95}) were consistent with each other and with
earlier independent findings \cite{Kra93}. A similar exponent, $\nu =
1.35 \pm 0.10$, was found by Berkovits and Avishai \cite{Ber96a} in a
three--dimensional quantum bond percolation system. This suggests that
the Anderson and the quantum percolation model are in the same
universality class.

Analytical work by Kravtsov {\it et al.} \cite{Krv94} led to a first 
indication that the result (\ref{eq_4:2}) was not completely correct. 
{}From the analytical derivation of the asymptotic behavior of the two--level
correlation function
\begin{equation}
R(s) \propto s^{-2+\gamma} \quad (s \gg 1)
\label{eq_4:6}
\end{equation}
where $\gamma = 1-(\nu d)^{-1}$  it was concluded \cite{Krv94} that
\begin{equation}
\langle \delta N^2\rangle \propto \langle
N\rangle^\gamma
\label{eq_4:7}
\end{equation}
instead of Eq.~(\ref{eq_4:2}). The constants of proportionality in
Eqs.~(\ref{eq_4:6}) and (\ref{eq_4:7}) depend only on $d$ and the
symmetry class (unitary, orthogonal, or symplectic), thus confirming
the universality of the statistics at the MIT. The absence of a term
linear in $\langle N \rangle$ was attributed to the sum rule
$\int_{-\infty}^\infty R(s) ds = 0$ \cite{Krv94,Aro94} (see also
Refs.~\onlinecite{Fre88a,Ley90}), which together with
\begin{equation}
\frac{d\langle\delta N^2\rangle}{d\langle N\rangle} = 
\int_{-\langle N\rangle}^{\langle N\rangle} R(s) ds
\label{eq_4:8}
\end{equation}
implies that the coefficient of the linear term vanishes for sufficiently
large $\langle N \rangle$. With the help of a certain ``plasma'' model
\cite{Krv95}, which re--interprets the distribution of levels as the 
distribution of interacting classical particles in one 
dimension, cf. Eq.~(\ref{E2}), it
was possible \cite{Aro94} to deduce the asymptotic form  
of $P_M(s)$,
\begin{equation}
P_M(s) \propto \exp(- \tilde{c}  s^{2-\gamma}) \quad (s\gg
1) \ ,
\label{eq_4:9}
\end{equation}
with $\tilde{c}$ some constant.
This was again at variance with previous claims \cite{Alt88,Shk93} that 
$P_M(s)$ should be Poisson--like. Moreover, numerical work \cite{Sea94} 
indicated the presence of {\it both}  a linear term and a term proportional 
to $\langle N\rangle^\gamma$ in the number variance. The situation obviously 
called for clarification.

Progress was made when the above--mentioned sum rule was critically examined
\cite{Aro95,Krv95a}. Although valid (in the limit $\langle N\rangle \to 
\infty$) for any finite system size, the sum rule may be violated \cite{Krv95a}
in the thermodynamic limit. In other words, the limits $\langle N\rangle \to 
\infty$ and $L\to\infty$ {\it do not commute}. This insight invalidated the 
claim that there could be no linear term in $\langle\delta N^2\rangle$. The 
number variance at the critical point was therefore now expected to have the 
form
\begin{equation}
\langle\delta N^2\rangle = A \langle N\rangle + B \langle
N\rangle^\gamma \ .
\label{eq_4:10}
\end{equation}
But why had earlier investigations \cite{Alt88} missed the non--trivial 
exponent $\gamma$? Some light was shed on this question by a transparent 
derivation of the relation (\ref{eq_4:6}) by Aronov, Kravtsov and Lerner 
who combined a semiclassical approach with scaling ideas \cite{Aro95a}.
We consider the spectral form factor (cf. the closely related quantity
$b_2(t)$ in Eq.~(\ref{1rc38}))
\begin{equation}
K(t) = \frac{1}{2\pi}\int R(s) \exp(ist) ds \ .
\label{eq_4:11}
\end{equation}
This form factor can be  related to the return probability $P(t)$ for a 
classically diffusing particle \cite{Arg93},
\begin{equation}
K(t) = \frac{2tP(t)}{4\pi^2\beta}  \ .
\label{eq_4:12}
\end{equation}
This relation is quite powerful. We recall that $t_d = \hbar / E_C$ is
the diffusion time through the system. In the ergodic regime $t \gg
t_d$ we have $P(t) = {\rm const.}$, $K(t) \propto t$ and hence $R(s)
\propto 1/s^2$, which corresponds to Wigner--Dyson statistics. In the
diffusive regime $t < t_d$, on the other hand, $P(t) \propto
(Dt)^{-d/2}$, $K(t) \propto t^{1-d/2}/{\cal D}^{d/2}$ and
Eq.~(\ref{eq_4:08}), i.e. the result derived by Altshuler and
Shklovskii \cite{Alt86}, is recovered. As in Ref.~\onlinecite{Alt88},
the transition from the diffusive to the critical regime was made with
the help of the idea that the diffusion constant ${\cal D}$ becomes
scale--dependent. Close to the MIT, the dimensionless conductance
scales as $g(L) = g_M (1+(L/\xi)^{1/\nu})$ with some critical value
$g_M$ \cite{Lee85}. On the other hand, $g \propto {\cal D} L^{d-2}$
and, therefore,
\begin{equation}
{\cal D}(L) \propto g_M L^{2-d} \left[ 1+\left(L/\xi\right)^{1/\nu}\right] \ .
\label{eq_4:14}
\end{equation}
In Ref.~\onlinecite{Alt88} the approximation $g(L)\approx g_M$ was
used, and the square bracket in the relation (\ref{eq_4:14}) was
absent. This leads to a time--independent spectral form factor and,
asymptotically, to $R(s)=0$.  Therefore the predictions in
Ref.~\onlinecite{Alt88} were close to the Poisson case.  With the more
accurate expression (\ref{eq_4:14}) for ${\cal D}(L)$ both time
dependence and the non--trivial exponent $\nu$ enter the form factor
$K(t)$, and Eq.~(\ref{eq_4:6}) is recovered \cite{Aro95a}. In this
way, the controversial issues associated with the two terms in the
number variance (\ref{eq_4:10}) were resolved.

Several numerical investigations added new aspects to the discussion of the 
critical level distribution. Evangelou \cite{Eva94} and Varga {\it et al.} 
\cite{Var95} found that $P_M(s)$ can be parametrized accurately by the function
\begin{equation}
P_M(s) = c_1 s\exp\left( -c_2 s^{2-\gamma} \right) \ ,
\label{eq_4:15}
\end{equation}
with $\gamma = 1- (\nu d)^{-1}$ as above and $c_1$ and $c_2$ two
$\gamma$--dependent normalization constants. This was in good
agreement with both the expected short--range level repulsion
\cite{Shk93} and the asymptotic form (\ref{eq_4:9}). (We note,
however, that Zharekeshev and Kramer \cite{Zha95a} fitted numerical
data with $P_M(s) \propto \exp(-\tilde{c}s)$, with $\tilde{c}$ some
constant, and identified only the linear term in $\langle\delta
N^2\rangle$). Detailed direct numerical calculations of the spectral
two--point function $R(s)$ by Braun and Montambaux \cite{Bra95}
provided the probably most complete picture of the behavior of this
quantity. In the metallic phase both the random matrix regime and (for
the first time) the Altshuler--Shklovskii regime could be identified
(cf. Fig.~\ref{fig12}).  At the mobility edge, the short--range
correlations were found to be weakened and the ensuing power--law
behavior for larger $s$ was consistent with the picture of anomalous
(scale--dependent) diffusion discussed above. Some evidence was found,
however, that the linear term in the number variance might have a
rather small, or even vanishing, coefficient. In
Ref.~\onlinecite{Fei95} the asymptotic form (\ref{eq_4:9}) and the
relation $\gamma = 1- (\nu d)^{-1}$ were confirmed for the center of
the lowest Landau band in a quantum Hall system.

Does the breaking of time--reversal symmetry affect the critical distribution? 
This is a controversial issue. While no effect of a magnetic field has been 
seen by Hofstetter and Schreiber \cite{Hof94a}, a recent investigation by 
Batsch {\it et al.} \cite{Bat96} reported a universal (i.e. size--independent) 
crossover behavior from the critical orthogonal distribution discussed so far 
to a new and distinct critical unitary distribution. A third  type of critical 
distribution was found in two--dimensional systems with spin--orbit scattering
\cite{Sch95,Eva95}. These systems have symplectic symmetry and do exhibit a 
MIT, in contrast to two--dimensional unitary or orthogonal systems.

Chalker, Lerner, and Smith~\cite{Cha96} used a Brownian motion model
to describe the energy level statistics at the mobility edge.
For the model
\begin{eqnarray}
H(\tau) &=& H_0 + \int_0^\tau d\tau' V(\tau',\vec{r}) \ , \nonumber\\
H_0     &=& -\frac{\hbar^2}{2m}\nabla^2 + U(\vec{r}) \ ,
\label{eq_4:16}
\end{eqnarray}
where both $U(\vec{r})$ and $V(\tau,\vec{r})$ are random white noise
potentials, one can derive a Langevin equation governing the parametric
dependence of the eigenvalues $E_n$ on $\tau$ \cite{Cha96},
\begin{equation}
\frac{dE_n(\tau)}{d\tau} = v^2 \sum_{l\ne 0}
\frac{c_{n,n+l}(\tau)}{E_n(\tau)-E_{n+l}(\tau)} + \xi_n(\tau) \ .
\label{eq_4:17}
\end{equation}
Here, $\xi_n(\tau)$ is a random force and the coefficients 
$c_{nm}(\tau)$, which are given by
$c_{nm}(\tau) = L^d \int d^dr 
|\psi_n(\tau,\vec{r})|^2|\psi_m(\tau,\vec{r})|^2$, contain information on 
eigenfunction correlations. By solving Eq.~(\ref{eq_4:17}) approximately,
a generalization of the expression (\ref{eq_4:12}) for the spectral form
factor could be derived, linking spectral statistics with correlations
in the wave functions (through the coefficients $c_{nm}$). The two main
results derived within this approach were an asymptotic power law for
the parametric two--point function $R(s,\lambda \propto \tau^2)$ at the
MIT \cite{Cha96},
\begin{equation}
R(0,\lambda)\propto \lambda^{2/(1+\eta/d)}
\label{eq_4:18}
\end{equation}
and an exact result for the spectral compressibility \cite{Cha96a},
\begin{equation}
\frac{d\langle\delta N^2\rangle}{d\langle N\rangle} = \frac{\eta}{2d} 
\quad (\langle N\rangle \gg 1) \ ,
\label{eq_4:19}
\end{equation}
i.e. the coefficient of the linear term in Eq.~(\ref{eq_4:10}). Here,
$\eta=d-d_2$, where $d_2$ is the multifractal exponent governing the
behavior of the inverse participation ratio. These results demonstrate
that spectral statistics at the MIT are influenced by the multifractal
properties of critical eigenstates. 

In spite of considerable progress, the eigenvalue statistics at the mobility 
edge must, at least in its details, still be considered an open question. 

\subsection{Persistent currents}
\label{pc}

An isolated mesoscopic ring threaded by a magnetic flux $\phi$ carries a 
persistent current $I$ as a thermodynamic equilibrium property. This amazing 
fact comes about as follows. An electron moving once around the ring picks 
up a phase factor $\exp( 2 i \pi \phi / \phi_0)$ where $\phi_0 = h c / e$ is
the elementary flux quantum. Therefore, all observables depend periodically 
on $\phi$. This applies, in particular, to the free energy $F = \beta^{-1} \ln 
{\rm tr} \exp( - \beta H  )$, so that $\partial F / \partial \phi$ differs 
from zero. The equilibrium current $I$ is then given by the thermodynamic 
relation 
\begin{equation}
I = c \frac{\partial F}{ \partial \phi}.
\label{current}
\end{equation}  
The current does not decay in time, hence the name ``persistent current''.

Persistent currents have played a central role in mesoscopic physics, for 
several reasons. First, their existence obviously hinges on the condition 
that the wave function of the system be coherent over a length scale given 
by the circumference $L$ of the ring. This must be true even in the case of 
multiple scattering by impurities. At the beginning of the mesoscopic era, 
the existence of phase coherence in the diffusive regime was a controversial 
issue. The positive answer to this question paved the way to experimental 
and theoretical investigations of a number of interference phenomena. Second, 
Eq.~(\ref{current}) shows the close connection between the persistent current, 
the sensitivity of eigenvalues to boundary conditions, and energy level 
statistics, putting the study of the persistent current at the heart of 
mesoscopic physics. Third, the problem of persistent currents brought to the 
fore the important differences between open and isolated (closed) systems. It 
turned out, in particular, that the grand canonical ensemble commonly used in 
the calculation of transport properties is inappropriate to describe persistent
currents. The grand canonical ensemble has to be replaced by the canonical 
ensemble, and this caused technical problems. Fourth, experimental
values of the  
persistent current could not be explained in the framework of theories for 
independent electrons. This was unexpected since theories neglecting the 
electron--electron interaction had been very successful in modeling transport 
properties of mesoscopic systems. Seemingly, electron--electron interactions 
are more important in closed than in open systems. A satisfactory theory with 
interactions that both reproduces the measured properties of the persistent 
current and explains why interactions are less critical for open systems is 
still lacking. 
   
After explaining the early insights and ideas leading to the theoretical 
prediction of persistent currents, we describe the three experiments performed 
up to now. In a third part we review analytical calculations neglecting 
interactions. In this section, aspects of Random Matrix Theory come into play.
Therefore, we devote considerable space to this discussion. Finally, we briefly
summarize the present status of theories which include the interaction between 
electrons. A more comprehensive review of the topic can be found in 
Ref.~\onlinecite{Mon95a}.

\subsubsection{Early theory, and the three experiments}
\label{pc_1}

It has been realized a long time ago \cite{Lon37,Hun38} that persistent
(dissipationless) currents occur in {\it ideal} rings free of impurities and 
interactions. This fact was rediscovered and reformulated more than twenty 
years later \cite{Bye61,Blo65}. Expanding on our introductory remarks above,
we consider an ideal one--dimensional ring threaded by a magnetic flux $\phi$. 
The magnetic field can be gauged away from the Hamiltonian \cite{Bye61} at the
expense of introducing a twisted boundary condition for the wavefunction, 
$\psi(x+L) = \exp(2i\pi\phi/\phi_0)\psi(x)$, where $x$ is the coordinate along 
the ring. The eigenfunctions at zero field are given by $\exp(2i\pi mx/L)$, 
and the corresponding eigenvalues are $(\hbar^2/2m_e)(2\pi/L)^2 m^2$, where 
$m$ is the quantum number of the $z$ component of angular momentum 
(perpendicular to the plane of the ring). At finite field we have to replace 
$m$ by $m-\phi/\phi_0$. This leads to a quadratic dependence of the 
single--particle energies on magnetic flux. Equation~(\ref{current})
shows that at  
zero temperature, the persistent current is given by the sum over all occupied 
single--particle levels of the flux derivatives of the single--particle 
energies $E_m$, 
\begin{equation}
I = - \sum_m \frac{\partial E_m}{\partial\phi} \ .
\label{eq_3:01}
\end{equation}
This relation explicitly shows the close connection between persistent 
currents, energy level statistics, and the sensitivity of the spectrum to an 
external perturbation alluded to above. 

Interest in this topic was greatly enhanced by a seminal paper of B\"uttiker, 
Imry, and Landauer \cite{Bue83}. These authors claimed that persistent currents
exist in normal--metal rings even at finite temperature and in spite of 
multiple elastic scattering. Their work was followed by a series of papers on 
1d \cite{Bue85,Che88} and multichannel \cite{Che88a,Che89,Ent89} rings with 
non--interacting electrons. These investigations tried to assess the chances to 
actually observe persistent currents in experiments by considering the 
influence of channel number, temperature and disorder. The main results of 
these papers can be summarized as follows. In ideal 1d rings with a fixed 
number $N$ of particles, the persistent current is a sawtooth--shaped function 
of $\phi$ jumping from $-I_0 = -ev_F/L$ to $+I_0$ ($v_F$ being the Fermi 
velocity) at even ($N$ even) or odd ($N$ odd) multiples of $\pm \phi_0/2$. 
The typical current $I_{\rm typ}= \langle I^2 \rangle^{1/2}$ is given by
$I_0 \sqrt{M}$, i.e. it grows with the number $M$ of channels. The relevant 
energy scale for temperature effects is the single--particle (rather than the 
many--particle) level spacing $\Delta$. In the presence of disorder, the Fermi 
velocity $v_F$ in the expression for $I_0$ is replaced by the diffusion 
velocity $L / t_d$, and the dependence on the channel number vanishes, $I_{\rm 
typ} = I_0 \ell/L$. The average current, on the other hand, was found to decay 
exponentially with $L$ on the scale of the elastic mean free path $\ell$.

Bouchiat, Montambaux {\it et al.} \cite{Bou89} showed that this last
assertion is incorrect, for a very fundamental reason. For closed
systems in the mesoscopic regime, it is of vital importance to keep
the particle number in the system exactly constant as one averages
over the disorder configurations.  Thus it is necessary to employ the
canonical ensemble. The exponentially small currents found in
Ref.~\onlinecite{Che89} were shown to be an artifact produced by the
use of the grand canonical ensemble, i.e. by keeping the chemical
potential (rather than the particle number) fixed in the averaging
process. The thermodynamic limit does not apply to mesoscopic rings,
and the two ensembles are not equivalent. By a combination of
analytical arguments and extensive numerical simulations, the
following results for an ensemble of isolated rings were obtained
\cite{Bou89}. The typical current $I_{\rm typ}=(\langle
I\rangle)^{1/2}$ is independent of channel number $M$ while the
average current decreases with increasing disorder and channel number
like $\langle I\rangle = I_0 (\ell/ML)^{1/2}$. Furthermore, the
persistent current was found to be periodic in the flux with period
$\phi_0/2$. This period halving comes about because the ensemble
considered contained rings with even and odd particle numbers. The
most important result of Ref.~\onlinecite{Bou89} was the comparatively
large value for the average current. Experimental confirmation of the
theoretically predicted phenomenon of persistent currents suddenly
seemed to be within reach.

\begin{figure}
\centerline{
\psfig{file=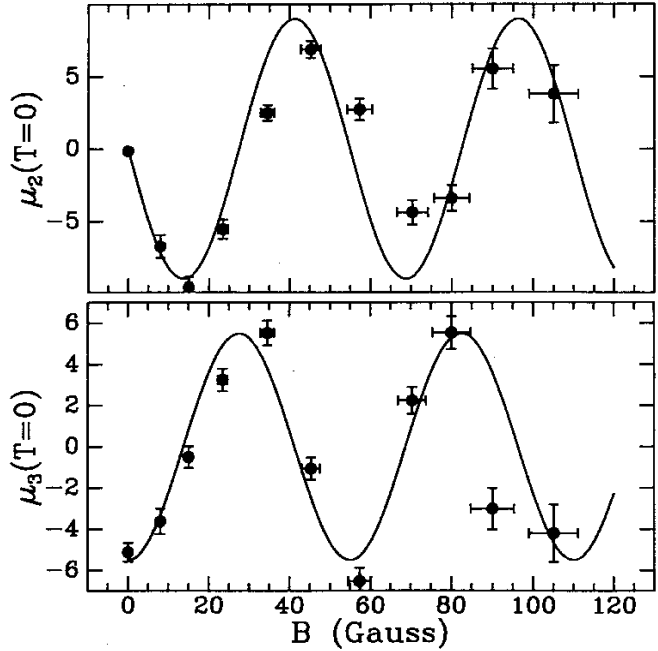,width=3.5in}
}
\caption{
  The second (top) and third (bottom) zero--temperature
  harmonic of the magnetic moment (induced by persistent currents in
  an ensemble of $10^7$ cop\-per rings) versus the dc magnetic 
  field. The Fourier decomposition was made with respect to the
  frequency of a small ac component in the field.
  One flux quantum through a single ring corresponds to 130 G. The
  magnitude of the oscillatory moment is estimated to be roughly
  $10^{-15}$ Am$^2$. Taken from Ref.~\protect\onlinecite{Lev90}.
}
\label{fig13}
\end{figure}

The first experiment on persistent currents was performed by Levy {\it
  et al.} \cite{Lev90} who used an ensemble of $10^7$ copper rings
mounted on a single wafer. For technical reasons, these rings were
actually squares, with a circumference of $L = 2.2 \mu$m. With $L \gg
\ell$, the experiment probed the diffusive regime. A magnetic flux
with a small, slowly varying ac component was applied to the sample
and the resulting magnetization measured. The ac component allowed for
an effective background rejection. The measured current was indeed
$\phi_0/2$--periodic, in agreement with the numerical simulations
\cite{Bou89}, see Fig.~\ref{fig13}.  The average current per ring was
$0.4 \ nA$, corresponding to $3 \times 10^{-3} ev_F/L$. This first
observation of persistent currents was a striking confirmation of the
theoretical ideas developed earlier. However, the current was roughly
one order of magnitude larger than the values estimated in
Ref.~\onlinecite{Bou89}.

Two more experiments followed \cite{Cha91,Mai93}. Both provided
interesting additional information that went beyond
Ref.~\onlinecite{Lev90}. Chandrasekhar {\it et al.} \cite{Cha91}
investigated three {\it single} gold loops. Two of these actually were
rings with diameters of $2.4$ $\mu$m and $4.0$ $\mu$m, respectively.
The third loop was a rectangle of dimensions $1.4$ $\mu$m $\times 2.6$
$\mu$m. Transport measurements for gold films fabricated like the
loops gave an elastic mean free path $\ell=70$ nm. Therefore, all
three probes were in the diffusive regime. The observed current was
flux--periodic with period $\phi_0$. The measured amplitudes at a
temperature of $4.5$ mK were in the range $0.2-3.0$ $ev_F/L$. The
absence of period halving is an immediate consequence of measuring the
response of single rings. No averaging over even and odd particle
numbers as in the previous experiment was involved. The large
amplitudes, however, pose a serious problem. Even when compared with
the theoretical estimates for the typical current cited above, these
values are larger by one to two orders of magnitude. In comparison
with the theoretical {\it average} current the discrepancy becomes
even worse.

The third persistent current experiment by Mailly, Chapelier, and Benoit 
\cite{Mai93} poses less of a problem for theory. In this experiment, the 
magnetic response of a single mesoscopic GaAlAs/GaAs ring was measured. This 
ring had an internal diameter of $2$ $\mu$m and an elastic mean free path 
$\ell = 11\mu$m. It was therefore in the ballistic regime. In contrast to the 
previous two experiments, where the channel number $M$ was of the order of 
$10^4$, $M$ was quite small, $M=4$. A very interesting feature of the 
semiconductor ring was the possibility to connect it via an electrostatic gate 
to two external leads. A second gate was placed on one of the arms of the ring,
making it possible to suppress all interference effects. With this 
unprecedented control over the ring it was possible to measure transport and 
thermodynamic properties in the same system. The current observed in the ring 
had the value $I_0 = ev_F/L$, exactly as expected for an almost clean sample 
with very few channels.

In summary, the persistent currents predicted theoretically were observed 
in three independent experiments. In the diffusive regime, the measured 
values exceeded theoretical expectations by at least one to two orders of 
magnitude. This poses a serious challenge for theory.

\subsubsection{Analytical theory for non--interacting electrons}
\label{pc_2}

The need to calculate the disorder average for the persistent current
within the canonical (rather than the grand canonical) ensemble
\cite{Bou89} led to three almost simultaneous publications by Schmid
\cite{Sch91}, v. Oppen and Riedel \cite{Opp91}, and Altshuler, Gefen,
and Imry \cite{Alt91}. The calculations for the average persistent
current used the impurity perturbation technique. In
Refs.~\onlinecite{Sch91,Alt91}, the constraint of constant particle
number $N$ was met by letting the chemical potential fluctuate, $\mu =
\bar{\mu} + \delta\mu$, and by choosing $\delta\mu$ in such a way that
-- to first order in $\delta \mu$ -- $N$ remained fixed. In
Ref.~\onlinecite{Opp91}, this approximation was avoided by relating the
average current to the typical fluctuation of a single level, and by
calculating the latter quantity using standard Green function
techniques. Strictly speaking, the perturbative results derived in
\cite{Sch91,Opp91,Alt91} are divergent at zero temperature $T$ and for
vanishing flux $\phi$. However, this divergence is regularized either
by finite values of $T$, or by the energy scale $1/\tau_\varphi$ set
by inelastic processes. Both of these scales can realistically be
assumed to be at least of the order of the single--particle level
spacing $\Delta$. With this assumption, all three investigations
yielded for the first even harmonics $I_m$ of the persistent current
the values
\begin{equation}
I_m \approx \frac{\Delta}{\phi_0} \approx \frac{ev_F}{LM} =
\frac{I_0}{M},
\label{eq_3:1}
\end{equation}
where $\Delta \approx h v_F/LM$. At first sight this seemed to
contradict earlier numerical results \cite{Bou89} according to which
the average current behaves as $\langle I\rangle \approx I_0
(\ell/LM)^{1/2}$. However, it was pointed out in
Ref.~\onlinecite{Alt91} that the first $\sqrt{E_C / \Delta}$ harmonics
contribute to the current provided the regularizing cutoffs discussed
above are of order $\Delta$. (Here, as always, $E_C =\hbar {\cal D} /
L^2$ is the Thouless energy and ${\cal D}$ the diffusion constant).
With $E_C / \Delta = g = M \ell / L$ and $g$ the dimensionless
conductance, these considerations led to a maximal amplitude of
\begin{equation}
I_{\rm max} \approx \frac{\sqrt{\Delta E_C}}{\phi_0} \approx
\frac{ev_F}{L} \sqrt{\frac{\ell}{ML}} \ ,
\label{eq_3:2}
\end{equation}
in good agreement with Ref.~\onlinecite{Bou89}. Unfortunately, these
analytical results did not remove the discrepancy between theory and
experiment: As stated above, the result (\ref{eq_3:2}) is too small by
one to two orders of magnitude.

Altland {\it et al.} \cite{All92} used Random Matrix Theory and the 
supersymmetry method to calculate the persistent current
non--perturbatively at  
zero temperature in the model of independent electrons. In this work, a novel
approach was used to calculate the impurity--ensemble average within the 
canonical ensemble. The condition that all rings in the ensemble carried an 
integer (but not the same) number of electrons was imposed. The resulting 
expressions turned out to be equivalent to those obtained from the first--order
expansion in $\delta\mu$ referred to above. The non--perturbative treatment 
removed the divergence typical for perturbative results and yielded the 
current at $T=0$, where all cutoffs lose their physical significance. This 
work \cite{All92} also contained the first calculation of the crossover from 
orthogonal to unitary symmetry within the supersymmetry formalism. The 
crossover occurs as the flux increases from $\phi = 0$ (where orthogonal 
symmetry applies) to non--zero values. In keeping with the scope of our 
review, we discuss this application of Random Matrix Theory in more detail 
than previous work.

The persistent current $I(\phi)$ can be expressed as follows,
\begin{equation}
I(\phi) = -2\sum_{n=1}^N \frac{\partial E_n}{\partial\phi} = 
-2 \int_0^E dE' \, \left[ \sum_{n=1}^\infty
\frac{\partial E_n}{\partial\phi} \delta(E'-E_n) \right].
\label{eq_3:3}
\end{equation}
The factor 2 accounts for spin degeneracy. To account for the constraint of 
fixed particle number $N$, the upper integration limit $E$ has to be adjusted 
as a function both of flux and of disorder realization. This is impossible in 
practice. The difficulty is avoided as follows. We recall that the first 
experiment \cite{Lev90} used a wafer with $10^7$ rings. It is natural to 
expect that the particle numbers on these rings, although integer, were 
different. This suggests averaging $I$ over a range $K$ of particle numbers 
centered at some value $N_0 \gg K$. The associated eigenvalues $E_N$ defining 
the upper limit of integration lie in an energy interval $S$,
\begin{equation}
I(\phi) = -\frac{2}{K} \int_S dE \int_0^E dE' \, 
\left[\sum_{n=1}^\infty
\frac{\partial E_n}{\partial\phi} \delta(E'-E_n) \right]
\left[\sum_{N=1}^\infty \delta(E-E_N-\epsilon) \right].
\label{eq_3:4}
\end{equation}
Here, $E_N$ is the energy for which there are exactly $N$ electrons on the 
ring. The second $\delta$ function with infinitesimal positive increment 
$\epsilon$ in Eq.~(\ref{eq_3:4}) guarantees the discreteness of the
particle number in the ensemble of rings. At first sight nothing has
been gained, the problem of adjusting $E$ has just been replaced by the
problem of adjusting the full interval $S$. However, as is explained
in detail in \cite{All92}, the dependence of $S$ on flux and disorder
can be neglected for $K \gg 1$. Expressing the $\delta$ functions in 
Eq.~(\ref{eq_3:4}) in terms of Green functions we get for the 
ensemble--averaged persistent current
\begin{eqnarray}
\langle I(\phi)\rangle &=& -\frac{2}{K} \int_S dE \int_0^E dE' \,
g(E,E') \nonumber\\
g(E,E') &=& 
\overline{
\frac{1}{4\pi^2} {\rm tr}
\left[ \frac{\partial H}{\partial \phi} \frac{1}{E^{\prime +} - H}
\right]
{\rm tr} 
\left[ \frac{1}{E^- - H} \right] + {\rm c.c.} 
}
\ .
\label{eq_3:5}
\end{eqnarray}
This is equivalent \cite{All92} to the expression $\langle I\rangle =
\Delta \partial_\phi \overline{\left( \int_0^\mu dE \, {\rm
    tr}[\delta(E-H)] \right)^2 }$ employed in Ref.~\onlinecite{Alt91}
(apart from a sign error in Ref.~\onlinecite{Alt91}).

In Ref.~\onlinecite{All92}, the model Hamiltonian (cf. Sec.~\ref{loc}),
\begin{eqnarray}
& &H(\phi) = \frac{1}{2m} \left( \vec{p} - e\vec{A} \right)^2 +
V(\vec{x}) \nonumber\\
& & \overline{ V(\vec{x}) } = 0; \quad \overline{ V(\vec{x})
V(\vec{y}) } = \frac{1}{2\pi\nu\tau} \
\delta (\vec{x}-\vec{y}) \ 
\label{eq_3:6}
\end{eqnarray}
was used. Here, $\nu = 1/\Delta V$ is the density of states (with $V$ the 
volume of the ring) and $\tau$ is the elastic scattering time. (Instead of 
a Gaussian random white noise potential as in Eq.~(\ref{eq_3:6}), the  
Hamiltonian could have been chosen as a chain of coupled GOE matrices, see
the IWZ model in Sec.~\ref{quasi1d_3}. Both models lead to the same
non--linear  
$\sigma$ model). Introducing a supersymmetric generating functional for
the Green functions as well as for the flux derivatives of $H$, one can write 
the average current as
\begin{eqnarray}
&&\langle I(\phi) \rangle = \frac{1}{8\pi^2 K} \int_S dE \int_0^E dE' \,
\Bigg[ \left( \partial_{\phi_\chi} - \partial_{\phi_s}\right) \partial_j
\overline{ Z[\phi_s,\phi_\chi,j] }
\bigg\vert_{j=0;\phi_s=\phi_\chi=\phi} \nonumber\\
&& \hspace{8cm} + \ {\rm c.c.} \Bigg] \ , \nonumber\\
&&Z[\phi_s,\phi_\chi,j] = \int d[\Psi] \,
\exp\left(\frac{i}{2} \int d^2x \, \Psi^\dagger(\vec{x})  L
(D+jI_2) \Psi(\vec{x}) \right) \ .
\label{eq_3:7}
\end{eqnarray}
The definitions of the supersymmetric quantities appearing in
Eq.~(\ref{eq_3:7}) can be found in Ref.~\onlinecite{All92}. For the
present purpose the following explanations hopefully suffice, cf.
Sec.~\ref{susy}. The inverse propagator $D$ is basically given by
$E-H$ with suitable additional matrix structure to account for (i) the
different energy arguments $E, E'$ and (ii) the necessary extension to
the $8$--dimensional superspace. The quantities $\phi_s,\phi_\chi$ are
the flux arguments of those ``copies'' of $H$ which are associated
with the Boson--Boson block and the Fermion--Fermion block,
respectively. The matrix $I_2$ is diagonal with elements $1$ and $-1$.
After the usual steps (averaging, Hubbard--Stratonovich
transformation, gradient expansion) we obtain the following
non--linear $\sigma$ model,
\begin{eqnarray}
\langle I(\phi) \rangle &=& -\frac{1}{8\pi^2 K}\frac{\pi^2\nu^2 {\cal D} \ e}
{8L} \int_S dE \int_0^E dE' \, \int d[Q] \, {\rm trg}[QI_2] \ 
{\rm trg}[\nabla Q \ [\hat{K},Q]] \nonumber\\
& & \quad\quad \times \
\exp\left[ \frac{\pi\nu}{8} \int \, 
 {\rm trg}[ {\cal D} \ (\nabla Q)^2 + 2\Delta E \ LQ ] d^3r
\right] \ ,
\label{eq_3:8}
\end{eqnarray}
where $\hat{K}$ is a suitable source matrix, where $\Delta E = E-E'$, and where
$\nabla = -i\vec{\partial} + (e\phi/L) [\tau_3,\cdot] e_\theta$ is a covariant 
derivative with the gradient $\vec{\partial}$ and the tangential unit vector 
$e_\theta$. We note that Eq.~(\ref{eq_3:8}) is of the form of 
Eq.~(\ref{eq_4:01}) except for the redefinition $Q \rightarrow -iQ$. The matrix
$\tau_3$ = diag $(1,-1)$ introduced at the end of Sec.~\ref{spec_1} breaks the 
symmetry between those parts of the $8\times 8$ supermatrix $Q$ that are 
connected by the operation of time reversal. As in the case of the spectral 
fluctuations discussed there, this matrix reflects the necessity to calculate 
the crossover from orthogonal to unitary symmetry for flux values near $\phi = 
0$. The divergence in the perturbative approaches discussed above originates 
from this crossover, more precisely from the ``zero mode'' in 
Eq.~(\ref{eq_3:8}) (i.e. from the contribution with $Q\equiv Q_0 =
{\rm const.})$ which cannot be treated adequately in a perturbative framework. 
Actually, symmetry breaking in this zero mode also poses a technical
problem in the  
context of the supersymmetric formulation. The problem arises from the term 
${\rm trg} ([\tau_3 Q_0]^2)$ in the covariant derivative, cf. the exponent in 
Eq.~(\ref{eq_3:8}). This problem was overcome by a trick which is a precursor  
of the parametrization of the $T$ matrix in Eq.~(\ref{eq_4:023}). Focussing 
attention on the zero mode, one gets
\begin{equation}
\langle I_0(\phi)\rangle = -\frac{1}{8\pi^{3/2}}
\frac{\sqrt{E_C\Delta}}{\phi_0} K(y)
\label{eq_3:9}
\end{equation}
with
\begin{eqnarray}
K(y) &=& \sqrt{y} \int_0^\infty dx \, \int d[Q_0] \,
\exp\left( y \ {\rm trg} \ [Q_0\tau_3Q_0\tau_3] +
\frac{x}{4} \ {\rm trg}\ [Q_0 L] \right)    \nonumber\\
& & \quad\quad
{\rm trg}\ [Q_0I_2] \left({\rm trg}\ [Q_0\tau_3 Q_0 \hat{K}] - 4 \right) +
{\rm c.c.} \ ,
\label{eq_3:10}
\end{eqnarray}
where $y=\pi(E_C/\Delta) \ (\phi/\phi_0)^2$ and $x = \pi \ \Delta
E/\Delta$.  After the final integration over $Q_0$ the result can be
well parametrized by the expression $\langle I\rangle =
(\sqrt{E_C\Delta}/\phi_0) \ (\kappa_1\sqrt{y}/(1+\kappa_2 y))$ with
$\kappa_1\approx 2.25$ and $\kappa_2\approx 12.40$. This gives a
maximal amplitude of $I_{\rm max} \approx 0.3
\sqrt{E_C\Delta}/\phi_0$, in agreement with the estimates given in
Ref.~\onlinecite{Alt91}. The modes with spatially varying $Q$ fields
can be treated perturbatively and added to the zero mode result to
obtain a full picture of the current \cite{All92}.

In a closely related study, Efetov and Iida \cite{Efe91} calculated the 
persistent current from a dynamic response. This point of view has the 
advantage that the dynamic response is not sensitive to the use of the  
canonical or the grand canonical ensemble. Within the model (\ref{eq_3:6}) of 
non--interacting electrons
in a ring threaded by a magnetic flux and with a slowly varying field
\begin{equation}
B(t) = B_0 + B_\omega \cos(\omega t),
\label{eq_3:11}
\end{equation}
the density of the oscillating current can be written in linear
response as
\begin{equation}
j_\omega = K(\omega) A_\omega \ .
\label{eq_3:12}
\end{equation}
Here, $K(\omega)$ is the current--current correlation function and
$A_\omega$ is the vector potential corresponding to the magnetic field
$B_\omega$. At first sight it is plausible that the current calculated from 
Eq.~(\ref{eq_3:12}) in the limit of vanishing frequency $\omega \to 0$ should 
be basically equivalent to the thermodynamic current (see, however, the 
discussion below). The quantity $K(\omega)$ is essentially given by the
correlator
\begin{equation}
\tilde{R}(\omega) = \overline{p_r G_E^+(r,r') p_{r'} G_{E-\omega}^-(r',r)
} \ ,
\label{eq_3:13}
\end{equation}
where $p_r = (1/m)[-i\nabla + eA]$ is the velocity operator and
$G_E^\pm$ the retarded (advanced) Green function. This correlator can
be expressed in the framework of the supersymmetry method as a
functional average $\langle \ldots \rangle_Q$ over a field of
$Q$ matrices \cite{Efe91},
\begin{equation}
\langle \ldots \rangle_Q = \int (\ldots ) \exp\left( -S[Q] \right) DQ
\label{eq_3:14}
\end{equation}
with
\begin{equation}
S[Q] = -\frac{\pi\nu}{8} \int {\rm trg} \left(
D [\nabla Q -ieA[Q,\tau_3]]^2 + 2i\omega L Q \right) dr \ .
\label{eq_3:15}
\end{equation}
This is, of course, absolutely analogous to Eq.~(\ref{eq_3:8}).
Without specifying the precise form of the terms in the angular
brackets in Eq.~(\ref{eq_3:14}) we confine ourselves to saying that
the procedure is analogous to the one explained above in the context
of the thermodynamic approach. Again, the calculation may be
restricted to the zero mode, i.e. to a spatially constant $Q$ field.
The final result for the current has a maximal amplitude of the order
of $\sqrt{E_C\Delta}/\phi_0$, in agreement with
Ref.~\onlinecite{All92}. However, the shape of the current as a
function of the magnetic flux differs considerably from the one found
in the thermodynamic calculation.

The relation between the thermodynamic and the dynamic approaches was
addressed by Kamenev and Gefen \cite{Kam93}. These authors
investigated a whole variety of averaging procedures, differing (i) in
the initial preparation of the system under study and (ii) in the
constraints imposed while the external field is varied adiabatically.
According to Ref.~\onlinecite{Kam93}, the thermodynamic calculation
corresponds to the situation where both the initial occupation of
levels and the evolution of the system with varying flux are
determined within the canonical ensemble. The dynamic response, on the
other hand, corresponds to a particular procedure: The initial
occupation of levels is defined grand--canonically but is then held
fixed as determined by the canonical ensemble. Hence there is no
reason why the persistent currents calculated in both cases should
coincide exactly. On physical grounds, however, one does not expect
drastic (order--of--magnitude) discrepancies.

The same authors also investigated the
magnetoconductance \cite{Kam95a} and the conductance distribution
\cite{Kam95b} of isolated rings in the quantum regime, where the level
widths are smaller than the level spacing.
Significant differences between the canonical
and the grand canonical ensemble were found.

\subsubsection{Theories with electron--electron interaction}
\label{pc_3}

After it had generally been recognized that the experimental values for the 
persistent current could not be explained in the framework of theories 
which neglect the electron--electron interaction, numerous investigations 
appeared which addressed the role of these interactions. A detailed discussion 
of this vast body of work is completely beyond the scope of the present review 
which focusses on Random Matrix Theory. We restrict ourselves to a brief 
enumeration of relevant work. The difficulty in attempts to treat both 
disorder {\it and} interaction lies in the fact that disorder alone poses
severe theoretical problems (witness the present section) while interactions 
add the full complexity of the non--relativistic many--body problem.

In an early paper by Ambegaokar and Eckern \cite{Amb90b} (actually
preceding the full solution of the non--interacting problem
\cite{Sch91,Opp91,Alt91,All92}), the grand potential of a mesoscopic
normal metal ring including electron--electron interactions was
calculated to first order in the interaction. Evaluating both the
Hartree and Fock diagrams, the authors found that the average
persistent current is parametrically given by $E_C/\phi_0$, in
apparent agreement with Ref.~\onlinecite{Lev90}. The same result was
later derived in Ref.~\onlinecite{Sch91} from the condition of local
charge neutrality.  Also, the temperature dependence of the result,
see Ref.~\onlinecite{Amb90a}, was closer to experiment than obtained
in theories without interactions. However, diagrams of higher order in
the interaction renormalize the Cooper channel and were estimated to
reduce the result to roughly one order of magnitude below measured
values \cite{Eck91}. A calculation of the typical current along
similar lines \cite{Eck92} resulted in a very large value, $I_{\rm
  typ} \approx ev_F/L$, but was later corrected \cite{Smi92} to
$I_{\rm typ}\approx ev_F\ell/L^2 = E_C/\phi_0$. Work using density
functional theory \cite{Arg93a} also led to relatively large currents
of order $E_C/\phi_0$, but the absence of renormalizing, higher--order
corrections could again not be established.  Recently, the results of
Ref.~\onlinecite{Amb90b} could be reproduced within a simple
(non--diagrammatic) Hartree--Fock picture \cite{Mon95}.

In a series of papers, the persistent current problem was approached
from the point of view of Luttinger liquid theory. From very early
work \cite{Mat74} it was known that in a Luttinger liquid the effect
of a single impurity is exponentially enhanced (suppressed) for
repulsive (attractive) interactions.  It was possible to generalize
the formalism to include the case of twisted boundary conditions
necessary to describe a clean mesoscopic ring in a magnetic field
\cite{Los92} (see also Ref.~\onlinecite{Sch68}).  Nevertheless the
Luttinger liquid, predicting enhanced impurity scattering for
interacting electrons, did not seem to be a promising candidate for
explaining the discrepancy between theory and experiment. Interacting
Fermions in clean, one--dimensional rings were also investigated in
Ref.~\onlinecite{Kus92} by means of the Bethe ansatz. In a refined
treatment \cite{Gog94} of the effect of impurities which improved on
the work of Ref.~\onlinecite{Mat74} by going beyond simple
perturbation theory for the disorder, spinless Fermions in one
dimension and with repulsive interaction were investigated. An
algebraic (rather than exponential) decay of the current with the ring
circumference was found.  On the other hand, Monte--Carlo simulations
\cite{Mor94} indicated that for a strong impurity potential, the
current may be enhanced. A significant new aspect was contributed to
the discussion by Giamarchi and Shastry \cite{Gia95}, who took into
account the spin degree of freedom. In earlier work \cite{Gia87} on
the localization properties of one--dimensional disordered electron
systems a combination of the replica trick (to average over disorder)
and renormalization group methods was used. The renormalization group
equations derived in Ref.~\onlinecite{Gia87} were employed in
Ref.~\onlinecite{Gia95} to prove that repulsive interactions can {\it
  enhance} the persistent current, provided the carriers have spin
(see Ref.~\onlinecite{Kam95}, and also
Refs.~\onlinecite{Hae96,Chi96}). This work emphasized the important
role of the additional spin degree of freedom and (indirectly) gave
rise to the interesting question whether a second channel (i.e., a
model beyond the strictly one--dimensional Luttinger liquid) would
play a similar role. It was pointed out in Ref.~\onlinecite{Mor95}
that the enhancement just mentioned does not lead back to the value
for the clean ring.

Another approach \cite{Kop93} invoked the classical electromagnetic energy 
associated with charge fluctuations and long--range Coulomb forces to derive a 
very large average current of the order of $ev_F/L$. The critique \cite{Vig94} 
leveled against this work basically points out that the assumed coexistence 
of large fluctuations and long--range forces is highly implausible.

A new mechanism explaining why the Coulomb interaction might indeed
counteract the influence of the disorder potential was suggested and
investigated in Ref.~\onlinecite{Mue93}. The key point was a
phase--space argument: By mixing many--particle states in a rather
large energy interval, the Coulomb interaction redistributes the
strength of the impurity potential to high--lying configurations, thus
reducing the influence of impurity scattering on the ground state
relevant for the persistent current. Numerical simulations supported
this claim, but the relation of the model used to real physical
systems remained controversial.

Much important work used computer simulations of lattice models of
interacting particles. Small one--dimensional systems of spinless
Fermions could be treated by numerical diagonalization of the
Hamiltonian, but no enhancement of the persistent current was found
\cite{Abr93,Ber93,Bou94,Bou94a}. In Ref.~\onlinecite{Abr93} this fact
was (at least partly) attributed to the onset of the Mott--Hubbard
transition which sets in as the interaction strength increases.  These
results were confirmed in Ref.~\onlinecite{Kat94} in the framework of a
Hartree--Fock approach. Similar conclusions were also reached in a
recent study using the density--matrix renormalization--group
algorithm \cite{Scm96}.  Numerical investigations of one-- and
three--dimensional rings \cite{Ram95} which included first--order
corrections due to interactions, found (among other results) agreement
with Ref.~\onlinecite{Amb90b} in the diffusive regime.

Exact diagonalization is possible only when the system sizes are
severely restricted. It is an important task to devise reliable
numerical approximation schemes capable of dealing simultaneously with
interactions and disorder. One step in this direction was the
development of the self--consistent Hartree--Fock method (SHF)
\cite{Bou95,Kat95}. Through a self--consistency condition,
higher--order terms in the interaction are effectively resummed.
Therefore, the method goes beyond the analytical calculations in
Ref.~\onlinecite{Amb90b}. With this method, an enhancement of the
persistent current has been found in three--dimensional rings and for
long--range interactions \cite{Kat95}. A similar enhancement was
observed after exact diagonalization of small two--dimensional
cylinders with long--range interactions in Ref.~\onlinecite{Ber95}.
For short--range (on--site and nearest--neighbor) interactions in two
dimensions, on the other hand, spin again seems to be an important
factor \cite{Bou95a}: the current was found to be strongly suppressed
in the spinless case but to be significantly enhanced once spin was
included.

Why do interactions affect the persistent current but not the conductance? 
This very important question was addressed by Berkovits and Avishai 
\cite{Ber96}. The authors expressed the conductance as the compressibility of 
the system times the flux derivative of the ensemble--averaged persistent
current. This formula leads to the plausible scenario that an enhancement of 
the persistent current is compensated by a corresponding reduction of the 
compressibilty, with zero net effect on the conductance.

\subsection{Transport in quasi one--dimensional wires}
\label{quasi1d}

The main experimental information on mesoscopic systems is derived from 
transport studies: Each such system is connected to voltmeters and/or
external voltage differences by a number of leads, and both the
resistance tensor and, by implication, the conductance tensor are
measured. Historically, these experiments raised questions of a
conceptual nature. What is the meaning of a transport coefficient like
the conductance in a system where inelastic effects and, hence, energy
dissipation are absent by definition? This is not the place to review
the heated discussion of nearly a decade ago. An overview over this
subject is given in an article by Stone and Szafer \cite{Sto88}. 
Suffice it to say that there are two approaches to quantum transport
in mesoscopic systems, the Kubo formula \cite{Kub56} based on linear
response theory, and the scattering approach pioneered by Landauer  
\cite{Lan57}, see Eqs.~(\ref{eq_2:1}) and (\ref{eq_2:2}) below. The problem
of dissipation in mesoscopic transport was mainly discussed in the
framework of Landauer's ideas. The key issue here were the so--called
reservoirs, fictitious sources and sinks for electrons. These  
reservoirs were introduced to represent the macroscopic contacts attached to 
the mesoscopic probe. Energy dissipation occurs in, and irreversibility is 
caused by, these reservoirs. In this way, the conductance of a mesoscopic 
system could be meaningfully defined even though neither dissipation nor any 
irreversible processes occur in that system itself. This result can be seen 
as another manifestation of quantum coherence in mesoscopic systems: The 
wave function is coherent in the entire domain connecting the reservoirs.

The study of electron transport in quasi one--dimensional (1d) wires is 
particularly interesting, for several reasons. In contrast to strictly 
1d systems, these wires possess a genuine diffusive regime. At the same 
time, they share many properties with strictly 1d systems. This is true, 
in particular, for the applicability of non--perturbative analytical methods.
In the last decade, both the DMPK equation \cite{Dor82,Mel88} and the 
non--linear $\sigma$ model \cite{Efe83,Iid90} have been successfully used 
to describe such wires. This has led to an (almost) complete
understanding of the physical properties of these wires for all length
scales $L$.

By definition, the transverse dimensions of quasi 1d wires are large enough 
to accomodate a number ($\Lambda$) of transverse quantized modes (often 
referred to as channels) below the Fermi energy. Typically, $\Lambda$ is of 
the order $10^2$ or $10^3$, hence $\Lambda \gg 1$. At the same time, the 
transverse dimensions are small enough (i.e., of the order of the elastic 
mean free path $\ell$) so that no diffusion takes place in the transverse 
direction. The ballistic regime $L < \ell$ and the localized regime $L > \xi$
(with $\xi = \Lambda \ell$) are separated by the diffusive regime $\ell < L 
< \xi$ where Ohm's law applies.

In a first part we briefly discuss the phenomenon of universal conductance
fluctuations. This phenomenon triggered most of the interest in mesoscopic 
physics and quasi 1d systems. Subsequently, we describe the two 
non--perturbative approaches mentioned above. Both originate from or are at
least intimately related to the concept of Random Matrix Theory. Lastly, we 
review the development leading to the insight that these two approaches, in 
spite of their considerable technical dissimilarity, are in fact equivalent.

\subsubsection{Universal conductance fluctuations}
\label{quasi1d_1}

\begin{figure}
\centerline{
\psfig{file=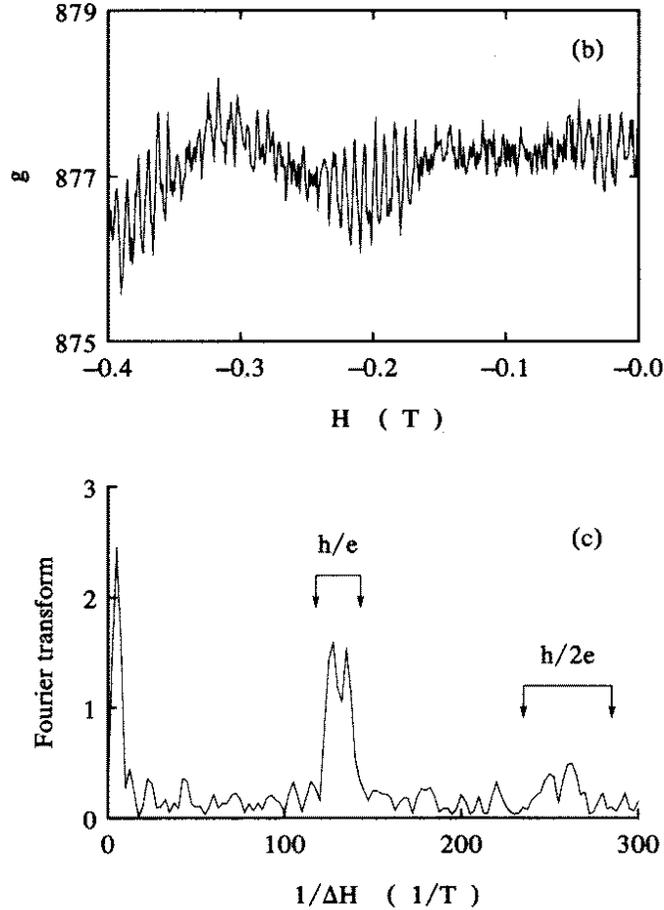,width=3.5in}
 }
\caption{
Top: The dimensionless conductance $g$ (i.e. the conductance in units
of $e^2/h$) of a mesoscopic gold loop versus magnetic field. Bottom:
The Fourier transform of the data in the top figure. Peaks in the
spectral weight are abserved near $1/\Delta H = 0$ (UCF), $130/T$
(principal Aharonov--Bohm periodicity), and $260/T$ (higher harmonic).
Taken from Ref.~\protect\onlinecite{Wash91}.
} 
\label{afig4}
\end{figure}

In 1984, peculiar fluctuations in the conductance of small metallic
wires \cite{Umb84,Blo84} and of small metallic rings
\cite{Web85,Was85,Cha85,Umb86} were observed experimentally, see
Fig.~\ref{fig7}. These fluctuations occurred as a function of an
external magnetic field. The same phenomenon was found in quasi 1d
MOSFETs \cite{Whe82,Lic85,How86,Kap86} as a function of gate voltage.
The phenomena are reviewed in Refs.~\onlinecite{Web86,Aro87}.  The
fluctuations occurred at Kelvin temperatures in systems that were much
larger than the elastic mean free path $\ell$, but smaller than the
phase coherence length $L_\phi$, i.e. in what became known as the
mesoscopic regime. Most strikingly, the amplitude of the fluctuations
at low $T$ was of the order of $e^2/h$, independent of the mean
conductance of the sample and of sample--specific details like length,
material or geometric shape. Therefore, these fluctuations were dubbed
``universal conductance fluctuations'' (UCF). The fluctuation pattern
turned out to be typical for a given sample and reproducible in
consecutive measurements at fixed temperature, hence the name
``magnetofingerprints''. This observation ruled out any explanation in
terms of time--dependent fluctuation processes. In rings like the one
shown in Fig.~\ref{afig1}, a superposition of UCF and periodic
oscillations of principal frequency $h/e$ was observed, see
Fig.~\ref{afig4}. This was a demonstration of the mesoscopic
Aharonov--Bohm effect~\cite{Web86}.

UCF can be qualitatively explained by estimating the level number
fluctuations, see Sec.~\ref{spec_1}, but
compared with expectations based on classical physics, the UCF are
abnormally large: From Ohm's law and in $d$ 
dimensions, we have $\langle g \rangle \propto L^{d-2}$ for the
dimensionless mean conductance $\langle g \rangle$. UCF implies ${\rm
  var}(g) \approx 
{\rm  constant}$. Therefore, the {\it relative} fluctuations ${\rm
  var}(g)/\langle g \rangle^2$ of the conductance behave as
$L^{2(2-d)}$. For a network of {\it classical} resistors of size $L$ in
$d$ dimensions, on the other hand, one can show \cite{Lee87} that the
relative fluctuations ${\rm var}(g)/\langle g \rangle^2$ of the
conductance behave as $L^{-d}$: The relative fluctuations vanish as
the system size goes to infinity. This is referred to as
``self--averaging''. The existence of UCF implies that there is no
self--averaging for $d = 1,2$.

Theoretical investigations by Altshuler, Lee, and Stone revealed that
these fluctuations were due to complex quantum interference effects
\cite{Sto85} typical for the diffusive regime. The amplitude was
indeed found to be universal and of order $e^2/h$
\cite{Alt85,Lee85a,Alt85a}. Shortly thereafter, Imry \cite{Imr86}
further elucidated the origin of UCF by assuming that the long--range
stiffness in the spectra of Gaussian Random Matrix Theory carries over
to certain parameters $x_j$ characterizing the transfer matrix of a
disordered segment. With this assumption the universality of the
conductance fluctuations followed immediately. While the assumption
was not quantitatively correct, it helped to develop a simple and
intuitive understanding of UCF (cf. the discussion of
Ref.~\onlinecite{Alt86} at the beginning of Sec.~\ref{spec_1}). We
mention that UCF have much in common with but are different from
Ericson fluctuations, cf. Sec.~\ref{qc3nu}.

The analytical modeling used an ensemble of impurity potentials of
the kind described in Sec.~\ref{loc}. A general problem of
stochastic modeling (first encountered in RMT in the context of
nuclear physics, see Sec.~\ref{early}) resurfaced in the present context:
To compare ensemble averages with data, an ergodic hypothesis had to
be invoked. It states that an ensemble average over the impurity
potential is equal to the running average (taken over a range of
values of the magnetic field strength or of the gate voltage) on a
fixed member of the ensemble (i.e., the physical sample). This
assumption is discussed in some detail in the comprehensive paper by
Lee, Stone, and Fukuyama \cite{Lee87}.

For a quantitative analysis, the multichannel Landauer
formula has been the most useful starting point. This formula has
acquired considerable importance in the interpretation of transport
properties of mesoscopic systems. It expresses the dimensionless
conductance $g$ in terms of the quantum--mechanical transmission
amplitudes through the mesoscopic device.

To derive this formula from linear response theory, i.e., from the
Kubo formula \cite{Kub56}, we consider a finite quasi 1d disordered
region in $d$ dimensions with volume $V=L_{\perp}^{d-1} L$. Infinite
ideal leads with identical cross sections $L_{\perp}^{d-1}$ are
attached to both ends of this rod of length $L$. In this model
geometry, the zero temperature Kubo formula for the conductance reads
\begin{eqnarray}
g = \left(\frac{2\pi}{L}\right)^2 \sum_{k,l} \left\vert
\int_V d^3r \frac{1}{2im} 
\left( \nabla \psi_k^* \psi_l - \psi_k^* \nabla\psi_l \right)
\right\vert^2 \delta(E_F - E_k) \delta(E_F-E_l) \ . \nonumber \\
\label{eq_2:1}
\end{eqnarray}
Here, $E_F$ is the Fermi energy, $\psi_k$ are the eigenstates of the
system (including the leads) and $E_k$ the corresponding
eigenvalues. The integral extends over the disordered region, and 
the derivatives are to be taken in the longitudinal direction of the 
quasi 1d rod. Lee and Fisher \cite{Fis81} have shown that 
Eq.~(\ref{eq_2:1}) is equivalent to
\begin{eqnarray}
g &=& \phantom{2} \sum_{a,b} (|S_{ab}^{RL}|^2 + |S_{ab}^{LR}|^2) \nonumber\\
  &=& 2           \sum_{a,b} |S_{ab}^{RL}|^2 \ ,
\label{eq_2:2}
\end{eqnarray}
where $S_{ab}^{RL}$ and $S_{ab}^{LR}$ are those elements of the
scattering matrix that connect the right lead ($R$) with the left one
($L$) and vice versa, respectively. Eq.~(\ref{eq_2:2}) is
known as the multichannel Landauer formula \cite{Lan70,Fis81}. 
It establishes a direct link between the conductance and scattering
theory, and has been extended to the case of many external leads by
B\"uttiker~\cite{Bue86}.
The proof of equivalence of the Kubo and the Landauer approach
was further generalized by Stone and Szafer \cite{Sto88} and by
Baranger and Stone \cite{Bar89}. (In the first of these papers, there
are some misleading statements about the lack of unitarity of the
scattering matrix which are due to an incorrect normalization of the
wave functions). 
 
At first sight, Eq.~(\ref{eq_2:2}) seems absolutely obvious: Aside
from a factor 2 for spin, $g$ is given by the total quantum--mechanical 
transmission probability through the mesoscopic device. At second sight, 
one begins to wonder, however: $g$ is expressed in terms of the scattering 
matrix. This quantity describes the dissipationless passage of
electrons through the sample. How is it possible, then, that $g$ is a
transport coefficient which accounts for dissipation? This is the
issue discussed at the beginning of this subsection. Here, we only
emphasize that Eq.~(\ref{eq_2:2}) is another manifestation of quantum
coherence in mesoscopic devices. Because of this coherence, it is 
problematic to define a {\it conductivity}, which would require local
dissipation in the probe. Rather, properties of the entire
device enter into the definition of the {\it conductance}.

The actual calculations employed impurity perturbation theory. An
early overview is given in Ref.~\onlinecite{Lee87}. This technique is
tailored towards the calculation of (products of) Green functions, and
the conductance must first be expressed in terms of such functions.
This can be done either by expressing the $S$ matrices in
Eq.~(\ref{eq_2:2}) in terms of such functions \cite{Iid90}, or by
using in Eq.~(\ref{eq_2:1}) the identity
\begin{equation}
\langle r| G^+(E)-G^-(E)| r'\rangle =
- 2\pi i\sum_k \psi^*_k(r)\psi_k(r') \delta(E-E_k) \ ,
\label{eq_2:3}
\end{equation}
where $G^\pm(E) = (E^\pm-H)^{-1}$. The resulting expression for $g$ is
bilinear in the Green functions. After averaging over the ensemble,
the only non--vanishing contribution has the form 
$\overline{ G^+(r,r';E_F)G^-(r,r';E_F)}$. In essence, the
calculation of such a term uses an expansion of the Green functions 
in powers of the impurity potential $V(r)$. (For the definition of 
$V(r)$, cf. Eq.~\ref{imp}). The ensuing perturbation series is well 
defined in the diffusive regime and can be reordered to yield a 
systematic expansion in powers of the disorder parameter 
$(k_F\ell)^{-1}$. To leading order in $(k_F\ell)^{-1}$, the Fourier 
representation of a single averaged Green function is given by \cite{Abr63}
\begin{equation}
G^+(p,p') = \frac{\delta(p-p')}{E_F - p^2/2m + i/2\tau}\ .
\label{eq_2:6}
\end{equation}
The average Green function acquires a finite lifetime $\tau =
\ell/v_F$ (the elastic mean free time). In coordinate space, the Green
function decays on the scale of the elastic mean free path. Again to
leading order in $(k_F\ell)^{-1}$ one can show \cite{Abr63} that the 
expression $\overline{G^+(r,r';E_F)G^-(r,r';E_F)}$ is given by the
diffusion propagator $\Pi(r,r')$, a solution of the equation 
\begin{equation}
\tau {\cal D} \Delta \Pi(r,r') = \delta(r-r').
\label{eq_2:7}
\end{equation}
Here, ${\cal D} = v_F^2\tau/3$ is the diffusion constant. Calculating 
higher moments of the conductance amounts to calculating averages of higher
products of Green functions and becomes increasingly complicated. For
the variance of the conductance it turns out \cite{Alt85,Lee85a} that
\begin{equation}
{\rm var}(g) \propto \int d^3r \int d^3r' \, \Pi(r,r') \Pi(r',r) =
{\rm tr} \, \Pi^2 \ .
\label{eq_2:8}
\end{equation}
For a quasi 1d wire without spin--orbit scattering and obeying 
time--reversal symmetry, Eq.~(\ref{eq_2:8}) yields the universal result
$8/15$ \cite{Lee85a}. For all three symmetry classes with $\beta
= 1,2,4$, the result is ${\rm var}(g) = 8/(15\beta)$. 

The impurity perturbation technique sketched above cannot give a
complete picture of the transport properties of quasi 1d systems
because it fails (as does any perturbative approach) in the localized
regime. It is therefore necessary to consider alternative,
non--perturbative approaches. This is the program for the remainder of
this section on quasi 1d transport.

\subsubsection{Dorokhov--Mello--Pereyra--Kumar equation}
\label{quasi1d_2}

A first non--perturbative approach deals directly with the transfer
matrix $M$ of a disordered conductor. This matrix is defined as
follows. We consider a disordered sample connected to two ideal
leads. The number of channels in each lead is denoted by $\Lambda$. 
The matrix $M$ expresses the $\Lambda$--dimensional vectors 
$\phi'_{\rm in}$ and $\phi'_{\rm out}$ in one lead (which describe the
incoming and outgoing wave amplitudes in that lead) in
terms of the corresponding $\Lambda$--dimensional vectors 
($\phi_{\rm in}$ and $\phi_{\rm out}$) in the other
lead,
\begin{equation}
\left[
\begin{array}{c}
\phi'_{\rm out} \\
\phi'_{\rm in}
\end{array}
\right]
=
M
\left[
\begin{array}{c}
\phi_{\rm in} \\
\phi_{\rm out}

\end{array}
\right]\ .
\label{eq_2:9}
\end{equation}
The matrix $M$ differs from the scattering matrix $S$. The latter
connects the outgoing wave amplituds in either lead with the incoming
ones, 
\begin{equation}
\left[
\begin{array}{c}
\phi_{\rm out} \\
\phi'_{\rm out}
\end{array}
\right]
=
S
\left[
\begin{array}{c}
\phi_{\rm in} \\
\phi'_{\rm in}
\end{array}
\right]\ .
\label{eq_2:9a}
\end{equation}
The two matrices $M$ and $S$ are, of course, not independent but are
connected by a non--linear equation not given here. Both matrices
yield equivalent descriptions of the scattering process. The virtue
and usefulness of the transfer matrix lies in a multiplicative
composition rule: The transfer matrix of a number of disordered
regions arranged in series and pairwise connected by ideal leads, is
the product of the transfer matrices of the individual regions. This
property is used in the construction of the transfer matrix in the
local approach described below. It relates directly to the scaling
theory of localization. 

Combining Eqs.~(\ref{eq_2:9}) and (\ref{eq_2:9a}) with the
Landauer formula (\ref{eq_2:2}) one can show that the
conductance is given in terms of the radial ``eigenparameters''
$\lambda_i$ (to be defined below) of $M$ as
\begin{equation}
g = 2 \sum_{i=1}^\Lambda \frac{1}{1+\lambda_i}
  \equiv 2 \sum_{i=1}^\Lambda T_i \ ,
\label{eq_2:9b} 
\end{equation}
where we have defined the transmission eigenvalues
$T_i=1/(1+\lambda_i)$. 
The task is then to determine $M$ and, thus, $\langle g \rangle$ or
higher moments of $g$.
 
Every microscopic realization of a disordered quasi 1d wire is
associated with a particular transfer matrix $M$. This fact suggests
the basic idea of the transfer matrix approach: To define an
appropriate ensemble of transfer matrices that correctly reflects the
transport properties of an ensemble of disordered wires. There are two
different ways to do so, the {\it global} and the {\it local}
approach.  In the global approach developed by Pichard, Stone {\it et
  al.} \cite{Mut87,Pic90,Pic90a}, the statistical properties of the
full transfer matrix are determined in a single step: The maximum
entropy approach is used in conjunction with physically motivated
constraints to determine an ensemble of transfer matrices. In the
local approach by Dorokhov \cite{Dor82}, Mello, Pereyra, and Kumar
\cite{Mel88}, the maximum entropy approach is employed to define an
ensemble of transfer matrices for a small (but still macroscopic)
piece of wire. The decomposition of the wire associated with this
model is similar to the one employed in Sec.~\ref{quasi1d_3} for the
IWZ model, see Fig.~\ref{afig7}.  The full ensemble of transfer
matrices for the entire system is then obtained by putting many of
these small pieces in series, i.e. by multiplying the corresponding
transfer matrices. This procedure defines a Brownian motion model for
the transfer matrix and leads to a Fokker--Planck equation for its
``eigenparameters'' $\lambda_i$. This is the
Dorokhov--Mello--Pereyra--Kumar (DMPK) equation. In the following we
focus attention on this latter theory. The global approach has been
reviewed in detail in Ref.~\onlinecite{Sto91}.

We briefly sketch the derivation of the DMPK equation. We follow a
formulation in Ref.~\onlinecite{Fra95b} which was applied by Frahm and
Pichard \cite{FrP95a,FrP95b} in a slightly different context, namely
for a Brownian motion model for the $S$ matrix. We restrict ourselves
to the case of unitary symmetry ($\beta=2$). The relevant symmetry of
$M$ is dictated by current conservation and reads \cite{Sto91}
\begin{equation}
M^\dagger \Sigma_z M = \Sigma_z,
\label{eq_2:10}
\end{equation}
where $\Sigma_z=$diag$(1,-1)$ in the block representation defined by
Eq.~(\ref{eq_2:9}). Putting $M = \exp(X)$ this implies
\begin{equation}
X = \left[
\begin{array}{cc}
a & b \\
b^\dagger & d \\
\end{array}
\right] \ '
\label{eq_2:11}
\end{equation}
with $a^\dagger = -a$ and $d^\dagger = -d$. 

Using this symmetry, we now consider $M(x)$ as a function of a
variable $x$ which describes the length of the wire. We wish to find a
differential equation for $M(x)$. We define a random process by 
\begin{equation}
M(x+\delta x) = e^{\delta M} M(x)
\label{eq_2:12}
\end{equation}
where $\delta M$ is a random matrix which must have the same symmetry
as $X$. For simplicity, we identify $\delta M$ with $X$ and introduce 
the following statistical assumptions (which can be justified in terms
of a maximum entropy approach \cite{Mel88}): $\delta M$ has a Gaussian
probability distribution with vanishing first moment and uncorrelated
matrix elements $a,b,d$. The second moments are given by
\begin{eqnarray}
\overline{ {\rm tr}(a^\dagger a) } =
\overline{ {\rm tr}(d^\dagger d) } &=& W_f \Lambda^2 \delta x \ ,
\nonumber\\
\overline{ {\rm tr}(b^\dagger b) } &=& W_b \Lambda^2 \delta x.
\label{eq_2:13}
\end{eqnarray}
We recall that $\Lambda$ is the number of channels. The quantities
$W_f$ and $W_b$ define the relative strength of the diagonal and
non--diagonal elements, respectively. 

It is useful to decompose the transfer matrix into radial and angular
degrees of freedom,
\begin{equation}
M(x) = U e^H V \ , \quad M(x+\delta x) = \tilde{U} e^{H+\delta H}
\tilde{V},
\label{eq_2:14}
\end{equation}
where $U={\rm diag}(u_1,u_2)$, $V={\rm diag}(v_1,v_2)$, and $H=\nu
\Sigma_x$ (and similarly for $\tilde{U}, \tilde{V}$, and $\delta H$).
Here $u_i$ and $v_i$ are unitary matrices, $\nu$ is a diagonal matrix
of ``radial'' parameters,  and $\Sigma_x$ is the first Pauli matrix.
The quantities $\nu_i$ are related to the ``eigenparameters''
$\lambda_i$ in Eq.~(\ref{eq_2:9b}) through
$\lambda_i=\sinh^2(\nu_i)$. 
It is the aim to express the variation $\delta H$ in the radial
paramters as a function of the statistical input (\ref{eq_2:13}) 
for the random process (\ref{eq_2:12}). This is achieved by an
expansion to second order in the small quantities $b$ and
$b^\dagger$. The quantities of central interest
are the increments $\delta \nu_i$ of the radial parameters. 
Their detailed properties follow from Eq.~(\ref{eq_2:13}).   
For $\beta=2$ the result reads
\begin{eqnarray}
\overline{\delta\nu_j} &=& \frac{1}{2} W_b \delta x
\left(\sum_{k(\neq j)} \left( \coth(\nu_j-\nu_k) + \coth(\nu_j+\nu_k)
\right) + 2\coth(2\nu_j) \right)\ , \nonumber\\
\overline{ \delta\nu_j\delta\nu_k } &=&  \frac{1}{2} W_b
\delta x \delta_{jk} \ .
\label{eq_2:19}
\end{eqnarray}
This defines a standard Brownian motion process leading immediately 
\cite{Dys62d} to a Fokker--Planck equation for the distribution
function $p(\nu,x)$ of the radial parameters (DMPK equation)
\begin{eqnarray}
\partial_x p(\nu,x) &=& \frac{W_b}{4} \sum_j
\partial_{\nu_j} \Bigg(
\partial_{\nu_j} - 2
\sum_{k(\neq j)} \left( \coth(\nu_j-\nu_k) + \coth(\nu_j+\nu_k)
\right)  \nonumber\\
& & \hspace{6cm} - 2\coth(2\nu_j) \Bigg) p(\nu,x) \ .
\label{eq_2:20}
\end{eqnarray}
The differential operator on the r.h.s. of Eq.~(\ref{eq_2:20}) is the radial
part of the Laplacian on the space of transfer matrices \cite{Hue90}. 
Introducing as new variables the ``eigenparameters'' $\lambda_i =
\sinh^2(\nu_i)$, we get an alternative form of the DMPK equation, 
\begin{equation}
\partial_x \hat{p}(\lambda,x) = 
\frac{2}{\xi} \sum_j \partial_{\lambda_j} J(\lambda)
\lambda_j(1+\lambda_j) \partial_{\lambda_j} J^{-1}(\lambda)
\hat{p}(\lambda,x) .
\label{eq_2:21}
\end{equation}
With $J(\lambda) = \prod_{j<k}|\lambda_j-\lambda_k|^\beta$, the
definition for the elastic mean free path $\ell =
1/(W_b(\Lambda+(2-\beta)/\beta))$ and the localization length $\xi =
\ell(\beta\Lambda + 2 -\beta)$, Eq.~(\ref{eq_2:21}) is valid for all three
symmetry classes.

The DMPK equation has been solved by Beenakker and Rejaei
\cite{Bee93,Bee94} in the unitary case ($\beta=2$). It is obviously of
considerable interest to have an exact solution for the distribution
function $\hat{p}(\lambda,x)$. The actual motivation for this work had
another source, however: A puzzling discrepancy between the global
approach \cite{Mut87,Pic90,Pic90a} and diagrammatic perturbation theory. 
In the global approach it is assumed that all correlations among the
eigenvalues $\lambda_i$ stem from the Jacobian $J(\lambda)$ introduced
above. (Formally, this is just the square of the Vandermonde
determinant discussed in Sec.~\ref{qc1rcb}).
Therefore the distribution function must have the form
\begin{equation}
\hat{p}_{\rm glob} = e^{-\beta ( \sum_{i<j} u(\lambda_i,\lambda_j) +
                     \sum_i V(\lambda_i))}
\label{eq_2:22}
\end{equation}
with the logarithmic two--body repulsion $u(\lambda_i,\lambda_j) =
-\ln|\lambda_i-\lambda_j|$ coming from the Jacobian. It was shown by
Beenakker \cite{Bee93a}, however, that Eq.~(\ref{eq_2:22}) leads to
${\rm var}(g) = 1/2\beta$ (instead of the correct value $8/15\beta$),
irrespective of the form of the confining potential $V(\lambda_i)$. 
Hence the logarithmic repulsion of eigenparameters in Eq.~(\ref{eq_2:22})
had to be incorrect. The correct interaction between the $\lambda_i$
was obtained with the help of a variant of the Sutherland
transformation \cite{Bee93,Bee94}. This transformation maps the DMPK
equation (\ref{eq_2:20}) onto a Schr\"odinger equation in
imaginary time. For $\beta = 2$, it has the form 
\begin{equation}
-\ell \frac{\partial \Psi}{\partial x} = ({\cal H} - U)\Psi
\label{eq_2:23}
\end{equation}
with
\begin{equation}
{\cal H} =
-\frac{1}{2\gamma}\sum_i\left(\frac{\partial^2}{\partial^2_{\nu_i}}
+\frac{1}{\sinh^22\nu_i} \right),
\label{eq_2:24}
\end{equation}
$\gamma = \beta\Lambda+2-\beta$, and $U = -\Lambda/2\gamma -
\Lambda(\Lambda-1)\beta/\gamma -
\Lambda(\Lambda-1)(\Lambda-2)\beta^2/6\gamma$. The Hamiltonian in
Eq.~(\ref{eq_2:24}) is the sum of single--particle Hamiltonians. The
corresponding single--particle Green functions can be explicitly
calculated and combined to give $G(\nu,x|\mu)$, the many--particle
Green function of Eq.~(\ref{eq_2:23}). From this function, one obtains
$p(\nu,x)$ \cite{Bee93,Bee94}. Simplifying this expression for the
metallic regime $\ell <x <\xi$, and transforming back to the
$\lambda_i$, one finds \cite{Bee93,Bee94},
\begin{eqnarray}
\hat{p}(\lambda,x) &\propto& e^{-\beta ( \sum_{i<j} u(\lambda_i,\lambda_j) +
                     \sum_i V(\lambda_i,x))} \nonumber\\
u(\lambda_i,\lambda_j) &=& -\ln|\lambda_j-\lambda_i|/2 - 
\ln\left\vert
{\rm arsinh}^2\sqrt{\lambda_j} - {\rm arsinh}^2\sqrt{\lambda_i}
\right\vert /2 \ ,
                     \nonumber\\ 
V(\lambda_i,x) &=& \frac{\Lambda\ell}{2x} {\rm arsinh}^2\sqrt{\lambda_i} \,
(1+{\cal O}(\Lambda^{-1})) \ .
\label{eq_2:25}
\end{eqnarray}
As the transmission eigenvalues $T_i = 1/(1+\lambda_i)$ go from $1$ to
$0$, the $\lambda_i$ go from $0$ to $1$, and the ``two--body
interaction'' term crosses over from $-\ln|\lambda_j - \lambda_i|$ to half
that value. This shows that it is the failure of the global approach
to correctly describe the behavior of weakly transmitting channels
which is responsible for the slight discrepancy in the UCF. 

The exact solution $\hat{p}(\lambda,x)$ was used by Frahm \cite{Fra95d}
to calculate all $m$--point correlation functions
\begin{equation}
R_m(\lambda_1,\ldots,\lambda_m,x) = \frac{\Lambda !}{(\Lambda - m)!}
\int_0^\infty d\lambda_{m+1} \ldots d\lambda_{\Lambda} \,
\hat{p}(\lambda,x) \ .
\label{eq_2:26}
\end{equation}
This was possible with the help of a generalization of the method of
orthogonal polynomials \cite{Meh91}. With $m,n,j = 1,\ldots,\Lambda$,
$\hat{p}(\lambda,x)$ could be written as the product of two
determinants of matrices of dimension $\Lambda$,
\begin{equation}
\hat{p}(\lambda,x) \propto {\rm det}[W_{n-1}(\lambda_j,x)]
                           {\rm det}[h_{m-1}(\lambda_j,x)],
\label{eq_2:27}
\end{equation}
The functions $W_n$ and $h_m$ are defined in Ref.~\onlinecite{Fra95d}
and fulfill the biorthogonality relation $\int_0^\infty d\lambda \,
W_{n-1}(\lambda,x) h_{m-1}(\lambda,x) = \delta_{nm}$. Defining the
functions
\begin{equation}
K_\Lambda(\lambda,\tilde{\lambda};x) = 
\sum_{m=0}^{\Lambda-1} W_m(\lambda,x) h_m(\tilde{\lambda},x)
\label{eq_2:28}
\end{equation}
one finds that the $m$--point functions have the form of a determinant, 
\begin{equation}
R_m(\lambda_1,\ldots,\lambda_m,x) = 
{\rm det}[K_\Lambda(\lambda_i,\lambda_j;x)_{1\le i,j \le m}] \ .
\label{eq_2:29}
\end{equation}
{}From $R_1$ and $R_2$, Frahm \cite{Fra95d} calculated explicitly the
first two moments of the conductance $g = \sum_i 1/(1+\lambda_i)$. The
results turned out to be equivalent to those obtained in the framework
of the non--linear $\sigma$ model described in the next section.

Extending this formalism, Frahm and M\"uller--Groeling \cite{Fra96b}
calculated in the unitary symmetry class the autocorrelation function
$\langle \delta g(L) \delta g(L+\Delta L) \rangle$ for all length
scales. Within the framework of the DMPK approach, this was the first
calculation of a four--point correlation function. Explicit analytical
results were obtained in the diffusive regime and in the strongly
localized regime. In the diffusive regime where $t=L/2\xi\ll 1$ and 
for large channel numbers $\Lambda \to \infty$ the correlation
function is given by a squared Lorentzian,
\begin{equation}
\langle \delta g(t)\delta g(t+\Delta t)\rangle = \frac{4}{15}
\frac{1}{(1+\frac{\Delta t}{t})^2} + {\cal O}(t) \ .
\label{eq_2:29_1}
\end{equation}
This result can be viewed as a generalization of the universality of
var$(g)$. Equation~(\ref{eq_2:29_1}) is independent of any absolute length
scale, and the scale on which the correlation function decays with
$\Delta t$ is set by the system size $t$ itself. In the strongly
localized regime $t,\Delta t \gg 1$, on the other hand, one gets
\begin{equation}
\langle \delta g(t)\delta g(t+\Delta t)\rangle \propto
(\Delta t)^{-3/2} t^{-3/2} e^{-\Delta t} e^{-t} \ .
\label{eq_2:29_2}
\end{equation}
In this limit the localization length $\xi$ sets an explicit scale for
the decay of the correlations. In the crossover regime between
metallic and localized behavior the analytical expressions derived in
Ref.~\onlinecite{Fra96b} had to be evaluated numerically. Examination
of these results explicitly confirmed the view that a wire of length
$L\gg\xi$ can be thought of as being composed of independently
fluctuating segments of length $\xi$.

\subsubsection{Non--linear $\sigma$ model for quasi 1d wires}
\label{quasi1d_3}

The second approach to transport in quasi one--dimensional systems uses
statistical properties of the {\it Hamiltonian} of the system under
study rather than of the transfer matrix. For this reason, it is
sometimes called {\it microscopic}, in contrast to the macroscopic
DMPK approach. We prefer another terminology. 
We will refer to the DMPK approach and the $\sigma$ model approach
also as to the random transfer matrix approach and the random
Hamiltonian approach, respectively. In
the latter approach, ensemble averages are calculated with the help of
Efetov's supersymmetry method \cite{Efe83,Ver85}, see Sec.~\ref{susy}. 
This is close to Efetov's original application of his method. In fact,
starting from a single--particle Hamiltonian with a random white noise
potential, Efetov \cite{Efe83} proved --- among other things --- 
that all wave functions in infinite quasi one--dimensional systems are
localized for all three symmetry classes. The first model, however, to
properly take into account the geometry discussed above including the
coupling to the two leads, was introduced by Iida, Weidenm\"uller, and
Zuk (IWZ) \cite{Iid90}. We briefly describe its main features. 

\begin{figure}
\centerline{
\psfig{file=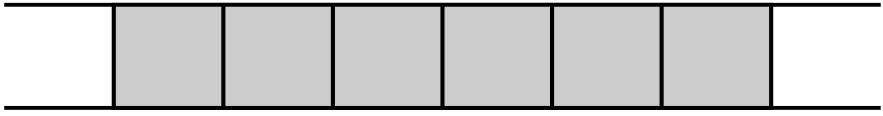,angle=90,width=4.5in}
 }
\vskip 0.5cm
\caption{
Schematic illustration of the IWZ model.  The disordered region
(shaded) is divided into $K$ segments. Ideal leads are
attached to the right and the left of the disordered region.
} 
\label{afig7}
\end{figure}

The disordered region is modeled as a chain of $K$ segments, each
roughly of 
the size of the mean free path $\ell$, see Fig.~\ref{afig7}.
The leads attached to the right
and the left of the disordered region support $\Lambda_1$ and
$\Lambda_2$ channels, respectively. Without the leads, the system is
described by the Hamiltonian $H^{ij}_{\mu\nu}$, where the indices
$i,j$ refer to the segments, while the indices $\mu,\nu$ denote the
$N$ states per segment. As usual, the limit $N \rightarrow \infty$ is
taken. The elements of $H$ are real ($\beta=1$), complex ($\beta =
2$), or quaternion ($\beta=4$) numbers. States within one segment
labelled $i$ are represented by the matrix $H^{ii}_{\mu\nu}$ with the
probability distribution  
\begin{equation}
P(H^{ii}) \propto \exp\left(-\frac{\beta N}{4v_1^2} {\rm tr}(H^{ii})^2
\right) \ .
\label{eq_2:29a}
\end{equation}
States in adjacent segments are coupled by the matrices
$H^{ij}_{\mu\nu}$ ($|i-j|=1$) with the probability distributions  
\begin{equation}
P(H^{ij}) \propto \exp\left(-\frac{\beta N^2}{2v_2^2} 
{\rm tr}(H^{ij}H^{ji}) \right) \ .
\label{eq_2:29b}
\end{equation}
Matrix elements $H^{ij}_{\mu\nu}$ with $|i-j|>1$ vanish identically. 
Matrix elements carrying different upper indices $(i,j) \neq (k,l)$
are uncorrelated. The Hamiltonian of the entire disordered region, a
matrix of dimension $K N$, is denoted by ${\cal H}$. The coupling to
the leads is effected by a fixed (non--random) rectangular matrix
$W=W_1+W_2$ of dimension $KN \times (\Lambda_1+\Lambda_2)$. It has
elements $W^i_{\mu n}$, where $i$ and $\mu$ identify a segment and a
state, respectively, in the disordered region and where $n$ refers to
the channels in the leads. It is assumed that only the first (the
last) segment couple to the left (the right) lead, repectively. Thus
the elements of $W_1$ are non--zero only for $i=1$ and $1\le n\le
\Lambda_1$, while the elements of $W_2$ are non--zero only for $i=K$ and
$\Lambda_1 < n\le \Lambda_1+\Lambda_2$. For this model, the scattering
matrix $S$ can be written down explicitly,
\begin{equation}
S = 1 - 2\pi i \, W^\dagger (E-{\cal H}+i\pi WW^\dagger)^{-1} W \ .
\label{eq_2:29c}
\end{equation}
The ensemble of Hamiltonians defines an ensemble of scattering
matrices. The latter determines the statistics of the conductance via
the Landauer formula (\ref{eq_2:2}).

The supersymmetric generating functional for a product of effective
propagators $(E-{\cal H} \pm i\pi WW^\dagger)^{-1}$ appearing in
Eq.~(\ref{eq_2:29c}) has the form 
\begin{eqnarray}
&&\overline{Z[J]} \ = \ \int d[{\cal H}] P({\cal H})
\label{eq_2:29x}
\\
&&
\int D\psi \, \exp\left( \frac{1}{2}i\psi^\dagger L (
E - {\cal H} + i\pi W_1W_1^\dagger  L + i\pi W_2W_2^\dagger L +
i\eta L + J)\psi \right) \ . \nonumber
\end{eqnarray}
After averaging, Hubbard--Stratonovich transformation, integration
over the $\psi$--fields and saddle--point approximation \cite{Iid90}, the
generating functional takes the form of a generalized non--linear
$\sigma$ model, 
\begin{eqnarray}
\overline{Z[J]} &=& \int D[Q] \,
\exp\Bigg(\frac{v_2^2}{2v_1^4}\sum_i {\rm trg}[Q_iQ_{i+1}]  \nonumber \\
& & \quad
-\frac{1}{2} {\rm Trg}\ln(E-Q + i\pi W_1W_1^\dagger L + i\pi
W_2W_2^\dagger L + J ) \Bigg)\ .
\label{eq_2:29e}
\end{eqnarray}
Setting $E=0$
the matrices $Q_i$ obey the constraint $Q_i^2 = -v_1^2$. The matrix $Q$ is
block--diagonal, with diagonal entries $Q_1, Q_2, \ldots, Q_K$. The
symbol ${\rm Trg}$ implies a trace over the segment label $i$ and the
state label $\mu$.

The functional (\ref{eq_2:29e}) has served as the starting point of
several papers dealing with the diffusive regime and/or the localized
regime of quasi one--dimensional mesoscopic systems. In the diffusive
regime, where the dimensionless conductance $g$ is much larger than
unity, the functional (\ref{eq_2:29e}) can be evaluated perturbatively
as follows \cite{Iid90}. The saddle--point manifolds of the matrices
$Q_i$, $i = 1, \ldots, K$ are parametrized in terms of a suitable set
of independent variables. Terms of second order in these variables are
proportional to $g$ \cite{Iid90} and are kept in the exponent. The
exponential containing all terms of higher order than the second is
expanded in a Taylor series. This reduces all integrations to an
application of Wick's theorem. The expansion generated in this way is
an asymptotic expansion in inverse powers of $g$. In this way,
conductance fluctuations were investigated in the orthogonal
\cite{Iid90}, unitary and symplectic \cite{All90} symmetry classes.
Phase breaking processes could be taken into account in a
phenomenological way \cite{All91}. The influence of a magnetic field
on the conductance was studied in ring structures by calculating the
weak localization correction \cite{Zuk92} as well as the
autocorrelation function versus magnetic field strength \cite{Mue93a}.
Resistance fluctuations were investigated in detail in three--
\cite{Iid91}, four-- \cite{Iid91a} and multi--lead \cite{Scm92}
devices. Some, but not all, of the results in these papers have been
obtained earlier by means of the impurity perturbation technique (see
Ref.~\onlinecite{Lee87} and references therein). The perturbative
evaluation of the supersymmetric functional for the IWZ model is
involved but conceptually relatively simple. In its results, it is
equivalent to the impurity diagram technique.

A non--perturbative treatment requires considerably more effort but
leads to results which go beyond the domain of applicability of the
impurity diagram technique. The appropriate technique, Fourier
analysis on supermanifolds, was developed by Zirnbauer \cite{Zir91}. 
This technique was used by Zirnbauer to calculate the mean
conductance \cite{Zir92} and by Mirlin, M\"uller--Groeling, and
Zirnbauer to calculate the variance of the conductance \cite{Mir94} in
quasi one--dimensional wires for all length scales and for all three
symmetry classes. We now sketch the calculation for the orthogonal
symmetry class.  

For $\Lambda_1=\Lambda_2=\Lambda$ and in the limit of large channel
number $\Lambda \gg 1$, the first two moments of the conductance can
be expressed as integrals over the coset space of the $Q$ matrices
(with the redefinition $Q\rightarrow Q'=iQ/v_1$ and omitting the
prime) \cite{Mir94},
\begin{eqnarray}
\mu_1 &=& \langle g \rangle = \frac{\Lambda^2}{2} 
\int D[Q] \,  Q_1^{51} Q_K^{51}  \nonumber\\
& & \ \exp\left(-\frac{\Lambda}{8} {\rm
  trg}[(Q_1+Q_K)L] \right) \exp\left(\frac{\xi}{32\ell} \sum_i {\rm
  trg} (Q_{i+1}-Q_i)^2 \right) \ ,   \nonumber\\
\mu_2 &=& \langle g^2 \rangle = \frac{\Lambda^4}{4}
\int D[Q] \, Q_1^{51} Q_1^{62} Q_K^{51} Q_K^{62} \nonumber\\
& & \  \exp\left(-\frac{\Lambda}{8} {\rm
  trg}[(Q_1+Q_K)L] \right) \exp\left(\frac{\xi}{32\ell} \sum_i {\rm
  trg} (Q_{i+1}-Q_i)^2 \right) \ .
\label{eq_2:30}
\end{eqnarray}
The superscripts on the preexponential terms indicate the appropriate
element of the $8\times 8$ supermatrix $Q$. We have introduced the
quantity $\xi = 8(v_1^2/v_2^2)\ell$.  By comparison with
Ref.~\onlinecite{Efe83} it can be identified with the localization
length for the orthogonal case. Integrating over all $Q$ matrices
except for those referring to the first and the last segment, $Q_1$
and $Q_K$, we obtain the {\it heat kernel}
\begin{eqnarray}
{\hat W}(Q_1,Q_K;L/2\xi) = \int dQ_2 \ldots dQ_{K-1} \,
\exp\left(\frac{\xi}{32\ell} \sum_i {\rm
  trg} (Q_{i+1}-Q_i)^2 \right) \ , \nonumber\\
\label{eq_2:31}
\end{eqnarray}
where $L=K\ell$ is the system size (not to be confused with the
supermatrix $L$ occuring, e.g., in Eqs.~(\ref{eq_2:29x}) and
(\ref{eq_2:29e})). In the continuum limit and with the metric defined
by the length element ${\rm trg}(dQdQ)/16$ one can show \cite{Mir94}
that ${\hat W}$ fulfills the {\it heat equation} 
\begin{equation}
\frac{\partial}{\partial s} {\hat W}(Q,Q';s) = \Delta_Q {\hat W}(Q,Q';s);
\quad \lim_{s\to 0} {\hat W}(Q,Q';s) = \delta(Q,Q').
\label{eq_2:32}
\end{equation}
Here, $s=L/2\xi$, $\delta(Q,Q')$ is the $\delta$ function and
$\Delta_Q$ the Laplace--Beltrami operator (Laplacian) on the coset
space. It turns out that all the information about the first two
conductance moments $\mu_1$ and $\mu_2$ is contained in the heat
kernel. Using for the $Q$ matrices the parametrization $Q = \exp(X)
 L \exp(-X)$, one obtains the following expansion of the heat
kernel in powers of $X$,
\begin{equation}
{\hat W}(e^X  L e^{-X},  L;s) = 
1- \frac{1}{2} \mu_1 \frac{{\rm trg} X^2}{4} + 
   \frac{1}{8} \mu_2 \left(\frac{{\rm trg} X^2}{4}\right)^2 +
   \ldots
\label{eq_2:33}
\end{equation}
We notice that the coefficients of the two leading terms in this
expansion are the first two moments of the conductance.

\begin{figure}
\centerline{
\psfig{file=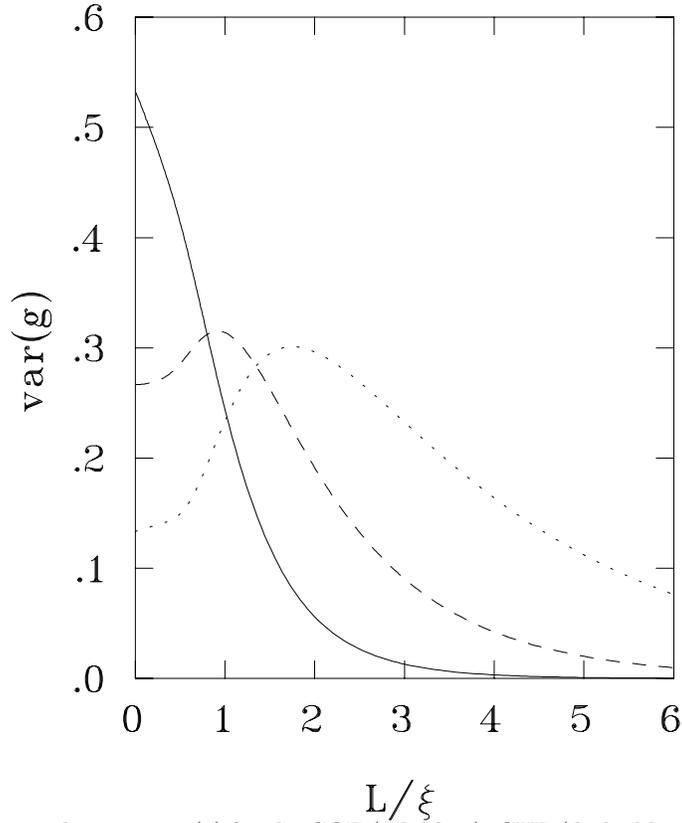,width=3.5in}
 }
\caption{
The variance of the conductance var$(g)$ for the GOE (solid line), GUE
(dashed line), and GSE (dotted line) versus $L/\xi$.
} 
\label{fig_varg}
\end{figure}

Fourier analysis was invented to solve the heat equation and is
expected to work here as well. Equation~(\ref{eq_2:32}) presents a
particularly complicated case, however, since it is defined on a
curved supermanifold. Nonetheless, the basic idea remains the same. 
Here, we only give the result. For GOE symmetry, one finds
\begin{eqnarray}
&&\mu_n(L) = \frac{\pi}{2} \int_0^\infty d\lambda \,
\tanh^2(\pi\lambda/2)(\lambda^2+1)^{-1} p_n(1,\lambda,\lambda)
T(1,\lambda,\lambda)  \nonumber\\
&&\quad + 2^4 \sum_{l\in 2{\bf N}+1} \int_0^\infty d\lambda_1d\lambda_2 \,
l(l^2-1)\lambda_1 \tanh(\pi\lambda_1/2)\lambda_2\tanh(\pi\lambda_2/2)
\nonumber\\ 
&&\quad\times p_n(l,\lambda_1,\lambda_2) \prod_{\sigma,\sigma_1,\sigma_2 =
  \pm 1}
(-1+\sigma l+i\sigma_1\lambda_1 + i\sigma_2\lambda_2)^{-1}
T(l,\lambda_1,\lambda_2),
\label{eq_2:39}
\end{eqnarray}
where $T(l,\lambda_1,\lambda_2) =
\exp(-2\tilde{\Delta}(l,\lambda_1,\lambda_2)  L/4\xi)$,
$\tilde{\Delta}(l,\lambda_1,\lambda_2) = l^2+\lambda_1^2 +
\lambda_2^2$ and
\begin{eqnarray}
p_1(l,\lambda_1,\lambda_2) &=& \tilde{\Delta}(l,\lambda_1,\lambda_2)
 \ , \nonumber\\
p_2(l,\lambda_1,\lambda_2) &=& (\lambda_1^4+\lambda_2^4+2l^4 +
3l^2(\lambda_1^2+\lambda_2^2) + 2l^2 - \lambda_1^2-\lambda_2^2-2)/2 \ .
\label{eq_2:40}
\end{eqnarray}
Corresponding results for the GUE and GSE were also calculated in
Ref.~\onlinecite{Mir94}. While var$(g)= \mu_2 - \mu_1^2$ showed
exponential localization as a function of system length for GOE and
GUE symmetry, the curve for the GSE case approached a finite limiting
value as $L/\xi\to\infty$. This suggested that for the GSE there is no
localization. A similar behavior had also been obtained for the mean
conductance \cite{Zir92}. However, as discussed below, this unusual
result turned out to be due to a subtle technical mistake
\cite{Bro96}. In Fig.~\ref{fig_varg} we show the correct behavior of
var$(g)$ for all three symmetry classes~\cite{Mir94,Bro96}.

\subsubsection{Equivalence of non--linear $\sigma$ model and DMPK
  equation}
\label{quasi1d_4}

The coincidence in both approaches of the first two moments of the
conductance for the GUE \cite{Fra95d} suggested at least a very close
connection between the two methods. Further evidence in this direction
was provided in Ref.~\onlinecite{Rej96}. Here the average density of
transmission eigenvalues (see Eq.~(\ref{eq_2:9b})) in quasi 1d wires,
\begin{equation}
\rho(T) = \langle \sum_i \delta(T-T_i) \rangle \ ,
\label{eq_2:41}
\end{equation}
was calculated for $\beta=2$ within the $\sigma$ model approach. The
result was identical with the one obtained from the DMPK equation
\cite{Rej96}. This effectively extended the equivalence of the two
approaches for the GUE to the average of all linear statistics,
i.e. of all quantities $A$ that can be written as $A=\sum_i
A(T_i)$. On the other hand, the surprising results for $g$ and
var$(g)$ obtained in the symplectic symmetry class within the
non--linear $\sigma$ model \cite{Zir92,Mir94} were in striking
disagreement with results obtained from the DMPK equation (which ruled
out any delocalization \cite{Fra95d}). Thus, a clarification of the
relation between the DMPK equation and the $\sigma$ model {\it for
all} $\beta$ was of considerable interest. This problem was solved
by Brouwer and Frahm \cite{Bro96} who proved the equivalence of the
two approaches for all three symmetry classes. We give a brief account
of the main steps in the proof.  

Before doing so we point out that the claimed equivalence holds only
for the distribution function of the conductance but not for all
parametric correlations of $g$ (like the autocorrelation function
versus an external magnetic field, or the weak localization correction). 
This is because the DMPK approach is more limited than the
$\sigma$ model approach, for the following reason. Within the random
Hamiltonian approach, the calculation of parametric correlations is
always possible, at least in principle, via a suitable modification of
the random Hamiltonian (the actual calculation may be very cumbersome,
of course). In contradistinction, the calculation of parametric
correlations within the random transfer matrix approach leads to
principal difficulties: In general it is not clear how to implement a
parametric dependence into the transfer matrix in a dynamically
meaningful way \cite{Rau95,FrP95a}.

The main idea of the proof consists in defining a suitable generating
functional $Z({\hat{\theta}},\lambda)$ which combines the key
quantities of both methods, namely the eigenparameters $\lambda_i$ of
the transfer matrix and the radial paramters $\theta_j$ of the
supersymmetric $Q$ matrix. The latter are defined by
\begin{eqnarray}
Q &=& T^{-1} L T   \ ,\nonumber\\
T &=& \left[ \begin{array}{cc}
              u^{-1}  &  0               \\
              0       &  v^{-1}
             \end{array}
      \right]
\exp\left[   \begin{array}{cc}
             0  &  \hat{\theta}/2  \\
            \hat{\theta}/2  &  0
             \end{array}
     \right]
\left[ \begin{array}{cc}
              u       &  0               \\
              0       &  v
             \end{array}
      \right] \ ,
\label{eq_2:42}
\end{eqnarray}
where $\hat{\theta}=$diag$(\hat{\theta}_a,\hat{\theta}_b)$ and
\begin{equation}
\hat{\theta}_a = \left[\begin{array}{cc}
                        \theta_1   &   \theta_2   \\
                        \theta_2   &   \theta_1
                       \end{array}
                 \right] \ ; \quad\quad
\hat{\theta}_b = \left[\begin{array}{cc}
                        i\theta_3   &   i\theta_4   \\
                        i\theta_4   &   i\theta_3
                       \end{array}
                 \right] \ .
\label{eq_2:42a}
\end{equation}
Additional symmetry restrictions are imposed on the $\theta_j$ for
$\beta = 1,2,4$ \cite{Bro96}. The proof of equivalence proceeds by
showing that $Z({\hat{\theta}},\lambda)$ obeys both, the DMPK
equation (\ref{eq_2:20}) with respect to the variables $\lambda$, and
the heat equation (\ref{eq_2:32}) with respect to the variables
${\tilde{\theta}}$. If in addition $Z({\hat{\theta}},\lambda)$
fulfills the proper initial conditions, the proof is complete. 

A variant of the supersymmetric generating functional for the Green
functions of the IWZ model serves this purpose,
\begin{eqnarray}
Z[Q] = 
\int D\psi \, \exp\left( \frac{1}{2}i\psi^\dagger L (
E - {\cal H} + i\pi W_1W_1^\dagger Q + i\pi W_2W_2^\dagger L +
i\epsilon L)\psi \right) \ . \nonumber\\
\label{eq_2:43}
\end{eqnarray}
The precise definitions of the quantities involved here can be found
in Ref.~\onlinecite{Bro96}. It can be shown \cite{Bro96} that $Z$
depends on $Q$ only through $\hat{\theta}$. Moreover, the dependence
of $Z$ on the Hamiltonian ${\cal H}$ and on the coupling matrix
elements $W_1,W_2$ can be replaced by a dependence on the transmission
eigenvalues $T_i = 1/(1+\lambda_i)$. Thus, $Z$ has the required form
$Z=Z(\hat{\theta},\lambda)$. Furthermore, all transport quantities
that can be obtained within the minimal $\sigma$ model (i.e. without
increasing the dimension of $Q$) can be generated from
Eq.~(\ref{eq_2:43}) by taking proper derivatives with respect to the
$\theta_j$.

We average $Z$ over the ensemble of Hamiltonians ${\cal H}$ and denote
the result by a bar ($\overline{Z}$). We then use the heat kernel ${\hat
  W}(Q_1,Q_K;s)$ defined in Eq.~(\ref{eq_2:31}), and we define the
functions 
\begin{equation}
f_{1(2)}(Q,Q_{1(K)}) = \exp\left(-\frac{d}{2}\sum_n {\rm
  trg}\ln(1+x_nQQ_{1 (K)} ) \right) \ ,
\label{eq_2:45}
\end{equation}
with $d=1 (1,2)$ for $\beta=1 (2,4)$. The quantities $x_n$ are
essentially the eigenvalues of $W_1W_1^\dagger$ (for $f_1$) and
$W_2W_2^\dagger$ (for $f_2$), respectively. As in Eq.~(\ref{eq_2:30}),
the functional $\overline{ Z }$ can be expressed as an integral
over the two $Q$ matrices which refer to the first and the last
segment, respectively,
\begin{equation}
\overline{ Z[Q] } = \int dQ_1\int dQ_K \,
 f_1(Q,Q_1) f_2( L,Q_K) {\hat W}(Q_1,Q_K;s) \ .
\label{eq_2:44}
\end{equation}
Because of the important symmetry property $\Delta_{Q'} f_1(Q,Q') =
\Delta_Q f_1(Q,Q')$, the averaged supersymmetric functional 
$\overline{ Z} $ 
also fulfills the heat equation with the radial part
$\Delta_{\hat{\theta}}$ of the full Laplacian $\Delta_Q$,
\begin{equation}
\frac{\partial}{\partial s} 
\overline{ Z(\hat{\theta},\lambda) } = 
\Delta_Q \overline{ Z(\hat{\theta},\lambda) } =
\Delta_{\hat{\theta}} \overline{ Z(\hat{\theta},\lambda) } \ .
\label{eq_2:46}
\end{equation}

We now onsider another average of the generating functional
$Z(\hat{\theta},\lambda)$, defined by the use of the probability
distribution $\hat{p}(\lambda,s)$, a solution of the DMPK equation
(\ref{eq_2:21}), as weight factor. This average is indicated by the
superscript $T$,
\begin{eqnarray}
\overline{Z(\hat{\theta},\lambda)}^T &=&
\int d\lambda_1 \ldots d\lambda_N \,
Z(\hat{\theta},\lambda) \hat{p}(\lambda,s).
\label{eq_2:47}
\end{eqnarray}
It is obvious that $\overline{Z}^T$ obeys the DMPK equation, 
\begin{eqnarray}
\frac{\partial}{\partial s} \overline{Z(\hat{\theta},\lambda)}^T &=&
(2/\gamma) \overline{D_\lambda Z(\hat{\theta},\lambda)}^T \ ,
\label{eq_2:47a}
\end{eqnarray}
where $D_\lambda$ is the differential operator on the r.h.s. of
Eq.~(\ref{eq_2:21}) (without the factor $2/\xi$). 

To prove the identity of both approaches, it is shown \cite{Bro96} that  
\begin{equation}
\Delta_{\hat{\theta}} Z(\hat{\theta},\lambda) \propto
D_\lambda Z(\hat{\theta},\lambda) \ . 
\label{eq_2:48}
\end{equation}
Substituting this result into Eq.~(\ref{eq_2:47a}) and interchanging
the average over $\lambda$ and the differentiation with respect to
${\hat{\theta}}$, one finds that $\overline{Z}$ and
$\overline{Z}^T$ obey the same differential
equation (\ref{eq_2:46}).  
Consideration of the initial conditions in both cases \cite{Bro96}
establishes the equivalence of both methods.

This insight implies, however, that the results for the GSE in
Refs.~\onlinecite{Zir92,Mir94} are incorrect. (We recall that these
results differ from those of the DMPK approach). Inspection shows that
the finite (i.e., non--zero) limiting values ($L \rightarrow \infty$)
for the first two moments of the conductance are technically due to a
zero mode of the Laplacean (\ref{eq_2:32}). In Ref.~\onlinecite{Bro96}
it was shown that in the GSE case, this zero mode is not a
single--valued function on the space of $Q$ matrices, and that it does
not exist when the Kramers degeneracy is correctly taken into account.
Therefore, the contribution due to this mode in
Refs.~\onlinecite{Zir92,Mir94} to the Fourier expansions and to the
expressions for the moments of the conductance, has to be omitted. In
this way, corrected expressions for the symplectic symmetry class were
obtained in Ref.~\onlinecite{Bro96}.  These expressions show
localization as expected from the DMPK equation.

\subsection{Random band matrices}
\label{rbm}

Originally, Random Matrix Theory was introduced by defining a number
of matrix ensembles which describe generic features of many
microscopically different systems. An important ingredient was the
postulated invariance (orthogonal, unitary or symplectic) in Hilbert
space. This concept was shown to have enormous power: Physical systems
are grouped into few universality classes characterized by very
general symmetry properties. Many examples were given in the previous
section where the three Gaussian ensembles played a prominent role. It
is natural to ask whether generalized ensembles of random matrices
exist from which the generic properties of {\it extended} systems can
likewise be generated. These ensembles must, of course, contain the
linear dimensions of the system in parametric form. The IWZ model
described in Sec.~\ref{quasi1d_3} suggests that such ensembles do
indeed exist, 
at least for quasi 1d wires, and that they differ from the classical
ensembles by relinquishing the requirement of (orthogonal, unitary or
symplectic) invariance. We now show that quasi 1d wires can be
modeled by an ensemble of random band matrices (RBMs). Early
investigations of RBMs are summarized in chapter 20 of Mehta's
book~\cite{Meh91}.

A band matrix is characterized by the dimension $N$ and by the
bandwidth $b$. The latter defines the distance from the main diagonal
beyond which the matrix elements become either zero or are negligibly
small. An ensemble of RBM's is obtained by assuming that the matrix
elements within the central band are uncorrelated Gaussian distributed
random varibles with zero mean and a variance which effectively
defines the energy scale. As long as $b\approx N$ the RBM ensembles
are clearly equivalent to Wigner's Gaussian ensembles, while for $b =
{\cal O}(1)$ we deal with an almost diagonal matrix with uncorrelated
eigenvalues and strongly localized eigenfunctions. This shows that
RBMs represent one way to interpolate between chaotic systems with
Wigner--Dyson statistics and Poisson regularity 
(see Refs.~\onlinecite{Guh96a,Guh96b,Guh96}
for another possibility). The wish to describe intermediate level
statistics was indeed the original motivation which led Seligman,
Verbaarschot, and Zirnbauer \cite{Sel84} to the investigation of RBM
ensembles. Many numerical investigations of these ensembles by Casati
{\it et al.} and Wilkinson {\it et al.} followed
\cite{Cas90b,Cas90,Cas90a,Wil91}. 

In Sec.~\ref{quasi1d_3}, we have introduced the Hamiltonian of the IWZ
model 
\cite{Iid90}. As mentioned above, this Hamiltonian, although endowed
with some fine structure, has the general form of a random band matrix. 
Another physical system that has been mapped
on a random band matrix is the kicked rotor
\cite{Boh84,Izr90,Cas90}. It has
been argued \cite{Boh84} that in a certain basis the evolution
operator $S$ connecting values of the wavefunction before and after a
single  kick, exhibits the structure of a RBM. 
For the kicked rotor, one classically expects unbounded
diffusion in angular momentum space once the strength of the kick
exceeds a critical value. Quantum mechanically, however, 
``dynamical localization'' sets in and the diffusion is eventually
suppressed, see Sec.~\ref{qc2dl}.
A first connection between dynamical and  Anderson localization was
established in Ref.~\onlinecite{Fis82}.
Recently, the correspondence between the kicked rotor and RBMs 
(and hence the correspondence between dynamical and Anderson
localization) has been made more
exact. Altland and  Zirnbauer \cite{All96a} have developed
a field theory for the time--dependent Hamiltonian
\begin{equation}
H(t) = \frac{\tilde{l}^2}{2} + k \cos(\tilde{\theta} +a)
\sum_{n=-\infty}^\infty \delta(n\tau - t) \ , 
\label{eq_add1}
\end{equation}
which defines the quantum kicked rotor. This approach is still under
discussion, see Ref.~\onlinecite{Cas97}.  The quantities $\tilde{l}$
and $\tilde{\theta}$ are the operators corresponding to the angular
momentum $l$ and the angular coordinate $\theta$, $k$ determines the
kick strength and $\tau$ its period, and $a$ is a symmetry breaking
parameter similar to a magnetic field. Using a novel variant of the
Hubbard Stratonovich transformation the authors of
Ref.~\onlinecite{All96a} claim that the long wave--length behavior of
Eq.~(\ref{eq_add1}) is described by the one--dimensional non--linear
$\sigma$ model. As we will see, the same is true for RBMs. These two
examples suffice to show how much insight can be gained by analysing
RBMs.

Numerical simulations \cite{Cas90a} have shown that the fluctuation
properties of observables derived from RBMs are governed by the
scaling parameter $\tilde{x} = b^2/N$. For completeness we mention
that a second scaling parameter arises if the diagonal elements of the
RBM are allowed to increase linearly in size with increasing index
\cite{Wil91}. Fyodorov and Mirlin \cite{Fyo91} studied RBMs
analytically and derived a non--linear $\sigma$ model which turned out
to be identical to the one found by Efetov \cite{Efe83} for
non--interacting electrons in a quasi 1d wire. This proved the close
correspondence between RBMs and random quasi 1d systems. If one
associates the matrix size $N$ with the length $L$ of the sample and uses
the result (see below) that the localization length $\xi$ scales like $b^2$,
it becomes clear that $\tilde{x}$ is proportional to the conductance
$g=\xi/L$. We now sketch the proof.

The ensemble of $N\times N$ real and symmetric band matrices $H$ has
independent Gaussian distributed random elements $H_{ij}=H_{ji}$ with
zero mean value and variance  
\begin{equation}
\overline{ H_{ij}^2 } = \frac{1}{2} A_{ij} [1+\delta_{ij}] \ .
\label{eq_5:1}
\end{equation}
Here, $A_{ij}$ decays (sufficiently fast) on the scale $b$. The
asymptotic behavior for large $|l-m|$ of the two--point correlation
function 
\begin{equation}
K_{lm}  = \overline{\left[\left(E+i\epsilon - H\right)^{-1}
\right]_{ll} \ \left[ \left(E-i\epsilon - H\right)^{-1}
\right]_{mm} }
\label{eq_5:2}
\end{equation}
defines the localization length $\xi$. With the help of the averaged
supersymmetric functional 
\begin{eqnarray}
\overline{ Z[J] } &=& \int d[\sigma] \, \exp\left(-S[\sigma,J]\right)
 \ , \nonumber\\ 
S[\sigma,J] &=& \frac{1}{2} {\rm trg} \sum_{ij} \sigma_i (A^{-1})_{ij}
\sigma_j + \frac{1}{2} {\rm trg} \sum_i \ln (E-\sigma_i + i\epsilon L +
J_i )
\label{eq_5:3}
\end{eqnarray}
the average of the correlation function $K_{lm}$ can be expressed as
\begin{equation}
K_{lm}  = \frac{1}{4} \frac{\partial^2 \overline{Z[J]}}{\partial J^1_l
  \partial J^2_m} \ .
\label{eq_5:4}
\end{equation}
In Eq.~(\ref{eq_5:3}), the $\sigma_i$ ($i=1,\ldots,N$) are a set of
$8\times 8$ supermatrices (not yet restricted to the saddle--point
manifold) and $J$ is a suitably defined \cite{Fyo91} source field with
elements $\pm J^{(1,2)}_l$ $(l=1,\ldots,N)$. The saddle--point
approximation is justified because $N\gg 1$ and $b\gg 1$. In the
continuum limit one defines $x/d=|l-m|$ where $d$ is some suitable
small length scale, e.g. the lattice constant and arrives at
\begin{eqnarray}
K_{lm}  &=& \left(\frac{\pi\nu d}{4}\right)^2
               \int d[Q] \, F(Q(0),Q(x))
               \exp\left(-S[Q]\right) \  , \nonumber\\
S[Q] &=& -\frac{\pi\nu}{8} \int {\rm trg} \left[
 {\cal D} (\nabla Q)^2 + 2i\omega LQ \right] dx.
\label{eq_5:5}
\end{eqnarray}
Here $\nu$ is the density of states per unit length. To define the
diffusion constant ${\cal D}$, we write $A_{ij} = a(|i - j|) =
a(x/d)$, define $B_2 = (1/2) \sum a(r) r^2 \propto b^2$ and have
${\cal D} = \pi\nu d B_2$.  The precise form of the preexponential
term $F(Q(0),Q(x))$ in Eq.~(\ref{eq_5:5}) can be found in
Ref.~\onlinecite{Fyo91}. The $8\times 8$ supermatrix $Q$ belongs to
the saddle--point manifold discussed in
Refs.~\onlinecite{Efe83,Ver85}. Thus, the one--dimensional $\sigma$
model of Eq.~(\ref{eq_5:5}) is identical to the one of
Ref.~\onlinecite{Efe83}. The localization length for this model has
been calculated in Ref.~\onlinecite{Efe83} and is given by
\begin{equation}
l_{\rm loc} \equiv \xi/d = 4\pi\nu D / d \propto b^2 \ .
\label{eq_5:6}
\end{equation}
Invoking the one--parameter scaling hypothesis \cite{Abr79} we see
that the scaling properties of a RBM can depend only on the ratio of
localization length and matrix size $N$. According to
Ref.~\onlinecite{Fyo91}, this explains the scaling parameter $b^2/N$
found numerically in \cite{Cas90a}.

In Refs.~\onlinecite{Fyo92,Fyo93,Mir93a,Mir93}, the investigation of
RBMs was extended to the statistical properties of the corresponding
eigenfunctions. The main results are summarized in
Ref.~\onlinecite{Mir94a}.  First, Fyodorov and Mirlin showed how to
formulate various statistical observables involving eigenfunctions in
the framework of the non--linear $\sigma$ model, i.e. for arbitrary
dimension of the disordered sample.  Second, they solved the $\sigma$
model in one dimension and calculated the distribution functions of
the following three quantities for arbitrary system length $L$: the
$n$th component $\psi_n^{(k)}$ of the $k$th eigenfunction, the inverse
participation ratio
\begin{equation}
P_2 = \sum_n |\psi_n^{(k)}|^4 \ ,
\label{eq_5:7}
\end{equation}
and the product $r(L)=|\psi_0^{(k)}\psi_L^{(k)}|^2$ of amplitudes
taken from opposite ends of the wire. The distribution of $r(L)$
turned out to be log--normal in the insulating regime. This implies a
Gaussian distribution for the localization lengths.

Technically, these results were derived by expressing the desired
observable, in particular its moments, in terms of arbitrarily high
products of Green functions. Employing the supersymmetry formalism
these moments could then be expressed in terms of (the discrete
variant of) the $\sigma$ model in Eq.~(\ref{eq_5:5}).
The one--dimensional
chainlike structure of the $Q$ matrices suggested the use of a method
\cite{Efe85} patterned like the transfer matrix approach. This led to
a recurrence relation and eventually to a comparatively simple
differential equation \cite{Fyo92}. In subsequent papers
\cite{Fyo93,Mir93,Mir93a} appropriate generalizations of this method
were used.

\subsection{Transport in higher dimensions}
\label{akl}

Many of the rigorous results reviewed so far in this section on
transport properties are restricted to quasi one--dimensional
systems. This is mainly due to the fact that the $\sigma$ model has
been well under control in the one--dimensional case but not in
higher dimensions. Very recently, however, important progress has
been made, and interesting information was obtained, from the
$\sigma$ model for $d=2$ and $d=3$. It is useful to briefly recall
the history of the non--linear $\sigma$ model and of scaling theory
prior to a discussion of this development. We thereby extend the
historical remarks of Sec.~\ref{loc}. For simplicity of notation, and
deviating from former usage, we use in the present subsection $g$ for
the average conductance and $\delta g$ for the difference between the
conductance and its average value. 

\subsubsection{Scaling theory and distributions}
\label{akl_1}

In Sec.~\ref{loc} it was mentioned that early scaling ideas by Thouless
\cite{Tho74} were developed into a one--parameter scaling theory by
Abrahams {\it et al.} \cite{Abr79}. This theory received strong
support from perturbative calculations by Anderson, Abrahams, and
Ramakrishnan \cite{And79}, Abrahams and Ramakrishnan \cite{Abr80},
Gorkov, Larkin, and Khmelnitskii \cite{Gor79}, and from
field--theoretical considerations involving a non--linear
$\sigma$ model by Wegner \cite{Weg79}. His field--theoretical
approach, subsequently developed in a series of papers
\cite{Efe80,Sca80,Hou80,Hik81,Sca82,Efe82}, marked the beginning of a 
development that led to the omnipresence of the $\sigma$ model (now in
its supersymmetric form \cite{Efe83}) in modern discussions of
disordered systems. 

Scaling theory determines the dependence of the dimensionless
conductance $g(L)$ on the linear dimension $L$ of the system. The
initial condition is set by the strength of the disorder as measured 
by the conductance $g_0 = g(\ell)$ at the scale of the elastic mean
free path $\ell$. The following general picture emerged. For large $g$
the $\beta$ function has the form
\begin{equation}
\beta(g) = \frac{d\ln g}{d\ln L} = \frac{L}{g}\frac{dg}{dL} = (d-2) -
\frac{a}{g} 
\label{eq_6:1}
\end{equation}
where the constant $a$ is of order unity. The
$\beta$ function is seen to depend on $g$ as the single parameter,
hence the term one--parameter scaling. For $\beta(g)>0$ the conductance
increases with increasing system size and we deal with a conductor
supporting extended electron states. For $\beta(g)<0$, however, the
conductance scales to zero with growing $L$, and we are in an insulating
regime with localized eigenfunctions. As long as $d\le 2$, $\beta(g)$
is always negative and one expects all states to be localized. In
three dimensions, however, the asymptotic value of $g$ depends on the
initial value $g_0$ of the conductance. We introduce (for $d=3$) the
critical conductance $g_c\equiv a$, which is defined by the condition
$\beta(g_c)=0$. For $g_0>g_c$ we have a
conductor and for $g_0<g_c$, an insulator. The point $g_0=g_c$ defines
the mobility edge or metal--insulator transition. This picture is
qualitative but in a nutshell represents today's view of disordered
systems. However, reasons have emerged to call into question the basic
tenets of one--parameter scaling. This has come about as follows.

Altshuler, Kravtsov, and Lerner \cite{Alt86a,Alt88a,Alt91a} emphasized
that the transport properties of a disordered sample cannot be
characterized by the mean conductance $g$ alone for which the
above--mentioned scaling theory had been formulated. Instead,
anomalously large sample--to--sample fluctuations have to be accounted
for even in the weak localization regime (see Sec.~\ref{quasi1d_1} on
universal conductance fluctuations, UCF). It was therefore proposed in
Ref.~\onlinecite{Alt86a} to study the whole distribution of the
conductance, and of the density of states, and to develop a scaling
theory for all moments of these quantities. Before we briefly outline
the method used to pursue this program we summarize the main results
\cite{Alt86a,Alt91a} on conductance fluctuations some of which
contradict the one--parameter scaling hypothesis.

We focus attention on the $n$th {\it cumulants} $\langle (\delta G)_c^n
\rangle$ of the conductance $G = (e^2 / \hbar) g$. For $n<g_0$ it was
found that  
\begin{equation}
\langle (\delta G)_c^n \rangle \propto g^{2-n} \left( \frac{e^2}{\hbar}
\right)^n\ .
\label{eq_6:2}
\end{equation}
The first cumulant is equal to $g$ itself and changes with $L$
according to Eq.~(\ref{eq_6:1}). The second cumulant, i.e. the
variance of $G$, does not depend on $g$ and therefore stays constant
in the metallic regime (UCF). Higher moments are small by powers of
$g^{-1}$ and can be neglected for $g\gg 1$. Hence, for not too large
fluctuations $\delta g$ and for $g\gg 1$, the distribution function
is almost Gaussian,
\begin{equation}
f(\delta g) \propto \exp\left[ - (\delta g)^2\right] \ ,
\label{eq_6:3}
\end{equation}
and one--parameter scaling applies: The distribution function is
completely determined by the mean conductance. Close to the localized
regime with $g\approx 1$, however, all higher moments are of
comparable size and the distribution function is strongly
non--Gaussian. It was emphasized in Refs.~\onlinecite{Alt86a,Alt91a}
that this picture is not entirely correct because it neglects certain
classes of perturbative corrections to the high--order cumulants of
the conductance. Thus, deviations from the Gausssian form
(\ref{eq_6:3}) must occur in the distant tails of the distribution
even in the regime of weak localization where $g\gg 1$. In these
tails, the distribution turned out to be logarithmically normal,
\begin{equation}
f(\delta g) \propto \frac{1}{\delta g} \exp\left[ -\frac{1}{4u} \ln^2
\left( \frac{\delta g}{g} \frac{1}{\tau\Delta} \right)
\right] 
\label{eq_6:4}
\end{equation}
where $u = \ln\left(\sigma_0 /\sigma \right)$, where $\sigma$ is
the conductivity of the system with linear dimension $L$, and where
$\sigma_0$ is the classical (Drude) conductivity. (We recall that
$\tau$ is the elastic mean free time and $\Delta$ is the average 
single--particle level spacing). In two dimensions, $\sigma$ is
identical to the conductance $g(L)$ while $\sigma_0$ corresponds to
$g_0$.  

The non--Gaussian tails in Eq.~(\ref{eq_6:4}) can be traced
back to additional contributions to the cumulants of $g$ arising from
the above--mentioned perturbative corrections,
\begin{equation}
\langle (\delta G)_c^n \rangle_{\rm add} \propto
g_0 \left(\frac{\ell}{L}\right)^{2n-d} e^{u(n^2-n)}
\left(\frac{e^2}{h}\right)^n  \ .
\label{eq_6:6}
\end{equation}
For $n<g_0$ these terms are insignificant but for $n>g_0$ they
dominate those in expression (\ref{eq_6:2}) and lead to the
log--normal tails in the expression (\ref{eq_6:4}). The tails dominate
the distribution function (i) in the metallic regime $g\gg 1$ for
fluctuations $\delta g \ge \sqrt{g_0}$ and (ii) in the region of
strong localization for fluctuations $\delta g \ge \sqrt{g_0}/\ln
g_0$. Exact results for the distribution function in one dimension
\cite{Mel81,Abr81,Alt87} coincide with formula (\ref{eq_6:4}). An
important point in this expression is the dependence on $u$, i.e. on
both $\sigma$ and $\sigma_0$. This introduces another microscopic
parameter in addition to the scale--dependent conductance $g = \sigma
L^{d-2}$. This is at variance with the hypothesis of one--parameter
scaling in its strict form. We briefly return to this issue later.
A formally similar violation of one--parameter scaling had earlier
been observed by Kravtsov and Lerner \cite{Krv85} in their
investigation of the frequency dependence of the diffusion constant
${\cal D}$. With the substitutions $\delta g \to \delta\nu $ and $g
\to \langle\nu\rangle$, the asymptotic form of the distribution of the
density of states is also given by Eq.~(\ref{eq_6:4}).

\subsubsection{Renormalization group analysis}
\label{akl_2}

First formal investigations employed the replica trick \cite{Edw75}
described in Sec.~\ref{loc}. We recall that this method circumvents
the problem of normalization of the generating functional for the
Green functions by considering $k$ replicas of the system under study.
At some point in the calculation the limit $k \rightarrow 0$ is taken.
This procedure is fully justified only within perturbative treatments
\cite{Ver85a}. This, however, suffices for the present purpose. In
Refs.~\onlinecite{Alt86a,Alt88a,Alt91a,Krv84a,Krv84b,Krv88,Krv89},
moments of both the conductance and the density of states were
considered. For the former, the replica functional reads
\begin{eqnarray}
\langle G^n \rangle &\propto& \prod_{i=1}^n \ 
{\rm tr} \left[ \frac{\partial^2}{\partial h_i^2}
  \frac{ \int d[Q] \exp(-S[Q,h])}{\int d[Q] \exp(-S[Q,h=0])}
         \right] \Bigg\vert_{h=0,k=0} \ ,   \nonumber\\
S[Q,h] &=& \frac{1}{t} \int {\rm tr} \left( \nabla Q - \frac{1}{L}
[h,Q] \right)^2 d^dr \ .
\label{eq_6:7}
\end{eqnarray}
Here, $t^{-1} = \pi\nu {\cal D}/8 \propto g$, $Q$ is a matrix field in
replica space containing commuting entries. The matrix $Q$ is
analogous to the supermatrix $Q$ in previous sections, and $h$ is a
source matrix with components $h_i$. The detailed structure of $Q$ and
$h$ is given in Ref.~\onlinecite{Alt86a}. The moments of the density
of states are represented analogously,
\begin{eqnarray}
\langle \nu^n(\epsilon)\rangle &\propto& \prod_{i=1}^n
\frac{\partial}{\partial\omega_i} 
\frac{\int d[Q] \exp(-S[Q,\omega])}{\int d[Q] \exp(-S[Q,\omega=0])}
\Bigg\vert_{\omega=0,k=0} \ , \nonumber\\
S[Q,\omega] &=& -\frac{1}{2} \int \, {\rm tr}(\omega LQ)
\frac{d^dr}{L^d} \ .
\label{eq_6:8}
\end{eqnarray}
The components $\omega_i$ of the source matrix $\omega$ are defined in
Ref.~\onlinecite{Alt86a}. The replica matrix $Q$ in
Eqs.~(\ref{eq_6:7}) and (\ref{eq_6:8}) is subject to the same
constraints as its supersymmetric counterpart, namely $Q^2 = 1$ and
${\rm tr}\, Q = 0$. The combination of the two actions in
Eqs.~(\ref{eq_6:7}) and (\ref{eq_6:8}) yields
\begin{equation}
S[Q,h,\omega] = \int \, {\rm tr} \left[
\frac{1}{t}\left(\nabla Q - \frac{1}{L}[h,Q] \right)^2 -
\frac{1}{2L^d}\omega LQ \right] d^dr
\label{eq_6:9}
\end{equation}
Both formally and in physical content the action $S$ in
Eq.~(\ref{eq_6:9}) corresponds to the supersymmetric action
(\ref{eq_4:022}) in the presence of a magnetic field. The source
matrix $h$ plays the role of the symmetry--breaking term $\propto
\tau_3$, and $\omega$ is analogous to the frequency in
Eq.~(\ref{eq_4:022}). 

Equations~(\ref{eq_6:7}) to (\ref{eq_6:9}) are based on the usual
gradient expansion (long wave--length limit). The authors of
Refs.~\onlinecite{Alt86a,Alt88a,Alt91a,Krv84a,Krv84b,Krv88,Krv89} made
the important observation that additional higher order terms, in
particular higher--order gradient terms, have to be included in the
action (\ref{eq_6:9}). To define the basic structure of these terms we
introduce the covariant derivative
\begin{equation}
{\cal D}_\alpha = \nabla_\alpha - \frac{1}{L}[h^\alpha,\cdot] \ ,
\label{eq_6:9a}
\end{equation}
where the index $\alpha$ refers to the direction in coordinate space. 
We suppress details of the spatial dependence. Then the additional
contributions to the action (\ref{eq_6:9}) are given by 
\begin{eqnarray}
S_n^{\rm add}[h] &=& z_n \int {\rm tr}({\cal D}Q)^{2n} d^dr
\label{eq_6:9b}\\
S_m^{add}[\omega] &=& Y_m \int {\rm tr}(\omega LQ)^m d^dr
\label{eq_6:10}
\end{eqnarray}
and
\begin{equation}
S_{n,m} \, = \, z_{n,m} \int {\rm tr} \left[ ({\cal D}Q)^{2n}(\omega
LQ)^l d^dr \right] 
\label{eq_6:11}
\end{equation}
with certain (bare) coupling constants $z_n$, $Y_m$, and $z_{n,m}$.

The perturbative analysis of the replica functional proceeds as
indicated after Eq.~(\ref{eq_2:29e}) for the supersymmetric functional. 
In the perturbation series generated from the action (\ref{eq_6:9})
{\it without} the additional terms in Eqs.~(\ref{eq_6:9b}) to
(\ref{eq_6:11}), a certain class of terms appears. In two dimensions,
these terms are logarithmically divergent (in $L/\ell$). In principle,
such terms can be summed effectively with the help of a
renormalization group analysis of the generating functional. This
would lead to a change of the prefactors of the vertices in
Eq.~(\ref{eq_6:9}). However, considerations involving particle number
conservation, the Einstein relation, and the continuity equation prove
that the action (\ref{eq_6:9}) {\it does not change its form under
renormalization }. It therefore
depends on the single parameter $t^{-1}\propto g$. This result is
consistent with the one--parameter scaling hypothesis \cite{Alt86a}.
However, higher moments of both the conductance and the density of
states are determined \cite{Alt86a,Alt91a} by the higher--order vertices
(\ref{eq_6:9b}), (\ref{eq_6:10}), and (\ref{eq_6:11}) which introduce
a large number of parameters, $z_n$, $Y_m$, and $z_{n,m}$. The
additional contributions (\ref{eq_6:6}) to the higher moments of the
conductance and, therefore, to the asymptotics (\ref{eq_6:4}) of the 
distribution function for g are due to such higher--order terms. The
number of diagrams originating from these terms is so high and their
structure so complicated that a direct summation appears to be hopeless. 
Therefore, one has to perform the renormalization group analysis for
the extended $\sigma$ model including the terms (\ref{eq_6:9b}),
(\ref{eq_6:10}), and (\ref{eq_6:11}). 

The principle of such an analysis is the following. In order to
distinguish between slow and fast variables, the $Q$ matrix is 
parametrized in the form
\begin{equation}
Q = \tilde{U}^\dagger Q_0 \tilde{U} = \tilde{U}^\dagger U_0^\dagger L
U_0 \tilde{U} \ .
\label{eq_6:11a}
\end{equation}
Here, the fast variables are given by $Q_0 = U_0^\dagger L U_0$ and
the slow ones by $\tilde{Q} = \tilde{U}^\dagger L
\tilde{U}$. Integration over the fast variables yields a functional of
the slow modes $\tilde{Q}$ alone which serves as a basis for the next
iteration of this procedure. Lowest--order perturbation theory for the
fully renormalized functional results in an effective summation of the
higher--order terms in the expansion of the ``bare'' functional. 

Using the results obtained in this way \cite{Alt86a,Alt88a,Alt91a},
several authors \cite{Alt86a,Alt88a,Alt91a,Kum86,Sha86,Sha87,Mut87}
have addressed the question whether or not one--parameter scaling
applies to mesoscopic fluctuations. To some extent this is a question
of definition \cite{Alt91a}. As we have seen, the asymptotic tails of
the distribution functions for $g$ and $\nu$ depend on the bare
conductivity $\sigma_0$ as an additional microscopic parameter. 
Therefore, these tails are not determined by $\sigma$ (or $g$) alone
and one might say that one--parameter scaling is violated. It turns
out, on the other hand, that renormalization of the parameters $z_n$,
$Y_m$, and $z_{n,m}$ in the higher--order terms does not influence the
coupling constant $1/t$ in Eq.~(\ref{eq_6:7}). One could argue that
{\it this} is the essence of one--parameter scaling. A more detailed
discussion can be found in Ref.~\onlinecite{Alt91a}. 

\begin{figure}
\centerline{
\psfig{file=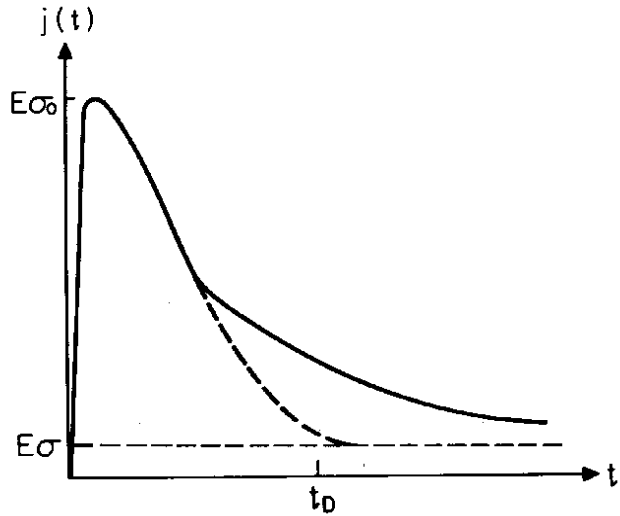,width=3.5in}
 }
\caption{
Schematic current response of a disordered conductor to
an electric field of the form $E(t) = E\Theta(t)$, where $\Theta(t)$
is the step function. For times greater than the diffusion time $t_d$,
higher quantum corrections lead to log--normal (rather than
exponential) decay towards the equilibrium value $E\sigma$.
Taken from Ref.~\protect\onlinecite{Altsh91}.
} 
\label{afig5}
\end{figure}

The mesoscopic fluctuations just discussed are intimately related to
current--relaxation processes \cite{Alt87a,Alt88a}. This is seen by
considering the distribution function of the time--dependent
conductance $G(t)$ which describes the response of the current to a
$\delta$ function pulse (in time) in the potential. This distribution
function also exhibits log--normal tails as in Eq.~(\ref{eq_6:4}), cf.
Fig.~\ref{afig5}. We
return to this point in the context of the supersymmetric
$\sigma$ model, see Sec.~\ref{akl_3}.

The {\it local} density of states $\nu(\vec{r},\epsilon) = -{\rm
Im}G^+(\vec{r},\vec{r};\epsilon)/\pi$, of particular interest for the
metal--insulator transition, was investigated by Lerner, Altshuler,
and Kravtsov \cite{Ler88,Alt91a}. In the metallic regime the
distribution of $\nu(\vec{r},\epsilon)$ turned out to be Gaussian with
log--normal tails, as discussed above for the conductance and the
density of states. In the vicinity of the metal--insulator transition,
however, the distribution was found to be completely log--normal
\cite{Ler88,Alt91a}. Recently, McCann and Lerner \cite{McC95} showed
that this log--normal distribution is linked to multifractal 
exponents for the cumulants of the local density of states. 
Multifractal analysis had been pioneered by Wegner \cite{Weg80} who
found a whole set of critical exponents for the moments of the inverse
participation ratio at the metal--insulator transition, and by 
Castellani and Peliti \cite{Cat86} who suggested that the electron
wave functions at the transition are multifractal. We now describe
recent work which has brought multifractal analysis and properties of
asymptotic distributions within reach of the non--linear $\sigma$ model.

\subsubsection{Supersymmetry approaches}
\label{akl_3}

Muzykantskii and Khmelnitskii \cite{Muz95} introduced the novel idea
that a non--trivial saddle--point configuration of the usual
$\sigma$ model in Eqs.~(\ref{eq_4:01}) and (\ref{eq_4:02}) 
determines the asymptotic tails of the distribution of mesoscopic
quantities in one, two, and three dimensions. Starting from Ohm's law
in its time--dependent form,
\begin{equation}
I(t) = \int_{-\infty}^t dt' \, G(t-t')V(t') \ ,
\label{eq_6:12}
\end{equation}
these authors expressed the average time--dependent conductance $G(t)$
in terms of the usual supersymmetric functional, 
\begin{equation}
G(t) = \int\frac{d\omega}{2\pi} \exp(-i\omega t)\int d[Q] P(Q)
\exp(-S[Q]) \ ,
\label{eq_6:13}
\end{equation}
where the action $S[Q]$ is given by Eq.~(\ref{eq_4:02}). The classical
(Drude) term has been suppressed. The next steps can probably be best
understood when viewed as arising from a saddle--point approximation
with respect to both $Q$ {\it and} $\omega$ of the combined action
\begin{equation}
S_{\rm comb}[Q,\omega] = -\frac{\pi\nu}{8} \int {\rm trg}[D(\nabla Q)^2
+ 2i\omega LQ] d^dr + i\omega t \ .
\label{eq_6:14}
\end{equation}
Variation of Eq.~(\ref{eq_6:14}) with respect to $\omega$ yields the
``self--consistency equation'' \cite{Muz95}
\begin{equation}
\int \frac{d^dr}{V} {\rm trg}[LQ] = \frac{4 t\Delta}{\pi} \ .
\label{eq_6:15}
\end{equation}
Variation with respect to $Q$ under the constraint $Q^2=1$ leads to an
equation reminiscent of the Eilenberger equation \cite{Eil68} for
dirty superconductors, 
\begin{equation}
2D\nabla(Q\nabla Q) + i\omega [L,Q] = 0 \ .
\label{eq_6:16}
\end{equation}
This saddle--point approximation should not be confused with the one
necessary to derive the $\sigma$ model of Eqs.~(\ref{eq_4:01}) and
(\ref{eq_4:02}). While the latter restricts the $Q$ matrices to the
coset space with $Q^2=1$, the present one fixes the spatial variation
of the $Q$ matrices within this coset space. The matrix equation
(\ref{eq_6:16}) as well as Eq.~(\ref{eq_6:15}) can be reduced to two
equations involving only one scalar function $\theta(\vec{r})$ 
\cite{Muz95}. Boundary conditions on $\theta$ determine the solutions
for $d=1,2$ and $3$. 

Two different regimes are distinguished by values of the parameter
$t\Delta$ which defines the size of the r.h.s. of the self--consistency
equation (\ref{eq_6:15}). Physically, $t\Delta$ determines whether
individual states in the mesoscopic probe can be resolved. For
$t\Delta \ll 1$ and with $t_D$ the diffusion time, $G(t)$ is
proportional to $\exp(-t/t_D)$, irrespective of the dimensionality. In
the opposite limit $t\Delta\gg 1$ the results depend on $d$. For $d=1$
one finds $G(t)\propto \exp(-g\ln^2(t\Delta))$ and for $d=2$,
$G(t) \propto (t\Delta)^{-g}$. Here $g$ is the dimensionless
conductance. These results do not apply for arbitrarily long times
because beyond a certain limit, $t>t^*$, the diffusion approximation
was found to break down \cite{Muz95}. In this extremely asymptotic
regime the function $\theta$ fluctuates on a scale exceeding the
elastic mean free path: The spatial derivative obeys $\theta' >
1/\ell$, and this contradicts the conditions under which $\theta$ is
constructed. Therefore, the ultra--long time limit in $d=1,2$
and the regime $t\Delta\gg1$ in $d=3$ seemed to be inaccessible and
the comparison with the results of Altshuler, Kravtsov, and Lerner
\cite{Alt91a} was inconclusive.

Mirlin \cite{Mir95} observed, however, that the regime $t>t^*$ can be
reached if the additional constraint $\theta'<1/\ell$ is taken into
account in solving Eqs.~(\ref{eq_6:15}) and (\ref{eq_6:16}). This
modifies the boundary conditions employed in Ref.~\onlinecite{Muz95}.
For a two--dimensional disk of radius $R$ Mirlin found the following
asymptotic behavior in time,
\begin{equation}
G(t) \propto \exp \left( -\frac{\pi\beta g}{4}
\frac{\ln^2(t/g\tau)}{\ln(R/\ell)} \right) \qquad\qquad (t\Delta \gg
(R/\ell)^2) \ .
\label{eq_6:17}
\end{equation}
This expression agrees exactly with the (slightly amended) result of
Ref.~\onlinecite{Alt91a}.

The calculation of eigenfunction statistics, in particular the
distribution of eigenfunction components, in the framework of the
supersymmetric non--linear $\sigma$ model has been pioneered by Fyodorov
and Mirlin \cite{Mir93,Mir94a,Mir94b,Mir95a,Mir96}. 
We have already mentioned some of their work in
Sec.~\ref{rbm}. The main results for the distribution function $f(u)$
of the eigenfunction amplitudes $u=|\psi(\vec{r}_0)|^2$ are the
following \cite{Mir96}. We consider the function
\begin{equation}
Y(Q_0) = \int_{Q(\vec{r}_0=Q_0} d[Q] \exp(-S[Q]) \ ,
\label{eq_6:17a}
\end{equation}
where $Q(\vec{r}_0)=Q_0$ is held fixed at the observation point
$\vec{r}_0$, and the action $S[Q]$ is essentially given by
Eq.~(\ref{eq_4:02}) (see \cite{Mir96} for the precise notational
conventions). In the limit of vanishing frequency $\omega$, 
and for both orthogonal ($\beta=2$) and unitary ($\beta=1$) symmetry,
this function depends on $Q_0$ only via the radial parameter
$\lambda_1$, $Y(Q_0) \rightarrow \tilde{Y}(-2\beta i \nu \omega
\lambda_1)$. In terms of $\tilde{Y}(z)$ the desired distribution
functions can be expressed as \cite{Mir93,Mir94b}
\begin{eqnarray}
f(u) &=& \frac{1}{V} \frac{d^2}{du^2} \tilde{Y}(u) \hspace{4.2cm}
(\beta=2) \ ,\nonumber\\
f(u) &=& \frac{2\sqrt{2}}{\pi V u^{1/2}} \frac{d^2}{du^2} \int_0^\infty
\frac{dz}{z^{1/2}} \tilde{Y}(z+u/2) \qquad (\beta=1) \ ,
\label{eq_6:17b}
\end{eqnarray}
where $V$ is the volume of the system.
These general formulas are valid for arbitrary dimension. In the limit
of a constant supermatrix $Q$ (zero mode approximation) the functional 
Eq.~(\ref{eq_6:17a}) becomes equivalent to Gaussian Random Matrix
Theory, see Sec.~\ref{spec_1}. In this case we obtain the
Porter--Thomas distribution discussed in detail in Sec.~\ref{qc1wwa},
\begin{eqnarray}
f(u) &=& V \exp(-uV) \hspace{2.4cm} (\beta = 2) \ , \nonumber\\
f(u) &=& \sqrt{\frac{V}{2\pi u}} \exp(-uV/2) \qquad\quad (\beta=1) \ .
\label{eq_6:17c}
\end{eqnarray}

For quasi 1d systems, the integral in Eq.~(\ref{eq_6:17a}) can be solved
exactly \cite{Mir93}. With $x=L/\xi \ll 1$ the ratio of system length and
localization length, and the coefficient $\alpha = 2(1-3L_-L_+/L^2)$,
where $L_-$ and $L_+$ are the distances from $\vec{r}_0$ to the sample
edges, the distribution function for $y=uV$ reads 
\cite{Mir93,Mir94a,Mir95a,Mir96}
\begin{eqnarray}
f^{(\beta=2)}(y) &=& \exp(-y)\left[1+\frac{\alpha x}{6}(2-4y+y^2)+
\ldots\right]  \hspace{2cm} (y\le x^{-1/2}) \ ,  \nonumber\\
f^{(\beta=1)}(y) &=& \frac{\exp(-y/2)}{\sqrt{2\pi y}}
 \left[ 1+ \frac{\alpha x}{6} (3/2 - 3y + y^2/2) + \ldots \right]
\hspace{0.7cm} (y\le x^{-1/2})  \ , \nonumber\\
f(y) &\propto&
\exp\left(\frac{\beta}{2}\left[-y+\frac{\alpha}{6}y^2x+\ldots
\right]\right)  \hspace{2.3cm} (x^{-1/2}\le y\le x^{-1}) \ ,
\nonumber\\
f(y) &\propto& \exp\left(-2\beta\sqrt{y/x}\right) \hspace{4.3cm}
(y\ge x^{-1}) \ .
\label{eq_6:17d}
\end{eqnarray}

In $d>1$ dimensions only approximate methods to evaluate
Eq.~(\ref{eq_6:17b}) are available. Let us define the parameter
$\kappa = \sum_{\vec{q}} (2\pi\nu{\cal D} \vec{q}^{\, 2} V)^{-1}$, where
the summation extends over the eigenmodes of the Laplace operator in
the sample. This sum depends strongly on dimensionality. A
perturbative treatment of the non--zero modes in Eq.~(\ref{eq_6:17b})
leads to results \cite{Mir95a} identical to the first two lines in
Eq.~(\ref{eq_6:17d}) with $\alpha x/6$ replaced by $\kappa$.

For large amplitudes $y\ge \kappa^{-1/2}$ perturbative methods are
no longer admissible. Falko and Efetov \cite{Fal95} employed a
saddle--point approximation similar to the one introduced in
Ref.~\onlinecite{Muz95} to derive the following result for $d=2,3$,
\begin{eqnarray}
f(y) &\propto& \exp\left(\frac{\beta}{2}(-y+\kappa y^2 +\ldots)
\right), \qquad\qquad (\kappa^{-1/2}\le y \le \kappa^{-1})
\nonumber\\ 
f(y) &\propto& \exp\left(-\frac{\beta}{8\kappa} \ln^d(\kappa
y)\right), \hspace{3.5cm} (y\ge\kappa^{-1}) \ .
\label{eq_6:17e}
\end{eqnarray}
Here, $\kappa = \ln(L/\ell)/4\pi^2\nu{\cal D}$ for $d=2$ and $\kappa
\propto (k_F\ell)^2$ for $d=3$, where $k_F$ is the Fermi momentum and
$\ell$ the elastic mean free path. In particular it was possible to
investigate the inverse participation numbers $u_n = \int_0^\infty u^n
f(u) du$. By studying the dependence of these quantities on system
size one can address the question of multifractality of the wave functions.
In two dimensions and for system sizes $L$ smaller than or of the
order of the localization length, the multifractal nature of the
wave functions had been established numerically \cite{Scr85,Poo91},
with the result 
\begin{equation}
t_n  \propto L^{-\tau(n)-2},  \qquad\qquad \tau(n) = (n-1)d^*(n) \ .
\label{eq_6:20}
\end{equation}
The fractal dimension $d^*(n) \neq 2$ had been found to depend on the
moment considered. In Ref.~\onlinecite{Fal95}, this fractal dimension
was derived analytically,
\begin{equation}
d^*(n) = 2 - \frac{1}{\beta} \frac{n}{2\pi^2\nu {\cal D}},
\label{eq_6:21}
\end{equation}
together with the corresponding power law decay in the tails of the
wave functions 
\begin{equation}
|\psi(r_0 + r)|^2 \propto r^{-2\mu} \ .
\label{eq_6:22}
\end{equation} 
In Eq.~(\ref{eq_6:21}) the order $n$ of the moment is restricted by the
requirement that $d^*(n)>0$. The exponent $\mu$ in Eq.~(\ref{eq_6:22})
is smaller than unity and varies from wavefunction to wavefunction.

We see from Eq.~(\ref{eq_6:17e}) that the 
asymptotic behavior of $f(y)$ is again
logarithmically normal, in close analogy to
the distribution functions for the conductance \cite{Alt91a,Muz95,Mir95}
and for the density of states \cite{Alt86a,Alt91a} discussed above.
What is the origin of all these log--normal tails? Already in the
metallic regime, they appear as an early manifestation of
localization, and they become more pronounced as one approaches the 
localized regime. This suggests the following interpretation. Even in
the metallic regime there exist rare states which are localized. As we
approach the localized regime, there occur rare states with abnormally
strong localization. The log--normal tails are due to the occurrence
of such ``atypical states'' at any value of $g$. This connection
between atypical states and asymptotic distribution functions is 
an active field of research \cite{Mir96,Mir96a,Smo96,Kra96}.

In conclusion it is satisfying to see how recent progress in the
evaluation of the $\sigma$ model helped to provide strong support for
earlier calculations \cite{Alt86a,Alt88a,Alt91a} of mesoscopic
distribution functions based on a combination of perturbation theory
and renormalization group methods.

\subsection{Interaction--assisted coherent transport}
\label{shp}

In 1990, Dorokhov \cite{Dor90} considered particles moving in a
strictly one--dimensional random potential. He showed that the
localization 
properties of a pair of particles bound tightly by a harmonic
potential differ from those of non--interacting particles. In
particular, he predicted a sizeable enhancement of the pair
localization length for the case where the pair energy is sufficient
to excite high--lying modes of the two--particle system. Four years
later, it was suggested by Shepelyansky \cite{She94} that the
wavefunction of two interacting particles in a one--dimensional
disordered chain might extend over a length scale $\xi_2$ much larger
than the one--particle localization length $\xi_1$, irrespective of the
sign of the two--particle interaction. Shepelyansky's result indicated
that interactions among electrons might significantly alter or modify
our present understanding of {\it transport} and {\it localization}
properties of mesoscopic systems. This important development should be
seen in connection with the persistent current problem in Sec.~\ref{pc}.
We recall that this problem has led to the general belief that
interactions are vital for the understanding of the {\it equilibrium}
(i.e. ground state) properties of mesoscopic rings. 

Technically, the problem of two interacting particles in a random
potential can be mapped onto an effective Hamiltonian which turns out
to be a random band matrix with strongly fluctuating diagonal entries.
Matrices of this type had not been studied
previously, and the two--particle problem is, therefore, also
interesting from the point of view of Random Matrix Theory.

\subsubsection{Idea and scaling picture}
\label{shp_1}

Shepelyansky \cite{She94} considered the following Schr\"odinger
equation for the two--particle wavefunction $\psi_{n_1n_2}$ in one
dimension,
\begin{eqnarray}
&&[ E_{n_1} + E_{n_2} + U\delta_{n_1n_2} ] \psi_{n_1n_2} \nonumber\\ 
&&\qquad + \  V [ 
\psi_{n_1+1,n_2} +
\psi_{n_1-1,n_2} +
\psi_{n_1,n_2+1} +
\psi_{n_1,n_2-1} 
]
= E \psi_{n_1n_2} \ .
\label{eq:1}
\end{eqnarray}
Here, $n_1$ und $n_2$ are the (discrete) coordinates of the first and
second electron, respectively, $E_{n_1}$ and $E_{n_2}$ are on--site
energies drawn randomly from the interval $[-W, W]$, $U$ characterizes
the strength of the Hubbard interaction, and $V$ is the hopping
integral. All lengths are measured in units of the lattice spacing and
are, therefore, dimensionless. For $U=0$, and not too strong disorder,
the one--particle
localization length $\xi_1$ in the middle of the energy band is
roughly given by $\xi_1 \approx 25 (V/W)^2$~\cite{Tho72,Czy81,Pic86}.
We define the two--particle
localization length $\xi_2$ as the typical scale on which the
two--particle wavefunction decays. It was predicted \cite{She94} that
in the interacting case,
\begin{equation}
\xi_2 \approx \frac{1}{32} \left( \frac{U}{V} \right)^2 \xi_1^2 \ ,
\label{eq:2}
\end{equation}
provided $\xi_1 \gg 1$ and $U/V = {\cal O}(1)$. This result
is noteworthy in two respects: First, for sufficiently large $\xi_1$
(i.e. sufficiently weak disorder) one can have a significant enhancement
of $\xi_2$ as compared to $\xi_1$. Second, since the interaction strength
$U$ enters the expression (\ref{eq:2}) quadratically, the effect does not
depend on the sign of the interaction.

\begin{figure}
\centerline{
\psfig{file=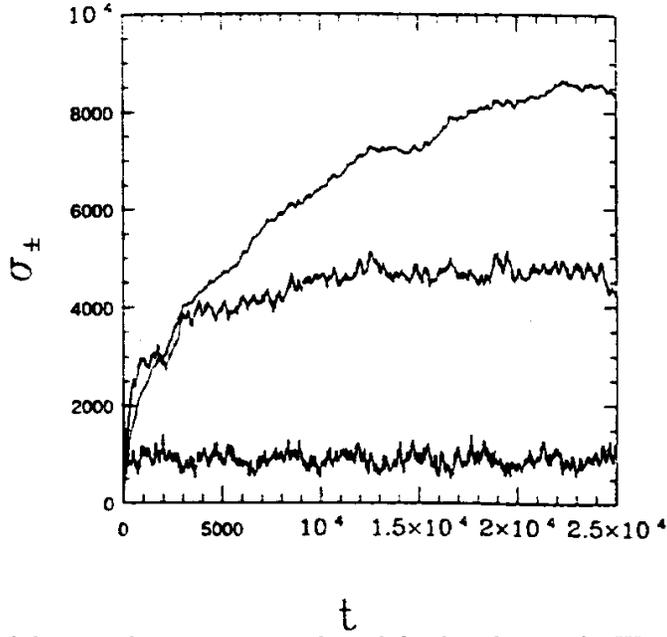,width=3.5in}
 }
\caption{
Time dependence of the second moments $\sigma_+$ and $\sigma_-$
defined in the text for $W=0.7$ and $V=1$ and (from top to bottom):
$U=1$ ($\sigma_+$), $U=1$ ($\sigma_-$), and $U=0$ ($\sigma_+$).
Taken from Ref.~\protect\onlinecite{She94}.
} 
\label{afig6}
\end{figure}

A first qualitative numerical confirmation of this effect came from
the study of the wave packet dynamics for Eq.~(\ref{eq:1}). The 
time--dependent quantities $\sigma_+ = \langle
(|n_1|+|n_2|)^2\rangle/4$ and  
$\sigma_- = \langle (|n_1|-|n_2|)^2\rangle$ (characterizing center
of mass and relative motion, respectively) are shown in
Fig.~\ref{afig6}. (The angular brackets denote an average over the
wave packet probability distribution at a given time $t$). An
enhancement effect is clearly visible.

The result (\ref{eq:2}) was obtained in two rather independent ways.
In Ref.~\onlinecite{She94} the problem was mapped onto the
above--mentioned random band matrix with strongly fluctuating diagonal
elements which was studied numerically.  The result (\ref{eq:2}) was
obtained from a fit to numerical data. Imry \cite{Imr95}, on the other
hand, invoked a generalization of Thouless' famous scaling block
picture to rederive Eq.~(\ref{eq:2}). We treat both approaches in
turn.

It is useful to rewrite Eq.~(\ref{eq:1}) in a basis defined by the
single--particle eigenfunctions of the non--interacting part of the system.
This yields
\begin{eqnarray}
H_{ij,kl} &=& (\epsilon_i + \epsilon_j) \delta_{ki}\delta_{jl} +
\tilde{Q}_{ij,kl} \ , \nonumber\\
\tilde{Q}_{ij,kl} &=& U \sum_n
\varphi_i(n)\varphi_j(n)\varphi_k(n)\varphi_l(n)
\ .
\label{eq:4}
\end{eqnarray}
In this representation the Hamiltonian is naturally decomposed into
two pieces: a diagonal matrix carrying the eigenvalues $\epsilon_i \in
[-B, B]$ of the non--interacting system with bandwidth $B \approx
2(2V+W)$, and a non--diagonal matrix $\tilde{Q}_{ij,kl}$ originating
from the interaction operator and defined by the overlap of four
eigenfunctions $\varphi_i$. A typical matrix element $\tilde{Q}$ can
be estimated \cite{She94} as $\tilde{Q} \approx U/\xi_1^{3/2}$.

Since the wave functions $\varphi_i$ are localized the Hamiltonian
(\ref{eq:4}) possesses band structure. Shepelyansky used the crucial
assumption that all elements of the matrix $\tilde{Q}_{ij,kl}$ in
Eq.~(\ref{eq:4}) can be chosen as independent random variables
centered at zero. In this way, the Hamiltonian in Eq.~(\ref{eq:4}) is
replaced by an effective random Hamiltonian. The latter is the sum of
two random matrices: (i) A random band matrix with bandwidth
$\xi_1^2$ and Gaussian distributed elements with variance
$U/\xi_1^{3/2}$. (ii) A random diagonal matrix with Gaussian
distributed elements having variance $B=2(2V+W)$. The addition of the
random diagonal matrix to the random band matrix defines a new class 
of random band matrices. This new class differs 
significantly from an ordinary random band matrix if the diagonal
elements fluctuate much more strongly than the off--diagonal
elements. This condition is met in the present case since $\xi_1 \gg
1$ and $U$ is typically of the order of $V$. Studying the localization
properties of the eigenfunctions of this effective Hamiltonian by
means of the transfer matrix method, Shepelyansky arrived at the
result (\ref{eq:2}).  
 
In an alternative approach, Imry \cite{Imr95} generalized the ideas
of Thouless \cite{Tho77}, see Sec.~\ref{loc}, to the two--body problem.
He divided the system into blocks of size $\xi_1^2$. The interaction
operator (\ref{eq:4}) couples the two--particle levels of one block
with those of its neighbor. The generalization of the Thouless
argument then says: If the ``two--particle Thouless energy''
$E_C^{(2)}$ is larger (smaller) than the two--particle level spacing
$\Delta_2$ in each of the blocks, the system is conducting
(localized), respectively. Using Fermi's golden rule and the estimate
$\tilde{Q} \approx U/\xi_1^{3/2}$ given above, Imry obtained
$E_C^{(2)} = \tilde{Q}^2/\Delta_2 \approx U^2 / B \xi_1$. Here, $B =
\Delta_2 \xi_1^2$ is the bandwidth which for $\xi_1 \gg 1$ is
proportional to $V$. In analogy to the one--electron case, Imry
defined a ``two--particle conductance'' $g_2$ by  
\begin{equation}
g_2(\xi_1) = \frac{E_C^{(2)}}{\Delta_2} = \frac{U^2}{B^2} \xi_1 .
\label{eq:6}
\end{equation}
For $\xi_1 \gg 1$ one can clearly have $g_2 \gg 1$ and the system is far
from being localized for electron pairs. Assuming that $g_2$ as a
function of the length $L$ of the system behaves like a proper
conductance, i.e. that $g_2(L) \propto L^{-1}$ in one dimension, and
using $g_2(\xi_2) \approx 1$ as the definition of $\xi_2$, Imry
obtained finally 
\begin{equation}
\frac{\xi_2}{\xi_1} = g_2(\xi_1) = \frac{U^2}{B^2}\xi_1
\ \ \Leftrightarrow \ \ \xi_2 = \left(\frac{U}{B}\right)^2\xi_1^2 . 
\label{eq:7}
\end{equation}
Apart from numerical factors which cannot be determined in this
approach, the result (\ref{eq:2}) has been reproduced in a
few lines.

\subsubsection{Microscopic approaches and refinements}
\label{shp_2}

Both approaches described above rest on rather important assumptions
and approximations. We mention, in particular, the statistical
assumptions of the effective random band matrix approach (which
are also implicit in the scaling picture, and which neglect all
correlations among matrix elements), and the assertion that $g_2$ 
has the scaling behavior of a proper conductance. Therefore, the work
of Shepelyansky and Imry was followed by a number of investigations
\cite{Fra95,Wei95,Opp96} all of which started from the microscopic
problem defined by Eq.~(\ref{eq:1}). 

Frahm {\it et al.} \cite{Fra95} employed the transfer matrix method to
solve the Schr\"odinger equation (\ref{eq:1}). These authors found
that in the parameter range considered, $V=U=1$ and $W = 0.7\ldots
2.0$, the two--body localization length behaved as $\xi_2 \propto
\xi_1^{\alpha}$ with $\alpha = 1.65$. This was the first direct
confirmation of the enhancement effect. The exponent $\alpha$ differed
from the predicted value $(\alpha=2)$, however. In
Ref.~\onlinecite{Fra95} this discrepancy was attributed to the
strongly non--Gaussian distribution of the interaction matrix elements
$\tilde{Q}_{ij,kl}$ in Eq.~(\ref{eq:4}). We will see later, however,
that the discrepancy may also be an artifact of fitting an exponent in
the crossover regime from $\xi_2 \propto \xi_1$ to $\xi_2 \propto
\xi_1^2$.

\begin{figure}
\centerline{
\psfig{file=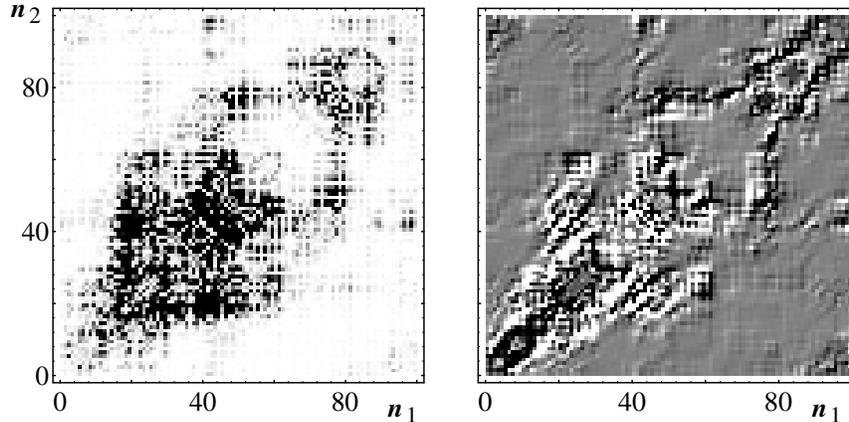,width=4.5in}
}
\caption{A particular two--electron wave function (left) of a
  one--dimensional, disordered ring and the
  corresponding local current distribution (right) demonstrating
  interaction--assisted coherent transport. In the left figure, dark
  and bright regions indicate high and low probability density,
  respectively. In the right figure, dark and bright regions indicate
  local currents in the direction and against the direction of
  increasing electron coordinates $n_1$, $n_2$, respectively. Taken
  from Ref.~\protect\onlinecite{Wei95}.
}
\label{fig14}
\end{figure}

A very direct demonstration of interaction--assisted coherent pair
propagation is due to Weinmann {\it et al.} \cite{Wei95}. These
authors considered a strongly localized, one--dimensional, disordered
ring threaded by a magnetic flux. The idea was to identify coherently
propagating electron pairs by the periodic flux dependence of their
eigenvalues. The period should be given by $h/2e$ instead of $h/e$,
the value for single electrons. Indeed, many states with $h/2e$
admixtures were found in Ref.~\onlinecite{Wei95}, some of them with
almost perfect $h/2e$ periodicity. The associated wave functions
showed the characteristic ``cigar shape'' expected when a pair of
typical size $\xi_1$ travels coherently over a distance $\xi_2$, cf.
Fig.~\ref{fig14}. This result provided compelling evidence for
Shepelyansky's effect.

A Green function approach introduced by v. Oppen {\it et al.}
\cite{Opp96} made it possible to investigate in a very efficient way
the dependence of the two--particle localization length on the
interaction strength $U$. For the problem defined by Eq.~(\ref{eq:1}),
a new scaling parameter could be identified. Irrespective of the
symmetry of the two--body wave function (Bosons or Fermions), it was
found numerically that the ratio $\xi_2/\xi_1$ depends exclusively on
the parameter combination $\mid U\mid \xi_1/V$,
\begin{equation}
\frac{\xi_2}{\xi_1} = \frac{1}{2} + c \frac{\mid U\mid}{V} \xi_1 ,
\label{eq:12}
\end{equation}
where $c$ is a number of the order $10^{-1}$. It is remarkable that
the dependence on $U$ is {\it linear} instead of quadratic as
predicted in Refs.~\onlinecite{She94,Imr95}. Attempts to explain this
result are discussed further below. Equation~(\ref{eq:12}) implies
that asymptotically, $\xi_2 \propto \xi_1^2$, in contrast to the
exponent $\alpha = 1.65$ found in Ref.~\onlinecite{Fra95}. The obvious
explanation suggested by Eq.~(\ref{eq:12}) is that the results of
Ref.~\onlinecite{Fra95} apply only in the crossover region where the
dependence of $\xi_2$ on $\xi_1$ changes from linear to quadratic.

It would be very desirable to have information on the enhancement
effect in $d > 1$ dimensions. Because of the metal--insulator
transition and the question how the mobility edge might be affected by
coherent pair propagation, data for $d>2$ dimensions would be of
particular interest. However, direct simulations for $d=3$ run into
severe numerical difficulties since the size of the basis grows with
the linear dimension $N$ like $N^6$. The only numerical information on
the delocalization effect for $d > 1$ comes from simulations of the
kicked rotor model by Borgonovi and Shepelyansky \cite{Bor95,Bor96}.
If in a $1d$ model, the time dependence of the perturbing kick is
modeled as the superposition of $\nu$ periodic oscillations with
incommensurate frequencies, the resulting problem corresponds to a
hopping model in $d=\nu+1$ dimensions. A similar correspondence for
the case of two interacting kicked rotors can be used to study two
interacting particles in higher dimensions. The gain in numerical
efficiency is impressive. It amounts to a factor $N^4$ \cite{Bor96}.
Two cases were investigated. They correspond to an interacting pair in
effective dimensions $d_{\rm eff}=2$ and $d_{\rm eff}=2.5$,
respectively.  The main results of Refs.~\onlinecite{Bor95,Bor96} can
be summarized in terms of the behavior of two quantities, $\sigma_+$
and $\sigma_-$, defined as the distance travelled by the pair and as
its typical size, respectively. For $d_{\rm eff} = 2$, a saturation of
$\sigma_+$ as function of time was observed. In the case $d_{\rm eff}
= 2.5$, however, $\sigma_+$ grew without saturation {\it even in a
  regime where single particles are localized}. This was a first
numerical indication that pairs might be delocalized earlier than
single particles. The numerical data in Ref.~\onlinecite{Bor96} also
confirmed qualitative arguments in Ref.~\onlinecite{Bor95}. It had
been predicted that the size of the diffusing electron pairs grows
logarithmically with time, while their motion is, correspondingly,
slightly subdiffusive: the diffusion constant is inversely
proportional to $\ln^{\mu}t$, with $\mu$ of the order of 1.

The linear dependence (\ref{eq:12}) of $\xi_2$ on the interaction
strength $U$ was addressed again in a recent study by Weinmann and
Pichard \cite{Wei96}. Equations~(\ref{eq:6})
and (\ref{eq:7}) show that it suffices to investigate the
$U$--dependence of the two--particle Thouless energy
$E_C^{(2)}$. Analytical arguments and numerical simulations
\cite{Wei96} show that for system sizes of the order of $\xi_1$, there
are two regimes for $E_C^{(2)}$: a linear ($E_C^{(2)} \propto U$) and a
quadratic one ($E_C^{(2)} \propto U^2$). The crossover occurs for
$E_C^{(2)}\approx\Delta_2$. 
Unfortunately, this result does not directly lead to the appropriate
scaling parameter $|U|\xi_1/V$, and additional statistical assumptions
are necessary to explain Eq.~(\ref{eq:12}) \cite{Wei96}.
Analytical investigations of this problem in the framework of a
suitable random matrix model were performed by
Guhr and M\"uller--Groeling \cite{Guh96}, who derived a novel
analytical expression for the spectral two--point correlation
function. Jacquod, Shepelyansky, and Sushkov~\cite{Jac96} have pointed
out that the linear $U$--dependence should only apply close to the
band center of the two--particle problem. In all other regions of the
spectrum they predict an essentially quadratic dependence of $\xi_2$ on
$U$.

We have seen that the problem of two interacting particles in a random
potential naturally leads to a random band matrix with strongly
fluctuating diagonal elements.
This class of random matrices generalizes the ordinary random band
matrices studied in the context of chaotic quantum systems and transport
in quasi 1d wires \cite{Fyo91,Fyo92,Fyo93,Mir93}. It has led to
partly \cite{Jac95} and fully \cite{Fyo95,Fra95a} analytical work on
the properties of the following ensemble of matrices, 
\begin{equation}
H_{nn'} = \eta_n\delta_{nn'} + \zeta_{nn'}/\sqrt{2b+1} \quad\quad (n,n' =
1,...,N) \ .
\label{eq:15}
\end{equation}
Here, $b$ is the bandwidth and $\eta_n$ and $\zeta_{nn'}$ are
uncorrelated random variables. In Ref.~\onlinecite{Jac95} the
$\eta_n$'s were drawn from the interval $[-W_b, W_b]$ and the
$\zeta_{nn'}$'s from $[-1,1]$ provided that $|n - n'| \leq b$.
Otherwise, $\zeta_{nn'} = 0$. In Refs.~\onlinecite{Fyo95,Fra95a} the
distribution of the $\eta_n$'s was rather arbitrary but characterized
by a strength $W_b$ while the $\zeta_{nn'}$'s had a zero--centered
Gaussian distribution with variance $\overline{|\zeta_{nn'}|^2} =
(1+\delta_{n n'}(\beta-2))A_{n n'}/2$. Here $A_{n n'}$ differs from
zero essentially only in the interval $|n - n'| \leq b$ where it has
the typical value unity. In both cases, the typical size of the
variables is given by the parameter $W_b$ and by unity, respectively.

In Ref.~\onlinecite{Jac95}, two main results for the regime $1\ll
W_b<\sqrt{b}$ were established by analytical arguments and numerical
simulations: (i) For a fixed unperturbed state with energy $\eta_n$,
the local density of states $\rho_n(E) = (1/\pi)
\overline{G_{nn}^+(E)}$ is always of Breit--Wigner form,
\begin{equation}
\rho_n(E) = \frac{\Gamma/2\pi}{(E-\eta_n)^2 + \Gamma^2/4} ,
\label{eq:16}
\end{equation}
where $\Gamma = \pi/3W_b$. (ii) The scale defined by
the localization length $\xi = b^2/2W_b^2$ differs from the one set by
the inverse participation ratio $\xi_{\rm IPR} =
\langle(\sum_n|\psi_{\lambda}(n)|^4)^{-1}\rangle_{\lambda} = b^2/4W_b^4$,
where $\psi_{\lambda}(n)$ is an eigenfunction of the Hamiltonian
(\ref{eq:15}) and $\langle ... \rangle_{\lambda}$ denotes a spectral
average. This means that typical wave functions have very strong
site--to--site fluctuations (We recall that the ensemble (\ref{eq:15})
is of interest only if $W_b \gg 1$). 

Simultaneously and independently, Fyodorov and Mirlin \cite{Fyo95} and
Frahm and M\"uller--Groeling \cite{Fra95a} investigated the ensemble
(\ref{eq:15}) by means of the supersymmetric non--linear $\sigma$
model. In these papers, the findings in Ref.~\onlinecite{Jac95} were
analytically confirmed and extended. In addition to the localization
length $\xi$ and the local density of states $\rho_n(E)$, the full set
of generalized inverse participation ratios $P_q = \sum_n
|\psi_n^{(k)}|^{2q}$, and the distribution of the eigenvector
components were calculated. These latter quantities were shown to be
connected with those for ordinary random band matrices by an
appropriate rescaling with a factor $\rho_n(E)/\bar{\rho}$
\cite{Fyo95,Fra95a}, where $\bar{\rho}$ is the average density of
states. Specialization of these results to the cases considered in
Ref.~\onlinecite{Jac95} confirmed all major results obtained there.

A variant of the Hamiltonian ensemble (\ref{eq:15}) was used by
Frahm, M\"uller--Groeling, and Pichard \cite{Fra96,Fra96a} to
investigate the pair dynamics in arbitrary dimension. Generally
speaking, the main result of these papers was the analytical proof
that the concepts familiar from transport theory of non--interacting
electrons also apply {\it mutatis mutandis} to the case of electron
pairs. A supersymmetric non--linear $\sigma$ model for the two--point 
function of the effective Hamiltonian was derived. Under omission of
source terms, the result is 
\begin{eqnarray}
Z_{\rm pair}&=& \int d[Q] \exp(-S_{\rm pair}[Q]) \ ,
\label{eq:21a}  \\
S_{\rm pair}[Q]&=& -\frac{\Gamma^2B_2}{8B_0^2} \int dR \,
{\rm trg}(\nabla_RQ)^2 -
i\frac{\pi}{4}\omega \, h\left(\frac{\Gamma}{\omega}\right)
\rho_0(E)\int dR \, {\rm trg}(Q L) \ .
\nonumber
\end{eqnarray}
Here, $B_0$ and $B_2$ are numbers characterizing the band structure of
the Hamiltonian, $\rho_0(E)$ is the density of unperturbed
(non--interacting) states, and $\Gamma$ is the interaction--induced
spreading width of these states at the scale $\xi_1$. The function
$h(\Gamma/\omega)$ is defined in Ref.~\onlinecite{Fra96,Fra96a}.
Equation~(\ref{eq:21a}) must be compared with the corresponding
expression for transport of non--interacting electrons in metals
\cite{Efe83} introduced in Sec.~\ref{spec_1},
\begin{eqnarray}
Z &=& \int d[Q] \exp(-S[Q]) \ ,   \nonumber\\
S[Q] &=& -\frac{\pi\nu}{8}  \int {\rm trg} \left[
{\cal D} (\nabla Q)^2 + 2i\omega LQ \right] dr \ .
\label{eq:22}
\end{eqnarray}
Here, the scaling parameter of localization theory is the conductivity
$\sigma = \nu {\cal D}$. It is clear that Eqs.~(\ref{eq:21a}) and
(\ref{eq:22}) are very closely related. This led to the following
statements: (i) The effective ``conductivity'' $\sigma_{\rm eff} =
\nu_{\rm eff}(\omega)D_{\rm eff}(\omega) \approx U^2\xi_1^2/W_b^2$
of the pairs plays exactly the same role as the one--particle quantity
$\sigma$. This rigorously justifies Imry's scaling approach
\cite{Imr95}. For $d=1$ one has $\sigma_{\rm eff} \propto \xi_2$ so that 
Eq.~(\ref{eq:2}) is recovered. (ii) The mean quadratic displacement of
the pairs increases weaker than linearly, $\langle R^2(t) \rangle
\propto t/ 
\ln(\Gamma t)^d$ for some time scale so that the pairs move
subdiffusively. This confirms numerical findings \cite{Bor95,Bor96}
discussed above. (iii) It is possible to define a pair Thouless energy
$E_C^{(2)}$ below which the spectral statistics of the diffusing pair
states is of Wigner--Dyson type. For energy scales larger than
$E_C^{(2)}$ one recovers the Altshuler--Shklovskii regime \cite{Alt86}.

In summary, we have seen that in its essentials, Shepelyansky's
discovery, the phenomenon of interaction--assisted coherent transport
of pairs of particles, has been corroborated by a wealth of subsequent
work. 

\subsubsection{Finite particle density}
\label{shp_3}

The problem of two interacting particles addressed in the previous
paragraph is academic, at least for electrons. Indeed, in the
many--body case the exclusion principle may prevent virtual
transitions into states which are essential for coherent pair
propagation. Hence it is necessary to study coherent pair propagation
at finite particle density. Only two papers have so far (partly)
addressed this important issue \cite{Imr95,Opp95}. In
Ref.~\onlinecite{Opp95}, von Oppen and Wettig considered two
interacting quasiparticles above the Fermi surface in one dimension.
Formally, the only difference to previous work was the blocking of all
single--particle states up to the Fermi energy $E_F$. The interaction
of particles within the Fermi sea with each other and with the two
quasiparticles was not taken into consideration. It is argued
\cite{Opp95} that this procedure amounts to a low--density
approximation. It was found that slightly above $E_F$, the
quasiparticle pair was localized on the scale $\xi_1$. With increasing
excitation energy the enhancement effect was eventually recovered, but
the energy scale for this to happen is given by the bandwidth. This
result appears to rule out the possibility to observe coherent pair
propagation in an experiment at low temperature in one dimension. Imry
\cite{Imr95} has emphasized that the situation might be more favorable
in higher dimensions. He argued that for $d=3$ the mobility edge for
pairs could be located {\it below} the mobility edge for single
particles. Let $\epsilon_{m1}$ and $\epsilon_{m2}$ be the excitation
energies of the mobility edges for single particles and pairs,
respectively.  Then, according to Ref.~\onlinecite{Imr95},
\begin{equation}
\epsilon_{m2} \propto \epsilon_{m1}^{\tilde{\nu} d/2} \ ,
\label{eq:24}
\end{equation}
where $\tilde{\nu}$ is the critical exponent defined by 
$\xi_1\propto \epsilon_{m1}^{-\tilde{\nu}}$.
Invoking the Harris criterion \cite{Mot76}, $\tilde{\nu} d/2 > 1$, it is
clear that $\epsilon_{m2} \to 0$ faster than $\epsilon_{m1}$, as
claimed.
 
Much more work is necessary to clarify the phenomenon of
interaction--assisted coherent transport in many--body systems. 
The whole field is rather new and progresses rapidly. The most
recent developments include a detailed study of the enhancement
mechanism in 1d~\cite{Pon96}, a numerical investigation~\cite{Jac96b}
of two quasiparticles in 2d and 3d, and an unfortunate controversy
concerning the role of finite size effects~\cite{Roe97,Fra97}.

\subsection{RMT in condensed matter physics --- a summary}
\label{sum}

To conclude our overview over developments in condensed matter physics
where RMT has played a significant role we now try to summarize and
judge the importance and prospects of RMT in this field. 

The most direct and therefore also historically first application of
RMT in solid state theory was to spectral fluctuations. Roughly, we
can distinguish two periods: One in which research focussed on proving
the {\it validity} of Wigner--Dyson (WD) statistics for some range of
parameters, and a second one where efforts were concentrated on
identifying {\it deviations} from this universal limit. Thus, WD
statistics was first established as a valid means of description and
then used as a standard of reference. Identifying the limits of
validity of WD statistics in the description of spectral fluctuations
has turned out to be very fruitful. It has led to a better
understanding of the various regimes (clean, ballistic, ergodic, 
diffusive, critical, and localized) in mesoscopic physics. In this
context the $\sigma$ model has served at least two purposes. First,
it provided the proof that an ergodic regime with WD type of spectral
correlations exists and, second, it turned out to be the proper tool
to calculate a number of corrections to this result. We have seen that
such corrections were also discovered with the help of diagrammatic
perturbation theory and/or scaling arguments. The discovery of a third
class of spectral statistics at the mobility edge (different from both
RMT and Poisson statistics) poses an interesting challenge: The
construction of a stochastic Hamiltonian which reproduces critical
spectral fluctuations. Although experimental verification of the
theoretical predictions seems to be out of reach for years to come,
finding such an ensemble would undoubtedly improve our understanding
of the mobility edge. An interesting recent step in this direction is
due to Kravtsov and Muttalib~\cite{Krv97}.

Persistent currents in mesoscopic rings are closely connected with
spectral fluctuations. It is therefore natural to expect RMT to play
an important role in this problem, too. However, the contribution of
RMT to the persistent current problem has been comparatively moderate
so far, for several reasons. Initially, one of the main problems was a
conceptual one: How to calculate averages in the canonical ensemble. 
When this question had been answered and first perturbative results
had become available, non--perturbative calculations based on RMT
did provide the complete and singularity--free solution of the persistent
current problem for non--interacting electrons at zero temperature. This
was an interesting technical achievement, but it did not alter the
discrepancy between theory and experiment. Moreover it is now widely
held that the important effect to be understood is the interplay
between disorder and electron--electron interactions. Unfortunately
this task is at present beyond the scope of RMT. One may speculate
that the embedded ensembles investigated some 20 years ago in nuclear 
physics could one day prove useful.

Applications of RMT to transport in quasi 1d systems have been
particularly successful. The paradigm ``quasi 1d wire'' has united
three formerly separate research activities, the random transfer
matrix approach, the random Hamiltonian approach of the IWZ model, and
random band matrices. Within well--defined limits these three
approaches are indeed equivalent. The random transfer matrix method
illustrates that RMT is not restricted to modeling the Hamiltonian of
a system. This fact has been known at least since Dyson introduced the
circular ensembles of $S$ matrices. For {\it spectral} problems the
distribution of eigenvalues and, thus, a modeling of the Hamiltonian
are of central importance. In {\it transport} problems, on the other
hand, it is the distribution of transmission coefficients that matters. 
These coefficients are directly related to the eigenparameters of the
transfer matrix. The ensemble of random transfer matrices corresponding
to an ensemble of disordered quasi 1d wires therefore appears to offer
the most direct approach to transport properties. However, an important
and sometimes decisive advantage of the Hamiltonian formulation is its
great flexibility. Any modification of the system under study is
{\it defined} in terms of its Hamiltonian and there is absolutely no
question about how to incorporate this change into the Hamiltonian
formalism. It is far less obvious how the transfer matrix has to be
adapted to such a modification. The autocorrelation function of the
conductance versus magnetic field strength illustrates this point.
This quantity can easily be formulated within the Hamiltonian approach
and has been calculated in the diffusive regime. A similar treatment
is, unfortunately, still missing in the framework of the transfer
matrix approach. 

Techniques of the $\sigma$ model have proven to be particularly
valuable. Both the equivalence of the transfer matrix approach and the
IWZ model, and the relation between random band matrices and quasi 1d
wires were established with this technique. Random band matrices are
found to play the same role for quasi 1d systems as do the Gaussian
ensembles for zero--dimensional systems like atoms, molecules, nuclei,
or quantum dots. It is of interest to ask whether generic ensembles of
random matrices exist which model properties of higher--dimensional
systems. 

Problems in $d > 1$ dimensions have taken center stage in recent work
on the non--linear $\sigma$ model. A systematic improvement of 
techniques has led from the zero--dimensional case 
(with perturbative corrections) to 
exact solutions in quasi 1d systems and, via
the recently discovered non--trivial saddle--point solutions, to
substantial progress in the understanding of two-- and
higher--dimensional systems. It has been possible to establish contact
with, and to reproduce results of, earlier perturbative
renormalization group calculations. This is the most recent success of
a program which goes ever further beyond the universal limit of the
Gaussian ensembles.

Another step towards more general matrix ensembles has been taken in
the context of interaction--assisted coherent pair propagation. The
effective Hamiltonian describing two particles in a 1d random
potential turned out to be a random band matrix with strong
fluctuations on the diagonal.
The necessity to introduce a new matrix ensemble originated 
from the presence of both disorder and interactions. The disorder
defines the diagonal entries (localized states) while the
interaction induces couplings between these basis states.  Again, the
$\sigma$ model turned out to be the appropriate tool to investigate
the newly defined ensemble. Differences and similarities between
ordinary band matrices and those with strongly fluctuating diagonal
elements could be 
identified, leading to an improved understanding of coherent pair
propagation. It would be desirable to extend this work and to
construct a matrix ensemble which correctly accounts for generic
features of Hamiltonians with one-- and two--body operators. The
improved understanding of the interplay between disorder and
interactions might also be helpful in the persistent current problem. 

We have seen that RMT has found wide applications in condensed matter
physics. These applications go far beyond the scope of the original
Gaussian ensembles. In this development, the supersymmetry method has
been an invaluable tool. Mapping a random matrix model or a
microscopic, effective Hamiltonian onto a non--linear $\sigma$ model
reduces the physical content to its very essence: Seemingly dissimilar
models turn out to be equivalent, at least in certain regimes. Several
examples were mentioned above. The supersymmetry approach is also
useful to establish the correspondence between dynamical localization
in the kicked rotor and Anderson localization in a thick metallic wire,
or to show that interaction--assisted propagation of pairs can
be described in terms similar to one--particle diffusion. The list can
be continued. Still, there are many open questions. We have mentioned
phenomena for which no adequate matrix model has been constructed up
to now. Doing so remains a challenge for the future.
\setcounter{equation}{0}
\section{Field theory and statistical mechanics}
\label{field}

In previous sections (Secs.~\ref{intro} to \ref{thasp}) of our review
we have dealt with the history and with mathematical aspects of RMT.
With this insight into the foundations and the scope of RMT the
subsequent applications to many body physics (Sec.~\ref{mbs}), quantum
chaos (Sec.~\ref{chaos}), and disordered systems (Sec.~\ref{disorder})
probably came as no surprise. In the present section we discuss the
importance of RMT for fields of physics which at first sight seem to
have little in common with the more obvious fields of applications
just mentioned. In particular we summarize the relation of RMT to
interacting Fermions in one dimension (Sec.~\ref{cs}), to QCD
(Sec.~\ref{chiral}), and to field theory and quantum gravity
(Sec.~\ref{ftqg}). The existence of these ``exotic'' relations once
again demonstrates the universality and ubiquity of RMT.

\subsection{RMT and interacting Fermions in one dimension}
\label{cs}

A deep-lying connection exists between classical Random Matrix Theory
and special systems of interacting Fermions in one dimension. This
fact was first demonstrated in work by Calogero \cite{Cal69} and by
Sutherland \cite{Sut71}. Recently, Simons, Lee, and Altshuler
\cite{Sim93c,Sim93e} have extended this result to the case of correlation
functions. This important work has led to a whole network of
connections between different branches of physics. Moreover, it has
made it possible to calculate the (hitherto unknown) time--dependent
correlation functions of such Fermionic systems using RMT.

We consider the $N$--particle Hamiltonian 
\begin{equation}
H_C = -\frac{\hbar^2}{2m} \left[ \sum_{i=1}^N
\frac{\partial^2}{\partial\lambda_i^2} - \frac{\beta}{2} (\beta-2)
\sum_{i<j} \frac{1}{(\lambda_i-\lambda_j)^2} \right] + \sum_i
V(\lambda_i)
\label{eq_cs_1}
\end{equation}
where the position of particle $i$ with $i = 1, \ldots,N$ in one
dimension is given by $\lambda_i$. It was shown by Calogero
\cite{Cal69} that $H_C$ is integrable provided the potential is
Gaussian, $V(\lambda_i)\propto \lambda_i^2$. Moreover, for this choice
of $V$ the ground state of $H_C$ defines a probability distribution
$P_{N\beta}(\{\lambda_i\})$ of the particle positions $\lambda_i$
which coincides \cite{Cal69} with the joint probability density
$P_{N\beta}(\{\epsilon_i\})$ for the rescaled eigenvalues $\epsilon_i$
(see below) with $i = 1, \ldots, N$ of the three Gaussian ensembles
with $\beta=1,2,4$. The positions $\lambda_i$ have to be measured in
units of the mean interparticle distance. There is a certain freedom
in defining the statistics of the particles described by
Eq.~(\ref{eq_cs_1}). Without entering into a detailed discussion we
will think of these particles as spinless Fermions.

Putting the confining potential $V(\lambda_i)$ equal to zero and
constraining the motion of the $N$ particles to a circle leads to the
Sutherland Hamiltonian \cite{Sut71}  
\begin{equation}
H_S = -\frac{\hbar^2}{2m} \left[ \sum_{i=1}^N
\frac{\partial^2}{\partial\lambda_i^2} - \frac{\beta}{2} (\beta-2)
\left(\frac{\pi}{N}\right)^2
\sum_{i<j} \frac{1}{\sin^2[\pi(\lambda_i-\lambda_j)/N]} \right] \ .
\label{eq_cs_2}
\end{equation}
The square of the ground--state wave function of $H_S$ is equal to the
equilibrium distribution of Dyson's circular ensembles. In the limit
$N\to\infty$ the correlation functions of the Hamiltonians
(\ref{eq_cs_1}) and (\ref{eq_cs_2}) are expected to coincide. We focus
attention on the Sutherland Hamiltonian. 

\begin{figure}
\centerline{
\psfig{file=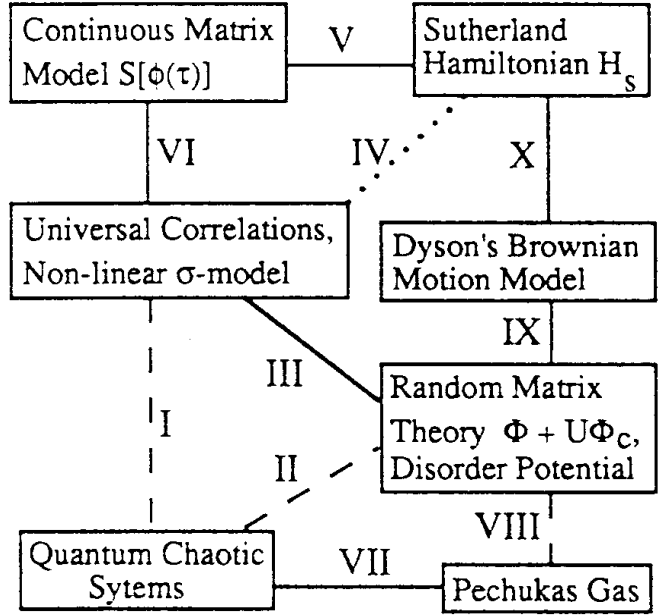,width=3.5in}
 }
\caption{
Connections between various branches of physics. Taken from
Ref.~\protect\onlinecite{Sim94}.
} 
\label{fig_cs}
\end{figure}

To formulate the conjecture made in Ref.~\onlinecite{Sim93c,Sim93e} we
recall some results concerning parametric correlation functions of
RMT, see Sec.~\ref{qc1pcb}. We consider the random Hamiltonian
\begin{equation}
H(U) = H_1 + UH_2 \ ,
\label{eq_cs_3}
\end{equation}
where $H_1$ belongs to one of the three Gaussian ensembles, and where
$H_2$ is a fixed traceless matrix drawn from the same ensemble. The
dimensionless parameter $U$ measures the strength of the perturbation
$H_2$. The eigenvalues $E_i$ of $H$ are obviously functions of this
strength, $E_i=E_i(U)$. To display universal aspects of the correlation
between the eigenvalues, it is necessary to rescale both $U$ and
the $E_i$. This is done by defining the rescaled eigenvalues
$\epsilon_i = E_i/\Delta$ where $\Delta$ is the mean level spacing,
and the rescaled strength 
\begin{equation}
u^2 = \left\langle\left(\frac{\partial\epsilon_i(U)}{\partial U}
\right)^2\right\rangle U^2.
\label{eq_cs_4}
\end{equation}
The resulting universal expressions for the level--density
autocorrelation functions have been given in Sec.~\ref{qc1pcb}, see
Eq.~(\ref{1pc15}). 
Simons, Lee, and Altshuler \cite{Sim93c,Sim93e} conjectured
that these 
universal functions coincide with the time--dependent particle density
correlation functions of $H_S$. To establish such a connection, one
must identify differences in imaginary time $\tau$ and position
$\lambda$ with those in perturbation strength $u$ and energy $\omega$
according to the rules  
\begin{equation}
u^2 = 2\tau,  \qquad\qquad \omega = \lambda \ .
\label{eq_cs_5}
\end{equation}
The first evidence for the conjecture came from expansions of the
time--dependent particle density correlation functions $k_\beta(
\lambda, \tau)$. To leading order in $\tau$ and $1/\tau$,
respectively, these functions were shown to agree with the 
corresponding limits of the level--density autocorrelators
(\ref{1pc15}). Subsequent
investigations by the same authors \cite{Sim93f,Sim94} and by Narayan and
Shastry \cite{Nar93} established the equivalence of both types of
correlation functions exactly. At the same time, Simons and coworkers
\cite{Sim93f,Sim94} set up and completed a network of interrelations
between 
different branches of physics that is schematically shown in
Fig.~\ref{fig_cs}.  
We explain and  summarize the various connections in the sequel.

The link between the Sutherland Hamiltonian and the universal
parametric correlations of Random Matrix Theory has been established
in three different ways. First, there are the asymptotic expansions of
$k_\beta(\lambda,\tau)$ just mentioned (connection IV in
Fig.~\ref{fig_cs}). These expansions provide a very direct link
between the two fields but are not exact and, in addition, restricted
to those correlators for which explicit expressions are known. In
practice, these are the two--point correlators. Second, it can be
shown that both $H_S$ and the non--linear $\sigma$ model for the
universal correlators are related to a continuous matrix model
introduced in Refs.~\onlinecite{Sim93f,Sim94} (connections V and VI).
Third, Dyson's Brownian motion model can be invoked to relate $H_S$
with the random matrix Hamiltonian (\ref{eq_cs_3}) and consequently
also with the universal correlations (connections III, IX, and X). We
will treat the two exact approaches in turn. In a final paragraph we
briefly summarize important new results for 1d Fermions.

\subsubsection{Continuous matrix model}
\label{cs_1}

Following Refs.~\onlinecite{Sim93f,Sim94} we consider the partition
function
\begin{eqnarray}
Z &=& \int D\Phi(\tau) \exp\left(-S[\Phi(\tau)]\right) \nonumber\\
S[\Phi(\tau)] &=& \frac{1}{\hbar} \int_{-\infty}^\infty d\tau \,
{\rm tr}\left[ \frac{m}{2} \left( \frac{\partial\Phi(\tau)}{\partial\tau}
\right)^2 + V(\Phi) \right]
\label{eq_cs_6}
\end{eqnarray}
describing the propagation of $N\times N$ random matrices $\Phi$ from
one of the Gaussian ensembles in a potential $V(\Phi)$. The
Hamiltonian corresponding to the path integral (\ref{eq_cs_6}) reads
\begin{equation}
H = {\rm tr} \left[ -\frac{\hbar^2}{2m} \left(
\frac{\delta}{\delta\Phi} \right)^2 + V(\Phi) \right] \ .
\label{eq_cs_7}
\end{equation}
With the help of the diagonalization $\Phi(\tau)=\Omega(\tau)
\Lambda(\tau)\Omega(\tau)^\dagger$ the ``Laplacian'' $\delta /
\delta\Phi$ in Eq.~(\ref{eq_cs_7}) can be separated into radial
($\Lambda$) and angular ($\Omega$) degrees of freedom,
\begin{equation}
\left(\frac{\delta}{\delta\Phi}\right)^2 = \sum_i 
\frac{1}{J(\lambda_i)} \frac{\partial}{\partial\lambda_i} J(\lambda_i)
\frac{\partial}{\partial\lambda_i} + Y(\Omega,\lambda_i) \ .
\label{eq_cs_8}
\end{equation}
Here $J(\lambda_i) = \prod_{i<j} |\lambda_i-\lambda_j|^\beta$ is the
Jacobian known from Random Matrix Theory. In the limit $N\to\infty$
the angular part $Y(\Omega,\lambda_i)$ can be neglected, and we arrive
at the Hamiltonian $H_C$ of Eq.~(\ref{eq_cs_1}) provided (i) the
potential $V(\Phi)$ does not depend on angular degrees of freedom and
(ii) a factor $[J(\lambda_i)]^{1/2}$ is absorbed into the wavefunctions.
Such a connection between an $N$--Fermion Hamiltonian and a continuous
matrix model had originally been discussed by Brezin, Itzykson,
Parisi, and Zuber \cite{Bre78}. In particular, ground--state
expectation values of $H_C$ can be expressed in terms of the path
integral 
(\ref{eq_cs_6}),
\begin{equation}
\langle 0|\ldots|0\rangle = Z^{-1} \int D\Phi(\tau) \ldots
\exp(-S[\Phi(\tau)]) \ . 
\label{eq_cs_9}
\end{equation}
The proof of equivalence between the time--dependent $m$--point
ground--state correlators of $H_C$ (or, equivalently, $H_S$) and the
$m$--point level--density autocorrelation functions of the random
matrix model of Eq.~(\ref{eq_cs_3}) is carried out as follows
\cite{Sim93f,Sim94}. Supersymmetric non--linear $\sigma$ models are
derived 
for both types of quantities in the large--$N$ limit. The two
non--linear $\sigma$ models are shown to coincide.  

The $m$--point ground--state correlators of $H_C$ are given by
\begin{equation}
\langle 0|f_m(\{\lambda_\alpha,\tau_\alpha\})|0\rangle = 
\left\langle 0\left\vert
\prod_{\alpha=1}^m {\rm tr} 
\left( \frac{1}{\lambda_\alpha^\pm - \Phi(\tau_\alpha)} \right)
\right\vert 0\right\rangle.
\label{eq_cs_10}
\end{equation}
The quantities $f_m$ can be expressed in a standard way in terms of a
supersymmetric generating function,
\begin{equation}
Z_m(\{\lambda_\alpha,\tau_\alpha,J_\alpha\}) = 
\int d[\psi] \exp\left[i\psi^\dagger(\hat{\lambda} - \hat{\Phi}(\tau)
+ \hat{J}I)\psi\right] \ .
\label{eq_cs_11}
\end{equation}
According to Eq.~(\ref{eq_cs_9}), calculating the ground--state
expectation value of $Z_m$ amounts to averaging over $\Phi(\tau)$ with
the probability density $\exp(-S[\Phi(\tau)])$. 
Since we are intersted in the universal large--$N$ limit of the
correlation functions we can restrict ourselves to the Gaussian case
and choose $V(\Phi)\propto {\rm tr}\, \Phi^2$.
Then the average can be trivially
performed. As in other derivations of a non--linear $\sigma$ model,
subsequent steps involve \cite{Sim93f,Sim94} a Hubbard--Stratonovich
transformation and a saddle--point approximation. The final expression
in terms of $4m\times 4m$ supermatrix fields reads
\begin{eqnarray}
\langle 0|Z_m|0\rangle &=& \int d[Q] \exp(-F[Q]) \nonumber\\
F[Q] &=& i\frac{\pi}{2} \sum_{\alpha} {\rm trg}[(\lambda_\alpha^\pm +
J_\alpha I)Q_{\alpha\alpha}] +
\frac{\pi^2\beta}{8\hbar} \sum_{\alpha\beta} |\tau_\alpha -
\tau_\beta| {\rm trg}[Q_{\alpha\beta}Q_{\beta\alpha}] \ , 
\label{eq_cs_12}
\end{eqnarray}
where the mean interparticle distance and $\hbar^2/2m$ have been set
equal to unity. In a similar way, the correlators of the random matrix
Hamiltonian (\ref{eq_cs_3}) 
\begin{equation}
f_m(\{\Omega_\alpha,U_\alpha\}) = \prod_{\alpha=1}^m {\rm tr} \left[
\frac{1}{\Omega_\alpha^\pm - H_1 - U_\alpha H_2} \right]
\label{eq_cs_13}
\end{equation}
can also be represented in terms of a non--linear $\sigma$ model
involving the appropriately rescaled parameters $\omega_\alpha =
\Omega_\alpha/\Delta$ and $u_\alpha$ (see above). For $m = 2$, this
$\sigma$ model can be shown to coincide with that of
Eq.~(\ref{eq_cs_12}) after the formal replacements
\begin{equation}
\omega_\alpha \leftrightarrow \lambda_\alpha, \qquad\qquad
(u_\alpha - u_\beta)^2 \leftrightarrow 2|\tau_\alpha -
\tau_\beta|/\hbar \ .
\label{eq_cs_14}
\end{equation}
For $m=2$, this establishes the correspondence between the two types
of correlation functions. For $m>2$ the replacements (\ref{eq_cs_14})
cannot be realized by a single unique set of transformations $u_\alpha
= u_\alpha(\{\tau_\beta\})$. Instead a slightly more complicated
argument has to be invoked, see Refs.~\onlinecite{Sim93f,Sim94}. This
completes the proof in the general case.

\subsubsection{Use of Dyson's Brownian motion model}
\label{cs_2}

To establish the link between one--dimensional interacting Fermions
and universal parametric correlations via Dyson's Brownian motion
model, Narayan and Shastry \cite{Nar93} started from the Calogero
Hamiltonian (\ref{eq_cs_1}) with $V(\lambda_i) = \lambda_i^2/4a^2$. 
For imaginary time $\tau = it$, the time--dependent Schr\"odinger
equation involving $H_C$ yields for the quantity $P(\{\lambda_i\};\tau)
= \psi_0(\{\lambda_i\})\psi(\{\lambda_i\};t)$ the Fokker--Planck equation
\begin{equation}
\frac{\partial P(\{\lambda_i\};\tau)}{\partial\tau} = \sum_i
\frac{\partial}{\partial\lambda_i} \left[
\frac{\partial P}{\partial\lambda_i} + \frac{\lambda_i}{a^2} P \right]
-\sum_{i\neq j} \frac{\partial}{\partial\lambda_i}
\left[\frac{\beta}{\lambda_i-\lambda_j} P \right] \ .
\label{eq_cs_15}
\end{equation}
Here, $\psi_0$ is the ground--state wave function of $H_C$. To
interpret Eq.~(\ref{eq_cs_15}), we may consider the $\lambda_i$
as the position coordinates of $N$ classical particles. Then,
Eq.~(\ref{eq_cs_15}) describes the dynamics of the Wigner--Dyson
Coulomb gas \cite{Dys62a}, the $N$ particles being confined by a
harmonic potential and interacting via a pairwise logarithmic
repulsion (connection X).

In a next step, Narayan and Shastry considered the evolution of the
eigenvalues of a perturbed random matrix model similar to
Eq.~(\ref{eq_cs_3}), 
\begin{equation}
H(x) = H_0 \cos(\Omega x) + V\sin(\Omega x)/\Omega \ ,
\label{eq_cs_16}
\end{equation}
where $H_0$ and $V$ are drawn from one of the Gaussian ensembles.
For any $x$, $H(x)$ is perturbed by $H'(x) = \partial H/\partial x$
and the evolution of  the eigenvalues $\epsilon_i(x)$ can be expressed
as a set of first--order differential equations,
\begin{eqnarray}
\frac{d\epsilon_i}{dx} &=& H_{ii}' \ , \nonumber\\
\frac{dH_{ii}'}{dx} &=& -\Omega^2\epsilon_i + \sum_{i\neq j}
\frac{2|H_{ij}'|^2}{\epsilon_i-\epsilon_j} \ , \nonumber\\
\frac{dH_{ij}'}{dx} &=& \sum_{k\neq i,j} H_{ik}'H_{kj}' \left(
\frac{1}{\epsilon_i-\epsilon_k} + \frac{1}{\epsilon_j-\epsilon_k}
\right) + H_{ij}'(H_{jj}' - H_{ii}') \frac{1}{\epsilon_i-\epsilon_j} \ .
\label{eq_cs_17}
\end{eqnarray}
Interpreted classically, these equations describe the dynamics of $N$
point par\-ticles at the positions $\epsilon_i$ ($i=1,\ldots N$) and
constitute the so--called Pechukas gas \cite{Pec83}. With certain
assumptions about the properties of irregular systems, in particular
their eigenfunctions, Pechukas derived \cite{Pec83} linear level
repulsion from the above equations (connection VII).  Moreover, the
classical system corresponding to Eqs.~(\ref{eq_cs_17}) has been shown
to be integrable \cite{Pec83,Nak86}.  In Ref.~\onlinecite{Nar93},
infinitesimal increments $\delta \epsilon_i$ of the eigenvalues
$\epsilon_i$ were considered. With the identification $x^2=2\tau$, the
following statistical properties of the $\delta \epsilon_i$ could be
derived in the large--$N$ limit,
\begin{eqnarray}
\langle\delta\epsilon_i\rangle &=& -\frac{\epsilon_i}{a^2}\delta\tau + 
\beta\sum_{i\neq j} 
\frac{\delta \tau}{\epsilon_i-\epsilon_j} \ ,\nonumber\\
\langle\delta\epsilon_i^2\rangle &=& 2\delta\tau \ .
\label{eq_cs_18}
\end{eqnarray}
After setting $\lambda_i=\epsilon_i$ these two equations imply the
Fokker--Planck equation (\ref{eq_cs_15}) for the distribution function
$P(\{\epsilon_i\};\tau)$. This completes connection IX in
Fig.~\ref{fig_cs}. Taking into account the close relation between the
Hamiltonians (\ref{eq_cs_3}), (\ref{eq_cs_16}) and the non--linear
$\sigma$ model \cite{Efe83,Sim93a,Sim93b,Sim93d} (connection III), we
see that the second exact link between universal correlations and the
Sutherland Hamiltonian has been established. Prior to the publication
of Ref.~\onlinecite{Nar93}, Beenakker \cite{Bee93b} already noticed
the relation between the universal parametric correlators and Dyson's
Brownian motion model within the hydrodynamic limit.

Finally, it remains to comment on the connections I, II, and VIII in
Fig.~\ref{fig_cs}. The corresponding lines are dashed because one has
to assume ergodicity, i.e. the equality of the ensemble average in
Random Matrix Theory and the running average over energy in a real
microscopic system.  In many cases this hypothesis has been checked
numerically. Numerical confirmations for connection I include the case
of quantum billiards \cite{Sim93a,Sim93b,Sim93d}, hydrogen in a
magnetic field \cite{Sim93g}, and strongly correlated many--particle
systems \cite{Mon93}. Connection II is also based on numerical
evidence, and connection VIII is discussed in Ref.~\onlinecite{Haa91}.

\subsubsection{Further results}
\label{cs_3}

The calculation of the dynamical density--density correlators for
$\beta = 1,2$ and $4$ and the discovery of the various relations
discussed in the preceding Secs.~\ref{cs_1} and \ref{cs_2} have
stimulated important further progress in the understanding of
interacting Fermions in one dimension. Using the information provided
by Simons {\it et al.}, Haldane \cite{Hal94} conjectured the form of
the density--density correlator for the Calogero--Sutherland model
{\it for all rational values of} $\beta$. In these cases there is, of
course, no known relation to Random Matrix Theory. Independently,
Minahan and Polychronakos \cite{Min94} used group--theoretical
arguments to guess the two--point density correlator for integral
values of $\beta/2$. Even earlier, Haldane and Zirnbauer \cite{Hal93}
calculated the retarded part of the single--particle Green function
for $\beta=4$ and later \cite{Zir95} extended their results to $\beta
= 1$ and to the advanced part. Haldane's conjecture was proven by Ha
\cite{Ha94} with the help of the theory of Jack polynomials
\cite{Sta89}. At the same time Ha extended the results derived in
Refs.~\onlinecite{Hal93,Zir95} to arbitrary rational $\beta/2$, thus
confirming another conjecture by Haldane \cite{Hal94_a}. Forrester
\cite{For93} seems to have been the first author to introduce methods
involving Jack polynomials in the present context. He, too, used such
methods \cite{For95} to prove Haldane's second conjecture.

This shows how new results in Random Matrix Theory have eventually
led to novel dynamical correlation functions for a whole class of
interacting Fermion systems.

\subsection{QCD and chiral Random Matrix Theory}
\label{chiral}

During the last few years, RMT has been applied successfully to
certain aspects of quantum chromodynamics (QCD). More precisely, some
properties of data generated from lattice gauge calculations have been
shown to agree with RMT predictions. This is true, in particular, for
level correlation functions and for low--energy sum rules. Other data
like those on the chiral phase transition have been semiquantitatively
reproduced by RMT. These discoveries are important, and have generated
much interest, because they show that QCD possesses generic features.
RMT is expected to be the ideal tool to identify and quantify such
features. Thus, application of RMT to lattice QCD data will hopefully
help to separate the generic from the system--specific properties, and
to identify the latter. The work was pioneered by Verbaarschot and his
colleagues at Stony Brook. Our references are incomplete. Fairly
complete lists can be found in Refs.~\onlinecite{Ver96b,Wet96a}.

Lack of space forces us to treat this topic but briefly, without paying 
attention to the historic development, and without attempting to put these
developments into the overall context of QCD. We confine ourselves to
a description of those aspects which are most relevant for RMT. We show
evidence that lattice data do have generic features (Sec.~\ref{gen}),
\begin{figure}
\centerline{
\psfig{file=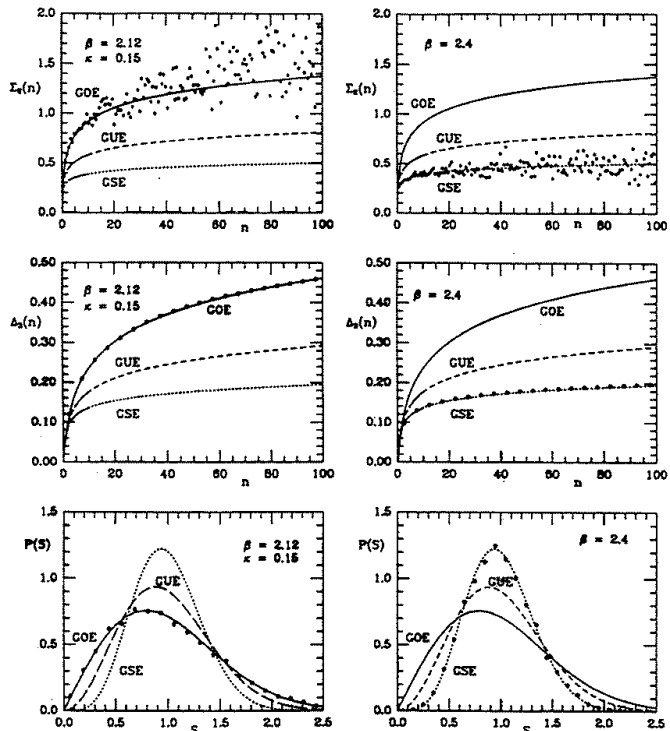,width=3.5in}
 }
\caption{
Comparison of lattice data and RMT predictions for the number variance
$\Sigma_2$, for the $\Delta_3$ statistic, and for the nearest neighbor
spacing distribution $p(s)$. Left: Wilson Fermions, right:
Kogut-Suskind Fermions. Taken from
Ref.~\protect\onlinecite{Hal95}. 
} 
\label{fig9}
\end{figure}
we sketch the connection between QCD and RMT and introduce the chiral 
ensembles (Sec.~\ref{conn}), we describe the application of chiral
Random Matrix Theory (chRMT) to the chiral phase transition
(Sec.~\ref{models}), and we mention some current lines of interest and
open problems (Sec.~\ref{further}).

\subsubsection{Evidence for generic features}
\label{gen}

The first piece of evidence comes from studies of the spectrum of the 
Euclidean (and Hermitean) Dirac operator 
\begin{equation}
iD\hspace{-0.27cm}/ = i\partial\hspace{-0.22cm}/ + g \frac{\lambda^a}{2}
A\hspace{-0.22cm}/
\label{dir}
\end{equation}
for massless Fermions. Here, $g$ is the coupling constant, the
matrices $\lambda^a$ are the generators of the gauge group, and
$A\hspace{-0.22cm}/$ are the gauge fields.

Kalkreuter \cite{Kal95} has calculated the spectrum of
$iD\hspace{-0.27cm}/$ for an $SU(2)$ gauge theory on the lattice, both
for Wilson and for staggered Fermions. These data have been used to
calculate the NNS distribution $p(s)$ and the $\Delta_3$ statistic
(cf. Eqs.~(\ref{1ba12}) and (\ref{1ba16})). The results are shown and
compared with chRMT predictions in Fig.~\ref{fig9}, taken from
Ref.~\onlinecite{Hal95}. (The symmetry of the Dirac operator on the
lattice differs from the case of the continuum. Wilson and staggered
Fermions correspond to the chGOE and chGSE, respectively, both defined
in Sec.~\ref{conn}). The agreement is impressive. It shows that the
spectral fluctuation properties of the Dirac operator are generic.
Somewhat unfortunately, these properties have virtually no bearing on
physical observables in QCD.

Another universal property (absent in classical RMT) occurs because
the spectrum of $iD\hspace{-0.27cm}/$ is symmetric about zero. The
operator $iD\hspace{-0.27cm}/$ anticommutes with $\gamma_5$ and the
eigenvalues $\lambda_n$ therefore occur in pairs $\pm \lambda_n$ with
opposite signs. Thus, level repulsion should affect directly the
average spectral density $\rho(\lambda)$ near zero energy ($\lambda =
0$) in a generic way. Here, $\rho(\lambda)$ is defined as
\begin{equation}
\rho(\lambda) =  \langle \sum_n \delta( \lambda - \lambda_n) \rangle. 
\label{rho}
\end{equation}
The average (indicated by angular brackets) is taken over all gauge
field configurations. The weight factor is the exponential of the
Euclidean action. The expected generic dependence of $\rho(\lambda)$
on $\lambda$ near $\lambda = 0$ has direct consequences for QCD
because the value of $\rho(0)$ is related to the vacuum expectation
value $\langle \overline{q} q \rangle$ of the chiral condensate by the
Banks--Casher formula $\langle \overline{q} q \rangle = - \pi
\rho(0)/V$ where $V$ is the space--time volume. Chiral symmetry is
broken if $\langle \overline{q} q \rangle$ is finite (non--zero). In
this case, $\rho(0) \sim V$, and the spacing of eigenvalues is $\sim
1/V$ rather than $\sim 1/V^{1/4}$ as would be the case for a
non--interacting system. Chiral symmetry is restored for $\rho(0) =
0$.

\begin{figure}
\centerline{
\psfig{file=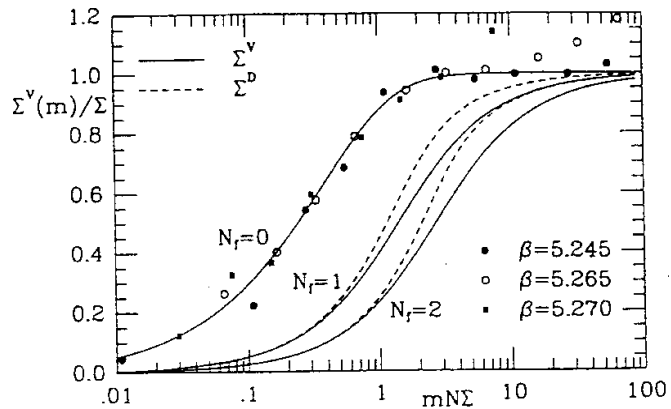,width=3.5in}
 }
\caption{
Dependence of the chiral condensate on the quark mass. Dots: Lattice
data. Solid curve: RMT prediction. Taken from
Ref.~\protect\onlinecite{Ver96}. 
} 
\label{fig10}
\end{figure}
First evidence in favor of this assertion (universality of
$\rho(\lambda)$ near $\lambda = 0$) came from the Leutwyler--Smilga
sum rules \cite{Leu92} for the eigenvalues $\lambda_n$. For any
integer $k$, the sum rules yield closed expressions for $V^{-2k}
\sum_n \lambda_n^{-2k}$. This sum can in turn be written as an
integral over $\rho(\lambda)$. It was shown \cite{Shu93} that the sum
rules can be derived from chRMT. Moreover, it was conjectured that the
microscopic limit $\rho_S(z) = \rho(z / (V | \langle \overline{q} q
\rangle|)) / (V | \langle \overline{q} q \rangle|)$ of the spectral
density, obtained by rescaling the argument by $1 / (V | \langle
\overline{q} q \rangle |) $, should in the thermodynamic limit be a
universal function. This conjecture was validated when it was shown
\cite{Ver96} that lattice results for the dependence of the chiral
condensate on the mass of the valence quarks can be reproduced using
the chRMT result for $\rho_S$ as the only input, see Fig.~\ref{fig10}.

Perhaps the most compelling evidence so far for the universality of
$\rho(\lambda)$ near $\lambda = 0$ is obtained from a comparison of
lattice data for staggered Fermions in the quenched approximation with
chRMT predictions. The lattice calculations were performed for strong
coupling ($\beta = 4 / g^2 = 2.0$) and for four different lattice
\begin{figure}
\centerline{
\psfig{file=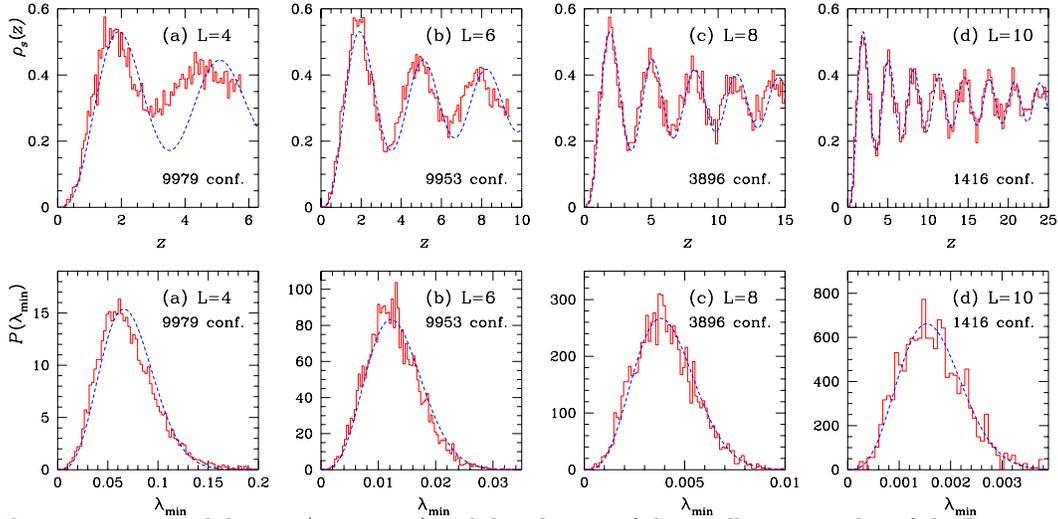,width=5.5in}
 }
\caption{
Microscopic spectral density (upper row) and distribution of the
smallest eigenvalue of the Dirac operator for four different lattice
sizes. The histograms represent lattice data, the dashed curves
correspond to chRMT predictions. Taken from
Ref.~\protect\onlinecite{Berbe97}. 
} 
\label{fig11}
\end{figure}
sizes $V = L^4$ with $L = 4,6,8$, and $10$, resulting in 9979, 9953,
3896, and 1416 configurations, respectively \cite{Berbe97}. The
comparison is shown in Fig.~\ref{fig11}. Higher correlation functions
also display universal features.

\subsubsection{Chiral random matrix ensembles}
\label{conn}

The chiral random matrix ensembles can best be introduced by writing 
the Euclidean Dirac operator of Eq.~(\ref{dir}) in matrix form. We choose 
a chiral basis (consisting of eigenstates of $\gamma_5$). In this basis,
$iD\hspace{-0.27cm}/$ 
only couples states of opposite chirality and has the form  
\begin{equation}
  \label{mat1}
  \left[\matrix{0&iD\hspace{-0.27cm}/\cr(iD\hspace{-0.27cm}/)^\dagger&0}
  \right] \ .   
\end{equation}
This suggests defining a random matrix model with matrices of the form
\begin{equation}
  \label{mat2}
  \left[\matrix{0 & W \cr W^\dagger & 0}\right] 
\end{equation}
where $W$ is a random matrix. In the chiral random matrix models, the 
average over gauge fields with the exponential of the Euclidean action 
as weight factor is replaced by the average over the matrices $W$ with 
a suitable weight factor $P(W)$ as in Sec.~\ref{qc1rca}. Normally, $P(W)$
is chosen as a Gaussian. The limit $N \rightarrow \infty$ of the 
dimension $N$ of $W$ is taken.

Depending on the symmetry of $iD\hspace{-0.27cm}/$ (which must be
reflected in the symmetry of $W$), there are \cite{Ver94} three chiral
ensembles, in complete analogy to Sec.~\ref{qc1rca}. They depend on
the number $N_c$ of colors and on the representation (fundamental [f]
or adjoint [a]) of the Fermions. The chiral Gaussian unitary,
orthogonal and symplectic ensembles are defined, respectively, and
apply as follows: (i) chGUE ($W$ complex): for [f] and $N_c \geq 3$,
(ii) chGOE ($W$ real): for [f] and $N_c = 2$, and (iii) chGSE ($W$
real quaternion): for [a] and $N_c \geq 2$. In the case of lattice
calculations with $SU(2)$ gauge symmetry, the chGOE (the chGSE)
applies for Wilson Fermions (for staggered Fermions, respectively).

The chiral random matrix ensembles differ from the classical ensembles
introduced in Sec.~\ref{qc1rcb} mainly in their behavior near $\lambda
= 0$. As a consequence of the chiral symmetry of
$iD\hspace{-0.27cm}/$, in these ensembles the point $\lambda = 0$
plays a special role, and the translational symmetry of the classical
ensembles is broken. Near $\lambda = 0$ there is repulsion of pairs of
eigenvalues of opposite sign. This causes $\rho_S(x)$ to become
generic. A very similar mechanism operates in the unitary ensemble
describing Andreev scattering, see Sec.~\ref{spec_2}. By way of
example, we give here the form of $\rho_S(x)$ for the chGUE. It
depends on the number $N_f$ of flavors and on the topological charge
$\nu$,
\begin{equation}
\rho_S(x) = \frac{x}{2}\left(J_{N_f+\nu}^2(x)-
            J_{N_f+\nu+1}(x)J_{N_f+\nu-1}(x)\right) \ .
\label{dens}
\end{equation}
The special role of the point $\lambda = 0$ does not, however, affect
the form of the NNS distribution $p(s)$, or of the $\Delta_3$
statistic which in the bulk are the same as for the classical
ensembles, see Fig.~\ref{fig9}.

\subsubsection{The chiral phase transition}
\label{models}

As shown in Sec.~\ref{gen}, for zero quark masses chiral random matrix
models yield a non--zero value of the chiral condensate. This
describes the spontaneous breaking of chiral symmetry at zero
temperature.  Two models have been put forward which modify
Eq.~(\ref{mat2}) in such a way that temperature is included in a
schematic way \cite{Jac96a,Janik97,Wetti97,Wet96}.  These models aim
at a description of the chiral phase transition. In both models, a
temperature--dependent diagonal matrix $D$ is added to $W$ in
Eq.~(\ref{mat2}).

What happens qualitatively to $\rho(\lambda)$ as $D$ is increased,
starting from $D = 0$? We focus attention on the {\it global}
properties of $\rho$ and disregard the local suppression of $\rho$ at
$\lambda = 0$ displayed in Sec.~\ref{gen}. In the simplest case where
$D$ is a multiple of the unit matrix, $D = D_0 {\bf 1}_N$, two
limiting cases are obvious. At $D_0 = 0$, $\rho(\lambda)$ has the
shape of a semicircle. For $W = 0$ and $D_0 > 0$, on the other hand,
all eigenvalues have values $\pm D_0$. Keeping $D_0$ fixed and
increasing $W$ from $W = 0$ will lift the degeneracy of eigenvalues of
fixed sign and generate two semicircles with identical radii and with
centers at $\pm D_0$. The two semicircles do (do not) overlap if their
radii are smaller (bigger) than $D_0$. We now apply these insights to
the chiral phase transition. We use the fact that according to the
Banks--Casher formula, the chiral condensate is proportional to
$\rho(0)$.  We keep $W$ fixed and increase $D_0$, starting from
$D_0=0$. The semicircle becomes deformed. At the value of $D_0$ where
it splits into two semicircles, $\rho(0)$ vanishes, and the chiral
phase transition occurs. This is basically the mechanism employed in
both Refs.~\onlinecite{Jac96a,Wet96}.

We have discussed this point in some detail because it shows that in
describing the chiral phase transition in terms of chRMT, use is made
of the semicircle law, not one of the generic properties of chRMT. And
the critical exponents for the phase transition are determined by the
shape of the semicircle at the end of the spectrum. Is this procedure
trustworthy? Several authors have investigated non--Gaussian
probability distributions $P$ for classical RMT, where $P(H)$ has the
form $\exp(-{\rm tr}V(H))$ with $V(H)$ some polynomial, see
Sec.~\ref{univ}. It was found that in the limit of infinite dimension,
the spectra are generically confined to a finite interval, and that
the behavior of the mean level density near the end points is always
of semicircle type. More precisely: Near the end point $a$ of the
spectrum, the mean level density $\rho(E)$ behaves like $(E -
a)^{\alpha}$ with $\alpha$ a rational number between zero and one.
This lends support to the models of Refs.~\onlinecite{Jac96a,Wet96}.
Still, it must be kept in mind that RMT does not necessarily give a
correct prediction of the details of the chiral phase transition.

The two models mentioned above differ in detail. In
Ref.~\onlinecite{Jac96a}, the temperature is included schematically in
the form of the lowest Matsubara frequency, and $D$ is accordingly
chosen as $\pi T {\bf 1}$. With $T_c$ the critical temperature, the
temperature dependence of the chiral condensate $\langle \overline{q}
q \rangle (T)$ is given by $\langle \overline{q} q \rangle (T) =
\langle \overline{q} q \rangle (0) \sqrt{1 - (T/T_c)^2}$. The chiral
phase transition is of second order with mean field critical
exponents. The inclusion of all Matsubara frequencies
\cite{Janik97,Wetti97} yields very similar results. In the model of
Ref.~\onlinecite{Wet96} (which was inspired by the instanton liquid
model), a fraction $\alpha$ of the diagonal elements of $D$ vanishes
(with $0 \leq \alpha \leq 1$ trivially), while the rest has identical
values $d$. Both $\alpha$ and $d$ are (unknown) monotonically
increasing functions of $T$. For a phase transition to occur, $\alpha$
must be equal to one, and $d$ must exceed a critical value $d_c$. The
order as well as the critical exponents of the chiral phase transition
depend on the way in which $\alpha$ and $d$ approach $1$ and $d_c$,
respectively. A comparison between these results and lattice and/or
instanton model calculations may shed light on the
temperature--dependence of $\alpha$ and $d$.

\subsubsection{Current lines of research and open problems}
\label{further}

The field is still quite young. Yet, a fair number of papers has
appeared. For recent reviews see Refs.~\onlinecite{Ver96b}.  Aside
from the topics mentioned above, these papers have addressed the
following points. (a) The spectrum of the Dirac operator at finite
temperature~\cite{Ste96}; (b) the calculation of $k$--point functions
at finite $T$ (with $k > 2$)~\cite{Guh97a} (c) the finite volume QCD
partition function for finite quark masses~\cite{Jur96}; (d) the
spectrum of the lattice Dirac operator for Wilson Fermions; (e)
inclusion of the chemical potential in
chRMT~\cite{Ste96a,Jan96,Halas97a,Halas97b}. The last point is
probably the most difficult but also the most challenging because on
the lattice, unquenched calculations with a chemical potential seem
beyond reach. As mentioned above, this list is incomplete.

\subsection{Random matrices in field theory and quantum gravity}
\label{ftqg}

After some preliminary remarks (Sec.~\ref{ftqg1}) we briefly explain
the use of topological concepts and RMT in general field theory
(Sec.~\ref{ftqg2}) before we discuss the special example of
two--dimensional quantum gravity in Sec.~\ref{ftqg3}.

\subsubsection{Preliminary remarks}
\label{ftqg1}

In the stochastic models discussed through most of this review, random
mat\-ri\-ces were used successfully to simulate generic properties of the
Hamiltonian or of a scattering matrix in a quantum problem described
by the Schr\"odinger equation. In Sec.~\ref{chiral} we have seen
that this substitution also works for certain aspects of relativistic
quantum mechanics where the Dirac equation applies. Demonstrative
experiments present strong evidence for the usefulness of this same
procedure in the study of classical wave phenomena, see
Sec.~\ref{qc2de}. This holds true,
in particular, for electromagnetic waves which, in three dimensions,
are described by a vectorial Helmholtz equation, and for
elastomechanical waves which are described by Navier's equation. 

We now turn to stochastic models where the use of random matrices has
a different origin. The representative examples given below stand for
a whole class of problems where it is the fields --- rather than the
Hamiltonian or another linear operator --- which are replaced by an
ensemble of random matrices. Often, the replacement has topological
reasons. Cases in point are field theory (Sec.~\ref{ftqg2}), 
and quantum gravity in two dimensions (Sec.~\ref{ftqg3}). Although
differently motivated, these random matrix models are, at least on the
formal and on the mathematical level, often closely related to the
ones discussed in other sections of this review. Because of this 
connection, a fruitful transfer of insights and methods between both
areas exists. We expect that in future years this transfer will
become even stronger in both directions. 

\subsubsection{Planar approximation to field theory}
\label{ftqg2}

Consider a non--Abelian gauge theory with internal symmetry group
$SU(N)$. We treat $N$ as a parameter and study the limit of large $N$.
Such a study is motivated by quantum chromodynamics where the gauge
group is $SU(3)$ because quarks come in three colors. In the absence
of exact solutions for $SU(3)$, it is hoped that the results for large
$N$ might be of help.

A useful classification of the diagrams generated by the perturbation
expansion of the field theory in powers of the coupling constant uses
topological concepts. Surfaces can be classified topologically in
terms of their genus. For instance, a sphere has genus $h=0$, a torus
has genus $h=1$, etc. The diagrams are classified by the minimum genus
of the surface needed to draw them. Diagrams with genus zero are
called planar. It was shown by 't Hooft~\cite{Hoo74} that the
asymptotic expansion of the field theory in inverse powers of $N$ 
yields terms which correspomd exactly to this topological classification.
Only planar diagrams survive in the limit $N \rightarrow \infty$,
provided the coupling constant is properly scaled with $N$. The higher
order terms in $1/N$ correspond to non--planar diagrams which have to
be drawn on surfaces of genus $h$ higher than zero.

The generality of this result suggests that a model study of the
expansion of field theories might be very useful. Brezin et
al.~\cite{Bre78} studied models for the planar approximations of
$\varphi^3$ and $\varphi^4$ theories. We focus here on the $\varphi^4$
theory. In a $d$-dimensional space $x_\mu, \mu=1, \ldots, d$ such a
theory is defined by a Lagrangean of the type
\begin{equation} 
\label{ft0} 
{\cal L} = \partial_\mu\varphi\partial_\mu\varphi + 
                    \frac{1}{2}\varphi^2 + \tilde{g}\varphi^4 
\end{equation} 
where $\tilde{g}$ is the coupling constant. A model for this theory is
obtained by replacing the field $\varphi(x)$ with a Hermitean $N
\times N$ matrix $M(x)$. The model Lagrangean reads 
\begin{eqnarray}
\label{ft1} 
{\cal L} &=& {\rm tr}\left(\partial_\mu M \partial_\mu M\right) 
                   + V(M,g) 
                                \nonumber\\ 
V(M,g) &=& \frac{1}{2}{\rm tr}M^2 + \frac{g}{N} {\rm tr}M^4 \ .  
\end{eqnarray} 
The global invariance group of this Lagrangean is the group $SU(N)$.
Thus, it is possible to study the planar limit by taking the limit $N
\rightarrow \infty$ while the parameter $g$ is held fixed. The Feynman
rules for the diagramatic expansion of this matrix model in powers of
$g$ involve the two matrix indices and are a straightforward extension
of those for scalar or vector fields. To make plausible the explicit
occurrence of the factor $1/N$ in the coupling constant $g/N$, we
refer to the discussion in Sec.~\ref{univ} where more general
non--Gaussian weight factors with a polynomial potential in the
exponent will be considered. Within this model one can generically
solve the so-called counting problem, i.e.~one can find the number
$E^{(d)}(g)$ of connected planar vacuum diagrams of a given order in
$g$ in $d$ dimensions. We discuss this for $d = 0$ and $d = 1$.

In zero dimensions, each diagram is unity, apart from an overall
weight. This implies \cite{Bre78} that, up to a $g$-independent
normalization constant, $E^{(0)}(g)$ can be obtained from the 
large--$N$ limit of the expression 
\begin{equation}
  \label{ft2}
  \exp\left(-N^2E^{(0)}(g)\right) = \int d[M] \exp\left(-V(M,g)\right)
  \qquad (N\to\infty)
\end{equation}
with the usual flat measure $d[M]$. To connect to standard random
matrix theory, the integrand can be viewed as a non--Gaussian
probability density. For large $N$, the normalization integral yields
the counting function versus $g$. It is shown in
Ref.~\onlinecite{Bre78} that this counting function coincides with the
one of the $\varphi^4$ theory. In the large--$N$ limit, there are no
contributions due to the matrix structure of the fields $M$.

Using the $U(N)$ invariance of the integrand, we reduce the problem to
integrations over the eigenvalues $\lambda_n, n=1,\ldots,N$ where the
squared Vandermonde determinant $\Delta_N^2(\lambda)$ appears in the
integrand. To compute the large--$N$ limit of Eq.~(\ref{ft2}), the
method of steepest descent is used and yields saddle--point equations
for the eigenvalues $\lambda_n$. The counting function $E^{(0)}(g)$ is
approximated by the logarithm of the integrand of the right hand side
of Eq.~(\ref{ft2}) taken at these saddlepoints. In a proper continuum
limit, the resulting sums are replaced by integrals. This yields
\begin{eqnarray}
  E^{(0)}(g) = \int_{-2a}^{+2a} d\lambda u(\lambda) 
                  \left(\frac{1}{2}\lambda^2 + g\lambda^4\right) -
               \int_{-2a}^{+2a} d\lambda u(\lambda)
               \int_{-2a}^{+2a} d\lambda^\prime u(\lambda^\prime)
                         \ln|\lambda-\lambda^\prime| \nonumber\\
\label{ft3}
\end{eqnarray}
where $u(\lambda)$ is the mean level density, and where $-2a$ and
$+2a$ with $a = a(g)$ are the end points of the spectrum. The first
term on the right hand side of Eq.~(\ref{ft3}) is due to the
potential $V(M,g)$, the second one, due to the Vandermonde
determinant.  The factor $1/N$ in the coupling constant in
Eq.~(\ref{ft1}) has disappeared. This is because the continuum limit
involves a rescaling of the eigenvalues by a factor of
$\sqrt{N}$. Brezin {\it et al.}~\cite{Bre78} find an explicit result for
$u(\lambda)$. In the limit $g=0$, this result reduces to the Wigner
semicircle law. Hence, a closed expression for the counting function
$E^{(0)}(g)$ is obtained. In the planar limit, the coefficients of an
expansion of $E^{(0)}(g)$ in powers of $g$ give the desired
combinatorical factors.

For $d=1$ the planar approximation provides an interesting connection
to the Calogero--Sutherland Model and related theories, see
Sec.~\ref{cs}. We replace the
Lagrangean of Eq.~(\ref{ft1}) by the Hamiltonian and quantize the
latter. This yields
\begin{equation}
  \label{ft4}
  {\cal H} = -\frac{1}{2}\Delta + V(M,g) 
\end{equation}
where the kinetic term is the usual Laplacian in matrix space,
\begin{equation}
  \label{ft5}
  \Delta = \sum_{n=1}^N \frac{\partial^2}{\partial M_{nn}^2} +
             \frac{1}{2}\sum_{n<m}
             \left(\frac{\partial^2}{\partial{\rm Re}^2M_{nm}}+
                   \frac{\partial^2}{\partial{\rm Im}^2M_{nm}}\right)
                   \ .
\end{equation}
It can be shown~\cite{Bre78} that for large $N$, the counting function
$E^{(1)}(g)$ is given by the lowest eigenvalue of ${\cal H}$, 
\begin{equation}
  \label{ft6}
  {\cal H}\psi = N^2 E^{(1)}(g) \psi \ .
\end{equation}
The value of $E^{(1)}(g)$ is found by varying the expectation value of
${\cal H}$ with respect to the functions $\psi$ in the limit of large
$N$. The ground--state wave function $\psi$ can only
depend on the radial degrees of freedom.  This is so because the
$U(N)$ invariance of the potential $V(M,g)$ implies the 
condition $\psi(M) = \psi(U M U^{\dagger})$.  Hence, $\psi$ is a
symmetric function of the eigenvalues $\lambda_n$ of $M$.  To
construct such symmetric wave functions $\psi$, Brezin {\it et al.}
make the ansatz
\begin{equation}
  \label{ft7}
  \psi(\lambda_1,\ldots,\lambda_N) = 
        \frac{\omega(\lambda_1,\ldots,\lambda_N)}{\Delta_N(\lambda)} \ .
\end{equation}
The Vandermonde determinant $\Delta_N(\lambda)$ is antisymmetric in
the eigenvalues $\lambda_n$. Since $\psi$ is symmetric, $\omega$ has
to be antisymmetric as well. This ansatz reduces the kinetic part of
Eq.~(\ref{ft4}) to the Euclidean Laplacian of the $N$ matrix
eigenvalues $\lambda_n, n=1,\ldots,N$. If, moreover, the interaction
is of $\varphi^4$ or related type, the Hamiltonian acquires the form
of $N$ non--interacting Fermions, described by the antisymmetric
wavefunction $\omega(\lambda_1,\ldots,\lambda_N)$. This construction
yields explicit results for the counting function. It also shows the
connection to the Calogero--Sutherland model, see Sec.~\ref{cs}. 

By adding to the potential the terms ${\rm tr}M_kM_{k^\prime}$,
Itzykson and Zuber~\cite{Itz80} extended these investigations to $K$
coupled matrix fields $M_k, k=1,\ldots,K$. In the resulting model, the
$U(N)$ invariance is broken. In the case $K=2$, it was possible to
integrate analytically over the diagonalizing groups. The result, the
famous Itzykson-Zuber integral, was later identified as a special case
of a more general formula due to Harish-Chandra~\cite{Har58}. 

\subsubsection{RMT and two-dimensional quantum gravity}
\label{ftqg3}

As in field theory, the use of Random Matrix Theory in
two--dimensional quantum gravity is motivated topologically. Again,
the basic idea derives from the planar approximation. The starting
point is string theory, providing a link to the theory of random
surfaces where further applications exist. A review was recently given
by Abdalla {\it et al.}~\cite{Abd94}. Here, we only try to motivate
the use of RMT in quantum gravity.

We consider a $d$-dimensional space in which functions
$X^\mu(\xi_0,\xi_1), \ \mu=1,\ldots,d$ parametrize the position of a
string on the two-dimensional world sheet $(\xi_0,\xi_1)$. For the
description of the string dynamics, it is convenient to use the
Polyakov string action~\cite{Pol81}, 
\begin{equation}
  \label{qg1}
  S = \frac{1}{2} \int d^2\xi \sqrt{|g|} g^{\alpha\beta}
                  \partial_\alpha X^\mu\partial_\beta X_\mu \ .
\end{equation}
Here $g_{\alpha\beta}(\xi_0,\xi_1)$ with $\alpha,\beta=0,1$ is
the two--dimensional gravitational field. Minimization of the
functional $S$ with respect to the $X^\mu$ and to the fields
$g_{\alpha\beta}$ gives the correct equations of motion for the string
with all required constraints. However, further contributions like
a cosmological term have to be added to the action $S$. The partition
function of a string action of this type is then given by the
functional integral 
\begin{equation}
  \label{qg2}
  Z = \int D[X] \int D[g] \exp(-S) \ .
\end{equation}
The integration over the gravitational field amounts to a summation
over all topologies of the two-dimensional surface described by the
string. Therefore, the partition function can be expanded in terms of
the genus $h$ and the area $A$ of the surface,
\begin{equation}
  \label{qg3}
  Z = \sum_{h=0}^\infty \kappa_0^{2h} \int_0^\infty dA 
                 \exp\left(-\mu_BA\right) Z_h(A) \ .
\end{equation}
Here $\mu_B$ is the bare cosmological constant and $\mu_BA$ is the
cosmological term of the action. The expansion proceeds in powers of
the bare string coupling constant $\kappa_0$. The expansion
coefficients $Z_h(A)$ are the partition functions for fixed genus $h$ 
and area $A$. At this point, the connection to the previous section
becomes apparent where the genus expansion was motivated by the
topology of diagrams. Hence, the $h=0$ term in Eq.~(\ref{qg3})
corresponds to a planar approximation. 

A way to proceed in actual calculations is to discretize the surfaces
of given genus $h$, i.e.~the sphere, the torus and so on, using
polygons with a fixed number $n$ of edges. Discretized models for
\begin{figure}
\centerline{
\psfig{file=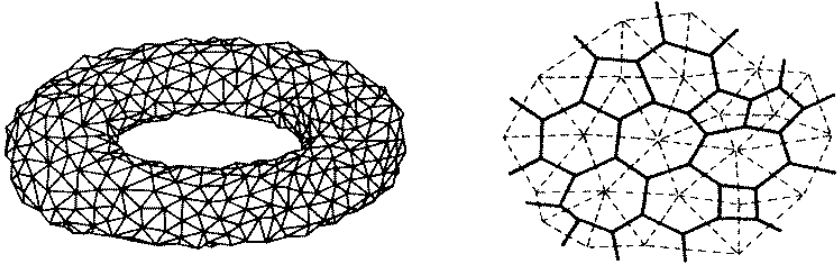,width=4.5in,angle=-90}
}
\caption{
  Left hand side: A triangulated torus. Right hand side: A piece of a 
  triangulated surface (dashed lines) superposed with its dual
  (solid lines). Taken from Ref.~\protect\onlinecite{Abd94}.
}
\label{figfs1}
\end{figure}
random surfaces were introduced by David~\cite{Dav85},
Kazakov~\cite{Kaz85} and Ambj\o rn, Durhuus and
Fr\"ohlich~\cite{Amb85}. Apparently, these authors were inspired by
Regge's work~\cite{Reg61} of 1961. Explicit calculations are often
easier in such discretized models than in the continuum version.

The dual graph of a discretization of a surface is defined by
connecting the centers of the covering $n$-gons with lines. These dual
graphs are typical Feynman graphs of given genus $h$ of a
zero-dimensional $\varphi^n$ field theory. Thus, the partition
functions $Z$ can be approximated in terms of a model of the
$\varphi^n$ type. A triangulated torus and the dual of a piece of a
triangulated surface are shown in Fig.~\ref{figfs1}. However, certain
constraints must be observed. Most importantly, the partition
functions $Z_h(A)$ in Eq.~(\ref{qg3}) show a double scaling relation.
It turns out that only Hermitean $N\times N$ matrices $M$ meet all the
requirements. The resulting matrix model is of the type
\begin{eqnarray}
  \label{qg4}
  Z^\prime &=& \int d[\varphi] \exp\left(-N{\rm tr}V(M)\right)
                                            \nonumber\\
  V(M) &=& \frac{1}{2}M^2 + 
                \sum_{n>2} (g_n\Lambda^{(n-2)/2}) M^n \ .
\end{eqnarray}
The $g_n$ are coupling constants and $\Lambda$ is associated with the
cosmological constant. As in the previous section, an asymptotic
expansion in inverse powers of the dimension $N$ effectively yields a
diagrammatic genus expansion. The coefficient functions have the
desired scaling properties and are identified with the functions
$Z_h(A)$. The expansion parameter $1/N$ plays the role of the string
coupling constant $\kappa_0$ in the discretized version of
Eq.~(\ref{qg3}). 


\setcounter{equation}{0}
\section{Universality}
\label{univ}

The overwhelming majority of investigations involving invariant random
matrix ensembles deals with the Gaussian case (GRMT). However, the
requirement that the ensemble should be invariant under appropriate
(unitary, orthogonal, or symplectic) transformations in matrix space,
admits a much wider class of matrix ensembles with a probability
density $P_{N\beta}(H)$ of the form
\begin{equation}
P_{N\beta}(H) = Z^{-1}\exp\left( - {\rm tr}V(H) \right) \ .
\label{eq_uni1}
\end{equation}
Here, $Z^{-1}$ is the normalization constant or inverse partition
function. The potential $V(H)$ is arbitrary provided the existence of
$Z$ is guaranteed. The Gaussian case corresponds to the choice
\begin{equation}
V(H) = \frac{\beta}{4v^2} H^2,
\label{eq_uni2}
\end{equation}
where $v$ and the elements of the $N\times N$ matrix $H$ have the
dimension energy. For other forms of the Gaussian
potential typically used in the literature we refer to the discussion
at the end of Sec.~\ref{qc1rcb}. It has been mentioned in
the Introduction of this review that the choice (\ref{eq_uni2}) is
dictated by mathematical convenience rather than by physical
principles. Under these circumstances it is necessary and important to
ask which statistical properties are generic, i.e. common to all or
very many ensembles defined in Eq.~(\ref{eq_uni1}), and which are
specific for the Gaussian case. This is the issue referred to as
``universality''.

First convincing numerical evidence in favor of universality was
provided by Porter and Rosenzweig~\cite{Por60}, see Appendix A.1 of
Mehta's book~\cite{Meh91}. These authors calculated the eigenvalue
density and 
the nearest neighbor spacing distribution for the Gaussian
probability density and for the densities $P(H_{ij})=
(1/2)\Theta(1-|H_{ij}|)$ and
$P(H_{ij})=(1/2)(\delta(H_{ij}+1)+\delta(H_{ij}-1))$. In all these
cases they found convergence to the semicircle law (eigenvalue
densities)  and to a curve close to the Wigner surmise (spacing
distribution), respectively. For obtaining the semicircle law,
statistical independence of the matrix elements is the essentil
ingredient. This investigation greatly enhanced the confidence in the
usefulness of RMT. From the mathematical point of view, an interesting
connection to the theory of free random variables was found by
Voiculescu \cite{Voicu85}. 

Orthogonal polynomials \cite{Meh91} provide a powerful analytical tool
to investigate Gaussian as well as non--Gaussian ensembles of random
matrices. We briefly outline this method. Upon diagonalization of the
matrices $H$ one obtains the joint probability density function for
the eigenvalues,
\begin{equation}
P_{N\beta}^{(E)}(E_1,\ldots,E_N) = Z^{-1} \prod_{i<j} |E_i-E_j|^\beta
\exp\left(-\sum_iV(E_i)\right),
\label{eq_uni3}
\end{equation}
where $Z$ is now defined without the integration over angles.
We have chosen the energy eigenvalues $E_i$ as arguments instead of
the dimensionless quantities $x_i$, see Sec.~\ref{qc1rcb}.
Equation~(\ref{eq_uni3}) involves the Vandermonde determinant
$\prod_{i<j} (E_i-E_j) = {\rm det}[(E_i)^{j-1}]$. The value of this
determinant remains unchanged if the monomials $(E_i)^{j-1}$ are
replaced by arbitrary (but linearly independent) polynomials $p_n(E)$
with degrees $n=0,1,\ldots,N-1$. If one chooses these polynomials to
be orthogonal in the following sense,  
\begin{equation}
\int_{-\infty}^\infty dE \exp\left(-V(E)\right) p_n(E) p_m(E) =
\delta_{nm} \ ,
\label{eq_uni4}
\end{equation}
and if one defines the functions
\begin{equation}
\psi_n(E) = p_n(E) \exp(-V(E)/2)
\label{eq_uni5}
\end{equation}
the joint probability density in the case $\beta=2$, to which we
restrict ourselves here, has the form
\begin{equation}
P_{N2}^{(E)}(E_1,\ldots,E_N) = {\rm det}[K(E_i,E_j)].
\label{eq_uni6}
\end{equation}
We have defined the kernel
\begin{equation}
K(E,E') = \sum_{i=0}^{N-1} \psi_i(E) \psi_i(E') \ .
\label{eq_uni7}
\end{equation}
In the Gaussian case Eq.~(\ref{eq_uni4}) defines the Hermite polynomials.
All correlation functions 
\begin{equation}
R_{\beta k}(E_1,\ldots,E_k) \propto \int dE_{k+1}\ldots
dE_N \ P_{N\beta}^{(E)}(E_1,\ldots,E_N)
\label{eq_uni7a}
\end{equation}
can now be expressed in terms of this
kernel. For the mean density of eigenvalues 
$\rho(E)\equiv R_{(\beta=2) 1}(E)$ 
and for the
connected two--point correlation function 
$T_{\beta 2}(E,E') = \rho(E)\rho(E') - R_{\beta 2}(E,E')$
we have, in particular,
\begin{eqnarray}
\rho(E) &=& K(E,E) \nonumber \\
T_{(\beta=2) 2} &=& [K(E,E')]^2.
\label{eq_uni8}
\end{eqnarray}

Fox and Kahn \cite{Fox64}, Leff \cite{Lef64}, and Bronk \cite{Bro65}
investigated certain ``classical'' cases of non--Gaussian potentials
$V(E)$, see also chapter 19 in Ref.~\onlinecite{Meh91}.
The cases studied explicitly include the potentials
corresponding via Eq.~(\ref{eq_uni4}) to the Legendre \cite{Lef64},
Laguerre \cite{Fox64,Bro65}, and Jacobi \cite{Fox64,Lef64}
polynomials. In these particular examples the kernel $K(E,E')$ was
shown to be universally given by 
\begin{equation}
K(E,E') = \frac{1}{\pi} \frac{\sin(\rho(\hat{E})(E-E'))}{E-E'}.
\label{eq_uni9}
\end{equation}
in the large--$N$ limit and for $\beta =2$. Performing the limit
$N\to\infty$ was made possible by the well--known Christoffel--Darboux
formula, see Eq.~(\ref{eq_uni16}).
Given $\hat{E}$, Eq.~(\ref{eq_uni9}) holds
for values of $E$ and $E'$ where the mean level density is essentially
constant, $\rho(\hat{E}) \simeq \rho(E) \simeq \rho(E')$. The result
(\ref{eq_uni9}) implied that after proper rescaling (``unfolding'') of
the energy variable, two-- and higher--point correlation functions
were identical for the classical ensembles, in contrast to the mean
level density. As a result, the conjecture \cite{Lef64} that
correlations should be ``relatively insensitive'' to the form of the
potential $V(E)$, became highly plausible.

Further significant progress in the universality problem had to wait
until 1990. Motivated by problems in two--dimensional quantum gravity
(see Sec.~\ref{ftqg}), Ambj{\o}rn, Jurkiewicz, and Makeenko
\cite{Amb90} then discovered a very general and far--reaching result.
(We follow the summary in Ref.~\onlinecite{Amb96}). We consider a
unitary random matrix ensemble with a general polynomial potential
\begin{equation} 
V(H) = \sum_k \frac{g_k}{k} H^k \ .
\label{eq_uni10}
\end{equation}
The coefficients $g_k$ are essentially arbitrary except that for
$N\to\infty$ the mean level density should be non--zero in only {\it
one} finite interval. Using an expansion of the resolvents in powers
of $H$, i.e. a perturbative method, and keeping only terms of leading
order in $N^{-1}$, Ambj{\o}rn {\it et al.} calculated the connected
part $G(E,E')$ of the two--point Green function in the large--$N$
limit. We remind the reader that a $1/N$ expansion is equivalent to a
topological expansion in the genus of the surfaces needed to draw the
diagrams of the field theory defined by Eq.~(\ref{eq_uni10}), see
Sec.~\ref{ftqg2}. 
While Brezin {\it et al.} \cite{Bre78} had found that the mean level
density $\rho(E)$ does depend on the details of the potential $V(H)$
it was now shown by Ambj{\o}rn {\it et al.}
that $G(E,E')$ depends on the potential only through
the spectral end points $a$ and $b$. In the special case of a
symmetric potential $V(H) = V(-H)$ with $b=-a$, $G(E,E')$ is given by
\cite{Amb90}
\begin{eqnarray}
&&G(E,E') = \overline{ {\rm tr} (E-H)^{-1} \ {\rm tr} (E'-H)^{-1}
} - \overline{ {\rm tr} (E-H)^{-1} } \ \
\overline{ {\rm tr} (E'-H)^{-1} } =  \nonumber\\
&&\frac{1}{4(E-E')^2}\left( -2+\frac{(E^2-a^2)+(E^{\prime 
    2}-a^2)}{\sqrt{(E^2-a^2)(E^{\prime 2}-a^2)}} \right) - \frac{1}{4}
    \frac{1}{\sqrt{(E^2-a^2)(E^{\prime 2}-a^2)}} \ . \nonumber\\
\label{eq_uni12}
\end{eqnarray}
The general formula for non--symmetric potentials and an extension to
complex rather than Hermitean matrices is also given in
Ref.~\onlinecite{Amb90}.  The general connected n--point Green
functions
\begin{equation}
G(E_1,\ldots,E_n) = N^{2n-2} \overline{ N^{-1} {\rm tr}
(E_1-H)^{-1} \ldots N^{-1} {\rm tr} (E_n-H)^{-1}
}^{\, c}
\label{eq_uni13}
\end{equation} 
were shown \cite{Amb90} to be also universal in the sense stated
above. In addition to the results in leading order, a systematic
$1/N^2$ expansion of the $n$--point correlators and of the free energy
$F = {\rm ln}Z/N^2$ was developed in Ref.~\onlinecite{Amb93},
\begin{eqnarray}
G(E_1,\ldots,E_n) &=& \sum_{h=0}^{\infty} \frac{1}{N^{2h}}
G_h(E_1,\ldots,E_n) \nonumber\\
F &=& \sum_{h=0}^\infty \frac{1}{N^{2h}} F_h \ .
\label{eq_uni14}
\end{eqnarray}
It was shown that these $1/N^2$ corrections are again universal. For
symmetric potentials, the corrections to the Green functions are of
the form 
\begin{equation}
G_h(E_1,\ldots,E_n) =
\frac{R_h\left(\{E_i^2-a^2\}, \{M_a^{(l)}\},
  a\right)}{\sqrt{\prod_{i=1}^n(E_i^2-a^2)} } \ .
\label{eq_uni15}
\end{equation}
Here, the $R_h$ are rational functions, and the $M_a^{(l)}$ are
additional parameters, cf. Ref.~\onlinecite{Amb96}. Analogous results
have also been derived for non--symmetric potentials \cite{Amb93} and
complex matrices \cite{Amb92}. Moreover, the statements made above for
the two--point Green functions could be generalized \cite{Amb96,Ake96}
to the case where, in the large--$N$ limit, the level density
$\rho(E)$ has support in $s$ distinct intervals. For each $s$ a new
universality class arises in which the two--point correlators are
universal and depend on the potential $V$ only through the end points
of the $s$ intervals.

For small distances $|E - E'|$, the two--point correlator of
Eq.~(\ref{eq_uni12}) does not yield the universal form (\ref{eq_uni9})
of the kernel $K(E,E')$. The reason is that the two expressions apply in
two different limits. This became particularly transparent in a paper
by Brezin and Zee \cite{Bre93} who considered potentials $V(H)$ in the
form of even polynomials in $H$. They made an {\it ansatz} for the
asymptotic (in $n$) form of the associated orthogonal polynomials
$p_n(E)$ and, hence, for $\psi_n(E)$. Knowledge of this asymptotic
form is helpful. Indeed, by means of the Christoffel--Darboux identity
the kernel $K(E,E')$ defined in Eq.~(\ref{eq_uni7}) can be written as
\begin{equation}
K(E,E') = c_N \frac{\psi_N(E)\psi_{N-1}(E') -
  \psi_{N-1}(E)\psi_N(E')}{E-E'} 
\label{eq_uni16}
\end{equation}
with an $N$--dependent constant $c_N$. In the large--$N$ limit the
{\it ansatz} in Ref.~\onlinecite{Bre93} and the ensuing expression for
$K(E,E')$ become exact. Two limiting cases were considered in
Ref.~\onlinecite{Bre93}. (i) The difference $E-E'$ is of order $1/N$
and both energy arguments are at a finite distance from the end points
of the spectrum. Then, the universal form (\ref{eq_uni9}) holds. Local
universality (which of course applies equally to all higher
correlation functions) was thus proven for all symmetric potentials of
polynomial form. For an extension of this result to more complicated
potentials see Ref.~\onlinecite{Bre93a}. (ii) The difference $|E -
E'|$ is so large that the number of levels in the interval $E-E'$ is
of order $N$. Because of this large number, the kernel $K(E,E')$
oscillates heavily with changes of $E$ or $E'$. Averaging over a
finite number of such oscillations in the vicinity of both $E$ and
$E'$, Brezin and Zee obtained
\begin{equation}
T_{(\beta=2) 2}^{\rm smooth}(E,E') = \frac{1}{2N^2\pi^2}
\frac{1}{(E-E')^2} 
\frac{a^2 - E E'}{\left[(a^2 - E^2)(a^2-E^{\prime 2}) \right]^{1/2}}.
\label{eq_uni12a}
\end{equation}
Thus, they rediscovered the universal correlation function
(\ref{eq_uni12}) of Ambj{\o}rn {\it et al.} \cite{Amb90}.  (The
relation between $G(E,E')$ and $T_{(\beta=2) 2}(E,E')$ is simple and
explained in Ref.~\onlinecite{Amb96}). Apparently, the perturbative
calculation of Ref.~\onlinecite{Amb90} cannot account for the
oscillatory behavior of the level correlations. In
Ref.~\onlinecite{Bre94}, this fact was attributed to an interchange of
limits. In a perturbative calculation $N$ is sent to infinity before
the poles in the resolvents are placed on the real axis. Therefore the
individual poles merge to form a cut and the fine structure is lost.
In the method of orthogonal polynomials and in the supersymmetry
method (see below), on the other hand, no such problem arises.

We digress and mention some related work although it is not directly
related to the subject of this section. In Ref.~\onlinecite{Bre94},
``time'' dependent correlations in a matrix ensemble defined by
\begin{equation}
P[H] = Z^{-1} \exp\left( -\int_{-T}^T dt \, {\rm tr}\left[
       \frac{1}{2}\left(\frac{dH}{dt}\right)^2 + V[H]\right]
       \right)
\label{eq_uni17}
\end{equation}
were calculated with diagrammatic techniques. 
Here, $H=H(t)$ is explicitly a function of time. In particular the
(smoothed) current--current correlator $\langle J(E,t) J(E',0)\rangle$
for arbitrary $t$ was derived, thus generalizing earlier results by Szafer
and Altshuler \cite{Sza93} and by Beenakker \cite{Bee93b}. 
The current operator was defined as $J(E,t)=N^{-1}\sum_i (dE_i/dt)
\delta(E-E_i(t))$. In subsequent work \cite{Bre94a} these
time--dependent correlations were related to the cross correlations
between the eigenvalues of two separate (but not independent) Gaussian
ensembles. The relationship is very similar to the one described in
Sec.~\ref{cs_1} between parametric correlations and a continuous matrix
model. In Ref.~\onlinecite{Bre95} the interesting relation 
\begin{equation}
G(E,E') = \frac{\partial^2}{\partial E\partial E'} {\rm ln}
\left(\left(\frac{G(E)-G(E')}{E-E'} \right) + \ldots \right)
\label{eq_uni18}
\end{equation}
was established for arbitrary polynomial $V[H]$. The dots indicate
terms that do not contribute to the connected two--point level density
correlation function $T_{\beta 2}(E,E')$. The result (\ref{eq_uni18}) is
rather surprising since it connects a universal quantity like
$G(E,E')$ with the non--universal one--point function $G(E)$. The
validity of Eq.~(\ref{eq_uni18}) could be extended \cite{Bre95}
to the case where $H=H_1+H_2$ with $P[H] = P_1[H_1]P_2[H_2]$.

Beenakker \cite{Bee94a} gave a particularly simple and elegant
derivation of the form (\ref{eq_uni12a}) of the universal ``wide'' 
correlator. At the same time, he could generalize the result to the
orthogonal and symplectic cases, and to non--polynomial
potentials. Using the functional derivative method which he had
developed \cite{Bee93a} and applied to the case of the positive
spectrum of the transfer matrix, he calculated directly the connected 
two--point correlation function $T_{\beta 2}(E,E')$. We briefly summarize
the main steps of the derivation. The two--point function is given by
the functional derivative of the eigenvalue density with respect to
the confining potential, 
\begin{equation}
T_{\beta 2}(E,E') = \frac{1}{\beta} \frac{\delta\rho(E)}{\delta
  V(E')} \ . 
\label{eq_uni19}
\end{equation}
Here the potential $V$ is arbitrary provided the support of the
spectrum remains compact as $N\to\infty$. In this limit, $V$ and
$\rho$ are connected by the integral relation 
\begin{equation}
{\bf P} \int_a^b dE \frac{\rho(E)}{E'-E} = \frac{d}{dE'} V(E')
\label{eq_uni20}
\end{equation}
where ${\bf P}$ denotes the principal value and $a$ and $b$ are the
end points of the spectrum. Variation of Eq.~(\ref{eq_uni20}) with
respect to $V$ and inversion of the result together with
Eq.~(\ref{eq_uni19}) immediately yields a universal expression for
$T_{\beta 2}(E,E')$. The result is valid for all $a$ and $b$, for
$\beta=1,2,4$, and for all $V$. For $a=-b$ and $\beta=2$ it reduces to
the known case given by Eq.~(\ref{eq_uni12a}). Extending the methods
used in Ref.~\onlinecite{Amb93}, Itoi \cite{Ito96} also derived
explicit expressions for two--point correlators in the orthogonal and
symplectic symmetry classes and established the universality of the
multi--point correlators.

Hackenbroich and Weidenm\"uller \cite{Hac95} showed that {\it all}
local correlation functions (i.e., functions on the scale of the mean
level spacing) are independent of the confining potential $V(H)$.
This statement holds under the following two provisos: The support of
the spectrum must remain finite in the limit $N \rightarrow \infty$,
and the arguments of the correlators must be scaled correctly (energy
differences are to be expressed in units of the local mean level
spacing). The proof applies to all three symmetry classes, and to all
types of correlation functions involving level densities, $S$ matrix
elements, and/or parametric correlations of arbitrary order. Starting
point is the supersymmetric generating functional   
\begin{equation}
I = \int d[\Psi] \int d[H] P_{N\beta}(H) \,
 \exp\left(\frac{i}{2}\Psi^\dagger
    L^{1/2}(H-E+M)L^{1/2}\Psi \right)
\label{eq_uni21}
\end{equation}
for a product of Green functions of $H$. Energy differences, source
terms and, if present, the coupling to external channels are all
contained in the matrix $M$ \cite{Hac95}. From the fact that $M$ is of
order $N^{-1}$ it follows that to leading order in $N^{-1}$, $I$ is a
function of the unitary invariants $A_{\alpha \beta} = \sum_{\mu}
(L^{1/2} \Psi)_{\alpha \mu} (\Psi^{\dagger} L^{1/2})_{\beta \mu}$
only. Before averaging, these invariants are introduced via an
additional double integration over a delta function. In a sense, this
step replaces the Hubbard--Stratonovich transformation which applies
in the Gaussian case only. Then, the integration over the variables
$\Psi$ is carried out. For large $N$, all remaining integrations
(including the ensemble average, i.e. the integration over the
distribution of $H$) are then evaluated with the help of the
saddle--point approximation. Except for the occurrence of the local
mean level density $\rho(E)$, the final expression for $I$ coincides
with the result for the Gaussian case. This proof derives its elegance
from the fact that one never leaves the level of the generating
functional. At the same time this is its drawback: The proof reduces
the form of the correlators to the Gaussian case but is of no help in
actual calculations of the latter.

Freilikher, Kanzieper, and Yurkevich \cite{Frei96,Frei96a} addressed
the issue of universality for wide classes of confining potentials
(for which the spectrum is bounded). The authors used orthogonal
polynomials and mathematical results \cite{Lub93,Sze21,Sze67,Lub89}
which partly are quite recent. In the ``$\alpha$--ensemble''
\cite{Che94,Can95} with $V(E)\propto |E|^\alpha$, the orthogonal
polynomials are defined \cite{Frei96} with respect to the ``Freud
weights'' \cite{Lub93}. Rigorous results for the kernel $K(E,E')$, see
Eq.~(\ref{eq_uni7}), led to the following statements. As $\alpha$
approaches unity from above, the mean level density $\rho(E)$ develops
a sharp peak in the center of the spectrum. In contradistinction, all
two--point correlators, local as well as smoothed ones, retain their
universal form down to and including the value $\alpha = 1$. In
Ref.~\onlinecite{Frei96a} the rigorous treatment was extended to the
``Erd\"os--type'' of confining potentials which at infinity grow
faster than any polynomial, and to the presence of a hard edge in the
spectrum of the Hamiltonian. This last condition is motivated by
applications involving the (non--negative) eigenparameters of random
transfer matrices, see Sec.~\ref{quasi1d_2}. Again, in all cases
considered, the universality of both global (smoothed) and local
two--point correlators could be shown.

Explicit and interesting examples for potentials $V(H)$ leading to
non--universal statistics were given by Nagao and Slevin \cite{Nag93},
and by Muttalib and coworkers \cite{Che92,Mut93}. In the spectrum of
the generalized Gaussian and Laguerre ensembles (which involve an
additional logarithmic term in the potential $V(H)$), there exist
domains with non--universal correlations \cite{Nag93}.  In
Ref.~\onlinecite{Mut93}, the $q$--dependent potential
\begin{equation}
V(H) = \sum_{n=0}^\infty {\rm ln} [1+2q^{n+1}\cosh(2\chi) + q^{2n+2}],
\quad H = \sinh\chi
\label{eq_uni18a}
\end{equation}
with $q\in[0,1]$ was considered. The spectrum is not bounded (the
eigenvalues range from $-\infty$ to $+\infty$). As a function of $q$,
the nearest neighbor spacing distribution interpolates between
Wigner--Dyson (WD) and Poisson statistics \cite{Mut93}. The orthogonal
polynomials corresponding to the choice (\ref{eq_uni18a}) are the $q$
Hermite polynomials. Potentials related to the $q$ Laguerre
polynomials exhibit similar spectral properties \cite{Che92}. It was
conjectured \cite{Mut93} that these properties are shared by the
entire family of potentials related to $q$ polynomials \cite{Gas90}. 

Since they interpolate between WD and Poisson statistics, potentials
of the type (\ref{eq_uni18a}) might be important for the description
of the crossover from metallic to localized behavior in mesoscopic
systems, see Sec.~\ref{spec_2}. Another possibility to describe this
transition is provided by random matrix ensembles which break the
rotational invariance in Hilbert space assumed in Eq.~(\ref{eq_uni1}). 
Examples are ensembles of random band matrices (see Sec.~\ref{rbm}) and
other ensembles \cite{Pic94,Mos94} with a preferential basis. 

Why do rotationally invariant potentials of the type (\ref{eq_uni18a})
deviate from universal behavior? Canali and Kravtsov \cite{Can95a}
gave an interesting answer: As one goes from the $\alpha$ ensemble
($\alpha\ge 1$) to potentials like (\ref{eq_uni18a}) which grow only 
logarithmically with $H$, the $U(N)$ invariance is spontaneously
broken. This seems the common feature of all approaches which describe
the localization transition in terms of random matrix ensembles.  

\setcounter{equation}{0}
\section{Common concepts}
\label{coco}

In previous sections of this review, 
we have encountered numerous examples of successful applications of
RMT to physical systems. We have also listed many open
problems. Perhaps the most impressive aspect of this discussion is the
wide range of topics involved. It suggests that RMT is a universal tool.
Indeed, RMT correctly describes
fluctuation properties of nonrelativistic and relativistic quantum
systems. It applies to chaotic quantum systems with few degrees of
freedom, to disordered quantum systems, to classical wave propagation
in disordered media, and to many--body quantum systems. RMT
also describes spectral fluctuation properties and parametric
correlations of several very different wave equations. And RMT
connects to quantum gravity and to the Calogero--Sutherland 
type models in one dimension. In this final section, we focus
attention on the ubiquity of RMT. 

To this end we first recall the foundation, success and
limitations of RMT as well as some of the open problems in the various
fields of application.

\subsection{Synopsis}
\label{coco_1}

{\it (i) Many--body systems.}
RMT started with Wigner's attempt to formulate a generic model for the
spectral fluctuation properties of complex many--body systems. The
Hamiltonian of every such system is assumed to be a member of an
ensemble of random matrices. The ensemble is characterized completely
by symmetry properties. Apart from appropriate mean values, no
detailed knowledge of the system at hand 
enters the definition of the ensemble. Thus, system--specific features
(beyond the mean values needed to define RMT properties) cannot be
reproduced and are assumed to be irrelevant for fluctuation properties.
This model has been extremely successful in the description of
spectral fluctuation properties and of cross--section fluctuations in
complex atoms, molecules, and atomic nuclei. This is true in spite of
the fact that the interactions between the constituents in these
systems are vastly different: Nuclei are bound by the short--range
nuclear force which is repulsive at very small distances, atoms and
molecules are governed by long--range Coulomb forces. Only states near
the ground state seem to display system--specific features. At higher
excitation energies, collective and/or single--particle properties may
still be present but seem to appear only as gross features. For
instance, the giant dipole resonance in nuclei appears as a broad
resonance in the $\gamma$--absorption cross section. The spreading
width of the resonance is very much larger than the mean level spacing
of the many--body system. The resonance does not correspond to a
single eigenstate of the nucleus. Similar statements apply to other
collective modes. Isobaric analogue resonances in heavy nuclei provide
a good example. Hence it now appears that, within each set of
conserved quantum numbers and sufficiently far above the ground state,
any strongly interacting quantum system with many degrees of freedom
must be expected to display fluctuation properties of RMT type of the
appropriate symmetry.

Does an energy scale exist in such systems (analogous to the Thouless
energy in disordered systems) beyond which RMT ceases to apply? Can we
expect to find the analog of localization in the eigenfunctions? As
long as the origin of stochastic behavior in such many--body systems
is unclear, questions such as these have no easy answer. It is
conceivable that stochasticity arises from classical chaos. This is
the case in systems with few degrees of freedom. In that case,
limitations of RMT similar to the ones to be mentioned under (ii) are
expected to apply. It is equally conceivable, however, that
stochasticity is an emergent property of strongly interacting systems
with many degrees of freedom irrespective of whether the underlying
classical dynamics is fully or only partially chaotic. After all, an
assumption of similar type (equal {\it a priori} occupation
probability of 
all accessible parts of phase space or Hilbert space) has been used
with amazing success as the foundation of classical and quantum
statistical mechanics. In this case, possible limitations of RMT would
have a different origin, and would probably be of different form. The
available experimental evidence (covering sequences of up to several
hundred levels) does not point to any such limitations at present. 

{\it (ii) Classically chaotic systems with few degrees of freedom.}
The fundamental property of these systems is the instability of the
classical trajectories. It leads to stochastic behavior on both the
classical and the quantum level. For fully chaotic systems with a
single intrinsic time scale, the situation is fairly well understood. 
We have discussed three attempts, based on semiclassics, the non--linear
$\sigma$ model, and structural invariance, respectively, to rigorously
establish the Bohigas conjecture. Although a complete proof does not yet
exist, the conditions for the occurrence of spectral fluctuations of
RMT type are known in their essentials. The energy interval within
which fluctuations of RMT type occur, is determined by the shortest
periodic orbits. Fully chaotic systems with several intrinsic time
scales, such as, e.g., a chain of coupled chaotic billiards, have not
been considered in detail in this review. They
show fluctuations similar to those of disordered quasi
one--dimensional mesoscopic systems and are described by random band
matrices. Here, localization phenomena may occur. Generic systems with
mixed phase space pose a major challenge. The idea of chaos--assisted
tunneling has been very successful. Closer inspection has shown,
however, that the structure of classical phase space is quite complex,
especially at the interface between regular and chaotic domains. This
fact does influence the fluctuation properties. We have emphasized the
conceptual importance of model systems (like two coupled harmonic
oscillators) for the investigation of many of these questions.
Another challenge
consists in extending the semiclassical results obtained for systems
with two or three degrees of freedom to higher dimensions. 

{\it (iii) Disordered systems.}
Both the origin of stochastic behavior and the limitations of RMT are
quite clear in disordered systems. The random distribution of impurity
scatterers together with a suitable ergodic theorem furnish an
immediate justification for modeling these systems in a
stochastic fashion. Such a justification is much more subtle in the
cases discussed under (i) and (ii), where stochasticity appears as an
emergent property of the many--body dynamics or the classically
chaotic dynamics, respectively. For disordered systems,
spectral
fluctuations and the fluctuations of transport coefficients are --
within known limits -- of classical RMT type. The applicability of
classical RMT is limited to the diffusive regime
(if we disregard the extreme ballistic limit where the system becomes
a quantum billiard)  and, within this
regime, to an energy interval given by the Thouless energy. Outside of
this interval, and in the crossover to either the ballistic or the
localized regime as well as  at the
mobility edge, other universal laws govern the fluctuation properties. 
Except for the behavior at the mobility edge, these laws can be
derived from suitable generalizations of classical RMT. In this sense,
RMT has been found to apply universally to disordered systems, too. We
have not been able to dwell on the passage of classical light waves,
or of radar waves, through a medium with a randomly varying index of
refraction. However, much of what has been said about disordered
mesoscopic systems applies here, too. In particular, weak localization
occurs and is thus seen to be not a quantum, but a wave phenomenon.

The phenomenon of localization is not limited to disordered systems. 
As ``dynamical localization'', it occurs likewise in chaotic quantum
systems. The kicked rotor is the standard example. This system can be
mapped onto a one--dimensional non--linear $\sigma$ model. This shows
the connection to quasi one--dimensional disordered wires and
to RMT.

{\it (iv) Wave equations.}
Spectral fluctuation properties of RMT type are not linked to a
specific linear second--order differential equation (the Schr\"odinger
equation). The spectra of acoustic and elastomechanical waves in
irregularly shaped solids show RMT fluctuations, and so do the
eigenmodes of Maxwell's equations in three--dimensional cavities. 
These phenomena are governed by rather different linear wave
equations. In the case of elastomechanical waves even the boundary
conditions (vanishing stress tensor) are completely different from those
for  Schr\"odinger waves. Another example of a linear wave equation,
the Dirac equation, will be mentioned under (v). Moreover, very
different interaction potentials in the Schr\"odinger equation also
give rise to RMT fluctuations. We
are not aware of any attempts to explain this general applicability of
RMT to various wave phenomena in a generic fashion.    

{\it (v) Field theory.}
We have considered two examples, QCD and a 
$\phi^4$ theory involving matrix fields $\phi$.
In general, local fluctuation properties are of little interest in
these field theories. Nevertheless, it is noteworthy that the spectral
fluctuations of the Dirac operator in QCD coincide with RMT
predictions, for two reasons. First, here is another example of
universality: A first--order differential operator coupled to gauge
fields possesses RMT fluctuation properties. Second, the existence of
RMT fluctuations shows that the data generated by lattice gauge
calculations contain generic features. This fact poses a challenge: To
separate the generic features from the physical content, in order both
to obtain a better understanding of the problem, and to simplify the
calculations.   

Certain {\it global} features of RMT are also universal. This is true,
for instance, for the shape of the global RMT spectrum near the end
points, or for the structure of diagrams generated through a loop
expansion of the above--mentioned $\phi^4$ theory. This is why RMT applies to
two--dimensional gravity and thereby establishes a conncection with
conformal field theory, and why RMT can 
be used to model the chiral phase transition of QCD.

\subsection{Discussion}
\label{coco_2}

This tour d'horizon shows that most applications of RMT to physical
systems predict local fluctuations in terms of a system--specific
input. The input consists of suitable mean values: Spectral
fluctuations require the mean level spacing as input. For cross
section fluctuations, the average scattering matrix serves the same
purpose. For conductance fluctuations, the mean conductance takes this
role. In this sense, RMT predictions are not parameter free: They
relate fluctuations to mean values. This statement can be given a
different and more interesting form. The mean values can be absorbed
into properly rescaled local variables, and RMT predictions for
fluctuations and correlation functions acquire universal form. This 
applies not only to level correlations, but very generally to the
dependence of correlation functions on any external parameter
(parametric correlations). We have presented many examples where such
universal RMT predictions have been tested successfully. 

By construction, RMT predictions are based on ensemble averages. 
On the other hand, variances or correlation functions extracted from a
data set are almost always obtained from a running average over a
given system. The equality of the results of both averaging procedures
is guaranteed by ergodicity. Although not proved in full generality,
ergodicity has been shown to hold in a number of cases. 

During the last fifteen years, the non--linear $\sigma$ model has
become an 
invaluable tool in RMT. Without this model, the universal features of
RMT might never have been discovered to their full extent. Through the
medium of field theory, the common statistical properties of chaotic
and of disordered systems, and the universality of parametric
correlations, have been displayed, and have been cast into compact
form. The model has been indispensable also in other contexts. For
instance, many aspects of stochastic scattering would not have been
accessible without it. And the insight that local RMT fluctuation
properties do not depend on the Gaussian distribution of matrix
elements finds perhaps the simplest and most general proof through
the non--linear $\sigma$ model. The need to calculate some RMT
correlation functions has led to a number of extensions of
supersymmetry. We think here of new and unconventional saddle points,
of Fourier analysis on graded spaces, and of extensions of the
Itzykson--Zuber integral. Due to lack of space we could not do justice
to some of these developments. These examples show that modern
applications of RMT 
have led to mathematical research which has meanwhile become a field
in its own right. The situation is reminiscent of the early days of
(classical) RMT where progress was eventually possible due to
mathematical input from the theory of orthogonal polynomials.

Are the success and the ubiquity of RMT really surprising? One might
argue that RMT is trivial because it only displays the consequences of
von Neumann--Wigner level repulsion, i.e. of the simple fact that two
states connected by a non--vanishing matrix element repel each other.
But this argument fails to take into account all other predictions of
RMT which go far beyond simple level repulsion. These predictions
derive from the symmetry of the Hamiltonian, and from the postulated
invariance properties of RMT which in turn embody the generic aspects
of the entire approach. The form of the Vandermonde determinant
appearing in the invariant measure of RMT is entirely due to the
postulated rotational invariance of the random matrix ensemble in
Hilbert space. It is true, of course, that this determinant implies
(a particular form of) level repulsion. The degree of repulsion is
determined by the symmetry of the Hamiltonian. A simple counting
argument leads directly to the exponent $\beta =1,2,4$ in the typical
factor $|E_\mu-E_\nu|^\beta$ in the Vandermonde determinant. But the
forms of the two--point function $Y_2$ and of higher correlation
functions, of the $\Delta_3$ statistic measuring the spectral
stiffness, and of the nearest neighbor spacing distribution are not
determined by level repulsion alone and require for their derivation
the full rotational invariance of the ensemble. The same is true of
the universal form of correlations of observables which depend upon an
external parameter. With regard to these observables, RMT possesses
genuine predictive power and leads to highly non--trivial results. The
results are generic, not based on dynamical principles, and derive
only from the underlying symmetry and invariance requirements. Even on
a global scale (as opposed to the local scale of the mean level
spacing), spectral correlations still display some degree of
universality. The extension of RMT to random band matrices and random
transfer matrices, important for the understanding of extended systems
and localization, breaks rotational invariance. The transition to
localization and the ensuing correlation functions are determined by
the way in which the influence of the Vandermonde determinant in the
invariant measure is gradually neutralized. But even in this domain of
application of RMT, generic non--dynamical features remain and are of
central interest. They are embodied in the effective Lagrangian of the
non--linear $\sigma$ model.

In the end, the universal features of RMT appear to correspond
remarkably well to generic properties of quantum (and certain
classical) systems. 
We recall Balian's proof that among all rotation invariant ensembles
Gaussian RMT maximizes the information entropy.
These facts seem to imply that in the absence of any further
knowledge about the system, Gaussian RMT yields the best approximation
to the physical properties of that system. 

In spite of this argument, we feel that the universal applicability of
RMT remains an amazing fact. It signals the emergence of a theory of
wave phenomena, both classical and quantal, which does not relate to
dynamical properties. RMT is 
valid essentially on the local scale defined by the mean level spacing
of the system. On larger scales, correlation functions are often used
to determine system--specific properties. The validity of RMT on the
local scale shows that as the resolution is improved and the local
scale is reached, such information is lost, and is replaced by
universal features. 

To us, the universal validity of RMT is reminiscent of the success of
Thermodynamics in the last century. Thermodynamics was built on a few
universal principles without recourse to any dynamical theory. With
the help of a few system--specific parameters, the thermodynamic
behavior of widely different systems could be described successfully. 
The universal applicability of RMT to local fluctuation properties of 
systems governed by a wave equation seems to mirror this development.

\bibliographystyle{axel}
\bibliography{review}

\end{document}